\documentclass[a4paper,11pt,twoside,onecolumn,final,openright]{book}

\makeatletter
\title{Approaching nuclear interactions with lattice QCD}\let\Title\@title
\author{Marc Illa Subi\~{n}a}\let\Author\@author

\renewcommand\@chapapp{\chaptername}
\newcommand{\chapternamefont}{\scshape\LARGE}% Chapter name font
\newcommand{\chaptertitlefont}{\Huge\bfseries}% Chapter title font
\def\@makechapterhead#1{% Header for \chapter
  \vspace*{50\p@}%
  {\parindent \z@ \raggedright \normalfont
    \ifnum \c@secnumdepth >\m@ne
        {\chapternamefont\@chapapp\space \thechapter\par\nobreak}% \huge\bfseries
        \vskip 20\p@
    \fi
    \interlinepenalty\@M
    {\chaptertitlefont #1\par\nobreak}
    \vskip 40\p@
  }}
\def\@makeschapterhead#1{% Header for \chapter*
  \vspace*{50\p@}%
  {\parindent \z@ \raggedright
    \normalfont
    \interlinepenalty\@M
    {\chaptertitlefont#1\par\nobreak}
    \vskip 40\p@
  }}
\makeatother

\usepackage[width=160mm, top=25mm, bottom=25mm, bindingoffset=15mm, headheight=13.6pt]{geometry} 

\usepackage[utf8]{inputenc}
\usepackage[T1]{fontenc}
\usepackage[english]{babel}

\usepackage{amsmath,amssymb,amsthm,mathtools}
\usepackage{cancel}
\usepackage{dsfont}
\usepackage{bm}
\usepackage{simplewick}
\usepackage{blkarray} %write outside matrices
\usepackage{mathabx} %big asterisk
\usepackage[low-sup]{subdepth} %unify subscript depths
\usepackage{scalerel}

\usepackage{graphicx}
\usepackage[dvipsnames]{xcolor}
\usepackage{float}
\usepackage[export]{adjustbox}
\usepackage{tikz}
\usepackage{pgf} %Update references on figures
\usepackage{ytableau}
\ytableausetup{centertableaux, smalltableaux}

\usetikzlibrary{decorations.markings,decorations.pathmorphing,decorations.pathreplacing,positioning,arrows.meta,calc}

\tikzset{->-/.style={decoration={markings,mark=at position #1 with {\arrow{stealth}}},postaction={decorate}}}

\tikzset{
particle/.style={thin,draw=black, postaction={decorate},
decoration={markings,mark=at position .55 with {\arrow[scale=1.5,black]{stealth}}}},
particlecross1/.style={thin,draw=black, postaction={decorate},
decoration={markings,mark=at position .8 with {\arrow[scale=1.5,black]{stealth}}}},
particlecross2/.style={thin,draw=black, postaction={decorate},
decoration={markings,mark=at position .3 with {\arrow[scale=1.5,black]{stealth}}}}
}

\definecolor{myblue}{RGB}{57,83,164}
\definecolor{mygreen}{RGB}{106,189,69}
\definecolor{myred}{RGB}{237,32,36}
\definecolor{myablue}{RGB}{247,236,19}
\definecolor{myagreen}{RGB}{185,82,159}
\definecolor{myared}{RGB}{111,204,221}

\definecolor{cc0}{HTML}{3F5878}
\definecolor{cc1}{HTML}{E07A5F}
\definecolor{cc2}{HTML}{81B29A}
\definecolor{cc3}{HTML}{F2CC8F}
\definecolor{cc4}{HTML}{B8B49A}
\definecolor{cc5}{HTML}{A76571}
\definecolor{cc6}{HTML}{6EB964}
\definecolor{cc7}{HTML}{FF8C98}

\usepackage{scalerel}
\def\shrinkage{-2.4mu}
\def\vecsign#1{\rule[1.388\LMex]{\dimexpr#1-2.5pt}{.36\LMpt}%
  \kern-6.0\LMpt\mathchar"017E}
\def\dvecsign#1{\rule{0pt}{7\LMpt}\smash{\stackon[-1.989\LMpt]{%
  \SavedStyle\mkern-\shrinkage\vecsign{#1}}%
  {\rotatebox{180}{$\SavedStyle\mkern-\shrinkage\vecsign{#1}$}}}}
\def\dvec#1{\ThisStyle{\setbox0=\hbox{$\SavedStyle#1$}%
  \def\useanchorwidth{T}\stackon[-4.2\LMpt]{\SavedStyle#1}{\,\dvecsign{\wd0}}}}
\usepackage{stackengine,amsmath}
\stackMath

\usepackage{multirow}
\usepackage{array}
\usepackage{enumitem}
\usepackage{tabu}
\usepackage{booktabs} %for \toprule, \midrule, and \bottomrule macros

\newlist{checks}{itemize}{1}
\setlist[checks]{label={--},wide=\parindent}

\usepackage{hyperref}
\hypersetup{
    pdfauthor={\Author},
	pdftitle={\Title},
    colorlinks=true,
    linkcolor=Blue,
    citecolor=Blue,
    filecolor=Blue,
    urlcolor=Blue,
    }
\usepackage{url}

\usepackage[english, capitalise]{cleveref} %reference chapter, sections...
\usepackage[nottoc]{tocbibind} %Bib appears in the TOC
\usepackage{cite} %for multiple citations (instead of [1,2,3,4], now [1-4]).

\usepackage{fancyhdr} 
\usepackage{rotating}
\usepackage{adjustbox}
\usepackage[final]{microtype} %better spacing between words
\usepackage{ragged2e}
\usepackage{afterpage} %For making figures occupy an entire page

\newcommand\blankpage{%
    \null
    \thispagestyle{empty}%
    \addtocounter{page}{-1}%
    \newpage}

\newcommand{\imag}{\mathrm{i}} %roman imag i
\newcommand{\Tr}{\operatorname{Tr}} %trace function
\newcommand{\Det}{\operatorname{det}} %trace function
\newcommand{\Ree}{\operatorname{Re}} %Real part
\newcommand{\arccosh}{\operatorname{arccosh}} %arccosh
\newcommand{\Dag}{\scalerel*{\dag}{X}} %shorter dagger

\def\1s0{^1 \hskip -0.03in S_0}
\def\3s1{^3 \hskip -0.025in S_1}

\theoremstyle{definition}
\newtheorem{definition}{Definition}

\makeatletter
\renewcommand*\env@matrix[1][c]{\hskip -\arraycolsep
  \let\@ifnextchar\new@ifnextchar
  \array{*\c@MaxMatrixCols #1}}
\makeatother

\begin{document}

%Front pages *******************************************
%\newpage
\thispagestyle{empty}
%\begin{titlepage}
    \begin{center}
        \vspace*{3\baselineskip}

        {\Huge \textbf{Approaching nuclear interactions \\[.3cm] with lattice QCD}}
        
        \vspace{4\baselineskip}
        {\huge Marc Illa Subiña}
        
        \vspace{4\baselineskip}
        PhD advisor:
        
        \vspace{0.2cm}
        {\Large Dra. Assumpta Parreño García}

        \vfill

        \begin{figure}[b]
            \centering
            \includegraphics[width=0.45\textwidth]{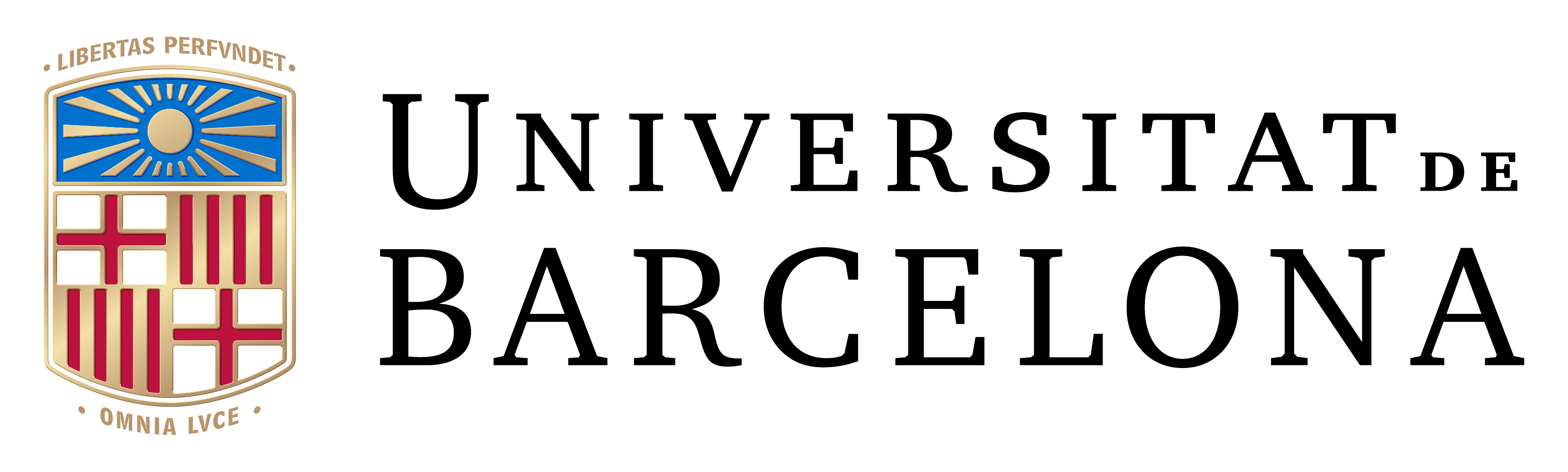}
        \end{figure}
    \end{center}
%\end{titlepage}

\afterpage{\blankpage}

\newpage
\thispagestyle{empty}
\begin{center}
    \vspace*{3\baselineskip}
    {\Huge \textbf{Approaching nuclear interactions \\[.3cm] with lattice QCD}}

    \vspace{0.6cm}
    {\large Memòria presentada per optar al grau de doctor per la Universitat de Barcelona}
    
    \vspace{1.2cm}
    {\scshape\LARGE Programa de Doctorat en Física}

    \vspace{1.5cm}
    {\Large \itshape Autor:}
    
	\vspace{0.1cm}
	{\LARGE Marc Illa Subiña}
    
    \vspace{0.6cm}
    {\Large \itshape Directora:}
    
	\vspace{0.1cm}
	{\LARGE Dra. Assumpta Parreño García}
    
    \vspace{0.6cm}
    {\Large \itshape Tutor:}
    
	\vspace{0.1cm}
	{\LARGE Dr. Joan Soto i Riera}

    \vspace{4cm}
    {\Large Departament de Física Quàntica i Astrofísica\\
            Institut de Ciències del Cosmos\\
            Universitat de Barcelona\\}

    \begin{figure}[b]
        \centering
        \includegraphics[width=0.45\textwidth]{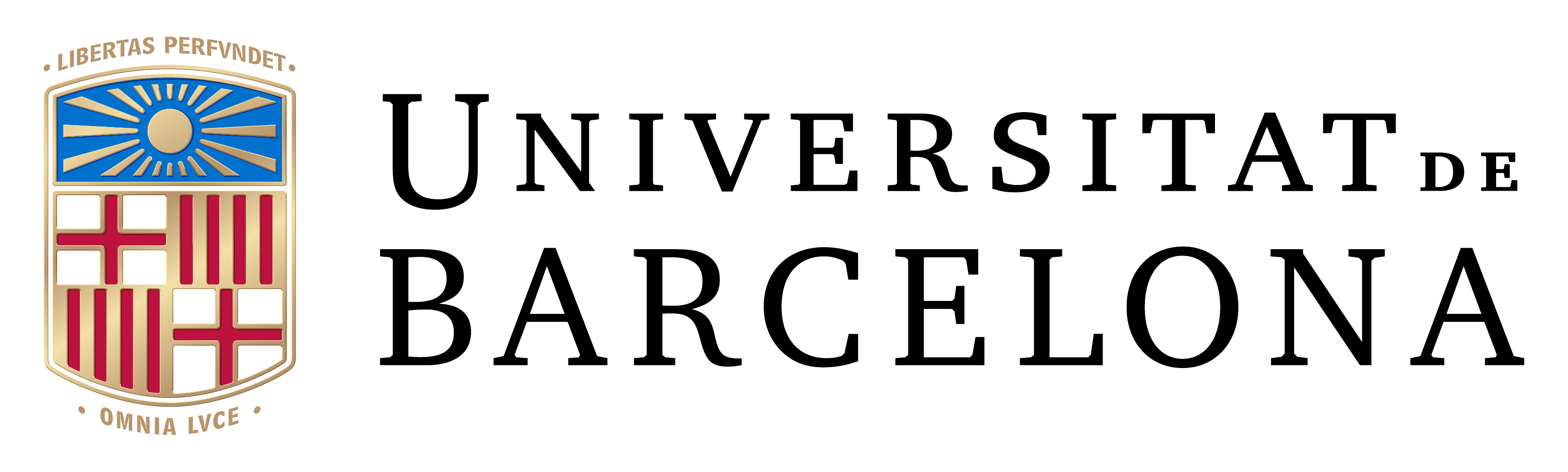}
    \end{figure}
\end{center}

\afterpage{\blankpage}

%%%%%%%%%%%%%%%%%%%%%%%%%%%%%%%%%%%%%%%%%%%%%%%%%%%%
\frontmatter

\cleardoublepage

{\pagestyle{plain}
% Resum (in catalan) *******************************************

\chapter*{Resum}
\addcontentsline{toc}{chapter}{Resum}

La descripció de les propietats bàsiques dels nuclis a partir dels seus constituents més fonamentals, els quarks i els gluons, és un dels principals objectius de la física nuclear, però degut al comportament singular de QCD a baixes energies, solucions teòriques en aquest rang han estat impossibles durant molts anys. La finalitat d'aquesta tesi doctoral és l'estudi de les interaccions entre dos barions, incloent aquells d'estranyesa diferent de zero, els anomenats hiperons, a partir de la teoria fonamental de la interacció forta, la Cromodinàmica Quàntica (QCD). Donat que a baixes energies la constant d'acoblament de QCD adquireix un valor molt gran, no és possible aplicar les tècniques pertorbatives, i s'han d'utilitzar altres mètodes alternatius. En el nostre cas, fem servir \textit{lattice} QCD (LQCD), proposat per K.~G.~Wilson l'any 1974~\cite{Wilson:1974sk}. 

Si ens fixem només en el sector dels quarks \textit{up} i \textit{down}, els únics quarks estables que formen protons i neutrons, podem trobar una immensa quantitat de dades experimentals provinents de l'estudi de la dispersió de dos nucleons o dels nivells d'energia de nuclis atòmics. Això permet construir models fenomenològics que, juntament amb tècniques de sistemes de molts cossos, s'utilitzen per estudiar una gran varietat de problemes. El problema amb aquests models és que no hi ha una connexió directe amb QCD, fet que va motivar que S.~Weinberg, l'any 1990, introduís el concepte de les teories de camp efectives (EFT)~\cite{Weinberg:1990rz}, que permeten descriure processos en el règim energètic no-pertorbatiu a partir de les simetries inherents de la teoria fonamental, QCD. Tot i aquest avenç, els graus de llibertat que s'utilitzen són components efectius i no els fonamentals (és a dir, hadrons i no quarks i gluons). El Lagrangià efectiu es construeix a partir d'operadors que reflecteixen aquelles simetries, acompanyats de coeficients de baixa energia (LEC), que encapsulen tota la física que no es té en compte de forma explícita, i s'han de determinar ajustant els càlculs fets utilitzant EFTs a les dades experimentals corresponents.

Si anem més enllà dels sistemes nuclears convencionals i considerem barions que contenen quarks \textit{strange}, observem que són inestables i es desintegren mitjançant processos febles. 
Un dels àmbits científics on els hiperons juguen un paper important és el de l'astrofísica nuclear, ja que aquests són determinants a l'hora d'estudiar l'estructura i dinàmica de les estrelles de neutrons. Nombrosos estudis teòrics demostren que quan s'introdueixen hiperons a l'equació d'estat de l'estrella, la seva massa màxima es situa per sota del valor observat (al voltant de dues masses solars), llevat que s'introdueixi una interacció repulsiva entre hiperons i nucleons.

Aquest problema, conegut com \textit{hyperon puzzle}, ha motivat diverses propostes teòriques per a la seva solució~\cite{Chatterjee:2015pua, Vidana:2018bdi, Tolos:2020aln}, i està lligat, per una banda, a la falta de dades experimentals de dispersió entre hiperó-nucleó i hiperó-hiperó que ajudin a determinar amb millor precisió les interaccions entre barions en el sector estrany, ja que entren necessàriament en la resolució de l'equació d'estat, i per altra, al desconeixement de la força a tres cossos en presència d'hiperons.

Degut a aquestes limitacions, tots els models teòrics fan servir la simetria de sabor $SU(3)$ que permet relacionar quantitats de les quals tenim dades experimentals a canals menys o totalment desconeguts. Per exemple, les dades dels sistemes amb estranyesa $0$ i $-1$ es poden utilitzar per fer prediccions pels canals amb estranyesa $-2$, $-3$ i $-4$. Com que aquesta simetria és aproximada (els tres quarks no tenen la mateixa massa), la EFT també ha d'incorporar termes que contribueixen al trencament de $SU(3)$, però degut a la poca quantitat de dades experimentals, només un LEC s’ha pogut determinar~\cite{ Haidenbauer:2014rna}. El coneixement insatisfactori d'aquestes interaccions fa necessari el desenvolupament i aplicació de mètodes alternatius, més directes, com és el de LQCD. Aquest formalisme ens permet solucionar les equacions de QCD fent servir un espai-temps discret i utilitzant mètodes numèrics a gran escala. 
La peculiaritat que fa que LQCD sigui l’eina ideal per investigar la interacció hipernuclear forta és que, a diferència del que passa a la natura, podem ``desconnectar'' la interacció feble i programar únicament el Lagrangià fort, fent que els hiperons es converteixin en partícules estables i eludint la principal complicació en l’estudi experimental d’aquests sistemes. No obstant això, també hi ha obstacles en estudis numèrics d'aquest tipus, com és per exemple la degradació del senyal en sistemes de més d'un barió, fet que comporta realitzar càlculs amb valors de les masses dels quarks lleugers per sobre dels valors físics. 

En aquesta tesi demostrem la viabilitat d'aquests tipus de càlculs i la seva importància a l'hora d'estudiar sistemes de dos barions (malgrat fer-ho amb unes masses dels quarks \textit{up} i \textit{down} que donen lloc a una massa del pió de 450 MeV), i així determinar les propietats de la interacció, com poden ser els desfasatges de dispersió, els paràmetres de dispersió a baixa energia (longitud de dispersió i rang efectiu), les energies de lligam o els LECs que descriuen la interacció. D’aquesta manera, LQCD pot proveir d'informació que pugui complementar la que obtenim directament de les dades experimentals, i ajudar a delimitar millor els models fenomenològics i teories efectives de les forces hipernuclears.

L’estructura de la tesi és la següent. Al Capítol~\ref{chap:2}, hi ha una introducció al formalisme de LQCD, passant primer per la teoria fonamental en el continu, QCD, per després posar-la en el reticle i fer les modificacions necessàries per tal de poder utilitzar les tècniques de Monte Carlo i extreure'n observables. N’hi ha de dos classes: podem calcular les energies d’un sistema (a partir de funcions de correlació de dos punts) i també calcular la interacció del sistema amb un corrent extern (a partir de funcions de correlació de tres punts). Per acabar, aquest capítol repassa el mètode de Lüscher~\cite{Luscher:1986pf, Luscher:1990ux}, que ens ajuda a calcular els desfasatges de dispersió i l’energia de lligam a partir de les energies extretes quan tenim un sistema dins d’un volum finit.

Al Capítol~\ref{chap:3}, ens centrem en l’estudi estadístic de les funcions de correlació de dos punts en relació a l'obtenció dels nivells d’energia. Primer descrivim detalladament l'algoritme que s’ha desenvolupat específicament per ajustar les dades de LQCD a una suma d’exponencials~\cite{Beane:2020ycc}. Per tal de fer una estimació dels errors sistemàtics, fem un estudi exhaustiu variant la quantitat de dades que s’inclouen en l’ajust, així com el nombre d’exponencials. En aquest capítol també discutim altres mètodes que ajuden a reduir la contaminació dels estats excitats, i finalitzem descrivim diferents mètodes per a l'estimació dels errors de les funcions de correlació.

Al Capítol~\ref{chap:4}, comencem amb un resum de la situació actual sobre el coneixement, tant experimental com teòric, de la interacció de dos barions. També repassem tots els càlculs de LQCD realitzats, i comparem els diferents mètodes utilitzats (es poden dividir en dos, el mètode directe i el mètode del potencial).
A continuació passem a descriure les diferents EFTs que volem estudiar. Donat que estem interessants en el règim de baixa energia, aquestes teories només contenen operadors de contacte, sense cap intercanvi de mesons (\textit{pionless} EFTs).
Estudiem dos casos: suposant que hi ha simetria de sabor $SU(3)$~\cite{Savage:1995kv,Petschauer:2013uua}, o que hi ha simetria de spin-sabor $SU(6)$~\cite{Kaplan:1995yg}. La primera treballa amb valors iguals de les masses dels tres quarks \textit{up}, \textit{down} i \textit{strange} (fet que es pot justificar davant de la gran diferència amb la massa del següent quark més massiu, el \textit{charm}, $\sim 1$ GeV per sobre de la del \textit{strange}), i la segona és una predicció en el límit d’un gran número de colors (QCD assumeix l'existència de tres càrregues de color). 
L’última part d’aquest capítol presenta els resultats principals de la tesi. En concret, estudiem sistemes amb estranyesa entre 0 i $-4$, i són $NN$, $\Sigma N$ ($I=3/2$) i $\Xi\Xi$ amb spin singlet i triplet, $\Sigma \Sigma$ ($I=2$) i $\Xi \Sigma$ ($I=3/2$) amb spin triplet, i $\Xi N$ ($I=0$) amb spin singlet. Els càlculs s’han realitzat treballant amb tres volums diferents (en la direcció espacial, van des de $2.8$ fm fins a $5.6$ fm) i amb un sol valor de l’espaiat del reticle (0.1167 fm)~\cite{Illa:2020nsi}. 
Els nivells d’energia de cada volum es poden fer servir per determinar els desfasatges de dispersió utilitzant el formalisme de Lüscher, revelant trets interessants sobre la naturalesa de les forces entre dos barions quan les masses dels quarks prenen valors no físics.
Concretament, els paràmetres de dispersió obtinguts ens permeten determinar els LECs de les EFTs, i en particular els coeficients relacionats amb el trencament de la simetria de sabor $SU(3)$.
Malgrat la diferència en massa reflectida en el trencament de simetria, els coeficients obtinguts resulten ser compatibles amb zero, possibilitant l'estudi de la simetria spin-sabor, i observem que les interaccions entre dos barions presenten simetria $SU(6)$. Aquesta simetria ja es va observar en un estudi previ~\cite{Wagman:2017tmp}, on les tres masses dels quarks prenien els mateixos valors, generant un pió amb una massa de $\sim 806$ MeV.
Mentre que l'estudi a 806 MeV va posar de manifest la simetria accidental $SU(16)$, és a dir, que amb un sol LEC es van poder descriure tots els canals d'interacció barió-barió amb estranyesa $0$ fins a $-4$, en el present estudi a 450 MeV no l'observem amb tanta claredat. 
Serà interessant veure com evoluciona la manifestació d'aquestes simetries a mesura que ens acostem al punt físic. En aquest capítol també es discuteixen canals pels quals no ha estat possible extreure els paràmetres de dispersió directament de les dades de LQCD. En aquests casos, hem utilitzat els valors dels LECs determinats prèviament per a determinar els valors corresponents.
Dins d'aquest capítol, també presentem les energies de lligam dels sistemes, i juntament amb els resultats a 806 MeV, les extrapolem fins al valor físic de les masses dels quarks utilitzant dues dependències funcionals molt simples per a poder comparar amb les prediccions dels models fenomenològics o EFTs, i també observar quina és la tendència a mesura que reduïm la massa dels quarks. Per exemple, s’observa el caràcter repulsiu dels canals $\Sigma N \; (\3s1)$ i $\Xi \Xi \; (\3s1)$, tal i com prediuen la majoria de models, com també l'atracció en els canals $\Xi \Sigma \; (\1s0)$ i $\Xi \Xi \; (\1s0)$. La dispersió observada entre les diferents prediccions teòriques, així com les conclusions contradictòries a què arriben diferents models, posen de manifest la necessitat de realitzar estudis de LQCD a prop del punt físic en el futur immediat.

Finalment, les conclusions de la tesi es presenten al Capítol~\ref{chap:summary}, seguides d’un conjunt d’apèndixs amb taules i figures que s’han omès en el text principal per facilitar la seva lectura.

\cleardoublepage
% Abstract *******************************************

\chapter*{Abstract}
\addcontentsline{toc}{chapter}{Abstract}

Nuclei make up the majority of the visible matter in the Universe; obtaining a first principles description of the nuclear properties and interactions between nuclei directly from the underlying theory of the strong interaction, Quantum Chromodynamics (QCD), is one of the main goals of the nuclear physics community. Although the theory was established nearly fifty years ago, the complexities of QCD at low energies precludes analytical solutions of the simplest hadronic systems, let alone the features of the nuclear forces.

Until the beginning of the century, the only way to overcome this handicap in the low-energy regime was to use phenomenological descriptions of nuclei or effective field theories (EFTs). While they have been very successful, these approaches rely heavily on experimental data. In contrast to what happens in the study of nucleon-nucleon interactions, where the amount of experimental data is overwhelming, the study of hadronic systems beyond the up-down quarks sector becomes more limited. This is because hyperons (baryons containing the next lightest quark, the strange quark) are unstable against weak interaction processes, making the experimental study of the interaction between hyperons and nucleons, and among hyperons, very difficult.

In this thesis we follow the lattice QCD (LQCD) approach, according to which QCD is solved non-perturbatively in a discretized space-time via large-scale numerical calculations. Specifically, the interactions between two octet baryons are studied at low energies with larger-than-physical quark masses corresponding to a pion mass of $m_{\pi}\sim 450$ MeV and a kaon mass of $m_{K}\sim 596$ MeV. The two-baryon systems that are analyzed have strangeness ranging from $S=0$ to $S=-4$ and include the spin-singlet and triplet $NN$, $\Sigma N$ ($I=3/2$), and $\Xi\Xi$ states, the spin-singlet $\Sigma \Sigma$ ($I=2$) and $\Xi \Sigma$ ($I=3/2$) states, and the spin-triplet $\Xi N$ ($I=0$) state.

Due to the inherent large noise in multi-baryon calculations (mitigated by the use of unphysical quark masses), the finite-volume energies are extracted using a robust fitting methodology, where in order to reliably estimate the systematic uncertainties, both the fitting form and the fitting range are varied. Then, the corresponding $S$-wave scattering phase shifts, low-energy scattering parameters, and binding energies when applicable, are extracted using Lüscher's formalism. While the results are consistent with most of the systems being bound at this pion mass, the interactions in the spin-triplet $\Sigma N$ and $\Xi \Xi$ channels are found to be repulsive and do not support bound states. Using results from previous studies of these systems at a larger pion mass, an extrapolation of the binding energies to the physical point is performed and is compared with available experimental values and phenomenological predictions.

The low-energy coefficients in pionless EFT relevant for two-baryon interactions, including those responsible for $SU(3)$ flavor-symmetry breaking, are constrained. The $SU(3)$ flavor symmetry is observed to hold approximately at the chosen values of the quark masses, as well as the $SU(6)$ spin-flavor symmetry, predicted at large $N_c$. A remnant of an accidental $SU(16)$ symmetry found previously at a larger pion mass is further observed. The $SU(6)$-symmetric EFT constrained by these LQCD calculations is used to make predictions for two-baryon systems for which the low-energy scattering parameters could not be determined within the present LQCD study, and to constrain the coefficients of all leading $SU(3)$ flavor-symmetric interactions, demonstrating the predictive power of two-baryon EFTs matched to LQCD.

\cleardoublepage
% Acknowledgements *******************************************

\chapter*{Acknowledgements}
\addcontentsline{toc}{chapter}{Acknowledgements}

Primer de tot vull donar les gràcies a la meva directora, l'Assumpta Parreño, per haver-me introduit en el món màgic de la física hipernuclear, per haver-me inculcat totes les pràctiques que ha de seguir un bon investigador, i per haver-me guiat durant més de sis anys, ja des del grau i passant pel màster. No saps que bé que m'ho he passat fent física amb tu.

I want to thank all the members of the NPLQCD Collaboration, with special mention to Silas Beane, Zohreh Davoudi, William Detmold, Martin Savage, Phiala Shanahan and Mike Wagman. I have learned a lot from you, and I am really grateful for your hospitality during my brief visits to your institutions. I also want to thank the collaboration for providing me with all the lattice data that is analyzed in this thesis, as well as for the permission to show the figures and results that are already published.

També vull agraïr a la gent del departament de Física Quàntica i Astrofísica i a l'Institut de Ciències del Cosmos, en especial a l'Àngels Ramos, en Volodymir Magas, en Bruno Julià, l'Artur Polls, en Javier Menéndez, en Joan Soto, en Federico Mescia i la María Concepción González, juntament amb la Laura Tolós i l'Isaac Vidaña, per ajudar-me sempre que ho he necessitat, i per aportar el seu gra de sorra en el meu desenvolupament.

No puc oblidar-me de la persona que encara no entenc d'on ha tret tanta paciència per aguantar-me al despatx, la Glòria, com també d'en Pere i en Jordi. A tots els companys que he conegut durant el doctorat, l'Adrià, l'Albert, l'Alejandro, l'Andreu, en Chiranjib, la Clàudia, l'Iván, en Javi, els Joseps i en Marc, com també als companys de grau i màster, en especial a l'Adrià, les Anes, la Caterina, la Clara, l'Elena, la Gemma, en Guillem, en Jordi, la Maria, en Manel, les Núries, en Pau, en Pere, en Pol, la Sara, en Sergi, en Xavi i la Xènia. Moltes gràcies a tots vosaltres per fer-me riure cada dos per tres.

Finalment vull agraïr el suport incondicional que he rebut de la meva família. Sort n'he tingut de vosaltres.

Aquesta tesi s'ha realitzat amb el suport de l'ajut APIF de la Universitat de Barcelona, del projecte MDM-2014-0369 de l'ICCUB (Unidad de Excelencia “María de Maeztu”) del Ministerio de Economía y Competitividad (MINECO), del contracte FIS2017-87534-P provinent dels fons europeus FEDER i del projecte EU STRONG-2020 del programa H2020-INFRAIA-2018-1 (grant agreement No.\ 824093).

\cleardoublepage
\tableofcontents
\cleardoublepage}

\pagestyle{fancy}
\fancyhead{} %Clear everything
\fancyhead[RE]{\nouppercase{\leftmark}}
\fancyhead[LO]{\nouppercase{\rightmark}}
\fancyhead[LE,RO]{\thepage}
\cfoot{}

%%%%%%%%%%%%%%%%%%%%%%%%%%%%%%%%%%%%%%%%%%%%%%%%%%%%
\mainmatter

% Chapter 1 - Introduction *******************************************

\chapter{Introduction}\label{chap:1}

The description of the basic properties of nuclei from their fundamental constituents, quarks and gluons, is one of the key objectives of nuclear physics. 
At the most fundamental level, the strong interaction binds quarks together forming nucleons ($N$), according to the rules of Quantum Chromodynamics (QCD), which combined with Quantum Electrodynamics (QED), the theory of the weak interaction, and the much weaker gravity, dictates how elementary particles interact with each other. For many years, QCD has been elusive to theoretical solutions in the energy regime characterizing nuclear processes. The strength of the interaction between quarks and gluons increases as the characteristic energy of a process decreases~\cite{Gross:1973id,Politzer:1973fx}, and at nuclear scales, perturbative techniques based on coupling expansion cannot be applied to find solutions of the theory starting from the elementary degrees of freedom. 

Since the up and down quarks are the only stable quarks, a vast amount of data from scattering experiments and spectroscopy are available~\cite{NNonline} to constrain theoretical studies of nuclear properties and interactions. These have proceeded by combining many-body techniques with phenomenological models describing the interaction of point-like nucleons~\cite{Machleidt:2020vzm}. Examples of such successful approaches are the Urbana $v_{14}$~\cite{Lagaris:1981mm} and Argonne $v_{14}$~\cite{Wiringa:1984tg} and $v_{18}$~\cite{Wiringa:1994wb} potentials, or boson-exchange models inspired in Yukawa's meson theory~\cite{Yukawa:1935xg}, like the Nijmegen~\cite{Nagels:1978sc,Stoks:1993tb}, the CD-Bonn~\cite{Machleidt:2000ge}, and the Stadler-Gross~\cite{Gross:2008ps} potentials.

Aiming at a model-independent description of the strong interaction, S.~Weinberg introduced at the beginning of the 1990s a new formulation~\cite{Weinberg:1990rz} which, over the years, has become established as an efficient and systematic way of studying nuclear systems from first principles. This effective field theory (EFT) approach is especially useful when different energy scales can be identified in the physical problem under study, and it is based on retaining only those degrees of freedom that appear explicitly below the largest energy scale that characterizes the process. A small parameter can be then formed from the ratio of the given scales. For example, in the study of $NN$ interaction at low energies, a convenient parameter is constructed from the ratio of the typical momentum carried by the nucleons to the chiral symmetry breaking scale (approximately the nucleon mass). The effective Lagrangian is then constructed by incorporating all allowed operators respecting the symmetries of the underlying QCD interactions and organized in increasing order according to an expansion in the small parameter, and therefore, as a power expansion on the momentum. Each term of the expansion is accompanied by a low-energy coefficient (LEC) that encapsulates the physics that is not explicitly retained, corresponding to energies beyond the largest scale identified in the problem, and which is determined by fitting EFT calculations to experimental data. Therefore, the predictive power of the method relies mainly on two things: the presence of sufficiently separated energy scales and the availability of precise experimental data for a given physics process. For example, two groups, the Bochum~\cite{Epelbaum:2014sza,Reinert:2017usi} and Idaho-Salamanca~\cite{Entem:2017gor} groups, have precisely extracted the phase shifts for the lowest partial waves and the low-energy scattering parameters of $NN$ up to fifth order using chiral effective field theory ($\chi$EFT), and the sixth order is being explored~\cite{Entem:2015xwa}.

Beyond nucleons we find hyperons ($Y$), particles with at least one strange quark, which are expected to appear in the interior of neutron stars~\cite{1960SvA4187A}. The main problem found when dealing with hyperons is that unless the strong interactions between hyperons and nucleons are sufficiently repulsive, the equation of state (EoS) of dense nuclear matter will be softer than for purely non-strange matter, leading to correspondingly lower maximum values for neutron star masses. 
While experimental data on scattering cross sections in the majority of the $YN$ channels are scarce, there are reasonably precise constraints on the interactions in the $\Lambda N$ channel from scattering and hypernuclear spectroscopy experiments~\cite{Feliciello:2015dua,Gal:2016boi}, and they indicate that the interactions in this channel are attractive. 
Given that the $\Lambda$ baryon is lighter than the other hyperons, it is likely the most abundant hyperon in the interior of neutron stars. 
However, models of the EoS including $\Lambda$ baryons and attractive $\Lambda N$ interactions~\cite{Lonardoni:2014bwa} predict a maximum neutron star mass that is below the maximum observed mass at $2M_{\odot}$~\cite{Champion:2008ge, Freire:2010tf,Demorest:2010bx,Antoniadis:2013pzd, Cromartie:2019kug}.\footnote{Very recently, the gravitational wave signal GW190814, originated from the merger of a $23 M_{\odot}$ black hole and a $2.6M_{\odot}$ compact object, was reported~\cite{Abbott:2020khf}, where the nature of the compact object is a subject of discussion. If this compact object was a neutron star, it would have been the most massive one known, imposing a mass-limit constraint very difficult to fulfill for the majority of existing nuclear EoS models.}
Several remedies have been suggested to solve this problem, known in the literature as the ``hyperon puzzle''~\cite{Chatterjee:2015pua, Vidana:2018bdi, Tolos:2020aln}. For example, if hyperons other than the $\Lambda$ baryon (such as $\Sigma$ baryons) are present in the interior of neutron stars and the interactions in the corresponding $YN$ and $YY$ channels are sufficiently repulsive, the EoS would become more stiff~\cite{Vidana:2010ip, Petschauer:2015nea}.
Another suggestion is that repulsive interactions in the $YNN$, $YYN$, and $YYY$ channels may render the EoS stiff enough to produce a $2M_{\odot}$ neutron star~\cite{Vidana:2010ip, Yamamoto:2013ada, Furumoto:2014ica, Lonardoni:2014bwa, Logoteta:2019utx}.
Repulsive density-dependent interactions in systems involving the $\Lambda$ and other hyperons have also been suggested, along with the possibility of a phase transition to quark matter in the interior of neutron stars; see Refs.~\cite{Chatterjee:2015pua,Vidana:2018bdi, Tolos:2020aln} for recent reviews. 
Given the scarcity or complete lack of experimental data on $YN$ and $YY$ scattering and all three-body interactions involving hyperons, $SU(3)$ flavor symmetry ($SU(3)_f$) is used to constrain EFTs and phenomenological meson-exchange models of hypernuclear interactions. In this way, quantities in channels for which experimental data exist can be related via symmetries to those in channels which lack such phenomenological constraints.
For example, the lowest-order effective interactions in several channels with strangeness $S\in\{-2,-3,-4\}$ were constrained using experimental data on $pp$ phase shifts and the $\Sigma^+p$ cross section in the same $SU(3)_f$ representation in the framework of $\chi$EFT in Refs.~\cite{Haidenbauer:2015zqb,Haidenbauer:2009qn,Haidenbauer:2014rna}. However, only a few of the $SU(3)_f$-breaking LECs of the EFT could be constrained~\cite{Haidenbauer:2014rna}. To date, the knowledge of these interactions in nature remains unsatisfactory, demanding more direct theoretical approaches.

During the last twenty years major formal, technological, and algorithmic advances have enabled rigorous exploration of the low-energy regime of QCD using large-scale numerical calculations. By performing a numerical evaluation of the equations of QCD in a discretized space-time, lattice QCD (LQCD) has been used to compute hadronic properties with high precision, in exceptional cases with more accuracy than that given by experiments~\cite{Aoki:2019cca}. Specific to nuclear physics, it has allowed a wealth of observables, from hadronic spectra and structure to nuclear matrix elements~\cite{Detmold:2019ghl, Drischler:2019xuo, Davoudi:2020ngi}, to be calculated directly from interactions of quarks and gluons, albeit with uncertainties that are yet to be fully controlled. In the context of constraining hypernuclear interactions, LQCD is a powerful theoretical tool because the lowest-lying hyperons are stable when only strong interactions are included in the computation, circumventing the limitations faced by experiments on hyperons and hypernuclei. Nonetheless, LQCD studies in the multi-baryon sector require large computing resources as there is an inherent signal-to-noise degradation present in the correlation functions of baryons~\cite{Parisi:1983ae, Lepage:1989hd, Beane:2009kya, Beane:2009gs, Beane:2009py, Wagman:2016bam}, among other issues as discussed in a recent review~\cite{Davoudi:2020ngi}.
Consequently, most studies of two-baryon systems to date~\cite{Beane:2006mx, Beane:2006gf, Beane:2009py, Beane:2010hg, Beane:2011zpa, Beane:2011iw, Beane:2012ey, Beane:2012vq, Beane:2013br, Orginos:2015aya, Wagman:2017tmp, Illa:2020nsi, Fukugita:1994ve, Yamazaki:2011nd, Yamazaki:2012hi, Yamazaki:2015asa, Nemura:2008sp, Inoue:2010es, Inoue:2011ai, Ishii:2013cta, Berkowitz:2015eaa, Horz:2020zvv, Francis:2018qch, Green:2021qol,Amarasinghe:2021} have used larger-than-physical quark masses to expedite computations, and only recently have results at the physical values of the quark masses emerged~\cite{Gongyo:2017fjb, Iritani:2018sra,Sasaki:2019qnh}, making it possible to directly compare with experimental data~\cite{Acharya:2020asf}.
The existing studies are primarily based on two distinct approaches. In one approach, the low-lying spectra of two baryons in finite spatial volumes are determined from the time dependence of Euclidean correlation functions computed with LQCD, and are then converted to scattering amplitudes at the corresponding energies through the use of Lüscher's formula~\cite{Luscher:1986pf, Luscher:1990ux} or its generalizations~\cite{Rummukainen:1995vs, Beane:2003da, Bedaque:2004kc, Feng:2004ua, Kim:2005gf, He:2005ey, Luu:2011ep, Davoudi:2011md, Leskovec:2012gb, Hansen:2012tf, Briceno:2012yi, Gockeler:2012yj, Briceno:2013lba, Briceno:2013bda, Briceno:2013hya, Lee:2017igf,  Briceno:2017max}.
In another approach, non-local potentials are constructed based on the Bethe-Salpeter wavefunctions determined from LQCD correlation functions, and are subsequently used in the Lippmann-Schwinger equation to solve for scattering phase shifts~\cite{Aoki:2020bew}.

While LQCD studies at unphysical values of the quark masses already shed light on the understanding of (hyper)nuclear and dense-matter physics, a full account of all systematic uncertainties, including precise extrapolations to the physical quark mass, is required to further impact phenomenology.
Additionally, LQCD results for scattering amplitudes can be used to better constrain the low-energy interactions within given phenomenological models and applicable EFTs. In the case of exact $SU(3)_f$ symmetry and including only the lowest-lying octet baryons, there are six two-baryon interactions at leading order (LO) in pionless EFT~\cite{vanKolck:1998bw,Chen:1999tn} that can be constrained by the $S$-wave scattering lengths in two-baryon scattering~\cite{Savage:1995kv}.
LQCD has been used in Ref.~\cite{Wagman:2017tmp} to constrain the corresponding LECs of these interactions by computing the $S$-wave scattering parameters of two baryons at an $SU(3)$ flavor-symmetric point with $m_{\pi}\sim 806$ MeV. Strikingly, the first evidence of a long-predicted $SU(6)$ spin-flavor symmetry in nuclear and hypernuclear interactions in the limit of a large number of colors ($N_c$)~\cite{Kaplan:1995yg} was observed in that study, along with an accidental $SU(16)$ symmetry. This extended symmetry has been suggested in Ref.~\cite{Beane:2018oxh} to support the conjecture of entanglement suppression in nuclear and hypernuclear forces at low energies, pointing to intriguing aspects of strong interactions in nature. 

The objective of this thesis is to extend the previous studies to quark masses that are closer to their physical values, corresponding to a pion mass of $\sim 450$ MeV and a kaon mass of $\sim 596$ MeV, and further to study these systems in a setting with broken $SU(3)_f$ symmetry as is the case in nature. Therefore, it provides new constraints that allow preliminary extrapolations to physical quark masses to be performed, and complements previous independent LQCD studies at nearby quark masses~\cite{Beane:2006mx,Beane:2006gf,Beane:2009py,Beane:2010hg,Beane:2011iw,Beane:2012ey,Orginos:2015aya,Yamazaki:2012hi,Yamazaki:2015asa,Inoue:2011ai,Ishii:2013cta}.
The LQCD results presented here are used to constrain the leading $SU(3)_f$ symmetry-breaking coefficients in pionless EFT. This EFT matching enables the exploration of large-$N_c$ predictions, pointing to the validity of $SU(6)$ spin-flavor symmetry at this pion mass as well, and revealing a remnant of an accidental $SU(16)$ symmetry that was observed at a larger pion mass in Ref.~\cite{Wagman:2017tmp}. Strategies to make use of the QCD-constrained EFTs to advance the \emph{ab initio} many-body studies of larger hypernuclear isotopes and dense nuclear matter are beyond the scope of this work. Nevertheless, the methods applied in Refs.~\cite{Barnea:2013uqa,Contessi:2017rww,Bansal:2017pwn} to connect the results of LQCD calculations to higher-mass nuclei can also be applied in the hypernuclear sector using the results presented.

The structure of this theses is organized as follows.
\cref{chap:2} gives a brief introduction of QCD, followed by a description of the LQCD method. For that, the discretization of QCD is explained, together with the observables that can be extracted (energies and matrix elements). Finally, the study of scattering processes in finite volume is detailed, with an appropriate summary of the necessary group-theoretical tools.
\cref{chap:3} is devoted to the main tools to analyze the correlation functions, including several more sophisticated methods to reduce excited-state contamination.
\cref{chap:4} is focused on the baryon-baryon interaction. First, a summary of the present status of the field, both experimentally and theoretically, is presented. After the EFT Lagrangians that will be constrained are explained, the main LQCD results are showed, which are the lowest-lying energies, the $S$-wave scattering parameters, and the binding energies (with a preliminary extrapolation to the physical point) of several two-baryons channels, followed by the constraints that these results impose on the LECs of the EFTs.
To conclude the thesis, \cref{chap:summary} summarizes the work.
Several appendices follow to supplement the thesis, omitted from the main body for clarity of presentation.
\cref{appen:GTtables} shows all the relevant group-theoretical relations between the point (and double)  and continuum angular momentum groups relevant for the states studied in this work.
\cref{appen:Zfun} presents the derivation of the exponentially-accelerated version of the $\mathcal{Z}$-function.
\cref{appen:BBsummary} tabulates the scattering parameters predicted by the available theoretical models for the baryon-baryon channels studied in this thesis, as well as the binding energies extracted from fully-dynamical LQCD calculations.
\cref{appen:SU3chann} explicitly states the full $SU(3)_f$ decomposition of all octet baryon-baryon channels.
\cref{appen:NLOEFT} contains the partial-wave decomposition of all the next-to-leading order (NLO) terms that appear in the Lagrangian of Ref.~\cite{Petschauer:2013uua}.
\cref{appen:LECsEFT} includes relations among the LECs of the three-flavor EFT Lagrangian of Ref.~\cite{Petschauer:2013uua} and the ones used in the present work, as well as a recipe to access the full set of leading symmetry-breaking coefficients from future studies of a more complete set of two-baryon systems.
\cref{appen:vs2015} presents an exhaustive comparison between the results obtained in this work and previous results presented in Ref.~\cite{Orginos:2015aya} for the two-nucleon channels using the same LQCD correlation functions, as well as with the predictions of the low-energy theorems analyzed in Ref.~\cite{Baru:2016evv}.
\cref{appen:figtab} contains additional figures and tables related to the LQCD results presented in~\cref{sec:450results}.

% Chapter 2 - LQCD *******************************************

\chapter{QCD on the computer}\label{chap:2}

\section{Quantum Chromodynamics}

More than two centuries have passed since the beginning of nuclear physics, with the accidental discovery of radioactivity by H.~Becquerel in 1896. A great deal of experiments and theoretical breakthroughs were needed to pinpoint the fundamental forces behind very distinct processes and elaborate what we know today as the Standard Model (SM). Some of these milestones, relevant for this thesis, are the first proposal for the description of the strong force by H.~Yukawa~\cite{Yukawa:1935xg}, the discovery of the first strange particles, the $K^0$ meson\footnote{That is the reason why the strangeness quantum number is negative, since the kaon was given $S=1$ although it carries an anti-strange quark.} by G.~D.~Rochester and C.~C.~Butler~\cite{Rochester:1947mi} and the $\Lambda$ baryon by V.~D.~Hopper and S.~Biswas~\cite{Hopper:1950}. In order to understand and organize the large number of particles discovered, the eightfold way was proposed by M.~Gell-Mann~\cite{GellMann:1961ky} and Y.~Ne'eman~\cite{Neeman:1961jhl}, followed by the more fundamental quark model by M.~Gell-Mann~\cite{GellMann:1964nj} and Z.~Zweig~\cite{Zweig:1964}. The proposal of a new quantum number, later on called color charge, by W.~Greenberg~\cite{Greenberg:1964pe}, M.~Y.~Han, and Y.~Nambu~\cite{Han:1965pf}, was one of the last steps before the definition of the QCD Lagrangian by H.~Fritzsch, M.~Gell-Mann, and H.~Leutwyler~\cite{Fritzsch:1973pi}.

From that point forward, it is known that the degrees of freedom of QCD are quarks and gluons. Mathematically, the quarks are spin-$\frac{1}{2}$ Dirac spinors $\alpha\in\{1,2,3,4\}$ that carry color $a\in\{1,2,3\}$ and flavor $q\in\{$up, down, strange, charm, bottom, top$\}$ indices, $\psi^a_{q,\alpha}$, and transform under the fundamental (triplet) representation of $SU(3)_c$ as
\begin{equation}
    \psi(x) \rightarrow \psi^\prime(x)=\Omega(x) \psi(x) = e^{-\imag g \theta_\mathfrak{a}(x) T_\mathfrak{a}} \psi(x)\, , \quad \text{with }\; \Omega(x)\in SU(3)_c\, ,
\end{equation}
where $g$ is the strong coupling constant, $\theta_\mathfrak{a}(x)$ is the parameter of the transformation that depends on the position (to account for the local gauge invariance), and $T_\mathfrak{a}$ are the generators of the $SU(3)_c$ Lie algebra, with $\mathfrak{a}\in\{1,\ldots,8\}$ (the number of $SU(N)$ generators equals the dimension of the adjoint representation, $N^2-1=8$ for $N=3$), which can be written as $T_\mathfrak{a}=\lambda_\mathfrak{a}/2$, with $\lambda_\mathfrak{a}$ being the Gell-Mann matrices~\cite{GellMann:1962xb}. These generators are traceless Hermitian matrices, normalized such that $\Tr(T_\mathfrak{a} T_\mathfrak{b}) = \frac{1}{2}\delta_{\mathfrak{a}\mathfrak{b}}$, obeying the commutation relation $[T_\mathfrak{a},T_\mathfrak{b}]=\imag f_{\mathfrak{a}\mathfrak{b}\mathfrak{c}}T_\mathfrak{c}$, where $f_{\mathfrak{a}\mathfrak{b}\mathfrak{c}}$ are the structure constants of $SU(3)_c$.

The mediators of the interaction, the gluons, are spin-$1$ gauge bosons that are usually written as $A_{\mu}=T_\mathfrak{a}A^\mathfrak{a}_{\mu}$. They transform under the adjoint (octet) representation of $SU(3)_c$ as
\begin{equation}
    A_{\mu}(x) \rightarrow A^\prime_{\mu}(x) = \Omega(x) A_{\mu}(x) \Omega^{\Dag}(x)+\frac{\imag}{g} \left[\partial_{\mu}\Omega(x)\right] \Omega^{\Dag}(x)\, .
\end{equation}
The Lagrangian of QCD has to be invariant under these local gauge transformations, and it can be written as
\begin{equation}
    \mathcal{L}_{QCD}=\sum_q\smash{\bar{\psi}}_q \left(\imag\gamma^{\mu}D_{\mu}-m_q\right)\psi_q-\frac{1}{4}G^{\mathfrak{a}}_{\mu\nu}G^{\mathfrak{a},\mu\nu}\, ,
\label{eq:QCDLag}
\end{equation}
where $\gamma^{\mu}$ are the Dirac matrices, $m_q$ are the masses of the quarks, and the covariant derivative $D_{\mu}$ contains the term that couples the quark and gluons, $D_{\mu}=\partial_{\mu}+\imag g T_\mathfrak{a} A^\mathfrak{a}_{\mu}$.

The purely gluonic part is written in terms of the gluon field strength tensor $G^\mathfrak{a}_{\mu\nu}=\partial_\mu A^\mathfrak{a}_\nu-\partial_\nu A^\mathfrak{a}_\mu - g f_{\mathfrak{a}\mathfrak{b}\mathfrak{c}}A^\mathfrak{b}_\mu A^\mathfrak{c}_\nu$, where the last term is characteristic of non-Abelian theories (no such term appears in the QED Lagrangian), and is responsible for the three- and four-gluon self-interactions. The reason why these types of interactions appear is due to the fact that the gluon is charged with color (the photon does not have electric charge), so it is able to interact with other charged particles, like quarks and other gluons. Since a term of the form $m_g A_{\mu}A^{\mu}$ is not gauge invariant, gluons are massless particles.

Another term that we have not included in Eq.~\eqref{eq:QCDLag} but is allowed by gauge invariance is one proportional to $\theta \epsilon^{\mu\nu\rho\omega} G^\mathfrak{a}_{\mu\nu} G^\mathfrak{a}_{\rho\omega}$, known as the $\theta$-term. This term, unlike the others, violates CP-symmetry, and the value of $\theta$ (specifically, the combination $\theta'=\theta+\arg\Det m_q$) has been constrained experimentally with the electric dipole moment of the neutron, giving an upper limit of $|\theta'| \lesssim 10^{-10}$~\cite{Graner:2016ses}. The reason why the value of $\theta$ is so small is still not understood, and it is known as the strong CP problem. There are additional terms in the Lagrangian of Eq.~\eqref{eq:QCDLag}, such as the gauge fixing term (with the fictitious Faddeev-Popov ghosts) and the corresponding counterterms, but they are not relevant for the subject of this thesis, LQCD~\cite{Gupta:1997nd}.

One of the most striking features of QCD is how the value of $g$ depends on the energy scale of the process. This is known as asymptotic freedom, and it was discovered by D.~J.~Gross, F.~Wilczek~\cite{Gross:1973id}, and H.~D.~Politzer~\cite{Politzer:1973fx} (the three of them were awarded the Nobel Prize in Physics in 2004). At very high energies (or very small distances) the coupling constant becomes small, so the quarks and gluons interact very weakly, and perturbation theory can be used to study processes in this energy regime. However, at low energies the situation is the opposite, with $g$ increasing in value as the energy decreases to the point where $g\sim\mathcal{O}(1)$ (around $\Lambda_{QCD}$) and perturbation techniques are no longer adequate. In this regime, the quarks and gluons are bound inside color-singlet hadrons, known as confinement. The most common hadrons are the mesons (pair of quark-antiquark) and baryons (three quarks), although more exotic ones, like pentaquarks or glueballs, are not prohibited by QCD.

Since perturbation theory is no longer applicable at low energies, several alternative methods and models have been developed to circumvent this problem. Examples are the use of phenomenological models and EFTs for the nuclear sector as mentioned in~\cref{chap:1}. The one we will focus on in this thesis is LQCD, the only non-perturbative method in which quantities are computed directly using quarks and gluons, and is systematically improvable.

\section{Discretization of QCD}

The formalism of LQCD was first introduced by K.~G.~Wilson~\cite{Wilson:1974sk}, and it is based on the path integral formalism of R.~P.~Feynman~\cite{Feynman:1948}, where observables are computed as vacuum expectation values of operators,
\begin{equation}
    \langle \hat{O}\rangle = \frac{1}{Z}\int \mathcal{D}\psi \mathcal{D}\bar{\psi} \mathcal{D}A_{\mu}\, \hat{O} \, e^{\imag S_{QCD}}\, ,
\label{eq:pathintMink}
\end{equation}
with $Z$ being the QCD partition function, $Z = \int \mathcal{D}\psi \mathcal{D}\bar{\psi} \mathcal{D}A_{\mu} \, e^{\imag S_{QCD}}$, and $S_{QCD}$ the QCD action, $S_{QCD}=\int d^4x \, \mathcal{L}_{QCD}$. 

Notice that this formalism resembles the one used in statistical mechanics (see Ref.~\cite{Gupta:1997nd}) except the imaginary unit in the exponential, which renders an oscillatory factor, troublesome for numerical evaluations. A solution to this problem is to perform a Wick rotation~\cite{Wick:1954}, which transforms the $(3+1)$-dimensional Minkowski field theory to a 4-dimensional Euclidean field theory. Under this rotation,
\begin{equation}
\begin{aligned}
    \eta_{\mu\nu}=\operatorname{diag}(+1,-1,-1,-1) &\rightarrow \eta^{(E)}_{\mu\nu}=\delta_{\mu\nu}=\operatorname{diag}(+1,+1,+1,+1) \\
    t=x_0 \rightarrow -\imag x^{(E)}_4 = -\imag \tau \,, &\qquad x_i \rightarrow x^{(E)}_i\,, \quad \text{with }\; i\in\{1,2,3\} \, ,\\
    \partial_0 \rightarrow \imag \partial^{(E)}_4\, , &\qquad \partial_i \rightarrow \partial^{(E)}_i \, ,\\
    \gamma^{0} \rightarrow \gamma^{4(E)}\, ,&\qquad \gamma^{i} \rightarrow \imag\gamma^{i(E)} \, , \\
    A_0 \rightarrow \imag A_4^{(E)}\, ,&\qquad A_i \rightarrow A_i^{(E)} \, .
\end{aligned}
\end{equation}
Note that, with the new metric tensor $\eta^{(E)}_{\mu\nu}$, in Euclidean space-time one does not need to worry about the position of the indices, since there will be no extra $-1$ factors when raising or lowering them.
If we apply these changes to Eq.~\eqref{eq:QCDLag}, the QCD Lagrangian becomes
\begin{equation}
\begin{aligned}
    \mathcal{L}_{QCD}=&\sum_q\smash{\bar{\psi}}_q \left( \imag \gamma^\mu D_\mu - m_q \right) \psi_q - \frac{1}{4} G^\mathfrak{a}_{\mu\nu} G^{\mathfrak{a},\mu\nu}\\
    =&\sum_q\smash{\bar{\psi}}_q \left(\imag \gamma^0 D_0 + \imag \gamma^i D_i - m_q \right) \psi_q - \frac{1}{4} \left(-G^\mathfrak{a}_{0i} G^\mathfrak{a}_{0i} + G^\mathfrak{a}_{ij} G^\mathfrak{a}_{ij} \right)\\
    \rightarrow & -\sum_q \smash{\bar{\psi}}_q \left( \gamma^{(E)}_4 D^{(E)}_4 + \gamma^{(E)}_i D^{(E)}_{i} + m_q \right) \psi_q - \frac{1}{4} \left( G^{(E)\mathfrak{a}}_{4i} G^{(E)\mathfrak{a}}_{4i} + G^{(E)\mathfrak{a}}_{ij} G^{(E)\mathfrak{a}}_{ij} \right)\\
    &= - \left[ \sum_q \smash{\bar{\psi}}_q \left( \gamma^{(E)}_\mu D^{(E)}_\mu + m_q \right) \psi_q + \frac{1}{4} G^{(E)\mathfrak{a}}_{\mu\nu} G^{(E)\mathfrak{a}}_{\mu\nu} \right] = - \mathcal{L}^{(E)}_{QCD}\, ,
\end{aligned}
\end{equation}
and the Euclidean QCD action is expressed as
\begin{equation}
    S_{QCD}=\int d^4x \, \mathcal{L}_{QCD} \rightarrow \int (-\imag)d\tau \int d^3\bm{x}\, (-\mathcal{L}^{(E)}_{QCD})=\imag\int d^4x^{(E)}\, \mathcal{L}^{(E)}_{QCD}=\imag S^{(E)}_{QCD}\, ,
\end{equation}
making the phase in Eq.~\eqref{eq:pathintMink} real. Therefore, and omitting the superscript $(E)$ for simplicity,
\begin{equation}
    \langle \hat{O}\rangle = \frac{1}{Z}\int \mathcal{D}\psi \mathcal{D}\bar{\psi} \mathcal{D}A_{\mu}\, \hat{O} \, e^{-S_{QCD}}\, .
\label{eq:pathintEucl}
\end{equation}
Similar changes occur in the partition function. With the current form, one can identify $e^{-S_{QCD}}$ as a probability distribution function and apply Monte Carlo methods to perform this multi-dimensional integral. Before we get to this point, we have to discretize $S_{QCD}$.\footnote{For a complete introduction and development of lattice gauge theories, see Refs.~\cite{Rothe:2005,DeGrand:2006,Gattringer:2010zz,Lin:2015dga,Knechtli:2017sna}.}

The simplest way to discretize QCD is by using an isotropic hypercubic lattice $\Lambda$,
\begin{equation}
    \Lambda = \Big\lbrace x_{\mu}=(x_1,x_2,x_3,x_4=\tau)\, \Big|\, 0\leq x_1,x_2,x_3< L\, ,\; 0\leq x_4< T \Big\rbrace\, ,
\end{equation}
where $L$ is the spatial extent and $T$ is the temporal extent (with total volume $V=L^3\times T$). The lattice spacing $b$ in this case is the same in both directions.\footnote{Other types of geometries are also used, like anisotropic lattices (where the temporal extent has a finer lattice spacing)~\cite{Burgers:1987mb,Klassen:1998ua} or asymmetric lattices~\cite{Li:2003jn,Detmold:2004qn}.} The discretization of QCD has two purposes: to make it amenable for computational calculations, and to introduce an ultraviolet cutoff (inverse of the lattice spacing), regularizing the theory. As will be discussed later, to make the connection to the physical world, the limits of zero lattice spacing $b\rightarrow 0$ and infinite volume $L\rightarrow \infty$ have to be taken. The calculations with non-zero $b$ and finite $L$ have to be chosen carefully: the mesh has to be fine enough so that it resolves the hadronic scale ($b \ll \Lambda_{QCD}$), and the spatial extent must be large compared to the typical range of the hadronic interactions under study, which is set by the Compton wavelength of the lightest particle exchanged (for the $NN$ interaction, this implies $L \gg m^{-1}_{\pi}$).

Within this formulation, the quarks, spin-$\frac{1}{2}$ objects with Dirac, color, and flavor indices, reside on the nodes of the lattice, $\psi(x)$. Due to the finite volume, boundary conditions (BC) are applied to both fields, quarks and gluons (discussed with more detail below). On the spatial direction, one typically applies periodic BC to both fields, although more sophisticated choices, like twisted BC~\cite{Byers:1961zz,Bedaque:2004kc}, are also possible. For the temporal direction, anti-periodic BC are imposed to the quarks (that are fermions) while periodic BC are imposed to gluons (bosons), so as to ensure the correct statistics.

To illustrate some of the problems inherent in the discretization method, we discuss below the simplest approximation, the so-called naive discretization, for the free quark case, for which the QCD action reads
\begin{equation}
    \int d^4x \, \bar{\psi}(x)(\gamma^{\mu}\partial_{\mu}+m)\psi(x)\rightarrow b^4\sum_{x\in\Lambda} \bar{\psi}(x)\left\lbrace\gamma^{\mu}\frac{1}{2b}\left[\psi(x+\hat{\mu})-\psi(x-\hat{\mu})\right]+m\psi(x)\right\rbrace\, ,
\end{equation}
where the integral is now a sum over the lattice sites, and the derivative has been discretized by a symmetric finite difference. We can further simplify this expression by using the Dirac operator $M(x,y)$,
\begin{equation}
    b^4\sum_{x,y\in\Lambda} \bar{\psi}(x)\left[\gamma^{\mu}\frac{1}{2b}\left(\delta_{x+\hat{\mu},y}-\delta_{x-\hat{\mu},y}\right)+m\delta_{x,y}\right] \psi(y)=b^4\sum_{x,y\in\Lambda} \bar{\psi}(x)M(x,y)\psi(y)\, .
\end{equation}
To look at the spectrum, it is easier to go to momentum space,
\begin{equation}
    \delta_{x,y}=\int^{\pi/b}_{-\pi/b}\frac{d^4p}{(2\pi)^4} e^{\imag p (x-y)}\, ,
\end{equation}
where the limits of the integration correspond to the first Brillouin zone (BZ), $(-\pi/b,\pi/b]$. The Dirac matrix can thus be written as 
\begin{equation}
    \tilde{M}(p)= \gamma^{\mu}\frac{1}{2b}\left(e^{\imag p_{\mu} b}-e^{-\imag p_{\mu} b}\right)+m = \frac{\imag}{b}\gamma^{\mu}\sin(p_{\mu}b)+m\, ,
\end{equation}
whose inverse is related to the quark propagator,
\begin{equation}
    \tilde{M}^{-1}(p)= \frac{1}{\frac{\imag}{b}\gamma^{\mu}\sin(p_{\mu}b)+m}=\frac{-\frac{\imag}{b}\gamma^{\mu}\sin(p_{\mu}b)+m}{\frac{1}{b^2}\sin^2(p_{\mu}b)+m^2}\, .
\end{equation}
In the limit $m\rightarrow 0$, the poles of the propagator one finds are the usual $p_{\mu}=(0,0,0,0)$, expected in the continuum theory, plus 15 unphysical poles at the corners of the BZ, $p_{\mu}\in\{(\frac{\pi}{b},0,0,0),\ldots,\allowbreak (\frac{\pi}{b},\frac{\pi}{b},\frac{\pi}{b},\frac{\pi}{b})\}$. These extra poles are the so-called doublers and are purely lattice artifacts~\cite{Wilson:1975id}. The reason these doublers appear is explained by the Nielsen-Ninomiya no-go theorem~\cite{Nielsen:1981hk}, which states that one cannot define an Hermitian, translational invariant, local, and chirally symmetric lattice regularized gauge theory without doublers.

There are several ways to remove these unwanted states. Since in~\cref{chap:4} the lattice results shown are computed using the Wilson approach~\cite{Wilson:1975id}, we will focus on this one, and the rest will only be mentioned. The proposed solution by Wilson consisted in adding the following irrelevant operator, $-\frac{1}{2}br\bar{\psi}\partial^{\mu}\partial_{\mu}\psi$ , with $r$ being the Wilson parameter, which is usually set to 1 (note that the added term vanishes in the limit $b\rightarrow 0$). The corresponding discretized version is
\begin{equation}
    -\frac{r}{2b}\bar{\psi}(x)\left(\delta_{x+\hat{\mu},y}-2\delta_{x,y}+\delta_{x-\hat{\mu},y}\right)\psi(y)\, ,
\end{equation}
with the momentum-space Dirac operator being
\begin{equation}
    \tilde{M}(p)= \frac{\imag}{b}\gamma^{\mu}\sin(p_{\mu}b)+\frac{r}{b}\sum_{\mu}[1-\cos(p_{\mu}b)]+m\, .
\end{equation}
If we compute the propagator and look at the poles, we see that the original pole is undisturbed, while the doublers acquire an extra factor proportional to $r/b$, which in the continuum limit will become infinitely massive and decouple from the theory. The usual way to write the (free) Wilson action is in terms of $\kappa=(2mb+8r)^{-1}$,
\begin{equation}
    S^W=b^4\sum_{x,y\in\Lambda} \bar{\psi}(x)M(x,y)\psi(y)\, , \quad M(x,y) = \delta_{x,y}-\kappa\left[(r-\gamma^{\mu})\delta_{x+\hat{\mu},y}+(r+\gamma^{\mu})\delta_{x-\hat{\mu},y}\right]\, ,
\end{equation}
where we have redefined $\psi \rightarrow \sqrt{2\kappa b}\,\psi$. 

We are ready now to introduce the gluons into the calculation in a gauge invariant way. Finite differences contain terms like $\bar{\psi}(x)\psi(x+\hat{\mu})$, which transform like
\begin{equation}
    \bar{\psi}(x) \psi(x+\hat{\mu}) \rightarrow \bar{\psi}^\prime(x) \psi^\prime(x+\hat{\mu}) = \bar{\psi}(x) \Omega^{\Dag}(x) \Omega(x+\hat{\mu}) \psi(x+\hat{\mu}) \neq \bar{\psi}(x) \psi(x+\hat{\mu})\, .
\end{equation}
For these terms to be invariant under a local gauge transformation, we need to introduce an additional field, $U_{\mu}(x)$, transforming as
\begin{equation}
    U_{\mu}(x)\rightarrow U^\prime_{\mu}(x) = \Omega(x) U_{\mu}(x) \Omega^{\Dag}(x+\hat{\mu})\, .
\end{equation}
Now $\bar{\psi}(x)U_{\mu}(x)\psi(x+\hat{\mu})$ is invariant under local gauge transformation. In the continuum, such object already exists, and is the path-ordered exponential integral of the gauge field $A_{\mu}$ along a curve $C$ connecting two points $x$ and $y$,
\begin{equation}
    U(x,y)=\mathcal{P}\, e^{\imag g\int_C A\cdot dl}\, .
\end{equation}
Therefore, we can interpret $U_{\mu}(x)$, named link variables, as the lattice version of the gauge transporter connecting the points $x$ and $x+\hat{\mu}$, taking the following form $U_{\mu}(x)=e^{\imag g b A_{\mu}(x)}$, with $U^{\Dag}_{\mu}(x)=U_{-\mu}(x-\hat{\mu})$. Then, the Dirac operator of the (gauge-invariant) Wilson action is
\begin{equation}
    M(x,y) = \delta_{x,y}-\kappa\left[(1-\gamma^{\mu})U_{\mu}(x)\delta_{x+\hat{\mu},y}+(1+\gamma^{\mu})U_{-\mu}(x)\delta_{x-\hat{\mu},y}\right]\, .
\end{equation}
An important property of this operator is the $\gamma^5$-hermiticity, which implies $\gamma^5 M(x,y)\gamma^5 = M^{\Dag}(y,x)$. This property will come in handy later, when dealing with propagators and correlation functions for mesons, but most importantly it forces the determinant of $M(x,y)$ to be real.

The Wilson action is only correct up to $\mathcal{O}(b)$ discretization errors. To improve the situation (avoiding calculations with very small $b$), one can introduce higher-dimensional operators to the action that cancel the $\mathcal{O}(b)$ errors. This is known as the Symanzik improvement program~\cite{Symanzik:1983dc}. For the Wilson action, this correction was computed by B.~Sheikholeslami and R.~Wohlert~\cite{Sheikholeslami:1985ij} by adding the following operator,
\begin{equation}
  S^{SW}=S^W-b^5 c_{SW} \sum_{x\in\Lambda} \bar{\psi}(x) \frac{1}{2} \kappa \sigma_{\mu\nu} G_{\mu\nu}(x) \psi(x)\, ,
\end{equation}
where the coefficient $c_{SW}$ has to be tuned so that it cancels the $\mathcal{O}(b)$ errors~\cite{Luscher:1996sc}, $\sigma_{\mu\nu} = [\gamma_{\mu},\gamma_{\nu}]/2$, and $G_{\mu\nu}(x)$ is the gluon field strength tensor. This term is usually called the clover term due to the way $G_{\mu\nu}(x)$ is discretized, resembling a clover leaf,
\begin{equation}
Q_{\mu\nu}(x)=
\begin{tikzpicture}[baseline={([yshift=-.5ex]current bounding box.center)}]
\draw[step=0.5cm,gray, thin] (-0.7,-0.7) grid (0.7,0.7);

\draw[thick,->-=.57] (0.05,0.05) -- (0.45,0.05);
\draw[thick,->-=.57] (0.45,0.05) -- (0.45,0.45);
\draw[thick,->-=.57] (0.45,0.45) -- (0.05,0.45);
\draw[thick,->-=.57] (0.05,0.45) -- (0.05,0.05);

\draw[thick,->-=.57] (-0.45,0.05) -- (-0.05,0.05);
\draw[thick,->-=.57] (-0.45,0.45) -- (-0.45,0.05);
\draw[thick,->-=.57] (-0.05,0.45) -- (-0.45,0.45);
\draw[thick,->-=.57] (-0.05,0.05) -- (-0.05,0.45);

\draw[thick,->-=.57] (-0.05,-0.05) -- (-0.45,-0.05);
\draw[thick,->-=.57] (-0.45,-0.05) -- (-0.45,-0.45);
\draw[thick,->-=.57] (-0.45,-0.45) -- (-0.05,-0.45);
\draw[thick,->-=.57] (-0.05,-0.45) -- (-0.05,-0.05);

\draw[thick,->-=.57] (0.45,-0.05) -- (0.05,-0.05);
\draw[thick,->-=.57] (0.45,-0.45) -- (0.45,-0.05);
\draw[thick,->-=.57] (0.05,-0.45) -- (0.45,-0.45);
\draw[thick,->-=.57] (0.05,-0.05) -- (0.05,-0.45);

\filldraw (0,0) circle (1pt);
\end{tikzpicture}\quad , \qquad \begin{matrix*}[l]
&G_{\mu\nu}(x)=\frac{1}{8}\left[Q_{\mu\nu}(x)-Q_{\nu\mu}(x)\right]\, ,\\[6pt]
&Q_{\mu\nu}(x)=P_{\mu\nu}(x)+P_{\nu-\mu}(x)+P_{-\nu\mu}(x)+P_{-\mu-\nu}(x)\, .
\end{matrix*}
\end{equation}
The objects $P_{\mu\nu}(x)$ are called the plaquettes, and will be discussed later, in the context of the discretization of the purely-gluonic action. 

This clover-improved action, although it removes the problem of the doublers, breaks chiral symmetry explicitly. Despite this, it is widely used in the LQCD community, resulting in some remarkable results. As an example, the BMW Collaboration has computed the mass-splittings between iso-multiplets~\cite{Borsanyi:2014jba} in total agreement with experimental data, and in some cases with better precision.

As mentioned before, there are alternative ways to remove the doublers besides the Wilson approach. These are summarized below:
\begin{itemize}[label={--}]
    \item Twisted-mass fermions~\cite{Aoki:1983qi,Frezzotti:2000nk} are a variant of the Wilson fermions, where the quarks are rotated in flavor space by some angle, which can be tuned to remove the $\mathcal{O}(b)$ lattice artifacts. However, it breaks isospin symmetry (see Ref.~\cite{Shindler:2007vp}). Using this formulation, the ETM Collaboration was able, for example, to make a full flavor decomposition of the spin and momentum fraction of the proton~\cite{Alexandrou:2020sml}.
    \item Staggered fermions~\cite{Kogut:1974ag} do not remove explicitly the doublers, but simply re-distribute them among lattice sites, leaving in the end 4 doublers (called tastes, similar to flavors but unphysical), which are removed using the so-called ``fourth-root procedure'' (see Ref.~\cite{Bazavov:2009bb}). These types of fermions are used, for example, to study thermodynamical properties, like the QCD equation of state~\cite{Borsanyi:2013bia,Bazavov:2014pvz}.
    \item Domain-wall~\cite{Kaplan:1992bt,Kaplan:1992sg,Shamir:1993zy,Furman:1994ky} and overlap fermions~\cite{Narayanan:1994gw,Neuberger:1997fp} are formulations whose main purpose is to maintain chiral symmetry (more specifically, a lattice version of it, known as the Ginsparg–Wilson equation~\cite{Ginsparg:1981bj}) at the cost, for example, of adding an extra dimension for the case of domain-wall. The main problem with these formulations is that they are $10-100$ times computationally more expensive than the rest. As an example, the RBC/UKQCD Collaboration used domain-wall fermions to compute the $K\rightarrow \pi\pi$ decay~\cite{Abbott:2020hxn}.
\end{itemize}

Now we can focus on the discretization of the gauge part of the action. To maintain gauge invariance, we have introduced the link variables $U_{\mu}(x)$, which are related to the $A_{\mu}$ fields. Working with only link variables, the only gauge invariant object is the trace of a path ordered closed loop, also called Wilson loop, $\Tr W_{\mu\cdots\nu}(x)= \Tr[U_{\mu}(x)\cdots U^{\Dag}_{\nu}(x)]$. The simplest case is the plaquette (introduced previously for the clover term), which has the following form,
\begin{equation}
P_{\mu\nu}(x)=
\begin{tikzpicture}[baseline={([yshift=-.5ex]current bounding box.center)}]
\draw[step=0.5cm,gray, thin] (-0.2,-0.2) grid (0.7,0.7);

\draw[thick,->-=.57] (0.05,0.05) -- (0.45,0.05);
\draw[thick,->-=.57] (0.45,0.05) -- (0.45,0.45);
\draw[thick,->-=.57] (0.45,0.45) -- (0.05,0.45);
\draw[thick,->-=.57] (0.05,0.45) -- (0.05,0.05);

\filldraw (0,0) circle (1pt);
\end{tikzpicture}\quad , \qquad P_{\mu\nu}(x)=U_{\mu}(x)U_{\nu}(x+\hat{\mu})U^{\Dag}_{\mu}(x+\hat{\nu})U^{\Dag}_{\nu}(x)\, .
\end{equation}
With this simple loop, we can write the action as
\begin{equation}
    \int d^4x \frac{1}{4}G^{\mathfrak{a}}_{\mu\nu}G^{\mathfrak{a},\mu\nu}\rightarrow \beta \sum_{x\in\Lambda} \sum_{\mu<\nu}\Ree\Tr\frac{1}{3}\left[1-P_{\mu\nu}(x)\right]\, ,
\end{equation}
where $\beta = 6/g^2$. It can be shown (using the Baker-Campbell-Hausdorff formula and performing a Taylor expansion of $A_{\mu}$ around $n$) that the discretized version is correct up to $\mathcal{O}(b^2)$. The Symanzik improvement program~\cite{Symanzik:1983dc} can also be applied here, where now the higher-dimensional operators correspond to larger Wilson loops, which for the case of the Lüscher-Weisz action~\cite{Luscher:1984xn} correspond to rectangular and parallelogram-shaped loops (besides the plaquette),
\begin{equation}
\begin{matrix*}[l]
&P_{pl}(x)=
\begin{tikzpicture}[baseline={([yshift=-.5ex]current bounding box.center)}]
\draw[step=0.5cm,gray, thin] (-0.15,-0.15) grid (0.65,0.65);

\draw[thick,->-=.57] (0.05,0.05) -- (0.45,0.05);
\draw[thick,->-=.57] (0.45,0.05) -- (0.45,0.45);
\draw[thick,->-=.57] (0.45,0.45) -- (0.05,0.45);
\draw[thick,->-=.57] (0.05,0.45) -- (0.05,0.05);

\filldraw (0,0) circle (1pt);
\end{tikzpicture}\\[8pt]
&R_{rt}(x)=
\begin{tikzpicture}[baseline={([yshift=-.5ex]current bounding box.center)}]
\draw[step=0.5cm,gray, thin] (-0.15,-0.15) grid (1.15,0.65);

\draw[thick,->-=.57] (0.05,0.05) -- (0.45,0.05);
\draw[thick,->-=.57] (0.45,0.05) -- (0.95,0.05);
\draw[thick,->-=.57] (0.95,0.05) -- (0.95,0.45);
\draw[thick,->-=.57] (0.95,0.45) -- (0.45,0.45);
\draw[thick,->-=.57] (0.45,0.45) -- (0.05,0.45);
\draw[thick,->-=.57] (0.05,0.45) -- (0.05,0.05);

\filldraw (0,0) circle (1pt);
\end{tikzpicture}\\[8pt]
&G_{pg}(x)=
\begin{tikzpicture}[baseline={([yshift=-.5ex]current bounding box.center)}]
\draw[step=0.5cm,gray, thin] (-0.15,-0.15) grid (0.65,0.65);
\draw[step=0.5cm,gray!60, thin, xshift=0.2cm, yshift=0.2cm] (-0.15,-0.15) grid (0.65,0.65);

\draw[thick,->-=.57] (0.05,0.05) -- (0.45,0.05);
\draw[thick,->-=.57] (0.45,0.05) -- (0.65,0.25);
\draw[thick,->-=.57] (0.65,0.25) -- (0.65,0.65);
\draw[thick,->-=.57] (0.65,0.65) -- (0.25,0.65);
\draw[thick,->-=.57] (0.25,0.65) -- (0.05,0.45);
\draw[thick,->-=.57] (0.05,0.45) -- (0.05,0.05);

\draw[step=0.5cm,gray, thin] (-0.15,-0.15) grid (0.65,0.65);

\filldraw (0,0) circle (1pt);
\end{tikzpicture}
\end{matrix*}\quad \qquad \begin{matrix*}[l]
S^{LW}=\beta \displaystyle\sum_{x\in\Lambda} \Big\lbrace & \hskip -0.1in \displaystyle\sum_{pl} c_{0} \Ree \Tr \dfrac{1}{3} \left[1-P_{pl}(x)\right] \\[8pt]
&+ \displaystyle\sum_{rt}c_{1}\Ree \Tr \dfrac{1}{3} \left[1-R_{rt}(x)\right] \\[8pt]
&+ \displaystyle\sum_{pg}c_{2}\Ree \Tr \dfrac{1}{3} \left[1-G_{pg}(x)\right]\Big\rbrace\, .
\end{matrix*}
\end{equation}
In order to recover the original action, a relation between the coefficients $c_i$ has to be satisfied: $c_{0}+8c_{1}+8c_{2}=1$. By choosing specific values for these coefficients, as it is the case of the Iwasaki action~\cite{Itoh:1984ym}, where $c_0=1-8c_1$, $c_1=-0.331$, and $c_2=0$, the error can be reduced to $\mathcal{O}(b^4)$.

\section{Extracting observables}

With the action discretized according to the previous section, we can now rewrite Eq.~\eqref{eq:pathintMink} as
\begin{equation}
    \langle \hat{O}\rangle = \frac{1}{Z}\int \mathcal{D}\psi \mathcal{D}\bar{\psi} \mathcal{D}U_{\mu}\, \hat{O}(\psi,\bar{\psi},U) \, e^{- S^{LW}(U)-S^{SW}(\psi,\bar{\psi},U)}\, ,
    \label{eq:pathintFull}
\end{equation}
where we have split the action into the gluonic and the quark parts. Since quarks are anticommuting variables (they are fermions), they are described by Grassmann numbers. As such, and given the form of the quark action $S^{SW}=\sum \bar{\psi}(x)M(x,y)\psi(y)$, one can perform the integral over $\psi$ and $\bar{\psi}$ analytically, and Eq.~\eqref{eq:pathintFull} is now written as
\begin{equation}
    \langle \hat{O}\rangle = \frac{1}{Z}\int \mathcal{D}U_{\mu}\, \hat{O}[M^{-1}(U),U] \det\left[M(U)\right]\, e^{- S^{LW}(U)}\, .
\label{eq:pathintDiscrt}
\end{equation}
All the dependence on $\psi$ and $\bar{\psi}$ has disappeared: the quark action in the exponential gives rise to the determinant of the Dirac operator (similar for $Z$), and the fields in the operator $\hat{O}$ have been contracted via the Wick theorem~\cite{Wick:1950ee} to quark propagators ($M^{-1}(U)$) that only depend on the link fields.

As mentioned before, the only possible way to compute this integral is via Monte Carlo methods. Summarized below are the steps in a typical LQCD calculation:
\begin{enumerate}[label=(\roman*)]
    \item If we want to perform the integral in a stochastic manner, we have to generate a set of gauge field configurations $\{U\}$ sampled from the distribution function $\frac{1}{Z} \det\left[M(U)\right]\, e^{- S^{LW}(U)}$. This is usually done via Markov chain Monte Carlo algorithms, such as the hybrid Monte Carlo algorithm~\cite{Duane:1987de,Gottlieb:1987mq}, but new ideas using machine-learning based methods are starting to appear that do not suffer from critical slowing down~\cite{Albergo:2019eim,Kanwar:2020xzo,Boyda:2020hsi}.
    \item Most of the observables require the computation of quark propagators (except the purely-gluoinc ones, like the study of the spectrum of glueballs). This is done by solving the following linear equation (where we have made all the indices explicit),
    \begin{equation}
        \sum_{y,b,\beta} M^{ab}_{\alpha\beta}(x,y) S^{bc}_{\beta\gamma}(y,z) = \phi^{ac}_{\alpha\gamma}(x,z)\, ,
    \label{eq:propagator_equation}
    \end{equation}
    where $S$ is the propagator, $M$ is the Dirac operator and $\phi$ is the source (usually a point-source written as a delta in position, color and spin-space). Due to the sparsity of the Dirac operator (there are only nearest-neighbour interactions), Krylov subspace solvers, such as the BiConjugate Gradient Stabilized (Bi-CGStab)~\cite{Vorst:1992} or the Generalized Minimum Residual Method (GMRES)~\cite{Saad:1986}, are used. The convergence of these methods can be related to the condition number of the Dirac matrix, which is the ratio between the largest and the smallest eigenvalue. Since the smallest eigenvalue is proportional to the mass of the lightest quark, as we approach the physical point the condition number increases rapidly, entering a region where these types of solvers are known for critical slowing down. In order to reduce the value of the condition number and increase the speed of convergence, preconditioners are used~\cite{Gattringer:2010zz,Knechtli:2017sna}. Examples are the even-odd preconditioning~\cite{DeGrand:1988vx}, domain decomposition~\cite{Luscher:2003qa}, and the multigrid method~\cite{Babich:2010qb,Frommer:2013fsa,Alexandrou:2016izb,Brower:2018ymy}, which is the most widely used in current LQCD calculations at (or near) the physical point.
    \item Once we have generated enough gauge-field configurations $\{U\}$, we can approximate the integral in Eq.~\eqref{eq:pathintDiscrt} by the mean value of the operator $\hat{O}$ over the set of configurations,
    \begin{equation}
        \langle \hat{O}\rangle \approx \frac{1}{N_{\text{cfg}}}\sum_{n=1}^{N_{\text{cfg}}} \hat{O}(\{U\}_n)\, .
    \end{equation}
    Since we have a finite number of $N_{\text{cfg}}$, an (statistical) uncertainty has to be assigned to the result. Other sources of uncertainty can be cast into the systematics, and can come from the extrapolation to $b\rightarrow 0$ and $L\rightarrow \infty$, or from the choice of fitting method used to extract $\hat{O}$.
    \item In order to compare the results of LQCD calculations to experimental values, we need to express them not in lattice units, but in physical units, for which we need to know the value of the lattice spacing, $b$. However, its value is not known a priori since the configurations in the first step are generated by fixing the gauge coupling $g$, and there is no analytical relation between the two (one can compute the running of $g$ with $b$, but only perturbatively). To compute $b$, a dimensionful quantity $X$ that is assumed to be insensitive to the quark masses is compared to its experimental value, so that $b=(b X_{\text{latt}})/X_{\text{exp}}$. The typical quantities used to determine the spacing are the mass of some heavy baryon (e.g., $\Omega$ or $\Upsilon$), the pseudoscalar decay constants (e.g., $f_{\pi}$ or $f_K$), or the Sommer scale $r_0$, related to the static quark potential (for a summary, see Ref.~\cite{Sommer:2014mea}).
\end{enumerate}
Several simplifications to these steps have been done in the past for computational purposes. The roughest one, known as quenching, consists in setting $\det\left[M(U)\right]=1$ during the gauge-field generation. Physically, this is equivalent to turning off the sea-quark effects. Halfway between the quenched and fully dynamical calculations are the partially quenched ones, where the masses of the sea quarks are different than the masses used for the valence quarks in the propagators. An analogous approach entails the use of a mixed action, where the action for the sea quarks is different than the one for valence quarks (e.g., in Ref.~\cite{Beane:2006gj}, staggered sea quarks and domain-wall valence quarks were used by the NPLQCD Collaboration to study $\pi K$ scattering).

\begin{figure}[hb]
\begin{equation*}
    \includegraphics[width=0.18\textwidth, valign=c]{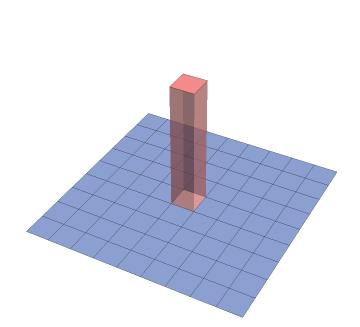} \overset{N=1}{\longrightarrow}
    \includegraphics[width=0.18\textwidth, valign=c]{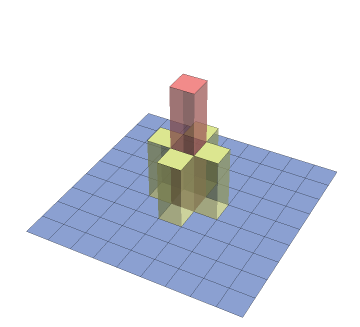} \overset{N=2}{\longrightarrow}
    \includegraphics[width=0.18\textwidth, valign=c]{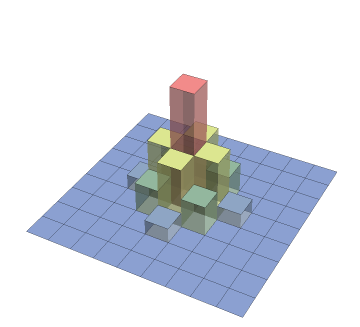} \cdots\; \overset{N}{\longrightarrow}
    \includegraphics[width=0.18\textwidth, valign=c]{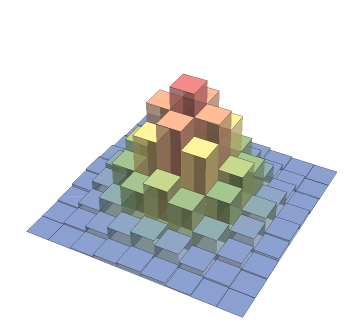}
\end{equation*}
\caption{Qualitative description of the iterative procedure to smear the source (or sink) in a two-dimensional lattice, with the height of the bars representing the relative weight. Starting with a delta function, as $N$ increases, the shape gets closer to a Gaussian.}
\label{fig:smearing}
\end{figure}

Besides the techniques described above, other improvements are available, like the tadpole improvement~\cite{Lepage:1992xa} or the APE~\cite{Albanese:1987ds}, HYP~\cite{Hasenfratz:2001hp}, and stout~\cite{Morningstar:2003gk} smearings of the gauge links. Another important improvement consists in using smeared quark sources and/or sinks. Since we know that hadrons are not point-like, to construct better operators with larger overlap to the ground-state and reduce contamination from excited states, smearing profiles can be applied to the quark source in Eq.~\eqref{eq:propagator_equation}. Among the several choices proposed, the Gaussian smearing~\cite{Gusken:1989qx}, which is the one used in~\cref{sec:450results}, takes the form
\begin{equation}
    \tilde{\phi}(x,z)=\sum_{y}[\delta_{x,y}+\alpha H(x,y)]\phi(y,z)\, ,
\label{eq:smearing}
\end{equation}
where $H(x,y)=\sum_{\mu=1}^3 U_\mu(x)\delta_{y,x+\hat{\mu}}+U^{\Dag}_\mu(x-\hat{\mu})\delta_{y,x-\hat{\mu}}$ is the hopping term, and $\alpha$ is a constant. If this procedure is repeated $N$ times, then the shape becomes closer to a Gaussian, with $\{\alpha , N\}$ being the parameters that determine the shape and size of the source. To understand how this procedure works, in Fig.~\ref{fig:smearing} we show the weight that each point (in a two-dimensional lattice) acquires after each iteration of Eq.~\eqref{eq:smearing}.

Another difficulty that is faced when trying to compute the properties of hadrons is the inclusion of electromagnetic (EM) interactions. That is because Gauss's law is not satisfied in a finite volume with PBC.
There are several proposals to circumvent this problem (see Ref.~\cite{Davoudi:2018qpl} for a comprehensive discussion). For the studies of nuclear systems ($A\geq 2$), the most common technique is the use of uniform EM background fields, where the EM fields are added to the gauge-field ensembles after its generation. For example, this technique was used by the NPLQCD Collaboration to study the magnetic moments~\cite{Beane:2014ora} and polarizabilities~\cite{Chang:2015qxa} of light nuclei, as well as the study of the first nuclear reaction cross section with direct input from LQCD, the radiative-capture process $np\rightarrow \gamma d$~\cite{Beane:2015yha}.
Very recently, a dynamical QCD$\,+\,$QED calculation studied two- and three-baryons systems (besides multi-pion and kaon systems)~\cite{Beane:2020ycc}, where instead of using EM background fields, the spatial zero mode of the photon is removed on every time slice.

The most important objects calculated on the lattice are two- and three-point correlation functions. The first ones are used for extracting the energy levels of the system, while the second ones to extract matrix elements. In the following subsections, we will discuss how to relate physical quantities with lattice objects.

\subsection{Two-point correlation functions}\label{subsec:2ptcorr}

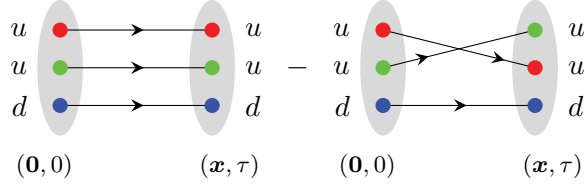
\begin{figure}[H]
\begin{equation*}
\begin{tikzpicture}[baseline={([yshift=1.5ex]current bounding box.center)}]
 \coordinate (A) at (1,1.8);
 \node at (A) [left = 3mm of A] {$u$};
 \coordinate (B) at (3,1.8);
 \node at (B) [right = 3mm of B] {$u$};
 \coordinate (C) at (1,1.3);
 \node at (C) [left = 3mm of C] {$u$};
 \coordinate (D) at (3,1.3);
 \node at (D) [right = 3mm of D] {$u$};
  \coordinate (H) at (1,0.8);
 \node at (H) [left = 3mm of H] {$d$};
 \coordinate (I) at (3,0.8);
 \node at (I) [right = 3mm of I] {$d$};
 
 \fill[gray!30] (1,1.3) ellipse (0.3 and 0.9);
 \fill[gray!30] (3,1.3) ellipse (0.3 and 0.9);

 \coordinate (F) at (1,0);
 \node at (F) [left = -3mm of F] {\footnotesize $(\bm{0},0)$};
 \coordinate (G) at (3,0);
 \node at (G) [right = -3mm of G] {\footnotesize $(\bm{x},\tau)$};

 \draw[particle] (A) -- (B);
 \draw[particle] (C) -- (D);
 \draw[particle] (H) -- (I);

 \fill[myred](A) circle (0.1);
 \fill[mygreen](C) circle (0.1);
 \fill[myblue](H) circle (0.1);
 \fill[myred](B) circle (0.1);
 \fill[mygreen](D) circle (0.1);
 \fill[myblue](I) circle (0.1);
\end{tikzpicture}\;-\;\begin{tikzpicture}[baseline={([yshift=1.5ex]current bounding box.center)}]
 \coordinate (A) at (1,1.8);
 \node at (A) [left = 3mm of A] {$u$};
 \coordinate (B) at (3,1.8);
 \node at (B) [right = 3mm of B] {$u$};
 \coordinate (C) at (1,1.3);
 \node at (C) [left = 3mm of C] {$u$};
 \coordinate (D) at (3,1.3);
 \node at (D) [right = 3mm of D] {$u$};
 \coordinate (H) at (1,0.8);
 \node at (H) [left = 3mm of H] {$d$};
 \coordinate (I) at (3,0.8);
 \node at (I) [right = 3mm of I] {$d$};
 
 \fill[gray!30] (1,1.3) ellipse (0.3 and 0.9);
 \fill[gray!30] (3,1.3) ellipse (0.3 and 0.9);

 \coordinate (F) at (1,0);
 \node at (F) [left = -3mm of F] {\footnotesize $(\bm{0},0)$};
 \coordinate (G) at (3,0);
 \node at (G) [right = -3mm of G] {\footnotesize $(\bm{x},\tau)$};

 \draw[particlecross1] (A) -- (D);
 \draw[particlecross2] (C) -- (B);
 \draw[particle] (H) -- (I);

 \fill[myred](A) circle (0.1);
 \fill[mygreen](C) circle (0.1);
 \fill[myblue](H) circle (0.1);
 \fill[mygreen](B) circle (0.1);
 \fill[myred](D) circle (0.1);
 \fill[myblue](I) circle (0.1);
\end{tikzpicture}
\end{equation*}
\vspace*{-5mm}
\caption{Wick contractions for the two-point correlation function of the proton.}
\label{fig:protoncontract}
\end{figure}

The two-point correlation functions are objects that give us the amplitude for the time evolution of a state, from its creation at a point in the lattice (source) to its annihilation at another point in the lattice (sink). Appropriate interpolation operators are used to: i) create a state out of the vacuum with specific quantum numbers (flavor, spin, parity, charge conjugation,...) so that it couples to the desired state ($\mathcal{X}_A$), ii) annihilate at the sink ($\mathcal{X}_B$). This object can be written as
\begin{equation}
    C_{2pt}(\tau,\bm{p})=\sum_{\bm{x}} e^{-\imag \bm{x}\cdot\bm{p}}\langle \mathcal{X}_B(\bm{x},\tau)\bar{\mathcal{X}}_A(\bm{0},0)\rangle\, ,
\end{equation}
where the sum over the position $\bm{x}$ projects the state to a definite momentum $\bm{p}$. 
There is a vast bibliography on operator construction~\cite{Basak:2005ir,Basak:2005aq,Peardon:2009gh,Dudek:2010wm,Morningstar:2011ka} to study both ground- and excited-states for mesons and baryons.
In order to understand how these functions are computed, we can use local operators of baryons (all the quarks are placed at the same point), which are the ones used to study the ground-state energy of systems (like in~\cref{sec:450results}). The simplest operator one can use for the proton is
\begin{equation}
    N(x)=\epsilon_{abc}\left[u^a(x)C\gamma_5 d^b(x)\right]u^c(x)\, ,
\label{eq:protOperator}
\end{equation}
where $\{a,b,c\}$ are color indices and $C=\gamma_2\gamma_4$ is the charge-conjugation matrix in Euclidean space-time, which together with $\gamma_5$ produces a spin-zero diquark object. The projected correlation function is then
expressed as (again, making all the spinor indices explicit)
\begin{equation}
\begin{aligned}
    \Gamma_{\beta\alpha}\langle & N_{\alpha}(\bm{x},\tau)\bar{N}_{\beta}(\bm{0},0)\rangle \\
    =& \epsilon_{abc}\epsilon_{def} \Gamma_{\beta\alpha}\langle \left\lbrace u_{\gamma}^a(\bm{x},\tau)[C\gamma_5]_{\gamma\delta} d^b_{\delta}(\bm{x},\tau)\right\rbrace u^c_{\alpha}(\bm{x},\tau) \bar{u}^d_{\beta}(\bm{0},0) \left\lbrace\bar{d}_{\eta}^e(\bm{0},0)[C\gamma_5]_{\eta\zeta} \bar{u}^f_{\zeta}(\bm{0},0)\right\rbrace\rangle \\
    =& \contraction[4ex]{\epsilon_{abc}\epsilon_{def} \Gamma_{\beta\alpha} [C\gamma_5]_{\gamma\delta} [C\gamma_5]_{\eta\zeta}\langle}{u}{{}_{\gamma}^a(\bm{x},\tau) d^b_{\delta}(\bm{x},\tau) u^c_{\alpha}(\bm{x},\tau) \bar{u}^d_{\beta}(\bm{0},0) \bar{d}_{\eta}^e(\bm{0},0)}{\bar{u}} \contraction[2.5ex]{\epsilon_{abc}\epsilon_{def} \Gamma_{\beta\alpha} [C\gamma_5]_{\gamma\delta} [C\gamma_5]_{\eta\zeta}\langle u_{\gamma}^a(\bm{x},\tau)}{d}{{}^b_{\delta}(\bm{x},\tau) u^c_{\alpha}(\bm{x},\tau) \bar{u}^d_{\beta}(\bm{0},0)}{\bar{d}} \contraction[2ex]{\epsilon_{abc}\epsilon_{def} \Gamma_{\beta\alpha} [C\gamma_5]_{\gamma\delta} [C\gamma_5]_{\eta\zeta}\langle u_{\gamma}^a(\bm{x},\tau) d^b_{\delta}(\bm{x},\tau)}{u}{{}^c_{\alpha}(\bm{x},\tau)}{\bar{u}} \bcontraction[4ex]{\epsilon_{abc}\epsilon_{def} \Gamma_{\beta\alpha} [C\gamma_5]_{\gamma\delta} [C\gamma_5]_{\eta\zeta}\langle}{u}{{}_{\gamma}^a(\bm{x},\tau) d^b_{\delta}(\bm{x},\tau) u^c_{\alpha}(\bm{x},\tau)}{\bar{u}} \bcontraction[3ex]{\epsilon_{abc}\epsilon_{def} \Gamma_{\beta\alpha} [C\gamma_5]_{\gamma\delta} [C\gamma_5]_{\eta\zeta}\langle u_{\gamma}^a(\bm{x},\tau)}{d}{{}^b_{\delta}(\bm{x},\tau) u^c_{\alpha}(\bm{x},\tau) \bar{u}^d_{\beta}(\bm{0},0)}{\bar{d}} \bcontraction[2ex]{\epsilon_{abc}\epsilon_{def} \Gamma_{\beta\alpha} [C\gamma_5]_{\gamma\delta} [C\gamma_5]_{\eta\zeta}\langle u_{\gamma}^a(\bm{x},\tau) d^b_{\delta}(\bm{x},\tau)}{u}{{}^c_{\alpha}(\bm{x},\tau)\bar{u}^d_{\beta}(\bm{0},0) \bar{d}_{\eta}^e(\bm{0},0)}{\bar{u}} \epsilon_{abc}\epsilon_{def} \Gamma_{\beta\alpha} [C\gamma_5]_{\gamma\delta} [C\gamma_5]_{\eta\zeta}\langle u_{\gamma}^a(\bm{x},\tau) d^b_{\delta}(\bm{x},\tau) u^c_{\alpha}(\bm{x},\tau) \bar{u}^d_{\beta}(\bm{0},0) \bar{d}_{\eta}^e(\bm{0},0) \bar{u}^f_{\zeta}(\bm{0},0)\rangle\\
    =&\epsilon_{abc}\epsilon_{def} \Gamma_{\beta\alpha} [C\gamma_5]_{\gamma\delta} [C\gamma_5]_{\eta\zeta}\left[ S^{af}_{u,\gamma\zeta}(\bm{x},\tau;\bm{0},0)S^{be}_{d,\delta\eta}(\bm{x},\tau;\bm{0},0)S^{cd}_{u,\alpha\beta}(\bm{x},\tau;\bm{0},0)\right.\\
    &\qquad\qquad\qquad\qquad\qquad\qquad\qquad\left.-S^{ad}_{u,\gamma\beta}(\bm{x},\tau;\bm{0},0)S^{be}_{d,\delta\eta}(\bm{x},\tau;\bm{0},0)S^{cf}_{u,\alpha\zeta}(\bm{x},\tau;\bm{0},0) \right]\, .
\label{eq:protcontract}
\end{aligned}
\end{equation}
In the previous expression we have introduced a projector $\Gamma$, which can take two forms,
\begin{equation}
    \Gamma_{\text{unpol}}=\frac{1}{2}(1+\gamma_4)\, , \quad \Gamma_{\text{pol}}=\frac{1}{4}(1+\gamma_4)(1+\imag \gamma_5(\bm{\gamma}\cdot\bm{s})\gamma_4)\, ,
\label{eq:projector_parityspin}
\end{equation}
where $\Gamma_{\text{unpol}}$ only projects to positive-parity states, while $\Gamma_{\text{pol}}$ additionally projects the spin of the state to the polarization direction $\bm{s}$, which is usually taken in the $z$-direction. The allowed Wick contractions (only two possibilities for the proton) are shown in Eq.~\eqref{eq:protcontract}, which are shown schematically in Fig.~\ref{fig:protoncontract}.
As the number of particles increases, the number of contractions grows. Naively, this number grows factorially with the number of quarks, but it can be reduced by using symmetries to remove duplicate and vanishing contributions. These techniques have been applied to multi-meson~\cite{Beane:2007es,Detmold:2008fn,Detmold:2008yn,Detmold:2010au,Detmold:2012wc} (up to 72 pions) and multi-baryon~\cite{Beane:2009gs,Yamazaki:2009ua,Doi:2012xd,Beane:2012vq,Detmold:2012eu} (up to $A=28$ nucleons) systems.

In order to relate $C_{2pt}(\tau,\bm{p})$ with the energy levels of the state, we need to write its spectral decomposition (we will drop the dependence on $\bm{p}$ and $\bm{x}$ for the moment). Using the Hamiltonian evolution operator,
\begin{equation}
    \langle \mathcal{X}_B(\tau)\bar{\mathcal{X}}_A(0)\rangle = \frac{1}{Z_T}\Tr\left[e^{-(T-\tau)\hat{H}} \mathcal{X}_B(0) e^{-\tau\hat{H}} \bar{\mathcal{X}}_A(0) \right]\, ,
    \label{eq:hamiltonianevolution}
\end{equation}
with the normalization factor $Z_T=\Tr e^{-T\hat{H}}$ and $T$ being the temporal extent of the lattice. The trace can be evaluated by inserting a complete set of states (with the normalization $\langle \mathfrak{n} | \mathfrak{n} \rangle = 1$),
\begin{equation}
\begin{aligned}
    \Tr\left[e^{-(T-\tau)\hat{H}} \mathcal{X}_B(0) e^{-\tau\hat{H}} \bar{\mathcal{X}}_A(0) \right] &= \sum_{\mathfrak{n},\mathfrak{m}} \langle \mathfrak{m} | e^{-(T-\tau)\hat{H}} \mathcal{X}_B(0)| \mathfrak{n} \rangle \langle \mathfrak{n} | e^{-\tau\hat{H}} \bar{\mathcal{X}}_A(0) | \mathfrak{m} \rangle \\
    &=\sum_{\mathfrak{n},\mathfrak{m}} e^{-(T-\tau)E_\mathfrak{m}} \langle \mathfrak{m} | \mathcal{X}_B(0)| \mathfrak{n} \rangle \, e^{-\tau E_\mathfrak{n}} \langle \mathfrak{n} | \bar{\mathcal{X}}_A(0) | \mathfrak{m} \rangle \, ,
\label{eq:2pcorrfun1}
\end{aligned}
\end{equation}
and similarly for $Z_T=\Tr e^{-T\hat{H}} = \sum_{\mathfrak{n}}\langle \mathfrak{n} |e^{-T\hat{H}}| \mathfrak{n} \rangle = \sum_{\mathfrak{n}} e^{-TE_\mathfrak{n}}$. The case $\mathfrak{n}=0$ represents the vacuum, which we take as a reference value. In particular, we choose $E_0=0$ and $\langle 0 | O | 0 \rangle=0$. Note that in the limit $T\rightarrow \infty$ the denominator in Eq.~\eqref{eq:hamiltonianevolution} tends to $1$. We can consider two limits for the sum in Eq.~\eqref{eq:2pcorrfun1}: one for $\tau$ fixed and $T\rightarrow \infty$ (forward propagation) and another for $T-\tau$ fixed and $T\rightarrow \infty$ (backward propagation),
\begin{equation}
    \sum_{\mathfrak{n}} \langle 0 | \mathcal{X}_B(0)| \mathfrak{n} \rangle \langle \mathfrak{n} | \bar{\mathcal{X}}_A(0) |0\rangle \, e^{-\tau E_\mathfrak{n}} + \sum_{\mathfrak{m}} \langle \mathfrak{m} | \mathcal{X}_B(0)| 0 \rangle \langle 0 | \bar{\mathcal{X}}_A(0) | \mathfrak{m} \rangle \, e^{-(T-\tau)E_\mathfrak{m}} + \cdots \, ,
\end{equation}
where the dots denote higher excited states contributions. It can be showed~\cite{Fucito:1982ip,Gattringer:2010zz} that the backward propagating state is the charge-conjugated version of the forward propagating state. For mesons, these are the same states, and taking the pion as an example, one can write
\begin{equation}
    C^\pi_{2pt}(\tau) = |\langle 0 | \mathcal{X}_\pi |\pi \rangle|^2 \left[ e^{-\tau m_{\pi}} + e^{-(T-\tau)m_{\pi}} \right] + \cdots\, .
\end{equation}
For baryons, the charge-conjugated state has opposite parity, giving for the case of the nucleon (with $N^+$ having positive parity and $N^-$ negative parity)
\begin{equation}
\begin{aligned}
    C^N_{2pt}(\tau) ={} & (1+\gamma_4)\left[A_+e^{-\tau M_{N^+}}+A_-e^{-(T-\tau)M_{N^-}}\right]\\
    &-(1-\gamma_4)\left[A_-e^{-\tau M_{N^-}}+A_+e^{-(T-\tau)M_{N^+}}\right]+\cdots\, ,
\end{aligned}
\end{equation}
where the constant $A_{\pm}$ is proportional to $|\langle 0 | O |N^{\pm} \rangle |^2$. It is easy to see that projecting with $\Gamma_{\text{unpol}}$, which picks only the positive forward propagating state, we can remove the contamination coming from the second line. The contamination from backward-propagating states is not relevant for baryonic systems, with negative parity state masses much heavier than their positive counterparts (in nature, $N^+\sim 939$ MeV while $N^-\sim 1535$ MeV), and where the analysis of correlation functions involves only $\tau < T/2$.

\begin{figure}[t]
\centering
\includegraphics[width=\textwidth]{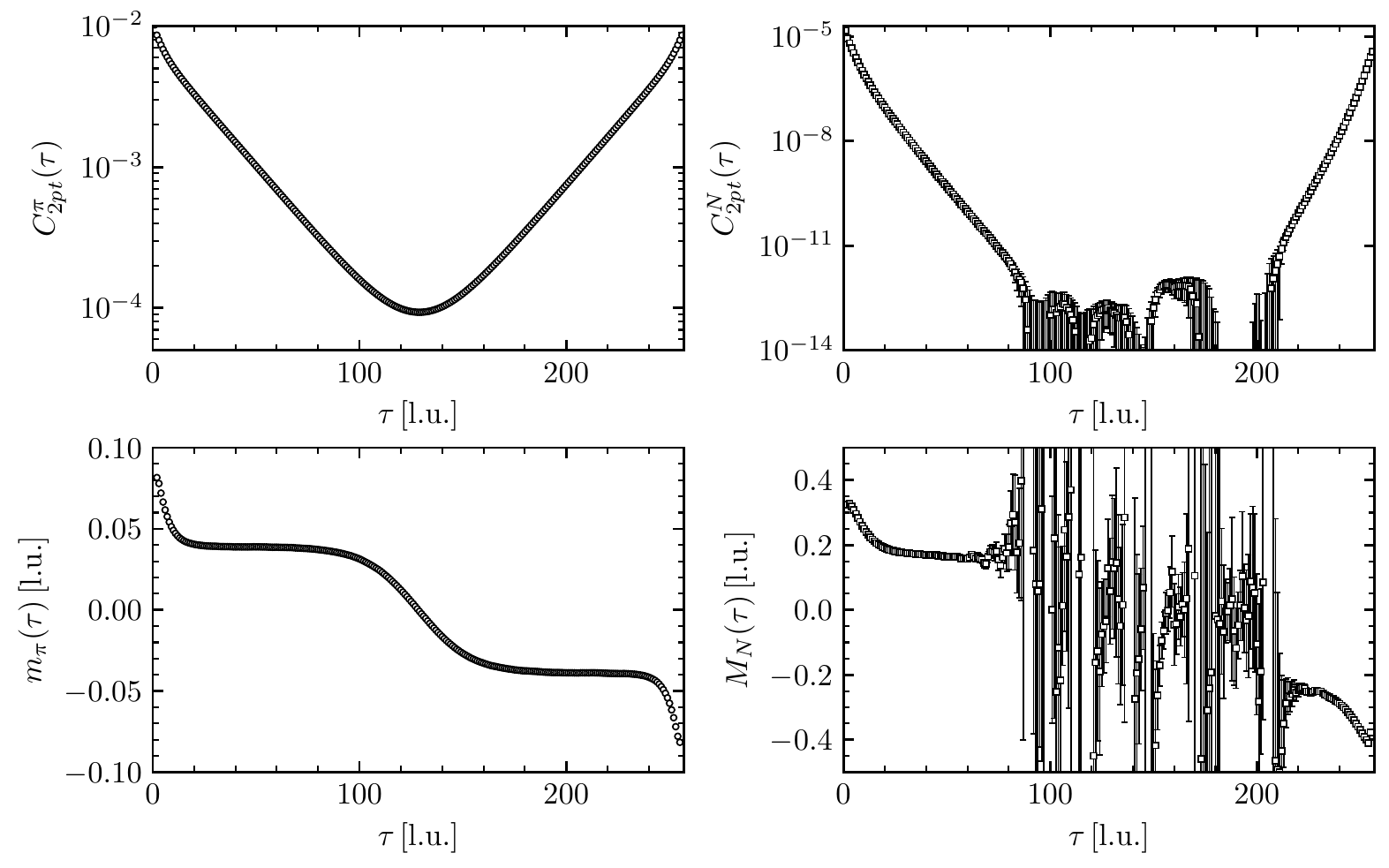}
\caption{Top row: two-point correlation functions for the pion (left) and proton (right). Bottom row: effective mass plots for the pion (left) and proton (right).}
\label{fig:corr_emp}
\end{figure}

The top-left panel of Fig.~\ref{fig:corr_emp} shows the correlation function of the pion, where the forward and backward propagating states have the same mass, giving rise to a symmetric function, with the same slope on both sides with respect to the center. The top-right panel shows the correlation function of the nucleon, where the backward state, associated to the opposite parity nucleon state, shows a steeper slope.

An alternative way to visualize the correlation functions is via the effective mass plot (EMP), which is defined as
\begin{equation}
    M(\tau) = \frac{1}{\tau_J} \ln \left[ \frac{C_{2pt}(\tau)}{C_{2pt}(\tau+\tau_J)} \right]\, ,
\label{eq:EMP}
\end{equation}
where $\tau_J$ is a non-zero integer that is introduced to stabilize the function (usual values for this parameter are $1\leq \tau_J \leq 5$~\cite{Beane:2009kya}). In the limit $\tau\rightarrow\infty$, this function plateaus to the ground state energy of the system, making it easier to read it off. In the bottom panels of Fig.~\ref{fig:corr_emp}, the effective mass plots for the pion (left) and nucleon (right) are shown.\footnote{For mesons, since both forward- and backward-propagating states have the same energy, more appropriate forms of the effective mass functions are available, as for example
\begin{equation}
    M(\tau)=\frac{1}{\tau_J}\arccosh\left[\frac{C_{2pt}(\tau+\tau_J)+C_{2pt}(\tau-\tau_J)}{2C_{2pt}(\tau)}\right]\, ,
\end{equation}
where now the change in sign at $\tau=T/2$ shown in the bottom-left panel of Fig.~\ref{fig:corr_emp} disappears.}

For multi-hadron systems, it is interesting to form the ratio of the correlation function describing the multi-hadron system with respect to the product of correlation functions of its constituents. For the case of two baryons, $B_1$ and $B_2$, this ratio reads
\begin{equation}
    R_{B_1B_2}(\tau)=\frac{C_{B_1B_2}(\tau)}{C_{B_1}(\tau)C_{B_2}(\tau)}\, .
\label{eq:energyshiftfun}
\end{equation}
We can construct an equivalent of the EMP for multi-baryons, called an effective energy-shift function,
\begin{equation}
    \Delta E_{B_1B_2}(\tau)=\frac{1}{\tau_J} \ln \left[ \frac{R_{B_1B_2}(\tau)}{R_{B_1B_2}(\tau+\tau_J)} \right]\, ,
\label{eq:effenergyshiftfun}
\end{equation}
which in the limit $\tau\rightarrow\infty$ it plateaus to $\Delta E_{B_1B_2} =E_{B_1B_2}-M_{B_1}-M_{B_2}$.

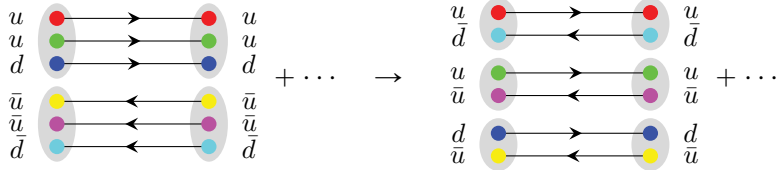
\begin{figure}[H]
\begin{equation*}
\begin{tikzpicture}[baseline={([yshift=0ex]current bounding box.center)}]
\coordinate (A) at (1,1.6);
 \node at (A) [left = 3mm of A] {\small $u$};
\coordinate (B) at (3,1.6);
 \node at (B) [right = 3mm of B] {\small $u$};
 \coordinate (C) at (1,1.3);
 \node at (C) [left = 3mm of C] {\small $u$};
 \coordinate (D) at (3,1.3);
 \node at (D) [right = 3mm of D] {\small $u$};
  \coordinate (H) at (1,1.0);
 \node at (H) [left = 3mm of H] {\small $d$};
 \coordinate (I) at (3,1.0);
 \node at (I) [right = 3mm of I] {\small $d$};
 
 \coordinate (J) at (1,0.5);
 \node at (J) [left = 3mm of J] {\small $\bar{u}$};
\coordinate (K) at (3,0.5);
 \node at (K) [right = 3mm of K] {\small $\bar{u}$};
 \coordinate (L) at (1,0.2);
 \node at (L) [left = 3mm of L] {\small $\bar{u}$};
 \coordinate (M) at (3,0.2);
 \node at (M) [right = 3mm of M] {\small $\bar{u}$};
  \coordinate (N) at (1,-0.1);
 \node at (N) [left = 3mm of N] {\small $\bar{d}$};
 \coordinate (O) at (3,-0.1);
 \node at (O) [right = 3mm of O] {\small $\bar{d}$};
 
 \fill[gray!30] (1,1.3) ellipse (0.25 and 0.5);
 \fill[gray!30] (3,1.3) ellipse (0.25 and 0.5);
 \fill[gray!30] (1,0.2) ellipse (0.25 and 0.5);
 \fill[gray!30] (3,0.2) ellipse (0.25 and 0.5);

 \draw[particle] (A) -- (B);
 \draw[particle] (C) -- (D);
 \draw[particle] (H) -- (I);
 \draw[particle] (K) -- (J);
 \draw[particle] (M) -- (L);
 \draw[particle] (O) -- (N);

 \fill[myred](A) circle (0.1);
 \fill[mygreen](C) circle (0.1);
 \fill[myblue](H) circle (0.1);
 \fill[myred](B) circle (0.1);
 \fill[mygreen](D) circle (0.1);
 \fill[myblue](I) circle (0.1);
 \fill[myablue](J) circle (0.1);
 \fill[myagreen](L) circle (0.1);
 \fill[myared](N) circle (0.1);
 \fill[myablue](K) circle (0.1);
 \fill[myagreen](M) circle (0.1);
 \fill[myared](O) circle (0.1);
\end{tikzpicture}
+\cdots \quad \rightarrow \quad \begin{tikzpicture}[baseline={([yshift=0ex]current bounding box.center)}]
\coordinate (A) at (1,1.7);
 \node at (A) [left = 3mm of A] {\small $u$};
\coordinate (B) at (3,1.7);
 \node at (B) [right = 3mm of B] {\small $u$};
 \coordinate (C) at (1,1.4);
 \node at (C) [left = 3mm of C] {\small $\bar{d}$};
 \coordinate (D) at (3,1.4);
 \node at (D) [right = 3mm of D] {\small $\bar{d}$};
 
  \coordinate (H) at (1,0.9);
 \node at (H) [left = 3mm of H] {\small $u$};
 \coordinate (I) at (3,0.9);
 \node at (I) [right = 3mm of I] {\small $u$};
 \coordinate (J) at (1,0.6);
 \node at (J) [left = 3mm of J] {\small $\bar{u}$};
\coordinate (K) at (3,0.6);
 \node at (K) [right = 3mm of K] {\small $\bar{u}$};
 
 \coordinate (L) at (1,0.1);
 \node at (L) [left = 3mm of L] {\small $d$};
 \coordinate (M) at (3,0.1);
 \node at (M) [right = 3mm of M] {\small $d$};
  \coordinate (N) at (1,-0.2);
 \node at (N) [left = 3mm of N] {\small $\bar{u}$};
 \coordinate (O) at (3,-0.2);
 \node at (O) [right = 3mm of O] {\small $\bar{u}$};
 
 \fill[gray!30] (1,1.55) ellipse (0.25 and 0.35);
 \fill[gray!30] (3,1.55) ellipse (0.25 and 0.35);
 \fill[gray!30] (1,0.75) ellipse (0.25 and 0.35);
 \fill[gray!30] (3,0.75) ellipse (0.25 and 0.35);
 \fill[gray!30] (1,-0.05) ellipse (0.25 and 0.35);
 \fill[gray!30] (3,-0.05) ellipse (0.25 and 0.35);

 \draw[particle] (A) -- (B);
 \draw[particle] (D) -- (C);
 \draw[particle] (H) -- (I);
 \draw[particle] (K) -- (J);
 \draw[particle] (L) -- (M);
 \draw[particle] (O) -- (N);

 \fill[myred](A) circle (0.1);
 \fill[myared](C) circle (0.1);
 \fill[myred](B) circle (0.1);
 \fill[myared](D) circle (0.1);
 
 \fill[mygreen](H) circle (0.1);
  \fill[myagreen](J) circle (0.1);
 \fill[mygreen](I) circle (0.1);
  \fill[myagreen](K) circle (0.1);
  
 \fill[myblue](L) circle (0.1);
 \fill[myablue](N) circle (0.1);
 \fill[myblue](M) circle (0.1);
 \fill[myablue](O) circle (0.1);
\end{tikzpicture}+\cdots
\end{equation*}
\vspace*{-5mm}
\caption{Wick contractions and quark re-organization for the quantity $\langle \mathcal{O}^2\rangle$ in the case of the proton correlation function. The dots denote additional contractions.}
\label{fig:protoncontract_2}
\end{figure}

From Fig.~\ref{fig:corr_emp}, we notice the different statistical behavior between meson and baryon correlation functions. This was first highlighted by G.~Parisi~\cite{Parisi:1983ae} and G.~P.~Lepage~\cite{Lepage:1989hd}, later on studied in detail for light-nuclei by the NPLQCD Collaboration~\cite{Beane:2009kya,Beane:2009gs,Beane:2009py,Beane:2010em} and also by M.~L.~Wagman and M.~J.~Savage~\cite{Wagman:2016bam,Wagman:2017xfh,Wagman:2017gqi}, motivated by previous works on the statistical properties of correlation functions~\cite{Endres:2011jm,Endres:2011er,Lee:2011sm,Endres:2011mm,DeGrand:2012ik,Grabowska:2012ik,Nicholson:2012xt,Beane:2014oea}. 
To understand this different behavior, we have to focus on the variance of the correlation function, which for an operator $\mathcal{O}$ is defined as $\text{Var}(\mathcal{O})=\langle \mathcal{O}^2\rangle - \langle \mathcal{O}\rangle^2$. As we have done in Fig.~\ref{fig:protoncontract}, where we have shown schematically the contractions leading to $\langle \mathcal{O}\rangle$ in the case of the proton, in Fig.~\ref{fig:protoncontract_2} we show the corresponding contractions corresponding to $\langle \mathcal{O}^2\rangle$.
As can be seen from Fig.~\ref{fig:protoncontract_2}, the long-time behavior of $\langle \mathcal{O}^2\rangle$ will be dominated not by the propagator of a proton and anti-proton, but by the lighter three-pion state ($3m_{\pi}<2M_N$). Then, the ratio between the mean value and the square root of the variance, also known as the signal-to-noise ratio (StN), is
\begin{equation}
    \text{StN}=\frac{\langle \mathcal{O} \rangle}{\sqrt{\text{Var}(\mathcal{O})}} \; \propto \; \frac{e^{-\tau M_N}}{\sqrt{e^{-\tau 3 m_\pi}}} = e^{-\tau (M_N-\frac{3}{2}m_{\pi})}\, ,
\end{equation}
where the explicit exponential degradation with time of the signal is manifested. If we compute the same quantity for the pion, we see that both $\langle \mathcal{O}^2\rangle$ and $\langle \mathcal{O}\rangle^2$ are dominated by two pions and StN ratio becomes a constant (no degradation). A similar degradation appears for the isovector mesons (like the $\rho$), with a decay that goes like $e^{-\tau(m_\rho-m_\pi)}$, and for states in higher partial-waves~\cite{Luu:2011ep}. If we go beyond the two-flavor sector and look at baryons with non-zero strangeness, we see that the degradation is less severe. For example, with the physical values of the meson masses, the StN of the cascade baryon ($\Xi=\{ssu/ssd\}$) degrades with the difference of $M_\Xi-\frac{1}{2}(2m_K+m_\eta)\sim 550$ MeV, smaller than the difference for the proton, which is $M_N-\frac{3}{2}m_\pi\sim 740$ MeV (see Refs.~\cite{Beane:2009kya,Beane:2009gs,Beane:2009py} for a detailed investigation of the noise scaling is one-, two-, and three-baryon systems including strangeness).

\begin{figure}[t]
\input{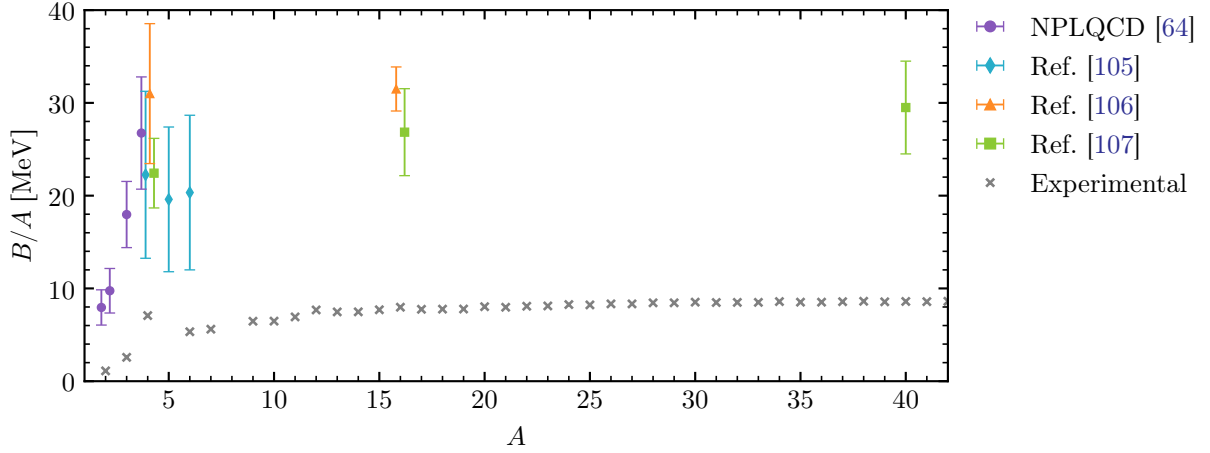}
\caption{Binding energy per nucleon $B/A$ as a function of the mass number $A$ from LQCD~\cite{Beane:2012vq} and many-body calculations~\cite{Barnea:2013uqa,Contessi:2017rww,Bansal:2017pwn} at $m_{\pi}\sim 806$ MeV, together with experimental values~\cite{Huang:2021nwk,Wang:2021xhn}.}
\label{fig:largenuclei}
\end{figure}

The situation worsens as we increase the number of baryons. In Ref.~\cite{Beane:2009kya} it was shown that the StN for a system with $A$ nucleons goes as $e^{-\tau \left[ A (M_N - \frac{3}{2} m_{\pi})\right]}$. This puts a limit on the atomic number of the nucleus that can be studied on the lattice for which a reasonable signal can be extracted.
The largest value has been computed by the NPLQCD Collaboration, with $A=4$ in Ref.~\cite{Beane:2012vq}, although at a heavier-than-physical pion mass, where the difference between $M_N$ and $m_\pi$ is smaller, reducing the degradation of the signal.
To reach larger systems, the energy levels of two- and three-body systems extracted directly from LQCD~\cite{Beane:2012vq,Yamazaki:2012hi} can be used to fix the coefficients of pionless EFT, and then compute the binding energy with many-body methods (the potentials extracted with LQCD have also been used to study larger nuclei in Refs.~\cite{Inoue:2014ipa,McIlroy:2017ssf}, however, the three-nucleon force was neglected in these studies). The use of this simpler EFT, compared to $\chi$EFT, is motivated by the unphysically large value of $m_{\pi}$ used in the lattice calculations, which allowed to consider pion exchange as a short-range effect and the nucleons as the only relevant degrees of freedom. This was first done in Ref.~\cite{Barnea:2013uqa}, where effective interaction hyperspherical harmonics and auxiliary-field diffusion Monte Carlo techniques were used to study ${}^{4}$He, ${}^{5}$He, ${}^{5}$Li, and ${}^{6}$Li nuclei, followed by Ref.~\cite{Contessi:2017rww} reaching ${}^{16}$O. The doubly magic isotopes ${}^{4}$He, ${}^{16}$O, and ${}^{40}$Ca were studied in Ref.~\cite{Bansal:2017pwn} using a discrete variable representation of the EFT in the harmonic oscillator basis. In order to compare all these results, the binding energy per nucleon is plotted in Fig.~\ref{fig:largenuclei} for the systems computed at $m_{\pi}\sim 806$ MeV together with the physical values of $B/A$ for the most common stable isotopes in nature~\cite{Huang:2021nwk,Wang:2021xhn}. It is interesting to note that a similar behavior is seen, where for small nuclei a rapid increase in $B/A$ is observed, while for larger nuclei this value seems to stabilize. 
An alternative method to study larger systems is via nuclear lattice EFT~\cite{Lahde:2019npb}, where $\chi$EFT is regularized on a lattice (with similar techniques used as in LQCD), and calculations of nuclear systems can be performed after the two- and three-body LECs are fitted to experimental data.

Let us go back to the spectral decomposition of the two-point correlation function, and look more closely to the proton (and without considering the backward propagating state),
\begin{equation}
    C_{2pt}(\tau,\bm{p})=\sum_{\bm{x}} e^{-\imag \bm{x}\cdot\bm{p}} \Gamma_{\beta\alpha} \langle 0| \mathcal{X}_\alpha (\bm{x},\tau) \bar{\mathcal{X}}_\beta(\bm{0},0)|0\rangle\, .
\end{equation}
Again, inserting a complete set of states $|\mathfrak{n},\bm{k},\bm{s}\rangle$ (where now in addition to $\mathfrak{n}$ we have included the momentum $\bm{k}$ and the spin $\bm{s}$ indices) and using the Hamiltonian and translation operators to shift the interpolating operator $\mathcal{X}_\alpha(\bm{x},\tau)$ to $\mathcal{X}_\alpha(\bm{0},0)$,
\begin{equation}
    C_{2pt}(\tau,\bm{p})=\sum_{\bm{x}} \sum_{\mathfrak{n},\bm{k},\bm{s}} e^{-\imag \bm{x}\cdot\bm{p}} e^{\imag \bm{x}\cdot\bm{k}} \Gamma_{\beta\alpha} \langle 0| \mathcal{X}_\alpha (\bm{0},0) |\mathfrak{n},\bm{k},\bm{s}\rangle \langle\mathfrak{n},\bm{k},\bm{s}|\bar{\mathcal{X}}_\beta(\bm{0},0) |0\rangle \, e^{-\tau E_{\mathfrak{n}}(\bm{k})}\, .
\end{equation}
Defining the overlap factor $Z_{\mathfrak{n}}$ as $\langle 0| \mathcal{X}_\alpha (\bm{0},0)|\mathfrak{n},\bm{k},\bm{s}\rangle = Z_{\mathfrak{n}}(\bm{k})U_\alpha(\bm{p},\bm{s})$, with $U_\alpha(\bm{p},\bm{s})$ being a spinor, we have
\begin{equation}
    C_{2pt}(\tau,\bm{p})=\sum_{\mathfrak{n},\bm{k},\bm{s}} Z_{\mathfrak{n}}(\bm{k})Z^*_{\mathfrak{n}}(\bm{k})\Gamma_{\beta\alpha}\sum_{\bm{x}}e^{-\imag \bm{x}\cdot(\bm{p}-\bm{k})} U_\alpha(\bm{k},\bm{s})\bar{U}_\beta(\bm{k},\bm{s}) e^{-\tau E_{\mathfrak{n}}(\bm{k})}\, .
\end{equation}
If we perform the sum over $\bm{x}$ and $\bm{k}$, we project all momenta to $\bm{p}$. Finally, by summing over spin,
\begin{equation}
    \sum_{\bm{s}} U_\alpha(\bm{p},\bm{s})\bar{U}_\beta(\bm{p},\bm{s})=[E(\bm{p})\gamma_4-\imag\bm{p}\cdot\bm{\gamma}+M_N]_{\alpha\beta}\, ,
\end{equation}
we obtain
\begin{equation}
\begin{aligned}
    C_{2pt}(\tau,\bm{p})&=\sum_{\mathfrak{n}} Z_{\mathfrak{n}}(\bm{p})Z^*_{\mathfrak{n}}(\bm{p})\Gamma_{\beta\alpha} [E_{\mathfrak{n}}(\bm{p})\gamma_4-\imag\bm{p}\cdot\bm{\gamma}+M_N]_{\alpha\beta}\, e^{-\tau E_{\mathfrak{n}}(\bm{p})}\\
    &=\sum_{\mathfrak{n}} Z_{\mathfrak{n}}(\bm{p})Z^*_{\mathfrak{n}}(\bm{p})\Tr\{\Gamma [E_{\mathfrak{n}}(\bm{p})\gamma_4-\imag\bm{p}\cdot\bm{\gamma}+M_N]\} e^{-\tau E_{\mathfrak{n}}(\bm{p})}\\
    &=\sum_{\mathfrak{n}} Z_{\mathfrak{n}}(\bm{p})Z^*_{\mathfrak{n}}(\bm{p}) \mathcal{F}_{\mathfrak{n}}(\Gamma) e^{-\tau E_{\mathfrak{n}}(\bm{p})}\, ,
\end{aligned}
\end{equation}
where the trace, $\mathcal{F}_{\mathfrak{n}}(\Gamma)$, can be evaluated using the gamma-matrix properties, leading to $\mathcal{F}_{\mathfrak{n}}(\Gamma_{\text{unpol}})=2[E_{\mathfrak{n}}(\bm{p})+M_N]$ for the case where $\Gamma$ only projects to positive parity, and $\mathcal{F}_{\mathfrak{n}}(\Gamma_{\text{pol}})=[E_{\mathfrak{n}}(\bm{p})+M_N]$ when, in addition, the spin projection is made. Nevertheless, these factors are not explicitly shown and are usually absorbed into the overlap factors. In the next chapter, we will discuss how to fit this two-point correlation function to extract the energy levels of the system.

\subsection{Three-point correlation functions}

\begin{figure}[ht]
\begin{equation*}
\begin{tikzpicture}[baseline={([yshift=1ex]current bounding box.center)}]
\coordinate (A) at (1,1.8);
 \node at (A) [left = 3mm of A] {$u$};
\coordinate (B) at (3,1.8);
 \node at (B) [right = 3mm of B] {$u$};
 \coordinate (C) at (1,1.3);
 \node at (C) [left = 3mm of C] {$u$};
 \coordinate (D) at (3,1.3);
 \node at (D) [right = 3mm of D] {$u$};
  \coordinate (H) at (1,0.8);
 \node at (H) [left = 3mm of H] {$d$};
 \coordinate (I) at (3,0.8);
 \node at (I) [right = 3mm of I] {$d$};
 
 \coordinate (Q) at (2,1.8);
 \node at (Q) {$\otimes$};
 \node at (2.3,2.1) {$\mathcal{Q}$};
 
 \fill[gray!30] (C) ellipse (0.3 and 0.9);
 \fill[gray!30] (D) ellipse (0.3 and 0.9);

 \coordinate (F) at (1,0);
 \node at (F) [left = -3mm of F] {\footnotesize $(\bm{0},0)$};
  \coordinate (FG) at (2,0);
 \node at (FG) [right = -5mm of FG] {\footnotesize $(\bm{y},\tau')$};
 \coordinate (G) at (3,0);
 \node at (G) [right = -3mm of G] {\footnotesize $(\bm{x},\tau)$};

 \draw[particle] (A) -- (Q);
 \draw[particle] (Q) -- (B);
 \draw[particle] (C) -- (D);
 \draw[particle] (H) -- (I);

 \fill[myred](A) circle (0.1);
 \fill[mygreen](C) circle (0.1);
 \fill[myblue](H) circle (0.1);
 \fill[myred](B) circle (0.1);
 \fill[mygreen](D) circle (0.1);
 \fill[myblue](I) circle (0.1);
\end{tikzpicture}\;+\;\begin{tikzpicture}[baseline={([yshift=-1.4ex]current bounding box.center)}]
\coordinate (A) at (1,1.8);
 \node at (A) [left = 3mm of A] {$u$};
\coordinate (B) at (3,1.8);
 \node at (B) [right = 3mm of B] {$u$};
 \coordinate (C) at (1,1.3);
 \node at (C) [left = 3mm of C] {$u$};
 \coordinate (D) at (3,1.3);
 \node at (D) [right = 3mm of D] {$u$};
  \coordinate (H) at (1,0.8);
 \node at (H) [left = 3mm of H] {$d$};
 \coordinate (I) at (3,0.8);
 \node at (I) [right = 3mm of I] {$d$};
 
 \coordinate (Q) at (2,2.8);
 \node at (Q) {$\otimes$};
  \node at (2.4,2.9) {$\mathcal{Q}$};
  
 \draw[decoration={markings, mark=at position 0.5 with {\arrow[scale=1.5,black]{stealth}}},postaction={decorate}] (2,2.5) circle (0.3);
 
 \fill[gray!30] (C) ellipse (0.3 and 0.9);
 \fill[gray!30] (D) ellipse (0.3 and 0.9);

 \coordinate (F) at (1,0);
 \node at (F) [left = -3mm of F] {\footnotesize $(\bm{0},0)$};
  \coordinate (FG) at (2,0);
 \node at (FG) [right = -5mm of FG] {\footnotesize $(\bm{y},\tau')$};
 \coordinate (G) at (3,0);
 \node at (G) [right = -3mm of G] {\footnotesize $(\bm{x},\tau)$};

 \draw[particle] (A) -- (B);
 \draw[particle] (C) -- (D);
 \draw[particle] (H) -- (I);

 \fill[myred](A) circle (0.1);
 \fill[mygreen](C) circle (0.1);
 \fill[myblue](H) circle (0.1);
 \fill[myred](B) circle (0.1);
 \fill[mygreen](D) circle (0.1);
 \fill[myblue](I) circle (0.1);
\end{tikzpicture}\;+\cdots
\end{equation*}
\vspace*{-5mm}
\caption{Possible Wick contractions for the three-point correlation function of the proton interacting with an external current of the form $\mathcal{Q}=\bar{u}\Phi u$.}
\label{fig:proton_ME_contract}
\end{figure}
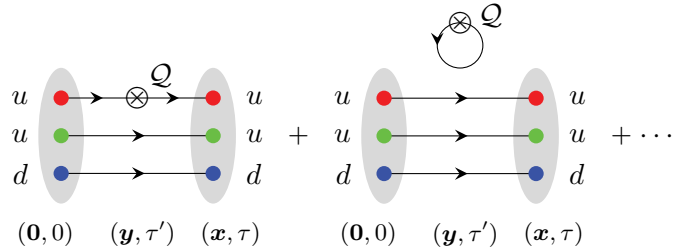

In order to extract matrix elements (MEs), we need to couple the quarks (or gluons) fields to external currents by calculating a three-point correlation function. The most common definition is the following,
\begin{equation}
    C_{3pt}(\tau,\tau',\bm{p},\bm{p}')=\sum_{\bm{x},\bm{y}} e^{-\imag \bm{x}\cdot\bm{p}'}e^{-\imag \bm{y}\cdot(\bm{p}'-\bm{p})} \Gamma_{\beta\alpha} \langle 0 | \mathcal{X}_\alpha(\bm{x},\tau) \mathcal{Q}(\bm{y},\tau') \bar{\mathcal{X}}_\beta(\bm{0},0)| 0\rangle\, ,
\end{equation}
with $\mathcal{Q}$ being the current inserted at time $\tau'$, and $\bm{p}$ and $\bm{p}'$ the initial and final state momenta (with the transferred momentum being $\bm{q}=\bm{p}'-\bm{p}$).
Focusing on bilinear operators, $\mathcal{Q}=\bar{q}\Phi q$, and for currents that do not change the quark flavor, we can draw schematically the contractions in Fig.~\ref{fig:proton_ME_contract} (again for the case of the proton).

Now we see that besides the usual connected diagrams, there is also the possibility of having disconnected diagrams, which originate in the Wick contraction of the quarks in the operator $\mathcal{Q}$. As we can see in Fig.~\ref{fig:proton_ME_contract}, these disconnected diagrams start and finish at the same spatial point, meaning that we will have to compute all-to-all propagators, which are more expensive that the point-to-all propagators from Eq.~\eqref{eq:propagator_equation}, since the whole Dirac matrix has to be inverted (stochastic methods are used to evaluate these types of contributions, e.g., see Ref.~\cite{Gambhir:2016uwp}). These disconnected contributions can be cancelled if certain combinations of $\mathcal{Q}$ are made, as we will see below. 

Among the possible $\mathcal{Q}$ operators, hadronic form factors (more specific, their corresponding charges), tell us about the internal structure of the hadrons as well as their interaction with other particles. For example, the electromagnetic form factors tell us about the charge and magnetic distribution of the hadron, and they are fairly simple to access experimentally using electron-proton scattering~\cite{Perdrisat:2006hj}. On the other hand, if we want to study the spin and gluon structure of the hadrons, nuclear reactions like double $\beta$-decay, or the interaction with scalar particles (dark matter candidates), QCD plays an important role, and for some of these quantities there is no clean extractions from experiments. Related to these quantities we have three different currents giving scalar, axial, and tensor MEs.

\subsubsection{Scalar ME}

Weakly interacting massive particles (WIMPs) are possible dark matter candidates, an extension of the SM. One of the problems when trying to study experimentally the interaction between WIMPs and nuclei is the large uncertainty in the spin-independent interaction, the scalar ME or sigma terms, which are defined as the ME of the scalar quark currents between hadronic $h$ states,
\begin{equation}
\begin{aligned}
    \sigma_{\pi h}&=m_{ud}\, \langle h|\bar{u}u+\bar{d}d|h\rangle\, , \\
    \sigma_{sh}&=m_s\, \langle h|\bar{s}s|h\rangle\, ,
\end{aligned}
\label{eq:ME_S}
\end{equation}
where $m_{ud}$ is the average of the up and down quark masses. What is interesting is that, in a LQCD calculation at $m_{\pi}\sim 806$ MeV, the scalar MEs for multi-hadronic states were found to be different from the naive estimation assuming a sum of free nucleons~\cite{Chang:2017eiq}, so quantifying these differences for light nuclei with LQCD with high accuracy will be needed for experiments using big elements (like xenon, germanium or argon) as detectors~\cite{Undagoitia:2015gya}.

Also, scalar MEs are used to access the strange content of the hadron. In principle, if we only consider hadrons with no strange valence quarks (like the proton or deuteron), the net strangeness is zero. However, the valence quarks are surrounded by a sea of $q\bar{q}$ pairs, and computing the vacuum contributions of this sea to observables is pertinent.

Experimentally, the $\sigma_{\pi N}$ has been extracted from pion-nucleon scattering experiments, $\sigma_{\pi N}\simeq 45$ MeV~\cite{Gasser:1990ce}, but the $\sigma_{sN}$ is not directly accessible. Instead, the flavour-singlet quantity $\sigma_0 = m_{ud}\, \langle N|\bar{u}u+\bar{d}d-2\bar{s}s|N \rangle$ can be extracted from octet baryon mass splittings, giving a value of $\sigma_0=36(7)$ MeV~\cite{Borasoy:1996bx}. However, when trying to use these two quantities to extract $\sigma_{sN}$, one gets large uncertainties. Therefore, a direct LQCD calculation can be very illuminating.

\subsubsection{Axial ME}

The hadron axial structure is characterized by the hadron axial form factors found in the ME of the axial-vector quark current, $A^{\mathfrak{a},\mu}=\bar{q}\gamma^{\mu}\gamma^5\frac{\tau^\mathfrak{a}}{2}q$~\cite{Lin:2015dga},
\begin{equation}
\begin{aligned}
    \langle h(\bm{p}',\bm{s}')|A^{\mathfrak{a},\mu}|h(\bm{p},\bm{s})\rangle = \bar{U}_h(\bm{p}',\bm{s}')&\left[\gamma^{\mu}\gamma^5 G^{\mathfrak{a}}_A(Q^2)+\frac{q^{\mu}\gamma^5}{2m_h}G^{\mathfrak{a}}_P(Q^2)\right.\\
    &\; \left.+\ \imag\frac{\sigma^{\mu\nu}q_{\nu}\gamma^5}{2m_h}G^{\mathfrak{a}}_T(Q^2)\right]U_h(\bm{p},\bm{s})\, ,
\end{aligned}
\label{eq:ME_A}
\end{equation}
where $Q^2=-q^2$, $\tau^{\mathfrak{a}}$ is the isospin Pauli matrix,\footnote{For this example, since we work in $SU(2)_f$, $q=( u\, d)^\top$ and we use the Pauli matrices $\tau^{\mathfrak{a}}$, but for $SU(3)_f$, $q=(u\, d\, s)^\top$ and we shall use the Gell-Mann matrices.} and the form factors $G^{\mathfrak{a}}_A(Q^2)$, $G^{\mathfrak{a}}_P(Q^2)$, and $G^{\mathfrak{a}}_T(Q^2)$ are the axial, induced pseudoscalar, and tensor form factors, respectively.

In the forward limit $Q^2=0$, the axial form factor gives the hadronic axial charge $G_A(0)=g_A$. Experimentally, the nucleon axial charge has been measured with great accuracy, with a value of $g_A=g^{u-d}_A=1.2723(23)$~\cite{Tanabashi:2018oca}. 
This axial current also allows us to study $\beta$-decay processes, although it requires a current that changes quark flavor. On the lattice, we can simplify this calculation by using isospin symmetry with the raising and lowering operators $\hat{T}_\pm$,
\begin{equation}
    \langle p|\bar{u}\gamma^{\mu}\gamma^5 d|n\rangle = \langle p|\bar{u}\gamma^{\mu}\gamma^5 d (\hat{T}_- \hat{T}_+) |n\rangle = \langle p|\bar{u}\gamma^{\mu}\gamma^5 d \,\hat{T}_- |p\rangle = \langle p|\bar{u}\gamma^{\mu}\gamma^5 u -\bar{d}\gamma^{\mu}\gamma^5 d |p\rangle \, ,
\end{equation}
where we have transformed the ME for the neutron $\beta$-decay to an isovector axial proton ME (in which the disconnected diagrams cancel with each other since the masses of $u$ and $d$ are set to be the same).

\subsubsection{Tensor ME}

The ME of the tensor quark current, $T^{\mathfrak{a},\mu\nu}=\bar{q}\,\imag\sigma^{\mu\nu}\frac{\tau^{\mathfrak{a}}}{2}q$, can be parametrized by three form factors~\cite{Ledwig:2010tu},
\begin{equation}
\begin{aligned}
    \langle h(\bm{p}',\bm{s}')|T^{\mathfrak{a},\mu\nu}|h(\bm{p},\bm{s})\rangle = \bar{U}_h(\bm{p}',\bm{s}')&\left[\imag \sigma^{\mu\nu} H^{\mathfrak{a}}_T(Q^2)+\frac{\gamma^{\mu} q^{\nu}-q^{\mu} \gamma^{\nu}}{2m_h}E^{\mathfrak{a}}_T(Q^2)\right.\\
    &\; \left.+\ \frac{n^{\mu} q^{\nu}-q^{\mu} n^{\nu}}{2m^2_h}\tilde{H}^{\mathfrak{a}}_T(Q^2)\right]U_h(\bm{p},\bm{s})\, ,
\label{eq:ME_T}
\end{aligned}
\end{equation}
where $n^{\mu}=p'^{\mu}+p^{\,\mu}$. Again, in the forward limit, the tensor charge is defined as $H_T(0)=g_T$. Tensor charges are involved in the quark electric dipole moment, a new source of CP violation in some beyond-SM models~\cite{Pospelov:2005pr,Pitschmann:2014jxa}, since the two sources in the SM (the $\theta$-term and the complex phase in the quark mixing matrix) are too small to explain the unbalance between matter and antimatter.

This quantity is accessible through the quark transversity distribution $h_1(x)$, which measures the number of quarks with transverse polarization parallel to that of the hadron minus that of quarks with antiparallel polarization. The first moment of $h_1(x)$ is related to $g_T$~\cite{Jaffe:1991ra},
\begin{equation}
    \int_0^1\, dx [h_1(x)-\overline{h}_1(x)]=g_T \, ,
\end{equation}
where $\overline{h}_1(x)$ is the antiquark transversity distribution and $x$ is the Bjorken variable (related to deep inelastic scattering). The most recent values obtained using this approach are given in Ref.~\cite{Kang:2015msa} using data from COMPASS, HERMES, and JLab, with $g_T=g^{u-d}_T=0.61^{(+0.26)}_{(-0.51)}$ for the isovector nucleon tensor charge, and $g^0_T=g^{u+d}_T=0.17^{(+0.47)}_{(-0.30)}$ for the isoscalar one (both at $90\%$ C.L.). It should be said that the future upgrade of JLab to 12 GeV is expected to improve the precision up to one order of magnitude~\cite{Ye:2016prn}. Given the challenging experimental extraction of the tensor charge, its estimation with LQCD calculations is extremely helpful.

\vspace{1em}

All these charges have been computed in the single-baryon sector, as compiled in the most recent FLAG summary~\cite{Aoki:2019cca}, as well for two- and three-nucleon states~\cite{Chang:2017eiq,Parreno:2021ovq}. One can extract these charges from the spectral decomposition of $C_{3pt}$,
\begin{equation}
\begin{aligned}
    C_{3pt}(\tau,\tau') =&\ \sum_{\bm{x},\bm{y}}\sum_{\mathfrak{n},\bm{k},s}\sum_{\mathfrak{m},\bm{k}',s'} e^{\imag\bm{x}\cdot\bm{k}'}e^{-\imag\bm{y}\cdot(\bm{k}'-\bm{k})} \Gamma_{\beta\alpha}\\
    &\times\langle 0| \mathcal{X}_{\alpha}(\bm{0},0)|\mathfrak{m},\boldsymbol{k}',\bm{s}'\rangle \langle\mathfrak{m},\boldsymbol{k}',\bm{s}' | \mathcal{Q}(\bm{0},0)|\mathfrak{n},\boldsymbol{k},\bm{s}\rangle \langle\mathfrak{n},\boldsymbol{k},\bm{s} | \bar{\mathcal{X}}_{\beta}(\bm{0},0)|0\rangle\\
    &\times e^{-(\tau-\tau')E_{\mathfrak{n}}(\boldsymbol{k})}e^{-\tau'E_{\mathfrak{m}}(\boldsymbol{k}')}\, ,
\end{aligned}
\end{equation}
where we have taken $\bm{p}=\bm{p}'=\bm{0}$ (and dropped the $\bm{p}$ and $\bm{p}'$ labels) for simplicity. The operator bracket gives us $\langle\mathfrak{m},\boldsymbol{k}',\bm{s}' | \mathcal{Q}(\bm{0},0)|\mathfrak{n},\boldsymbol{k},\bm{s}\rangle=\bar{U}_\gamma(\boldsymbol{k}',\bm{s}')\Phi_{\gamma\delta}U_\delta(\boldsymbol{k},\bm{s}) g_\Phi$, with $g_\Phi$ being the charge. Therefore,
\begin{equation}
\begin{aligned}
    C_{3pt}(\tau,\tau') =&\ \sum_{\mathfrak{n},\mathfrak{m}} Z_{\mathfrak{n}}Z^*_{\mathfrak{m}} \Tr[\Gamma(E_{\mathfrak{m}}\gamma_4+M_N)\Phi(E_{\mathfrak{n}}\gamma_4+M_N)] e^{-(\tau-\tau')E_{\mathfrak{n}}}e^{-\tau'E_{\mathfrak{m}}}g_\Phi \\
    =&\ \sum_{\mathfrak{n},\mathfrak{m}} Z_{\mathfrak{n}}Z^*_{\mathfrak{m}} \tilde{\mathcal{F}}_{\mathfrak{n}\mathfrak{m}}(\Gamma,\Phi) e^{-(\tau-\tau')E_{\mathfrak{n}}}e^{-\tau'E_{\mathfrak{m}}}g_\Phi\, .
\label{eq:c3pt_spectral}
\end{aligned}
\end{equation}
We note that while for scalar currents, $2\tilde{\mathcal{F}}_{\mathfrak{n}\mathfrak{m}}(\Gamma_{\text{unpol}},1)=\tilde{\mathcal{F}}_{\mathfrak{n}\mathfrak{m}}(\Gamma_{\text{pol}},1)=(E_{\mathfrak{m}}+M_N)(E_{\mathfrak{n}}+M_N)$, the axial and tensor currents have to be computed using $\Gamma_{\text{pol}}$, since $\tilde{\mathcal{F}}_{\mathfrak{n}\mathfrak{m}}(\Gamma_{\text{unpol}},\gamma^\mu\gamma^5)=\tilde{\mathcal{F}}_{\mathfrak{n}\mathfrak{m}}(\Gamma_{\text{unpol}},\sigma^{\mu\nu})=0$. Thus, using $\Gamma_{\text{pol}}$, we see that for the axial current $\tilde{\mathcal{F}}_{\mathfrak{n}\mathfrak{m}}(\Gamma_{\text{pol}},\gamma^i\gamma^5)\,\propto\, \bm{s}_i$, and for the tensor current $\tilde{\mathcal{F}}_{\mathfrak{n}\mathfrak{m}}(\Gamma_{\text{pol}},\sigma^{ij})\,\propto\, \epsilon^{ijk4}\bm{s}_k$. This means that if we polarize our state in the $z$-direction, we have to pick $\gamma^3\gamma^5$ and $\sigma^{12}$ in order to get a non-zero measurement.

Explicitly writing the first terms of the sums in Eq.~\eqref{eq:c3pt_spectral}, we see that (and again absorbing $\tilde{\mathcal{F}}_{\mathfrak{n}\mathfrak{m}}(\Gamma,\Phi)$ into the $Z$ factors),
\begin{equation}
\begin{aligned}
    C_{3pt}(\tau,\tau') = g_\Phi &\left[ Z_{\mathfrak{0}} Z^*_{\mathfrak{0}} e^{-\tau E_{\mathfrak{0}}} + Z_{\mathfrak{1}} Z^*_{\mathfrak{1}} e^{-\tau E_{\mathfrak{1}}}\right. \\
    &\; \left.+\ Z_{\mathfrak{0}} Z^*_{\mathfrak{1}} e^{-(\tau-\tau')E_{\mathfrak{0}}} e^{-\tau'E_{\mathfrak{1}}} + Z_{\mathfrak{1}} Z^*_{\mathfrak{0}} e^{-(\tau-\tau')E_{\mathfrak{1}}} e^{-\tau'E_{\mathfrak{0}}} +\cdots \right]\, ,
\end{aligned}
\end{equation}
which tells us that in order to reduce exited state contamination, $\tau'$ and $\tau$ have to satisfy $0 \ll \tau' \ll \tau$. Appropriate ratios of $C_{3pt}(\tau,\tau')$ over $C_{2pt}(\tau)$ can help cancel higher-excited state contributions~\cite{Hagler:2003jd}.

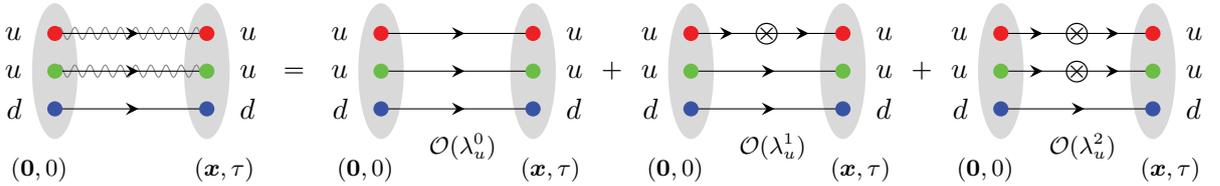
\begin{figure}[ht]
\begin{equation*}
\begin{tikzpicture}[baseline={([yshift=1.5ex]current bounding box.center)}]
\coordinate (A) at (1,1.8);
 \node at (A) [left = 3mm of A] {$u$};
\coordinate (B) at (3,1.8);
 \node at (B) [right = 3mm of B] {$u$};
 \coordinate (C) at (1,1.3);
 \node at (C) [left = 3mm of C] {$u$};
 \coordinate (D) at (3,1.3);
 \node at (D) [right = 3mm of D] {$u$};
  \coordinate (H) at (1,0.8);
 \node at (H) [left = 3mm of H] {$d$};
 \coordinate (I) at (3,0.8);
 \node at (I) [right = 3mm of I] {$d$};
 
 \fill[gray!30] (1,1.3) ellipse (0.3 and 0.9);
 \fill[gray!30] (3,1.3) ellipse (0.3 and 0.9);

 \coordinate (F) at (1,0);
 \node at (F) [left = -3mm of F] {\footnotesize $(\bm{0},0)$};
 \coordinate (G) at (3,0);
 \node at (G) [right = -3mm of G] {\footnotesize $(\bm{x},\tau)$};

 \tikzset{decoration={snake,amplitude=.8mm,segment length=2mm,
                       post length=0mm,pre length=0mm}}
 \draw[gray, decorate] (A) -- (B);
 \draw[gray, decorate] (C) -- (D);

 \draw[particle] (A) -- (B);
 \draw[particle] (C) -- (D);
 \draw[particle] (H) -- (I);

 \fill[myred](A) circle (0.1);
 \fill[mygreen](C) circle (0.1);
 \fill[myblue](H) circle (0.1);
 \fill[myred](B) circle (0.1);
 \fill[mygreen](D) circle (0.1);
 \fill[myblue](I) circle (0.1);
\end{tikzpicture}\;=\;\begin{tikzpicture}[baseline={([yshift=1.5ex]current bounding box.center)}]
\coordinate (A) at (1,1.8);
 \node at (A) [left = 3mm of A] {$u$};
\coordinate (B) at (3,1.8);
 \node at (B) [right = 3mm of B] {$u$};
 \coordinate (C) at (1,1.3);
 \node at (C) [left = 3mm of C] {$u$};
 \coordinate (D) at (3,1.3);
 \node at (D) [right = 3mm of D] {$u$};
  \coordinate (H) at (1,0.8);
 \node at (H) [left = 3mm of H] {$d$};
 \coordinate (I) at (3,0.8);
 \node at (I) [right = 3mm of I] {$d$};

 \fill[gray!30] (1,1.3) ellipse (0.3 and 0.9);
 \fill[gray!30] (3,1.3) ellipse (0.3 and 0.9);

 \coordinate (F) at (1,0);
 \node at (F) [left = -3mm of F] {\footnotesize $(\bm{0},0)$};
 \coordinate (FG) at (2,0.3);
 \node at (FG) [right = -5mm of FG] {\footnotesize $\mathcal{O}(\lambda^0_u)$};
 \coordinate (G) at (3,0);
 \node at (G) [right = -3mm of G] {\footnotesize $(\bm{x},\tau)$};

 \draw[particle] (A) -- (B);
 \draw[particle] (C) -- (D);
 \draw[particle] (H) -- (I);

 \fill[myred](A) circle (0.1);
 \fill[mygreen](C) circle (0.1);
 \fill[myblue](H) circle (0.1);
 \fill[myred](B) circle (0.1);
 \fill[mygreen](D) circle (0.1);
 \fill[myblue](I) circle (0.1);
\end{tikzpicture}\;+\;\begin{tikzpicture}[baseline={([yshift=1.5ex]current bounding box.center)}]
\coordinate (A) at (1,1.8);
 \node at (A) [left = 3mm of A] {$u$};
\coordinate (B) at (3,1.8);
 \node at (B) [right = 3mm of B] {$u$};
 \coordinate (C) at (1,1.3);
 \node at (C) [left = 3mm of C] {$u$};
 \coordinate (D) at (3,1.3);
 \node at (D) [right = 3mm of D] {$u$};
  \coordinate (H) at (1,0.8);
 \node at (H) [left = 3mm of H] {$d$};
 \coordinate (I) at (3,0.8);
 \node at (I) [right = 3mm of I] {$d$};
 
 \coordinate (Q) at (2,1.8);
 \node at (Q) {$\otimes$};
 
 \fill[gray!30] (1,1.3) ellipse (0.3 and 0.9);
 \fill[gray!30] (3,1.3) ellipse (0.3 and 0.9);

 \coordinate (F) at (1,0);
 \node at (F) [left = -3mm of F] {\footnotesize $(\bm{0},0)$};
 \coordinate (FG) at (2,0.3);
 \node at (FG) [right = -5mm of FG] {\footnotesize $\mathcal{O}(\lambda^1_u)$};
 \coordinate (G) at (3,0);
 \node at (G) [right = -3mm of G] {\footnotesize $(\bm{x},\tau)$};

 \draw[particle] (A) -- (Q);
 \draw[particle] (Q) -- (B);
 \draw[particle] (C) -- (D);
 \draw[particle] (H) -- (I);

 \fill[myred](A) circle (0.1);
 \fill[mygreen](C) circle (0.1);
 \fill[myblue](H) circle (0.1);
 \fill[myred](B) circle (0.1);
 \fill[mygreen](D) circle (0.1);
 \fill[myblue](I) circle (0.1);
\end{tikzpicture}\;+\;\begin{tikzpicture}[baseline={([yshift=1.5ex]current bounding box.center)}]
\coordinate (A) at (1,1.8);
 \node at (A) [left = 3mm of A] {$u$};
\coordinate (B) at (3,1.8);
 \node at (B) [right = 3mm of B] {$u$};
 \coordinate (C) at (1,1.3);
 \node at (C) [left = 3mm of C] {$u$};
 \coordinate (D) at (3,1.3);
 \node at (D) [right = 3mm of D] {$u$};
  \coordinate (H) at (1,0.8);
 \node at (H) [left = 3mm of H] {$d$};
 \coordinate (I) at (3,0.8);
 \node at (I) [right = 3mm of I] {$d$};
 
  \coordinate (Q) at (2,1.8);
 \node at (Q) {$\otimes$};
  \coordinate (QQ) at (2,1.3);
 \node at (QQ) {$\otimes$};
 
 \fill[gray!30] (1,1.3) ellipse (0.3 and 0.9);
 \fill[gray!30] (3,1.3) ellipse (0.3 and 0.9);

 \coordinate (F) at (1,0);
 \node at (F) [left = -3mm of F] {\footnotesize $(\bm{0},0)$};
  \coordinate (FG) at (2,0.3);
 \node at (FG) [right = -5mm of FG] {\footnotesize $\mathcal{O}(\lambda^2_u)$};
 \coordinate (G) at (3,0);
 \node at (G) [right = -3mm of G] {\footnotesize $(\bm{x},\tau)$};

 \draw[particle] (A) -- (Q);
 \draw[particle] (Q) -- (B);
 \draw[particle] (C) -- (QQ);
 \draw[particle] (QQ) -- (D);
 \draw[particle] (H) -- (I);

 \fill[myred](A) circle (0.1);
 \fill[mygreen](C) circle (0.1);
 \fill[myblue](H) circle (0.1);
 \fill[myred](B) circle (0.1);
 \fill[mygreen](D) circle (0.1);
 \fill[myblue](I) circle (0.1);
\end{tikzpicture}
\end{equation*}
\vspace*{-5mm}
\caption{Contributions that appear when using the compound propagator $S^{(u)}_{\Phi,\lambda_u}$ in a proton.}
\label{fig:proton_ME_contract_2}
\end{figure}

The background field method, used in studies of the magnetic properties of nuclear systems, can also be applied to compute matrix elements. However, in this case, instead of modifying the gauge-field configurations, the propagator is modified by including the insertion of the operator $\mathcal{Q}$, leading to the compound propagator~\cite{Savage:2016kon},
\begin{equation}
    S^{(q)}_{\Phi,\lambda_q}(x,y)=S^{(q)}_{\alpha\beta}(x,y)+\lambda_q \sum_z S^{(q)}_{\alpha\gamma}(x,z)\Phi_{\gamma\delta} S^{(q)}_{\delta\beta}(z,y)\, ,
\end{equation}
where $\lambda_q$ is a constant number. The use of this compound propagator implies a sum over the time $\tau'$ where the operator is inserted, and multiple insertions are possible. To understand the difference with the previous method, let us draw the contractions depicted in Fig.~\ref{fig:proton_ME_contract_2} where we replace the up-quark propagators by the corresponding compound ones.

When constructing correlation functions with these compound propagators, $C_{\Phi,\lambda_q}$, we will have contributions from the two-point correlation function as well as from three-point correlation functions with as many insertions of the operator as number of valence quarks of flavor $q$ has the hadron (the disconnected diagrams have to be computed separately). This means that we can write $C_{\Phi,\lambda_q}$ as a polynomial in $\lambda_q$,
\begin{equation}
\begin{aligned}
    C_{\Phi,\lambda_q}(\tau)=\Gamma_{\beta\alpha}\sum_{\bm{x}} & \bigg[ \langle 0| \mathcal{X}_{\alpha}(\bm{x},\tau) \bar{\mathcal{X}}_{\beta}(\bm{0},0)|0\rangle \\
    &\; +\ \lambda_q \sum_{\boldsymbol{y}}\sum_{\tau'=0}^\tau \langle 0| \mathcal{X}_{\alpha}(\bm{x},\tau) \{\bar{q}\Phi q\} (\bm{y},\tau')\bar{\mathcal{X}}_{\beta}(\bm{0},0) |0\rangle \bigg] + \mathcal{O}(\lambda^2_q)\, .
\label{eq:corr_ex}
\end{aligned}
\end{equation}
For example, for the case of the proton, $C_{\Phi,\lambda_q}$ is a polynomial of maximum order $\lambda^2_u$ and $\lambda_d$. In order to disentangle each contribution, multiple $\lambda_q$ are used in the calculation, and a system of equations with a Vandermonde matrix for the coefficients is obtained, which has an exact analytical solution.
This technique has been extensively used by the NPLQCD Collaboration for the study of multi-baryon systems (for a detailed discussion, see Ref.~\cite{Davoudi:2020ngi}), where the usual propagators are replaced by the compound ones.
The most recent results obtained with this method correspond to the calculation of the momentum fraction of ${}^3\text{He}$~\cite{Detmold:2020snb} and of the axial charge of the triton~\cite{Parreno:2021ovq}, which only need the evaluation of the $\mathcal{O}(\lambda_q)$ contribution. 
Processes that require the $\mathcal{O}(\lambda^2_q)$ components are, for example, double-$\beta$ decays~\cite{Shanahan:2017bgi,Tiburzi:2017iux}.

Looking at the spectral decomposition of $\left. C_{\Phi,\lambda_q}\right|_{\mathcal{O}(\lambda_q)}$ we see that, compared to Eq.~\eqref{eq:c3pt_spectral}, the sum over $\tau'$ yields different time-dependence for the ground state as well as different excited-state contamination,
\begin{equation}
\begin{aligned}
    \left. C_{\Phi,\lambda_q}\right|_{\mathcal{O}(\lambda_q)} &= \sum_{\mathfrak{n},\mathfrak{m}} Z_{\mathfrak{n}}Z^*_{\mathfrak{m}} \,g_\Phi\, e^{-\tau E_{\mathfrak{n}}}\sum_{\tau'=0}^{\tau} e^{-\tau'(E_{\mathfrak{m}}-E_{\mathfrak{n}})}\\
    &= \sum_{\mathfrak{n},\mathfrak{m}} Z_{\mathfrak{n}}Z^*_{\mathfrak{m}} \,g_\Phi \left[\frac{e^{-\tau E_{\mathfrak{n}}}}{1-e^{E_{\mathfrak{n}}-E_{\mathfrak{m}}}}+\frac{e^{-\tau E_{\mathfrak{m}}}}{1-e^{E_{\mathfrak{m}}-E_{\mathfrak{n}}}}\right]\\
    &=\sum_{\mathfrak{n}} Z_{\mathfrak{n}}Z^*_{\mathfrak{n}} \,g_\Phi (1+\tau)e^{-\tau E_{\mathfrak{n}}}+\sum_{\mathfrak{n}}\sum_{\mathfrak{m}\neq\mathfrak{n}} Z_{\mathfrak{n}}Z^*_{\mathfrak{m}} \,g_\Phi \left[\frac{e^{-\tau E_{\mathfrak{n}}}}{1-e^{E_{\mathfrak{n}}-E_{\mathfrak{m}}}}+\frac{e^{-\tau E_{\mathfrak{m}}}}{1-e^{E_{\mathfrak{m}}-E_{\mathfrak{n}}}}\right]\, .
\end{aligned}
\end{equation}
Again, appropriate combinations and ratios of $\smash{\left. C_{\Phi,\lambda_q}\right|_{\mathcal{O}(\lambda_q)}}$ and $C_{2pt}$ can help extract $g_\Phi$ more cleanly~\cite{Detmold:2020snb}. 

All these matrix elements obtained from the lattice are computed in the ``lattice'' scheme. What this means is that the lattice itself is a regularization scheme (the lattice spacing imposes an ultraviolet cutoff, $\Lambda_{\text{latt}}=b^{-1}$), so the operators computed in the lattice will be regularized in this particular scheme. These are what we call bare operators. If we want to compare them to their continuum counterparts, we need to switch to an appropriate renormalization scheme, which is usually the modified subtraction scheme ($\overline{\text{MS}}$) at a scale $\mu=2$ GeV~\cite{Gattringer:2010zz} (typical renormalization scale used when dealing with experimental data),
\begin{equation}
    \mathcal{O}^{R}_\Phi (\mu) = Z_\Phi (\mu,b) \mathcal{O}^{\text{latt}}_\Phi (b)\, ,
\end{equation}
where $Z_{\Phi}$ is called the renormalization constant, which depends on the lattice action used and the operator, but not on the external states in the ME. To be precise, the renormalization procedure is divided in two steps, since one cannot apply the $\overline{\text{MS}}$ scheme directly on the lattice. First, a non-perturbative calculation of $Z_{\Phi}$ is done on the lattice using a specific scheme, such as the RI-MOM scheme (Regularization Independent MOMentum subtraction scheme~\cite{Martinelli:1994ty}). Then, a perturbative matching between the RI-MOM and $\overline{\text{MS}}$ schemes is performed.

Since the lattice version of QCD is written in Euclidean space-time, the original Lorentz group is replaced by the orthogonal group $O(4)$, which is further reduced due to the discretization of space-time to the hypercubic group $H(4)\subset O(4)$~\cite{Gockeler:1996mu}. Just as it happens with the angular momentum (we will give more details in the next section), since $H(4)$ is a finite group, mixing between different operators is possible, and $Z_{\Phi}(\mu,b)$ is, in general, a matrix with non-zero off-diagonal elements, rather than just a number.

\section{Scattering in finite volume}\label{sec:scatteringFV}

Computing the relevant parameters in the description of the interaction of two (and more) hadrons directly from QCD is one of the goals of nuclear physics. Typically, phase shifts are determined from scattering experiments by parametrizing the scattering amplitude at low-energies as a function of the energy in the center-of-mass (c.m.) frame. For baryon-baryon processes, a common parametrization below the $t$-channel cut is the effective range expansion (ERE)~\cite{Schwinger:notes,Blatt:1949zz,Bethe:1949yr}, which for the case of $S$ wave can be written as
\begin{equation}
    k^*\cot\delta(k^*)=-\frac{1}{a}+\frac{1}{2}rk^{*2}+Pk^{*4}+\mathcal{O}(k^{*6})\, ,
\label{eq:ERE}
\end{equation}
where $a$ is the scattering length, $r$ is the effective range, and $P$ is the leading shape parameter. For other types of processes, like the study of resonances, different parametrizations are more suitable (e.g., see Ref.~\cite{Dudek:2012xn,Briceno:2017qmb}).
When trying to study scattering processes with LQCD, we have to confront the Maiani-Testa no-go theorem~\cite{Maiani:1990ca}, which states that scattering matrix elements cannot be extracted from infinite-volume Euclidean correlation functions except at kinematic thresholds. An easy way to circumvent this problem is by computing correlation functions at finite volume. The method was formalized by M.~Lüscher~\cite{Luscher:1986pf,Luscher:1990ux}, who extracted $S$-wave scattering elements from the discrete spectrum of the two particle state in a finite three-dimensional box below inelastic thresholds. Before diving into the formalism, we need to understand how do the spin, orbital, and total momentum of the system get modified when it is put in a box.

\subsection{Angular momentum group theory}

The study of angular momentum on the lattice requires some knowledge of group theory. A summary of the main concepts (taken from Refs.~\cite{koster1963properties,ATKINS2008tables,weissbluth1978atoms,elliott1979symmetry,Johnson:1982yq,dresselhaus2007group}) and useful definitions are given below.

\subsubsection{Elements of group theory}

\begin{definition} 
A collection of elements $A,B,C,\ldots$ form a \textit{group} when the following four conditions are satisfied: 
\begin{enumerate}[label=(\roman*)]
 \item The product of any two elements of the group is also a member of the group.
 \item The associative law applies to the product of three elements, i.e., $(AB)C=A(BC)$.
 \item There exists a unit element $E$ (or identity element) such that $EA=AE=A$.
 \item Each element has an inverse element which is also a member of the group, i.e., for every element $A$ there exists an inverse element $A^{-1}$ such that $A^{-1}A=AA^{-1}=E$.
 \end{enumerate}
\end{definition}

\begin{definition} 
The \textit{order of a group} $h$ is the number of elements in the group.
\end{definition}

\begin{definition} 
A \textit{subgroup} is a collection of elements within a group that by themselves satisfy the group postulates.
\end{definition}

\begin{definition} 
Two groups $G$ and $G'$ are \textit{isomorphic} when to each element of $G$ there corresponds one, and only one, element of $G'$, and conversely (one-to-one correspondence).
\end{definition}
 
\begin{definition} 
Two groups $G$ and $G'$ are \textit{homomorphic} when to each element of $G$ there corresponds one, and only one, element of $G'$, but to each element of $G'$ there corresponds at least one and possibly more than one element of $G$ (many-to-one correspondence).
\end{definition}

\begin{definition} 
If a group $G$ contains two subgroups $G_a$ and $G_b$ whose elements commute, and if every element of $G$ can be written uniquely as a product of the elements of $G_a$ and $G_b$, then $G$ is the \textit{direct product} of $G_a$ and $G_b$ and is written $G=G_a \otimes G_b$.
\end{definition}
 
\begin{definition} 
An element $B$ \textit{conjugate} of $A$ is by definition $B=XAX^{-1}$, where $X$ is some member of the group.
\end{definition}
 
\begin{definition} 
A \textit{class} is the totality of elements which can be obtained from a given group element by conjugation. For an Abelian group, since all members are selfconjugate, the number of classes is equal to the order $h$ of the group.
\end{definition}

\begin{definition} 
A \textit{representation} of an abstract group is a substitution group (matrix group with square matrices with non-vanishing determinants) such that the substitution group is homomorphic (or isomorphic) to the abstract group. We assign a matrix $\Pi(A)$ to each element $A$ of the abstract group such that $\Pi(AB)=\Pi(A)\Pi(B)$. These representations are \textit{not unique}.
\end{definition}

\begin{definition} 
The \textit{dimensionality} of a representation is equal to the dimensionality of each of its matrices.
\end{definition}

\begin{definition} 
If by means of the same equivalence transformation, all the matrices in the representation of a group can be made to acquire the same block form, then the representation is said to be \textit{reducible}; otherwise it is \textit{irreducible}. An irreducible representation (irrep) cannot be expressed in terms of representations of lower dimensionality.
\end{definition}

If $\Pi$ is a reducible representation, then the reduction is expressed by writing
\begin{equation}
\Pi=\bigoplus_{i}m_i \Pi^{(i)} \, ,
\label{eq:reduction}
\end{equation}
where each $m_i$ is a positive integer that indicates how many times the representation matrix $\Pi^{(i)}(R)$ appears along the main diagonal of $\Pi(R)$. The number of non-equivalent irreps is equal to the number of classes. Furthermore, there is a restriction on the dimensions of the irreps. If $l_i$ is the dimension of the $i$th irrep and $h$ the order of the group, then
\begin{equation}
\sum_i l_i^2=h \, .
\end{equation}

\begin{definition} 
The \textit{character} of a representation matrix $\Pi^{(j)}(R)$ is the trace of the matrix. The character for each element in a class is the same.
\end{definition}
The usual way to summarize the information on the characters of the representations of a group is using the \textit{character table}. In a character table we list the irreps as columns and the class as rows. Looking at Eq.~\eqref{eq:reduction}, it follows that the character $\chi$ of $\Pi$ is related to the irreducible characters by a similar equation. If $\chi(R)$ denotes the character of $\Pi$ for an element in a class $R$, then
\begin{equation}
\chi(R)=\sum_{i}m_i \chi^{(i)}(R)\, .
\end{equation}
It is obviously of interest, given $R$, to be able to deduce the $m_i$, which can be done if the irreducible characters $\chi^{(i)}(R)$ are known,
\begin{equation}
m_i=\frac{1}{h}\sum_R n_R\, \chi^{(i)}(R)^*\chi(R)\, ,
\label{eq:reduction_2}
\end{equation}
where $n_R$ denotes the number of elements in class $R$.

\subsubsection{Angular momentum in the continuum}

In the continuum, the relevant group is the $O(3)$ group, which is the full three-dimensional rotation group. This group includes $3\times 3$ real matrices with determinant $\pm 1$. This means that $O(3)$ does not only include the \textit{proper} rotations (matrices with $\Det=+ 1$), but also the \textit{improper} rotations (with $\Det=- 1$). These improper rotations correspond to proper rotations combined with a spatial inversion (also known as parity). Therefore, $O(3)=SO(3) \otimes C_i$, where the group $C_i$ contains just two elements: the identity element $E$ and the inversion element $I$.

Since $O(3)$ is the direct product of $SO(3)$ and $C_i$, each irrep of $O(3)$ can be regarded as the direct product of an irrep of $SO(3)$ with an irrep of $C_i$. We already know the irreps of $SO(3)$, there are an infinite number of irreps with dimension $(2L+1)$, which are labeled by integer angular momentum $L\in\{0,1,2,\ldots\}$. For the group $C_i$, since it is an Abelian group with two elements, there will be two 1-dimensional irreps, which separates states between proper and improper states of parity. Then, the character of the irreps of $O(3)$ is given by
\begin{equation}
\chi^{(L)}(\theta,\text{proper})=\frac{\sin\left[\left( L+\frac{1}{2}\right)\theta\right]}{\sin\left(\frac{\theta}{2}\right)}\, , \quad\quad \chi^{(L)}(\theta,\text{improper})=(-1)^L\frac{\sin\left[\left( L+\frac{1}{2}\right)\theta\right]}{\sin\left(\frac{\theta}{2}\right)}\, .
\label{eq:S03_char}
\end{equation}

If we want to include half-integer angular momentum (required for fermions), we need to consider the double cover representation of $SO(3)$, which is $SU(2)$. This can be argued as follows. If we look at the first expression in Eq.~\eqref{eq:S03_char}, under the rotation $\theta+2\pi$,
\begin{equation}
\chi^{(J)}(\theta+2\pi)=\frac{\sin\left[\left( J+\frac{1}{2}\right)(\theta+2\pi)\right]}{\sin\left(\frac{\theta+2\pi}{2}\right)}=\frac{\sin\left[\left( J+\frac{1}{2}\right)\theta\right] \cos\left[\left( J+\frac{1}{2}\right)2\pi\right]}{\sin\left(\frac{\theta}{2}\right) \cos(\pi)}\, .
\end{equation}
For integer values of $J$, $\cos\left[\left( J+\frac{1}{2} \right) 2\pi \right]=-1$, while for half-integer values of $J$ it takes the value $+1$. Therefore, we have the important relation
\begin{equation}
\chi^{(J)}(\theta+2\pi)=(-1)^{2J}\chi^{(J)}(\theta)\, ,
\end{equation}
which implies that for integer $J$, a rotation by $\theta, \theta\pm 2\pi, \theta\pm 4\pi,\ldots$ yields identical characters, but for half-integer values of $J$, we have
\begin{equation}
\chi^{(J)}(\theta+2\pi)=-\chi^{(J)}(\theta)\, , \quad\quad \chi^{(J)}(\theta+4\pi)=\chi^{(J)}(\theta)\, .
\end{equation}
The need to rotate by $4\pi$ rather than by $2\pi$ to generate the identity operation leads to the concept of double groups. For the improper rotations with half-integer $J$, the change of sign in Eq.~\eqref{eq:S03_char} depends on the parity ($P$) of the system. If $P=+$, no modification is required, but if $P=-$, one needs to multiply by $(-1)$. The same will happen when $J$ comes from coupling orbital angular momentum $L$ to spin $s$. For the rest of the discussion and for simplicity, we will use $J=L$ ($s=0$) for $J$ integer (and parity $(-1)^L$).

An interesting property of $\chi^{(J)}(\theta)$ that will be used later on is to see what happens when $J\rightarrow J+12n$,
\begin{equation}
\chi^{(J+12n)}(\theta)=\frac{\sin\left[\left( J+\frac{1}{2}\right)\theta\right] \cos (12n\theta) + \cos\left[\left( J+\frac{1}{2}\right)\theta\right] \sin (12n\theta)}{\sin\left(\frac{\theta}{2}\right)}\, .
\end{equation}
The rotations relevant for the lattice are only $\theta\in \{\pi,\pi/2,\pi/3\}$ (and multiples). Therefore, $\cos (12n\theta) = 1 \; \forall \theta$, and
\begin{equation}
\chi^{(J+12n)}(\theta)=\chi^{(J)}(\theta)+\frac{\cos\left[\left( J+\frac{1}{2}\right)\theta\right] \sin (12n\theta)}{\sin\left(\frac{\theta}{2}\right)}=\begin{dcases}
    \chi^{(J)}(\theta)+24n(-1)^{2mJ} , & \theta= 2m\pi \\
    \chi^{(J)}(\theta) , & \text{otherwise}
\end{dcases}\, .
\label{eq:j12_rep}
\end{equation}

\subsubsection{Angular momentum on the lattice}

The lattice is symmetric with respect to a set of rotations that form the octahedral group $O$, which is a subgroup of the continuum rotational group $SO(3)$ (if we boost the system, or use an asymmetrical lattice, the symmetry will be broken down to a more restrictive one). The octahedral group consists of 24 group elements, each corresponding to a discrete rotation that leaves invariant a cube, which are enumerated in the following 5 classes:
\begin{itemize}[label={--}]
\item $E$: identity.
\item $8C_3$: rotations of $\pm 2\pi/3$ about the four body diagonals.
\item $3C_2$: rotations of $\pi$ about the coordinate axes.
\item $6C_4$: rotations of $\pm \pi/2$ about the coordinate axes.
\item $6C'_2$: rotations of $\pi$ about the six axes parallel to the six face diagonals.
\end{itemize}
If we consider spatial inversion, we get the octahedral group $O_h$, where $O_h=O\otimes C_i$. With the inclusion of inversion, we get 24 more group elements, which are also organized into 5 classes:
\begin{itemize}[label={--}]
\item $I$: inversion
\item $8IC_3 =8S_6$: rotations of $\pm 2\pi/3$ about the four body diagonals + inversion.
\item $3IC_2 =3\sigma_h$: rotations of $\pi$ about the coordinate axes + inversion.
\item $6IC_4 =6S_4$: rotations of $\pm \pi/2$ about the coordinate axes + inversion.
\item $6IC'_2 =6\sigma_d$: rotations of $\pi$ about the six axes parallel to the six face diagonals + inversion.
\end{itemize}
Therefore, $O_h$ has 48 elements divided into 10 classes. This will correspond to 10 irreps (5 irreps for each parity), which are denoted as $A_1^{\pm}, A_2^{\pm}, E^{\pm}, T_1^{\pm}, T_2^{\pm}$, with respective dimensions $1, 1, 2, 3, 3$.

When the objects that are rotated involve half-integer values of the angular momentum, the group elements double, forming the double octahedral group $O^D_h$. The elements of this group are labeled with a horizontal line over the name of the element. The doubling of the elements does not mean that the classes also double (e.g., if it is a class of rotation by $\pi$, the original elements and the elements of the double group are members of the same class). Therefore, there will only be 6 more classes, with 6 more irreps, $G_1^{\pm}, G_2^{\pm}, H^{\pm}$, with respective dimensions $2,2,4$.

When we boost the system inside the lattice, the Lorentz transformation deforms the cubical volume and only some subgroups of the original $O^D_h$ group survive. Since we only work with boosts of the form $\bm{d}=(0,0,n)$, the relevant symmetry groups are the tetragonal point group $D_{4h}$ (if both particles have the same mass) and $C_{4v}$ (if both particles have different masses). For a more general boost, the relevant point and double groups are summarized in Refs.~\cite{Rummukainen:1995vs,Moore:2005dw}. The tetragonal point group $D_{4h}$ has $8\times 2$ (including parity) elements grouped into 10 classes, and the corresponding irreps are denoted as $A_1^{\pm}, A_2^{\pm}, B_1^{\pm}, B_2^{\pm}, E^{\pm}$, with respective dimensions $1, 1, 1, 1, 2$. For the double tetragonal group $D^D_{4h}$, there are only 4 more classes, so 4 more irreps, $G^{\pm}_1, G_2^{\pm}$, with respective dimensions $2,2$. The tetragonal point group $C_{4v}$ has 8 elements (the spatial inversion is not an element) grouped into 5 classes, with the irreps denoted as $A_1, A_2, B_1, B_2, E$, with respective dimensions $1, 1, 1, 1, 2$. For the double tetragonal group $C^D_{4v}$ (also known as $\text{Dic}_4$), there are only 2 more classes, so 2 more irreps, $G_1, G_2$, with respective dimensions $2,2$.

The character tables of $O^D_h$, $D^D_{4h}$, and $C^D_{4v}$ are shown in Tables~\ref{tab:Oh_char},~\ref{tab:D4_char}, and~\ref{tab:C4_char}, respectively, from~\cref{appen:GTtables}, where the values of $\theta$ are also included.

\subsubsection{Decomposition of the angular momentum}

In order to determine the decomposition of the angular momentum in the continuum into the corresponding lattice group, we have to extract the coefficients $m_i$ from the combination of the characters of $SO(3)$ (see Eq.~\eqref{eq:S03_char}) and the characters of the corresponding point (double) group (see Tables~\ref{tab:Oh_char},~\ref{tab:D4_char}, and~\ref{tab:C4_char}) into Eq.~\eqref{eq:reduction_2}. For example, let us try to compute the case $J=0$ for the group $O^D_h$. Starting with $A^+_1$,
\begin{equation}
\begin{aligned}
    m_{A^+_1}&=\frac{1}{h}\sum_R n_R\, \chi^{A^+_1}(R)\chi^{(J=0)}(R)\\
    &=\frac{1}{96}\left[ \chi^{A^+_1}(E)\chi^{(0)}(4\pi)+\chi^{A^+_1}(\overline{E})\chi^{(0)}(2\pi)+8\chi^{A^+_1}(C_3)\chi^{(0)}(2\pi/3) +\cdots \right] \\
    &=\frac{1}{96}\left[ 1\cdot 1+1\cdot 1+8(1\cdot 1) +\cdots \right] =1\, ,
\end{aligned}
\end{equation}
while for the rest of the irreps we get $\smash{m_{A^+_2}}=m_{E^+}=\cdots=m_{H^-}=0$. Therefore, $J=0$ is only described by the $A^+_1$ irrep. Doing the same for angular momentum up to $J<12$, we get Table~\ref{tab:spin_reduction_o} for the group $O^D_h$, Table~\ref{tab:spin_reduction_d} for the group $D^D_{4h}$, and Table~\ref{tab:spin_reduction_c} for the group $C^D_{4v}$. Having calculated the values for $J<12$, Eq.~\eqref{eq:j12_rep} allows us to find the $J+12n$ values:
\begin{table}[H]
\centering
\centering
\renewcommand{\arraystretch}{1.2}
\begin{tabu}{c @{\hskip 0.3in} c @{\hskip 0.5in} c}
 & $J+12n$ integer & $J+12n$ half-integer \\ \cmidrule(r{3em}){2-2}\cmidrule{3-3}
$O^D_h$ & $\Gamma^{(J)} \oplus n (A^{\pm}_1\oplus A^{\pm}_2\oplus 2E^{\pm} \oplus 3T^{\pm}_1\oplus 3T^{\pm}_2)$ & $\Gamma^{(J)} \oplus 2n(G^{\pm}_1\oplus G^{\pm}_2 \oplus 2 H^{\pm})$ \\
$D^D_{4h}$ & $\Gamma^{(J)} \oplus 3n (A^{\pm}_1\oplus A^{\pm}_2\oplus B^{\pm}_1\oplus B^{\pm}_2\oplus 2E^{\pm})$ & $\Gamma^{(J)}\oplus 6n(G^{\pm}_1\oplus G^{\pm}_2)$ \\
$C^D_{4v}$ & $\Gamma^{(J)} \oplus 3n (A_1\oplus A_2\oplus B_1\oplus B_2\oplus 2E)$ & $\Gamma^{(J)}\oplus 6n(G_1\oplus G_2)$ \\
\end{tabu}
\end{table}
\noindent
where $\Gamma^{(J)}$ is the corresponding decomposition of the angular momentum $J$.

However, when working on the lattice, we have the inverse problem: we want to know the momentum content of a lattice energy eigenstate transforming as a single irrep of the corresponding point (or double) group (the operators are build to transform correctly under the discrete groups, and not the full rotation group). In other words, we want the inverse of the correspondence given by Eq.~\eqref{eq:reduction_2} and Table~\ref{tab:spin_reduction_o}. The results for the group $O^D_h$ are given in Table~\ref{tab:spin_reduction_o_2}, for the group $D^D_{4h}$ in Table~\ref{tab:spin_reduction_d_2}, and for the group $C^D_{4v}$ in Table~\ref{tab:spin_reduction_c_2}. It is interesting to see that the groups that do not have inversion as an element, mix partial waves with opposite parity.

\subsection{Lüscher's formalism}\label{subsec:LuscherQCD}

When two hadrons are put inside a box with periodic BC, the energy levels of the system, now quantized, will shift from the free-particle values due to their mutual interactions. The connection between these finite-volume levels and the infinite-volume scattering parameters, below three-particle production, was first presented in Refs.~\cite{Luscher:1986pf,Luscher:1990ux}, and later generalized to two-hadrons with non-zero momenta and non-zero spin~\cite{Rummukainen:1995vs,Beane:2003da,Bedaque:2004kc,Feng:2004ua,Kim:2005gf,He:2005ey,Christ:2005gi,Ishizuka:2009bx,Luu:2011ep,Davoudi:2011md,Leskovec:2012gb,Hansen:2012tf,Briceno:2012yi,Gockeler:2012yj,Briceno:2013lba,Briceno:2013bda,Briceno:2013hya,Briceno:2014oea,Lee:2017igf,Briceno:2017max} and even three-particle states~\cite{Hansen:2015zga,Hammer:2017kms,Mai:2017bge,Briceno:2018mlh,Doring:2018xxx,Mai:2018djl,Blanton:2019igq}.
Throughout this section as well as in~\cref{sec:450results}, where the results will be discussed, we will assume that the box is larger than the range of the interaction of the system~\cite{Luscher:1985dn}, given by the lightest particle that can be exchanged, which for all the baryon-baryon channels studied in this work is the pion, so $m_\pi L \gg 1$ (corrections that take into account these effects have been computed in Refs.~\cite{Bedaque:2006yi,Sato:2007ms}).

We will follow the notation of Refs.~\cite{Luu:2011ep,Briceno:2013lba,Briceno:2013bda}, where two-baryon systems are labeled by the spin $s$ and orbital momentum $L$, both coupled to give a total angular momentum $J$. The Lüscher's quantization condition (QC) is given by
\begin{equation}
    \det\left[ (\mathcal{M}^{\infty})^{-1} + \delta\mathcal{G}^V \right] = 0 \, ,
\label{eq:QC}
\end{equation}
where $\mathcal{M}^{\infty}$ is the infinite volume scattering amplitude. In the non-relativistic limit, for channels with no partial-wave mixing (like in the $NN$ spin-singlet state), it has the following form,
\begin{equation}
    (\mathcal{M}^{\infty})^{LL';s}_{JM_J}=\frac{2\pi}{\tilde{M}k^*}\frac{e^{2\imag\delta^{Ls}_{J}}-1}{2\imag}\delta_{L,L'}\delta_{L,J}=\frac{2\pi}{\tilde{M}k^*}\frac{1}{\cot\delta^{Ls}_{J}-\imag}\delta_{L,L'}\delta_{L,J} \,,
\label{eq:ampl_nomixing}
\end{equation}
where $\tilde{M}$ is the reduced mass of the system, $k^*$ is the momentum of each baryon in the c.m.\ frame, and $\delta^{Ls}_{J}$ is the scattering phase shift. When partial-wave mixing is possible (like in the $NN$ spin-triplet state), some paramerization of the scattering amplitude has to be used, like the Stapp-Ypsilantis-Metropolis (or barred)~\cite{Stapp:1956mz}, the Blatt–Biedenharn~\cite{Blatt:1952zza,Blatt:1952zz}, the Bryan-Klarsfeld-Sprung~\cite{Bryan:1981cc,Klarsfeld:1983iph,Bryan:1984sq,Sprung:1985zz}, or the Kabir-Kermode~\cite{Kabir:1987} parametrization. In this work we will use the Blatt–Biedenharn one, which has the following form,
\begin{equation}
\begin{aligned}
    (\mathcal{M}^{\infty})^{J\pm 1 J\pm 1;s}_{JM_J} &= \frac{2\pi}{\tilde{M}k^*} \frac{\cos^2(\epsilon_J)e^{2\imag\delta^{Ls}_{J}}+\sin^2(\epsilon_J)e^{2\imag\delta^{L^\prime s}_{J}}-1}{2\imag} \delta_{L,J-1} \delta_{L^\prime,J+1} \,, \\
    (\mathcal{M}^{\infty})^{J\pm 1 J\mp 1;s}_{JM_J} &= \frac{2\pi}{\tilde{M}k^*} \frac{\cos(\epsilon_J)\sin(\epsilon_J)(e^{2\imag\delta^{Ls}_{J}}-e^{2\imag\delta^{L^\prime s}_{J}})}{2\imag} \delta_{L,J-1} \delta_{L^\prime,J+ 1} \,,
\label{eq:ampl_mixing}
\end{aligned}
\end{equation}
where now we have two scattering phase shifts and a mixing angle $\epsilon_J$. This parametrization is usually written in terms of the $\mathcal{S}$ matrix,
\begin{equation}
\begin{aligned}
    \mathcal{S} &= 1+\imag\frac{\tilde{M}k^*}{\pi}\mathcal{M}^{\infty}\\
    &=\setlength\arraycolsep{5pt}\def\arraystretch{1.2}\begin{pmatrix*}[r] \cos(\epsilon_J) & -\sin(\epsilon_J) \\ \sin(\epsilon_J) & \cos(\epsilon_J) \end{pmatrix*}\begin{pmatrix} e^{2\imag\delta^{J-1,s}_{J}} & 0 \\ 0 & e^{2\imag\delta^{J+1,s}_{J}} \end{pmatrix}\begin{pmatrix*}[r] \cos(\epsilon_J) & -\sin(\epsilon_J) \\ \sin(\epsilon_J) & \cos(\epsilon_J) \end{pmatrix*}\, .
\label{eq:smat_mixing}
\end{aligned}
\end{equation}

The second term, $\delta\mathcal{G}^V$, is a function of the finite-volume energy levels, and has the following form,
\begin{equation}
    [\delta\mathcal{G}^V]_{JM_J,Ls;J'M_{J'},L' s'} = \frac{\imag\tilde{M}k^*}{2\pi} \delta_{ss'} \left[ \delta_{JJ'} \delta_{M_JM_{J'}} \delta_{LL'} + \imag F^{(FV)}_{JM_J,L;J'M_{J'},L'} \right]\, ,
\end{equation}
with
\begin{equation}
\begin{aligned}
    F^{(FV)}_{JM_J,L;J'M_{J'},L'} & = \sum_{m,m',\nu}\langle L m; s \nu | J M_J \rangle \langle L' m'; s \nu | J' M_{J'} \rangle \ \overline{F}^{(FV)}_{L m;L' m'}\, , \\
    \overline{F}^{(FV)}_{L m;L' m'} & = \frac{(-1)^{m'}}{q \gamma \pi^{3/2}}\ \sqrt{(2L+1)(2L'+1)}\ \sum_{\overline{l}=|L-L'|}^{|L+L'|} \frac{\sqrt{2\overline{l}+1}}{ q^{\overline{l}}} \\
    & \qquad\quad \times\sum_{\overline{m}=-\overline{l}}^{\overline{l}} \begin{pmatrix} L & \overline{l} & L' \\ 0 & 0 & 0 \end{pmatrix} \begin{pmatrix} L & \overline{l} & L' \\ -m & -\overline{m} & m' \end{pmatrix}\, \mathcal{Z}^{\bm{d}}_{\overline{l},\overline{m}}(1;q^2) \, , 
\end{aligned}
\end{equation}
where $q=\frac{k^* L_s}{2\pi}$ (here $L_s$ is the spatial size of the lattice, to avoid confusion with the angular momentum $L$), $\langle L' m'; s \nu | J' M_{J'} \rangle$ are Clebsch-Gordan coefficients, and $\begin{pmatrix} l_1 & l_2 & l_3 \\ m_1 & m_2 & m_3\end{pmatrix}$ are Wigner $3j$ symbols~\cite{Wigner1959}. The generalized $\mathcal{Z}$-function is defined through the equation
\begin{equation}
    \mathcal{Z}^{\bm{d}}_{lm}(s;q^2)=\sum_{\bm{r}\in P_{d}} \frac{\mathcal{Y}_{lm}(\bm{r})}{\left(|\bm{r}|^2-q^2\right)^s}\, ,
\label{eq:zeta_fun}
\end{equation}
where $\mathcal{Y}_{lm}(\bm{r})=|\bm{r}|^l Y_{lm}(\hat{\bm{r}})$, with $Y_{lm}(\hat{\bm{r}})$ being the spherical harmonics, and the summation is over the set
\begin{equation}
    P_{d}=\left\lbrace \bm{r} \left| \; \bm{r} = \hat{\gamma}^{-1}\left(\bm{n}-\alpha \bm{d} \right) \right. , \; \bm{n}\in \mathbb{Z}^3\right\rbrace\, .
\label{eq:QC_nonint}
\end{equation}
In order to understand this relation and the new quantities that appear, we need to introduce some new definitions. Let us consider a system of two hadrons (with masses $m_1$ and $m_2$) in a finite box with spatial dimensions $L_s^3$ with periodic BC. The system, in the laboratory frame (LAB), is given a non-zero total three-momentum $\bm{P}$, which has to satisfy the following condition (in the non-interacting case),
\begin{equation}
    \bm{P} = \bm{p}_1+\bm{p}_2 = \frac{2\pi}{L_s}\bm{d}\, , \quad \bm{d}\in\mathbb{Z}^3\, .
\label{eq:P_mom}
\end{equation}
Then, the energy eigenvalues for this system are $E = \sqrt{m_1^2+|\bm{p}_1|^2} + \sqrt{m_2^2+|\bm{p}_2|^2}$.
In the c.m.\ frame (denoted with an asterisk $*$), the total momentum is zero, and the relative one is $\bm{p}^* = \bm{p}_1^* = -\bm{p}_2^*$. Knowing that the total four-momentum squared is invariant, the relation between the LAB and c.m.\ frames is
\begin{equation}
    {E^*}^2=E^2-\bm{P}^2\, .
\end{equation}
Since the c.m.\ frame moves with a velocity $\bm{v}=\bm{P}/E$ with respect to the LAB frame, the momenta $\bm{p}_i$ and $\bm{p}^*$ are related via the standard Lorentz transformation,
\begin{equation}
    \bm{p}_1 = \hat{\gamma} (\bm{p}^*+\bm{v}E^*_1)\, ,\quad \bm{p}_2 = \hat{\gamma} (-\bm{p}^*+\bm{v}E^*_2)\, ,
\label{eq:p_lorentz}
\end{equation}
where the boost factor $\hat{\gamma}$ acts in the direction of $\bm{v}$ (which is the same direction as $\bm{d}$),
\begin{equation}
    \hat{\gamma} \bm{n} = \gamma \bm{n}_{\parallel} + \bm{n}_{\perp} \, , \quad \hat{\gamma}^{-1} \bm{n} = \gamma^{-1} \bm{n}_{\parallel} + \bm{n}_{\perp} \, , \quad \bm{n}_{\parallel} = \frac{\bm{n}\cdot \bm{d}}{|\bm{d}|^2} \bm{d}, \quad \bm{n}_{\perp} = \bm{n} - \bm{n}_{\parallel}\, ,
\end{equation}
and $E^*_1$ and $E^*_2$ are the energy eigenvalues of the particles 1 and 2 in the c.m.\ frame, respectively. To extract them, knowing that $E^*_i=\sqrt{m_i^2+|\bm{p}^*|^2}$, we can write $\bm{p}^*$ in terms of $E^*$ and $m_{1,2}$, and then substitute it back to $E^*_i$,
\begin{equation}
    E^*_1 = \sqrt{m_1^2+\frac{1}{4{E^*}^2} \left[ ({E^*}^2-m_1^2-m_2^2)^2 - 4 m_1^2 m_2^2 \right]} = \frac{E^*}{2} \left( 1 + \frac{m_1^2-m_2^2}{{E^*}^2} \right) \, .
\label{eq:CM_energy1}
\end{equation}
For $E^*_2$, we get the same as for $E^*_1$, but with $m_1$ and $m_2$ interchanged. To know the values of the allowed momenta $\bm{p}^*$ in the c.m.\ frame for the non-interacting case, we use Eqs.~\eqref{eq:P_mom},~\eqref{eq:p_lorentz} and~\eqref{eq:CM_energy1},
\begin{equation}
\begin{aligned}
    \bm{p}^* & = \hat{\gamma}^{-1} \bm{p}_1 - \bm{v} E^*_1 = \hat{\gamma}^{-1} (\bm{p}_1-\gamma \bm{v} E^*_1) = \hat{\gamma}^{-1} \left[\bm{p}_1 - \frac{E}{E^*} \frac{\bm{P}}{E} \frac{E^*}{2} \left( 1 + \frac{m_1^2-m_2^2}{{E^*}^2} \right) \right] \\ 
    &=\hat{\gamma}^{-1} \left[\bm{p}_1 - \frac{\bm{P}}{2} \left( 1 + \frac{m_1^2-m_2^2}{{E^*}^2} \right) \right] = \hat{\gamma}^{-1} \left( \bm{p}_1 - \alpha\bm{P} \right)\, ,
\end{aligned}
\end{equation}
where we have defined $\alpha=\frac{1}{2}\left( 1 + \frac{m_1^2-m_2^2}{{E^*}^2} \right)$.
Thus, we see that $\bm{p}^*$ is quantized to the values $\bm{p}^* = \frac{2\pi}{L} \bm{r}$, with $\bm{r}\in P_d$, shown in Eq.~\eqref{eq:QC_nonint}.

The sum in~\eqref{eq:zeta_fun} converges when $\Ree(2s)>l+3$, but can be analytically continued to the whole complex plane, since we are interested in evaluating the $\mathcal{Z}$-function at $s=1$. This is shown in~\cref{appen:Zfun}, where a convenient expression has been derived to exponentially accelerate the numerical evaluation of the function. The symmetry properties of $\mathcal{Z}^{\bm{d}}_{lm}$ follow directly from the definition in Eq.~\eqref{eq:zeta_fun} and the properties of the spherical harmonic functions under symmetry operations:
\begin{enumerate}[label=(\roman*)]
\item Under the interchange $m_1\leftrightarrow m_2$, it can be shown that
\begin{equation}
    \left\lbrace \bm{r} \left|\; \bm{r} = \hat{\gamma}^{-1} \left[\bm{n}-\alpha(m_1 ,m_2) \bm{d} \right] \right. \right\rbrace =\left\lbrace -\bm{r} \left|\; \bm{r} = \hat{\gamma}^{-1} \left[\bm{n}-\alpha(m_2,m_1) \bm{d}\right] \right. \right\rbrace \, ,
\end{equation}
which, using $Y_{lm}(-\hat{\bm{r}})=(-1)^l Y_{lm}(\hat{\bm{r}})$, results in
\begin{equation}
    \mathcal{Z}^{\bm{d}(m_1, m_2)}_{lm}(s;q^2) = (-1)^l \mathcal{Z}^{\bm{d}(m_2, m_1)}_{lm}(s;q^2)\, .
\end{equation}
\item A consequence of the previous result is that for odd $l$ the $\mathcal{Z}$-function must vanish when $m_1=m_2$ or $\bm{d}=\bm{0}$,
\begin{equation}
    \mathcal{Z}_{lm}(s;q^2)=0 \quad \text{for $l$ odd}\, .
\end{equation}
\item The property $Y_{l-m}=(-1)^m Y^*_{lm}$ translates directly to
\begin{equation}
    \mathcal{Z}^{\bm{d}}_{l-m}(s;q^2) = (-1)^m [\mathcal{Z}^{\bm{d}}_{lm}(s;q^2)]^* \, .
\end{equation}
\item For $\bm{d}=\bm{0}$ or $\bm{d}=(0,0,n)$, the system is symmetric under rotations around the $z$-axis by $\pi/2$, $\varphi\rightarrow\varphi+\pi/2$, so $Y_{lm}(\theta,\varphi+\pi/2)=f(\theta)e^{\imag m (\varphi + \pi/2)}=Y_{lm} e^{\imag m \pi/2}$,
\begin{equation}
    \mathcal{Z}^{\bm{d}}_{lm}(s;q^2)=\imag^m \mathcal{Z}^{\bm{d}}_{lm}(s;q^2)\; \Rightarrow \; \mathcal{Z}^{\bm{d}}_{lm}(s;q^2) = 0 \quad \text{for }m\neq 0,4,8,\ldots
\end{equation}
\item For $\bm{d}=\bm{0}$ or $\bm{d}=(0,0,n)$, the system is symmetric under a mirror reflection about the $xz$-plane, $\varphi\rightarrow 2\pi-\varphi$, so $Y_{lm}(\theta,\varphi)=Y_{lm}(\theta,2\pi-\varphi)=Y^*_{lm}(\theta,\varphi)$,
\begin{equation}
    \mathcal{Z}^{\bm{d}}_{lm}(s;q^2)=\mathcal{Z}^{\bm{d}}_{l-m}(s;q^2) \, .
\end{equation}
\item Additional relations between non-zero $\mathcal{Z}_{lm}$ can be found by looking at the symmetries of the spherical harmonics. For any rotation matrix $R$ corresponding to each symmetry operation of the discrete group,
\begin{equation}
    \mathcal{Y}_{lm}(R\bm{r})=\sum_{m'=l}^{l}\mathcal{D}^{(l)}_{mm'}(R)\mathcal{Y}_{lm'}(\bm{r})\, ,
\end{equation}
where $\mathcal{D}^{(l)}_{mm'}(R)$ is the Wigner $\mathcal{D}$-matrix~\cite{Luscher:1990ux} (this matrix is written in terms of the Euler angles of the rotation, which can be read off from, e.g., Ref.~\cite{Lee:2017igf}). Then, by looking at Eq.~\eqref{eq:zeta_fun}, we see that
\begin{equation}
    \mathcal{Z}^{\bm{d}}_{lm}(s;q^2)=\sum_{m'=l}^{l}\mathcal{D}^{(l)}_{mm'}(R)\mathcal{Z}^{\bm{d}}_{lm'}(s;q^2)\, .
\end{equation}
For example, if we look at the group $O^D_h$ (with $\bm{d}=\bm{0}$), we obtain that $\mathcal{Z}_{20}(s;q^2)=0$. Other possible relations are
\begin{equation}
\begin{aligned}
    \mathcal{Z}_{44}(s;q^2) &= \sqrt{\frac{5}{14}} \mathcal{Z}_{40}(s;q^2)\, , &\quad \mathcal{Z}_{64}(s;q^2) &= -\sqrt{\frac{7}{2}} \mathcal{Z}_{60}(s;q^2)\, , \\
    \mathcal{Z}_{84}(s;q^2) &= \frac{1}{3}\sqrt{\frac{14}{11}} \mathcal{Z}_{80}(s;q^2)\, , &\quad \mathcal{Z}_{88}(s;q^2) &= \frac{1}{3}\sqrt{\frac{65}{22}} \mathcal{Z}_{80}(s;q^2)\, .
\end{aligned}
\end{equation}
For the relations between $\mathcal{Z}_{lm}$ with higher values of $l$ and other groups, see Refs.~\cite{Luu:2011ep,Gockeler:2012yj,Briceno:2013lba,Morningstar:2017spu}.
\end{enumerate}
A useful short-hand notation that we will use when writing the QCs is the following,
\begin{equation}
    z_{lm}=\frac{1}{q^{l+1}\gamma\pi^{3/2}\sqrt{2l+1}}\mathcal{Z}^{\bm{d}}_{lm}(1;q^2)\, .
\label{eq:zlm_shorthandnot}
\end{equation}

The QC in Eq.~\eqref{eq:QC} is written in the basis $|J M_J L\rangle$, not suitable for lattice calculations, where we work in the basis of the irreps $\Gamma$ of the point (or double) group. This basis can be written as $|\Gamma \sigma \nu J L \rangle$, with $\sigma\in\{1,\ldots,\dim\Gamma\}$ and $\nu\in\{1,\ldots,N(\Gamma,J)\}$, where $N(\Gamma,J)$ denotes the multiplicity of $J$ in the irrep $\Gamma$. The basis vectors $|\Gamma \sigma \nu J L \rangle$ can be decomposed in terms of the $|J M_J L\rangle$ basis,
\begin{equation}
    |\Gamma \sigma \nu J L \rangle = \sum_{M_J} c^{\Gamma\sigma\nu}_{J M_J} |J M_J L\rangle \, ,
\end{equation}
where the coefficients $c^{\Gamma\sigma\nu}_{J M_J}$ can be read off directly from Tables~\ref{tab:vectors_o},~\ref{tab:vectors_d}, and~\ref{tab:vectors_c}, in which the basis vectors of $O_{h}$, $D_{4h}$, and $C_{4h}$ are listed (only for integer values of total angular momentum). The matrix elements of $\delta \mathcal{G}^V$ in the new basis are given by
\begin{equation}
    \langle \Gamma \sigma \nu J L |\delta\mathcal{G}^V |\Gamma' \sigma' \nu' J' L' \rangle=\sum_{M_J,M_{J'}} \left( c^{\Gamma\sigma\nu}_{J M_J} \right)^* c^{\Gamma'\sigma'\nu'}_{J' M_{J'}} [\delta\mathcal{G}^V]_{JM_JL;J'M_{J'}L'}\, .
\end{equation}
A similar block-diagonalization procedure is presented in Refs.~\cite{Luu:2011ep,Briceno:2013lba}. According to Schur's lemma~\cite{Bernard:2008ax}, $\delta\mathcal{G}^V$ is partially diagonalized in this new basis,
\begin{equation}
    \langle \Gamma \sigma \nu J L |\delta\mathcal{G}^V |\Gamma' \sigma' \nu' J' L' \rangle=\delta_{\Gamma\Gamma'}\delta_{\sigma\sigma'} [\delta\mathcal{G}^V]^{\Gamma}_{JL\nu;J'L'\nu'}\, ,
\end{equation}
and Eq.~\eqref{eq:QC} can thus be rewritten as
\begin{equation}
    \prod_\Gamma \det\left\lbrace\left([\mathcal{M}^{\infty}]^{LL'}_J\right)^{-1}\delta_{JJ'}\delta_{\nu\nu'} + [\delta\mathcal{G}^V]^{\Gamma}_{JL\nu;J'L'\nu'} \right\rbrace=0\, ,
\end{equation}
which is independent of $\sigma$. For practical reasons, the matrices are truncated in the space of total angular momentum $J$ and orbital momentum $L$. Now let us look in detail at the different QCs for baryon-baryon systems in both spin-singlet and spin-triplet states. Since we will study the systems at very low energy, we will only need the relevant irrep that couple to the lowest $J$ value (like the $A_1$ for $J=0$). A full list of the QCs for all the irreps in both $NN$ spin channels can be found in Refs.~\cite{Briceno:2013lba,Briceno:2013bda}.

\subsubsection{Spin-singlet states}

For spin-singlet states, no mixing between different partial waves is possible, and the parametrization of the scattering amplitude in Eq.~\eqref{eq:ampl_nomixing} can be used, simplifying the QC,
\begin{equation}
    \det\left[\delta_{JJ'}\delta_{\nu\nu'} \cot\delta_{J} - F^{(FV,\Gamma)}_{J\nu;J'\nu'} \right]=0\, .
    \label{eq:QC_spinsinglet}
\end{equation}
For the $O_h$ group, $A^+_1$ is the only irrep that couples to $J=0$. Evaluating Eq.~\eqref{eq:QC_spinsinglet} up to $J=4$, and using the short-hand notation $z_{lm}$ from Eq.~\eqref{eq:zlm_shorthandnot},
\begin{equation}
    \det\left[\begin{pmatrix}\cot\delta_{\1s0} & 0 \\ 0 & \cot\delta_{^1G_4} \end{pmatrix}- \begin{pmatrix}z_{00} & \sqrt{\frac{108}{7}}z_{40} \\ \sqrt{\frac{108}{7}}z_{40} & z_{00}+\frac{324}{143}z_{40}+\frac{80}{11}z_{60}+\frac{560}{143}z_{80} \end{pmatrix}\right]=0\, .
\end{equation}
Assuming that the phase shift from higher partial waves (like $L=4$) vanish, the QC reduces to
\begin{equation}
    \cot\delta_{\1s0}-z_{00}=0\quad \rightarrow \quad k^*\cot\delta_{\1s0} = \frac{2}{\sqrt{\pi} L} \mathcal{Z}_{00}(1;q^2)\, .
\label{eq:QC_Od_A1+}
\end{equation}

Boosts of the form $\bm{d}=(0,0,2n)$ will be used in the present work. Ref.~\cite{Briceno:2013lba} showed that, in the non-relativistic limit, systems with equal masses and with boost given by $\bm{d}=(0,0,2n)$ behave as if they were boosted with $\bm{d}=(0,0,0)$. While this is no longer true when masses are unequal, the same QC, given by Eq.~\eqref{eq:QC_Od_A1+}, can be used when higher partial waves are neglected, a reasonable assumption in the low-energy regime. Therefore,  
\begin{equation}
    k^*\cot\delta_{\1s0} = \frac{2}{\sqrt{\pi}\gamma L} \mathcal{Z}^{\bm{d}}_{00}(1;q^2)\, .
\label{eq:QC_A1+_boost}
\end{equation}

\subsubsection{Spin-triplet states}

For spin-triplet states, the strong interaction mixes different partial waves for a given $J$ value, $L\in\{J-1,J+1\}$, and some type of parametrization of the scattering amplitude has to be used, like Eq.~\eqref{eq:ampl_mixing}. Now, for the $O_h$ group, $T^+_1$ is the only irrep that couples to $J=1$. Let us analyze the mixing between the $S$ and $D$ waves, which can be done by taking $J=1$ as the maximum value. Rewriting the QC in terms of the $\mathcal{S}$ matrix (see Eq.~\eqref{eq:smat_mixing}),
\begin{equation}
    \det\left[ \mathcal{S}\left(\imag - F^{(FV,T^+_1)}\right)+\left(\imag + F^{(FV,T^+_1)}\right)\right]=0\, , 
\end{equation}
where $F^{(FV,T^+_1)}$ only has diagonal entries, both being $z_{00}$. The two phase shifts from the Blatt–Biedenharn parametrization are usually called the $\alpha$-wave $\delta_{\alpha}$ and the $\beta$-wave $\delta_{\beta}$. While the first one is predominantly $S$ wave with a small admixture of $D$ wave, the second one is predominantly $D$ wave with a small admixture of $S$ wave. For this case, if we do not consider contamination from the $\delta_{\beta}$ since it is expected to be small, we get
\begin{equation}
    \cot\delta_{\alpha}-z_{00}=0\quad \rightarrow \quad k^*\cot\delta_{\alpha} = \frac{2}{\sqrt{\pi} L} \mathcal{Z}_{00}(1;q^2)\, ,
\label{eq:QC_Od_T1+}
\end{equation}
and we recover Eq.~\eqref{eq:QC_Od_A1+}. The same argument made for the spin-singlet states with boosts of the form $\bm{d}=(0,0,2n)$ applies here too.

In order to have access to the mixing parameter $\epsilon_1$, one has to go to a boosted frame. In Ref.~\cite{Briceno:2013bda} it was shown that the QC for the irreps $A^+_2$ and $E^+$ from the group $D_{4h}$ gives us access to both $\delta_{\alpha}$ and $\epsilon_1$. In Ref.~\cite{Orginos:2015aya}, the NPLQCD Collaboration attempted to extract this mixing parameter by building correlation functions of the deuteron with boost $\bm{d}=(0,0,1)$. The $A^+_2$ irrep was accessed by projecting the spin in the perpendicular direction of $\bm{d}$ ($j_z=0$), while the $E^+$ irrep was obtained from the corresponding parallel spin projections ($j_z=\pm 1$). However, since the energy difference between these two irreps was statistically consistent with zero, no useful bound was placed on $\epsilon_1$.

For clarity, the $\delta_\alpha$ will be denoted as $\delta_{\3s1}$ for the rest of the thesis.

\subsection{Binding energy extraction}\label{subsec:QC_binding}

As first explored in Ref.~\cite{Beane:2003da}, being later on expanded in Refs.~\cite{Sasaki:2006jn,Konig:2011nz,Bour:2011ef,Davoudi:2011md}, Lüscher's formalism can be applied to extract the binding momenta from the finite-volume ground-state energies.

The bound states in the infinite-volume limit correspond to poles of the scattering amplitude. Since we are dealing with the region $k^{*2}<0$, we define the binding momentum as $k^* \equiv \imag\kappa^{(\infty)}$ (so that $\kappa^{(\infty)2}>0$), and the pole position can be obtained from
\begin{equation}
    \left. k^*\cot\delta \right|_{k^*=\imag\kappa^{(\infty)}}+\kappa^{(\infty)}=0\, .
\label{eq:pole_binding}
\end{equation}
However, in finite-volume this relation gets modified. We can start by expanding Eq.~\eqref{eq:zeta_fun} for the $l=m=0$ case in the region where $q^2<0$. This expansion was first derived in Ref.~\cite{Elizalde:1997jv}, and it uses the Poisson summation formula given in~\cref{appen:Zfun} (Eq.~\eqref{eq:Poissformula}),
\begin{equation}
    \frac{1}{\sqrt{4\pi}}\sum_{\bm{r}\in P_{d}} \frac{1}{\left(|\bm{r}|^2-q^2\right)^s} = \frac{\gamma}{\sqrt{4\pi}} \sum_{\bm{m}\in \mathbb{Z}^3} \int d^3\bm{r} \frac{1}{\left(|\bm{r}|^2-q^2\right)^s} e^{\imag 2\pi\bm{m}(\hat{\gamma}\bm{r}-\alpha\bm{d})}\, .
\end{equation}
To compute this integral, we separate the contribution in the sum coming from $\bm{m}=\bm{0}$ and the rest. For the first one, we keep $s$ different from $1$, since the integral is only convergent for $s>\frac{3}{2}$, but in the end we analytically continue it to $s=1$,
\begin{equation}
    \frac{\gamma}{\sqrt{4\pi}}\int d^3\bm{r} \frac{1}{\left(|\bm{r}|^2-q^2\right)^s}=\frac{\gamma}{\sqrt{4\pi}}(-q^2)^{3/2-s}\pi^{3/2}\frac{\Gamma(s-\frac{3}{2})}{\Gamma(s)}\;\overset{s\rightarrow 1}{\longrightarrow}\;-\gamma\pi^{3/2}\sqrt{-q^2}\, .
\end{equation}
For the second one, we do not have this problem and we can set $s=1$ from the beginning,
\begin{equation}
    \frac{\gamma}{\sqrt{4\pi}} \sum_{\bm{m}\neq \bm{0}} \int d^3\bm{r} \frac{1}{|\bm{r}|^2-q^2} e^{\imag 2\pi\bm{m}(\hat{\gamma}\bm{r}-\alpha\bm{d})}=\frac{\gamma}{\sqrt{4\pi}} \sum_{\bm{m}\neq \bm{0}} \frac{\pi}{|\hat{\gamma}\bm{m}|} e^{-\imag 2\pi\alpha\bm{m}\cdot\bm{d}} e^{-\imag 2\pi|\hat{\gamma}\bm{m}|\sqrt{-q^2}}\, .
\end{equation}
Combining both expressions we get\footnote{The different sign in the first exponential compared to Ref.~\cite{Davoudi:2011md} comes from the different sign in the definition of $P_d$ in Eq.~\eqref{eq:QC_nonint}.}
\begin{equation}
    \mathcal{Z}^{\bm{d}}_{00}(1;q^2)\;\overset{q^2<0}{\longrightarrow}\;\frac{\gamma}{\sqrt{4\pi}}\left(-2\pi^2\sqrt{-q^2}+\sum_{\bm{m}\neq \bm{0}} \frac{\pi}{|\hat{\gamma}\bm{m}|} e^{-\imag 2\pi\alpha\bm{m}\cdot\bm{d}} e^{-2\pi|\hat{\gamma}\bm{m}|\sqrt{-q^2}}\right)\, .
\end{equation}
Inserting this relation into either Eq.~\eqref{eq:QC_Od_A1+} or~\eqref{eq:QC_Od_T1+}, and setting $k^*=\imag\kappa$, we obtain the following QC,
\begin{equation}
    \left. k^*\cot\delta \right|_{k^*=\imag\kappa} + \kappa = \frac{1}{L} \sum_{\bm{m}\neq \bm{0}} \frac{1}{|\hat{\gamma}\bm{m}|} e^{-\imag 2\pi\alpha\bm{m}\cdot\bm{d}} \, e^{-|\hat{\gamma}\bm{m}| \kappa L} = \frac{1}{L} F^{\bm{d}}(\kappa L)\, ,
\label{eq:pole_binding_FV}
\end{equation}
where the infinite-volume relation~\eqref{eq:pole_binding} is recovered when taking the limit $L\rightarrow \infty$. In this limit, a perturbative solution to Eq.~\eqref{eq:pole_binding_FV} can be found~\cite{Beane:2003da,Davoudi:2011md},
\begin{equation}
    |k^*|=\kappa^{(\infty)}+\frac{Z^2}{L}F^{\bm{d}}(\kappa^{(\infty)} L)+\mathcal{O}(e^{-2\kappa^{(\infty)} L}/L)\, ,
\label{eq:mom_binding_expression}
\end{equation}
where $Z^2$ is the residue of the scattering amplitude
at the bound-state pole. This expression can be used to extract $\kappa^{(\infty)}$ by fitting the finite-volume values $|k^*|$ obtained with different boosts $\bm{d}$.

An alternative way to extract the binding energies in lattice calculations is by first constraining $k^*\cot\delta$ using higher energies besides the ground-state value, and then solving Eq.~\eqref{eq:pole_binding} to extract $\kappa^{(\infty)}$. This approach, therefore, requires an intermediate step compared with the first method, but does not require a truncation of the sum in Eq.~\eqref{eq:mom_binding_expression}. These two approaches will be discussed in~\cref{subsec:bindingLQCD}.

% Chapter 3 - BB interaction *******************************************
\chapter{Analysis of two-point correlation functions}\label{chap:3}

\section{Extraction of the ground-state energy}

As we have discussed in the previous chapter, the two-point correlation functions constructed have the following spectral representation in Euclidean spacetime,
\begin{equation}
    C(\tau) = \sum_{\mathfrak{n}} \mathcal{Z}_{\mathfrak{n}}\mathcal{Z}^*_{\mathfrak{n}} e^{-\tau E_{\mathfrak{n}}},
\label{eq:correlator_chap3}
\end{equation}
where all quantities are expressed in lattice units. $E_{\mathfrak{n}}$ is the energy of the $\mathfrak{n}$th eigenstate $| E_{\mathfrak{n}} \rangle$ and $\mathcal{Z}_{\mathfrak{n}}$ is an overlap factor.
The lowest-lying energies of the one- and two-baryon systems required for the subsequent analyses can be extracted by fitting the correlation functions to this form. To reliably discern the first few exponents given the discrete $\tau$ values and the finite statistical precision of the computations is a challenging task.
In particular, as mentioned in~\cref{subsec:2ptcorr}, a well-known problem in the study of baryons with LQCD is the exponential degradation of the signal-to-noise ratio in the correlation function as the source-sink time separation increases---an issue that worsens as the masses of the light quarks approach their physical values. Another problem that complicates the study of multi-baryon systems is the small energy separation between the excited states in the finite-volume spectrum, that leads to significant excited-state contributions to correlation functions. To overcome these issues, sophisticated methods have been developed to analyze the correlation functions, such as the generalized pencil-of-function~\cite{Aubin:2010jc} and matrix-Prony~\cite{Beane:2009kya} techniques, as well as signal-to-noise optimization techniques~\cite{Detmold:2014hla}.
Ultimately, a large set of single- and multi-baryon interpolating operators with the desired quantum numbers must be constructed to provide a reliable variational basis to isolate the lowest-lying energy eigenvalues via solving a generalized eigenvalue problem~\cite{Michael:1982gb,Luscher:1990ck} (all these methods are related, as shown in Ref.~\cite{Fischer:2020bgv}). 
Such an approach is not yet widely applied to the study of two-baryon correlation functions, given the large amount of computational resources that are required, but progress is being made. In Refs.~\cite{Francis:2018qch,Horz:2020zvv,Green:2021qol}, a partial set of two-baryon scattering interpolating operators were used to study the two-nucleon and $H$-dibaryon channels at the $SU(3)$ flavor-symmetric point, with results that, in general, disagreed with previous works at similar pion masses~\cite{Beane:2012vq, Berkowitz:2015eaa,Beane:2010hg,Wagman:2017tmp}. An extension of these studies has been carried out~\cite{Amarasinghe:2021}, where additional operators, like hexaquark and quasi-local operators, are included. Investigations continue to understand and resolve the observed discrepancies~\cite{Iritani:2017rlk, Beane:2017edf, Wagman:2017tmp, Davoudi:2017ddj, Yamazaki:2017jfh, Drischler:2019xuo, Davoudi:2020ngi}.

\subsection{Exponential fitting}\label{subsec:2ptfitting}

The straightforward fitting strategy consists in using Eq.~\eqref{eq:correlator_chap3} to fit the calculated correlation functions. Given that correlation functions are only evaluated at a finite number of times and with finite precision, the spectral representation has to be truncated to a relatively small number of exponentials and in a given time range. These choices introduce systematic uncertainties that have to be included in the final result in order to have a good estimation of the total uncertainty. The approach that we will describe, and use later in~\cref{sec:450results}, was used for the first time in Ref.~\cite{Beane:2020ycc} and then applied to other investigations~\cite{Detmold:2020snb,Illa:2020nsi,Parreno:2021ovq}.

The first step in this approach is fixing the time range $[\tau_{\text{min}},\tau_{\text{max}}]$ to be used in the fit. Due to the StN problem with multi-baryon correlation functions, the points at large Euclidean time do not significantly contribute to the $\chi^2$ as compared to the points at earlier times. Therefore, the result should be quite insensitive to the value of $\tau_{\text{max}}$, which can be fixed to a single value during the whole fitting process. In order to find the optimal $\tau_{\text{max}}$ value, several conditions have to be satisfied. The first is that the chosen $\tau_{\text{max}}$ is such that the StN has not degraded to the point where estimations of the correlation functions are not reliable. To achieve this, we fix a threshold value for the noise tolerance, the inverse of the StN, $\text{Var}[C(\tau)]/\overline{C}(\tau)$. In our calculations, this value is fixed to $tol_{\text{noise}}=0.1$. Another fact that we have to consider is that due to limited statistics, the mean value of the correlation function might become negative, points that must be avoided in the fitting procedure. Finally, we could have the situation where very large values for $\tau$ are reached, like $\tau > T/2$, where contributions from the opposite parity state become important. Therefore, we also have to impose a limit on $\tau$, fixed by $tol_{\text{temp}}$, which is set to be smaller than $T/2$ (in practice, when working with baryonic systems, this last condition is not needed since the StN goes above $tol_{\text{noise}}$ before reaching $T/2$). The value of $\tau_{\text{min}}$ does have a big impact in the calculation, since excited states contamination is more noticeable at short times. In order to explore its effects, we sample $\tau_{\text{min}}$ over a range of values. The smallest value is fixed by the temporal non-locality in the lattice action, and given that for the improved action used in~\cref{sec:450results} the transfer matrix involves fields on two adjacent timeslices~\cite{Luscher:1984xn}, $\tau_{\text{min}} \geq 2$ is required. As for the largest value, the minimum length of the fitting window, $\tau_{\text{plateau}}$, is introduced as a free parameter, so $\tau_{\text{min}} \leq \tau_{\text{max}}-\tau_{\text{plateau}}$. In this study, we choose $\tau_{\text{plateau}}=5$, which means that a minimum of five points are fitted. Therefore, for each correlation function to be fitted, $\tau_{\text{min}}$ is randomly sampled between $[2,\tau_{\text{max}}-\tau_{\text{plateau}}]$ for a maximum of $N_{\text{fits}}=200$ times or after all possible choices have been explored.

The next step is to compute the covariance matrix $\mathcal{C}^{ij}_{\tau\tau'}$ for the time range selected previously, $\{\tau,\tau'\}\in\{\tau_{\text{min}},\ldots,\tau_{\text{max}}\}$, and for as many different operators $\{i,j\}\in\{1,\ldots,N_{\text{op}}\}$ available (as it will be explained in~\cref{sec:450results}, different smearings are used, producing two different correlation functions) from the $n\in\{1,\ldots,N_{\text{meas}}\}$ number of correlation functions computed $C^i_n(\tau)$. This matrix will be used during the $\chi^2$ minimization step, and its calculation will be discussed in detail in the next section. However, it is appropriate to note that fits with a large number of points may render the covariance matrix ill-defined if not enough statistical ensembles are used during the evaluation due to finite sample-size fluctuations, making it impossible to compute the inverse, needed for the $\chi^2$. Shrinkage techniques~\cite{stein1956,ledoit2004well,Rinaldi:2019thf} are available to better estimate the underlying covariance matrix in these situations. For that, the estimator $\tilde{\mathcal{C}}^{ij}_{\tau\tau'}$ is defined as a combination of the original covariance matrix $\mathcal{C}^{ij}_{\tau\tau'}$ and a well-conditioned matrix, which is taken to be a diagonal matrix, with its elements taken from $\text{diag}(\mathcal{C}^{ij}_{\tau\tau'})$~\cite{Rinaldi:2019thf}, 
\begin{equation}
   \tilde{\mathcal{C}}^{ij}_{\tau\tau^\prime}(\lambda) = (1-\lambda)\mathcal{C}^{ij}_{\tau\tau'} +  \lambda\sqrt{\mathcal{C}^{ii}_{\tau\tau\phantom{'}}\mathcal{C}^{jj}_{\tau'\tau'}}\delta_{ij}\delta_{\tau\tau'}\, ,
   \label{eq:shrinkcovariance}
\end{equation}
where $\lambda$ is the shrinkage parameter. The two extreme values of this parameter are $\lambda=0$, where we recover the original matrix, and $\lambda=1$, where all the off-diagonal elements are set to zero and we have an uncorrelated covariance matrix. In the infinite-statistics limit, the optimal value of $\lambda$ is zero, but for a finite ensemble, a new parameter $\lambda^*$ is defined, which reduces the average mean-squared difference between $\tilde{\mathcal{C}}^{ij}_{\tau\tau'}(\lambda)$ and the underlying covariance matrix~\cite{Rinaldi:2019thf},
\begin{equation}
    \lambda^*=\min \left(\frac{\overline{b}^2}{d^2},1\right)\, ,
\end{equation}
where the quantities $\overline{b}^2$ and $d^2$ are defined as
\begin{equation}
\begin{aligned}
    \overline{b}^2 &= \frac{1}{N_{\text{meas}}} \sum_{n=1}^{N_{\text{meas}}} \sum_{\tau,\tau'=\tau_{\text{min}}}^{\tau_{\text{max}}} \sum_{i,j=1}^{N_{\text{op}}} \left[y_n^i(\tau)y_n^j(\tau')-\rho^{ij}_{\tau\tau'}\right]^2\, , \\
    d^2 &= \sum_{\tau,\tau'=\tau_{\text{min}}}^{\tau_{\text{max}}} \sum_{i,j=1}^{N_{\text{op}}} \left(\rho^{ij}_{\tau\tau'}-\delta_{ij}\delta_{\tau\tau'}\right)^2\, ,\\
    y_n^i(\tau) &=\frac{C_n^i(\tau)-x^i(\tau)}{\sqrt{S^{ii\phantom{j}}_{\tau\tau\phantom{'}}}}\, ,\quad x^i(\tau) = \frac{1}{N_{\text{meas}}} \sum_{n=1}^{N_{\text{meas}}}C_n^i(\tau)\, ,\\
    \rho^{ij}_{\tau\tau'} &= \frac{S^{ij}_{\tau\tau'}}{\sqrt{S^{ii\phantom{j}}_{\tau\tau\phantom{'}}S^{jj}_{\tau'\tau'}}}\, , \quad S^{ij}_{\tau\tau'} = \frac{1}{N_{\text{meas}}-1} \sum_{n=1}^{N_{\text{meas}}}\left[C_n^i(\tau)-x^i(\tau)\right]\left[C_n^j(\tau')-x^j(\tau')\right]\, .
\end{aligned}
\end{equation}
Therefore, the covariance matrix with the optimal shrinkage parameter is given by $\tilde{\mathcal{C}}^{ij}_{\tau\tau'}(\lambda^*)$. With this matrix, we can build the $\chi^2$ to be minimized and obtain the fit parameters.

With the spectral decomposition from Eq.~\eqref{eq:correlator_chap3}, it is evident that the function $f$ to be fitted can be written as $f(\tau;\bm{E},\bm{Z})=\sum_{\mathsf{n}=0}^\mathsf{e} Z_\mathsf{n} e^{-\tau E_\mathsf{n}}$, where $\mathsf{e}$ is the number of total excited states included, and $\{\bm{E},\bm{Z}\}$ are the parameters (energies and overlap factors) to fit. While the values of $\bm{Z}$ are different for different operators, the values of $\bm{E}$ are set to be the same (different operators might have different overlap to the ground- and excited-states, but the energy of these levels has to be the same). Then, 
\begin{equation}
    \chi^2= \sum_{\tau,\tau'=\tau_{\text{min}}}^{\tau_{\text{max}}} \sum_{i,j=1}^{N_{\text{op}}}\left[\overline{C}^i(\tau)-f(\tau;\bm{E},\bm{Z})\right]\left[\tilde{\mathcal{C}}^{-1}(\lambda^*)\right]^{ij}_{\tau\tau'} \left[\overline{C}^j(\tau')-f(\tau';\bm{E},\bm{Z})\right]\, .
    \label{eq:chi2_2pt}
\end{equation}
Again, depending on the error estimation procedure used, the average value of the correlation function $\overline{C}^i(\tau)$ will be computed differently (as explained in~\cref{sec:errestimation}). In particular, for the analysis presented here, we choose the bootstrap method. Given the linear $\bm{Z}$ dependence of $f$, variable projection (VarPro) techniques~\cite{Golub:2003,Olearly:2013} can be used so that only the $\bm{E}$ parameters enter in the non-linear minimization procedure. For a given value of $\bm{E}$, the $\bm{Z}$ factors can be obtained by solving a system of linear equations. This can be seen if we write Eq.~\eqref{eq:chi2_2pt} in matrix form, and imposing that the partial derivative with respect to $\bm{Z}$ is zero,
\begin{equation}
    \chi^2=[\bm{C}-\bm{X}\bm{Z}]^\top\tilde{\bm{\mathcal{C}}}^{-1}[\bm{C}-\bm{X}\bm{Z}]\;\rightarrow \;\frac{\partial \chi^2}{\partial \bm{Z}}=0\; \Rightarrow \;\bm{Z}=(\bm{X}^\top \tilde{\bm{\mathcal{C}}}^{-1} \bm{X})^{-1} \bm{X}^\top\tilde{\bm{\mathcal{C}}}^{-1}\bm{C}\, ,
\end{equation}
where $\bm{X}$ is the matrix with components $e^{-\tau E_{\mathsf{n}}}$, whose columns contain the different energy levels (different $\mathsf{n}$) and rows the different times $\tau$ and operators (as said before, given $\mathsf{n}$, the value $E_{\mathsf{n}}$ is the same for different operator choices). Another technicality is that to ensure a positive spectrum and avoid degenerate levels, the parameters that enter into the optimization procedure are not $E_\mathsf{n}$ directly but $\log E_0$ for the ground state and $\log(E_\mathsf{k}-E_{\mathsf{k}-1})$ for the excited states, with $1\leq \mathsf{k} \leq \mathsf{e}$.

In order to fix the number of excited states $\mathsf{e}$ to include in the function $f$, the first fit is performed with only one exponential for the ground state ($\mathsf{e}=0$). Then, additional exponentials are added until there is no further improvement according to some information criteria. In our case we use the Akaike information criterion ($AIC$)~\cite{Akaike:1074}, but others are available, like the Bayesian information criterion~\cite{Schwarz:1978}. When comparing two models, the $AIC$ value for each one has to be computed. For a model with $\mathsf{e}$ excited states,
\begin{equation}
    AIC(\mathsf{e})=\chi^2(\mathsf{e})+2 N_{\text{param}}(\mathsf{e})\, ,
\end{equation}
where $N_{\text{param}}$ is the number of parameters fitted. Then, to penalize overfitting the data, we impose that $\Delta AIC =AIC(\mathsf{e})-AIC(\mathsf{e}-1)< -\mathcal{A} N_{\text{dof}}(\mathsf{e})$, where $\mathcal{A}$ is a free parameter and $N_{\text{dof}}(\mathsf{e})=N_{\text{pts}}-N_{\text{param}}(\mathsf{e})$ is the number of degrees of freedom, with $N_{\text{pts}}$ being the number of points included in the fit. Let us assume the following criterion: we want the reduced $\chi^2$, $\chi^2_{\text{dof}}=\chi^2/N_{\text{dof}}$, to improve by $\mathcal{O}(1)$ when adding an extra exponential to the calculation. Then, for this condition to hold, $\mathcal{A}$ must fulfill the following relation,
\begin{equation}
    \chi^2_{\text{dof}}(\mathsf{e}) - \chi^2_{\text{dof}}(\mathsf{e}-1) - \underbrace{\frac{N_{\text{dof}}(\mathsf{e})-N_{\text{dof}}(\mathsf{e}-1)}{N_{\text{dof}}(\mathsf{e})}\left[2-\chi_{\text{dof}}^2(\mathsf{e}-1)\right]}_{\ll 1} < -\mathcal{A}\, ,
\end{equation}
which leads us to $\chi^2_{\text{dof}}(\mathsf{e}) \approx \chi^2_{\text{dof}}(\mathsf{e}-1) - \mathcal{O}(1)$ when $\mathcal{A}\approx \mathcal{O}(1)$, as we required. In our case, we set $\mathcal{A}=0.5$, but other values were explored and consistent results were obtained. Therefore, if $\Delta AIC < -\mathcal{A}N_{\text{dof}}(\mathsf{e})$, we accept the model with $\mathsf{e}$ excited states and compute the one with $\mathsf{e}+1$, otherwise, we reject it and keep the one with $\mathsf{e}-1$ excited states.

After the model has been selected, with fitted parameters $\{\bm{E}^{f},\bm{Z}^{f}\}$, and before the confidence intervals of these parameters are computed, we conduct two tests: i) we impose that the value of $\chi^2_{\text{dof}}$ has to be smaller than some specific tolerance $tol_{\chi^2}$ (in our case set to 2), ii) we require that fits resulting from different optimization algorithms must agree with the energies within the same absolute tolerance ($tol_{\text{sol}}=10^{-5}$ in the present study). For this last test, the Nelder-Mead (NM) and gradient-based Newton solver (CG) are compared.

Once these two checks have been passed, the confidence intervals of the parameters are computed using bootstrap resampling methods (as it is explained in the next section), and $N_{\text{boot}}$ bootstrap resampled ensembles are generated, $\{\bm{E}^{b,f},\bm{Z}^{b,f}\}$. Afterwards, the 67\% confidence intervals for the energies are extracted by using the $\tfrac{5}{6}$th and $\tfrac{1}{6}$th quantiles (Q) of the difference between $\{\bm{E}^{b,f},\bm{Z}^{b,f}\}$ and the central values $\{\bm{E}^{f},\bm{Z}^{f}\}$,
\begin{equation}
    \delta \bm{E}^f=\frac{Q_{5/6}(\bm{E}^{b,f}-\bm{E}^{f})-Q_{1/6}(\bm{E}^{b,f}-\bm{E}^{f})}{2}\, ,
\end{equation}
with a similar expression for $\delta \bm{Z}^f$. With the bootstrap ensemble and uncertainty computed, two additional checks are performed: i) the results of a given uncorrelated $\chi^2$ minimization (this can be done by setting $\lambda=1$ in Eq.~\eqref{eq:shrinkcovariance}) must agree with $\{\bm{E}^{f},\bm{Z}^{f}\}$ within a tolerance of $tol_{\text{corr}}=5\sigma$ (with $\sigma$ being $\delta \bm{E}^f$ or $\delta \bm{Z}^f$), and ii) the bootstrap median (computed with $Q_{1/2}$) must reproduce the mean within a tolerance $tol_{\text{med}}=2\sigma$. 

\afterpage{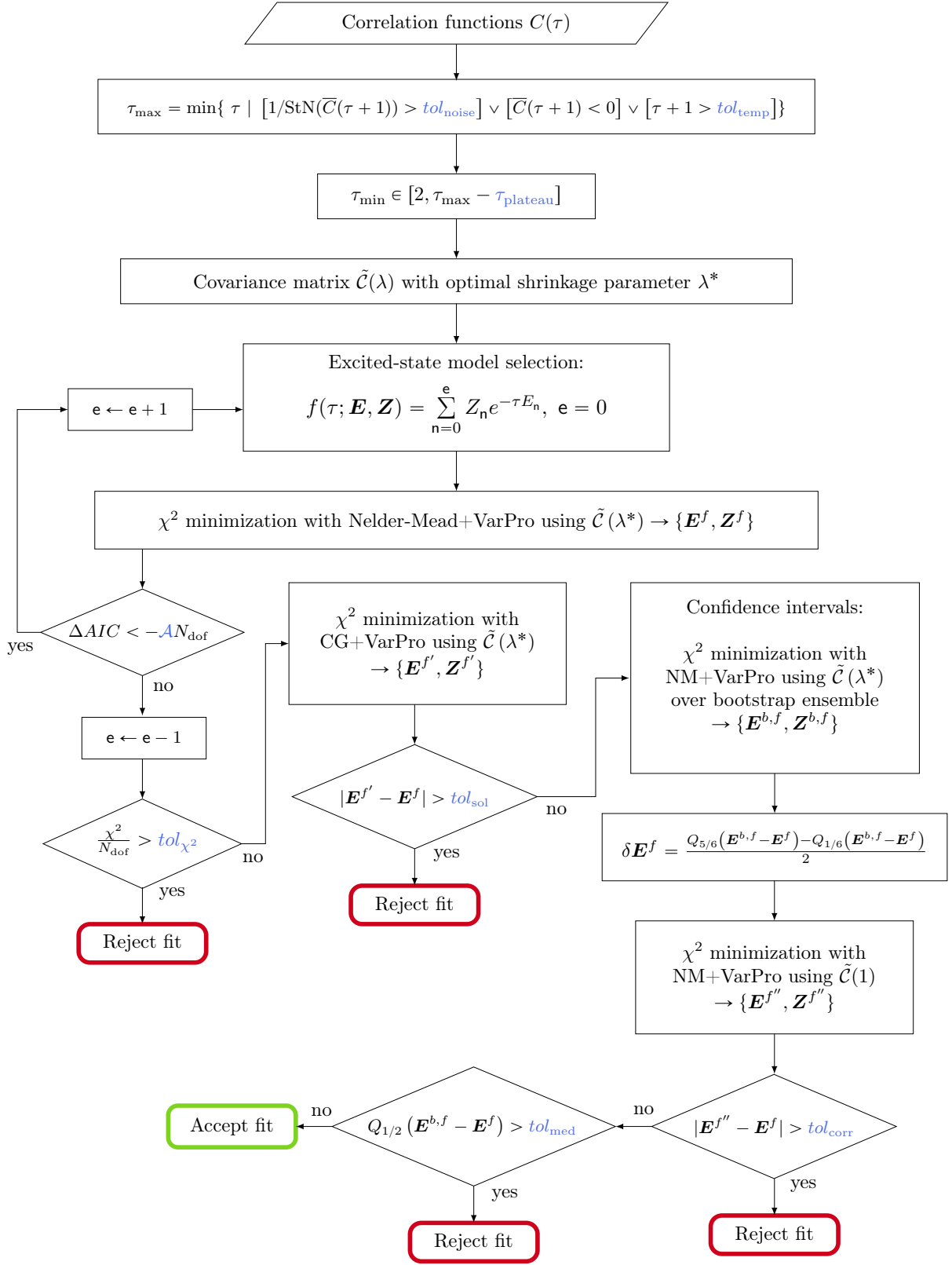
\begin{figure}[ht!]
\centering
\resizebox{\linewidth}{!}{\begin{tikzpicture}[scale=1.1,x=0.75pt,y=0.75pt,yscale=-1,xscale=1,>=latex, every picture/.style={line width=0.75pt}]

%uncomment if require: \path (0,877); %set diagram left start at 0, and has height of 877

\draw   (239.9,1) -- (482,1) -- (462.1,27) -- (220,27) -- cycle ;

\draw[->]    (351,27) -- (351,50) ;

\draw   (265,108) -- (437,108) -- (437,136) -- (265,136) -- cycle ;

\draw   (129,49) -- (573,49) -- (573,83) -- (129,83) -- cycle ;

\draw   (143,161) -- (559,161) -- (559,189) -- (143,189) -- cycle ;

\draw[->]    (351,136) -- (351,161) ;

\draw[->]    (351,189) -- (351,214) ;

\draw   (219,214) -- (483,214) -- (483,281) -- (219,281) -- cycle ;

\draw[->]    (351,281) -- (351,306) ;

\draw   (127,306) -- (575,306) -- (575,342) -- (127,342) -- cycle ;

\draw[->]    (156.5,342) -- (156.5,367) ;

\draw   (156.5,367) -- (219.5,394) -- (156.5,421) -- (93.5,394) -- cycle ;

\draw   (119,447) -- (196,447) -- (196,473) -- (119,473) -- cycle ;

\draw   (156.44,498) -- (217.88,525.88) -- (156.44,553.76) -- (95,525.88) -- cycle ;

\draw   (247.26,363) -- (419,363) -- (419,439) -- (247.26,439) -- cycle ;

\draw[->]    (325.13,439) -- (325.13,464) ;

\draw   (325.13,464) -- (401,496.5) -- (325.13,529) -- (249.26,496.5) -- cycle ;

\draw   (459,363) -- (637.76,363) -- (637.76,482) -- (459,482) -- cycle ;

\draw[->]    (548.13,482) -- (548.13,507) ;

\draw   (441.26,507) -- (655,507) -- (655,549) -- (441.26,549) -- cycle ;

\draw[->]    (548,549) -- (548,574) ;

\draw   (462,574) -- (634,574) -- (634,644.5) -- (462,644.5) -- cycle ;

\draw   (547.75,669.75) -- (623.5,702.25) -- (547.75,734.75) -- (472,702.25) -- cycle ;

\draw[->]    (472,702.25) -- (449.5,702.25) ;

\draw[->]    (548,644.5) -- (548,669.5) ;

\draw  [color={rgb, 255:red, 208; green, 2; blue, 27 }  ,draw opacity=1 ][line width=2.25]  (508.5,764.79) .. controls (508.5,760.92) and (511.63,757.79) .. (515.5,757.79) -- (580.5,757.79) .. controls (584.37,757.79) and (587.5,760.92) .. (587.5,764.79) -- (587.5,774.75) .. controls (587.5,778.62) and (584.37,781.75) .. (580.5,781.75) -- (515.5,781.75) .. controls (511.63,781.75) and (508.5,778.62) .. (508.5,774.75) -- cycle ;

\draw[->]    (548,734.75) -- (548,757.79) ;

\draw   (361.75,665.13) -- (449.5,702.25) -- (361.75,739.38) -- (274,702.25) -- cycle ;

\draw  [color={rgb, 255:red, 126; green, 211; blue, 33 }  ,draw opacity=1 ][line width=2.25]  (172.5,695.25) .. controls (172.5,691.94) and (175.19,689.25) .. (178.5,689.25) -- (245.5,689.25) .. controls (248.81,689.25) and (251.5,691.94) .. (251.5,695.25) -- (251.5,709.25) .. controls (251.5,712.56) and (248.81,715.25) .. (245.5,715.25) -- (178.5,715.25) .. controls (175.19,715.25) and (172.5,712.56) .. (172.5,709.25) -- cycle ;

\draw  [color={rgb, 255:red, 208; green, 2; blue, 27 }  ,draw opacity=1 ][line width=2.25]  (322.25,769.41) .. controls (322.25,765.55) and (325.38,762.41) .. (329.25,762.41) -- (394.25,762.41) .. controls (398.12,762.41) and (401.25,765.55) .. (401.25,769.41) -- (401.25,779.38) .. controls (401.25,783.24) and (398.12,786.38) .. (394.25,786.38) -- (329.25,786.38) .. controls (325.38,786.38) and (322.25,783.24) .. (322.25,779.38) -- cycle ;

\draw[->]    (361.75,739.38) -- (361.75,762.41) ;

\draw[->]    (274,702.25) -- (251.5,702.25) ;

\draw   (110.5,242) -- (187.5,242) -- (187.5,268) -- (110.5,268) -- cycle ;

\draw[<-]    (219,255) -- (187.5,255) ;

\draw[<-]    (110.5,255) -- (79,255) -- (79,394) -- (93.5,394) ;

\draw[->]    (351,83) -- (351,108) ;

\draw[->]    (156.5,421) -- (156.5,446) ;

\draw[->]    (156.5,473) -- (156.5,498) ;

\draw  [color={rgb, 255:red, 208; green, 2; blue, 27 }  ,draw opacity=1 ][line width=2.25]  (117,583.04) .. controls (117,579.17) and (120.13,576.04) .. (124,576.04) -- (189,576.04) .. controls (192.87,576.04) and (196,579.17) .. (196,583.04) -- (196,593) .. controls (196,596.87) and (192.87,600) .. (189,600) -- (124,600) .. controls (120.13,600) and (117,596.87) .. (117,593) -- cycle ;

\draw[->]    (156.5,553) -- (156.5,576.04) ;

\draw  [color={rgb, 255:red, 208; green, 2; blue, 27 }  ,draw opacity=1 ][line width=2.25]  (285.63,559.04) .. controls (285.63,555.17) and (288.76,552.04) .. (292.63,552.04) -- (357.63,552.04) .. controls (361.5,552.04) and (364.63,555.17) .. (364.63,559.04) -- (364.63,569) .. controls (364.63,572.87) and (361.5,576) .. (357.63,576) -- (292.63,576) .. controls (288.76,576) and (285.63,572.87) .. (285.63,569) -- cycle ;

\draw[->]    (325.13,529) -- (325.13,552.04) ;

\draw[<-]    (247.26,401) -- (232.75,401) -- (232.75,525.88) -- (217.88,525.88) ;

\draw[<-]    (458.13,422.5) -- (437.25,422.5) -- (437.25,496.5) -- (401,496.5) ;

\draw (351,14) node  [font=\small]  {$\text{Correlation functions} \ C(\tau)$};

\draw (351,122) node  [font=\small]  {$\tau_{\text{min}} \in [ 2,\tau_{\text{max}} -\textcolor[rgb]{0.29,0.41,0.89}{\tau_{\text{plateau}}}]$};

\draw (351,67) node  [font=\footnotesize]  {$\tau_{\text{max}} = \text{min}\{\ \tau\ |\ \left[ 1/\text{StN}( \overline{C}(\tau + 1))  >\textcolor[rgb]{0.29,0.41,0.89}{tol_{\text{noise}}} \right] \lor \left[ \overline{C}(\tau + 1) < 0 \right] \lor \left[\tau + 1  >\textcolor[rgb]{0.29,0.41,0.89}{tol_{\text{temp}}} \right] \}$ };

\draw (351,225) node  [font=\small] [align=left] {Excited-state model selection:};

\draw (351,255) node    {$f( \tau;\bm{E} ,\bm{Z}) =\sum\limits ^{\mathsf{e}}_{\mathsf{n}=0} Z_{\mathsf{n}} e^{-\tau E_{\mathsf{n}}} ,\ \mathsf{e}=0$};

\draw (351,324) node  [font=\small]  {$\chi ^{2} \ \text{minimization with Nelder-Mead+VarPro using} \ \tilde{\mathcal{C}} \left( \lambda ^{*}\right)\rightarrow \{\bm{E}^{f} ,\bm{Z}^{f}\}$};

\draw (154,394) node  [font=\footnotesize]  {$\Delta AIC< -\mathcal{\textcolor[rgb]{0.29,0.41,0.89}{A}} N_{\text{dof}}$};

\draw (325.13,496.5) node  [font=\footnotesize]  {$|\bm{E}^{f'} -\bm{E}^{f} | >\textcolor[rgb]{0.29,0.41,0.89}{tol_{\text{sol}}}$};

\draw (81,404) node  [font=\small] [align=left] {yes};

\draw (149,255) node  [font=\footnotesize]  {$\mathsf{e}\leftarrow \mathsf{e}+1$};

\draw (157.5,460) node  [font=\footnotesize]  {$\mathsf{e}\leftarrow \mathsf{e}-1$};

\draw (156.5,588.02) node  [font=\small] [align=left] {Reject fit};

\draw (159.5,523.5) node  [font=\small]  {$\frac{\chi ^{2}}{N_{\text{dof}}}  >\textcolor[rgb]{0.29,0.41,0.89}{tol_{\chi ^{2}}}$};

\draw (212,702.25) node  [font=\small] [align=left] {Accept fit};

\draw (169.5,428) node  [font=\small] [align=left] {no};

\draw (333.13,401) node  [font=\small] [align=center] {$\displaystyle \chi ^{2}$ minimization with\\CG+VarPro using $\displaystyle \tilde{\mathcal{C}}\left( \lambda ^{*}\right)$\\$\displaystyle \rightarrow \{\bm{E}^{f'} ,\bm{Z}^{f'}\}$};

\draw (548.38,378) node  [font=\small] [align=left] {Confidence intervals:};

\draw (548.38,429) node  [font=\small] [align=center] {$\displaystyle \chi ^{2}$ minimization with\\NM+VarPro using $\displaystyle \tilde{\mathcal{C}}\left( \lambda ^{*}\right)$\\over bootstrap ensemble\\$\displaystyle \rightarrow \{\bm{E}^{b,f} ,\bm{Z}^{b,f}\}$};

\draw (325.13,564.02) node  [font=\small] [align=left] {Reject fit};

\draw (225,535) node  [font=\small] [align=left] {no};

\draw (175,557) node  [font=\small] [align=left] {yes};

\draw (345,533) node  [font=\small] [align=left] {yes};

\draw (413,505) node  [font=\small] [align=left] {no};

\draw (548,769.77) node  [font=\small] [align=left] {Reject fit};

\draw (361.75,774.39) node  [font=\small] [align=left] {Reject fit};

\draw (548.13,528) node  [font=\small]  {$\delta \bm{E}^f =\frac{Q_{5/6}\left( \bm{E}^{b,f} - \bm{E}^{f}\right) -Q_{1/6}\left( \bm{E}^{b,f} - \bm{E}^{f}\right)}{2}$};

\draw (547.75,702.25) node  [font=\footnotesize]  {$|\bm{E}^{f''} -\bm{E}^{f} | >\textcolor[rgb]{0.29,0.41,0.89}{tol_{\text{corr}}}$};

\draw (548,609.25) node  [font=\small] [align=center] {$\displaystyle \chi ^{2}$ minimization with\\NM+VarPro using $\displaystyle \tilde{\mathcal{C}}( 1)$\\$\displaystyle \rightarrow \{\bm{E}^{f''} ,\bm{Z}^{f''}\}$};

\draw (566,739) node  [font=\small] [align=left] {yes};

\draw (467,692) node  [font=\small] [align=left] {no};

\draw (360.75,702.25) node  [font=\footnotesize]  {$Q_{1/2}\left(\bm{E}^{b,f} -\bm{E}^{f}\right)  >\textcolor[rgb]{0.29,0.41,0.89}{tol_{\text{med}}}$};

\draw (381,745) node  [font=\small] [align=left] {yes};

\draw (267.5,693) node  [font=\small] [align=left] {no};

\draw (351,175) node  [font=\small] [align=left] {Covariance matrix $\displaystyle \tilde{\mathcal{C}}(\lambda)$ with optimal shrinkage parameter $\displaystyle \lambda ^{*}$};

\end{tikzpicture}}
\caption{Flowchart representing the steps of the fitting algorithm for a fixed time range. After a fit is accepted (green box) or rejected (red box), the procedure is repeated for a different $\tau_{\text{min}}$. The blue variables are the free parameters of the algorithm (see text for details).}
\label{fig:flowchart}
\end{figure}
\clearpage}

In Fig.~\ref{fig:flowchart}, as a summary, a flowchart of the fitting algorithm for a specific $[\tau_{\text{min}},\tau_{\text{max}}]$ range is shown. This algorithm has been implemented with the \texttt{Julia} language~\cite{Bezanson:2017}, together with the \texttt{Optim} package~\cite{Mogensen:2018}.

Once all the possible values of $\tau_{\text{min}}$ have been used (or $N_{\text{fits}}$ fits have been performed), a list of $N_{\text{acc}}$ accepted fits is obtained. The final step is to combine them and give a reliable estimation of the associated statistical and systematic uncertainties. The only quantity that will be extracted,  being the one we are interested in, is the ground-state energy $E_0$. For that, we have to define some weights for each result, $\omega_f$, that takes into account the quality of the fit and its accuracy (see Ref.~\cite{Jay:2020jkz} for a Bayesian framework). The quality will be given by the $p$-value,
\begin{equation}
    p_f \equiv \text{Prob}(\chi^2_{f,\text{dof}}\leq \chi^2)=\frac{\Gamma(N_{\text{dof}}/2,\chi^2_f/2)}{\Gamma(N_{\text{dof}}/2)}\, ,
\end{equation}
where $\Gamma(x,y)$ is the upper incomplete gamma function, $\Gamma(x,y)=\int_y^{\infty}dt\, t^{x-1} e^{-t}$, and $\chi^2_f$ is computed inserting the $\{\bm{E}^f,\bm{Z}^f\}$ values in Eq.~\eqref{eq:chi2_2pt}. The statistical uncertainty is chosen to be the 67\% confidence region $\delta E^f_0$. In terms of these quantities, the weights and the mean energy value read~\cite{Rinaldi:2019thf}
\begin{equation}
    \omega_f=\frac{p_f \left(\delta E^f_0\right)^{-2}}{\sum_{f'=1}^{N_{\text{acc}}} p_{f'} \left(\delta E^{f'}_0 \right)^{-2}}\, , \qquad \overline{E}_0=\sum_{f=1}^{N_{\text{acc}}} \omega_f E^f_0\, .
    \label{eq:weights}
\end{equation}
In the present work, we define the statistical uncertainty as that of the fit with the highest weight, while the systematic uncertainty is defined as the average difference between the weighted mean value and each of the accepted fits,
\begin{equation}
    \delta \overline{E}_{\text{stat}}=\delta E_0^{f:\max[\{\omega_f\}]}\, ,\qquad \delta \overline{E}_{\text{sys}}=\sqrt{\sum_{f=1}^{N_{\text{acc}}} \omega_f \left(E_0^f-\overline{E}_0\right)^2}\, .
\label{eq:weights_fitting}
\end{equation}

An alternative way to extract the energy of the system is by using the EMP as defined in Eq.~\eqref{eq:EMP}, based in the assumption that the overlap to the ground state is stronger than the one to excited states, which allows us to use as a fitting function a constant, $E_0$, plus exponentials to account for excited states contributions.

In previous works (as in Ref.~\cite{Orginos:2015aya}), the energy shift in the finite-volume energies of two interacting baryons, $\Delta E_0=E_0-M_{B_1}-M_{B_2}$, was obtained by fitting a constant to the effective energy-shift function given by Eq.~\eqref{eq:effenergyshiftfun}. In these calculations, the correlation functions for both the single- and two-baryon systems are required to be described by a single-state fit within the fitting time range. This is due to the fact that otherwise, cancellations may occur between excited states (including the finite-volume states that would correspond to elastic scattering states in the infinite volume), either in the correlation function or in ratios of correlation functions, producing a ``mirage plateau''~\cite{Iritani:2017rlk} (see further discussion in~\cref{sec:presenstatus}). To minimize this problem, here the single- and two-baryon correlation functions are fitted separately to multi-exponential forms within the same fitting range, and afterwards the energy shifts are computed at the bootstrap level, in such a way that correlations between the different correlation functions are taken into account.
The use of correlated differences of multi-state fit results is particularly convenient for automated fit-range sampling, since the number of excited states can be varied independently for one- and two-baryon correlation functions, unlike fits to the ratio in Eq.~\eqref{eq:energyshiftfun}. Despite this fact, consistent results were obtained with fits to the ratio in Eq.~\eqref{eq:energyshiftfun} in the allowed time regions.

\subsection{Generalized pencil-of-functions method}

\begin{figure}[t]
\centering
\includegraphics[width=\textwidth]{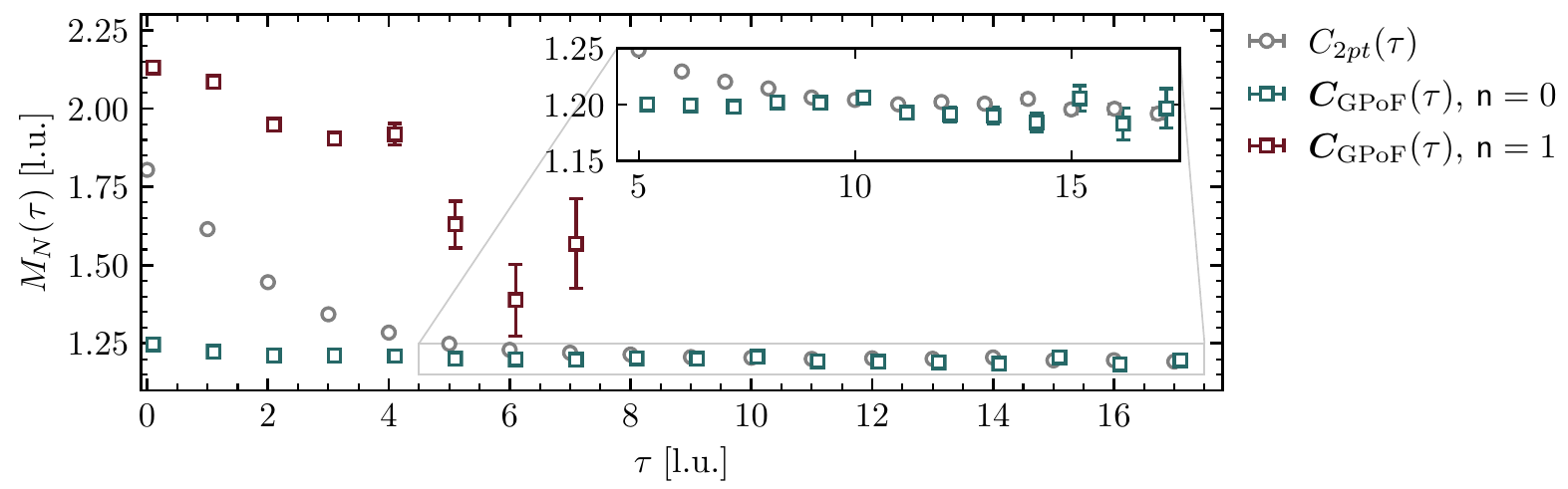}
\caption{Effective mass plot for the proton, comparing the results obtained either using the single correlation function $C_{2pt}(\tau)$ or with a $2\times 2$ GPoF matrix $\bm{C}_{\text{GPoF}}(\tau)$ (using $\delta\tau=2$) showing the ground-state ($\mathsf{n}=0$) and excited-state ($\mathsf{n}=1$) energy levels.}
\label{fig:proton_GPoF}
\end{figure}

Before discussing the variational method, there are a few improvements that we can do in order to remove excited-state contributions, which would allow us to extend the time range at early times. When only a single type of operator is computed, such method is the generalized pencil-of-functions method (GPoF)~\cite{Hua:1989,Sarkar:1995,Aubin:2010jc}, where a set of linearly-independent operators are created via the time evolution operator,
\begin{equation}
    \mathcal{X}_{\delta \tau}(\tau)= e^{\delta \tau \hat{H}}\mathcal{X}(\tau) e^{-\delta \tau \hat{H}} =\mathcal{X}(\tau+\delta \tau)\, .
\end{equation}
Then, a correlation function matrix can be built with the two operators, $\mathcal{X}$ and $\mathcal{X}_{\delta \tau}$,
\begin{equation}
\begin{aligned}
    \bm{C}_{\text{GPoF}}(\tau)&=\begin{pmatrix}
    \langle \mathcal{X}(\tau) \bar{\mathcal{X}}(0) \rangle & \langle \mathcal{X}(\tau) \bar{\mathcal{X}}_{\delta \tau}(0) \rangle \\
    \langle \mathcal{X}_{\delta \tau}(\tau) \bar{\mathcal{X}}(0) \rangle & \langle \mathcal{X}_{\delta \tau}(\tau) \bar{\mathcal{X}}_{\delta \tau}(0) \rangle 
    \end{pmatrix} \\
    &= \begin{pmatrix}
    \langle \mathcal{X}(\tau) \bar{\mathcal{X}}(0) \rangle & \langle \mathcal{X}(\tau) \bar{\mathcal{X}}(-\delta \tau) \rangle \\
    \langle \mathcal{X}(\tau+\delta \tau) \bar{\mathcal{X}}(0) \rangle & \langle \mathcal{X}(\tau+\delta \tau)\bar{\mathcal{X}}(-\delta \tau) \rangle
    \end{pmatrix} = \begin{pmatrix}
    C(\tau) & C(\tau + \delta \tau) \\
    C(\tau + \delta \tau) & C(\tau + 2\delta \tau)
    \end{pmatrix}\, ,
\end{aligned}
\end{equation}
where in the last step we have used the time-translation invariance of the correlation functions. This procedure can be generalized to construct a larger matrix by using multiple $\delta \tau$ shifts, $\mathcal{X}_{n\delta \tau}$, and it can also be applied to three-point correlation functions~\cite{Aubin:2010jc}. Once this matrix is built, the generalized eigenvalue problem (GEVP) can be solved (discussed in~\cref{subsec:variational}) and the energies can be extracted. In order to illustrate this method (as well as the following ones), the data from Ref.~\cite{Amarasinghe:2021} will be used. In Fig.~\ref{fig:proton_GPoF}, the EMP for the proton is shown, comparing the raw $C_{2pt}(\tau)$ correlation function data with the results from a $2 \times 2$ GPoF analysis with $\delta \tau=2$. It can be seen that while the excited state contamination for small $\tau$ is reduced for the ground state, the errors increase faster at earlier times than the raw correlator. It can be noticed that the excited state cannot be extracted reliably, and more suitable methods, like the variational method, are better when dealing with these states.

\subsection{Prony method}

Another method that can be used to separate the different energy levels of the correlation functions is the Prony method~\cite{Prony:1795,Hildebrand:1987}, which was first applied to LQCD data in Refs.~\cite{Fleming:2004hs,Lin:2007iq}, and later on generalized for multiple correlation functions in Ref.~\cite{Beane:2009kya}, known as the matrix-Prony method.

Starting with the case where we only work with a single type of correlation function, and assuming that only $\mathsf{e}$ excited states contribute to $C(\tau)$, we can write $C(\tau)= \sum_{\mathsf{n}=0}^{\mathsf{e}} Z_\mathsf{n} e^{-\tau E_\mathsf{n}} \equiv \sum_{\mathsf{n}=0}^{\mathsf{e}} Z_\mathsf{n} \xi_\mathsf{n}^\tau$. We have two ways to proceed, the analytical approach and the linear prediction. Within the analytical approach~\cite{Fleming:2004hs,Lin:2007iq}, we extract the $\mathsf{e}+1$ energies $\xi_\mathsf{n}$ analytically from $2(\mathsf{e}+1)$ time-shifted correlation functions. For example, if we assume $\mathsf{e}=0$, then we have to solve the following system of equations,
\begin{equation}
    \begin{pmatrix} C(\tau) \\ C(\tau+1) \end{pmatrix} = \begin{pmatrix} \xi^{\tau\hphantom{+1}}_0 \\ \xi^{\tau+1}_0 \end{pmatrix} Z_0\, .
\end{equation}
The solutions are $Z_0=\xi^\tau_0/C(\tau)$ and $\xi_0=C(\tau+1)/C(\tau)$, and we see that we recover the definition of the effective mass function~\eqref{eq:EMP}. If we now set $\mathsf{e}=1$, the system of equations to solve is
\begin{equation}
    \begin{pmatrix} C(\tau) \\ C(\tau+1) \\C(\tau+2) \\ C(\tau+3) \end{pmatrix} = \begin{pmatrix} \xi^{\tau\hphantom{+1}}_0 & \xi^{\tau\hphantom{+1}}_1 \\ \xi^{\tau+1}_0 & \xi^{\tau+1}_1 \\ \xi^{\tau+2}_0 & \xi^{\tau+2}_1 \\ \xi^{\tau+3}_0 & \xi^{\tau+3}_1 \end{pmatrix} \begin{pmatrix} Z_0 \\ Z_1 \end{pmatrix}\, .
\end{equation}
For this case, the solutions are (only for $\xi_0$ and $\xi_1$)
\begin{equation}
\begin{aligned}
    \xi_{0,1}=\smash[b]{\frac{-y\pm\sqrt{y^2-4xz}}{2x}}\, , \quad\text{with }\; x &= C(\tau+1)^2 + C(\tau)C(\tau+2)\, , \\
    y &= C(\tau)C(\tau+3)-C(\tau+1)C(\tau+2)\, ,\\
    z &= C(\tau+2)^2 - C(\tau+1)C(\tau+3)\, .
\end{aligned}
\end{equation}
Due to the Abel–Ruffini theorem, this is the maximum number of states $\mathsf{e}$ for which a general algebraic solution can be found. For $2(\mathsf{e}+1)\geq 5$, numerical methods have to be employed to find the solution, and as it can be seen in Ref.~\cite{Lin:2007iq}, the procedure is not very stable numerically.

The linear prediction approach~\cite{Fleming:2004hs,Lin:2007iq,Beane:2009kya} starts by forming an $(\mathsf{e}+1)$th order polynomial, with $\xi_{\mathsf{n}}$ as the roots,
\begin{equation}
    p(\xi)=\prod_{\mathsf{n}=0}^{\mathsf{e}}(\xi-\xi_{\mathsf{n}})=\sum_{\mathsf{k}=0}^{\mathsf{e}+1}p_{\mathsf{k}}\,\xi^{\mathsf{e}+1-\mathsf{k}}\, ,\quad \text{with }\; p_0=1\, .
\end{equation}
Since $\xi_{\mathsf{n}}$ are the roots of this polynomial ($p(\xi_{\mathsf{n}})=0$), we can rewrite it as
\begin{equation}
    \xi^\mathsf{m}_{\mathsf{n}}=-\sum_{\mathsf{k}=1}^{\mathsf{e}+1}p_{\mathsf{k}}\,\xi^{\mathsf{m}-\mathsf{k}}_{\mathsf{n}}\, ,\quad \text{with }\; \mathsf{m}\geq\mathsf{e}+1\, .
\end{equation}
With this last expression, we can write the correlation function as
\begin{equation}
    C(\tau)=\sum_{\mathsf{n}=0}^{\mathsf{e}}Z_\mathsf{n}\xi_\mathsf{n}^\tau = -\sum_{\mathsf{n}=0}^{\mathsf{e}}Z_\mathsf{n}\sum_{\mathsf{k}=1}^{\mathsf{e}+1}p_{\mathsf{k}}\,\xi^{\tau-\mathsf{k}}_{\mathsf{n}}=-\sum_{\mathsf{k}=1}^{\mathsf{e}+1} p_{\mathsf{k}}\, C(\tau-\mathsf{k})\, ,\quad \text{with }\; \tau\geq\mathsf{e}+1\, .
    \label{eq:linearpred}
\end{equation}
This is the reason why this method is called linear prediction, since we predict $C(\tau)$ using the values of the correlation function at earlier times. Once the values of $p_{\mathsf{k}}$ have been obtained by solving Eq.~\eqref{eq:linearpred}, they can be used to solve for the roots of the polynomial $p(\xi)$. An easy example to see how this works is by setting $\mathsf{e}=0$, in which case we get
\begin{equation}
    C(\tau+1)+p_1 C(\tau)=0 \quad \text{and} \quad p(\xi)=\xi+p_1\, ,
\end{equation}
 again recovering the effective mass function.

For the matrix-Prony method~\cite{Beane:2009kya}, we assume we have computed $N_{\text{op}}$ different correlation functions, so we can write them in a vector form,
\begin{equation}
    \bm{c}(\tau)=\begin{pmatrix} C^1(\tau) & \cdots & C^N_{\text{op}}(\tau) \end{pmatrix}^\top\, .
\end{equation}
Since correlation functions can be written as a sum of exponentials, we want to build an operator such that $\bm{c}(\tau+\tau_t)=\bm{T} \bm{c}(\tau)$ (similar to the linear prediction approach). Defining $\bm{T}=\bm{M}^{-1} \bm{V}$, we get the following relation,
\begin{equation}
    \bm{M}\bm{c}(\tau+\tau_t) = \bm{V}\bm{c}(\tau)\, .
\end{equation}
A simple solution for the matrices $\bm{M}$ and $\bm{V}$ is
\begin{equation}
    \bm{M} = \left[ \sum_{\tau'=\tau}^{\tau+\tau_1} \bm{c}(\tau'+\tau_t) \bm{c}^\top(\tau') \right]^{-1}\, , \quad \bm{V} = \left[ \sum_{\tau'=\tau}^{\tau+\tau_1} \bm{c}(\tau') \bm{c}^\top(\tau') \right]^{-1}\, .
\end{equation}
The value of $\tau_1$ has to be large enough to make the matrices inside the brackets full rank (this happens for $\tau_1 \geq N_{\text{op}}$). With these matrices, a GEVP can be solved,
\begin{equation}
    \bm{M} \bm{u}^\alpha = (\lambda_\alpha)^{\tau_t} \bm{V}\bm{u}^\alpha\, ,
    \label{eq:GEVP_prony}
\end{equation}
with $\bm{u}^\alpha$ being the general eigenvectors and $\lambda_\alpha=e^{E_\alpha}$ the general eigenvalues (again, this system will be discussed in~\cref{subsec:variational}). The parameter $\tau_t$ can be varied to improve stability. To illustrate the method, we use two correlation functions (so $\tau_1 \geq 2$) for the proton, corresponding to two different smearings: a narrow ($N$) and a wide ($W$) Gaussian profile at both source and sink. Then, according to Eq.~\eqref{eq:GEVP_prony}, these correlation functions can be written as
\begin{equation}
    \begin{pmatrix} C^{N}(\tau) \\ C^{W}(\tau) \end{pmatrix} = Z'_0 \begin{pmatrix} u^0_{N} \\ u^0_{W} \end{pmatrix} (\lambda_0)^{-\tau}+Z'_1 \begin{pmatrix} u^1_{N} \\ u^1_{W} \end{pmatrix} (\lambda_1)^{-\tau}\, .
\end{equation}
In order to extract the energies $\lambda_\alpha$, we need to find the correct combination of $C^{N/W}(\tau)$  with the components of the eigenvectors $u^\alpha_{N/W}$ (the remaining overlap factors $Z'_\alpha$ are not relevant in the present discussion, since they cancel out in the ratios which define the EMPs). With a bit of algebra, we arrive at
\begin{equation}
    (\lambda_0)^{-\tau}\;\propto \; \frac{u^1_{N}C^{W}(\tau)-u^1_{W}C^{N}(\tau)}{u^0_{N}u^1_{W}-u^1_{N}u^0_{W}}\, , \quad (\lambda_1)^{-\tau}\;\propto \; \frac{u^0_{N}C^{W}(\tau)-u^0_{W}C^{N}(\tau)}{u^0_{N}u^1_{W}-u^1_{N}u^0_{W}}\, .
\end{equation}

\begin{figure}[t]
\centering
\includegraphics[width=\textwidth]{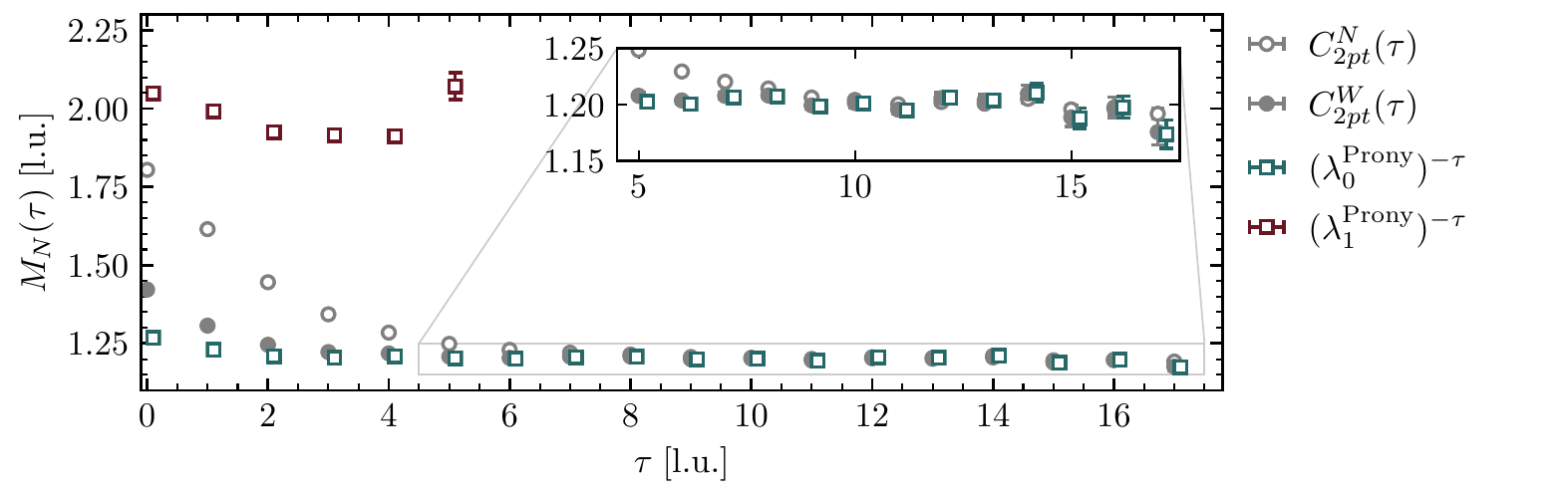}
\caption{Effective mass plot for the proton, comparing the results obtained using two different correlation functions $C^{N/W}_{2pt}(\tau)$ as discussed in the text, and using the matrix-Prony method, showing the ground-state ($\lambda^{\text{Prony}}_0$) and excited-state ($\lambda^{\text{Prony}}_1$) eigenvalues.}
\label{fig:proton_prony}
\end{figure}

In Fig.~\ref{fig:proton_prony}, the EMP for the proton is shown, comparing the raw $C_{2pt}(\tau)$ correlation function data with the results from a matrix-Prony analysis with $\{\tau_t,\tau_1\}=\{2,3\}$. Like in the GPoF case, the effective mass for the ground-state reaches the plateau at small times, but the excited state cannot be reliably extracted.

\subsection{Variational method}\label{subsec:variational}

In order to make a reliable extraction of the spectrum of a system, a variational study~\cite{Michael:1982gb,Luscher:1990ck} with an Hermitian correlation function matrix should be constructed and then diagonalized to obtain ground-state and excited-state energies. The advantage of using an Hermitian matrix is that the energies extracted are an upper limit of the underlying energy levels, removing the issue of the ``mirage'' plateau.

The idea here is that starting with a set of operators $\mathcal{X}_i$, with $i\in\{1,\ldots,N_{\text{op}}\}$, we can produce a new basis $\mathcal{E}^{\alpha}$, with $\alpha\in\{1,\ldots,N_{\text{op}}\}$, which couple to a single energy eigenstate, such that $\langle 0 | \mathcal{E}^{\alpha} | \mathfrak{n} \rangle = \delta_{\alpha \mathfrak{n}}$.
These new operators can be build using the left and right eigenvectors of the correlation function matrix $\bm{C}(\tau)=C_{ij}(\tau)=\langle 0 | \mathcal{X}_i(\tau) \bar{\mathcal{X}}_j(0) |0 \rangle$, which we assume are the same, since $C_{ij}$ is a real and symmetric matrix. Labeling the eigenvectors as $\bm{u}^{\alpha}$,
\begin{equation}
    \mathcal{E}^{\alpha}=\sum_{i=1}^{N_{\text{op}}} u^{\alpha}_i \mathcal{X}_i\, .
\end{equation}
To understand how to extract these eigenvectors, we can look at $C_{ij}(\tau)u^{\alpha}_j$,
\begin{equation}
\begin{aligned}
    C_{ij}(\tau)u^{\alpha}_j &= \langle 0 | \mathcal{X}_i(\tau) \bar{\mathcal{X}}_j(0) | 0 \rangle u^{\alpha}_j = \langle 0 | \mathcal{X}_i(\tau) (\bar{\mathcal{X}}_j(0)u^{\alpha}_j ) | 0 \rangle = \langle 0 | \mathcal{X}_i(\tau) \bar{\mathcal{E}}^{\alpha}(0) | 0 \rangle \\
    &= \sum_{\mathfrak{n}} e^{-\tau E_{\mathfrak{n}}} \langle 0 | \mathcal{X}_i | \mathfrak{n} \rangle \langle \mathfrak{n} | \bar{\mathcal{E}}^{\alpha} | 0 \rangle = \sum_{\mathfrak{n}} e^{-\tau E_{\mathfrak{n}}} \langle 0 | \mathcal{X}_i | \mathfrak{n} \rangle \delta_{\alpha\mathfrak{n}} = e^{-\tau E_{\alpha}} Z_{\alpha}\\
    & = e^{-(\tau-\tau_0)E_{\alpha}} e^{-\tau_0 E_{\alpha}} Z_{\alpha} = e^{-(\tau-\tau_0)E_{\alpha}} \sum_{\mathfrak{m}} e^{-\tau_0 E_{\mathfrak{m}}} \langle 0 | \mathcal{X}_i | \mathfrak{m} \rangle \langle \mathfrak{m} | \bar{\mathcal{E}}^{\alpha} | 0 \rangle \\
    &  = e^{-(\tau-\tau_0) E_{\alpha}} \langle 0 | \mathcal{X}_i(\tau_0) \bar{\mathcal{X}}_j(0) | 0 \rangle u^{\alpha}_j = e^{-(\tau-\tau_0)E_{\alpha}}C_{ij}(\tau_0)u^{\alpha}_j\, .
\end{aligned}
\label{eq:GEVP}
\end{equation}
After some manipulation, we have ended up with a GEVP problem, and we can identify the general eigenvalues as $\lambda_{\alpha}\equiv e^{-(\tau-\tau_0)E_{\alpha}}$.\footnote{Note that there are contributions from higher-energy states we are not accounting for given the limited size of the truncated Hilbert space we are working with.}
There are several strategies when trying to solve this system, since if we transform $\bm{C}(\tau)\bm{u}^\alpha=\lambda_\alpha \bm{C}(\tau_0)\bm{u}^\alpha$ to $\bm{C}^{-1}(\tau_0)\bm{C}(\tau)\bm{u}^\alpha=\lambda_\alpha \bm{u}^\alpha$ to be a conventional eigenvalue problem, the matrix $\bm{C}^{-1}(\tau_0)\bm{C}(\tau)$ is no longer Hermitian, and we could get imaginary eigenvalues, with eigenvectors that may not be orthogonal.
One possible solution is to use the Cholesky decomposition of $\bm{C}(\tau_0)$~\cite{Dudek:2007wv}, $\bm{C}(\tau_0)=\bm{L}\bm{L}^{\Dag}$, with $\bm{L}$ being a lower triangular matrix. Then,
\begin{equation}
    \bm{C}(\tau) \bm{u}^{\alpha} = \lambda_{\alpha} \bm{L} \bm{L}^{\Dag} \bm{u}^{\alpha}\; \Rightarrow \;\bm{L}^{-1} \bm{C}(\tau) \bm{u}^{\alpha} = \lambda_{\alpha} \bm{L}^{\Dag} \bm{u}^{\alpha}\; \Rightarrow \;\bm{L}^{-1} \bm{C}(\tau) \bm{L}^{\Dag -1} \bm{L}^{\Dag} \bm{u}^{\alpha} = \lambda_{\alpha} \bm{L}^{\Dag} \bm{u}^{\alpha}\,  .
\end{equation} 
Redefining the eigenvectors to be $\bm{\tilde{u}}^{\alpha}\equiv\bm{L}^{\Dag}\bm{u}^{\alpha}$, the matrix that needs to be diagonalized is $\bm{L}^{-1}\bm{C}(\tau) \bm{L}^{\Dag -1}$, which is Hermitian. Another solution is to use $\bm{C}^{-1/2}(\tau_0)\bm{C}^{1/2}(\tau_0)=\bm{1}$~\cite{Lin:2007iq},
\begin{equation}
\begin{gathered}
    \bm{C}(\tau) \bm{C}^{-1/2}(\tau_0) \bm{C}^{1/2}(\tau_0) \bm{u}^{\alpha} = \lambda_{\alpha} \bm{C}(\tau_0) \bm{u}^{\alpha}\\
    \bm{C}^{1/2}(\tau_0)\bm{C}(\tau) \bm{C}^{-1/2}(\tau_0)\bm{C}^{1/2}(\tau_0) \bm{u}^{\alpha} = \lambda_{\alpha} \bm{C}^{1/2}(\tau_0) \bm{u}^{\alpha}\, ,
\end{gathered}
\end{equation}
where now the eigenvectors are redefined to be $\bm{\tilde{u}}^{\alpha}\equiv\bm{C}^{1/2}(\tau_0)\bm{u}^{\alpha}$, and the matrix that needs to be diagonalized is $\bm{C}^{1/2}(\tau_0)\bm{C}(\tau) \bm{C}^{-1/2}(\tau_0)$, which is also Hermitian.
While some works extract the energies directly from the eigenvalues $\lambda_{\alpha}$, recent works on two-baryon systems~\cite{Francis:2018qch,Horz:2020zvv,Green:2021qol,Amarasinghe:2021} reconstruct a diagonalized correlation function matrix~\cite{Bulava:2016mks},
\begin{equation}
    \hat{C}_{\alpha\alpha}(\tau)=\bm{u}^{\alpha\Dag}(\tau_{\text{ref}},\tau_0)\bm{C}(\tau)\bm{u}^{\alpha}(\tau_{\text{ref}},\tau_0)\, ,
\end{equation}
where the eigenvectors $\bm{u}^{\alpha}(\tau_{\text{ref}},\tau_0)$ are extracted from the GEVP with a fixed $\{\tau=\tau_{\text{ref}},\tau_0\}$. A detailed discussion on the variational method can be found in Refs.~\cite{Blossier:2009kd,Amarasinghe:2021}.

\begin{figure}[t]
\centering
\includegraphics[width=\textwidth]{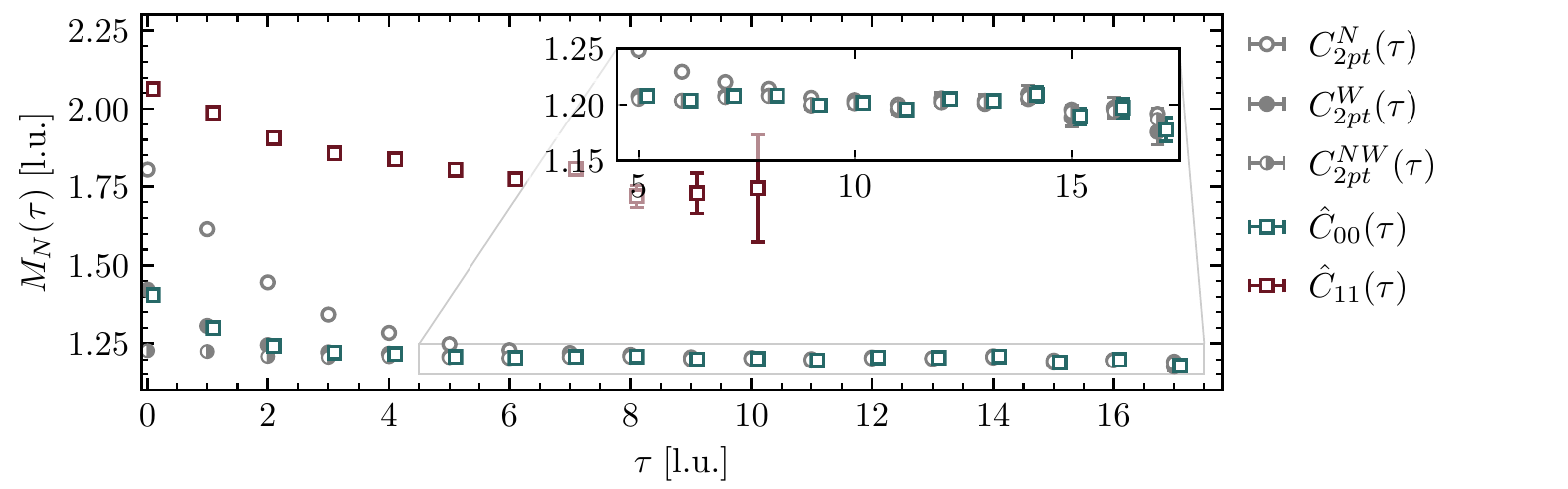}
\caption{Effective mass plot for the proton, comparing the results obtained using correlation functions from the matrix $\bm{C}_{2pt}(\tau)$ and from the variational analysis, showing the ground-state $\hat{C}_{00}(\tau)$ and excited-state $\hat{C}_{11}(\tau)$ diagonalized correlation functions.}
\label{fig:proton_var}
\end{figure}

In Fig.~\ref{fig:proton_var}, the EMP for the proton is shown, comparing the raw $C_{2pt}(\tau)$ correlation function data (in this case, a matrix where at the source or sink we can have narrow $N$ or wide $W$ Gaussian smearing) with the results from computing the diagonalized correlation function matrix $\hat{\bm{C}}(\tau)$ with $\{\tau_{\text{ref}},\tau_0\}=\{4,2\}$. Now, comparing with Figs.~\ref{fig:proton_GPoF} and~\ref{fig:proton_prony}, the excited state can be extracted, although with much larger uncertainty than the ground state.\footnote{Other options for the operators, such as displaced sources, are more suitable to study excited states~\cite{Dudek:2007wv}.} Also, the uncertainty in the ground state does not grow as fast as with the use of GPoF or matrix-Prony methods.

A tricky issue with the variational method is that one should include all possible operators that can couple to the states under study, otherwise some levels will be missed. This has already been seen in the mesonic sector, in the study of the $\rho$ resonance~\cite{Dudek:2012xn}, where if only bilinear $\overline{q}\Phi q$ operators or two-pion operators are included, some levels are missed or misplaced, and it is not until both types are used in a variational study that the resonance is observed.

\section{Error estimation}\label{sec:errestimation}

The analysis of statistical uncertainties is one of the key steps in LQCD, since only a finite number of gauge-field configurations are produced to extract observables. Moreover, these configurations are not completely independent, showing correlations among them.

Assuming we have computed $N$ samples of some variable $x_i$, with $n\in\{1,\ldots,N\}$ (the usual starting point is the correlation function in LQCD), the sample mean, variance, and covariance with another variable $y_n$ are given by
\begin{equation}
    \overline{x}=\frac{1}{N}\sum_{n=1}^N x_n\, , \quad \sigma^2_{\overline{x}}=\frac{1}{N(N-1)}\sum_{n=1}^N (x_n-\overline{x})^2\, , \quad \mathcal{C}(x,y)=\frac{1}{N-1}\sum_{n=1}^N(x_n-\overline{x})(y_n-\overline{y})\, .
    \label{eq:meanvarcov}
\end{equation}
As we have seen in the previous sections, the correlation functions are usually manipulated and inserted into functions, such as the effective mass. Then, when we look at some function $f$, the average value, labeled as $f(X)$, can be computed by using $f(\overline{x})$ (and not $\overline{f(x)}=\sum_n f(x_n)/N$, since it would be a biased estimator~\cite{Young:notes}), and the uncertainty can be obtained via error propagation,
\begin{equation}
    \sigma_{f(\overline{x})} = \sigma_{\overline{x}}\left| \frac{d f}{d x} \right|_{x=\overline{x}} \, .
\end{equation}
However, as the number of variables in $f$ increases, and for correlated data, this formula is no longer practical, and alternative methods have to be used. In our case, such methods are resampling methods, the jackknife and bootstrap methods, both widely used in the LQCD community (for an extended and detailed discussion, see Refs.~\cite{Young:notes,Efron:1982,Beane:2014oea}).

\subsection{Jackknife method}

The starting point is to construct the jackknife samples $x^J_n$, which are defined by taking the average of the variable $x$ without including the $n$th sample,
\begin{equation}
    x^J_n=\frac{1}{N-1}\sum_{m\neq n}x_m=\frac{N}{N-1}\overline{x}-\frac{1}{N-1}x_n\, .
\end{equation}
Then, the mean value of the function $f$ and the estimation of its uncertainty is
\begin{equation}
    f(X)\simeq \overline{f}^J_x=\frac{1}{N}\sum_{n=1}^N f(x^J_n)\, , \quad \sigma^2_{f(X)}\simeq \frac{N-1}{N}\sum_{n=1}^N\left[f(x^J_n)-\overline{f}_x^J\right]^2\, .
\end{equation}
An easy check that we can perform is to see that if $f(x)=x$, we recover the expressions from Eq.~\eqref{eq:meanvarcov},
\begin{equation}
\begin{aligned}
    \overline{x} = &\ \frac{1}{N}\sum_{n=1}^N x^J_n=\frac{1}{N}\sum_{n=1}^N\left(\frac{N}{N-1}\overline{x}-\frac{1}{N-1}x_n\right)=\frac{N}{N-1}\overline{x}-\frac{1}{N-1}\overline{x}=\overline{x}\, , \\
    \sigma^2_{\overline{x}} = &\ \frac{N-1}{N}\sum_{n=1}^N\left(x^J_n-\overline{x}\right)^2=\frac{N-1}{N}\sum_{n=1}^N\left(\frac{N}{N-1}\overline{x}-\frac{1}{N-1}x_n-\overline{x}\right)^2\\
     = &\ \frac{1}{N(N-1)}\sum_{n=1}^N\left(x_n-\overline{x}\right)^2=\sigma^2_{\overline{x}}\, .
\end{aligned}
\end{equation}

As pointed out in Ref.~\cite{Young:notes}, a better estimator for $f(X)$ that includes $1/N$ bias correction is given by
\begin{equation}
    f(X)\simeq Nf(\overline{x})-(N-1)\overline{f}^J_x\, .
\end{equation}
Also, this method can be used to compute the covariance matrix, needed when performing $\chi^2$ minimizations for the fitting of the correlation function. It is given by
\begin{equation}
    \mathcal{C}^J[f(x),f(y)]=\frac{N-1}{N}\sum_{n=1}^N\left[f(x^J_n)-\overline{f}^J_x\right]\left[f(y^J_n)-\overline{f}^J_y\right]\, .
\end{equation}

A generalization of the jackknife method is based on not only subtracting one sample from the set, but $k$ samples, ending with $N_k=N/k$ jackknife samples.

\subsection{Bootstrap method}

The bootstrap resampling method, compared to the jackknife one, creates $N_{\text{boot}}$ bootstrap samples $x^B_\alpha$, with $\alpha\in\{1,\ldots,N_{\text{boot}}\}$, where each one is obtained from a random selection from the original sample $N$ points (with repetitions allowed),
\begin{equation}
    x^B_\alpha=\frac{1}{N}\sum_{n=1}^N x_{\text{rand}(1,N)}=\frac{1}{N}\sum_{n=1}^N r_n^{\alpha}x_n\, ,
\end{equation}
where $r_n^{\alpha}$ is the number of times $x_n$ appears in the $\alpha$th bootstrap sample, with the constrain that $\sum_n r^{\alpha}_n = N$, which follows a binomial distribution~\cite{Young:notes},
\begin{equation}
    P(r_n^{\alpha})=\frac{N!}{r_n^{\alpha}!(N-r_n^{\alpha})!}p^{r_n^{\alpha}}(1-p)^{N-r_n^{\alpha}}\, ,
\end{equation}
where $p=1/N$ is the probability that $x_n$ is chosen. With this method, we can also write how to compute the mean value of the function $f$ and the estimation of its uncertainty,
\begin{equation}
    f(X)\simeq \overline{f}^B_x=\frac{1}{N_{\text{boot}}}\sum_{\alpha=1}^{N_{\text{boot}}} f(x^B_\alpha)\, , \quad \sigma^2_{f(X)}\simeq \frac{N}{N-1}\frac{1}{N_{\text{boot}}}\sum_{\alpha=1}^{N_{\text{boot}}}\left[f(x^B_\alpha)-\overline{f}^B_x\right]^2\, .
\end{equation}
In the case of $f(x)=x$, we recover the known formulas. For example, for the mean value,
\begin{equation}
    \overline{x} = \frac{1}{N_{\text{boot}}} \sum_{\alpha=1}^{N_{\text{boot}}} x^B_\alpha = \frac{1}{N} \sum_{n=1}^N \left(\frac{1}{N_{\text{boot}}}\sum_{\alpha=1}^{N_{\text{boot}}} r_n^{\alpha}\right)x_n = \frac{1}{N} \sum_{n=1}^N x_n = \overline{x}\, ,
\end{equation}
where we have assumed that $N_{\text{boot}}$ is very large, so we can use the mean value of a binomial variable, which is $Np$ ( with $p=1/N$, we get $Np=1$).

A better estimator for $f(X)$, which reduces the bias from order  $1/N$ to $1/N^2$~\cite{Young:notes}, is
\begin{equation}
    f(X)\simeq 2f(\overline{x})-\overline{f}^B_x\, .
\end{equation}

The covariance matrix can also be estimated with bootstrap,
\begin{equation}
    \mathcal{C}^B[f(x),f(y)]=\frac{N}{N-1}\frac{1}{N_{\text{boot}}}\sum_{\alpha=1}^{N_{\text{boot}}}\left[f(x^B_\alpha)-\overline{f}^B_x\right]\left[f(y^B_\alpha)-\overline{f}^B_y\right]\, .
\end{equation}

What is interesting with the bootstrap resampling method is that we can look at the distribution function of $f(x)$ and compute the errors with the quantiles (as it is done in the~\cref{subsec:2ptfitting}). Also, the confidence regions of the parameters in some non-linear fits can be extracted easily.

\subsection{Hodges–Lehmann estimator}\label{subsec:HL}

One possible issue that we might encounter with LQCD data is the presence of outliers in the correlation functions due to fluctuations in a finite-size sample. For these cases there exists what are called robust estimators, which are more resilient to these outliers than the usual estimators, like Eq.~\eqref{eq:meanvarcov}. The Hodges–Lehmann (HL) is an example of such estimators~\cite{Hodges:1963}.
The starting point, similar to the bootstrap method, is to generate $N_{\text{HL}}$ HL samples $x^{\text{HL}}_{\mu}$, $\mu\in\{1,\ldots,N_{\text{HL}}\}$, taking $N$ random points from the original sample, compute the pairwise averages (known as the Walsh averages~\cite{Walsh:1949}) and take the median,
\begin{equation}
    x^{\text{HL}}_{\mu}=\text{Median}\left[\left\lbrace\frac{\tilde{x}_n+\tilde{x}_m}{2}\right\rbrace\right]\, ,\quad \text{with  }\; \tilde{x}_n=x_{\text{rand}(1,N)}\; \text{ and }\; 1\leq n,m\leq N\, .
\end{equation}
Then, a robust estimator of the median is defined as
\begin{equation}
    f(X)\simeq \overline{f}^{\text{HL}}_x=\text{Median}\left[\left\lbrace f(x^{\text{HL}}_{\mu}) \right\rbrace \right]\, .
\end{equation}
For the uncertainty associated to this estimator and the computation of covariance matrices, the median absolute deviation (MAD) is used,
\begin{equation}
\begin{aligned}
    \sigma_{f(X)}&=\frac{1}{\Phi^{-1}(\frac{3}{4})}\text{Median}[\{|f(x^{\text{HL}}_{\mu})-\overline{f}^{\text{HL}}_x|\}]\, ,\\
    \mathcal{C}^{\text{HL}}[f(x),f(y)]&=\text{Median}[\{(f(x^{\text{HL}}_{\mu})-\overline{f}^{\text{HL}}_x)(f(y^{\text{HL}}_{\mu})-\overline{f}^{\text{HL}}_y)\}]\, ,
\end{aligned}
\end{equation}
where $\Phi^{-1}(\frac{3}{4})=0.67449$ is the inverse of the normal cumulative distribution function (this factor is needed to convert to the usual error definition of $\sim 68\%$ coverage, since the MAD only covers $\sim 50\%$). In~\cref{appen:vs2015}, a detailed comparison with the bootstrap method is performed, and for additional discussions applied to LQCD, see Refs.~\cite{Beane:2014oea,Orginos:2015aya}.

% Chapter 4 - BB interaction *******************************************
\chapter{The strong baryon-baryon interaction}\label{chap:4}

\section{Present status}\label{sec:presenstatus}

In order to see what the current knowledge of the strong interaction between two baryons at low energies is, let us summarize the present status in the three main fronts: experimental, theoretical, and lattice. For a review on some of these topics, see Refs.~\cite{Feliciello:2015dua,Gal:2016boi,Hiyama:2018lgs,Tolos:2020aln}.

\subsubsection{Experimental front}

\begin{figure}[hbt!]
\input{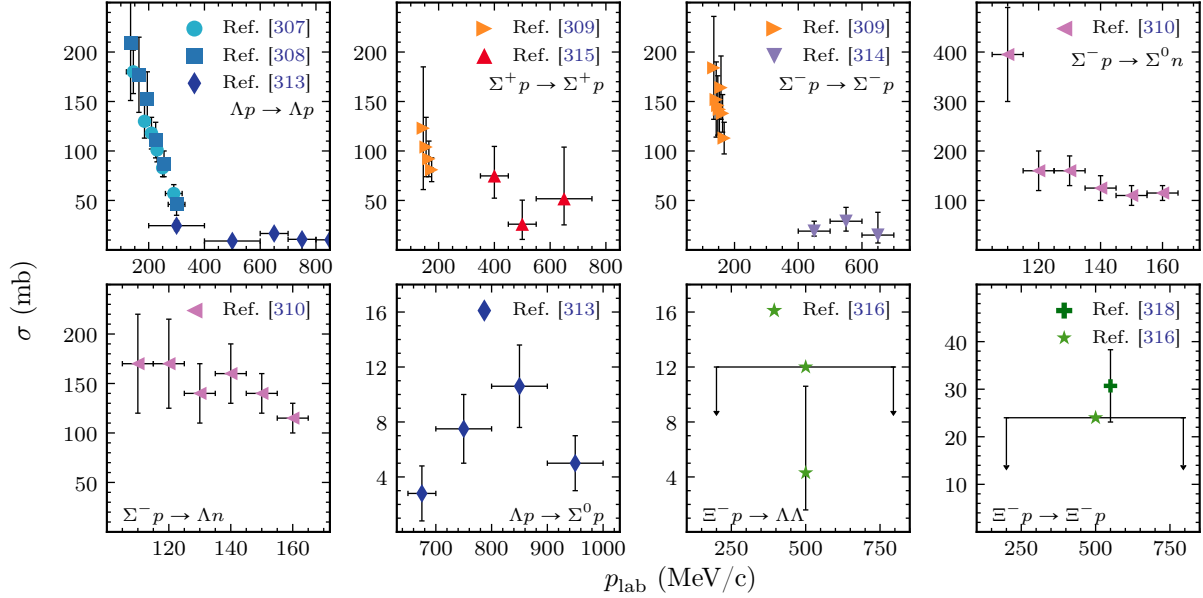}
\caption{Total cross section as a function of $p_{\text{lab}}$ for the baryon-baryon channels in the strangeness sector where experimental data is available. The arrows for the points from Ref.~\cite{Ahn:2005jz} represent the upper limits on the cross section at the 90\% confidence level.}
\label{fig:cross_section}
\end{figure}

Up until the last decade, the main sources of experimental information on the low-energy strong interaction between two baryons were scattering experiments and the study of hypernuclei. Regarding scattering data in the strange sector, the standard set of data used for constraining theoretical models is composed of 36 points, 35 of which are total cross sections: 12 for $\Lambda p \rightarrow \Lambda p$~\cite{Alexander:1969cx,SechiZorn:1969hk}, 4 for $\Sigma^+ p \rightarrow \Sigma^+ p$, 7 for $\Sigma^- p \rightarrow \Sigma^- p$~\cite{Eisele:1971mk}, 6 for $\Sigma^- p \rightarrow \Sigma^0 n$, and 6 for $\Sigma^- p \rightarrow \Lambda n$~\cite{Engelmann1966}. In addition, the inelastic capture rate of $\Sigma^- p$ at rest~\cite{Hepp:1968zza,Stephen:1970} is included.
It is interesting to notice that all the experimental data used was obtained in the early 70s (older data is also available, see~\cite{NNonline}) with quite large uncertainties, and only for systems with only one strange quark (none for $S\leq -2$).
Additional points at slightly higher energies are also included: 7 for $\Lambda p \rightarrow \Lambda p$, 4 for $\Lambda p \rightarrow \Sigma^0 p$~\cite{Kadyk:1971tc}, 3 for $\Sigma^- p \rightarrow \Sigma^- p$~\cite{Kondo:2000hn}, and 3 for $\Sigma^+ p \rightarrow \Sigma^+ p$~\cite{Ahn:2005gb}, bringing the total number of points to 53 (differential cross sections have also been reported in Refs.~\cite{Engelmann1966,Eisele:1971mk,Kondo:2000hn,Ahn:2005gb}, but they are not usually included in this fitting set).
For systems with $S=-2$, there are some upper limits on the total cross section for the channels $\Xi^- p \rightarrow \Xi^- p$ and $\Xi^- p \rightarrow \Lambda \Lambda$~\cite{Ahn:2005jz}, and other in-medium values~\cite{Aoki:1998sv,Tamagawa:2001tk}. All these points are plotted in Fig.~\ref{fig:cross_section}, an updated version of a similar figure appearing in Ref.~\cite{Dover:1990id}.
Additional experiments are running to increase this database, such as the E40 at J-PARC~\cite{Nakada:2019jzs,J-PARCE40:2021qxa}, which focuses on $\Sigma^{\pm}p$ elastic and inelastic channels.
For the sake of comparison, a total of 6713 points have been collected in the $NN$ sector (2996 for $pp$ and 3717 for $np$) for energies below 350 MeV~\cite{Perez:2013mwa,Perez:2013jpa}, including total and differential cross sections, analyzing powers, depolarizations, etc.

An alternative way to learn about two-baryon interactions is by means of the femtoscopy technique, in which two-particle correlation functions in momentum space are measured in high-energy collisions~\cite{Lisa:2005dd}. This correlation function can be computed theoretically, so comparisons with theoretical models~\cite{Lednicky:1981su,Mihaylov:2018rva} as well as extraction of scattering parameters are possible.
The first application of this method to hyperon physics was performed by the STAR Collaboration using heavy-ion collisions, where the $p$--$\Lambda$ correlation function was extracted~\cite{Adams:2005ws}, although with large uncertainties. Several studies have followed by other collaborations, mainly HADES and ALICE.
The channels that have been studied are $p$--$\Lambda$~\cite{Adams:2005ws,Agakishiev:2010qe,Adamczewski-Musch:2016jlh,Acharya:2018gyz}, $\Lambda$--$\Lambda$~\cite{Adamczyk:2014vca,Acharya:2018gyz,Acharya:2019yvb}, $p$--$\Sigma^0$~\cite{Acharya:2019kqn}, $p$--$\Xi^-$~\cite{Acharya:2019sms}, and $p$--$\Omega^-$~\cite{STAR:2018uho,Acharya:2020asf}.
The most important contribution of this method is the possibility to access experimentally systems with $S\leq -2$, very challenging for scattering experiments.
Although some constraints on the scattering parameters for $\Lambda\Lambda$ have been obtained (with quite large uncertainties)~\cite{Acharya:2019yvb}, at the moment the data extracted from femtoscopy is used to compare with other theoretical or lattice predictions.
In the future, the upcoming Runs 3 and 4 of the LHC will allow ALICE to increase statistics for the $p$--$\Lambda$ and $p$--$\Sigma^0$ channels, to access new systems, such as the coupled channel  $N\Lambda$-$N\Sigma$ (the first experimental evidence has been reported in Ref.~\cite{Acharya:2021fdf}) or $\Omega$--$\Omega$, and to study three-body forces~\cite{Fabbietti:2020bfg}.

Other observational means to constrain these interactions, such as radius measurements of neutron stars, their thermal and structural evolution, and the emission of gravitational waves in hot and rapidly rotating newly born neutron stars, can be used to indirectly probe the strangeness content of dense matter and provide complementary constraints on models of hypernuclear interactions~\cite{Chatterjee:2015pua}.

\subsubsection{Theoretical front}

The theoretical approaches that have been used to describe the strong baryon-baryon interaction can be classified into three different types: meson-exchange models, quark models, and effective field theories. For all the cases, since scattering data for systems containing a single strange quark is used (information for systems with $S<-1$ can be obtained, for example, from hypernuclei spectroscopy), the parameters of the theory for more strange systems are related via the use of symmetries, like $SU(3)$ flavor for meson-exchange and EFTs, or $SU(6)$ spin-flavor for the quark models.

Meson-exchange models were developed in the early 70s to approach the problem. The Nijmegen ND~\cite{Nagels:1973rq,Nagels:1976xq} and NF~\cite{Nagels:1978sc} potential models included a short-distance hard core and the exchange of mesons belonging to the pseudoscalar and vector ground-state octets, as well as the two-pion exchange mechanism. Scalar mesons were also included in the NF model.
These models were improved by introducing a soft-core potential, giving rise to the NSC89 model~\cite{Maessen:1989sx} and the five versions of the widely used NSC97 model~\cite{Rijken:1998yy,Stoks:1999bz}. More recent versions of the meson-exchange models are the extended soft-core ESC03~\cite{RIJKEN200527}, ESC04~\cite{Rijken:2006ep,RijkenIII:2006}, ESC08~\cite{Rijken:2010zzb,Rijken:2013wxa,Nagels08II:2015,Nagels08III:2015}, and ESC16~\cite{Nagels:2015lfa,Nagels:2020oqo} models, which contain two-meson exchange contributions. A second group of models, which focused only on $S=-1$ systems, had released three versions~\cite{Holzenkamp:1989tq,Reuber:1993ip,Haidenbauer:2005zh} based on the $NN$ Bonn model~\cite{Machleidt:1987hj}, where $SU(6)$ symmetry relations were used to constrain coupling constants. Finally, hypernuclear spectroscopy data was also included in the fit to constrain the Ehime potential~\cite{Tamotsu:1998,Tominaga:1998,Yamaguchi:2001ip}, an extension of the meson-exchange model of Refs.~\cite{PhysRev.174.1304,PhysRevC.17.1763}, which supplemented the contribution of pseudoscalar, scalar, and vector mesons with the isoscalar scalar part of the uncorrelated and correlated two-pion exchanges.

The second type of approach is non-relativistic quark models, which are based on the resonating-group method~\cite{Oka:1980ax,Oka:1981ri,Oka:1981rj}. While the repulsive short-distance physics is well described via the quark-quark interaction (color-magnetic interaction and Pauli effect), the medium- and long-range effects have to be included by hand via the exchange of mesons. The most known versions are FSS~\cite{Fujiwara:1995fx,Fujiwara:1996qj,Fujita:1998sg} and fss2~\cite{Fujiwara:2001xt,Fujiwara:2006yh}, although other versions like RGM-F and RGM-H are also available~\cite{Nakamoto:1995,Fujiwara:1995td,Fujiwara:1995tc,Fujiwara:1996qj,Fujita:1998sg}.

Lastly, the EFT approach is a model-independent technique based only on the symmetries of the fundamental theory, QCD. Using a power counting scheme, the approach allows us to write the Lagrangian as an expansion, typically in powers of the (low) momenta of the particles, and estimate the error of a calculation at a given order produced by the neglected higher-order terms. This approach was first applied in Ref.~\cite{Savage:1995kv} to the study of the modification of the baryon masses in nuclear matter. Later on, the first study of hyperon-nucleon scattering was performed in Ref.~\cite{Korpa:2001au} within the EFT formalism of Ref.~\cite{Savage:1995kv}. Following the studies in the $NN$ sector, $\chi$EFT calculations at LO~\cite{Polinder:2006zh,Polinder:2007mp,Haidenbauer:2009qn} and NLO~\cite{Haidenbauer:2013oca,Haidenbauer:2014rna,Haidenbauer:2015zqb,Haidenbauer:2018gvg,Haidenbauer:2019boi} were performed for baryon-baryon channels up to $S=-4$. In these studies, several values of the cutoff used to regularize the resulting potential were used to assess the uncertainty, and the difference between the LO and NLO contributions was used to address convergence issues (more detailed studies of uncertainty quantification in EFTs are available for the $NN$ sector~\cite{Epelbaum:2014efa,Melendez:2017phj,Wesolowski:2018lzj}).

\begin{figure}[ht!]
\centering \includegraphics[width=\textwidth]{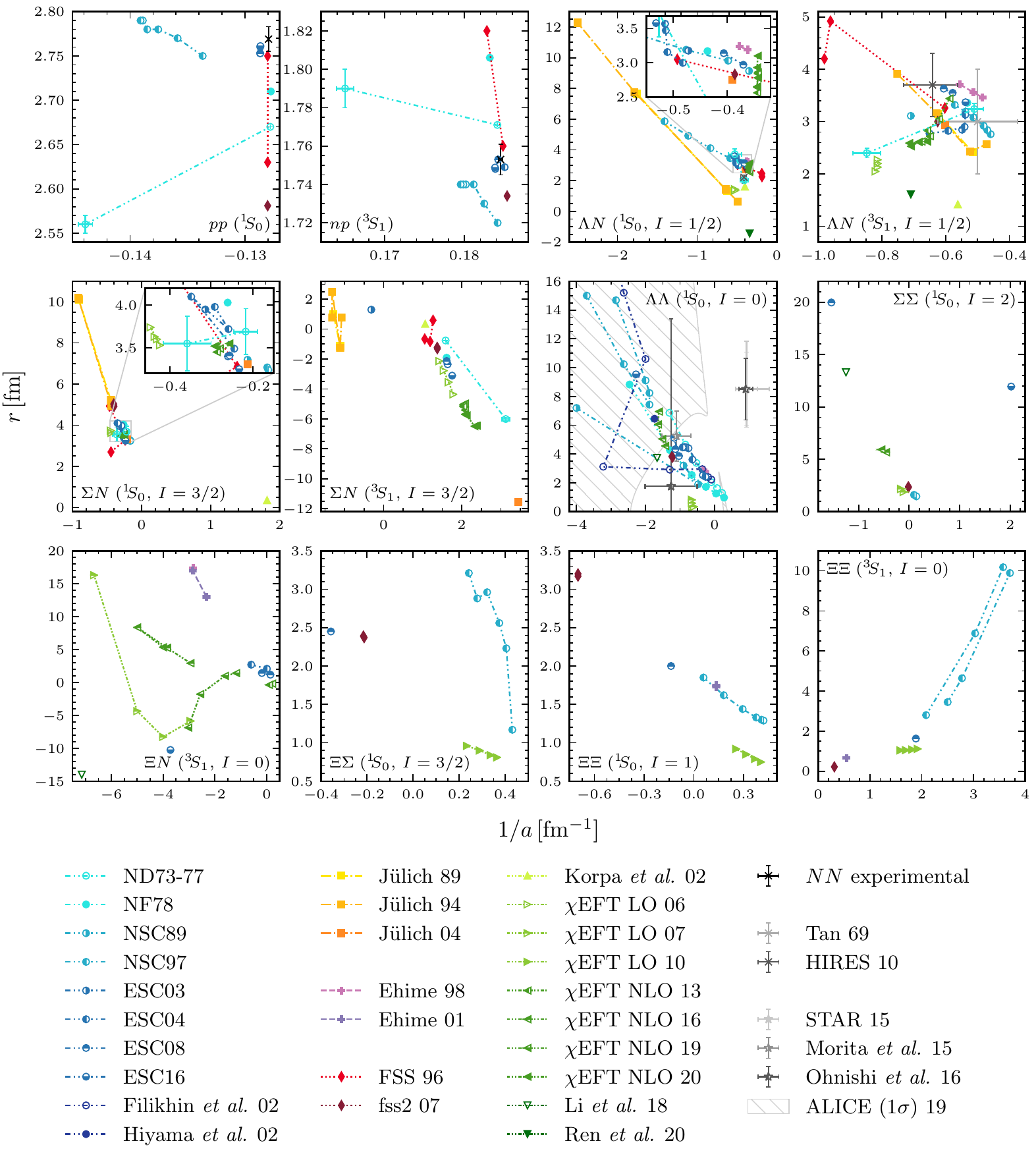}
\caption{Inverse scattering length and effective range predicted by different theoretical models as well as constraints from experiments for different two-baryon systems. The points connected with lines correspond to variations of the same model (like cutoff variations for EFT calculations).}
\label{fig:ar_BB_summary}
\end{figure}

Before discussing the lattice state-of-the-art calculations, it is interesting to see what are the predictions for the scattering parameters obtained with all the above-mentioned theoretical models and compare them to the few available experimental extractions.
All the values are tabulated in Tables~\ref{tab:LNscatt}-\ref{tab:XXscatt} of~\cref{appen:BBsummary}, and plotted in Fig.~\ref{fig:ar_BB_summary} for the cases where both the scattering length and effective range are calculated.
Moreover, the first two panels in Fig.~\ref{fig:ar_BB_summary} are for the $NN$ channels, showing the scattering parameters computed with the same models as the ones used for the hyperon channels (not specific for $NN$)~\cite{Nagels:1973rq,Nagels:1975,Nagels:1978sc,Stoks:1999bz,Rijken:2006en,Rijken:2010zzb,Nagels:2014,Nagels:2014qqa,Fujiwara:1996qj,Fujiwara:2001jg} as well as the experimental values~\cite{Machleidt:2000ge}, where one can see the impact of the quality and quantity of data on the final values.

\subsubsection{Lattice front}

The search for complementary inputs besides experimental data on baryon-baryon scattering has led LQCD to be an important tool to help constrain theoretical models. The first study of the interaction between two baryons was performed by the CP-PACS Collaboration~\cite{Fukugita:1994na,Fukugita:1994ve} for the $NN$ sector, using a quenched calculation with unphysical quark masses. 
This seminal work motivated more studies of the baryon-baryon interaction, by several groups and collaborations, using two different methods: the direct and the potential method. The differences and issues between these two methods will be discussed after presenting a summary of the work and results published to date. 

Ten years later after the appearance of the CP-PACS paper, the NPLQCD Collaboration studied the $NN$ systems in finite volume in detail~\cite{Beane:2003da} considering both scattering and bound states, and also presented the advantages of using LQCD to approach hypernuclear physics~\cite{Beane:2003yx}.
The same collaboration made the first fully-dynamical calculations for the nucleon-nucleon~\cite{Beane:2006mx} and hyperon-nucleon~\cite{Beane:2006gf} systems with a pion mass of around $350$, $490$, and $590$ MeV (earlier investigations on the $\Lambda p$ system were performed by other authors using the quenched approximation in Ref.~\cite{Muroya:2004fz}).
These works were followed by a high statistics analysis of several baryon-baryon systems with strangeness ranging from $0$ to $-4$~\cite{Beane:2009py,Beane:2011iw} at a pion mass of $350$ MeV (the first fully-dynamical three-baryon calculations were also part of this analysis~\cite{Beane:2009gs}), with special interest in the $H$-dibaryon~\cite{Beane:2010hg,Beane:2011zpa} (some previous quenched calculations on this system are summarized in Ref.~\cite{Beane:2011zpa}).
The $\Sigma N$ system was studied in Ref.~\cite{Beane:2012ey}, where the authors extrapolated the interaction to the physical point via the assumption of mass-independence of the LECs, and applied the result to estimate the $\Sigma^-$ energy shift in dense nuclear matter.
In Refs.~\cite{Beane:2012vq,Beane:2013br,Wagman:2017tmp}, the baryon-baryon interaction was studied at the $SU(3)_f$ symmetric point ($m_{\pi}\sim 806$ MeV) with a different action compared to the previous calculations, and in Refs.~\cite{Orginos:2015aya,Illa:2020nsi} a similar study was conducted but at a lower pion mass, $m_{\pi}\sim 450$ MeV, which provided some insights into $SU(3)_f$ symmetry breaking effects. This last work~\cite{Illa:2020nsi} will be thoroughly presented in~\cref{sec:450results}. In a recent preprint~\cite{Amarasinghe:2021}, a detailed variational analysis of two nucleons has been performed. Also, in a joint work with the QCDSF Collaboration, the two- and three-nucleon channels (besides multi-pion and kaon systems) where studied in a LQCD+QED calculation for the first time~\cite{Beane:2020ycc}.

Another collaboration which has worked in the two-baryon sector is the PACS-CS Collaboration, with several works focused only on the $NN$ interaction: first a quenched calculation~\cite{Yamazaki:2011nd}, which came after the first quenched LQCD calculation of ${}^3\text{He}$ and ${}^4\text{He}$~\cite{Yamazaki:2009ua}, followed by fully-dynamical ones with lower quark masses, around $m_\pi \sim 510$~\cite{Yamazaki:2012hi} and $300$ MeV~\cite{Yamazaki:2015asa}.

The other two groups which use the direct method to study two baryons are the CalLat Collaboration and the Mainz group. The former, besides performing an exhaustive study of the $NN$ interaction at higher partial waves, with the extraction of the corresponding scattering parameters~\cite{Berkowitz:2015eaa,Berkowitz:2019yrf}, they also implemented the variational method to the $NN$ system~\cite{Horz:2020zvv}. Nevertheless, the first variational study of two baryons with LQCD calculations was performed by the Mainz group~\cite{Francis:2018qch,Hanlon:2018yfv}, which focused on the $H$-dibaryon channel. Additionally, a continuum extrapolation was performed in Ref.~\cite{Green:2021qol}.

Finally, the HAL QCD Collaboration has pursued the study of two-baryon systems with a different method, known as the potential method.
Following a first set of calculations of $NN$ and $YN$ systems within the quenched approximations~\cite{Ishii:2006ec,Aoki:2008hh,Nemura:2008sp,Aoki:2009ji}, fully-dynamical studies were performed for several two-baryon systems at the $SU(3)_f$-symmetric point~\cite{Inoue:2010hs,Inoue:2010es,Inoue:2011ai}, as well as including $SU(3)_f$ symmetry breaking~\cite{Ishii:2013cta,Murano:2013xxa}. 
This collaboration has also presented the results for two-baryon systems with the quark masses closest to their physical values ($m_\pi\sim 146$ MeV). While some preliminary results were presented in several conferences (e.g., Ref.~\cite{Doi:2015oha,Doi:2017cfx,Doi:2017zov}), the published ones are for the $\Lambda\Lambda$ and $N\Xi$ systems~\cite{Sasaki:2019qnh}. Several decuplet systems have been studied the same ensembles, mainly $\Omega\Omega$~\cite{Yamada:2015cra,Gongyo:2017fjb} and $N\Omega$~\cite{Etminan:2014tya,Iritani:2018sra}). The quark-mass dependence of the $\Lambda N$ system was studied in a joint calculation performed by the HAL QCD and PACS-CS collaborations~\cite{Nemura:2009kc,Nemura:2010nh}. 

\begin{figure}[b!]
\input{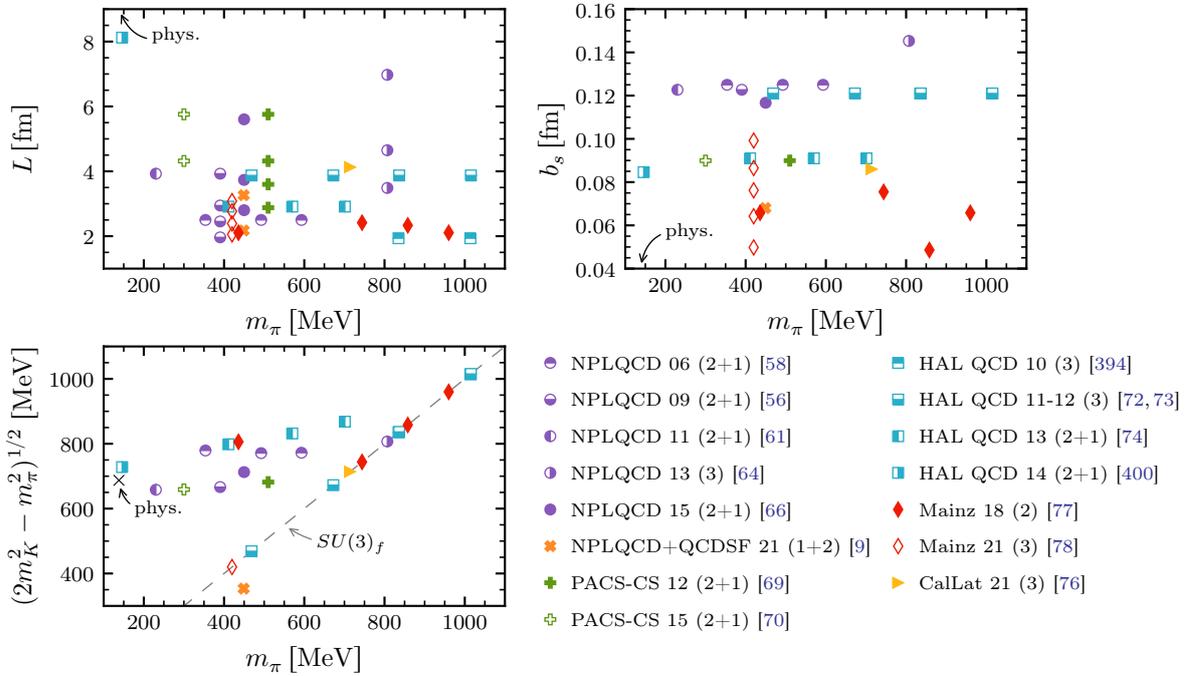}
\caption{Summary of the gauge-field ensemble parameters of fully-dynamical LQCD calculations for two-baryon systems. These include the lattice spatial extent (top left), the lattice spatial spacing (top right), and the strange quark mass (bottom left) plotted as a function of the pion mass.}
\label{fig:LQCD_parameters_allcollabs}
\end{figure}

In Fig.~\ref{fig:LQCD_parameters_allcollabs}, a summary of the parameters of the gauge-field ensembles used in fully-dynamical calculations made by all the collaborations is shown (similar to the ones in Ref.~\cite{Hoelbling:2011kk} for single-hadron physics).
These parameters are the pion mass $m_{\pi}$, spatial extent $L$, lattice spacing in the spatial direction $b_s$ (this clarification is needed for the anisotropic calculations), and the quantity $(2m^2_K-m^2_{\pi})^{1/2}$, which is proportional to the strange quark mass (see Eq.~\eqref{eq:quarkmassmatrix}).
In all of these plots, the physical point is indicated with a cross or an arrow labeled with ``phys.'', and in the left-bottom panel the $SU(3)_f$ case ($m_u=m_d=m_s$) with a dashed line. Some of the ensembles have been used more than one time (e.g., ``NPLQCD 15'' was first used in 2015 in Ref.~\cite{Orginos:2015aya} and later on re-analyzed in 2020 in Ref.~\cite{Illa:2020nsi}) and by more than one collaboration (e.g., ``NPLQCD 13'' was first used by the NPLQCD Collaboration in Refs.~\cite{Beane:2012vq,Beane:2013br} and later one re-used by the CalLat Collaboration in Ref.~\cite{Berkowitz:2015eaa}), so they are labeled by the first publication (collaboration and year). In Tables~\ref{tab:NNLQCD_summary}-\ref{tab:XXLQCD_summary} of~\cref{appen:BBsummary}, the binding energies computed by all these collaborations (only for fully dynamical calculations) are summarized.

Discrepancies in the spectrum of baryon-baryon systems obtained with the two methods exist (as seen in the tables in~\cref{appen:BBsummary}), in particular whether there are bound-states or not at heavier-than-physical quark masses. A summary of these can be found in the following reviews~\cite{Drischler:2019xuo,Davoudi:2020ngi}, and we recapitulate them here:
\begin{itemize}[label={--}]
    \item The direct (or Lüscher) method relies on extracting the finite-volume energy levels of the system directly from two-point correlation functions, and then applying Lüscher QCs, which are model independent, to extract phase shifts and binding energies.
    The principal criticism~\cite{Iritani:2016jie,Iritani:2017rlk,Aoki:2017byw} is that two-baryon correlation functions are contaminated by small elastic excitation gaps (few MeV) at $\sim 1$ fm (typical time which the region of the plateau, where the fits are performed, starts), and cancellations between these excited states produce a ``mirage'' plateau, requiring the extension of the fitting region to include larger times ($\sim 8$ fm, entering the noise region) to correctly extract the ground-state energy. This incorrect identification is then observed in the phase shift extracted, with some unphysical behaviors.
    These criticisms have been refuted in Refs.~\cite{Wagman:2017tmp,Beane:2017edf,Davoudi:2017ddj,Illa:2020nsi}, where one of the arguments is the volume-independence of the bound ground-state energies, indicating that it is unlikely that the contamination from excited states is the same in all the volumes studied (since this contamination comes from scattering states, which are volume dependent).
    In Ref.~\cite{Yamazaki:2017jfh}, the authors showed that with high-enough statistics, their correlation functions are operator independent, and that optimized operators can reduce excited state contamination and produce a plateau at early times.
    One possible way to face this issue is via the use of the variational method, as it is done in Refs.~\cite{Francis:2018qch,Horz:2020zvv,Green:2021qol} with the inclusion of dibaryon operators, and Ref.~\cite{Amarasinghe:2021}, where a more complete set, including hexaquark and quasi-local operators, is used. Another possible explanation for the discrepancies may be related to discretization errors, not studied until Ref.~\cite{Green:2021qol}, where quite large lattice spacing effects on the binding energy of the $H$-dibaryon are seen.
    \item The potential (or HAL QCD) method first computes the Bethe-Salpeter wavefunction of the two-baryon system from two-point correlation functions, then extracts the non-local potential, parametrized in a truncated velocity (derivative) expansion, by solving the Schrödinger equation in the infinite volume. Then, using this potential, the phase shifts and binding energy of the system can be computed.
    This method has two variants: the time-independent~\cite{Ishii:2006ec} and time-dependent~\cite{HALQCD:2012aa} one. The difference between both variants is that the latter, which is the one currently used by HAL QCD, allegedly does not require ground-state saturation (in contrast to the former and the direct method).
    Several criticisms for unaccounted systematic uncertainties in this method are collected and discussed in Refs.~\cite{Beane:2010em,Walker-Loud:2014iea,Savage:2016egr,Yamazaki:2017gjl,Yamazaki:2018qut}, and answered in Refs.~\cite{Aoki:2017yru,Iritani:2018zbt}. On the one hand, the potential (and phase shift) computed is only defined at the energy extracted from the correlation function, not for all values of the momentum. Then, the velocity expansion has momentum-dependent coefficients, and its convergence is not clear.
    On the other hand, the potential is not a physical quantity~\cite{Birse:2012ph} and is operator-dependent in its short-distance behavior. Also, its long-distance behavior is qualitatively different from the expected one from meson-exchange models in some channels, like $\Omega N$ and $\Omega\Omega$~\cite{Haidenbauer:2019utu}.
\end{itemize}
Progress is being made to try to understand the differences, as mentioned before. Regarding the lattice results that will be shown later on in this chapter, all possible consistency checks examined confirm the robustness of the extracted energies (see~\cref{subsec:validityLQCD}).

\section{Effective description of two baryons at low energies}

In this section we describe the two EFTs that will be used to describe the interaction between two octet baryons. One based on $SU(3)_f$ symmetry, with the inclusion of $SU(3)_f$ breaking terms, and the other based on $SU(6)$ spin-flavor symmetry, predicted in the limit of large number of colors $N_c$.

\subsection{Assuming \texorpdfstring{$SU(3)$}{SU(3)} flavor symmetry}\label{subsec:SU3EFT}

The Lagrangian for the low-energy interactions of two octet baryons under the assumption of $SU(3)_f$ was first constructed in Ref.~\cite{Savage:1995kv} using the heavy-baryon chiral EFT (HB$\chi$EFT) formalism~\cite{Jenkins:1990jv}, and consists of two-baryon contact operators at LO. Ten years later, Refs.~\cite{Polinder:2006zh,Haidenbauer:2007ra} studied these interactions using chiral perturbation theory ($\chi$PT), where in addition to the momentum-independent operators at LO, the pseudoscalar-meson exchanges were included in the interacting potential. At LO, all terms in both HB$\chi$EFT and $\chi$PT are $SU(3)_f$ symmetric.
At NLO, there are two types of contributions: the $SU(3)_f$-symmetric interactions, obtained by the addition of derivative terms to the LO Lagrangian, and the $SU(3)_f$ symmetry-breaking interactions, denoted by $\cancel{SU(3)}_f$ hereafter, that arise from the inclusion of the quark-mass matrix. Higher partial waves were included in the NLO extension of the two-baryon potential in $\chi$PT, which first appeared in Refs.~\cite{Petschauer:2013uua,Haidenbauer:2013oca}.

In this thesis, two-baryon systems are analyzed at low energies; therefore only $S$-wave interactions are considered. Also, pionless EFT~\cite{vanKolck:1998bw,Chen:1999tn} in the hypernuclear sector is considered, so only contact terms are used. The octet-baryon fields are introduced as a $3\times 3$ matrix in flavor space,
\begin{equation}
    B=\begin{pmatrix}
    \frac{\Sigma^0}{\sqrt{2}}+\frac{\Lambda}{\sqrt{6}} & \Sigma^+ & p \\
    \Sigma^- & -\frac{\Sigma^0}{\sqrt{2}}+\frac{\Lambda}{\sqrt{6}} & n \\
    \Xi^- & \Xi^0 & -\sqrt{\frac{2}{3}}\Lambda
    \end{pmatrix},
\label{eq:Bmatrix}
\end{equation}
that transforms under the chiral symmetry group $SU(3)_L\times SU(3)_R$ as $B \rightarrow U B U^{\Dag}$, where $U$ depends on $L,R\in SU(3)_{L,R}$ and the pseudoscalar meson matrix~\cite{carruthers1966}. At LO, we need two baryon fields $B$ and two adjoint baryon fields $B^{\Dag}$ (with no external fields), which can be arranged in seven different ways,
\begin{equation}
\begin{aligned}
    \mathcal{O}_1 &= \Tr(B^{\Dag}_i B_i B^{\Dag}_j B_j)\, , &\quad \mathcal{O}_2 &= \Tr(B^{\Dag}_i B_j B^{\Dag}_j B_i)\, , \\
    \mathcal{O}_3 &= \Tr(B^{\Dag}_i B^{\Dag}_j B_i B_j)\, , &\quad \mathcal{O}_4 &= \Tr(B^{\Dag}_i B^{\Dag}_j B_j B_i)\, , \\
    \mathcal{O}_5 &= \Tr(B^{\Dag}_i B_i) \Tr(B^{\Dag}_j B_j)\, , &\quad \mathcal{O}_6 &= \Tr(B^{\Dag}_i  B_j) \Tr(B^{\Dag}_j B_i)\, , \\
    \mathcal{O}_7 &= \Tr(B^{\Dag}_i B^{\Dag}_j) \Tr(B_i B_j)\, ,
\end{aligned}
\end{equation}
where the indices $i$ and $j$ denote spin indices (up and down). The last term, $\mathcal{O}_7$, can be eliminated using the Cayley-Hamilton identity~\cite{Urech:1994hd}, which relates the product of traces involving four $3\times 3$ matrices, $A_k$ with $k\in\{1,2,3,4\}$, as
\begin{equation}
\begin{aligned}
    0=&\sum_{6 \text{ perm}} \Tr(A_1A_2A_3A_4) -\sum_{8 \text{ perm}} \Tr(A_1A_2A_3) \Tr(A_4)- \sum_{3 \text{ perm}} \Tr(A_1A_2) \Tr(A_3 A_4) \\
    & + \sum_{6 \text{ perm}} \Tr(A_1A_2) \Tr(A_3) \Tr(A_4) - \Tr(A_1) \Tr(A_2) \Tr(A_3) \Tr(A_4) \, .
\end{aligned}
\end{equation}
Since in our case the baryon matrix is traceless, as it can be seen in Eq.~\eqref{eq:Bmatrix}, $\Tr(A_k)=0$, the previous relation becomes
\begin{equation}
    \sum_{6 \text{ perm}} \Tr(A_1A_2A_3A_4) = \sum_{3 \text{ perm}} \Tr(A_1A_2) \Tr(A_3 A_4)\, .
\end{equation}
Taking $A_1=B^{\Dag}_i$, $A_2=B^{\Dag}_j$, $A_3=B_i$, and $A_4=B_j$, we get
\begin{equation}
\begin{aligned}
    &\Tr(B^{\Dag}_i B^{\Dag}_j B_i B_j) + \Tr(B^{\Dag}_i B_i B^{\Dag}_j B_j) + \Tr(B^{\Dag}_j B^{\Dag}_i B_i B_j) + \Tr(B^{\Dag}_j B_i B^{\Dag}_i B_j) + \Tr(B_i B^{\Dag}_i B^{\Dag}_j B_j) \\ 
    &+ \Tr(B_i B^{\Dag}_j B^{\Dag}_i B_j) = \Tr(B^{\Dag}_i B^{\Dag}_j) \Tr(B_i B_j) + \Tr(B^{\Dag}_i B_i) \Tr(B^{\Dag}_j  B_j) + \Tr(B^{\Dag}_i B_j) \Tr(B^{\Dag}_j B_i)\, .
\end{aligned}
\label{eq:Cayley_expand}
\end{equation}
We notice that on the left-hand side of Eq.~\eqref{eq:Cayley_expand} there are some terms that are equal (using the cyclic property of the trace),
\begin{equation}
\begin{aligned}
 \Tr(B^{\Dag}_i &B^{\Dag}_j B_i B_j) +\Tr(B^{\Dag}_j B^{\Dag}_i B_i B_j) + \Tr(B_i B^{\Dag}_i B^{\Dag}_j B_j) + \Tr(B_i B^{\Dag}_j B^{\Dag}_i B_j) \\
&= \Tr(B^{\Dag}_i B^{\Dag}_j B_i B_j) +\underbrace{ \Tr(B^{\Dag}_j B^{\Dag}_i B_i B_j)}_{i\leftrightarrow j} + \Tr(B^{\Dag}_i B^{\Dag}_j B_j B_i) + \underbrace{\Tr(B^{\Dag}_j B^{\Dag}_i B_j B_i)}_{i\leftrightarrow j} \\
&= 2 \Tr(B^{\Dag}_i B^{\Dag}_j B_i B_j) + 2\Tr(B^{\Dag}_i B^{\Dag}_j B_j B_i) \, ,
\end{aligned}
\end{equation}
leading to
\begin{equation}
 \Tr(B^{\Dag}_i B^{\Dag}_j) \Tr(B_i B_j) = \mathcal{O}_7 = \mathcal{O}_1 + \mathcal{O}_2 + 2\mathcal{O}_3 + 2\mathcal{O}_4 - \mathcal{O}_5 - \mathcal{O}_6\, .
\end{equation}
Therefore, the LO Lagrangian can be written as
\begin{equation}
\begin{aligned}
    \mathcal{L}^{(0),\, SU(3)_f}_{BB}=&-c_1 \Tr(B^{\Dag}_i B_i B^{\Dag}_j B_j)-c_2 \Tr(B^{\Dag}_i B_j B^{\Dag}_j B_i)-c_3 \Tr(B^{\Dag}_i B^{\Dag}_j B_i B_j) \\
    &-c_4 \Tr(B^{\Dag}_i B^{\Dag}_j B_j B_i)-c_5 \Tr(B^{\Dag}_i B_i )\Tr(B^{\Dag}_j B_j)-c_6\Tr(B^{\Dag}_i B_j )\Tr(B^{\Dag}_j B_i)\, . \label{eq:LagLO}
\end{aligned}
\end{equation}
The coefficients $c_i$ are known as Savage-Wise coefficients~\cite{Savage:1995kv}. We can compare this expression with the non-relativistic LO Lagrangian that follows from Ref.~\cite{Petschauer:2013uua},
\begin{equation}
\begin{aligned}
     \mathcal{L}^{(0)}=&\, a_{1,1} \Tr(\overline{\mathcal{B}}_1\mathcal{B}_1 \overline{\mathcal{B}}_2\mathcal{B}_2) + a_{2,1} \Tr(\overline{\mathcal{B}}_1 \overline{\mathcal{B}}_2 \mathcal{B}_1\mathcal{B}_2)+ a_{3,1} \Tr(\overline{\mathcal{B}}_1\mathcal{B}_1) \Tr(\overline{\mathcal{B}}_2\mathcal{B}_2)\\
     &+ a_{1,2} \Tr[\overline{\mathcal{B}}_1(\gamma_5\gamma^{\mu}\mathcal{B})_1 \overline{\mathcal{B}}_2(\gamma_5\gamma_{\mu}\mathcal{B})_2] + a_{1,3} \Tr[\overline{\mathcal{B}}_1(\sigma^{\mu\nu}\mathcal{B})_1 \overline{\mathcal{B}}_2(\sigma_{\mu\nu}\mathcal{B})_2]\\
    &+ a_{2,2} \Tr[\overline{\mathcal{B}}_1 \overline{\mathcal{B}}_2(\gamma_5\gamma^{\mu}\mathcal{B})_1(\gamma_5\gamma_{\mu}\mathcal{B})_2] + a_{2,3} \Tr[\overline{\mathcal{B}}_1 \overline{\mathcal{B}}_2(\sigma^{\mu\nu}\mathcal{B})_1(\sigma_{\mu\nu}\mathcal{B})_2]\\
    &+ a_{3,2} \Tr[\overline{\mathcal{B}}_1(\gamma_5\gamma^{\mu}\mathcal{B})_1] \Tr[\overline{\mathcal{B}}_2(\gamma_5\gamma_{\mu}\mathcal{B})_2] + a_{3,3} \Tr[\overline{\mathcal{B}}_1(\sigma^{\mu\nu}\mathcal{B})_1] \Tr[\overline{\mathcal{B}}_2(\sigma_{\mu\nu}\mathcal{B})_2]\, ,
\end{aligned}
\end{equation}
where here $\mathcal{B}$ is the baryon matrix with the baryon fields being full four-component spinors. 
We can obtain an equivalent expression to Eq.~\eqref{eq:LagLO} and reduce the number of terms to six by considering only the large components of the spinor (the two upper ones), and the terms $a_{i,2}$ and $a_{i,3}$ are proportional to each other. These terms lead to expressions involving Pauli matrices, and we can get rid of them by using the Majorana exchange operator $(\sigma_k B)_i (\sigma_k B)_j = 2B_jB_i - B_iB_j$~\cite{Miller:2006jf}.

The reason why there are only six terms can be understood from group-theoretical considerations. In the spin-flavor decomposition of the product of two octet baryons with $J^P=\frac{1}{2}^+$, the 64 existing channels can be grouped into six different irreps,
\begin{equation}
    \mathbf{8}\otimes \mathbf{8} = \mathbf{27}\oplus \mathbf{8}_s \oplus \mathbf{1} \oplus \overline{\mathbf{10}} \oplus \mathbf{10} \oplus \mathbf{8}_a\, .
\label{eq:8x8SU3}
\end{equation}
There are several ways to find out which two-baryon states belong to which irrep. One possibility is to use the $SU(3)_f$ Clebsch-Gordan coefficients~\cite{RevModPhys.35.916,RevModPhys.36.1005}, to use the quadratic and cubic $SU(3)$ Casimir operators, as well as the $V$-spin, $U$-spin, and isospin raising and lowering operators~\cite{Beane:2012vq}, or to compute the coefficients using the Lagrangian in Eq.~\eqref{eq:LagLO} for all the two-baryon systems, and diagonalize the flavor matrix, giving directly the combination of baryons and $c_i$ for each irrep. This last procedure was used in Refs.~\cite{Wagman:2017tmp,Illa:2020nsi}, and as an example, let us look at the $\Sigma\Sigma$ channels. For this case, we have to consider the following flavor matrix,
\begin{equation}
   \begin{blockarray}{cccccccccc}
     & \Sigma^+ \Sigma^+ & \Sigma^+ \Sigma^0 & \Sigma^+ \Sigma^- & \Sigma^0 \Sigma^+ & \Sigma^0 \Sigma^0 & \Sigma^0 \Sigma^- & \Sigma^- \Sigma^+ & \Sigma^- \Sigma^0 & \Sigma^- \Sigma^- \\
    \begin{block}{c(ccccccccc)}
     \Sigma^+ \Sigma^+ & X & 0 & 0 & 0 & 0 & 0 & 0 & 0 & 0 \\
     \Sigma^+ \Sigma^0 & 0 & X & 0 & X & 0 & 0 & 0 & 0 & 0 \\
     \Sigma^+ \Sigma^- & 0 & 0 & X & 0 & X & 0 & X & 0 & 0 \\
     \Sigma^0 \Sigma^+ & 0 & X & 0 & X & 0 & 0 & 0 & 0 & 0 \\
     \Sigma^0 \Sigma^0 & 0 & 0 & X & 0 & X & 0 & X & 0 & 0 \\
     \Sigma^0 \Sigma^- & 0 & 0 & 0 & 0 & 0 & X & 0 & X & 0 \\
     \Sigma^- \Sigma^+ & 0 & 0 & X & 0 & X & 0 & X & 0 & 0 \\
     \Sigma^- \Sigma^0 & 0 & 0 & 0 & 0 & 0 & X & 0 & X & 0 \\
     \Sigma^- \Sigma^- & 0 & 0 & 0 & 0 & 0 & 0 & 0 & 0 & X \\
    \end{block}
  \end{blockarray}\, ,
\end{equation}
where the $X$ denote the non-zero entries $\Sigma_1 \Sigma_2\rightarrow \Sigma'_1 \Sigma'_2$. Focusing only in the spin-singlet case, we compute the matrix elements $\langle \Sigma'_1 \Sigma'_2 | \mathcal{L}^{(0),\, SU(3)_f}_{BB} | \Sigma_1 \Sigma_2 \rangle$ where the states have the correct spin wavefunction, which in this case is $\frac{1}{\sqrt{2}}(\Sigma_1^\uparrow \Sigma_2^\downarrow - \Sigma_1^\downarrow \Sigma_2^\uparrow)$, leading to the following coefficient matrix,
\begin{equation}
    \begin{pmatrix}
     2 c_A & 0 & 0 & 0 & 0 & 0 & 0 & 0 & 0 \\
     0 & c_A & 0 & c_A & 0 & 0 & 0 & 0 & 0 \\
     0 & 0 & c_B & 0 & c_C & 0 & c_B & 0 & 0 \\
     0 & c_A & 0 & c_A & 0 & 0 & 0 & 0 & 0 \\
     0 & 0 & c_C & 0 & c_D & 0 & c_C & 0 & 0 \\
     0 & 0 & 0 & 0 & 0 & c_A & 0 & c_A & 0 \\
     0 & 0 & c_B & 0 & c_C & 0 & c_B & 0 & 0 \\
     0 & 0 & 0 & 0 & 0 & c_A & 0 & c_A & 0 \\
     0 & 0 & 0 & 0 & 0 & 0 & 0 & 0 & 2 c_A \\
    \end{pmatrix}\, ,
\end{equation}
where $c_A=c_1-c_2+c_5-c_6$, $c_B=-c_3+c_4+c_5-c_6$, $c_C=-c_1+c_2-c_3+c_4$, and $c_D=c_1-c_2-c_3+c_4+2 c_5-2 c_6$. Diagonalizing this matrix, and writing the eigenvectors in terms of the basis $\{\Sigma^+\Sigma^+,\Sigma^+\Sigma^0,\ldots,\Sigma^-\Sigma^-\}$, we obtain the following eigenvalues,
\begin{equation}
\begin{aligned}
    &\begin{Bmatrix}
    \Sigma^+\Sigma^+, \Sigma^-\Sigma^-, \\
    \sqrt{\frac{1}{2}}(\Sigma^0\Sigma^+ +\Sigma^+\Sigma^0), \sqrt{\frac{1}{2}}(\Sigma^-\Sigma^0 +\Sigma^0\Sigma^-), \\
    \sqrt{\frac{1}{6}}(\Sigma^+\Sigma^- -2\Sigma^0\Sigma^0 + \Sigma^-\Sigma^+) \\
    \end{Bmatrix} &:\quad & 2 (c_1-c_2+c_5-c_6) \, ,\\[0.2cm]
    &\begin{Bmatrix}
    \sqrt{\frac{1}{2}}(\Sigma^0\Sigma^+ -\Sigma^+\Sigma^0), \sqrt{\frac{1}{2}}(\Sigma^-\Sigma^0 -\Sigma^0\Sigma^-), \\
    \sqrt{\frac{1}{2}}(\Sigma^-\Sigma^+ -\Sigma^+\Sigma^-) \\
    \end{Bmatrix} &:\quad & 0 \, ,\\[0.2cm]
    &\qquad\qquad\quad\;\begin{Bmatrix}
    \sqrt{\frac{1}{3}}(\Sigma^+\Sigma^- + \Sigma^0\Sigma^0 + \Sigma^-\Sigma^+) \\
    \end{Bmatrix} &:\quad & -c_1+c_2-3c_3+3c_4+2c_5-2c_6 \, .
\end{aligned}
\end{equation}
Looking at the results we see that the diagonalization has produced the three isospin structures: $I=2$ (top), $1$ (middle), and $0$ (bottom), and that the flavor-antisymmetric wavefunction (in the center) has a null eigenvalue since it cannot have spin zero, as expected (if we do the same calculation but for spin-triplet states, we get the opposite results). The isospin $I=2$ channels belong to the $\mathbf{27}$ irrep: when we apply this same procedure to all the baryon-baryon channels, we find that there are 27 channels with the coefficients $2(c_1-c_2+c_5-c_6)$. The $I=0$ channel is somewhat more intricate because it can couple to other flavor channels (mainly $\Lambda \Lambda$ and $\Xi N$), so we have to repeat this procedure again, considering all the possible $I=0$ states, and see that we get the combinations for the $\mathbf{27}$, $\mathbf{8}_s$, and $\mathbf{1}$ irreps. This procedure can be automated using \texttt{Mathematica}~\cite{Mathematica} together with the \texttt{grassmann} package~\cite{grassmann}, which leads to the following combination of Savage-Wise coefficients for each irrep,
\begin{equation}
\begin{aligned}
    c^{(27)}&= 2(c_1-c_2+c_5-c_6)\, ,& c^{(\overline{10})}&= 2(c_1+c_2+c_5+c_6)\, ,\\
    c^{(8_s)}&= \frac{1}{3}(-4c_1+4c_2-5c_3+5c_4+6c_5-6c_6),& c^{(10)}&= 2(-c_1-c_2+c_5+c_6)\, ,\\
    c^{(1)}&= \frac{2}{3}(-c_1+c_2-8c_3+8c_4+3c_5-3c_6)\, ,& c^{(8_a)}&= 3c_3+3c_4+2c_5+2c_6\, . \label{eq:SU3eq}
\end{aligned} 
\end{equation}
As we will see later on, this set of combinations is the one that we can access with LQCD data, and since we do not study all irreps, in principle we cannot invert this relation and compute each $c_i$ separately. In~\cref{appen:SU3chann}, the full flavor decomposition of all two-baryon channels in their $SU(3)_f$ irreps is shown.

Let us focus now on the NLO terms. The non-relativistic $SU(3)_f$-symmetric Lagrangian obtained from Ref.~\cite{Petschauer:2013uua} contains 19 terms with different derivative structures. However, here we are only interested in systems that are in the $S$ wave, so we need to find out which terms contribute to which partial wave. After a detailed analysis, found in~\cref{appen:NLOEFT}, only the second derivatives contribute to the $S$ wave, so the $SU(3)_f$ NLO Lagrangian, following the organization of the LO terms in Eq.~\eqref{eq:LagLO}, has the form
\begin{equation}
    \begin{aligned}
    \mathcal{L}^{(2),\, SU(3)_f}_{BB} = &-\tilde{c}_{1} \Tr( B^{\Dag}_i \nabla^2 B_i B^{\Dag}_j B_j+\text{h.c.})
    -\tilde{c}_{2}\Tr(B^{\Dag}_i \nabla^2 B_j B^{\Dag}_j B_i+\text{h.c.})\\
    &-\tilde{c}_{3}\Tr(B^{\Dag}_i B^{\Dag}_j \nabla^2B_i B_j+\text{h.c.})
    -\tilde{c}_{4}\Tr(B^{\Dag}_i B^{\Dag}_j \nabla^2 B_j B_i+\text{h.c.})\\
    &-\tilde{c}_{5}[\Tr(B^{\Dag}_i \nabla^2 B_i)\Tr(B^{\Dag}_j B_j)+\text{h.c.}]
    -\tilde{c}_{6}[\Tr(B^{\Dag}_i \nabla^2 B_j)\Tr(B^{\Dag}_j B_i)+\text{h.c.}]\, , \label{eq:LagNLO1}
    \end{aligned}
\end{equation}
where the same relations as in Eq.~\eqref{eq:SU3eq} hold for $\tilde{c}^{(\text{irrep})}$, just by replacing $c_i$ with $\tilde{c}_i$.

For the $\cancel{SU(3)}_f$ terms, as pointed out in Ref.~\cite{Petschauer:2013uua}, we need to introduce the quark mass matrix $\chi$,
\begin{equation}
    \chi=2B_0\begin{pmatrix} m_u & 0 & 0 \\ 0 & m_d & 0 \\ 0 & 0 & m_s \end{pmatrix}\approx \begin{pmatrix} m^2_\pi & 0 & 0 \\ 0 & m^2_\pi & 0 \\ 0 & 0 & 2m^2_K-m^2_\pi \end{pmatrix}\, ,
\label{eq:quarkmassmatrix}
\end{equation}
where $B_0$ is proportional to the quark condensate, and we have used the Gell-Mann–Oakes–Renner relation~\cite{GellMann:1968rz} to write this matrix in terms of the meson masses instead of the quark masses,
\begin{equation}
    m^2_\pi\approx(m_u+m_d)B_0=2m_lB_0\, , \quad m^2_K\approx(m_l+m_s)B_0\, ,
\end{equation}
with $m_l=m_u=m_d$. Since this part of the Lagrangian breaks $SU(3)_f$, we will not have six terms, and we need to find all possible combinations involving Eqs.~\eqref{eq:LagLO} and~\eqref{eq:quarkmassmatrix}. Again, this was first derived in Ref.~\cite{Petschauer:2013uua} (the relations between both sets of $\cancel{SU(3)}_f$ coefficients are presented in Table~\ref{tab:coeff_petshauervsme} of~\cref{appen:LECsEFT}), and we write it here with the terms following the organization as presented in the LO case,
\begin{equation}
    \begin{aligned}
    \mathcal{L}^{(2),\, \cancel{SU(3)}_f}_{BB} = &-c_1^{\chi}\Tr(B^{\Dag}_i \chi B_i B^{\Dag}_j B_j) - c_2^{\chi}\Tr(B^{\Dag}_i \chi B_j B^{\Dag}_j B_i)\\
    & - c_3^{\chi}\Tr(B^{\Dag}_i B_i \chi B^{\Dag}_j B_j) - c_4^{\chi}\Tr(B^{\Dag}_i B_j \chi B^{\Dag}_j B_i)\\
    & - c_5^{\chi}\Tr(B^{\Dag}_i \chi B^{\Dag}_j B_i B_j +\text{h.c.}) - c_6^{\chi}\Tr(B^{\Dag}_i \chi B^{\Dag}_j B_j B_i+\text{h.c.}) \\
    & - c_7^{\chi}\Tr(B^{\Dag}_i B^{\Dag}_j \chi B_i B_j ) - c_8^{\chi}\Tr(B^{\Dag}_i B^{\Dag}_j \chi B_j B_i)\\
    & - c_9^{\chi}\Tr(B^{\Dag}_i B^{\Dag}_j B_i B_j \chi) - c_{10}^{\chi}\Tr(B^{\Dag}_i B^{\Dag}_j B_j B_i \chi )\\
    & - c_{11}^{\chi}\Tr(B^{\Dag}_i \chi B_i ) \Tr(B^{\Dag}_j B_j ) - c_{12}^{\chi}\Tr(B^{\Dag}_i \chi B_j) \Tr( B^{\Dag}_j B_i)\, . \label{eq:LagNLO2}
    \end{aligned}
\end{equation}
Now, when writing the elements $\langle B'_1 B'_2 | \mathcal{L}^{(2),\, \cancel{SU(3)}_f}_{BB} | B_1 B_2 \rangle$, we will separate the contributions to the $c_i^{\chi}$ coefficients that are accompanied by $m^2_K-m^2_\pi$ from the ones that are accompanied by $m^2_\pi$. Let us look at an example to understand why,
\begin{equation}
    \begin{aligned}
    \langle \Sigma^+ p |\mathcal{L}^{(2),\, \cancel{SU(3)}_f}_{BB}| \Sigma^+ p \rangle &= 2(c_3^{\chi} - c_4^{\chi}) m^2_K + 2(c_1^{\chi}- c_2^{\chi}-c_{11}^{\chi}+ c_{12}^{\chi})m^2_{\pi}\\
    &=2(c_3^{\chi}- c_4^{\chi})(m^2_K-m^2_{\pi}) + 2(c_1^{\chi}- c_2^{\chi}+ c_3^{\chi}- c_4^{\chi}-c_{11}^{\chi}+ c_{12}^{\chi})m^2_{\pi}\, .
    \end{aligned}
\end{equation}
The coefficients accompanied by $m^2_K-m^2_{\pi}$ are the coefficients that contribute to the symmetry breaking, since they vary for different channels belonging to the same irrep (we will write them as $\bm{c}^{\chi}_i\equiv c^{\chi}_i (m^2_K-m^2_{\pi})$ for simplicity), while the rest (accompanied by $m^2_{\pi}$) are the same for all the channels in the same irrep, and therefore can be reabsorbed in the LO terms.

The final expression for the Lagrangian we will be using is
\begin{equation}
    \mathcal{L}_{BB}=\, \mathcal{L}^{(0),\, SU(3)}_{BB} + \mathcal{L}^{(2),\, SU(3)}_{BB} + \mathcal{L}^{(2),\, \cancel{SU(3)}}_{BB}\, ,
\label{eq:LEFTSU3}
\end{equation}
built from the expressions given in Eqs.~\eqref{eq:LagLO},~\eqref{eq:LagNLO1}, and~\eqref{eq:LagNLO2}.
In~\cref{subsec:EFTconstraints}, the specific LECs that we will have access to with the LQCD data available will be discussed.

\subsection{Assuming \texorpdfstring{$SU(6)$}{SU(6)} spin-flavor symmetry}\label{subsec:SU6EFT}

In the limit of $SU(3)_f$ symmetry and large $N_c$, two-baryon interactions are predicted to be invariant under an $SU(6)$ spin-flavor symmetry, with corrections that generally scale as $1/N_c$~\cite{Kaplan:1995yg} (this symmetry was previously discussed in Refs.~\cite{Gursey:1964,PhysRevLett.13.175,Dyson:1964xwa}, as the natural extension of the $SU(4)$ spin-flavor Wigner symmetry when including strangeness).
In the two-nucleon sector, this encompasses the $SU(4)$ spin-flavor Wigner symmetry~\cite{Wigner:1936dx, Wigner:1937zz, Wigner:1939zz}, with corrections that scale as $1/N^2_c$ (a recent study has proven again that this is an approximate symmetry at the physical point~\cite{Lee:2020esp}, but it is only seen at an optimal momentum resolution scale).
Under $SU(6)$ group transformations, the baryons transform as a three-index symmetric tensor $\Psi^{\mu\nu\rho}$, where each $SU(6)$ index $\mu$ is a pair of spin and flavor indices $(i \alpha)$. At LO, only two independent terms contribute to the two-baryon interacting Lagrangian,
\begin{equation}
    \mathcal{L}^{(0),SU(6)}_{BB}=-a(\Psi^{\Dag}_{\mu\nu\rho}\Psi^{\mu\nu\rho})^2-b\Psi^{\Dag}_{\mu\nu\sigma}\Psi^{\mu\nu\tau}\Psi^{\Dag}_{\rho\delta\tau}\Psi^{\rho\delta\sigma}\, ,
\label{eq:su6lag}
\end{equation}
where the baryon tensor can be expressed as a function of the octet $B$ and decuplet $T$ baryon matrices,
\begin{equation}
    \Psi^{\mu\nu\rho} = \Psi^{(i \alpha)(j \beta)(k \gamma)} = T^{\alpha\beta\gamma}_{ijk}+\frac{1}{\sqrt{18}}\left(B^{\alpha}_{\omega,i}\epsilon^{\omega \beta \gamma}\epsilon_{jk}+B^{\beta}_{\omega, j}\epsilon^{\omega \gamma \alpha}\epsilon_{ik}+B^{\gamma}_{\omega,k}\epsilon^{\omega \alpha \beta}\epsilon_{ij}\right) \, .
\label{eq:BaryonsSymmetricSU6}
\end{equation}
Here, $\alpha,\beta,\gamma,\omega$ are flavor indices, $i,j,k$ are spin indices, and the Levi-Civita tensor $\epsilon$ is in either flavor or spin space depending on the type and number of indices.

In order to understand why there are only two terms and why the baryons are written as a symmetric tensor, we will derive the $SU(6)$ spin-flavor symmetry of the nuclear (and hypernuclear) forces from group-theoretical arguments~\cite{Lichtenberg:1978pc,Quang:1998yw}, using the \texttt{Mathematica} package \texttt{LieArt}~\cite{Feger:2012bs,Feger:2019tvk} for the group-theory calculations.

The main idea behind is that if we assume $SU(3)_f$ symmetry together with $SU(2)_J$ spin symmetry, all the baryons will have the same mass regardless of the spin ($\tfrac{1}{2}$ or $\tfrac{3}{2}$) value. This means that all these objects remain invariant under transformations of the unitary group $SU(6)$, which has as subgroups the product $SU(3)_f\otimes SU(2)_J$. Therefore, now the quarks will be dimension-six objects, living in the fundamental representation,
\begin{equation}
    q^{i}=\begin{pmatrix}
    u^{\uparrow} & u^{\downarrow} & d^{\uparrow} & d^{\downarrow} & s^{\uparrow} & s^{\downarrow}
    \end{pmatrix}^\top\, .
\end{equation}
To identify the $SU(6)$ irrep where the octet and decuplet baryons live, we start by taking the product of three $\mathbf{6}$,
\begin{equation}
    \mathbf{6}\otimes\mathbf{6}\otimes\mathbf{6}=2\; \mathbf{70} \oplus \mathbf{56} \oplus \mathbf{20}\, ,
\end{equation}
and write it in terms of the irreps from the product of the subgroups $SU(3)_f\otimes SU(2)_J$,
\begin{equation}
    \mathbf{3}\otimes\mathbf{3}\otimes\mathbf{3}= \mathbf{10} \oplus 2\; \mathbf{8} \oplus \mathbf{1}\, , \qquad \mathbf{2}\otimes\mathbf{2}\otimes\mathbf{2}= \mathbf{4} \oplus 2\; \mathbf{2} \, .
\end{equation}
To continue, we can use the Young tableaux to study the symmetry of each irrep: totally symmetric (SS), totally antisymmetric (AA), and mixed symmetry (SA).\footnote{As a reminder, irreps with only a row of boxes are SS, with only a column of boxes are AA, and a mix of both are SA.} Therefore, for the $SU(6)$ irreps,
\begin{equation}
    \mathbf{70}\; \rightarrow \; \ydiagram{2,1} \;\; \text{(SA)}\, , \qquad \mathbf{56}\; \rightarrow \;  \ydiagram{3}\;\; \text{(SS)}\, ,\qquad \mathbf{20} \; \rightarrow \;   \ydiagram{1,1,1}\;\; \text{(AA)}\, ,
\end{equation}
and for the $SU(3)_f$ and $SU(2)_J$ irreps,
\begin{equation}
\begin{aligned}
    &\mathbf{10}\; \rightarrow \; \ydiagram{3}\;\; \text{(SS)} \, , \qquad \mathbf{8}\; \rightarrow \;  \ydiagram{2,1}\;\; \text{(SA)}\, ,\qquad \mathbf{1} \; \rightarrow \;  \ydiagram{1,1,1} \;\; \text{(AA)}\, ,\\
    &\mathbf{4\phantom{0}}\; \rightarrow \; \ydiagram{3}\;\; \text{(SS)} \, , \qquad \mathbf{2}\; \rightarrow \;  \ydiagram{1} \;\; \text{(SA)}\, .
\end{aligned}
\end{equation}

In order to obtain an SS state, we can take the product of (SS,SS), (AA,AA), and (SA,SA); for AA, we can do (SS,AA) or (SA,SA); and for SA, we can do (SS,AS), (AA,AS), or (AS,AS). Therefore,
\begin{equation}
    \mathbf{70}\rightarrow (\mathbf{10},\mathbf{2}) \oplus (\mathbf{8},\mathbf{2}) \oplus (\mathbf{8},\mathbf{4}) \oplus (\mathbf{1},\mathbf{2})\, , \quad \mathbf{56}\rightarrow (\mathbf{10},\mathbf{4}) \oplus (\mathbf{8},\mathbf{2})\, , \quad \mathbf{20}\rightarrow (\mathbf{8},\mathbf{2}) \oplus (\mathbf{1},\mathbf{4})\, .
\end{equation}
From these symmetry arguments, we see that the only irrep where the octet $(\mathbf{8},\mathbf{2})$ and decuplet $(\mathbf{10},\mathbf{4})$ live is the $\mathbf{56}$ irrep. Therefore, since $\mathbf{56}$ is completely symmetric, the baryons can be written as a completely symmetric tensor~\cite{Kaplan:1995yg,Quang:1998yw}, like in Eq.~\eqref{eq:BaryonsSymmetricSU6}.

When looking at the interaction of two baryons, we have to perform the tensor product of two $\mathbf{56}$ to see how many parameters will be needed to describe it,
\begin{equation}
    \mathbf{56}\otimes \mathbf{56} = \mathbf{462}\oplus \mathbf{490}\oplus \mathbf{1050}'' \oplus \mathbf{1134}'\, .
\end{equation}
Contrary to the arguments in Ref.~\cite{Kaplan:1995yg}, we see that in principle there should be four parameters, and not only two. To elucidate this point, we write these irreps in terms of the corresponding ones in $SU(3)_f\otimes SU(2)_J$. For two octet baryons, the flavor and spin decomposition is
\begin{equation}
    \mathbf{8}\otimes\mathbf{8}= \mathbf{27} \oplus \mathbf{10} \oplus \overline{\mathbf{10}} \oplus \mathbf{8}_s \oplus \mathbf{8}_a  \oplus \mathbf{1}\, , \qquad \mathbf{2}\otimes\mathbf{2}= \mathbf{1} \oplus \mathbf{3} \, .
\end{equation}
In order to identify the correct flavor-spin combinations for a baryon-baryon system, it is convenient to identify at least one system of two identical particles in each irrep,
\begin{equation}
\begin{aligned}
    \mathbf{27} & : \ pp & \mathbf{10} & :\ \tfrac{1}{\sqrt{2}}(\Xi^0\Xi^- -\Xi^-\Xi^0) \, ,\\
    \overline{\mathbf{10}} & : \ \tfrac{1}{\sqrt{2}}(np-pn)\, , & \mathbf{8}_s & :\ \Lambda \Lambda \, ,\\
    \mathbf{8}_a & : \ \tfrac{1}{\sqrt{2}}(\Sigma^-\Sigma^0 -\Sigma^0\Sigma^-)\, , & \mathbf{1} & :\ \Lambda \Lambda \, .
\end{aligned}
\end{equation}
We see that the irreps $\mathbf{27}$, $\mathbf{8}_s$, and $\mathbf{1}$ are symmetric with respect to the exchange of both baryons, while the irreps $\mathbf{10}$, $\overline{\mathbf{10}}$, and $\mathbf{8}_a$ are antisymmetric with respect to the exchange of both baryons. Taking into account that the spin-singlet state is antisymmetric and the spin-triplet state is symmetric with respect to the exchange of the spin of the baryons, the allowed (spin-flavor) combinations are
\begin{equation}
    BB\rightarrow BB:\ (\mathbf{27},\mathbf{1}),(\mathbf{8},\mathbf{1}),(\mathbf{1},\mathbf{1}),(\mathbf{10},\mathbf{3}),(\overline{\mathbf{10}},\mathbf{3}),(\mathbf{8},\mathbf{3})\, .
\end{equation}
In the case of having two baryons of the decuplet~\cite{Haidenbauer:2017sws}, the corresponding flavor and spin decompositions are
\begin{equation}
    \mathbf{10}\otimes\mathbf{10}= \mathbf{35} \oplus \mathbf{28}\oplus \mathbf{27} \oplus \overline{\mathbf{10}} \, , \qquad \mathbf{4}\otimes\mathbf{4}= \mathbf{1} \oplus \mathbf{3} \oplus \mathbf{5} \oplus \mathbf{7} \, .
\end{equation}
While the spin-0 and spin-2 wavefunctions are antisymmetric ($\mathbf{1}$ and $\mathbf{5}$ irreps), the spin-1 and spin-3 wavefunctions are symmetric ($\mathbf{3}$ and $\mathbf{7}$ irreps). Considering the following two-particle states for each $SU(3)_f$ irrep,
\begin{equation}
\begin{aligned}
    \mathbf{35}& : \ \tfrac{1}{\sqrt{2}}\left(\Delta^{++} \Delta^{+}-\Delta^{+}\Delta^{++} \right)\, , & \mathbf{28}& :\ \Omega^-\Omega^- \, ,\\
    \mathbf{27}& : \ \Xi^{*0}\Xi^{*0}\, , & \overline{\mathbf{10}}& :\ \tfrac{1}{\sqrt{2}}(\Sigma^{*0}\Sigma^{*+}-\Sigma^{*+}\Sigma^{*0}) \, ,
\end{aligned}
\end{equation}
the irreps $\mathbf{28}$ and $\mathbf{27}$ are symmetric and the irreps $\mathbf{35}$ and $\overline{\mathbf{10}}$ are antisymmetric, and the possible combinations are
\begin{equation}
    T T \rightarrow T T:\ (\mathbf{28},\mathbf{1}),(\mathbf{28},\mathbf{5}),(\mathbf{27},\mathbf{1}),(\mathbf{27},\mathbf{5}),(\mathbf{35},\mathbf{3}),(\mathbf{35},\mathbf{7}),(\mathbf{\overline{10}},\mathbf{3}),(\mathbf{\overline{10}},\mathbf{7})\, .
\end{equation}
Lastly, for the system composed of one baryon from the octet and one of the decuplet, the flavor and spin decompositions are
\begin{equation}
    \mathbf{8}\otimes\mathbf{10}= \mathbf{35} \oplus \mathbf{27}\oplus \mathbf{10} \oplus \mathbf{8} \, , \quad \mathbf{2}\otimes\mathbf{4}= \mathbf{3} \oplus \mathbf{5}  \, .
\end{equation}
Since in this case the two fermions are distinguishable, the Pauli principle does not apply, and any combination is allowed,
\begin{equation}
    B T \rightarrow B T:\ (\mathbf{35},\mathbf{3}),(\mathbf{27},\mathbf{3}),(\mathbf{10},\mathbf{3}),(\mathbf{8},\mathbf{3}),(\mathbf{35},\mathbf{5}),(\mathbf{27},\mathbf{5}),(\mathbf{10},\mathbf{5}),(\mathbf{8},\mathbf{5})\, .
\end{equation}
For the crossed channels ($BB\rightarrow TB$, $BB\rightarrow T T$ and $TB\rightarrow TT$), the possible irreps will be those that appear in both the initial and final states. As a summary,
\begin{equation}
\begin{aligned}
    BB\rightarrow BB:&\ (\mathbf{27},\mathbf{1}),(\mathbf{8},\mathbf{1}),(\mathbf{1},\mathbf{1}),(\mathbf{10},\mathbf{3}),(\overline{\mathbf{10}},\mathbf{3}),(\mathbf{8},\mathbf{3})\, , \\
    BB\rightarrow TB:&\ (\mathbf{10},\mathbf{3}),(\mathbf{8},\mathbf{3})\, , \\
    BB\rightarrow TT:&\ (\mathbf{27},\mathbf{1}),(\mathbf{\overline{10}},\mathbf{3})\, , \\
    TB\rightarrow TB:&\ (\mathbf{35},\mathbf{3}),(\mathbf{27},\mathbf{3}),(\mathbf{10},\mathbf{3}),(\mathbf{8},\mathbf{3}),(\mathbf{35},\mathbf{5}),(\mathbf{27},\mathbf{5}),(\mathbf{10},\mathbf{5}),(\mathbf{8},\mathbf{5})\, , \\
    TB\rightarrow TT:&\ (\mathbf{35},\mathbf{3}),(\mathbf{27},\mathbf{5})\, , \\
    TT\rightarrow TT:&\ (\mathbf{28},\mathbf{1}),(\mathbf{28},\mathbf{5}),(\mathbf{27},\mathbf{1}),(\mathbf{27},\mathbf{5}),(\mathbf{35},\mathbf{3}),(\mathbf{35},\mathbf{7}),(\mathbf{\overline{10}},\mathbf{3}),(\mathbf{\overline{10}},\mathbf{7})\, .
    \label{eq:BTirreps}
\end{aligned}
\end{equation}
From the $SU(6)$ tensor product, decomposing the four irreps into $SU(3)_f\otimes SU(2)_J$ multiplets gives
\begin{equation}
\begin{aligned}
    \mathbf{462}=&\ {\color{gray}(\overline{\mathbf{10}},\mathbf{1})} \oplus (\mathbf{27},\mathbf{3}) \oplus {\color{gray}(\mathbf{28},\mathbf{7})} \oplus (\mathbf{35},\mathbf{5})\, , \\
    \mathbf{490}=&\ (\mathbf{1},\mathbf{1}) \oplus (\mathbf{8},\mathbf{3}) \oplus (\mathbf{10},\mathbf{3}) \oplus (\overline{\mathbf{10}},\mathbf{3}) \oplus (\mathbf{8},\mathbf{5}) \oplus (\overline{\mathbf{10}},\mathbf{7}) \oplus (\mathbf{27},\mathbf{1}) \oplus (\mathbf{28},\mathbf{1})\\
    & \oplus (\mathbf{27},\mathbf{5}) \oplus (\mathbf{35},\mathbf{3})\, , \\
    \mathbf{1050}'' =&\ (\mathbf{8},\mathbf{1}) \oplus (\mathbf{8},\mathbf{3}) \oplus (\mathbf{10},\mathbf{3}) \oplus (\overline{\mathbf{10}},\mathbf{3}) \oplus (\mathbf{10},\mathbf{5}) \oplus (\mathbf{27},\mathbf{1}) \oplus (\mathbf{27},\mathbf{3}) \oplus (\mathbf{27},\mathbf{5})\\
    & \oplus (\mathbf{28},\mathbf{5}) \oplus (\mathbf{35},\mathbf{3}) \oplus (\mathbf{35},\mathbf{5}) \oplus (\mathbf{35},\mathbf{7})\, , \\
    \mathbf{1134}'=&\ {\color{gray}(\mathbf{1},\mathbf{3})} \oplus (\mathbf{8},\mathbf{1}) \oplus 2(\mathbf{8},\mathbf{3}) \oplus {\color{gray}(\mathbf{10},\mathbf{1})} \oplus {\color{gray}(\overline{\mathbf{10}},\mathbf{1})} \oplus (\mathbf{10},\mathbf{3}) \oplus (\mathbf{8},\mathbf{5}) \oplus {\color{gray}(\mathbf{10},\mathbf{5})}\\
    & \oplus {\color{gray}(\overline{\mathbf{10}},\mathbf{5})} \oplus 2(\mathbf{27},\mathbf{3}) \oplus {\color{gray}(\mathbf{28},\mathbf{3})} \oplus (\mathbf{27},\mathbf{5}) \oplus {\color{gray}(\mathbf{27},\mathbf{7})} \oplus {\color{gray}(\mathbf{35},\mathbf{1})} \oplus (\mathbf{35},\mathbf{3}) \oplus (\mathbf{35},\mathbf{5}) , 
\end{aligned}
\end{equation}
where the gray-colored numbers are non-physical states with symmetric wavefunctions, not appearing in Eq.~\eqref{eq:BTirreps}. For example, the combination $(\mathbf{28},\mathbf{7})$ is not possible because $\mathbf{28}$ has a symmetric flavor wavefunction and $\mathbf{7}$ also has a symmetric spin wavefunction, and with the combination of both a total symmetric wavefunction is prohibited by the Pauli principle. The only irreps that survive are the ones that contain only physical states, therefore we are left with two, $\mathbf{490}$ and $ \mathbf{1050}''$, meaning that we only need two coefficients to describe the baryon-baryon interaction, as in Eq.~\eqref{eq:su6lag}.

We can match the $SU(6)$ LO Kaplan-Savage coefficients $a$ and $b$ to the $SU(3)_f$ LO Savage-Wise coefficients $c_i$ by computing $\langle B'_1 B'_2 | \mathcal{L}^{(0),\, SU(6)}_{BB} | B_1 B_2 \rangle$,
\begin{equation}
\begin{aligned}
    c^{(27)}&=2a-\frac{2b}{27}+\mathcal{O}\left(\frac{1}{N^2_c}\right), &\qquad c^{(\overline{10})}&=2a-\frac{2b}{27}+\mathcal{O}\left(\frac{1}{N^2_c}\right), \\
    c^{(8_s)}&=2a+\frac{2b}{3}+\mathcal{O}\left(\frac{1}{N_c}\right), &\qquad c^{(10)}&=2a+\frac{14b}{27}+\mathcal{O}\left(\frac{1}{N_c}\right), \\
    c^{(1)}&=2a-\frac{2b}{3}+\mathcal{O}\left(\frac{1}{N_c}\right), &\qquad c^{(8_a)}&=2a+\frac{2b}{27}+\mathcal{O}\left(\frac{1}{N_c}\right).
\label{eq:SU6eq}
\end{aligned}
\end{equation}

\subsection{Matching the LECs to the scattering amplitude}

In order to constrain the LECs with the LQCD data, our approach is to match them to a momentum expansion of the scattering amplitude. We will follow the formalism from Refs.~\cite{Kaplan:1998tg,Kaplan:1998we}, where they work for simplicity with a toy model: two heavy spinless baryons $\tilde{B}_i$ in the non-relativistic limit (with masses $M_1$ and $M_2$), whose interaction is characterized by a scale $\Lambda$. Then, the EFT will describe scattering at a c.m.\ momentum $k^*\ll \Lambda$ using contact interactions in a derivative expansion,
\begin{equation}
\begin{aligned}
    \mathcal{L}=\, &\tilde{B}_1^{\Dag}\left(\imag\partial_0+\frac{\nabla^2}{2M_1}\right)\tilde{B}_1+\tilde{B}_2^{\Dag}\left(\imag\partial_0+\frac{\nabla^2}{2M_2}\right)\tilde{B}_2\\
    &+\left(\frac{\mu}{2}\right)^{4-D}\left\lbrace C_0\left(\tilde{B}_1^{\Dag}\tilde{B}_2\right)^2+C^{(i)}_2\left(\tilde{B}_1^{\Dag}\dvec{\nabla} \tilde{B}_2\right)^2+C^{(ii)}_2\left[\imag\vec{\nabla}\left(\tilde{B}_1^{\Dag} \tilde{B}_2\right)^2\right]+\cdots\right\rbrace \, ,
\end{aligned}
\label{eq:EFT_lag}
\end{equation}
where the ellipses denote higher derivative terms. The mass scale $\mu/2$ is introduced because we will use dimensional regularization for the loop integrals. Also, if the spacetime dimension is $D$, one can check that the couplings $C_{2n}$ will have the same dimensions regardless of the value of $D$,
\begin{equation}
    [S]=\text{adim.}\rightarrow [\mathcal{L}]=M^{D}\rightarrow [\tilde{B}]=M^{\frac{D-1}{2}}\rightarrow [C_{2n}]=M^{-2(1+n)} \, .
\end{equation}

From~\cref{sec:scatteringFV}, we know that the scattering amplitude $\mathcal{M}$ can be written in terms of the phase shift,
\begin{equation}
    \mathcal{M}=\frac{2\pi}{\tilde{M}}\frac{1}{k^*\cot \delta - \imag k^*} \, .
\label{eq:amp_phshift}
\end{equation}
We also know that $k^*\cot \delta$ has an expansion (ERE) for $k^*\ll \Lambda$. Writing Eq.~\eqref{eq:ERE} slightly differently,
\begin{equation}
    k^*\cot \delta = -\frac{1}{a}+\frac{1}{2}\Lambda^2\sum^{\infty}_{n=0}r_n \left(\frac{k^{*2}}{\Lambda^2} \right)^{n+1} \, .
\end{equation}
For the baryon-baryon interaction, all $r_n$ are of the order $\mathcal{O}(1/\Lambda)$, but $a$ can take any value. Therefore, the radius of convergence of the momentum expansion of $\mathcal{M}$ will depend on the size of the scattering length. We will study two cases: the first one will be for natural interactions, with $|a|\sim 1/\Lambda$, and the second one for unnatural interactions, with $|a|\gg 1/\Lambda$.\footnote{For a detailed discussion of naturalness in EFTs, see Ref.~\cite{vanKolck:2020plz}.}

\subsubsection{Scattering length of natural size}

For natural interactions, $|a|\sim 1/\Lambda$ and $|r_n| \sim 1/\Lambda$, and $\mathcal{M}$ has a simple momentum expansion,
\begin{equation}
\begin{aligned}
    \mathcal{M}(k^*)&=\, \mathcal{M}(0)+\mathcal{M}'(0)k^*+\frac{1}{2!}\mathcal{M}''(0)k^{*2}+\cdots \\
    &=-\frac{2\pi a}{\tilde{M}}+\frac{2\pi \imag a^2}{\tilde{M}}k^* +\frac{2\pi}{\tilde{M}}\left[(-a)^3(-\imag)^2-\frac{ra^2}{2}\right]k^{*2} +\mathcal{O}(k^{*3}/\Lambda^4)\\
    &=-\frac{2\pi a}{\tilde{M}}\left[1-\imag ak^* +\left(\frac{ra}{2}-a^2\right)k^{*2}+\mathcal{O}(k^{*3}/\Lambda^3)\right] \, .
\end{aligned}
\label{eq:amplitude_param}
\end{equation}
This expansion converges for momenta up to $k^*\sim \Lambda$,
\begin{equation}
    \mathcal{M}\sim \frac{1}{\Lambda}+ \frac{k^*}{\Lambda^2}+ \frac{k^{*2}}{\Lambda^3}+\cdots \, .
\end{equation}
This is the expansion we want to obtain starting from the EFT Lagrangian of Eq.~\eqref{eq:EFT_lag}.

The scattering amplitude $\mathcal{M}$ will be given by the sum of all the Feynman diagrams of this theory. At tree level, the $S$-wave amplitude is given by
\begin{equation}
    \imag\mathcal{M}_{\text{tree}}=-\imag\left(\frac{\mu}{2}\right)^{4-D}\sum_{n=1}^{\infty}C_{2n}(\mu)k^{*2n} \, ,
\label{eq:tree_amplitude}
\end{equation}
with $C_{2n}$ being linear combinations of the couplings in the Lagrangian. The full amplitude can be obtained from the following diagrammatic sum,
\begin{equation}
    \mathcal{M}=\includegraphics[scale=0.4, valign=c]{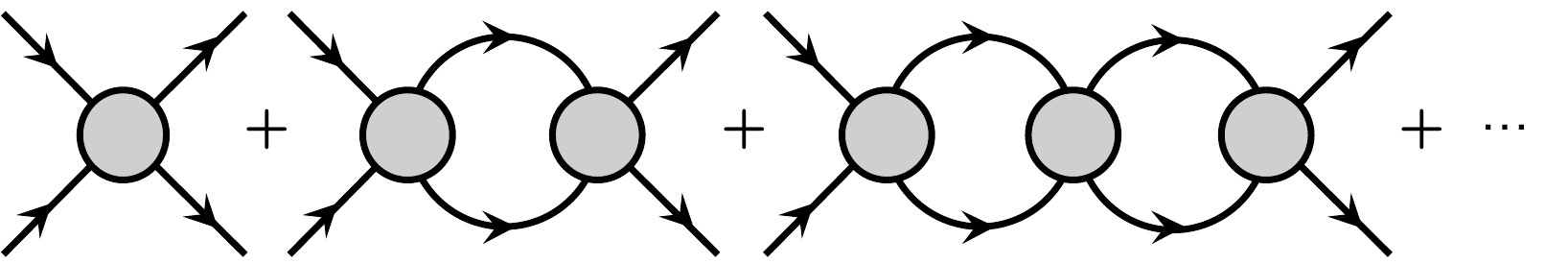} ,\quad \text{with }\; \includegraphics[scale=0.4, valign=c]{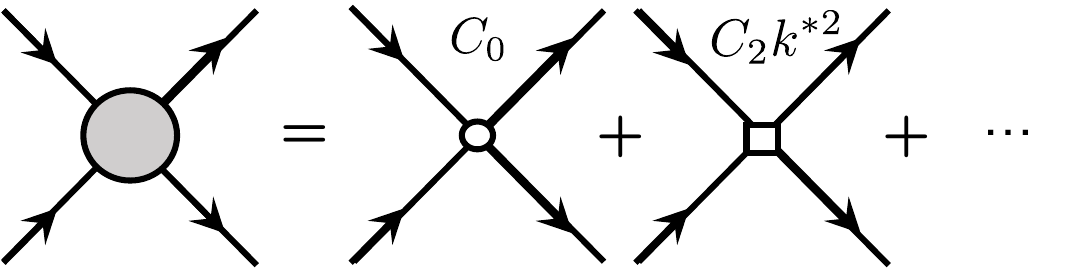}\, ,
\label{eq:Feynman_amplitude}
\end{equation}
where loop integrals are of the form
\begin{equation}
    I_n=-\imag\left(\frac{\mu}{2}\right)^{4-D}\int \frac{d^Dq}{(2\pi)^D}\bm{q}^{2n}\frac{\imag}{\frac{E}{2}-q_0-\frac{\bm{q}^2}{2M_1}+\imag\varepsilon}\frac{\imag}{\frac{E}{2}+q_0-\frac{\bm{q}^2}{2M_2}+\imag\varepsilon} \, ,
    \label{eq:int_loop}
\end{equation}
where $E$ is the energy of the incoming two-baryon system, and $q$ the loop momentum, which has to be integrated out. First, we can integrate $q_0$ by forming a contour closing the pole in the upper half of the complex plane and using the residue theorem,
\begin{equation}
\begin{aligned}
    I_n&=-\imag\left(\frac{\mu}{2}\right)^{4-D}\int \frac{d^{D-1}\bm{q}}{(2\pi)^D}\bm{q}^{2n}\int dq_0\frac{\imag}{\frac{E}{2}-q_0-\frac{\bm{q}^2}{2M_1}+\imag\varepsilon}\frac{\imag}{\frac{E}{2}+q_0-\frac{\bm{q}^2}{2M_2}+\imag\varepsilon} \\
    &=-\imag\left(\frac{\mu}{2}\right)^{4-D}\int \frac{d^{D-1}\bm{q}}{(2\pi)^D}\bm{q}^{2n} (2\pi \imag)\sum \text{Res}(q_0)\\
    &=-\imag\left(\frac{\mu}{2}\right)^{4-D}\int \frac{d^{D-1}\bm{q}}{(2\pi)^D}\bm{q}^{2n} (2\pi \imag) \frac{-\imag^2}{\frac{E}{2}+\left(\frac{E}{2}-\frac{\bm{q}^2}{2M_1}+\imag\varepsilon\right)-\frac{\bm{q}^2}{2M_2}+\imag\varepsilon} \\
    &=\left(\frac{\mu}{2}\right)^{4-D}\int \frac{d^{D-1}\bm{q}}{(2\pi)^{D-1}}\frac{\bm{q}^{2n}}{E-\frac{\bm{q}^2}{2}\left(\frac{1}{M_1}+\frac{1}{M_2}\right)+\imag\varepsilon} =\left(\frac{\mu}{2}\right)^{4-D}\int \frac{d^{D-1}\bm{q}}{(2\pi)^{D-1}}\frac{\bm{q}^{2n}}{E-\frac{\bm{q}^2}{2\tilde{M}}+\imag\varepsilon}\, .
\end{aligned}
\end{equation}
We can split $d^{D-1}\bm{q}$ in a radial and an angular part in $(D-1)$-spherical coordinates,
\begin{equation}
\begin{aligned}
    I_n&=\left(\frac{\mu}{2}\right)^{4-D}\int \frac{d^{D-1}\bm{q}}{(2\pi)^{D-1}}\frac{\bm{q}^{2n}}{E-\frac{\bm{q}^2}{2\tilde{M}}+\imag\varepsilon}=\left(\frac{\mu}{2}\right)^{4-D}\int d\Omega_{D-1}\int_0^{\infty} \frac{dq}{(2\pi)^{D-1}}q^{D-2}\frac{q^{2n}}{E-\frac{q^2}{2\tilde{M}}+\imag\varepsilon}\\
    &=\left(\frac{\mu}{2}\right)^{4-D}\frac{2\pi^{\frac{D-1}{2}}}{(2\pi)^{D-1}\Gamma\left(\frac{D-1}{2}\right)}\int_0^{\infty} dq\frac{q^{2n+D-2}}{E-\frac{q^2}{2\tilde{M}}+\imag\varepsilon}
    \, .
    \label{eq:Inintegral}
\end{aligned}
\end{equation}
The integral in the last line of Eq.~\eqref{eq:Inintegral} can rearranged as 
\begin{equation}
\begin{aligned}
    \int_0^{\infty} dq\frac{q^{2n+D-2}}{E-\frac{q^2}{2\tilde{M}}+\imag\varepsilon} &= 2\tilde{M}\int_0^{\infty} dq\frac{q^{2n+D-2}}{2\tilde{M}E-q^2+\imag\varepsilon} = -2\tilde{M}\int_0^{\infty} dq\frac{q^{2n+D-2}}{q^2-2\tilde{M}E-\imag\varepsilon}\\
    &=\frac{-2\tilde{M}}{-2\tilde{M}E-\imag\varepsilon}\int_0^{\infty} dq\frac{q^{2n+D-2}}{\frac{q^2}{-2\tilde{M}E-\imag\varepsilon}+1} \, ,
\end{aligned}
\end{equation}
which, after applying the following change of variables $p=q^2/(-2\tilde{M}E-\imag\varepsilon)$, takes the form
\begin{equation}
\begin{aligned}
    \frac{-2\tilde{M}}{-2\tilde{M}E-\imag\varepsilon} & \int_0^{\infty} dp \frac{\sqrt{-2\tilde{M}E-\imag\varepsilon}}{2\sqrt{p}} \frac{(-2\tilde{M}E-\imag\varepsilon)^{n+\frac{D-2}{2}} p^{n+\frac{D-2}{2}}}{p+1}\\
     & =-\tilde{M} (-2\tilde{M}E)^n (-2\tilde{M}E-\imag\varepsilon)^{\frac{D-3}{2}} \int_0^{\infty} dp \frac{p^{n+\frac{D-3}{2}}}{p+1} \, .
\end{aligned}
\end{equation}
This integral can be solved using the beta function,
\begin{equation}
    B(\alpha,\gamma)=\int_0^{\infty}dp\, p^{\alpha-1}(p+1)^{-\alpha-\gamma}=\frac{\Gamma\left(\alpha\right)\Gamma\left(\gamma\right)}{\Gamma\left(\alpha+\gamma\right)} \, ,
\end{equation}
with the following identification,
\begin{equation}
    p^{\alpha-1}(p+1)^{-\alpha-\gamma}=p^{n+\frac{D-3}{2}} (p+1)^{-1} \quad \Rightarrow	\quad \alpha=n+\frac{D-1}{2}\, ,\quad \gamma=\frac{3-D}{2}-n \, .
\end{equation}
Collecting all the terms, we get
\begin{equation}
\begin{aligned}
    I_n&=\left(\frac{\mu}{2}\right)^{4-D}\frac{2\pi^{\frac{D-1}{2}}}{(2\pi)^{D-1}\Gamma\left(\frac{D-1}{2}\right)}(-\tilde{M})(-2\tilde{M}E)^n (-2\tilde{M}E-\imag\varepsilon)^{\frac{D-3}{2}}\frac{\Gamma\left(n+\frac{D-1}{2}\right)\Gamma\left(\frac{3-D}{2}-n\right)}{\Gamma\left(1\right)} \\
    &=(-2\tilde{M})(-2\tilde{M}E)^n (-2\tilde{M}E-\imag\varepsilon)^{\frac{D-3}{2}}\frac{\Gamma\left(n+\frac{D-1}{2}\right)\Gamma\left(\frac{3-D}{2}-n\right)}{\Gamma\left(\frac{D-1}{2}\right)}\frac{\left(\mu/2\right)^{4-D}}{(4\pi)^{\frac{D-1}{2}}} \, .
\end{aligned}
\end{equation}
We can further simplify this expression by using the following relation,
\begin{equation}
    \Gamma(1+z)=z\Gamma(z) \quad \Rightarrow \quad \Gamma(z-1)=\frac{\Gamma(z)}{z-1} \, .
\end{equation}
Looking at specific values of $n$,
\begin{equation}
\begin{aligned}
    n=0 \; : \quad &\Gamma\left(\frac{D-1}{2}\right)\Gamma\left(\frac{3-D}{2}\right) \, , \\
    n=1 \; : \quad &\Gamma\left(1+\frac{D-1}{2}\right)\Gamma\left(\frac{3-D}{2}-1\right) =\frac{\frac{D-1}{2}}{\frac{3-D}{2}-1}\Gamma\left(\frac{D-1}{2}\right)\Gamma\left(\frac{3-D}{2}\right)\\
    &=\frac{\frac{D-1}{2}}{\frac{1-D}{2}}\Gamma\left(\frac{D-1}{2}\right)\Gamma\left(\frac{3-D}{2}\right)=(-1)\Gamma\left(\frac{D-1}{2}\right)\Gamma\left(\frac{3-D}{2}\right)\, , \\
    n=2 \; : \quad &\Gamma\left(2+\frac{D-1}{2}\right)\Gamma\left(\frac{3-D}{2}-2\right) =\frac{\frac{D-1}{2}+1}{\frac{3-D}{2}-2}\Gamma\left(1+\frac{D-1}{2}\right)\Gamma\left(\frac{3-D}{2}-1\right)\\
    &=\frac{\frac{D+1}{2}}{\frac{-1-D}{2}}(-1)\Gamma\left(\frac{D-1}{2}\right)\Gamma\left(\frac{3-D}{2}\right)=(-1)^2\Gamma\left(\frac{D-1}{2}\right)\Gamma\left(\frac{3-D}{2}\right) \, ,\\
    \vdots \quad \quad &\\
    n=n \; : \quad &\Gamma\left(n+\frac{D-1}{2}\right)\Gamma\left(\frac{3-D}{2}-n\right) =(-1)^n\Gamma\left(\frac{D-1}{2}\right)\Gamma\left(\frac{3-D}{2}\right) \, .
\end{aligned}
\end{equation}
Therefore,
\begin{equation}
\begin{aligned}
    I_n&=(-2\tilde{M})(-2\tilde{M}E)^n (-2\tilde{M}E-\imag\varepsilon)^{\frac{D-3}{2}}\frac{(-1)^n\Gamma\left(\frac{D-1}{2}\right)\Gamma\left(\frac{3-D}{2}\right)}{\Gamma\left(\frac{D-1}{2}\right)}\frac{\left(\mu/2\right)^{4-D}}{(4\pi)^{\frac{D-1}{2}}}\\
    &=(-2\tilde{M})(2\tilde{M}E)^n (-2\tilde{M}E-\imag\varepsilon)^{\frac{D-3}{2}}\Gamma\left(\frac{3-D}{2}\right)\frac{\left(\mu/2\right)^{4-D}}{(4\pi)^{\frac{D-1}{2}}}\, .
\label{eq:loop_int}
\end{aligned}
\end{equation}
In general, when taking the limit $D\rightarrow 4$, the loop integrals diverge. To avoid this issue, subtraction schemes are defined to avoid these divergences, and the minimal subtraction scheme (MS) is one of them, which accounts for removing the $1/(D-4)$ poles before taking the limit $D\rightarrow 4$. However, in this case, the integral does not have such poles, and we can take the limit directly,
\begin{equation}
\begin{aligned}
    I^{\text{MS}}_n &=(-2\tilde{M})(2\tilde{M}E)^n \sqrt{-2\tilde{M}E}\;\Gamma\!\left(-\frac{1}{2}\right)\frac{1}{(4\pi)^{\frac{3}{2}}}=\imag(-2\tilde{M})(2\tilde{M}E)^n \sqrt{2\tilde{M}E}\frac{(-2\sqrt{\pi})}{(4\pi)^{\frac{3}{2}}} \\
    &=\imag(2\tilde{M}E)^n \sqrt{2\tilde{M}E}\left(\frac{2\tilde{M}}{4\pi}\right) \, .
    \label{eq:loop_int_ms}
\end{aligned}
\end{equation}
Since we are in the non-relativistic limit, the energy of the system in the c.m.\ frame is
\begin{equation}
    E=\frac{k^{*2}}{2M_1}+\frac{k^{*2}}{2M_2}=\frac{k^{*2}}{2\tilde{M}}\quad \Rightarrow \quad k^{*2}=2\tilde{M}E \, ,
\end{equation}
and Eq.~\eqref{eq:loop_int_ms} becomes
\begin{equation}
    I^{\text{MS}}_n=\imag(2\tilde{M}E)^n \sqrt{2\tilde{M}E}\left(\frac{2\tilde{M}}{4\pi}\right)=\imag\left(\frac{\tilde{M}}{2\pi}\right) k^{*2n+1}\, .
\end{equation}
We see that the factors of $q$ in the loop integral~\eqref{eq:int_loop} have been replaced by factors of $k^*$. This means that to compute the Feynman diagrams in Eq.~\eqref{eq:Feynman_amplitude}, we can use the tree level amplitude~\eqref{eq:tree_amplitude} for the vertices, and for each loop,
\begin{equation}
    I^{\text{MS}}=\imag\left(\frac{\tilde{M}k^*}{2\pi}\right) \, ,
\end{equation}
giving for the sum of bubble diagrams in Eq.~\eqref{eq:Feynman_amplitude},
\begin{equation}
\begin{aligned}
    \mathcal{M}=&-\sum C_{2n}(\mu)k^{*2n}+\sum C_{2n}(\mu)k^{*2n} I^{\text{MS}}\sum C_{2n}(\mu)k^{*2n}-\cdots \\
    =&-\sum C_{2n}(\mu)k^{*2n}\left[1- I^{\text{MS}}\sum C_{2n}(\mu)k^{*2n}+\left(I^{\text{MS}}\sum C_{2n}(\mu)k^{*2n}\right)^2+\cdots\right]\\
    =&\frac{-\sum C_{2n}(\mu)k^{*2n}}{1+I^{\text{MS}}\sum C_{2n}(\mu)k^{*2n}}=\frac{-\sum C_{2n}(\mu)k^{*2n}}{1+\imag\left(\frac{\tilde{M}k^*}{2\pi}\right) \sum C_{2n}(\mu)k^{*2n}}\, .
\end{aligned}
\end{equation}
In this scheme, the power counting goes as follows: each vertex of $C_{2n}\nabla^{2n}$ counts as $k^{*2n}$, while each loop adds a $k^*$. Thus, we can expand the amplitude in powers of $k^*$,
\begin{equation}
    \mathcal{M}=\sum_{n=0}^{\infty}\mathcal{M}_{n}\, , \quad \mathcal{M}_n\sim  \mathcal{O}(k^{*n}) \, ,
\end{equation}
where each term, $\mathcal{M}_n$, is given by the diagrams with $L\leq n$ loops, and it can be matched to Eq.~\eqref{eq:amplitude_param} to find out the values of $C_{2n}$. So,
\begin{align}
    \mathcal{M}_0&=\includegraphics[scale=0.4, valign=c]{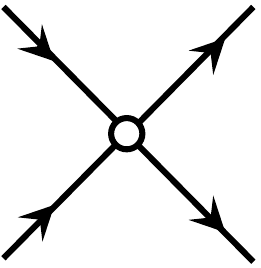}= -C_0 \, , \label{eq:nat_amp0} \\
    \mathcal{M}_1&=\includegraphics[scale=0.4, valign=c]{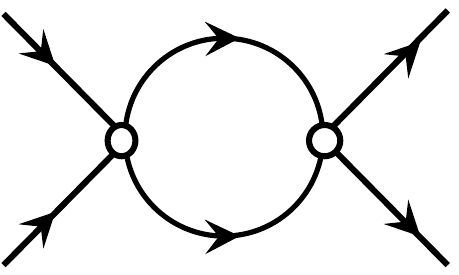} = \imag C^2_0\frac{\tilde{M}}{2\pi}k^* \, , \label{eq:nat_amp1}\\
    \mathcal{M}_2&=\includegraphics[scale=0.4, valign=c]{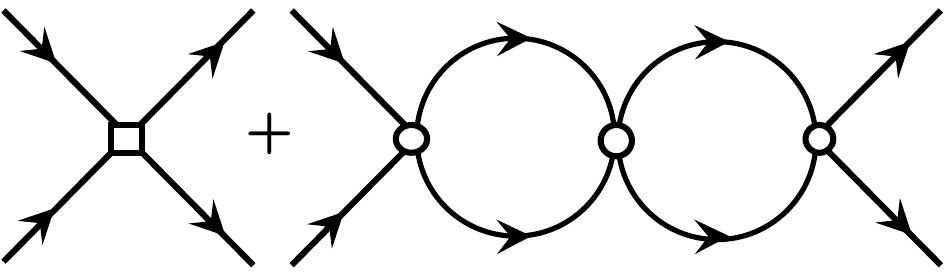} = \left[-C_2+C^3_0\left(\frac{\tilde{M}}{2\pi}\right)^2\right] k^{*2} \, .
\label{eq:nat_amp2}
\end{align}
Then, if we compare Eqs.~\eqref{eq:nat_amp0}-\eqref{eq:nat_amp2} with Eq.~\eqref{eq:amplitude_param}, 
\begin{equation}
    C_0=\frac{2\pi a}{\tilde{M}} \, , \quad C_2=\frac{2\pi}{\tilde{M}}\frac{r a^2}{2\vphantom{\tilde{M}}}=C_0\frac{r a}{2} \, .
\label{eq:coef_nat}
\end{equation}
In general, $C_{2n}$ will have a size of $(2\pi/\tilde{M}\Lambda)/\Lambda^{2n}$, which satisfy that each term is smaller than the immediately preceding one, giving rise to a completely perturbative EFT.

\subsubsection{Scattering length of unnatural size}

Now let us consider the case where $|a|\gg 1/\Lambda$ and $|r_n| \sim 1/\Lambda$. We can see that if we use the previous expansion of $\mathcal{M}$, Eq.~\eqref{eq:amplitude_param}, the radius of convergence has now decreased to $k^*\leq 1/|a| \ll \Lambda$, and also the couplings are very large, $C_{2n}\sim 2\pi a^{n+1}/\tilde{M}\Lambda^n$. This is not a problem of the EFT, but rather of the subtraction scheme used.

Instead of expanding the amplitude like Eq.~\eqref{eq:amplitude_param}, now we have to expand it in terms of $k^*/\Lambda$ but retaining $ak^*$ to all orders,
\begin{equation}
\begin{aligned}
    \mathcal{M}&=-\frac{2\pi a}{\tilde{M}}\left[1-\imag ak^*+\left(\frac{ra}{2}-a^2\right)k^{*2}+\mathcal{O}(k^{*3}/\Lambda^3)\right] \\
    &=-\frac{2\pi }{\tilde{M}}\left(a-\imag a^2k^*-a^3k^{*2}+\cdots \right)\left[1+\frac{r}{2}\left(a-\imag a^2k^*-a^3k^{*2}+\cdots \right)k^{*2}+\cdots\right]\\
    &=-\frac{2\pi}{\tilde{M}}\frac{1}{\frac{1}{a}+\imag k^*} \left[1+\frac{r/2}{\frac{1}{a}+\imag k^*}k^{*2}+\frac{(r/2)^2}{\left(\frac{1}{a}+\imag k^*\right)^2} k^{*4}+\cdots\right] \, .
\end{aligned}
\label{eq:amplitude_param_un}
\end{equation}
If $k^*>1/|a|$, we get
\begin{equation}
    \mathcal{M} \sim \frac{1}{k^*} + \frac{1}{\Lambda} + \frac{k^*}{\Lambda^2} + \frac{k^{*2}}{\Lambda^3} +\cdots \, .
\end{equation}
We can see that instead of scaling as $\{k^{*0},k^{*1},k^{*2},\ldots\}$, now it scales as $\{k^{*-1},k^{*0},k^{*1},\ldots\}$. Therefore, the amplitude computed via the EFT should take the form
\begin{equation}
    \mathcal{M}=\sum_{n=-1}^{\infty}\mathcal{M}_{n}\, , \quad \mathcal{M}_n \sim \mathcal{O}(k^{*n}) \, .
\end{equation}
So, for example, if we want to get $\mathcal{M}_{-1}$, we cannot have single diagrams, since we know from the previous derivation that they contribute with positive powers of $k^*$. In order to get $1/k^*$, we have to sum an infinite number of bubble diagrams,
\begin{equation}
    \mathcal{M}_{-1}=\includegraphics[scale=0.4, valign=c]{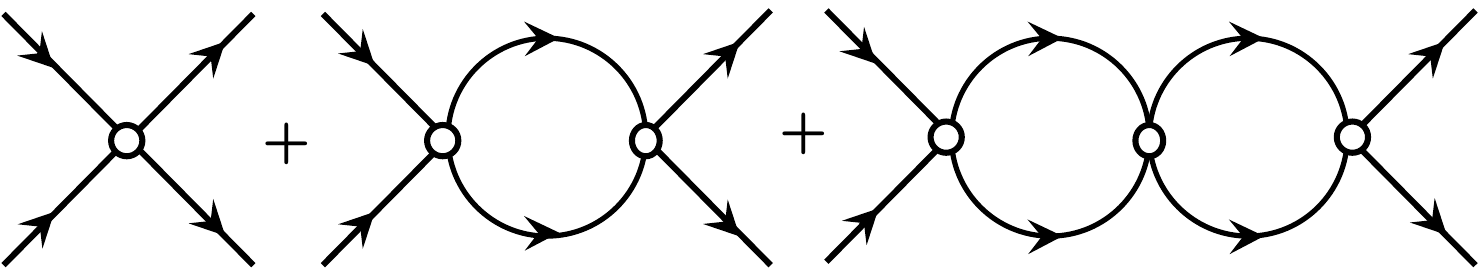}=\frac{-C_0}{1+\imag\frac{\tilde{M}k^*}{2\pi}C_0} \, .
\end{equation}
To find $C_0$, we can compare this expression with Eq.~\eqref{eq:amplitude_param_un}, and we also get $C_0=2\pi a/\tilde{M}$, like before. But if we look more closely, we can see that each diagram in the bubble expansion gets bigger and bigger, $(2\pi a/\tilde{M}) (\imag a k^*)^L$, with $L$ being the number of loops. Even if the total sum is a small number, each new term is bigger than the previous one. This is not a good expansion parameter for an EFT, since $|ak^*|>1$. Without a good expansion parameter, we cannot determine the size of a particular diagram, and we do not know which ones contribute at a particular order.

The size of the contact interactions depends on the renormalization scheme, so we have to revisit the MS scheme (not the dimensional regularization). Also, since $\mathcal{M}_{-1}$ scales as $1/k^*$, our new subtraction scheme must also produce a $C_0$ such that it scales as $1/k^*$. The one proposed by D.\ Kaplan, M.\ Savage, M.\ Wise~\cite{Kaplan:1998tg,Kaplan:1998we}, and U. van Kolck~\cite{vanKolck:1998bw} is called power divergence subtraction (or PDS) scheme (also known as KSW-vK scheme), which consist in removing from the dimensional regularized loop not only the divergences coming from $1/(D-4)$ poles (as in MS, which correspond to log divergences), but also poles from lower dimensions (which correspond to power law divergences at $D=4$). 

In our case, looking at the loop integral $I_n$ in Eq.~\eqref{eq:loop_int}, we see that it has a pole with $D=3+2\varsigma$, with $\varsigma$ being a small parameter,
\begin{equation}
I_n=(-2\tilde{M})(2\tilde{M}E)^n (-2\tilde{M}E-\imag\varepsilon)^{\varsigma} \Gamma\left(-\varsigma\right) \frac{\left(\mu/2\right)^{1-2\varsigma}}{(4\pi)^{1+\varsigma}} \; \overset{\varsigma \ll 1}{\longrightarrow} \, \frac{2\tilde{M}(2\tilde{M}E)^n}{\varsigma} \frac{\mu/2}{4\pi} + \mathcal{O}(\varsigma^0) \, .
\end{equation}
This means that we have to add the following counterterm,
\begin{equation}
    \delta I_n = -\frac{\tilde{M}(2\tilde{M}E)^n}{D-3}\frac{\mu}{2\pi} \, ,
\end{equation}
and the total integral becomes
\begin{equation}
    I_n^{\text{PDS}}=I_n+\delta I_n=(2\tilde{M}E)^n\left(\frac{\tilde{M}}{2\pi}\right)(\mu+ \imag k^*) \, .
\end{equation}
The PDS scheme maintains the nice feature of MS that powers of $q$ inside the loop integration are replaced by powers of the external momentum $k^*$. Also, setting $\mu=0$ we recover the result obtained by the MS scheme.

In this scheme, the full amplitude is
\begin{equation}
    \mathcal{M}=\frac{-\sum C_{2n}(\mu)k^{*2n}}{1+\frac{\tilde{M}}{2\pi}(\mu+\imag k^*) \sum C_{2n}(\mu)k^{*2n}} \, .
\end{equation}
There are several ways to find the value of $C_0$ in terms of the scattering parameters. One of them is by imposing that the amplitude is independent of the subtraction point $\mu$. But we can do the same as we did for the natural case, where we find which diagrams are the ones contributing at each order. To do that, we first need to see what is the size of $C_{2n}$ for $\mu\gg 1/|a|$. Using a rearranged version of Eq.~\eqref{eq:amp_phshift},
\begin{equation}
\begin{aligned}
    k^*\cot\delta &= \imag k^* +\frac{2\pi}{\tilde{M}\mathcal{M}}=-\frac{2\pi}{\tilde{M}\sum C_{2n}(\mu)k^{*2n}}-\mu \, .
\end{aligned}
\end{equation}
Expanding $k^*\cot\delta$ with the ERE and in the limit $\mu\gg 1/|a|$,
\begin{equation}
\begin{aligned}
    \mu+\frac{1}{2}\Lambda^2\sum^{\infty}_{n=0}r_n \left(\frac{k^{*2}}{\Lambda^2} \right)^{n+1} &=-\frac{2\pi}{\tilde{M}\sum C_{2n}(\mu)k^{*2n}}\\
    &=-\frac{2\pi}{\tilde{M}}\left(\frac{1}{C_0}-\frac{C_2}{C^2_0}k^{*2}+\frac{C_2^2-C_0C_4}{C^3_0}k^{*4}+\cdots \right) \, .
\end{aligned}
\end{equation}
Therefore, with $r_n\sim 1/\Lambda$,
\begin{equation}
    C_0\sim \frac{2\pi}{\tilde{M}\mu}\, , \quad C_2\sim\frac{2\pi}{\tilde{M}\Lambda\mu^2} \, ,\quad C_4\sim\frac{2\pi}{\tilde{M}\Lambda^2\mu^3} \quad \rightarrow \quad C_{2n}\sim \frac{2\pi}{\tilde{M}\Lambda^n\mu^{n+1}} \, .
\end{equation}
Hence, if we take $\mu \sim k^*$, then $C_{2n}\sim 1/k^{*n+1}$. We also have insertions of $\nabla^{2n}$, which count as $k^{*2n}$, so each vertex will count as $C_{2n}\nabla^{2n}\sim k^{*n-1}$, and each loop contributes a factor of $k^*$. Therefore, the leading order, $\mathcal{M}_{-1}$, which scales as $1/k^*$, is the sum of bubble diagrams with $C_0$ vertices, 
\begin{equation}
    \mathcal{M}_{-1}=\includegraphics[scale=0.4, valign=c]{figures/chapter4/unnat_a_-1_w.pdf}=\frac{-C_0}{1+\frac{\tilde{M}}{2\pi}(\mu+\imag k^*)C_0} \, ,
\label{eq:unnat_amp-1}
\end{equation}
and the rest will come from perturbative insertions of derivative interactions, dressed to all orders with $C_0$. These dressed propagators are (we will write $I\equiv I^{\text{PDS}}$ for simplicity)
\begin{equation}
\includegraphics[scale=0.4, valign=c]{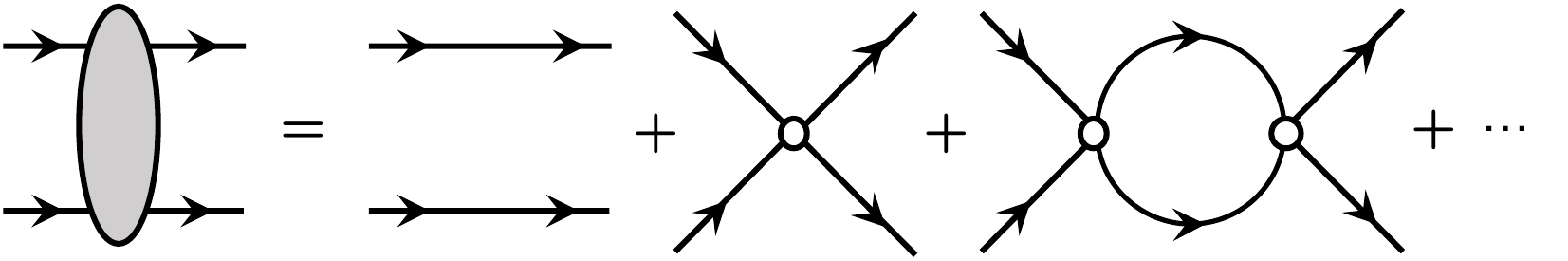} \; \rightarrow \quad 1 -C_0 + IC_0^2-I^2C_0^3+\cdots \, .
\end{equation}
We need to be careful because these dressed propagators will need to be connected with a derivative interaction, so in some cases we will have to add an extra loop,
\begin{equation}
\begin{aligned}
    \mathcal{M}_{0}=\includegraphics[scale=0.4, valign=c]{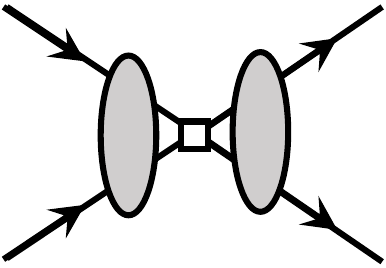}&=(1 -IC_0 + I^2C_0^2+\cdots)(-C_2k^{*2})(1 -IC_0 + I^2C_0^2+\cdots)\\
    &=\frac{1}{1+IC_0}(-C_2k^{*2})\frac{1}{1+IC_0} = \frac{-C_2k^{*2}}{\left[1+\frac{\tilde{M}}{2\pi}(\mu+\imag k^*)C_0\right]^2} \, . 
    \label{eq:unnat_amp0}
\end{aligned}
\end{equation}
Again, if we compare Eqs.~\eqref{eq:unnat_amp-1} and~\eqref{eq:unnat_amp0} with Eq.~\eqref{eq:amplitude_param_un} to write $C_{2n}$ in terms of $a$ and $r_n$,
\begin{equation}
\begin{aligned}
    \frac{-C_0}{1+\frac{\tilde{M}}{2\pi}(\mu+\imag k^*)C_0}=-\frac{2\pi }{\tilde{M}}\frac{1}{\frac{1}{a}+\imag k^*} \quad &\rightarrow\quad C_0=\frac{2\pi}{\tilde{M}}\frac{1}{\frac{1}{a}-\mu} \, , \\
    \frac{-C_2k^{*2}}{\left[1+\frac{\tilde{M}}{2\pi}(\mu+\imag k^*)C_0\right]^2} = -\frac{2\pi }{\tilde{M}}\frac{r/2}{(\frac{1}{a}+\imag k^*)^2}k^{*2} \quad &\rightarrow\quad C_2=\frac{2\pi}{\tilde{M}}\frac{r/2}{(\frac{1}{a}-\mu)^2}=C_0\frac{r/2}{\frac{1}{a}-\mu} \, .
\end{aligned}
\label{eq:coef_unnat}
\end{equation}

\vspace{1em}

Recapitulating, we have seen that for natural interactions, we can set $\mu=0$, corresponding to a tree-level expansion of the scattering amplitude. For the unnatural case, the expansion does not converge for momenta larger than $a^{-1}$, and in the KSW-vK scheme $\mu$ is introduced as a renormalization scale for the $s$-channel two-baryon loops appearing in the all-orders expansion of the amplitude with LO interactions.
Since a pionless EFT is used, a convenient choice is $\mu=m_{\pi}$, and the relations between the coefficients defined in~\cref{subsec:SU3EFT,subsec:SU6EFT} and the scattering parameters are
\begin{align}
    \left[-\frac{1}{a_{\scriptscriptstyle B_1B_2}}+\mu\right]^{-1}&= \frac{\tilde{M}_{\scriptscriptstyle B_1B_2}}{2\pi}(c^{(\text{irrep})}+\bm{c}^{\chi}_{\scriptscriptstyle B_1B_2})\, , \label{eq:scattparam1}\\
    \frac{r_{\scriptscriptstyle B_1B_2}}{2}\left[-\frac{1}{a_{\scriptscriptstyle B_1B_2}}+\mu\right]^{-2}&=\frac{\tilde{M}_{\scriptscriptstyle B_1B_2}}{2\pi}\tilde{c}^{(\text{irrep})} \, , \label{eq:scattparam2}
\end{align}
with Eq.~\eqref{eq:scattparam1} being the relation for the LECs that accompany momentum-independent operators, with contributions from LO and NLO $\cancel{SU(3)}_f$ terms in the Lagrangian, and Eq.~\eqref{eq:scattparam2} the LECs from momentum-dependent operators, with only contributions from NLO $SU(3)_f$ terms.

Although the KSW-vK power counting has been very useful to study the $NN$ and other nuclear systems, it suffers from convergence issues in the spin-triplet channels~\cite{Fleming:1999ee}, and alternative schemes have been developed~\cite{Beane:2001bc}, which combine the KSW-vK and Weinberg power counting~\cite{Weinberg:1990rz,Weinberg:1991um}. A different approach to learn about the LECs is to first build the potential of the two-baryon system and then solve the Lippmann-Schwinger equation to compute the scattering amplitude with the contact interactions regularized with a cutoff (e.g., see Ref.~\cite{Polinder:2006zh}).

\section{Lattice results for \texorpdfstring{$m_{\pi}\sim 450$}{mpi = 450} MeV}\label{sec:450results}

The present study continues, revisits, and expands upon the work of Ref.~\cite{Orginos:2015aya}. In particular, the same ensembles of QCD gauge-field configurations that were used in Ref.~\cite{Orginos:2015aya} to constrain the low-lying spectra and scattering amplitudes of spin-singlet and spin-triplet two-nucleon systems at a pion mass of $\sim 450$ MeV are used here. The same configurations have also been used to study properties of baryons and light nuclei at this pion mass, including the rate of the radiative capture process $np \to d \gamma$~\cite{Beane:2015yha}, the response of two-nucleon systems to large magnetic fields~\cite{Detmold:2015daa}, the magnetic moments of octet baryons~\cite{Parreno:2016fwu}, the gluonic structure of light nuclei~\cite{Winter:2017bfs}, and the gluon gravitational form factors of hadrons~\cite{Detmold:2017oqb,Shanahan:2018pib,Shanahan:2018nnv}.

\begin{table}[b!]
\centering
\caption{Parameters of the gauge-field ensembles used in this work. $L$ and $T$ are the spatial and temporal dimensions of the hypercubic lattice, $\beta$ is related to the strong coupling, $b$ is the lattice spacing, $m_{l(s)}$ is the bare light (strange) quark mass, $N_{\text{cfg}}$ is the number of configurations used and $N_{\text{src}}$ is the total number of sources computed. For more details, see Ref.~\cite{Orginos:2015aya}.}
\label{tab:gauge_param}
\renewcommand{\arraystretch}{1.5}
\resizebox{\columnwidth}{!}{
\begin{tabular}{ccccccccccc}
\toprule
$L^3\times T$ & $\beta$ & $bm_l$ & $bm_s$ & $b$ [fm] & $L$ [fm] & $T$ [fm] & $m_{\pi}L$ & $m_{\pi}T$ & $N_{\text{cfg}}$ & $N_{\text{src}}$ \\ \midrule 
$24^3\times 64$ & $6.1$ & $-0.2800$ & $-0.2450$ & $0.1167(16)$ & $2.8$ & $7.5$ & $6.4$ & $17.0$ & $4407$ & $1.16\times 10^6$ \\
$32^3\times 96$ & $6.1$ & $-0.2800$ & $-0.2450$ & $0.1167(16)$ & $3.7$ & $11.2$ & $8.5$ & $25.5$ & $4142$ & $3.95\times 10^5$ \\
$48^3\times 96$ & $6.1$ & $-0.2800$ & $-0.2450$ & $0.1167(16)$ & $5.6$ & $11.2$ & $12.8$ & $25.5$ & $1047$ & $6.8\times 10^4$\\\bottomrule
\end{tabular} }
\end{table}

For completeness, a summary of the technical details, widely discussed in Ref.~\cite{Orginos:2015aya}, is presented here. The LQCD calculations are performed with $n_f=2+1$ quark flavors, with the Lüscher-Weisz gauge action~\cite{Luscher:1984xn} and a clover-improved quark action~\cite{Sheikholeslami:1985ij} with one level of stout smearing ($\rho=0.125$)~\cite{Morningstar:2003gk}. The lattice spacing is $b=0.1167(16)$ fm, determined using quarkonium hyperfine splittings~\cite{Meinel:private}. The strange quark mass is tuned to its physical value, while the degenerate light (up and down)-quark masses produce a pion of mass $m_{\pi}=450(5)$ MeV and a kaon of mass $m_K=596(6)$ MeV.
Ensembles at these values of the parameters with three different volumes are used. Using the two smallest volumes with dimensions $24^3\times 64$ and $32^3\times 96$, two different sets of correlation functions were produced, with sink interpolating operators that are either point-like or smeared with 80 steps of a gauge-invariant Gaussian profile at the quark level with parameter $\rho=3.5$. In both cases, the source interpolating operators were smeared with the same values of the parameters. These two types of correlation functions are labeled SP and SS, respectively. For the largest ensemble with dimensions $48^3\times 96$, only SP correlation functions were produced for computational expediency. Table~\ref{tab:gauge_param} summarizes the parameters of these ensembles.

Correlation functions were constructed by forming baryon blocks at the sink~\cite{Detmold:2012eu},
\begin{equation}
    \mathcal{B}^{ijk}_B(\bm{p},\tau;x_0)=\sum_{\bm{x}}e^{\mathrm{i}\bm{p}\cdot\bm{x}}\, S^{(q_1),i'}_i \hskip -0.03in (\bm{x},\tau;x_0)\; S^{(q_2),j'}_j \hskip -0.03in (\bm{x},\tau;x_0)\; S^{(q_3),k'}_k \hskip -0.03in (\bm{x},\tau;x_0) \; w^{B}_{i'j'k'}\, ,
\label{eq:blocks}
\end{equation}
where $S^{(q),n'}_n$ is a quark propagator with flavor $q\in\{u,d,s\}$ and with combined spin-color indices $n,n'\in\{1,\ldots,N_sN_c\}$, where $N_s=4$ is the number of spin components and $N_c=3$ is the number of colors. The weights $w^{B}_{i'j'k'}$ are tensors that antisymmetrize and collect the terms needed to have the quantum numbers of the baryons $B\in\{N,\Lambda, \Sigma, \Xi \}$. 
The interpolating operators for the single-baryon systems studied in this work are local, i.e., include no covariant derivatives, and are similar to Eq.~\eqref{eq:protOperator}. Explicitly,
\begin{equation}
\begin{aligned}
    \hat{N}_{\alpha_1\alpha_2\alpha_3}(x)&=\epsilon^{abc}\frac{1}{\sqrt{2}}[u^a_{\alpha_1}(x)d^b_{\alpha_2}(x)-d^a_{\alpha_1}(x)u^b_{\alpha_2}(x)]u^c_{\alpha_3}(x)\, ,\\
    \hat{\Lambda}_{\alpha_1\alpha_2\alpha_3}(x)&=\epsilon^{abc}\frac{1}{\sqrt{2}}[u^a_{\alpha_1}(x)d^b_{\alpha_2}(x)-d^a_{\alpha_1}(x)u^b_{\alpha_2}(x)]s^c_{\alpha_3}(x)\, ,\\
    \hat{\Sigma}_{\alpha_1\alpha_2\alpha_3}(x)&=\epsilon^{abc}u^a_{\alpha_1}(x)u^b_{\alpha_2}(x)s^c_{\alpha_3}(x)\, ,\\
    \hat{\Xi}_{\alpha_1\alpha_2\alpha_3}(x)&=\epsilon^{abc}s^a_{\alpha_1}(x)s^b_{\alpha_2}(x)u^c_{\alpha_3}(x)\, ,
\end{aligned}
\end{equation}
where $\alpha_i$ denote spin indices and $a,b,c$ denote color indices~\cite{Basak:2005aq}. Only the upper-spin components in the Dirac spinor basis were used, requiring specific values for the $\alpha_i$ indices: $p_{\uparrow}(x) = \hat{N}_{121}(x)$, $\Lambda_{\uparrow}(x) = \hat{\Lambda}_{121}$, $\Sigma^+_{\uparrow}(x) = \sqrt{\frac{2}{3}}[\hat{\Sigma}_{112}(x)-\hat{\Sigma}_{121}(x)]$, and $\Xi^0_{\uparrow}(x) = \sqrt{\frac{2}{3}}[\hat{\Xi}_{112}(x)-\hat{\Xi}_{121}(x)]$. The neutron, $\Sigma^-$, and $\Xi^-$ operators were obtained by simply swapping $u\leftrightarrow d$ in the expressions above.
The sum over the sink position $\bm{x}$ in Eq.~(\ref{eq:blocks}) projects the baryon blocks to well-defined three-momentum value $\bm{p}$.
In particular, two-baryon correlation functions were generated with total momentum $\bm{P}=\bm{p}_1+\bm{p}_2$, where $\bm{p}_i$ is the three-momentum of the $i$th baryon taking the values $\bm{p}_i=\tfrac{2\pi}{L}\bm{n}$ with $\bm{n}\in\{ (0,0,0),(0,0,\pm 1) \}$. Therefore, $\bm{P}=\tfrac{2\pi}{L}\bm{d}$, with $\bm{d}\in\{ (0,0,0),(0,0,\pm 2)\}$.\footnote{For the rest of the thesis, $\bm{d}=(0,0,\pm 2)$ will be denoted as $\bm{d}=(0,0,2)$ for brevity.}
Additionally, two baryon correlation functions with back-to-back momenta were generated at the sink, $\bm{p}_1=-\bm{p}_2=\tfrac{2\pi}{L}\bm{n}$. This latter choice provides interpolating operators for the two-baryon system that primarily overlap with states that are unbound in the infinite-volume limit, providing a convenient means to identify excited states as well.
The next step in the construction of the correlation functions consists in forming a fully antisymmetrized local quark-level wavefunction at the location of the source, with quantum numbers of the two-baryon system of interest. Appropriate indices from the baryon blocks at the sink are then contracted with those at the source, in a way that is dictated by the quark-level wavefunction, see Refs.~\cite{Beane:2012vq,Detmold:2012eu} for more detail.
The contraction codes used to produce the correlation functions in this study are the same as those used to perform the contractions for the larger class of interpolating operators used in previous studies of the $SU(3)_f$-symmetric spectra of nuclei and hypernuclei up to $A= 5$~\cite{Beane:2012vq}, and two-baryon scattering~\cite{Beane:2013br,Orginos:2015aya,Wagman:2017tmp}.\footnote{The same code was generalized to enable studies of $np\rightarrow d\gamma$~\cite{Beane:2015yha}, proton-proton fusion~\cite{Savage:2016kon}, and other electroweak processes, as reviewed in Ref.~\cite{Davoudi:2020ngi}}

In this study, correlation functions for nine different two-baryon systems have been computed, ranging from strangeness $S=0$ to $-4$. Using the notation $(^{2s+1} L_J,\, I)$, where $s$ is the spin, $L$ the orbital momentum, $J$ the total angular momentum, and $I$ the isospin, the systems are
\begin{align*}
    S=\phantom{-}0\; &: \; NN \;(\1s0,\, 1)\, ,\ NN \;(\3s1,\, 0)\, , \\
    S=-1\; &: \; \Sigma N \;(\1s0,\, \tfrac{3}{2})\, ,\ \Sigma N (\3s1,\, \tfrac{3}{2})\, ,\\
    S=-2\; &: \; \Sigma \Sigma \;(\1s0,\, 2)\, ,\ \Xi N \;(\3s1,\, 0)\, , \\
    S=-3\; &: \; \Xi \Sigma \;(\1s0,\, \tfrac{3}{2})\, ,\\
    S=-4\; &: \; \Xi\Xi \;(\1s0,\, 1)\, ,\ \Xi\Xi \;(\3s1,\, 0)\, .
\end{align*}
Under strong interactions, these channels do not mix with other two-baryon channels or other hadronic states below three-particle inelastic thresholds. In the limit of exact $SU(3)$ flavor symmetry, the states belong to a single irrep of $SU(3)_f$: $\mathbf{27}$ (all the singlet states), $\overline{\mathbf{10}}$ (triplet $NN$), $\mathbf{10}$ (triplet $\Sigma N$ and $\Xi \Xi$), and $\mathbf{8}_a$ (triplet $\Xi N$). In the rest of this work, the isospin label will be dropped for simplicity.

\subsection{Low-lying finite-volume spectra of two baryons}

In order to extract the two-baryon energies, we first need to analyze the single baryon two-point correlation functions. The EMPs for the $N$, $\Lambda$, $\Sigma$, and $\Xi$ octet baryons are displayed in Fig.~\ref{fig:B1_EMP} of~\cref{appen:figtab} for each of the ensembles studied in the present work. The bands shown in the figures indicate the baryon mass which results from the fitting strategy explained in~\cref{subsec:2ptfitting}, with the statistical and systematic uncertainties included, and the corresponding numerical values listed in Table~\ref{tab:baryon_mass}. 

\begin{table}[t!]
\centering
\caption{The values of the masses of the octet baryons. The first uncertainty is statistical, while the second is systematic. Quantities are expressed in lattice units (l.u.).}
\label{tab:baryon_mass}
\renewcommand{\arraystretch}{1.2}
\setlength{\tabcolsep}{12pt}
\begin{tabular}{ccccc}
\toprule
Ensemble & $M_N$ [l.u.] & $M_{\Lambda}$ [l.u.] & $M_{\Sigma}$ [l.u.] & $M_{\Xi}$ [l.u.] \\
\midrule
$24^3\times 64$ & $0.7261(08)(15)$ & $0.7766(07)(13)$ & $0.7959(07)(10)$ & $0.8364(07)(08)$ \\
$32^3\times 96$ & $0.7258(05)(08)$ & $0.7765(05)(06)$ & $0.7963(05)(06)$ & $0.8362(05)(05)$ \\
$48^3\times 96$ & $0.7250(06)(12)$ & $0.7761(05)(09)$ & $0.7955(06)(07)$ & $0.8359(08)(08)$ \\ \midrule
$\infty$ & $0.7253(04)(08)$ & $0.7763(04)(06)$ & $0.7959(04)(05)$ & $0.8360(05)(05)$ \\\bottomrule
\end{tabular}
\end{table}

\begin{figure}[t]
\centering \includegraphics[width=0.9\textwidth]{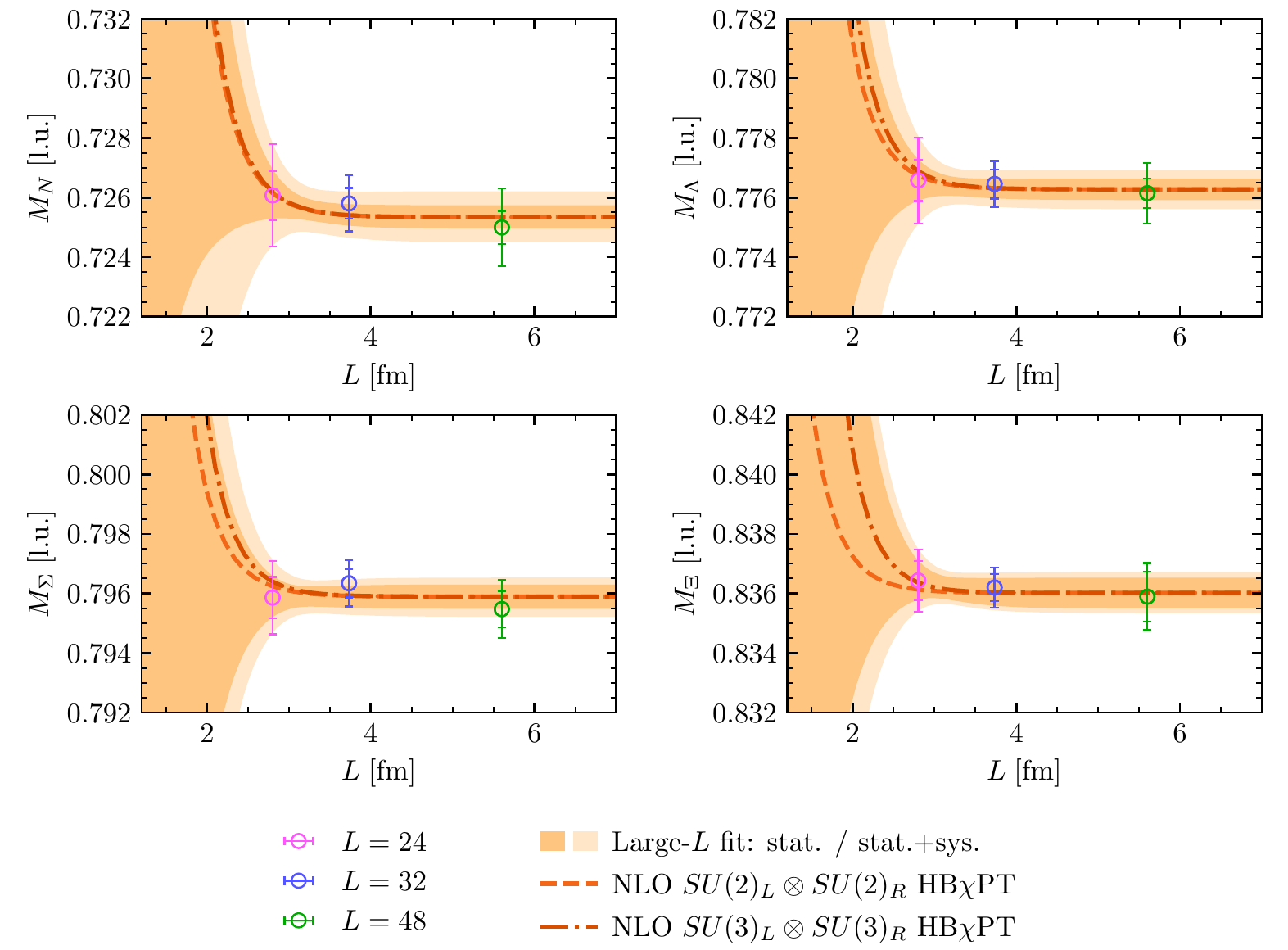}
\caption{The octet baryon masses as a function of the spatial extent $L$. The inner (outer) bands show the statistical (statistical and systematic) uncertainty from the fit to Eq.~\eqref{eq:mass_extrap}, and the lines show the expected behavior given by Eqs.~\eqref{eq:mass_extrapSU2} (dashed), with only the contribution $\delta M^{(\pi)}_B$, and~\eqref{eq:mass_extrapSU3} (dash-dotted), with both contributions $\delta M^{(\pi)}_B+\delta M^{(K,\eta)}_B$.}
\label{fig:baryonmass_extrap}
\end{figure}

To compute the binding energies as well as to constrain the LECs, we need the values of the masses in infinite volume. A very extensive study on finite-volume effects on baryons (as well as mesons) was performed by the NPLQCD Collaboration~\cite{Beane:2011pc} (see also Refs~\cite{AliKhan:2003ack,Beane:2004tw} for studies on finite-size effects on the nucleon mass). With our current values of $m_\pi$ and $L$ (listed in Table~\ref{tab:gauge_param}) such that $m_\pi L \gg 1$ (known as the $p$-regime), and noting the similarity of the values of the masses across the three different volumes, it is expected that there will be small (or imperceptible) finite volume effects. In the limit $m_\pi L \gg 1$, the volume dependence of the octet baryons can be described by
\begin{equation}
    M^{(V)}_{B}(L)=M^{(\infty)}_B+c_B\frac{e^{-m_{\pi}L}}{m_{\pi}L}\, ,
\label{eq:mass_extrap}
\end{equation}
where $M^{(\infty)}_B$ and $c_B$ are the two fit parameters. This form incorporates LO volume corrections to the baryon masses in heavy-baryon chiral perturbation theory (HB$\chi$PT). The values in the last row from Table~\ref{tab:baryon_mass} are obtained using this form, and all the values $c_B$ are of $\mathcal{O}(1)$ but consistent with zero within uncertainties (e.g., for the nucleon, $c_B=3(4)(7)$ l.u.). In physical units, $M_N\sim 1226$ MeV, $M_{\Lambda}\sim 1313$ MeV, $M_{\Sigma}\sim 1346$ MeV and $M_{\Xi}\sim 1414$ MeV. While the $\Lambda$ baryon is not relevant to subsequent analysis of the two-baryon systems (there is no system with one or two $\Lambda$s in the present study), the centroid of the four octet-baryon masses is used to define appropriate units for the EFT LECs, hence $M_{\Lambda}$ is reported for completeness. The extrapolation is shown in Fig.~\ref{fig:baryonmass_extrap}.

In order to check that higher-order terms proportional to $e^{-\sqrt{2}m_\pi L}/(\sqrt{2}m_\pi L)$ or $e^{-m_K L}/(m_K L)$, ignored in Eq.~\eqref{eq:mass_extrap}, do not contribute to this study, we can compare our fit with the explicit NLO expressions. For the case where only $\pi$ loops are considered (NLO $SU(2)_L\otimes SU(2)_R$ HB$\chi$PT), the finite volume corrections $\delta M_B=M^{(V)}_{B}(L)-M^{(\infty)}_B$ are~\cite{Beane:2011pc}
\begin{equation}
\begin{aligned}
    \delta M^{(\pi)}_N=&\ (D+F)^2\frac{9m^3_\pi}{8\pi f^2_\pi}F_N(m_\pi L)+C^2\frac{m^3_\pi}{\pi f^2_\pi}F_\Delta(m_\pi L,\Delta_{\Delta N}L)\, ,\\
    \delta M^{(\pi)}_\Lambda=&\ 4D^2\frac{3m^3_\pi}{8\pi f^2_\pi}F_\Delta(m_\pi L,\Delta_{\Sigma\Lambda}L) + \frac{C^2}{2}\frac{3m^3_\pi}{2\pi f^2_\pi}F_\Delta(m_\pi L,\Delta_{\Sigma^* \Lambda}L)\, ,\\
    \delta M^{(\pi)}_\Sigma=&\ 4D^2\frac{m^3_\pi}{8\pi f^2_\pi}F_\Delta(m_\pi L,\Delta_{\Lambda\Sigma}L) + 4F^2\frac{3m^3_\pi}{4\pi f^2_\pi}F_N(m_\pi L)\\
    &+\frac{C^2}{3}\frac{m^3_\pi}{2\pi f^2_\pi}F_\Delta(m_\pi L,\Delta_{\Sigma^* \Sigma}L)\, ,\\
    \delta M^{(\pi)}_\Xi=&\ (D-F)^2\frac{9m^3_\pi}{8\pi f^2_\pi}F_N(m_\pi L)+\frac{C^2}{3}\frac{3 m^3_\pi}{4 \pi f^2_\pi}F_\Delta(m_\pi L,\Delta_{\Xi^* \Xi}L)\, .
\end{aligned}
\label{eq:mass_extrapSU2}
\end{equation}
Including $K$ and $\eta$ loops (NLO $SU(3)_L\otimes SU(3)_R$ HB$\chi$PT), the corrections are~\cite{Beane:2011pc}
\begin{equation}
\begin{aligned}
    \delta M_N^{(K,\eta)} =&\ (D-F)^2 \frac{9 m_K^3}{8\pi f_K^2} F_\Delta (m_K L, \Delta_{\Sigma N} L) + (D+3F)^2 \frac{m_K^3}{8\pi f_K^2} F_\Delta (m_K L, \Delta_{\Lambda N} L)\\ 
    & + (D-3F)^2 \frac{m_\eta^3}{8\pi f_\eta^2} F_N (m_\eta L) + C^2 \frac{m_K^3}{4\pi f_K^2} F_\Delta (m_K L, \Delta_{\Sigma^* N} L)\, , \\
    \delta M_\Lambda^{(K,\eta)} =&\ (D+3F)^2 \frac{m_K^3}{4\pi f_K^2} F_\Delta( m_K L, \Delta_{N\Lambda} L) +(D-3F)^2 \frac{m_K^3}{4\pi f_K^2} F_\Delta( m_K L, \Delta_{\Xi\Lambda} L)\\
    & + D^2 \frac{m_\eta^3}{2\pi f_\eta^2} F_N (m_\eta L) + C^2 \frac{m_K^3}{2\pi f_K^2} F_\Delta( m_K L, \Delta_{\Xi^*\Lambda} L)\, ,\\
    \delta M_\Sigma^{(K,\eta)} =&\ (D-F)^2 \frac{3 m_K^3}{4\pi f_K^2} F_\Delta (m_K L, \Delta_{N\Sigma} L) + (D+F)^2 \frac{3 m_K^3}{4\pi f_K^2} F_\Delta (m_K L, \Delta_{\Xi\Sigma} L)\\
    & + D^2 \frac{m_\eta^3}{2\pi f_\eta^2} F_N (m_\eta L) + C^2 \frac{m_\eta^3}{4\pi f_\eta^2} F_\Delta (m_\eta L, \Delta_{\Sigma^*\Sigma} L)\\
    & + C^2 \frac{2 m_K^3}{3 \pi f_K^2} F_\Delta(m_K L, \Delta_{\Delta\Sigma} L) + C^2 \frac{m_K^3}{6\pi f_K^2} F_\Delta(m_K L, \Delta_{\Xi^*\Sigma} L)\, ,\\
    \delta M_\Xi^{(K,\eta)} =&\ (D+F)^2 \frac{9 m_K^3}{8\pi f_K^2} F_\Delta (m_K L, \Delta_{\Sigma\Xi} L) + (D-3F)^2 \frac{m_K^3}{8\pi f_K^2} F_\Delta (m_K L, \Delta_{\Lambda\Xi} L)\\
    & + (D+3 F)^2 \frac{m_\eta^3}{8\pi f_\eta^2} F_N (m_\eta L) + C^2 \frac{m_\eta^3}{4\pi f_\eta^2} F_\Delta (m_\eta L, \Delta_{\Xi^*\Xi} L)\\
    & + C^2 \frac{m_K^3}{4\pi f_K^2} F_\Delta (m_K L, \Delta_{\Sigma^*\Xi} L) + C^2 \frac{m_K^3}{2\pi f_K^2} F_\Delta (m_K L, \Delta_{\Omega\Xi} L) \, . 
\end{aligned}
\label{eq:mass_extrapSU3}
\end{equation}
The functions $F_N$ and $F_\Delta$ are defined as follows,~\cite{Beane:2011pc}
\begin{equation}
\begin{aligned}
    F_N(x) =&\ \frac{1}{6}\sum_{\bm{n}\neq \bm{0}}\frac{e^{-|\bm{n}|x}}{|\bm{n}|x}\, , \\
    F_\Delta(x,y) =&\ \frac{1}{3\pi}\sum_{\bm{n}\neq \bm{0}}\int_0^{\infty}dw\, \beta(w,\frac{y}{x})\left\lbrace\beta(w,\frac{y}{x})K_0[x|\bm{n}|\beta(w,\frac{y}{x})]-\frac{K_1[|\bm{n}|x\beta(w,\frac{y}{x})]}{|\bm{n}|x}\right\rbrace\, ,
\end{aligned}
\end{equation}
with $\beta(w,z)=\sqrt{w^2+2wz+1}$ and $K_n(z)$ are the modified Bessel functions of the second kind.
The other quantities used are the $SU(3)$-symmetric axial couplings $\{D,F,C\}=\{0.79,0.47,1.47\}$ fitted to reproduce the experimental results on hyperonic semileptonic decays~\cite{Jenkins:1991ts,Flores-Mendieta:1998tfv}, the physical meson decay constants $\{f_\pi,f_K,f_\eta\}\sim \{132,160,160\}$ MeV, $\Delta_{AB}=M_A-M_B$, and the masses of the mesons and decuplet baryons, taken from Ref.~\cite{Orginos:2015aya} ($m_\eta$ is fixed via the Gell-Mann–Okubo relation~\cite{GellMann:1962xb,Okubo:1961jc}, $3m^2_\eta=4m^2_K-m^2_\pi$). The masses of the octet baryons are taken from Table~\ref{tab:baryon_mass}.
The expected behavior given by Eqs.~\eqref{eq:mass_extrapSU2} and~\eqref{eq:mass_extrapSU3} is also plotted in Fig.~\ref{fig:baryonmass_extrap}, shown with a dashed line ($\delta M^{(\pi)}_B$) and a dash-dotted line ($\delta M^{(\pi)}_B+\delta M^{(K,\eta)}_B$). We can see that there is a complete agreement between Eqs.~\eqref{eq:mass_extrapSU2}-\eqref{eq:mass_extrapSU3} and their large-$L$ limit Eq.~\eqref{eq:mass_extrap}, and higher statistics, especially for the $L=24$ ensemble, would be needed to clearly identify the finite-volume effects.

\begin{figure}[t!]
\includegraphics[width=\textwidth]{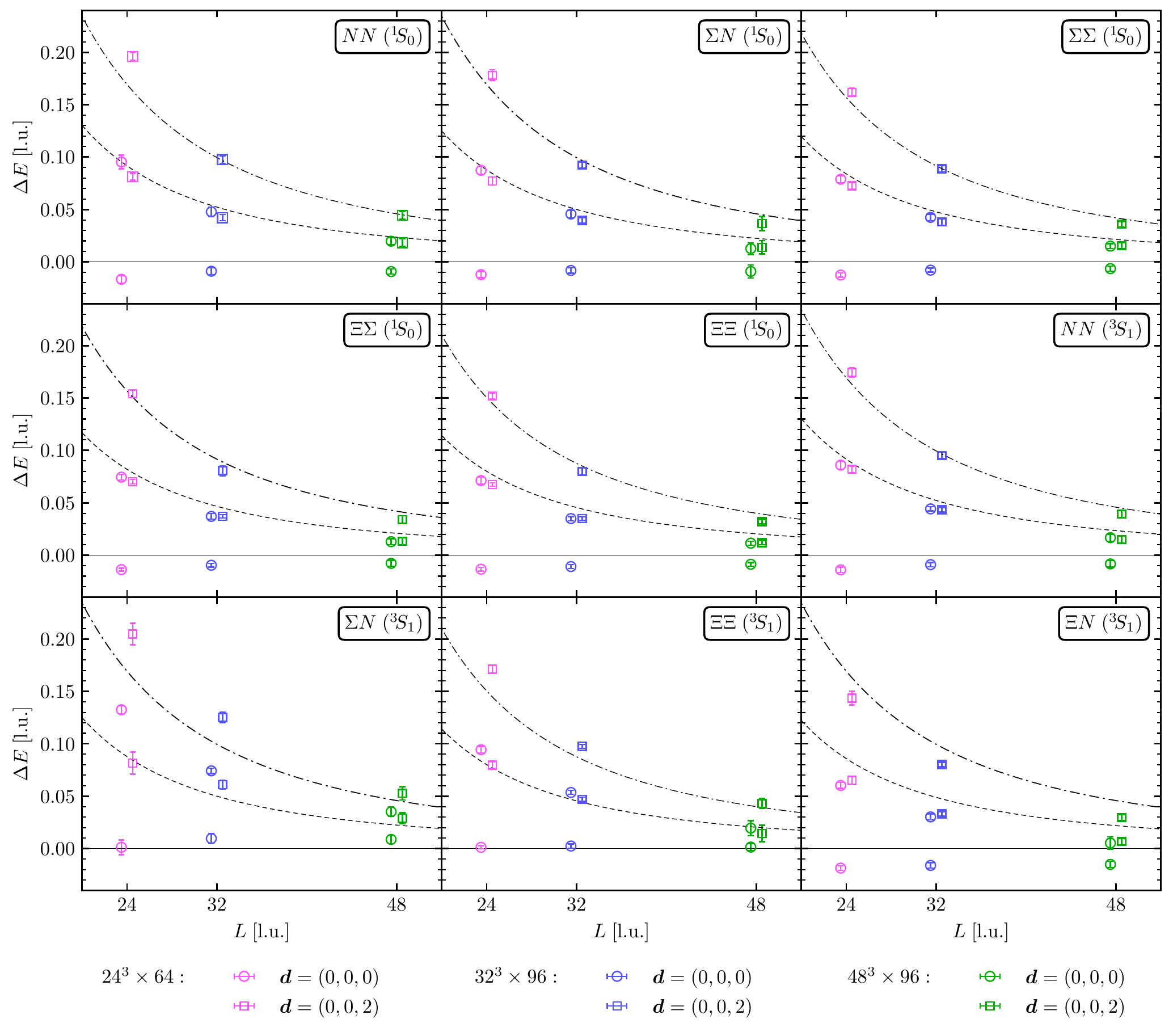}
\caption{Summary of the energy shifts extracted from LQCD correlation functions for all two-baryon systems studied in this work, together with the non-interacting energy shifts defined as $\Delta E = \sqrt{m_1^2+|\bm{p}_1|^2}+\sqrt{m_2^2+|\bm{p}_2|^2}-m_1-m_2$, where $|\bm{p}_1|^2=|\bm{p}_2|^2=0$ corresponds to systems that are at rest (continuous line), $|\bm{p}_1|^2=|\bm{p}_2|^2=(\frac{2\pi}{L})^2$ corresponds to systems which are either boosted or are unboosted but have back-to-back momenta (dashed line), and $|\bm{p}_1|^2=0$ and $|\bm{p}_2|^2=(\frac{4\pi}{L})^2$ corresponds to boosted systems where only one baryon has non-zero momentum (dashed-dotted line). The points with no boost have been shifted slightly to the left, and the ones with boosts have been shifted to the right for clarity. Quantities are expressed in lattice units.}
\label{fig:energy-levels}
\end{figure}

The results for the two-baryon energy shifts are shown in Fig.~\ref{fig:energy-levels},\footnote{The channels within the figures/tables are sorted according to the $SU(3)_f$ irrep they belong to in the limit of exact flavor symmetry, ordered as $\mathbf{27},~\overline{\mathbf{10}},~\mathbf{10},~\text{and}~\mathbf{8}_a$, and within each irrep according to their strangeness, from the largest to the smallest.} obtained using the fitting methodology described in~\cref{subsec:2ptfitting}.
For display purposes, the effective energy-shift functions, defined in Eq.~\eqref{eq:effenergyshiftfun}, are shown in Figs.~\ref{fig:NN1s0_EMP}-\ref{fig:XN3s1_EMP} of~\cref{appen:figtab}, along with the corresponding two-baryon effective energy functions, defined in Eq.~\eqref{eq:EMP}. The associated numerical values are listed in Tables~\ref{tab:eshift_ini}-\ref{tab:eshift_fin} of the same appendix.\footnote{Future studies with a range of values of the lattice spacing will be needed to extrapolate the results of two-baryon energies to the continuum limit. Nonetheless, the use of an improved lattice action in this study suggests that the discretization effects may be mild, and the associated systematic uncertainty, which has not been reported in the values in this thesis, may not be significant at the present level of precision. A recent work on the $H$-dibaryon has found quite large discretization effects~\cite{Green:2021qol} (although a small set of operators were included in the variational analysis), a system that is not studied here.}
In each subfigure of Figs.~\ref{fig:NN1s0_EMP}-\ref{fig:XN3s1_EMP}, two different correlation functions are displayed: the one yielding the lowest energy (labeled as $n=1$ in Tables~\ref{tab:eshift_ini}-\ref{tab:eshift_fin}) corresponds to having both baryons at rest or, if boosted, with the same value of the momentum, and the one yielding a higher energy (labeled as $n=2$ in the tables) corresponds to the two baryons having different momenta, e.g., having back-to-back momenta or one baryon at rest and the other with non-zero momentum. While the first case ($n=1$) couples primarily to the ground state, the latter ($n=2$) is found to have small overlap onto the ground state and gives access to the first excited state directly. 

As a final remark, it should be noted that the single-baryon masses and the energies extracted for the two-nucleon states within the present analysis are consistent within $1\sigma$ with the results of Ref.~\cite{Orginos:2015aya}, obtained with the same set of data but using different fitting strategies. Despite this overall consistency, the uncertainties of the two-nucleon energies in the present work are generally larger compared with those reported in Ref.~\cite{Orginos:2015aya} for the channels where results are available in that work. The reason lies in a slightly more conservative systematic uncertainty analysis employed here. The comparison between the results of this work and that of Ref.~\cite{Orginos:2015aya} is discussed extensively in~\cref{appen:vs2015}.

\subsection{Low-energy scattering phase shifts and effective range parameters}\label{subsec:scattparamLQCD}

\begin{figure}[t!]
\includegraphics[width=\textwidth]{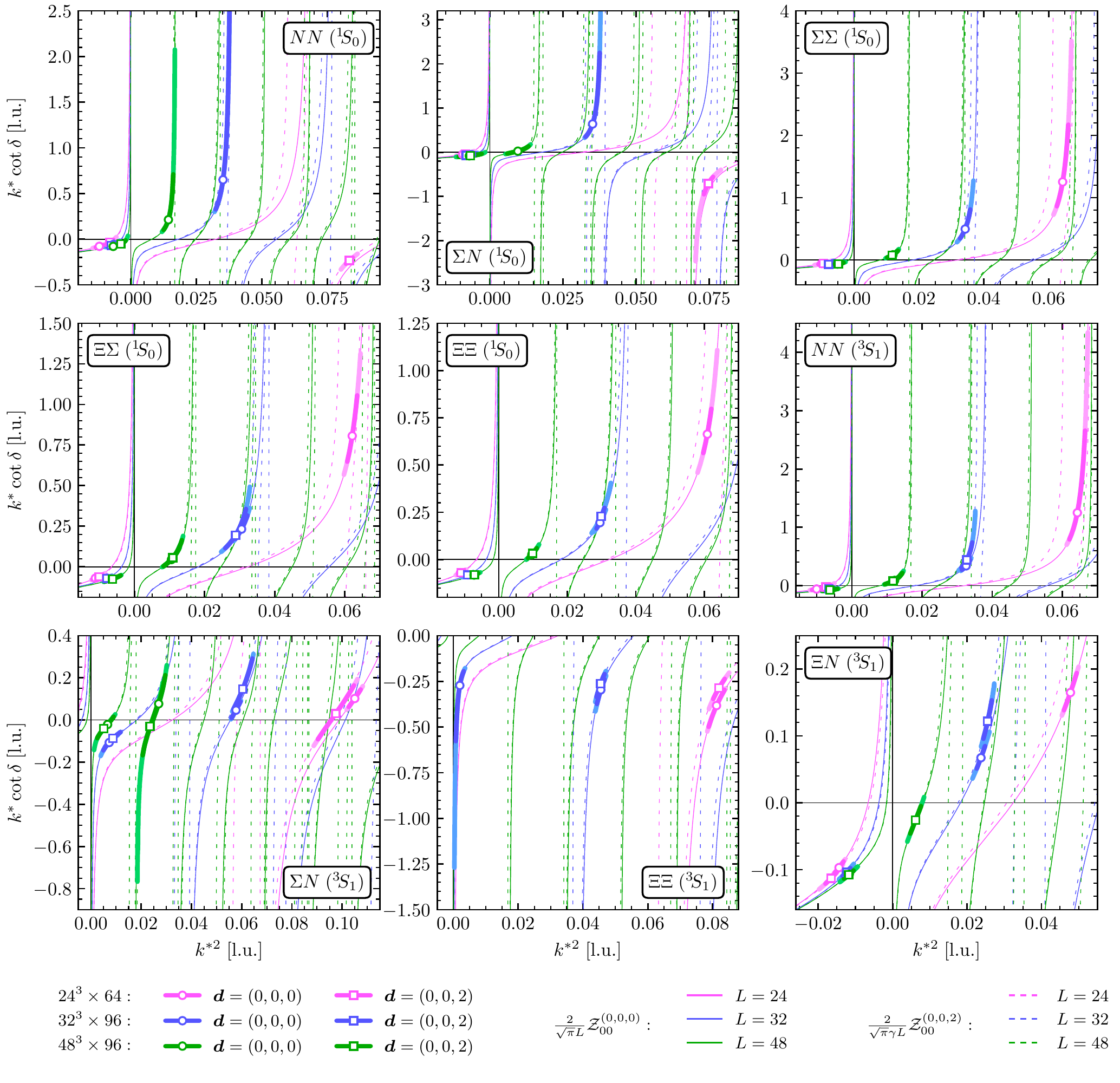}
\caption{$k^*\cot \delta$ values as a function of the squared c.m.\ momentum $k^{*2}$ for all two-baryon systems studied in this work. The darker uncertainty bands are statistical, while the lighter bands show the statistical and systematic uncertainties combined in quadrature. The different $\mathcal{Z}$-functions are shown as continuous and dashed lines. Quantities are expressed in lattice units.}
\label{fig:kcotdelta}
\end{figure}

The values of $k^* \cot\delta$ at given $k^{*2}$ values are shown for all two-baryon systems studied in this thesis in Fig.~\ref{fig:kcotdelta}, and the associated numerical values are listed in Tables~\ref{tab:eshift_ini}-\ref{tab:eshift_fin} of~\cref{appen:figtab}.  These values have been obtained using the QCs derived in~\cref{subsec:LuscherQCD}.
The validity of Lüscher's QC must be verified in each channel, in particular in those that have exhibited anomalously large ranges in previous calculations, such as $\Sigma N\; (\3s1)$.
The consistency between solutions to Lüscher's condition and the finite-volume Hamiltonian eigenvalue equation using a LO EFT potential was established in Ref.~\cite{Beane:2012ey} for the $\Sigma N\; (\3s1)$ channel and at values of the quark masses ($m_{\pi}\sim 389$ MeV) close to those of the current analysis. Based on the conclusion of the work in Ref.~\cite{Beane:2012ey}, we use the Lüscher's QC in the current work for this channel.

As mentioned before, the energy dependence of $k^* \cot\delta$ can be parametrized by an ERE below the $t$-channel cut, and since the pion is the lightest hadron that can be exchanged between any of the two baryons considered in the present study, $k^*_{t\text{-cut}}=m_{\pi}/2$. The scattering parameters can be constrained by fitting $k^* \cot\delta$ values obtained from the use of Lüscher's QC as a function of $k^{*2}$. To this end, one could use a one-dimensional choice of the $\chi^2$ function, minimizing the vertical distance between the fitted point and the function,
\begin{equation}
    \chi^2(a^{-1},r,P)=\sum_i \frac{[(k^*\cot\delta)_i-f(a^{-1},r,P,k_i^{*2})]^2}{\sigma_i^2}\, ,
\label{eq:chisquared1D}
\end{equation}
where\footnote{The inverse scattering length can be constrained far more precisely compared with the scattering length itself given that $a^{-1}$ samples can cross zero in the channels considered. As a result, in the following all dependencies on $a$ enter via $a^{-1}$.} $f(a^{-1},r,P,k^{*2})$ corresponds to the ERE parametrization given by the right-hand side of Eq.~\eqref{eq:ERE}, and the sum runs over all extracted pairs of $\{k_i^{*2}, (k^*\cot\delta)_i \}$, where the compound index $i$ counts data points for different boosts, $n$ values of the level, and different volumes.
Each contribution is weighted by an effective variance that results from the combination of the uncertainty in both $k_i^{*2}$ and $(k^*\cot\delta)_i$, $\sigma_i^2 =[\delta (k^*\cot\delta)_i]^2+[\delta k_i^{*2}]^2\,$, with $\delta x$ being the mid-68\% confidence interval of the quantity $x$. The uncertainty on the $\{k_i^{*2}, (k^*\cot\delta)_i \}$ pair can be understood by recalling that each pair is a member of a bootstrap ensemble with the distribution obtained in the previous step of the analysis.
To generate the distribution of the scattering parameters, pairs of $\{k_i^{*2}, (k^*\cot\delta)_i \}$ are randomly selected from each bootstrap ensemble and are used in Eq.~\eqref{eq:chisquared1D} to obtain a new set of $\{a^{-1},r,P\}$ parameters. This procedure is repeated $N$ times, where $N$ is chosen to be equal to the number of bootstrap ensembles for $\{k_i^{*2}, (k^*\cot\delta)_i \}$. This produces an ensemble of $N$ values of fit parameters $\{a^{-1},r,P\}$, from which the central value and the associated uncertainty in the parameters can be determined (median and mid-$68\%$ intervals are used for this purpose). 

Alternatively, one can use a two-dimensional choice of the $\chi^2$ function~\cite{Aoki:private}.
Knowing that the $k^*\cot\delta$ values must lie along the $\mathcal{Z}$-function, as can be seen from Eq.~\eqref{eq:QC} and Fig.~\ref{fig:kcotdelta}, one could take the distance between the data point and the point where the ERE crosses the $\mathcal{Z}$-function along this function (arc length) in the definition of $\chi^2$. Explicitly,
\begin{equation}
    \chi^2(a^{-1},r,P)=\sum_i \frac{D_{\mathcal{Z}}[\{k_i^{*2},(k^*\cot\delta)_i\},\{K_i^{*2},f(a^{-1},r,P,K_i^{*2})\}]^2}{\sigma_i^2}\, ,
\label{eq:newchi}
\end{equation}
where $\sigma_i^2$ is now defined as
\begin{equation}
    \sigma_i^2 =[\delta (k^*\cot\delta)_i]^2+\left(\left.\frac{\partial (k^*\cot\delta)_i}{\partial k^{*2}}\right|_{k^{*2}=k_i^{*2}}\right)^2 [\delta k_i^{*2}]^2\, ,
\label{eq:weights_chi2}
\end{equation}
and $D_{\mathcal{Z}}[\{x_a,y_a\},\{x_b,y_b\}]$ denotes the distance between the two points $\{x_a,y_a\}$ and $\{x_b,y_b\}$ along the $\mathcal{Z}$-function.
The quantity $K^{*2}$ is the point where the ERE ($f$ in Eq.~\eqref{eq:newchi}) crosses the $\mathcal{Z}$-function. To obtain this point, and given the large number of discontinuities present in the $\mathcal{Z}$-function, Householder's third order method can be used as a reliable root-finding algorithm~\cite{Householder:231838},
\begin{equation}
    \frac{2}{\sqrt{\pi}\gamma L} \mathcal{Z}^{\bm{d}}_{00}(1;Q^2)-f(a^{-1},r,P,K^{*2})\equiv F(K^{*2})=0 \; : \; K_{m+1}^{*2}=K_{m}^{*2}+3\left . \frac{(1/F)''}{(1/F)'''}\right |_{K_{m}^{*2}}\, ,
    \label{eq:Householder}
\end{equation}
where $Q=K^*L/2\pi$. The starting point is set to be $K^{*2}_0=k_i^{*2}$, and the number of primes over $1/F$ indicates the order of the derivative computed at the point $K_{m}^{*2}$. The stopping criterion is defined as $|K_{m+1}^{*2}-K_{m}^{*2}|<10^{-6}$, which occurs for $m\sim\mathcal{O}(10)$. Since the extraction of this point requires knowledge of the scattering parameters, the minimization must be implemented iteratively.

\begin{figure}[t!]
\includegraphics[width=\textwidth]{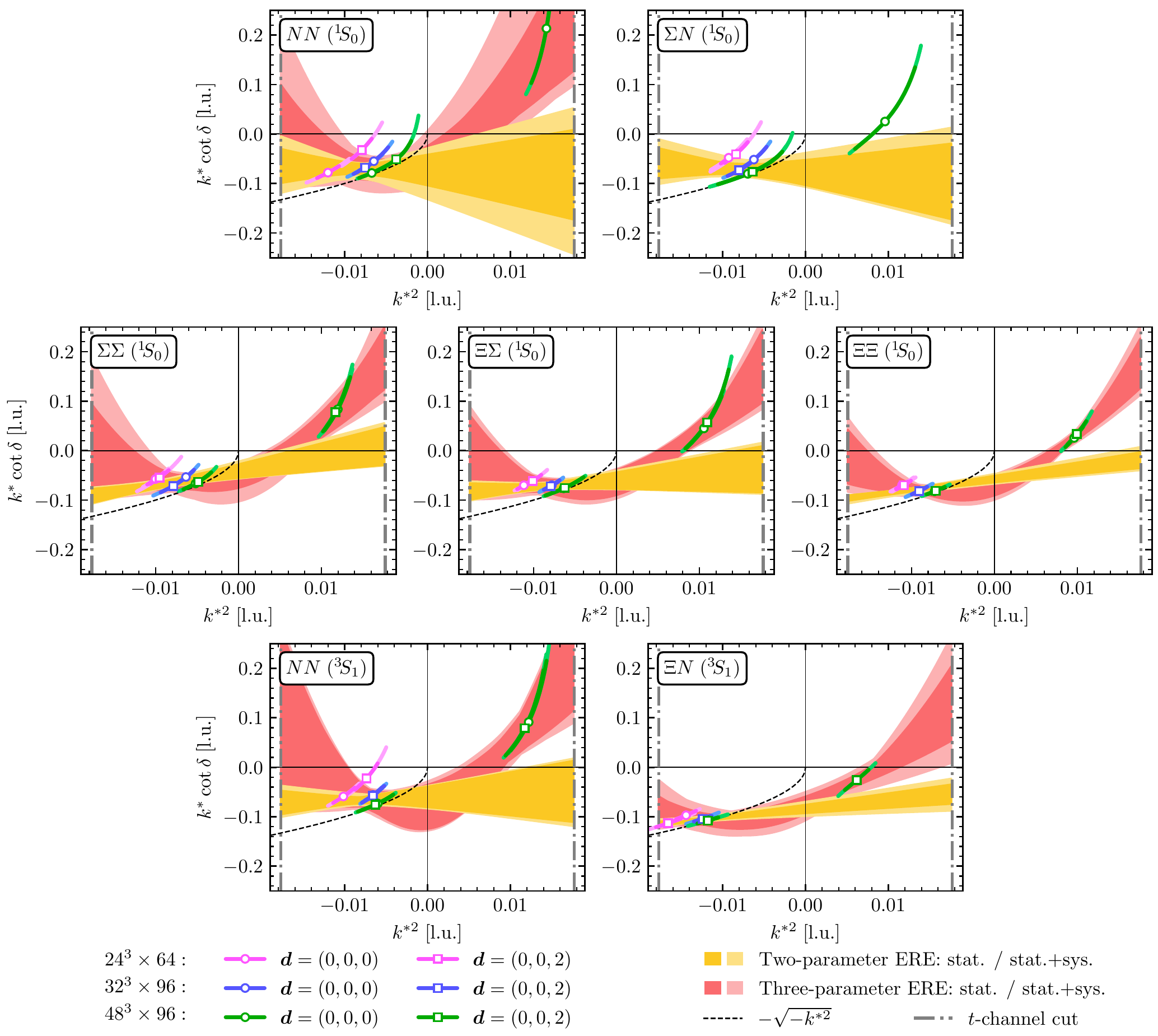}
\caption{$k^*\cot \delta$ values as a function of the c.m.\ momenta $k^{*2}$, along with the band representing the two- (yellow) and three-parameter (red) ERE for the two-baryon channels shown. The bands denote the 68\% confidence regions of the fits. Quantities are expressed in lattice units.}
\label{fig:ere-fit}
\end{figure}

\begin{figure}[t!]
\includegraphics[width=\textwidth]{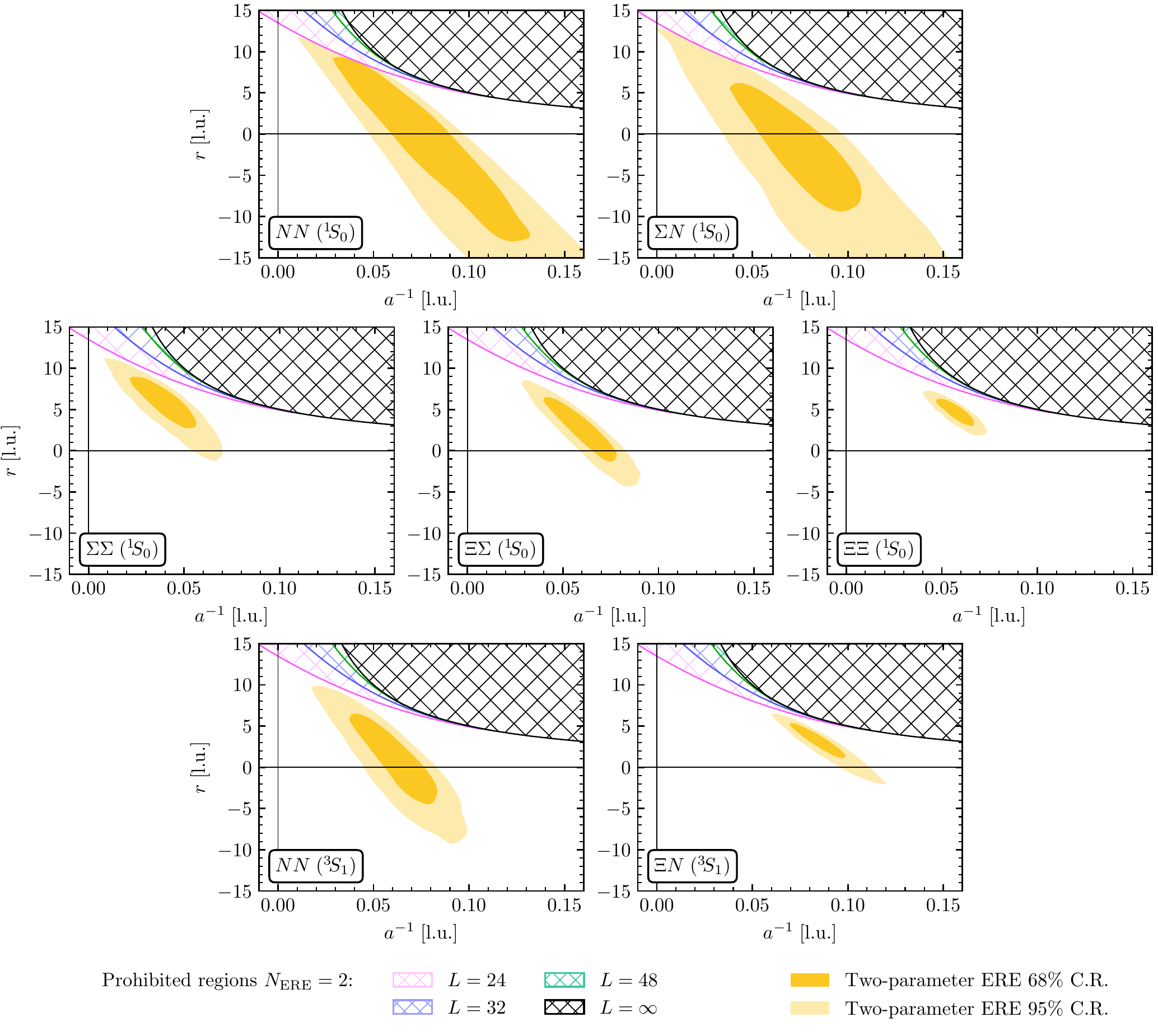}
\caption{The two-dimensional 68\% and 95\% confidence regions (C.R.) corresponding to the combined statistical and systematic uncertainty on the scattering parameters $a^{-1}$ and $r$ for all two-baryon systems that exhibit bound states, obtained from two-parameter ERE fits. The prohibited regions where the two-parameter ERE does not cross the $\mathcal{Z}$-function for given volumes (as well as the infinite-volume case) are denoted by hashed areas. Quantities are expressed in lattice units.}
\label{fig:ere-parameters}
\end{figure}

Computing the $\mathcal{Z}^{\bm{d}}_{00}$ function each time it is called during the root-finding algorithm may not be the best procedure for a fast evaluation and extraction of the scattering parameters. The solution proposed here is to precompute this function for all $k^{*2}$ possible values, in our case $k^{*2}\in[-0.05,0.2]$ l.u.\ with a step of $\Delta k^{*2}=2\cdot 10^{-6}$ l.u.\ (resulting with $1.25\cdot 10^5$ evenly distributed points). The value of the tolerance used in Householder's method is set to be the same as $\Delta k^{*2}$. Then, the distance function $D_{\mathcal{Z}}$ is computed by approximating the arc length as a sum of small straight lines,
\begin{equation}
    D_{\mathcal{Z}}[\{x_a,y_a\},\{x_b,y_b\}] = \int_{x_a}^{x_b}\sqrt{1+\left(\frac{dy}{dx}\right)^2} dx \approx \sum_{n=a}^{b-1}\sqrt{(x_n-x_{n+1})^2+(y_n-y_{n+1})^2}\, ,
\end{equation}
and the derivatives of the $\mathcal{Z}$-functions are computed via finite differences. For example, expanding Eq.~\eqref{eq:Householder},
\begin{equation}
    \frac{(1/F(x))''}{(1/F(x))'''}=\frac{F(x)^2 F''(x)-2 F(x)F'(x)^2}{F(x)^2 F'''(x)+6 F'(x)^3-6 F(x) F'(x) F''(x)}\, ,
\end{equation}
and only focusing on the $\mathcal{Z}$ part of $F$ (the derivatives of the ERE function $f$ can be computed analytically, since it is a polynomial in $k^{*2}$), we get
\begin{equation}
\begin{aligned}
    F'(x_0) &\approx \frac{1}{2}[F(x_{+1})-F(x_{-1})] \, , \\ F''(x_0) &\approx \frac{1}{2}[F(x_{+1})-2F(x_{0})+F(x_{-1})] \, , \\
    F'''(x_0) &\approx \frac{1}{2}[F(x_{+2})-2F(x_{+1})+2F(x_{-1})-F(x_{-2})]\, ,
\end{aligned}
\end{equation}
with $x_0$ being the evaluation point of $F$ (which is $K_{m}^{*2}$) and $\{x_{+2},\ldots,x_{-2}\}$ the shifts $\Delta k^{*2}$ away from $x_0$.

This second choice of $\chi^2$ function has been used in the main analysis of this work, however, the use of the one-dimensional $\chi^2$ function is shown to yield statistically consistent results (within $1\sigma$) for scattering parameters, as demonstrated in~\cref{appen:vs2015}. A similar approach was taken in Ref.~\cite{Horz:2020zvv}. 

For a precise extraction of the ERE parameters, a sufficient number of points below the $t$-channel cut must be available, for positive or negative $k^{*2}$.
In general, for the channels studied throughout this work, there are only a few points in the positive $k^{*2}$ region below the $t$-channel cut (starting at $k_{t\text{-cut}}^{*2}\sim 0.018 \text{ l.u.}$). 
For a non-interacting system, states above scattering threshold have c.m.\ energies $\sqrt{m_1^2+k^{*2}}+\sqrt{m_2^2+k^{*2}}$, with the c.m.\ momenta roughly scaling with the volume as $k^{*2}\sim(2\pi |\bm{n}|/L)^2$. With the minimum value of $|\bm{n}|^2$ used in this work ($|\bm{n}|^2=1$), only the states from the ensemble with $L=48$ are expected to lie below the $t$-channel cut ($4\pi^2/48^2<k_{t\text{-cut}}^{*2}$).
This behavior is consistent with the data. Comparing with the results of the analysis at $m_\pi \sim 806$ MeV of Ref.~\cite{Wagman:2017tmp}, where lattice configurations of comparable size (in lattice units) were used, the larger value of the pion mass resulted in the position of the $t$-channel cut being moved further away from zero, and the majority of the lowest-lying states extracted in that study remained inside the region where the ERE parametrization could be used.
Therefore, with only ground-state energies available for the analysis of the ERE in the ensembles with $L\in\{24,32\}$, the precision in the extraction of scattering parameters is noticeably reduced compared with the study at $m_{\pi} \sim 806$ MeV.
Inclusion of the shape parameter, $P$, does not improve the fits, and although the scattering lengths remain consistent with those obtained with a two-parameter fit, the effective ranges are larger in magnitude, and the uncertainties in the scattering parameters are increased. Moreover, the central values of the extracted shape parameters are rather large, bringing into question the assumption that the contribution of each order in the ERE should be smaller than the previous order. However, uncertainties on the shape parameters are sufficiently large that no conclusive statement can be made regarding the convergence of EREs.
In one case, i.e., the $\Sigma N\; (\1s0)$ channel, the three-parameter ERE fit is not performed given the large uncertainties. For these reasons, while the scattering parameters are reported for both the two- and three-parameter fits in this subsection, only those of the two-parameter fits will be used in the EFT study in the next subsection.

\begin{table}[t!]
\centering
\caption{The values of the inverse scattering length $a^{-1}$, effective range $r$, and shape parameter $P$ determined from the two- and three-parameter ERE fits to $k^*\cot\delta$ versus $k^{*2}$ for various two-baryon channels (see Fig.~\ref{fig:ere-fit}). Quantities are expressed in lattice units.}
\label{tab:scattPar}
\renewcommand{\arraystretch}{1.5}
\resizebox{\columnwidth}{!}{
\begin{tabular}{c cr ccr}
\toprule
 & \multicolumn{2}{c}{Two-parameter ERE fit} & \multicolumn{3}{c}{Three-parameter ERE fit} \\
 & $a^{-1}$ [l.u.] & \multicolumn{1}{c}{$r$ [l.u.]} & $a^{-1}$ [l.u.] & $r$ [l.u.] & \multicolumn{1}{c}{$P$ [l.u.]} \\
\cmidrule(r){2-3}\cmidrule(l){4-6}
$NN\;(\1s0)$	&	$0.084_{(-42)(-35)}^{(+20)(+44)}$	&	$-2.4_{(-5.5)(-9.0)}^{(+8.4)(+8.3)}$	&	$0.053_{(-29)(-52)}^{(+33)(+43)}$	&	$15.4_{(-6.2)(-5.7)}^{(+6.5)(+20.8)}$	&	$803_{(-570)(-190)}^{(+46)(+510)}$	\\
$\Sigma N\;(\1s0)$	&	$0.079_{(-27)(-31)}^{(+25)(+14)}$	&	$-2.8_{(-5.3)(-4.0)}^{(+6.7)(+6.0)}$	&	-	&	-	&	\multicolumn{1}{c}{-}	\\
$\Sigma\Sigma\;(\1s0)$	&	$0.040_{(-13)(-14)}^{(+15)(+06)}$	&	$5.8_{(-2.9)(-0.9)}^{(+2.8)(+1.5)}$	&	$0.059_{(-28)(-18)}^{(+17)(+41)}$	&	$10.0_{(-2.4)(-4.1)}^{(+3.8)(+3.8)}$	&	$563_{(-330)(-260)}^{(+200)(+490)}$	\\
$\Xi\Sigma\;(\1s0)$	&	$0.061_{(-17)(-12)}^{(+16)(+06)}$	&	$2.4_{(-3.4)(-1.6)}^{(+3.6)(+1.8)}$	&	$0.062_{(-22)(-11)}^{(+28)(+21)}$	&	$10.6_{(-2.1)(-0.9)}^{(+2.5)(+1.8)}$	&	$469_{(-280)(-140)}^{(+310)(+210)}$	\\
$\Xi\Xi\;(\1s0)$	&	$0.058_{(-07)(-08)}^{(+07)(+07)}$	&	$4.6_{(-1.4)(-0.8)}^{(+0.8)(+1.5)}$	&	$0.075_{(-22)(-16)}^{(+16)(+19)}$	&	$10.9_{(-1.0)(-1.0)}^{(+0.9)(+1.0)}$	&	$538_{(-250)(-180)}^{(+190)(+200)}$	\\
$NN\;(\3s1)$	&	$0.063_{(-24)(-09)}^{(+18)(+10)}$	&	$0.5_{(-4.1)(-2.9)}^{(+5.5)(+2.4)}$	&	$0.082_{(-47)(-26)}^{(+42)(+18)}$	&	$8.0_{(-5.1)(-1.9)}^{(+5.0)(+4.8)}$	&	$812_{(-560)(-340)}^{(+570)(+300)}$	\\
$\Xi N\;(\3s1)$	&	$0.086_{(-10)(-13)}^{(+07)(+11)}$	&	$3.0_{(-0.9)(-1.6)}^{(+1.7)(+1.7)}$	&	$0.080_{(-21)(-22)}^{(+14)(+23)}$	&	$12.2_{(-3.0)(-4.5)}^{(+3.5)(+4.8)}$	&	$307_{(-190)(-170)}^{(+220)(+310)}$	\\\bottomrule
\end{tabular}}
\end{table}

\begin{figure}[tb!]
\includegraphics[width=\textwidth]{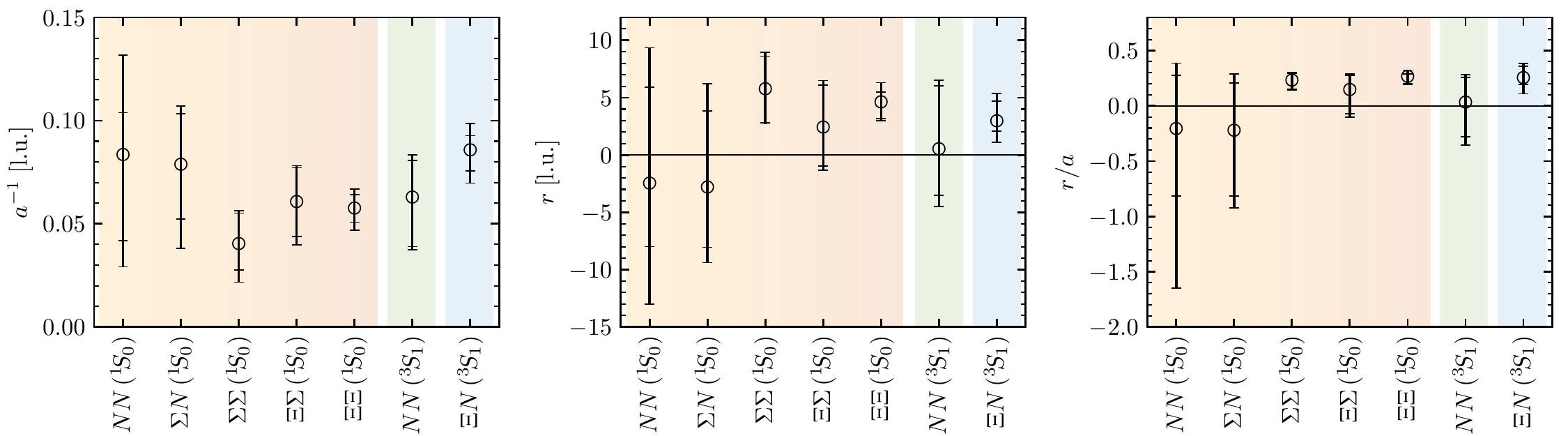}
\caption{Summary of the inverse scattering length $a^{-1}$ (left panel), effective range $r$ (middle panel), and ratio $r/a$ (right panel) determined from the two-parameter ERE fit for the two-baryon systems analyzed. The background color groups the channels by the $SU(3)_f$ irreps they would belong to if $SU(3_f)$ symmetry were exact (orange for $\mathbf{27}$, green for $\overline{\mathbf{10}}$, and blue for $\mathbf{8}_a$). Quantities are expressed in lattice units.}
\label{fig:scattPar}
\end{figure}

\begin{table}[tb!]
\caption{The values of the ratio of the effective range and scattering length, $r/a$, determined from the two-parameter ERE fit to $k^*\cot\delta$ values in each channel.}
\label{tab:scattPar-ratio}
\renewcommand{\arraystretch}{1.2}
\resizebox{\columnwidth}{!}{
\begin{tabular}{ccccccc}
\toprule
\multicolumn{7}{c}{$r/a$} \\
$NN\;(\1s0)$	&	$\Sigma N\;(\1s0)$	&	$\Sigma\Sigma\;(\1s0)$	&	$\Xi\Sigma\;(\1s0)$	&	$\Xi\Xi\;(\1s0)$	&	$NN\;(\3s1)$	&	$\Xi N\;(\3s1)$	\\ \midrule
$-0.2_{(-0.6)(-1.3)}^{(+0.5)(+0.3)}$	&	$-0.22_{(-60)(-37)}^{(+43)(+27)}$	&	$0.23_{(-08)(-04)}^{(+06)(+04)}$	&	$0.15_{(-22)(-11)}^{(+13)(+05)}$	&	$0.27_{(-07)(-03)}^{(+02)(+05)}$	&	$0.03_{(-31)(-23)}^{(+22)(+12)}$	&	$0.26_{(-06)(-13)}^{(+10)(+07)}$	\\\bottomrule
\end{tabular}}
\end{table}

Fits to $k^*\cot{\delta}$ as a function of $k^{*2}$ in various two-baryon channels are shown in Fig.~\ref{fig:ere-fit}, along with the correlation between inverse scattering length and effective range in each channel depicted in Fig.~\ref{fig:ere-parameters} using the 68\% and 95\% confidence regions of the parameters.
The areas in the parameter space that are prohibited by the constraints imposed by Eq.~\eqref{eq:QC} are also shown in Fig.~\ref{fig:ere-parameters}, highlighting the fact that the two-parameter ERE must cross the $\mathcal{Z}$-functions for each volume in the negative-$k^{*2}$ region. For fits including higher-order parameters, these constraints are more complicated and are not shown.
For the $\Sigma N \; (\3s1)$ and $\Xi \Xi \; (\3s1)$ channels, the ground-state energy is positively shifted, i.e., $\Delta E \gtrsim 0$, and only the values of $k^{*2}$ associated with the ground states are inside the range of validity of the ERE. As a result, no extraction of the ERE parameters is possible in these channels given the number of data points.
Results for the scattering parameters obtained using two- and three-parameter ERE fits in the other seven channels are summarized in Table~\ref{tab:scattPar} and are shown in Fig.~\ref{fig:scattPar} for better comparison in the case of two-parameter fits.

The inverse scattering lengths extracted for all systems are compatible with each other (albeit within rather large uncertainties), signaling that there may exist enhanced flavor symmetries at this pion mass at low energies, a feature that will be thoroughly examined in~\cref{subsec:EFTconstraints}.
The effective range in most systems is compatible with zero.
Furthermore, the ratio $r/a$ can be used as an indicator of the naturalness of the interactions; for natural interactions, $r/a\sim 1$, while for unnatural interactions $r/a \ll 1$.
At the physical point, both $NN$ channels are unnatural and exhibit large scattering lengths, with $|r/a|$ being close to $0.1$ for the spin-singlet channel, and $0.3$ for the spin-triplet channel.
As can be seen in Table~\ref{tab:scattPar-ratio}, the most constrained ratios are obtained for the $\Sigma \Sigma\; (\1s0)$, $\Xi\Xi\; (\1s0)$, and $\Xi N\; (\3s1)$ channels, for which $r/a\sim 0.2-0.3$, indicating unnatural interactions at low energies.
For other channels, the larger uncertainty in this ratio precludes drawing conclusions about naturalness.
Alternatively, naturalness can be assessed by considering the ratio of the binding momentum to the pion mass, as this quantity is better constrained in the present study, see Table~\ref{tab:binding_momenta} in the next subsection. The values for $\kappa^{(\infty)}/m_{\pi}$ in each of the bound two-baryon channels fall in the interval $[0.2,0.4]$, indicating that the interaction range associated to pion exchange is not the only characteristic scale in the system, suggesting unnaturalness.
However, at larger-than-physical quark masses, pion exchange may not be the only significant contribution to the long-range component of the nuclear force, as discussed in Ref.~\cite{Beane:2013br}.
For these reasons, both natural and unnatural interactions are considered in~\cref{subsec:EFTconstraints} when adopting a power-counting scheme in constraining the LECs of the EFT.
In~\cref{appen:vs2015}, the results for the $NN$ channels are compared with the previously obtained scattering parameters of Ref.~\cite{Orginos:2015aya}, using the same correlation functions, as well as with the predictions obtained from low-energy theorems~\cite{Baru:2016evv}. Through a thorough investigation, the differences observed between the results corresponding to the various methods are discussed and understood.

\begin{table}[tb!]
\centering
\caption{The values of the parameters $\tilde{a}^{-1}, \tilde{r},\tilde{P}$ from a two- or three-parameter polynomial fit for two-baryon channels that exhibit smooth and monotonic behavior in $k^*\cot\delta$ as a function of $k^{*2}$ beyond the $t$-channel cut. Quantities are expressed in lattice units.}
\label{tab:scattPar-beyond}
\renewcommand{\arraystretch}{1.5}
\resizebox{\columnwidth}{!}{
\begin{tabular}{c cr ccr}
\toprule
 & \multicolumn{2}{c}{Two-parameter polynomial fit} & \multicolumn{3}{c}{Three-parameter polynomial fit} \\
 & $\tilde{a}^{-1}$ [l.u.] & \multicolumn{1}{c}{$\tilde{r}$ [l.u.]} & $\tilde{a}^{-1}$ [l.u.] & $\tilde{r}$ [l.u.] & \multicolumn{1}{c}{$\tilde{P}$ [l.u.]} \\
\cmidrule(r){2-3}\cmidrule(l){4-6}
$\Sigma\Sigma\;(\1s0)$	&	$0.038_{(-16)(-05)}^{(+12)(+09)}$	&	$6.2_{(-2.6)(-0.7)}^{(+2.7)(+0.8)}$	&	$0.044_{(-12)(-07)}^{(+08)(+11)}$	&	$10.9_{(-2.6)(-0.6)}^{(+1.3)(+2.5)}$	&	$331_{(-120)(-80)}^{(+100)(+98)}$	\\
$\Xi\Sigma\;(\1s0)$	&	$0.043_{(-10)(-05)}^{(+08)(+07)}$	&	$6.3_{(-1.1)(-1.3)}^{(+1.7)(+0.5)}$	&	$0.052_{(-11)(-07)}^{(+11)(+07)}$	&	$7.7_{(-2.4)(-1.0)}^{(+1.6)(+1.8)}$	&	$173_{(-46)(-37)}^{(+43)(+25)}$	\\
$\Xi\Xi\;(\1s0)$	&	$0.047_{(-07)(-03)}^{(+03)(+08)}$	&	$6.9_{(-0.3)(-0.9)}^{(+0.9)(+0.5)}$	&	$0.053_{(-06)(-04)}^{(+03)(+06)}$	&	$8.9_{(-0.9)(-1.0)}^{(+0.7)(+1.3)}$	&	$149_{(-23)(-28)}^{(+23)(+31)}$	\\
$NN\;(\3s1)$	&	$0.038_{(-16)(-07)}^{(+12)(+07)}$	&	$7.2_{(-1.9)(-1.3)}^{(+2.3)(+1.0)}$	&	$0.051_{(-13)(-08)}^{(+12)(+09)}$	&	$8.3_{(-3.1)(-2.4)}^{(+2.2)(+2.0)}$	&	$265_{(-66)(-72)}^{(+89)(+62)}$	\\
$\Sigma N\;(\3s1)$	&	$0.073_{(-20)(-21)}^{(+22)(+16)}$	&	$3.5_{(-1.2)(-0.8)}^{(+1.2)(+0.9)}$	&	$0.085_{(-39)(-19)}^{(+23)(+31)}$	&	$5.2_{(-5.6)(-3.2)}^{(+2.9)(+4.3)}$	&	$-8_{(-14)(-15)}^{(+27)(+16)}$	\\
$\Xi\Xi\;(\3s1)$	&	$0.20_{(-09)(-18)}^{(+17)(+14)}$	&	$-2.6_{(-2.9)(-6.1)}^{(+4.3)(+1.0)}$	&	$0.25_{(-14)(-13)}^{(+22)(+29)}$	&	$1_{(-15)(-10)}^{(+14)(+22)}$	&	$-19_{(-54)(-88)}^{(+87)(+62)}$	\\
$\Xi N\;(\3s1)$	&	$0.059_{(-01)(-05)}^{(+05)(+02)}$	&	$6.9_{(-0.3)(-0.3)}^{(+0.2)(+0.4)}$	&	$0.066_{(-04)(-03)}^{(+02)(+04)}$	&	$7.1_{(-0.3)(-0.5)}^{(+0.5)(+0.3)}$	&	$36_{(-11)(-04)}^{(+08)(+08)}$	\\\bottomrule
\end{tabular}}
\end{table}

\begin{figure}[t!]
\includegraphics[width=\textwidth]{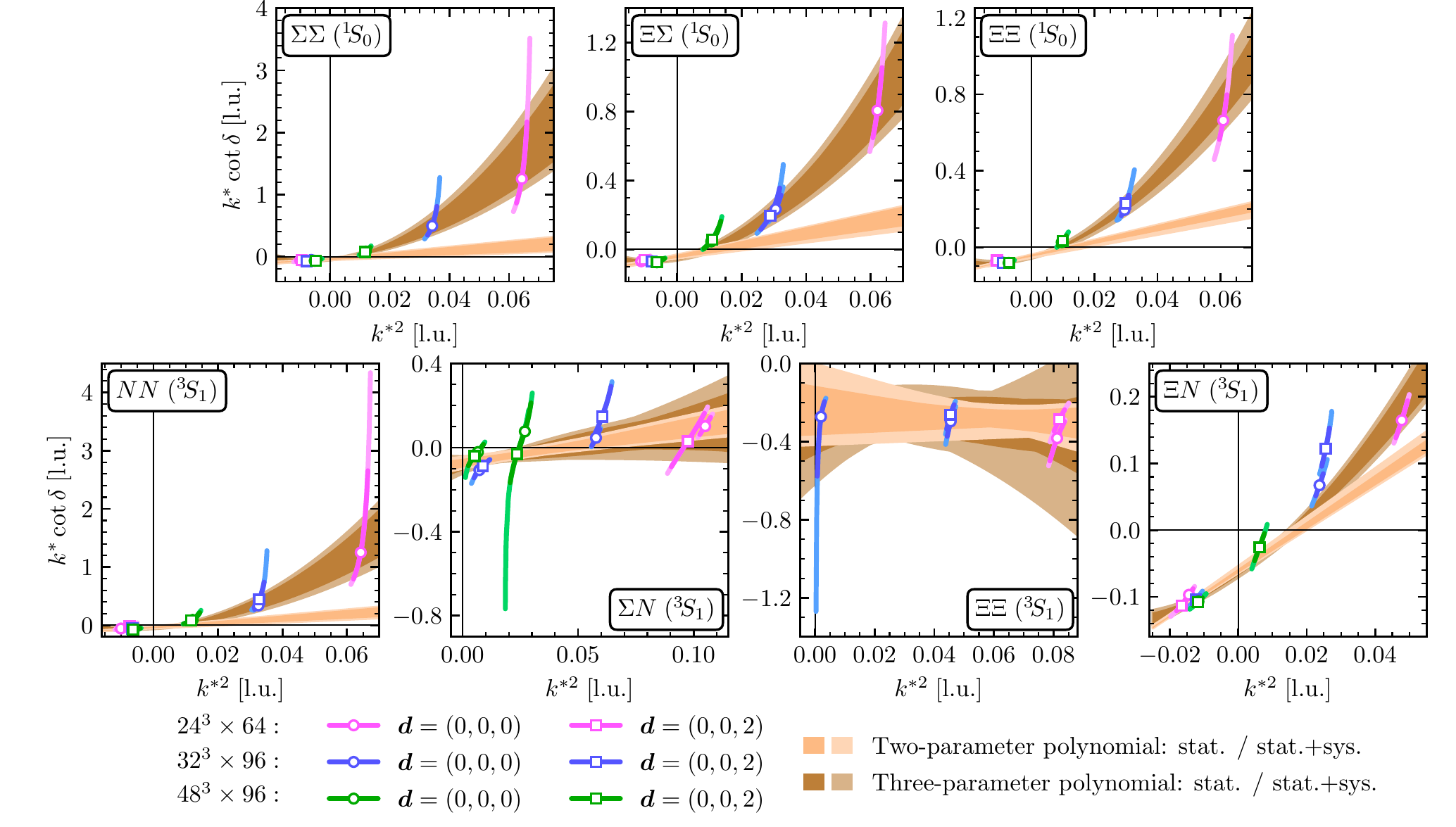}
\caption{$k^*\cot \delta$ values as a function of the c.m.\ momenta $k^{*2}$, along with the bands representing the two- and three-parameter polynomial fits for two-baryon systems under the assumption that there is a smooth and monotonic behavior in $k^*\cot\delta$ as a function of $k^{*2}$ beyond the $t$-channel cut. Quantities are expressed in lattice units.}
\label{fig:ere-fit-beyond}
\end{figure}

Although the ERE is only valid below the $t$-channel cut, one may still fit the $k^*\cot\delta$ values beyond this threshold using a similar polynomial form as the ERE in Eq.~\eqref{eq:ERE}.
To distinguish the fit parameters of this model from those obtained from the ERE, two- and three-parameter polynomials are characterized by two $\{\tilde{a}^{-1},\tilde{r}\}$ and three $\{\tilde{a}^{-1},\tilde{r},\tilde{P}\}$ parameters. Such forms are motivated by the fact that in most channels, $k^*\cot\delta$ values as a function of $k^{*2}$ exhibit smooth and monotonic behavior beyond the $t$-channel cut, as seen in Fig.~\ref{fig:kcotdelta}. The only exceptions are the spin-singlet $NN$ and $\Sigma N$ channels, for which such a polynomial fit will not be performed.
The results of this fit, using the same strategy as described above for ERE fits, are shown in Table~\ref{tab:scattPar-beyond} and in Fig.~\ref{fig:ere-fit-beyond}. In the next section, the EFTs and approximate symmetries of the interactions will be utilized to make predictions for the inverse scattering length in channels for which ERE fits could not be performed, i.e., $\Sigma N\;(\3s1)$ and $\Xi\Xi\;(\3s1)$ channels, and in those cases, the scattering length is found consistent with the $\tilde{a}^{-1}$ values obtained from this model analysis. It should be emphasized that such a polynomial fit beyond the $t$-channel cut is only one out of many applicable parametrizations of the amplitude, and a systematic uncertainty associated with multiple model choices and model-selection criteria needs to be assigned to reliably constrain the energy dependence of the amplitude at higher energies.
More precise LQCD results may be required to identify non-polynomial behavior in $k^{*2}$. This is analogous to the efforts to uniquely identify non-analytic terms in chiral expansions, such as in $\pi$-$\pi$ scattering, where very high-precision calculations are required to reveal the logarithmic dependence on $m_\pi$, e.g., see Ref.~\cite{Beane:2007xs}.

\subsection{Binding energies}\label{subsec:bindingLQCD}

A negative shift in the energy of two baryons in a finite volume compared with that of the non-interacting baryons may signal the presence of a bound state in the infinite-volume limit. However, to conclusively discern a bound state from a scattering state, a careful inspection of the volume dependence of the energies is required.
As discussed in~\cref{subsec:QC_binding}, Lüscher's QC can be used to identify the volume dependence of bound-state energies. Alternatively, LQCD eigenenergies in a finite volume can be matched to an EFT description of the system in the same volume to constrain the interactions. The constrained EFT can then be used to obtain the infinite-volume binding energy, e.g., see Ref.~\cite{Eliyahu:2019nkz,Detmold:2021oro}. This approach is more easily applicable to the multi-baryon sector; however, it relies on the validity of the EFT that is used.

Expanding Eq.~\eqref{eq:mom_binding_expression} for the boost vectors $\bm{d}$ used in this work, it reads,
\begin{align}
    \bm{d}=(0,0,0) \; : \; |k^*|=\kappa^{(\infty)}+\frac{Z^2}{L}&\left[ 6e^{-\kappa^{(\infty)}L}+6\sqrt{2}e^{-\sqrt{2}\kappa^{(\infty)}L}+\frac{8}{\sqrt{3}}e^{-\sqrt{3}\kappa^{(\infty)}L} \right] \, , \label{eq:kinf1}\\
    \bm{d}=(0,0,2) \; : \; |k^*|=\kappa^{(\infty)}+\frac{Z^2}{L}&\left[\, 4e^{\kappa^{(\infty)}L}+\frac{2}{\gamma}\cos(4\pi\alpha)e^{-\gamma\kappa^{(\infty)}L} \right. \nonumber\\
    &\;+2\sqrt{2}e^{-\sqrt{2}\kappa^{(\infty)}L}+\frac{8}{\sqrt{\gamma^2+1}}\cos(4\pi\alpha)e^{-\sqrt{\gamma^2+1}\kappa^{(\infty)}L} \nonumber\\
    &\;\left.+\frac{8}{\sqrt{\gamma^2+2}}\cos(4\pi\alpha)e^{-\sqrt{\gamma^2+2}\kappa^{(\infty)}L}\right]\, . \label{eq:kinf2}
\end{align}
The values of $\gamma$ deviate from one at the percent level.\footnote{The largest value of $\gamma$ is found in the $NN \; (\1s0)$ system with $L=24$, where $\gamma \sim 1.015$.}
Therefore, all systems considered are non-relativistic to a good approximation. Only the first few terms in the sum in Eq.~\eqref{eq:mom_binding_expression}, corresponding to $|\bm{m}|\in\{0,1,\sqrt{2}\}$, are considered in the volume extrapolation performed below, with corrections that scale as $\mathcal{O}(e^{-2\kappa^{(\infty)}L})$.

Alternatively, as mentioned in~\cref{subsec:QC_binding}, one can compute $\kappa^{(\infty)}$ by finding the pole location in the $S$-wave scattering amplitude,
\begin{equation}
    \left. k^* \cot\delta \right|_{k^*=i \kappa^{(\infty)}}+\kappa^{(\infty)}=0\, .
\label{eq:kinf3}
\end{equation}
To obtain $\kappa^{(\infty)}$, the scattering amplitude has to first be constrained using Lüscher's QC as discussed in the previous subsection, and then be expressed in terms of an ERE expansion.

Results for the infinite-volume binding momenta $\kappa^{(\infty)}$ are shown in Table~\ref{tab:binding_momenta}. The columns labeled as $\bm{d}=(0,0,0)$ and $\bm{d}=(0,0,2)$ correspond, respectively, to fitting separately the values of $k^*$ with no boost, or with boost $\bm{d}=(0,0,2)$, using Eqs.~\eqref{eq:kinf1} or~\eqref{eq:kinf2}.
The column labeled as $\bm{d}=\{(0,0,0),(0,0,2)\}$ is the result of fitting both sets of $k^*$ values simultaneously, i.e., imposing the same value for $\kappa^{(\infty)}$ and $Z^2$ in both fits.
The last column shows the $\kappa^{(\infty)}$ values obtained using Eq.~\eqref{eq:kinf3}, with the parameters listed in Table~\ref{tab:scattPar} as obtained with a two-parameter ERE fit to $k^*\cot\delta$.
The results obtained with the different extractions of $\kappa^{\infty}$ are seen to be consistent with each other within uncertainties. The largest difference observed is in the $\Xi\Xi\;(\1s0)$ channel, with a difference between the volume-extrapolation and pole-location results of around $1.5\sigma$. The agreement between the two approaches suggests that the higher-order terms neglected in the sum in Eq.~\eqref{eq:mom_binding_expression} are not significant.

\begin{table}[t!]
\centering
\caption{The infinite-volume binding momenta $\kappa^{(\infty)}$ for bound states obtained either by using the extrapolation in Eqs.~\eqref{eq:kinf1} and~\eqref{eq:kinf2} or from the pole location of the scattering amplitude as in Eq.~\eqref{eq:kinf3}. Quantities are expressed in lattice units.}
\label{tab:binding_momenta}
\renewcommand{\arraystretch}{1.5}
\begin{tabular}{ccccc}
\toprule
 & \multicolumn{4}{c}{$\kappa^{(\infty)}$ [l.u.]} \\
 & $\bm{d}=(0,0,0)$ & $\bm{d}=(0,0,2)$ & $\bm{d}=\{(0,0,0),(0,0,2)\}$ & $\left.-k^*\cot\delta\right|_{k^*=i\kappa^{(\infty)}}$ \\\cmidrule(r){2-4}\cmidrule(l){5-5}
$NN\;(\1s0)$	&	$0.077_{(-11)(-04)}^{(+08)(+06)}$	&	$0.072_{(-14)(-16)}^{(+10)(+08)}$	&	$0.075_{(-10)(-01)}^{(+05)(+06)}$	&	$0.076_{(-28)(-32)}^{(+06)(+12)}$	\\
$\Sigma N\;(\1s0)$	&	$0.073_{(-16)(-13)}^{(+13)(+05)}$	&	$0.083_{(-09)(-13)}^{(+09)(+06)}$	&	$0.080_{(-09)(-09)}^{(+08)(+02)}$	&	$0.072_{(-14)(-24)}^{(+12)(+09)}$	\\
$\Sigma\Sigma\;(\1s0)$	&	$0.068_{(-10)(-11)}^{(+08)(+08)}$	&	$0.072_{(-10)(-16)}^{(+11)(+07)}$	&	$0.069_{(-07)(-09)}^{(+07)(+06)}$	&	$0.047_{(-15)(-17)}^{(+15)(+07)}$	\\
$\Xi\Sigma\;(\1s0)$	&	$0.078_{(-09)(-09)}^{(+08)(+06)}$	&	$0.080_{(-08)(-11)}^{(+08)(+05)}$	&	$0.079_{(-05)(-07)}^{(+06)(+04)}$	&	$0.066_{(-14)(-14)}^{(+10)(+05)}$	\\
$\Xi\Xi\;(\1s0)$	&	$0.086_{(-05)(-06)}^{(+05)(+05)}$	&	$0.086_{(-05)(-09)}^{(+06)(+06)}$	&	$0.086_{(-03)(-05)}^{(+04)(+04)}$	&	$0.069_{(-08)(-09)}^{(+05)(+08)}$	\\
$NN\;(\3s1)$	&	$0.072_{(-11)(-08)}^{(+08)(+06)}$	&	$0.076_{(-09)(-08)}^{(+08)(+03)}$	&	$0.074_{(-07)(-05)}^{(+08)(+04)}$	&	$0.064_{(-20)(-08)}^{(+10)(+08)}$	\\
$\Xi N\;(\3s1)$	&	$0.108_{(-04)(-08)}^{(+04)(+06)}$	&	$0.106_{(-04)(-08)}^{(+05)(+06)}$	&	$0.107_{(-03)(-05)}^{(+03)(+05)}$	&	$0.101_{(-05)(-09)}^{(+05)(+06)}$	\\\bottomrule
\end{tabular}
\end{table}

\begin{table}[t!]
\centering
\caption{Binding energies for bound states in MeV. The values are obtained using $\kappa^{(\infty)}$ from the volume-extrapolation method with a combined fit to $\bm{d}=(0,0,0)$ and $\bm{d}=(0,0,2)$ data. The uncertainty from scale setting is an order of magnitude smaller than the statistical and systematic uncertainties quoted.}
\label{tab:binding_energy}
\renewcommand{\arraystretch}{1.2}
\resizebox{\columnwidth}{!}{
\begin{tabular}{ccccccc}
\toprule
\multicolumn{7}{c}{$B$ [MeV]} \\
$NN\;(\1s0)$	&	$\Sigma N\;(\1s0)$	&	$\Sigma\Sigma\;(\1s0)$	&	$\Xi\Sigma\;(\1s0)$	&	$\Xi\Xi\;(\1s0)$	&	$NN\;(\3s1)$	&	$\Xi N\;(\3s1)$	\\ \midrule
$13.1_{(-3.1)(-0.4)}^{(+2.0)(+2.3)}$	&	$14.3_{(-3.0)(-2.8)}^{(+3.1)(+0.9)}$	&	$10.2_{(-1.9)(-2.3)}^{(+2.1)(+2.0)}$	&	$12.8_{(-1.6)(-2.2)}^{(+2.1)(+1.6)}$	&	$14.9_{(-1.0)(-1.8)}^{(+1.5)(+1.4)}$	&	$12.7_{(-2.4)(-1.7)}^{(+2.4)(+1.5)}$	&	$25.3_{(-1.5)(-2.2)}^{(+1.5)(+2.2)}$	\\\bottomrule
\end{tabular}}
\end{table}

The binding energy, $B$, is defined in terms of the infinite-volume baryon masses and binding momenta as
\begin{equation}
B=M^{(\infty)}_1+M^{(\infty)}_2-\sqrt{M^{(\infty)2}_1-\kappa^{(\infty)2}}-\sqrt{M^{(\infty)2}_2-\kappa^{(\infty)2}}\, ,
\label{eq:bind_en}
\end{equation}
where $M_i^{(\infty)}$ is the infinite-volume mass of baryon $i$ obtained from Eq.~\eqref{eq:mass_extrap}. This quantity is computed for all systems that exhibit a negative c.m.\ momentum squared in the infinite-volume limit, i.e., those listed in Table~\ref{tab:binding_momenta}.
The binding energies in physical units are listed for these systems in Table~\ref{tab:binding_energy}. The binding energies of the two-nucleon systems computed here are consistent within $1 \sigma$ with the values published previously in Ref.~\cite{Orginos:2015aya} using the same LQCD correlation functions.
The same two-baryon systems studied here were also studied at $m_{\pi}\sim 806$ MeV in Ref.~\cite{Wagman:2017tmp} and were found to be bound albeit with larger binding energies. While the results at $m_{\pi} \sim 806$ MeV were inconclusive regarding the presence of bound states in the ${\bf 10}$ irrep, the $\Sigma N \; (\3s1)$ and $\Xi \Xi \; (\3s1)$ systems are found to be unbound at $m_{\pi} \sim 450$ MeV.
The results obtained in the present work can be combined with those of Ref.~\cite{Wagman:2017tmp} to perform a preliminary extrapolation of the binding energies to the physical pion mass.\footnote{The results that can be found in the literature for the binding energies of two-baryon systems obtained at larger-than-physical quark masses must be compared with the results of the current work with caution, as the use of different scale setting schemes makes a comparison in physical units meaningless, unless the physical limit of the quantities are taken, as can be seen in Ref.~\cite{Green:2021qol}.}
This enables a postdiction of binding energies in nature in cases where there are experimental data, and a prediction for the presence of bound states and their binding in cases where no experimental information is available.

At the beginning of this chapter, we mentioned that for systems with non-zero strangeness, experimental knowledge is notably limited in comparison to the nucleon-nucleon sector, and almost all phenomenological predictions are based on $SU(3)$ flavor-symmetry assumptions. For the systems analyzed here, the predictions from all these models are:
\begin{itemize}[label={--}]
\item The $\1s0$ and $\3s1$ $\Sigma N$ channels do not exhibit bound states. The spin-singlet state behaves in a similar way to $NN\; (\1s0)$, and the interactions are slightly attractive, while those in the spin-triplet channel are found to be repulsive.
\item For the $\Xi N\;(\3s1)$ system, almost all the models find that the interactions are slightly attractive, but only a few exhibit a bound state.\footnote{Since the binding energies are not explicitly computed in these references and only the $S$-wave scattering parameters are reported, binding energies are computed here using Eqs.~\eqref{eq:kinf3} and \eqref{eq:bind_en}, assuming a two-parameter ERE for $k^*\cot\delta$. These are marked with the symbol $^{\ddagger}$.}
Among the most recent calculations, the ESC08a model~\cite{Rijken:2010zzb} gives $B=0.9$ MeV$^{\ddagger}$, while the ESC08c1~\cite{Rijken:2013wxa} gives $B=0.5$ MeV$^{\ddagger}$. There is only one LQCD calculation of this system near the physical values of the quark masses, performed by the HAL QCD Collaboration~\cite{Sasaki:2019qnh}, and no bound state is observed.
\item The NSC97 model~\cite{Stoks:1999bz} finds a bound state for the $\Sigma \Sigma \;(\1s0)$ channel, with binding energies ranging from $1.53$ to $3.17$ MeV. $\chi$EFT at NLO~\cite{Haidenbauer:2014rna} finds a binding energy between $0$ and $0.01$ MeV (no bound state is found with ESC or quark models in this channel).
\item The $\Xi \Sigma \;(\1s0)$ system is found to be bound in the NSC97 model~\cite{Stoks:1999bz}, with a binding energy between $3.02$ and $16.5$ MeV, and by $\chi$EFT~\cite{Haidenbauer:2009qn,Haidenbauer:2014rna}, with a binding energy between $2.23$ and $6.18$ MeV at LO and $0.19$ and $0.58$ MeV at NLO. With the quark model fss2~\cite{Fujiwara:2006yh}, although the interaction in this system is found to be attractive, no bound state is predicted (similar to the ESC08c1 model~\cite{Rijken:2013wxa}).
\item Meson-exchange potentials give different results for the $\Xi \Xi \;(\1s0)$ system, depending on the model. The NSC97 model~\cite{Stoks:1999bz} predicts there is a bound state with a binding energy between $0.1$ and $15.8$ MeV, and with Ehime~\cite{Yamaguchi:2001ip} between $0.23$ and $0.71$ MeV, but no bound state is found with the ESC08c1 model~\cite{Rijken:2013wxa}. $\chi$EFT~\cite{Haidenbauer:2009qn,Haidenbauer:2014rna} also finds this state to be bound with a binding energy of $2.56-7.27$ MeV at LO and $0.40-1.00$ MeV at NLO. The quark model fss2~\cite{Fujiwara:2006yh} does not find a bound state. In the $\Xi \Xi \;(\3s1)$ channel, no bound state is found with meson-exchange potentials, except for Ehime that finds a deeply bound state with a binding energy of $9-15$ MeV. fss2~\cite{Fujiwara:2006yh} finds this channel to be repulsive.
\end{itemize}

The quark-mass dependences of multi-baryon spectra have not been studied extensively in the literature. For the octet-baryon masses, it was found that LQCD calculations performed with $2+1$ dynamical fermions are consistent with a linear dependence on the pion mass at unphysical values of the quark masses, compared to the quadratic dependence of the HB$\chi$PT prediction at LO~\cite{WalkerLoud:2008bp,WalkerLoud:2008pj,Walker-Loud:2014iea}.
Nonetheless, recent precision studies near the physical values of the quark masses appear to be more consistent with chiral predictions~\cite{Borsanyi:2020bpd}.
In the two-baryon sector the situation is more complicated. On the theoretical side, $\chi$EFT was used in Ref.~\cite{Haidenbauer:2011za} to extrapolate LQCD results to the physical point, assuming no dependence on the light quark masses for the LECs of the EFT (at a fixed order).
The same premise was taken in Ref.~\cite{Beane:2012ey} to determine the $I=3/2$ $\Sigma N$ interaction at LO, which was used to address the possible appearance of $\Sigma^-$ hyperons in dense nuclear matter.
In the absence of a conclusive form for the quark-mass extrapolation of two-baryon binding energies, two naive expressions with linear and quadratic $m_{\pi}$ dependence were used in Ref.~\cite{Beane:2011zpa} to extrapolate the binding energy of the $H$-dibaryon to its physical value, but no definitive statement concerning its nature could be made due to the large uncertainties.
In Refs.~\cite{Shanahan:2011su, Thomas:2011cg, Shanahan:2013yta}, under the assumption that the $H$-dibaryon is a compact 6 valence-quark state (and not a two-baryon molecule), $\chi$EFT was used to extrapolate the binding energies, resulting in an unbound state.

\begin{table}[t!]
\centering
\caption{Extrapolated binding energies (in MeV) at the physical quark masses for bound states using two different forms, linear and quadratic in $m_{\pi}$.}
\label{tab:extrap_bind}
\renewcommand{\arraystretch}{1.5}
\resizebox{\columnwidth}{!}{
\begin{tabular}{crrrrrrr}
\toprule
&	\multicolumn{1}{c}{$NN\;(\1s0)$}	&	\multicolumn{1}{c}{$\Sigma N\;(\1s0)$}	&	\multicolumn{1}{c}{$\Sigma\Sigma\;(\1s0)$}	&	\multicolumn{1}{c}{$\Xi\Sigma\;(\1s0)$}	&	\multicolumn{1}{c}{$\Xi\Xi\;(\1s0)$}	&	\multicolumn{1}{c}{$NN\;(\3s1)$}	&	\multicolumn{1}{c}{$\Xi N\;(\3s1)$}	\\ \cmidrule{2-8}
$B_{\rm lin}(m_{\pi}^{\rm phys})$	&	$6.4_{(-6.5)}^{(+6.3)}$	&	$8.4_{(-6.6)}^{(+7.8)}$	&	$1.0_{(-6.1)}^{(+6.1)}$	&	$5.9_{(-5.8)}^{(+5.7)}$	&	$9.6_{(-4.7)}^{(+4.5)}$	&	$-0.9_{(-6.1)}^{(+6.1)}$	&	$11.7_{(-6.2)}^{(+5.4)}$	\\
$B_{\rm quad}(m_{\pi}^{\rm phys})$	&	$9.9_{(-4.5)}^{(+4.6)}$	&	$11.5_{(-4.8)}^{(+5.7)}$	&	$5.8_{(-4.3)}^{(+4.2)}$	&	$9.5_{(-4.0)}^{(+3.8)}$	&	$12.4_{(-3.1)}^{(+3.0)}$	&	$6.3_{(-4.4)}^{(+4.3)}$	&	$18.9_{(-4.1)}^{(+3.8)}$	\\\bottomrule
\end{tabular}}
\end{table}

\begin{figure}[tb!]
\includegraphics[width=\textwidth]{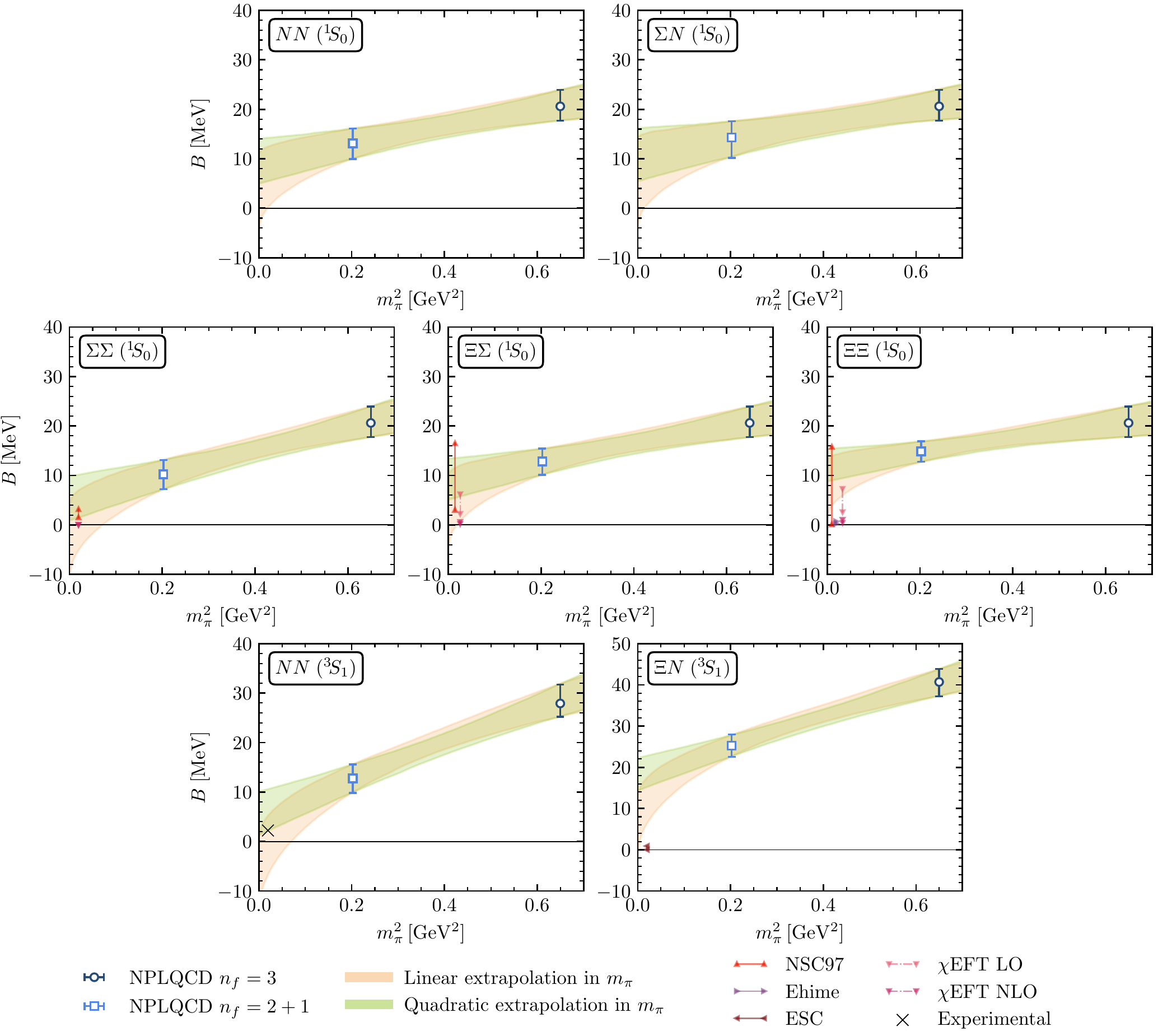}
\caption{Extrapolation of the binding energies of different two-baryon systems, using the results obtained in this work and those at $m_{\pi}\sim 806$ MeV from Ref.~\cite{Wagman:2017tmp}. For comparison, the results with values obtained using one-boson-exchange models or $\chi$EFTs are also shown (and where needed, are shifted slightly in the horizontal direction for clarity).}
\label{fig:extrap_bind}
\end{figure}

Two analytical forms with different $m_{\pi}$ dependence are used here to obtain the binding energies at the physical light-quark masses, using the results presented in Ref.~\cite{Wagman:2017tmp} at $m_\pi \sim 806$ MeV and those listed in Table~\ref{tab:binding_energy} for $m_\pi \sim 450$ MeV:
\begin{align}
    B_{\rm lin}(m_{\pi})&=B^{(0)}_{\rm lin}+B^{(1)}_{\rm lin}\, m_{\pi}\, , \label{eq:extrapLin}\\
    B_{\rm quad}(m_{\pi})&=B^{(0)}_{\rm quad}+B^{(1)}_{\rm quad} \, m^2_{\pi}\, ,\label{eq:extrapQuad}
\end{align}
where $B^{(0)}_{\rm lin}$, $B^{(1)}_{\rm lin}$, $B^{(0)}_{\rm quad}$, and $B^{(1)}_{\rm quad}$ are parameters to be constrained by fits to data. These fits are shown in Fig.~\ref{fig:extrap_bind}, along with the experimental values and predictions at the physical point. The binding energies extrapolated to the physical point, i.e., $B_{\rm lin}(m_{\pi}^{\rm phys})$ and $B_{\rm quad}(m_{\pi}^{\rm phys})$, are summarized in Table~\ref{tab:extrap_bind}.\footnote{Performing fits to dimensionless ratios of the binding energies to the baryon masses (to minimize the effects of non-zero lattice spacing) do not change the qualitative conclusions presented in the text (results agree within $1\sigma$).} 
It should be emphasized that given the lack of knowledge of the quark-mass dependence of binding energies, the results of the extrapolations shown should be treated with caution and are not intended to draw definitive conclusions about the value of the energies, but only to show an emerging trend towards the physical point

These extrapolations highlight some interesting features. The values obtained at the physical point are consistent with the experimental values for the $NN$ channels. For the other channels, the predictions show the same level of precision as the phenomenological results.
The $\Xi \Xi \; (\1s0)$ and $\Xi N \; (\3s1)$ channels are more consistent with being bound than the other channels, using both extrapolation functions. Moreover, the $\Sigma N \; (\3s1)$ channel was found not to support a bound state in this study, a conclusion that is in agreement with phenomenological models.
The same conclusion holds for $\Xi \Xi \; (\3s1)$, noting that only in one model, namely Ehime, a different conclusion is reached~\cite{Yamaguchi:2001ip}. The spread of results and some contradictory conclusions in the model calculations motivate the need for LQCD studies of these states at near-physical values of the quark masses in the upcoming years. 

\subsection{Validity of the extraction of the lowest-lying energies and the corresponding scattering amplitudes}\label{subsec:validityLQCD}

In Refs.~\cite{Iritani:2017rlk,Iritani:2018talk}, several criteria were presented to validate studies of two-baryon systems that rely on the extraction of finite-volume energies from Euclidean LQCD correlation functions to be used in Lüscher's formalism. The results of the present work are examined and validated with regard to these criteria. Similar investigations were performed in Refs.~\cite{Wagman:2017tmp,Beane:2017edf} within study performed by the NPLQCD Collaboration at $m_{\pi} \sim 806$ MeV.

\begin{figure}[b!]
\includegraphics[width=\textwidth]{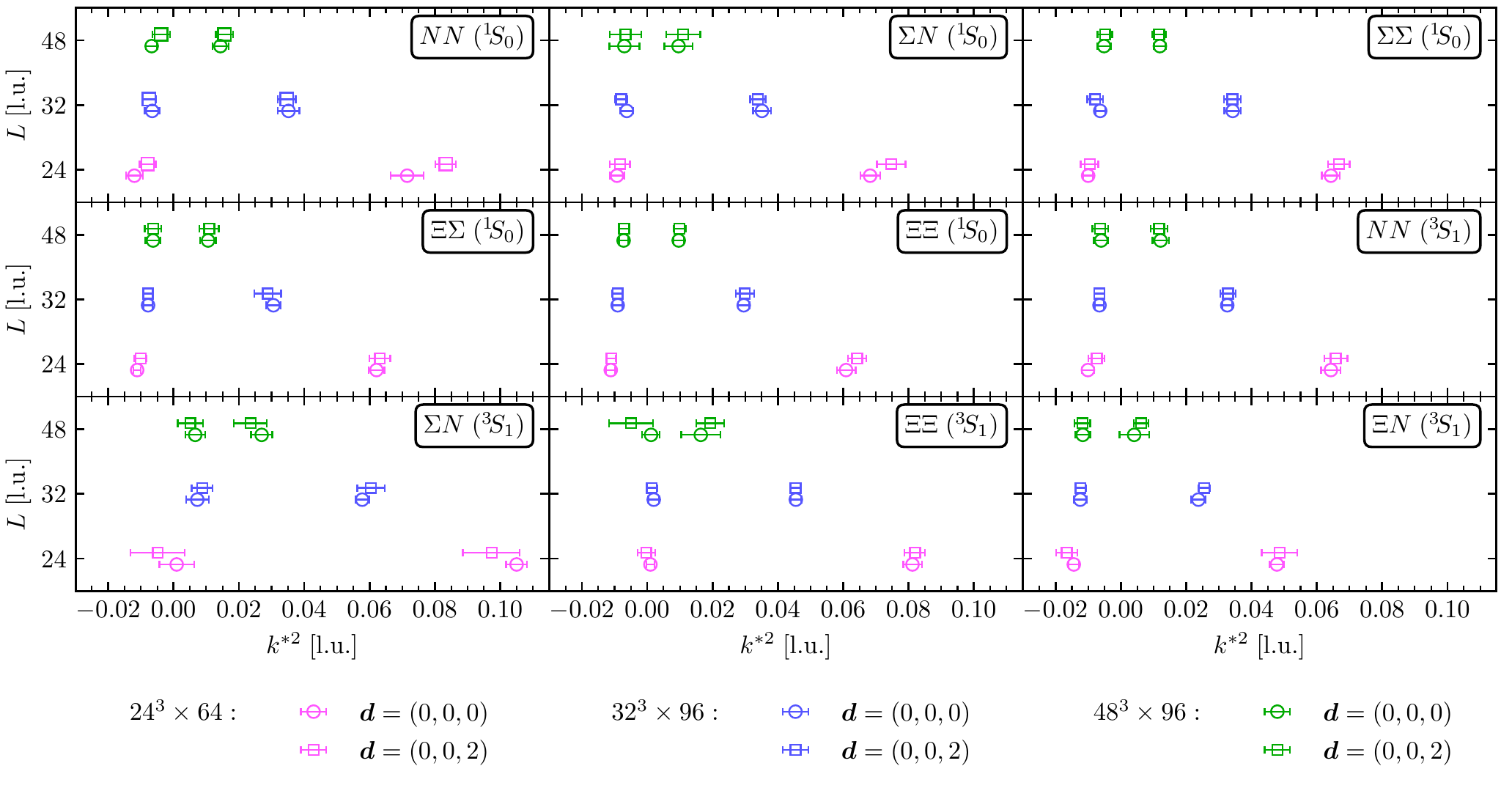}
\caption{The values of $k^{*2}$ for all systems analyzed in this work, using three different lattice volumes and different boost vectors. Quantities are expressed in lattice units.}
\label{fig:checks_k2}
\end{figure}

\begin{checks}
\item \textit{Interpolating-operator independence}: The two different source-sink operator structures used in the present work, denoted SP and SS, yield the same energies for both the ground and the first excited states. This consistency can be verified by examining the late-time behavior of the effective energy and effective energy-shift functions, shown in Figs.~\ref{fig:B1_EMP}-\ref{fig:XN3s1_EMP}, constructed from the SS and SP correlation functions. Moreover, the c.m.\ momenta $k^{*2}$ obtained from the correlation functions with $\bm{d}=(0,0,0)$ and $\bm{d}=(0,0,2)$ must be consistent, up to negligible relativistic and small $\mathcal{O}\left((m_1^2-m_2^2)/E^{*2}\right)$ corrections~\cite{Davoudi:2011md}, a feature that is observed in the results presented here, as shown in Fig.~\ref{fig:checks_k2}. The largest difference is seen in the $NN~(\1s0)$ channel for the $n=2$ level on the ensemble with $L=24$, for which the c.m.\ momenta in the unboosted and boosted cases exhibit a $\sim 2 \sigma$ difference. 

\begin{figure}[t]
\includegraphics[width=\textwidth]{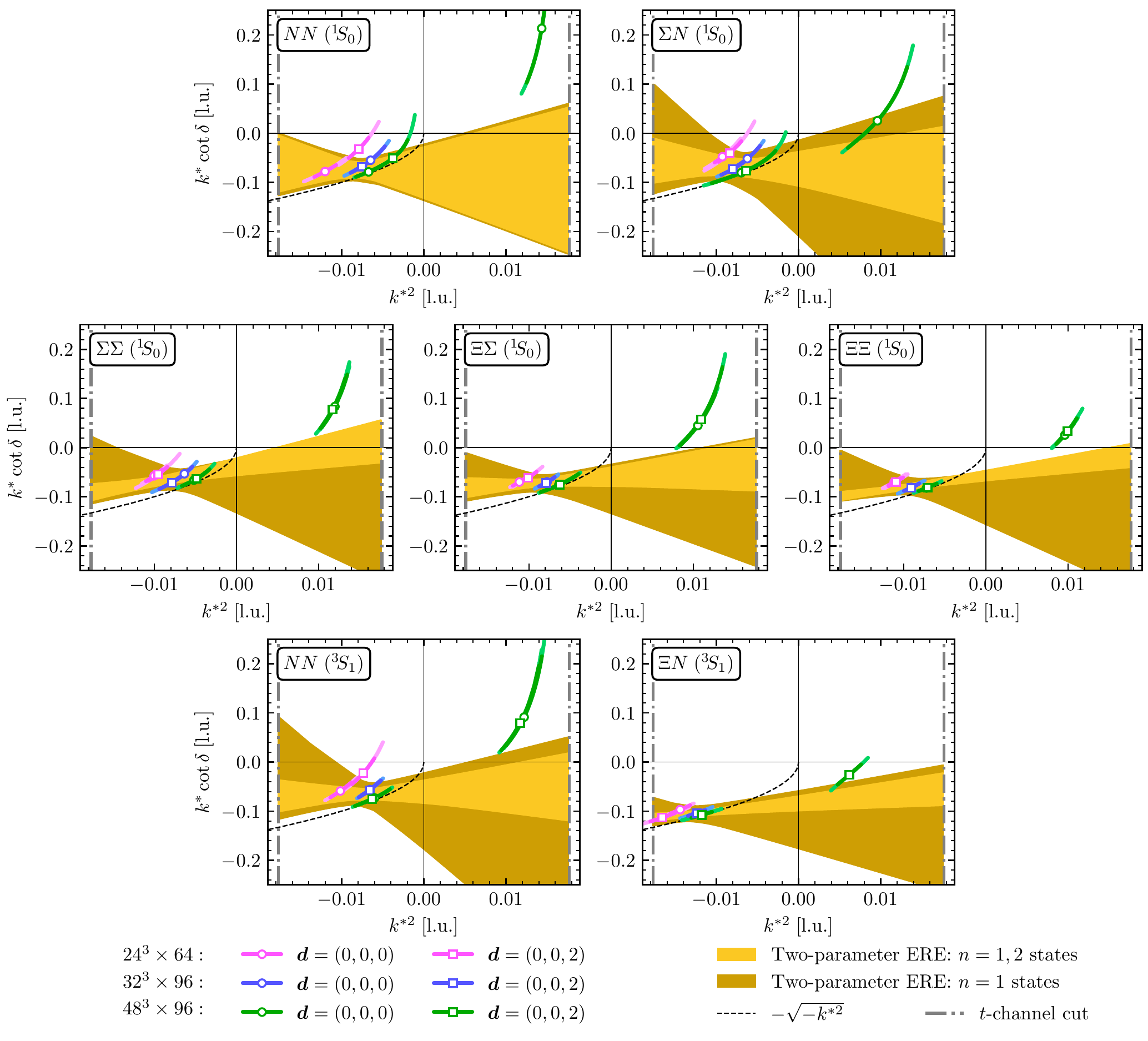}
\caption{$k^{*2}\cot\delta$ values as a function of the c.m.\ momenta $k^{*2}$, together with bands representing the two-parameter ERE using all the energy levels (ground state $n=1$ and excited states $n=2$) in lighter yellow, or using just the ground state in darker yellow. Quantities are expressed in lattice units.}
\label{fig:checks_n12}
\end{figure}

\item \textit{Consistency between ERE parameters for $k^{*2}<0$ and $k^{*2}>0$}: In the two-baryon channels studied in this work, there are not sufficient data points for $k^*\cot\delta$ below the $t$-channel cut to extract precise scattering parameters, as pointed out in~\cref{subsec:scattparamLQCD}. Nonetheless, in those cases where two sets of data at positive and negative values of $k^{*2}$ are available, the ERE fits obtained by either fitting to all $k^{*2}$ or only to $k^{*2}<0$ values are fully consistent with each other, as shown in Fig.~\ref{fig:checks_n12}.

\item \textit{Non-singular scattering parameters}: None of the scattering parameters extracted show singular behavior, as can be seen from the values in Table~\ref{tab:scattPar}.

\item \textit{Requirement on the residue for the scattering amplitude at the bound-state pole}: In order to support a physical bound state, the slope of the ERE as a function of $k^{*2}$ must be smaller than the slope of the $-\sqrt{-k^{*2}}$ at the bound-state pole. This can be understood with the following argument. Near this bound-state pole, the $\mathcal{S}$ matrix can be represented in the form~\cite{Sitenko:102667}
\begin{equation}
    \mathcal{S}(k^*\sim \imag\kappa^{(\infty)})\simeq\frac{-\imag\beta^2_b}{k^*-\imag\kappa^{(\infty)}}\, ,
\label{eq:Spole}
\end{equation}
where $\beta^2_b$ is real and positive for physical poles. Also, as we have seen in~\cref{subsec:LuscherQCD}, the $\mathcal{S}$ matrix can be written as
\begin{equation}
    \mathcal{S}(k^*)-1=e^{2\imag\delta}-1=\frac{2\imag k^*}{k^*\cot\delta-\imag k^*}\; \rightarrow	 \; k^* \cot\delta = \imag k^* \frac{\mathcal{S}(k^*)+1}{\mathcal{S}(k^*)-1}\, .
\label{eq:Sgen}
\end{equation}
Replacing Eq.~\eqref{eq:Spole} into Eq.~\eqref{eq:Sgen}, we get
\begin{equation}
    k^*\cot\delta\simeq \imag k^* \frac{-\imag\beta^2_b+k^*-\imag\kappa^{(\infty)}}{-\imag\beta^2_b-k^*+\imag\kappa^{(\infty)}}\, ,
\label{eq:Spole2}
\end{equation}
which allows us to put some constrains on the slope of the ERE near $\kappa^{(\infty)}$,
\begin{equation}
    \frac{d}{dk^{*2}}\left[k^*\cot\delta-(-\sqrt{-k^{*2}})\right]_{k^{*2}=-\kappa^{(\infty)2}}\, .
\end{equation}
By developing these derivatives, we get
\begin{align}
    \frac{d}{dk^{*2}}\left[k^*\cot\delta\right]_{k^{*2}=-\kappa^{(\infty)2}} &= \frac{dk^*}{dk^{*2}}\frac{d}{dk^*}\left[k^*\cot\delta\right]_{k^*=\imag\kappa^{(\infty)}}\nonumber\\
    &= \frac{1}{2\imag\kappa^{(\infty)}}\frac{d}{dk^*}\left[\imag k^*\frac{-\imag\beta^2_b+k^*-\imag\kappa^{(\infty)}}{-\imag\beta^2_b-k^*+\imag\kappa^{(\infty)}}\right]_{k^*=\imag\kappa^{(\infty)}} \\
    &=\frac{1}{2\imag\kappa^{(\infty)}}\left[\imag+\frac{2\kappa^{(\infty)}}{\imag\beta^2_b}\right]=\frac{1}{2\kappa^{(\infty)}}-\frac{1}{\beta^2_b}\, , \nonumber \\
\frac{d}{dk^{*2}}\left[\sqrt{-k^{*2}}\right]_{k^{*2}=-\kappa^{(\infty)2}}&=-\frac{1}{2\kappa^{(\infty)}}\, .
\end{align}
Collecting both terms, we get
\begin{equation}
    \frac{d}{dk^{*2}}\left[k^*\cot\delta-(-\sqrt{-k^{*2}})\right]_{k^{*2}=-\kappa^{(\infty)2}}=-\frac{1}{\beta^2_b}<0 \, ,
\end{equation}
which means that the slope of the ERE around the bound-state pole has to be smaller than the slope of $-\sqrt{-k^{*2}}$. The two slopes and associated uncertainty bands are depicted in Fig.~\ref{fig:checks_slope} for all two-baryon channels and the two-parameter EREs obtained, demonstrating that the requirement is satisfied. The values of the binding momenta used in this analysis are taken from Table~\ref{tab:binding_momenta} (the $\bm{d}=\{(0,0,0),(0,0,2)\}$ column).

\begin{figure}[t!]
\includegraphics[width=\textwidth]{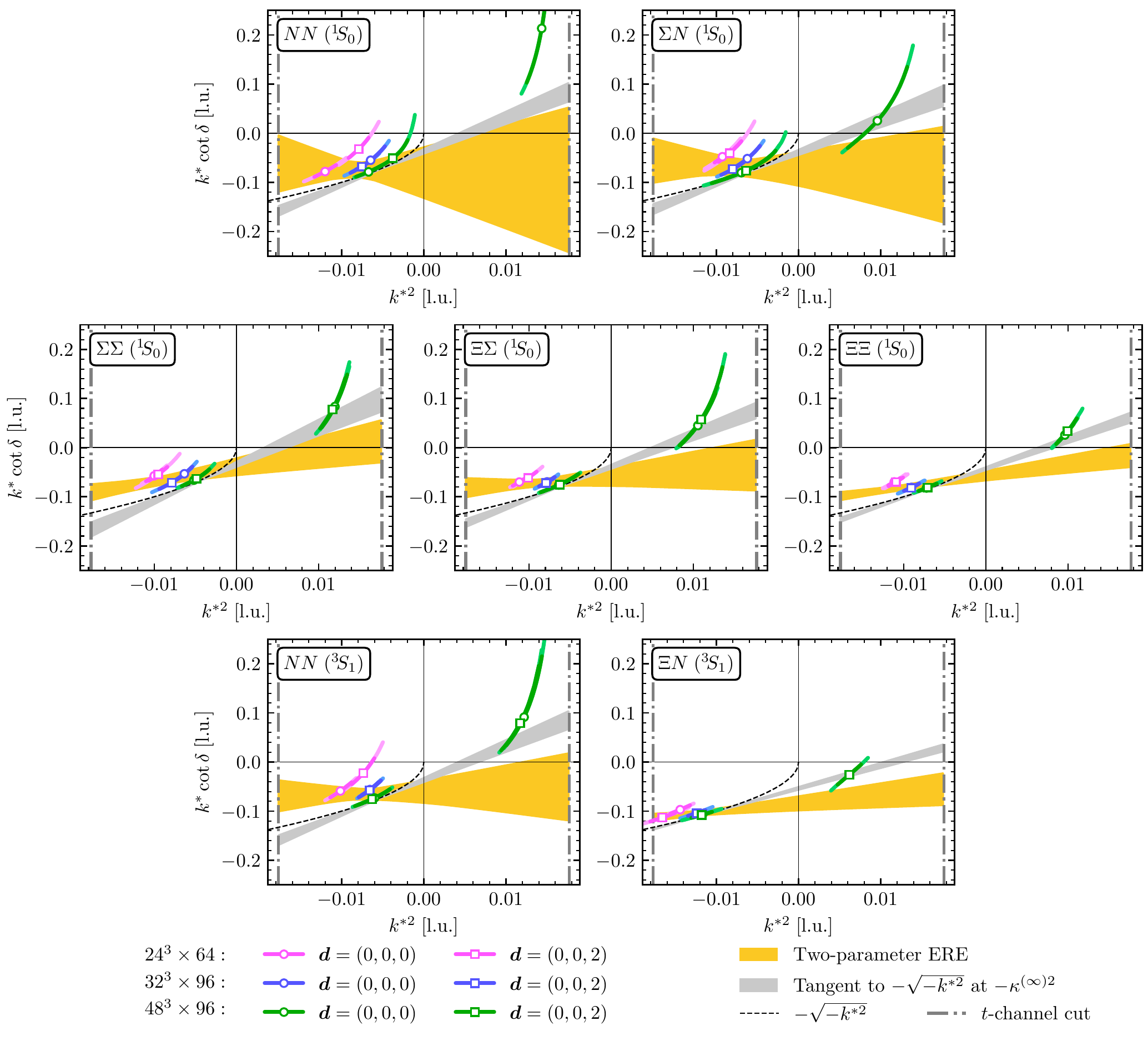}
\caption{Comparison between the two-parameter ERE and the slope of $-\sqrt{-k^{*2}}$ at $k^{*2}=-\kappa^{(\infty)2}$, where $\kappa^{(\infty)}$ is taken from the $\bm{d}=\{(0,0,0),(0,0,2)\}$ column of Table~\ref{tab:binding_momenta}. Quantities are expressed in lattice units.}
\label{fig:checks_slope}
\end{figure}

\item \textit{The absence of more than one bound state with an ERE parametrization of amplitudes}: None of the systems analyzed exhibit more than one bound state, i.e., the ERE does not cross the $-\sqrt{-k^{*2}}$ curve more than once. Therefore, applying the ERE parametrization of the $S$-wave scattering amplitude in all channels appears to be justified.

\item \textit{Constrained range for ERE parameters in the presence of a bound state}: The condition for the system to represent a physical bound state is that the ratio $r/a$ must be smaller than 1/2, so that the two-parameter ERE crosses the $-\sqrt{-k^{*2}}$ function once from below. This can be easily seen by equating both expressions,
\begin{equation}
    -\frac{1}{a}+\frac{1}{2}rk^{*2}=-\sqrt{-k^{*2}}\;\overset{k^{*2} =-\kappa^{(\infty)2}}{\longrightarrow} \, -\frac{1}{a}-\frac{1}{2}r\kappa^{(\infty)2}=-\kappa^{(\infty)}\; \Rightarrow\; \kappa^{(\infty)}=\frac{1\pm\sqrt{1-2\frac{r}{a}}}{r} \, .
\end{equation}
If we want to get a real solution, $r/a < 1/2$. Moreover, the ERE must cross the $\mathcal{Z}$-functions corresponding to different volumes to satisfy Lüscher's QC, introducing more constraints on scattering parameters. With the use of the two-dimensional $\chi^2$ in this work to fit the $k^*\cot\delta$ values, the confidence region of the ERE parameters does not cross these prohibited areas, as was demonstrated in Fig.~\ref{fig:ere-parameters}.
\end{checks}

\subsection{Constraints on the EFTs LECs}\label{subsec:EFTconstraints}

As mentioned in~\cref{subsec:scattparamLQCD}, given the large uncertainties in the scattering parameters (in particular in the effective range), the ratio $r/a$ shown in Table~\ref{tab:scattPar-ratio} is not well constrained, and does not conclusively prove unnaturalness in all channels.
Since in at least two channels the interactions seem unnatural, in the following both the natural and the unnatural cases will be considered in expressing relations between LECs and the scattering parameters, given by Eqs.~\eqref{eq:scattparam1} and~\eqref{eq:scattparam2}. The corresponding LECs for each two-baryon channel are given in Table~\ref{tab:LECtab}.

\begin{table}[b!]
\centering
\caption{The LECs of the LO and NLO pionless EFT that contribute to the scattering amplitude of the various two-baryon channel. The first three columns are total angular momentum ($J$), strangeness ($S$), and isospin ($I$).}
\label{tab:LECtab}
\renewcommand{\arraystretch}{1.2}
\resizebox{\columnwidth}{!}{
\begin{tabular}{ccccccc}
\toprule
$J$ & $S$ & $I$ & Channel & $SU(3)_f$ LO & $SU(3)_f$ NLO & $\cancel{SU(3)}_f$ NLO \\ \midrule
$0$ & $\phantom{-}0$ & $1$ & $NN$ & $c^{(27)}$ & $\tilde{c}^{(27)}$ & $4(c_3^{\chi}-c_4^{\chi})$\\
 & $-1$ & $\frac{3}{2}$ & $\Sigma N$ & $c^{(27)}$ & $\tilde{c}^{(27)}$ & $2(c_3^{\chi}-c_4^{\chi})$\\ 
 & $-2$ & $2$ & $\Sigma\Sigma$ & $c^{(27)}$ & $\tilde{c}^{(27)}$ & $0$\\ 
 & $-3$ & $\frac{3}{2}$ & $\Xi \Sigma$ & $c^{(27)}$ & $\tilde{c}^{(27)}$ & $2(c_1^{\chi}-c_2^{\chi}+c_{11}^{\chi}-c_{12}^{\chi})$\\ 
 & $-4$ & $1$ & $\Xi \Xi$ & $c^{(27)}$ & $\tilde{c}^{(27)}$ & $4(c_1^{\chi}-c_2^{\chi}+c_{11}^{\chi}-c_{12}^{\chi})$\\ 
$1$ & $\phantom{-}0$ & $0$ & $NN$ & $c^{(\overline{10})}$ & $\tilde{c}^{(\overline{10})}$ & $4(c_3^{\chi}+c_4^{\chi})$\\ 
 & $-1$ & $\frac{3}{2}$ & $\Sigma N$ & $c^{(10)}$ & $\tilde{c}^{(10)}$ & $-2(c_3^{\chi}+c_4^{\chi})$\\ 
 & $-4$ & $0$ & $\Xi \Xi$ & $c^{(10)}$ & $\tilde{c}^{(10)}$ & $-4(c_1^{\chi}+c_2^{\chi}-c_{11}^{\chi}-c_{12}^{\chi})$\\ 
 & $-2$ & $0$ & $\Xi N$ & $c^{(8_a)}$ & $\tilde{c}^{(8_a)}$ & $2(2c_5^{\chi}+2c_6^{\chi}+2c_7^{\chi}+2c_8^{\chi}+2c_9^{\chi}+2c_{10}^{\chi}+c_{11}^{\chi}+c_{12}^{\chi})$\\ \bottomrule
\end{tabular} }
\end{table}

\begin{figure}[t!]
\includegraphics[width=\textwidth]{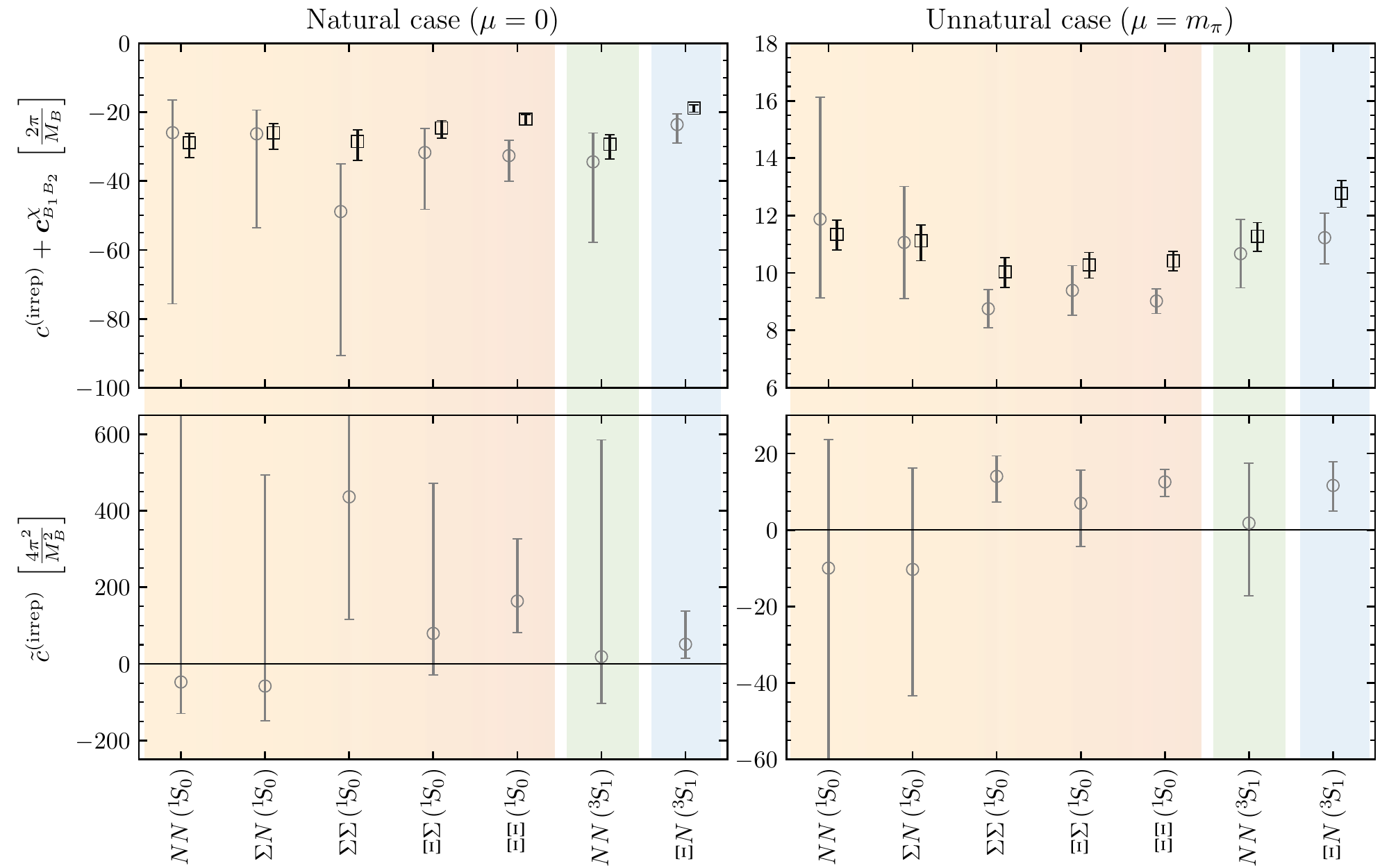}
\caption{LECs obtained by solving Eqs.~\eqref{eq:scattparam1} (upper panels) and~\eqref{eq:scattparam2} (lower panels) under the assumption of natural (left panels) and unnatural (right panels) interactions. The LECs of momentum-independent operators are in units of $[\frac{2\pi}{M_{B}}]$ and those of the momentum-dependent operators are in units of $[\frac{4\pi^2}{M^2_{B}}]$, where $M_B$ is the centroid of the octet-baryon masses. The gray-circle markers denote quantities that are extracted using the ERE parameters (method I), while black-square markers are those obtained from scattering lengths that are computed from binding momenta (method II).}
\label{fig:EFT_LONLO}
\end{figure}

Two sets of inputs can be used to constrain the numerical values of the LECs: 1) the scattering parameters $\{a^{-1},r\}$ obtained from two-parameter ERE fits, as discussed in~\cref{subsec:scattparamLQCD} and listed in Table~\ref{tab:scattPar}, can be used to compute the LECs of both momentum-independent and momentum-dependent operators (method~I), and 2) the binding momenta from~\cref{subsec:bindingLQCD} can be used to compute the corresponding scattering length, related at LO by \mbox{$-a^{-1}+\kappa^{(\infty)}=0$}, and this single parameter can be used to constrain the LECs of momentum-independent operators (method~II). 
This second method is motivated by the fact that $\kappa^{(\infty)}$ is extracted with higher precision than the parameters from the ERE fits, therefore enabling tighter constraints on the LECs of momentum-independent operators. The results for both sets of LECs are presented in Table~\ref{tab:LECtab1} and are depicted in Fig.~\ref{fig:EFT_LONLO}.
Results are presented in units of $2\pi/M_{B}$ for the momentum-independent operators and $4\pi^2/M^2_{B}$ for the momentum-dependent operators, where $M_B$ is the centroid of the octet-baryon masses, $M_B=\tfrac{1}{4}M_N+\tfrac{1}{8}M_{\Lambda}+\tfrac{3}{8}M_{\Sigma}+\tfrac{1}{4}M_{\Xi}=0.78583(23)(30)$ l.u.

\begin{table}[t!]
\centering
\caption{Numerical values of the LECs listed in Table~\ref{tab:LECtab}. Eq.~\eqref{eq:scattparam1} is used to obtain the ones from momentum-independent operators, given in units of $[\frac{2\pi}{M_{B}}]$, while Eq.~\eqref{eq:scattparam2} is used to obtain the ones from momentum-dependent operators, given in units of $[\frac{4\pi^2}{M^2_{B}}]$. $\tilde{c}^{(\text{irrep})}$ are only determined using method I. See text for more details.}
\label{tab:LECtab1}
\renewcommand{\arraystretch}{1.5}
\resizebox{\columnwidth}{!}{
\begin{tabular}{cccccccccc}
\toprule
LECs	&	$\mu$	&	Method	&	$NN\;(\1s0)$	&	$\Sigma N\;(\1s0)$	&	$\Sigma\Sigma\;(\1s0)$	&	$\Xi\Sigma\;(\1s0)$	&	$\Xi\Xi\;(\1s0)$	&	$NN\;(\3s1)$	&	$\Xi N\;(\3s1)$	\\\midrule
\multirow{4}{*}{$c^{(\text{irrep})}+\bm{c}^{\chi}_{\scriptscriptstyle B_1B_2}$}	&	\multirow{2}{*}{$0$}	&	I	&	$-26_{(-50)}^{(+9)}$	&	$-26_{(-27)}^{(+7)}$	&	$-49_{(-42)}^{(+14)}$	&	$-32_{(-17)}^{(+7)}$	&	$-33_{(-7)}^{(+5)}$	&	$-34_{(-23)}^{(+8)}$	&	$-24_{(-5)}^{(+3)}$	\\
	&		&	II	&	$-29_{(-4)}^{(+3)}$	&	$-26_{(-5)}^{(+3)}$	&	$-28_{(-5)}^{(+3)}$	&	$-25_{(-3)}^{(+2)}$	&	$-22_{(-2)}^{(+1)}$	&	$-29_{(-4)}^{(+3)}$	&	$-19_{(-1)}^{(+1)}$	\\
	&	\multirow{2}{*}{$m_{\pi}$}	&	I	&	$11.9_{(-2.7)}^{(+4.2)}$	&	$11.1_{(-2.0)}^{(+2.0)}$	&	$8.8_{(-0.7)}^{(+0.7)}$	&	$9.4_{(-0.9)}^{(+0.9)}$	&	$9.0_{(-0.4)}^{(+0.4)}$	&	$10.7_{(-1.2)}^{(+1.2)}$	&	$11.2_{(-0.9)}^{(+0.9)}$	\\
	&		&	II	&	$11.3_{(-0.5)}^{(+0.5)}$	&	$11.1_{(-0.7)}^{(+0.6)}$	&	$10.0_{(-0.5)}^{(+0.5)}$	&	$10.3_{(-0.5)}^{(+0.4)}$	&	$10.4_{(-0.3)}^{(+0.3)}$	&	$11.3_{(-0.5)}^{(+0.5)}$	&	$12.8_{(-0.5)}^{(+0.5)}$	\\ \midrule
\multirow{2}{*}{$\tilde{c}^{(\text{irrep})}$}	&	$0$	&	I	&	$-47_{(-82)}^{(+1600)}$	&	$-58_{(-91)}^{(+550)}$	&	$437_{(-320)}^{(+1800)}$	&	$80_{(-110)}^{(+390)}$	&	$164_{(-83)}^{(+160)}$	&	$19_{(-120)}^{(+570)}$	&	$51_{(-36)}^{(+86)}$	\\
	&	$m_{\pi}$	&	I	&	$-10_{(-84)}^{(+34)}$	&	$-10_{(-33)}^{(+27)}$	&	$14_{(-7)}^{(+5)}$	&	$7_{(-11)}^{(+9)}$	&	$13_{(-4)}^{(+3)}$	&	$2_{(-19)}^{(+16)}$	&	$12_{(-7)}^{(+6)}$	\\\bottomrule
\end{tabular}}
\end{table}

As can be seen from the values of the LECs that are obtained, the NLO $SU(3)_f$ coefficients have large uncertainties, and are mostly consistent with zero, because the effective ranges used to constrain them have rather large uncertainties.
Another feature of the results is that assuming the interactions to be unnatural leads to better-constrained parameters in general, as a non-zero scale $\mu$ in the left-hand side of Eqs.~\eqref{eq:scattparam1} and \eqref{eq:scattparam2} reduces the effect of uncertainties on the scattering lengths (this was also observed in Ref.~\cite{Wagman:2017tmp} for systems at $m_{\pi} \sim 806$ MeV).
Furthermore, as expected, the values obtained with method II have smaller uncertainties than the ones obtained from method I, given the more precise scattering lengths, although the method is limited to LO predictions.
Another anticipated feature is that in the cases where the effective range is resolved from zero within uncertainties (e.g., in the $\Xi\Xi\; (\1s0)$ channel), the values from method II are slightly different from those obtained from method I, indicating the non-negligible effect of the NLO effective range contributions that are neglected in this method.

\begin{table}[t!]
\centering
\caption{The values of the momentum-independent $SU(3)_f$ coefficient $c^{(27)}$ and specific linear combinations of the $\cancel{SU(3)}_f$ coefficients $\bm{c}^{\chi}_i$. Quantities are expressed in units of $[\frac{2\pi}{M_{B}}]$, where $M_B$ is the centroid of the octet-baryon masses.}
\label{tab:LECtab2}
\renewcommand{\arraystretch}{1.5}
\resizebox{\columnwidth}{!}{
\begin{tabular}{ccccccc}
\toprule
$\mu$	&	Method	&	$c^{(27)}$ $\{NN, \Sigma N\}$	&	$c^{(27)}$ $\{\Sigma\Sigma\}$	&	$c^{(27)}$ $\{\Xi\Sigma, \Xi\Xi\}$	&	$\bm{c}^{\chi}_3-\bm{c}^{\chi}_4$	&	$\bm{c}^{\chi}_1-\bm{c}^{\chi}_2+\bm{c}^{\chi}_{11}-\bm{c}^{\chi}_{12}$	\\ \midrule
\multirow{2}{*}{$0$}	&	I	&	$-27_{(-62)}^{(+58)}$	&	$-49_{(-42)}^{(+14)}$	&	$-31_{(-36)}^{(+21)}$	&	$0_{(-26)}^{(+18)}$	&	$0_{(-7)}^{(+10)}$	\\
	&	II	&	$-23_{(-12)}^{(+9)}$	&	$-28_{(-5)}^{(+3)}$	&	$-27_{(-7)}^{(+6)}$	&	$-1_{(-3)}^{(+4)}$	&	$1_{(-2)}^{(+2)}$	\\
\multirow{2}{*}{$m_{\pi}$}	&	I	&	$10.3_{(-7.7)}^{(+6.3)}$	&	$8.8_{(-0.7)}^{(+0.7)}$	&	$9.8_{(-2.1)}^{(+2.1)}$	&	$0.4_{(-2.2)}^{(+2.9)}$	&	$-0.2_{(-0.6)}^{(+0.6)}$	\\
	&	II	&	$10.9_{(-1.9)}^{(+1.6)}$	&	$10.0_{(-0.5)}^{(+0.5)}$	&	$10.1_{(-1.2)}^{(+1.2)}$	&	$0.1_{(-0.5)}^{(+0.6)}$	&	$0.1_{(-0.4)}^{(+0.4)}$	\\\bottomrule
\end{tabular}}
\end{table}

\begin{figure}[t!]
\centering
\includegraphics[width=0.8\textwidth]{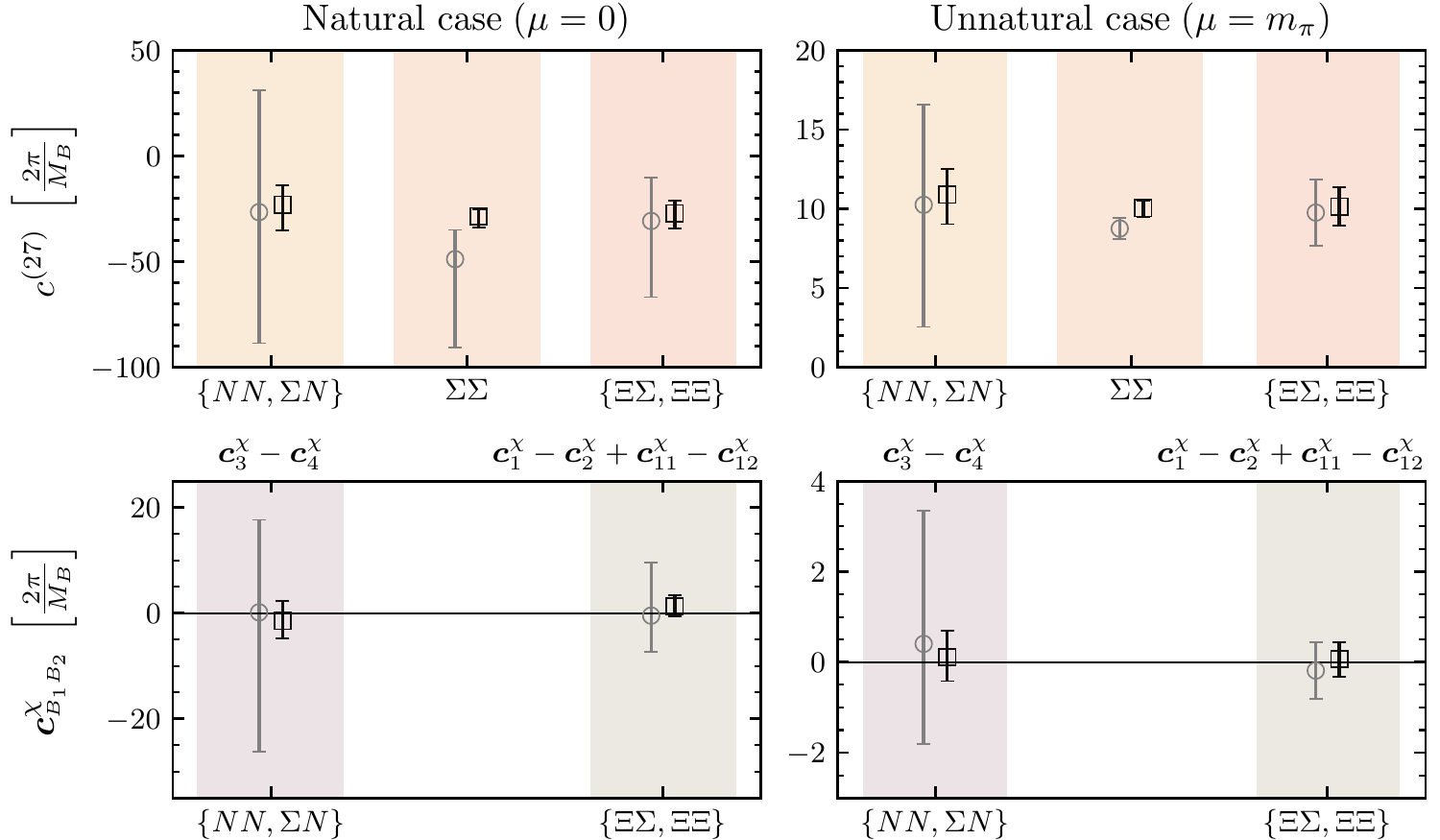}
\caption{The LO $SU(3)_f$ LEC $c^{(27)}$ (upper panels) and NLO $\cancel{SU(3)}_f$ LECs $\bm{c}^{\chi}_{\scriptscriptstyle B_1B_2}$ (lower panels) under the assumption of natural (left panels) and unnatural (right panels) interactions, in units of $[\frac{2\pi}{M_{B}}]$, where $M_B$ is the centroid of the octet-baryon masses. The gray-circle markers denote quantities that are extracted using method I, while black-square markers show results obtained from method II. See the text for further details.}
\label{fig:EFT_c27cchi}
\end{figure}

It should be noted that the input for scattering parameters is not sufficient to disentangle the LO $SU(3)_f$ and NLO $\cancel{SU(3)}_f$ coefficients in general, hence the $c^{(\text{irrep})}+\bm{c}^{\chi}_{\scriptscriptstyle B_1B_2}$ entry in Table~\ref{tab:LECtab1} and Fig.~\ref{fig:EFT_LONLO}.
For the systems that belong to the ${\bf 27}$ irrep, since the spin-singlet pairs $\{NN,\Sigma N\}$ and $\{\Xi \Sigma, \Xi \Xi\}$ depend on the same $SU(3)_f$ LO and $\cancel{SU(3)}_f$ NLO LECs but with different linear combinations of the coefficients, a system of equations can be formed to separate each contribution.
The results are shown in Table~\ref{tab:LECtab2} and Fig.~\ref{fig:EFT_c27cchi}, along with the result for the $\Sigma \Sigma$ channel for comparison purposes, as there is no contribution from $\cancel{SU(3)}_f$ interactions for this channel at this order.
From these results, it can be seen that the values of the symmetry-breaking coefficients $\bm{c}^{\chi}_3-\bm{c}^{\chi}_4$ and $\bm{c}^{\chi}_1-\bm{c}^{\chi}_2+\bm{c}^{\chi}_{11}-\bm{c}^{\chi}_{12}$ are compatible with zero. Together with the observation that the scattering lengths and binding energies in all of the systems are similar within uncertainties, it appears that the $SU(3)$ flavor symmetry remains an approximate symmetry at the quark masses used in this study.
These observations in the two-baryon sector are consistent with those in the single-baryon sector as presented in Ref.~\cite{Orginos:2015aya} at the same quark masses. There, the authors looked at the Gell-Mann-Okubo mass relation~\cite{GellMann:1962xb,Okubo:1961jc}, since its violation results from $SU(3)_f$ breaking transforming in the $\mathbf{27}$ irrep of $SU(3)_f$ symmetry~\cite{Beane:2006pt}, which can only arise from insertions of the light-quark mass matrix or from non-analytic meson-mass dependence induced by loops in $\chi$PT. The quantity they studied was $\delta_{\text{GMO}}=\frac{1}{M_B}(M_{\Lambda}+\frac{1}{3}M_{\Sigma}-\frac{2}{3}M_N-\frac{2}{3}M_{\Xi})$, which, using the values of the masses from Table~\ref{tab:baryon_mass}, gives $\delta_{\text{GMO}}=0.0008(13)(12)$, consistent with the value computed in Ref.~\cite{Orginos:2015aya} and an order of magnitude smaller than the experimental value, $\delta^{\text{exp}}_{\text{GMO}} \sim 0.0076$.

In Table~\ref{tab:full_eft_coeff} of~\cref{appen:LECsEFT}, the full list of relations needed to independently constrain all 24 different LECs that appear at LO and NLO are shown, demonstrating that the proper combinations of 18 two-baryon flavor channels are sufficient to extract all these LECs.
Future LQCD studies toward the physical values of the quark masses, and involving all these channels, will be instrumental in achieving this goal.

The $a$ and $b$ coefficients can also be matched to scattering amplitudes in a momentum expansion at LO. Since at least some of the $SU(3)_f$ symmetry-breaking LECs $\bm{c}^{\chi}_i$ were found to be consistent with zero in the present study, one can assume an approximate $SU(3)_f$ symmetry, and relate the $SU(6)$ LECs $a$ and $b$ directly to the LECs of the LO $SU(3)_f$-symmetric Lagrangian for given irreps using Eqs.~\eqref{eq:SU6eq}. A priori, the relative size of the Kaplan-Savage coefficients, $a$ and $b$, is unknown, and only experimental data or LQCD input can constrain these LECs.
As seen in Eqs.~\eqref{eq:SU6eq}, the contribution from the $b$ coefficient to the LO amplitude is parametrically suppressed compared with that of the coefficient $a$. As a result, if $b$ in Eq.~\eqref{eq:su6lag} is comparable to or smaller than $a$, there remains only one type of interaction that contributes significantly to the scattering amplitude, a situation that would realize an accidental $SU(16)$ symmetry of the nuclear and hypernuclear forces.
The first evidence for $SU(16)$ symmetry in the two-(octet)baryon sector was observed in a LQCD study at a pion mass of $\sim 806$ MeV~\cite{Wagman:2017tmp}, and the goal of the present study is to examine these predictions at smaller values of the light-quark masses.
Such a symmetry is suggested in Ref.~\cite{Beane:2018oxh} to be consistent with the conjecture of maximum entanglement suppression of the low-energy sector of QCD.

\begin{table}[t!]
\centering
\caption{The leading $SU(6)$ LECs, $a$ and $b/3$, obtained by solving a given pair of equations in Eqs.~\eqref{eq:SU6eq}. The last column shows the results of a constant fit to the LECs obtained in each case as described in Eqs.~\eqref{eq:averging}. The spin specifications are dropped from channel labels for brevity, but one clarification is necessary: in the first pair of two-baryon channels, $NN$ refers to the spin-singlet case, while in the last pair, it denotes the spin-triplet case. Quantities are expressed in units of $[\frac{2\pi}{M_{B}}]$, where $M_B$ is the centroid of the octet-baryon masses.}
\label{tab:SU6tab}
\renewcommand{\arraystretch}{1.5}
\resizebox{\columnwidth}{!}{
\begin{tabular}{cccccccccc}\toprule
LEC	&	$\mu$	&	Method	&	$\{NN, \Xi N\}$	&	$\{\Sigma N, \Xi N\}$	&	$\{\Sigma \Sigma, \Xi N\}$	&	$\{\Xi \Sigma, \Xi N\}$	&	$\{\Xi \Xi, \Xi N\}$	&	$\{NN, \Xi N\}$	&	Combined    \\ \cmidrule(r){1-9} \cmidrule(l){10-10}
\multirow{4}{*}{$a$}	&	\multirow{2}{*}{$0$}	&	I	&	$-12_{(-13)}^{(+3)}$	&	$-12_{(-8)}^{(+2)}$	&	$-18_{(-11)}^{(+4)}$	&	$-14_{(-5)}^{(+2)}$	&	$-14_{(-3)}^{(+2)}$	&	$-14_{(-7)}^{(+3)}$	&	$-15_{(-4)}^{(+4)}$	\\
	&		&	II	&	$-11.9_{(-1.3)}^{(+0.9)}$	&	$-11.2_{(-1.4)}^{(+0.9)}$	&	$-11.8_{(-1.6)}^{(+1.0)}$	&	$-10.8_{(-1.0)}^{(+0.8)}$	&	$-10.2_{(-0.7)}^{(+0.6)}$	&	$-12.0_{(-1.4)}^{(+0.9)}$	&	$-11.0_{(-1.0)}^{(+1.0)}$	\\
	&	\multirow{2}{*}{$m_{\pi}$}	&	I	&	$5.8_{(-0.9)}^{(+1.2)}$	&	$5.6_{(-0.7)}^{(+0.7)}$	&	$5.0_{(-0.4)}^{(+0.4)}$	&	$5.2_{(-0.4)}^{(+0.4)}$	&	$5.1_{(-0.3)}^{(+0.3)}$	&	$5.5_{(-0.5)}^{(+0.5)}$	&	$5.2_{(-0.4)}^{(+0.4)}$	\\
	&		&	II	&	$6.0_{(-0.3)}^{(+0.2)}$	&	$6.0_{(-0.3)}^{(+0.3)}$	&	$5.7_{(-0.3)}^{(+0.2)}$	&	$5.8_{(-0.2)}^{(+0.2)}$	&	$5.8_{(-0.2)}^{(+0.2)}$	&	$6.0_{(-0.2)}^{(+0.2)}$	&	$5.9_{(-0.2)}^{(+0.2)}$	\\ \midrule
\multirow{4}{*}{$\dfrac{b}{3}$}	&	\multirow{2}{*}{$0$}	&	I	&	$5_{(-31)}^{(+110)}$	&	$6_{(-26)}^{(+65)}$	&	$57_{(-42)}^{(+99)}$	&	$18_{(-27)}^{(+41)}$	&	$20_{(-22)}^{(+23)}$	&	$24_{(-29)}^{(+57)}$	&	$24_{(-36)}^{(+36)}$	\\
	&		&	II	&	$23_{(-9)}^{(+11)}$	&	$16_{(-8)}^{(+12)}$	&	$22_{(-10)}^{(+14)}$	&	$13_{(-7)}^{(+9)}$	&	$7_{(-6)}^{(+6)}$	&	$24_{(-9)}^{(+12)}$	&	$14_{(-9)}^{(+9)}$	\\
	&	\multirow{2}{*}{$m_{\pi}$}	&	I	&	$-1_{(-11)}^{(+7)}$	&	$0_{(-6)}^{(+6)}$	&	$6_{(-3)}^{(+3)}$	&	$4_{(-4)}^{(+4)}$	&	$5_{(-3)}^{(+3)}$	&	$1_{(-4)}^{(+4)}$	&	$4_{(-4)}^{(+4)}$	\\
	&		&	II	&	$3_{(-2)}^{(+2)}$	&	$4_{(-2)}^{(+2)}$	&	$6_{(-2)}^{(+2)}$	&	$6_{(-2)}^{(+2)}$	&	$5_{(-2)}^{(+2)}$	&	$3_{(-2)}^{(+2)}$	&	$5_{(-2)}^{(+2)}$	\\ \bottomrule
\end{tabular} }
\end{table}

\begin{figure}[t!]
\includegraphics[width=\textwidth]{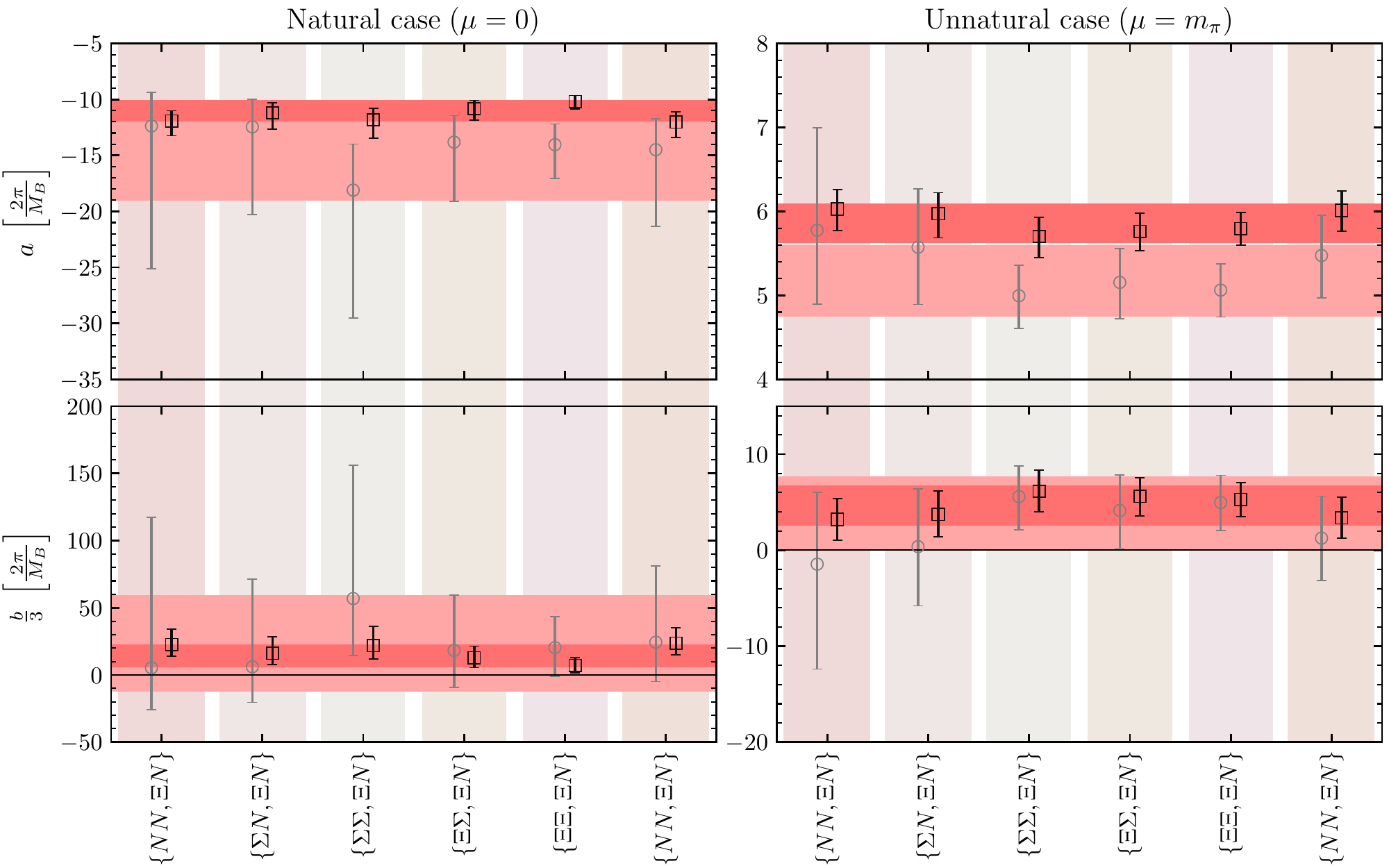}
\caption{The leading $SU(6)$ LECs, $a$ (upper panels) and $b/3$ (lower panels), under the assumption of natural (left panels) and unnatural (right panels) interactions, in units of $[\frac{2\pi}{M_{B}}]$, where $M_B$ is the centroid of the octet-baryon masses. The gray-circle markers denote quantities extracted using the ERE parameters (method I), with the light pink band showing the averaged value, while black-square markers show results obtained from scattering lengths that are constrained by binding momenta (method II), with the dark pink band showing the averaged value.}
\label{fig:SU6coeff}
\end{figure}

In order to extract $a$ and $b$, states in the $\mathbf{27}$ and $\overline{\mathbf{10}}$ irreps can be combined with those in the $\mathbf{8}_a$ irrep, allowing for six possible extractions.\footnote{Note that the ERE parameters were obtained in the previous section only for two-baryon channels belonging to the $\{\mathbf{27},\overline{\mathbf{10}},\mathbf{8}_a\}$ irreps.}
The results are shown in Table~\ref{tab:SU6tab} and Fig.~\ref{fig:SU6coeff}. As seen in Eqs.~\eqref{eq:SU6eq}, the contributions from the $b$ coefficient are suppressed by at least a factor of 3 compared with those from the $a$ coefficient, and thus the rescaled coefficient $b/3$ is considered.

Considering that the results presented should be valid only up to corrections that scale as $1/N_c$, individual values of the coefficients $a$ and $b/3$ obtained from different pairs of channels exhibit remarkable agreement, indicating that the $SU(6)$ spin-flavor symmetry is a good approximation at these values of the quark masses.
A correlated weighted average of the results is obtained, following the procedure introduced by Schmelling~\cite{Schmelling:1994pz} and used by the FLAG collaboration~\cite{Aoki:2013ldr}, and is shown as the pink bands in Fig.~\ref{fig:SU6coeff}. The average of a series of values $\{x_i\}$ with uncertainties $\{\sigma_i\}$ is computed as:
\begin{equation}
    x_{\text{average}}=\sum_i x_i w_i\, , \quad w_i=\frac{\sigma^{-2}_i}{\sum_j\sigma^{-2}_j}\, , \quad \sigma^2_{\text{average}}=\sum_{ij} w_i w_j C_{ij}\, , \quad C_{ij}=\sigma_i\sigma_j\, ,\label{eq:averging}
\end{equation}
where, since the different values of $x_i$ (and their uncertainties) are correlated, a 100\% correlation is assumed when computing $\sigma^2_{\text{average}}$. For asymmetric uncertainties in $x_i$, the following procedure is used to symmetrize them: a value $x_i=c^{(+u)}_{(-l)}$ is modified to $c+(u-l)/4$ with uncertainty $\sigma=\text{max}[(u+3l)/4,(3u+l)/4]$.

Given the uncertainty in $b/3$, no conclusion can be drawn about the relative importance of $a$ and $b/3$. We will return to the question of the presence of an accidental $SU(16)$ symmetry shortly.

\begin{figure}[hbt!]
\centering
\includegraphics[width=\textwidth]{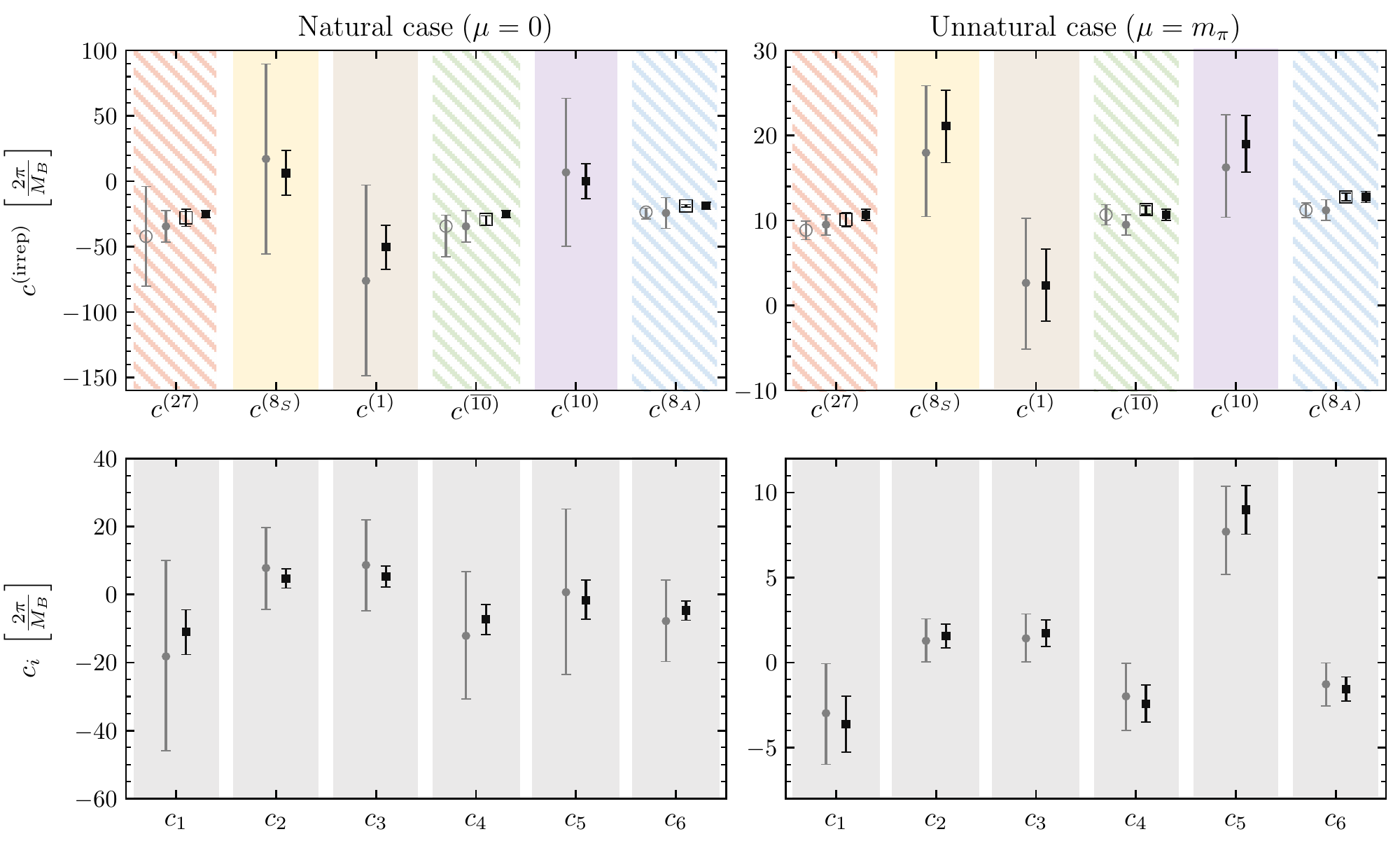}
\caption{The predicted (filled markers) LO $SU(3)_f$ coefficients $c^{(\text{irrep})}$ (upper panels), as well as Savage-Wise coefficients $c_i$ (lower panels) reconstructed from the $SU(6)$ relations, are compared with the directly-extracted LECs (empty markers) under the assumption of natural (left panels) and unnatural (right panels) interactions, in units of $[\frac{2\pi}{M_{B}}]$, where $M_B$ is the centroid of the octet-baryon masses. The gray-circle symbols denote quantities that have been extracted using the scattering parameters obtained from the ERE fit (method I), while black-square symbols denote those that are obtained from scattering lengths constrained by binding momenta (method II). The hashed background in the upper panels denotes coefficients whose values were used to constrain $a$ and $b$, and hence are not predictions.}
\label{fig:abSU3coeff}
\end{figure}

Given the extracted values of $a$ and $b/3$, several checks can be performed, and several predictions can be made. The simplest check is to compute all the LO $SU(3)_f$ LECs, $c^{(\text{irrep})}$, using the relations in Eqs.~\eqref{eq:SU6eq}.
The results are shown in the first rows of Table~\ref{tab:abSU3coeff} and in the upper panels of Fig.~\ref{fig:abSU3coeff}. Columns with hashed backgrounds are the coefficients whose values were used as an input to make predictions for other coefficients, presented in panels with solid-colored backgrounds.
These input coefficients ($c^{(27)}$, $c^{(\overline{10})}$, and $c^{(8_a)}$) can be reevaluated using the average values of $a$ and $b/3$, which therefore gives back consistent values but with different uncertainties (for $c^{(27)}$, the average of the values given in Table~\ref{tab:LECtab2} is computed).
The large uncertainties in the $c^{(8_s)}$, $c^{(1)}$, and $c^{(10)}$ coefficients are due to the fact that $b/3$, with a larger uncertainty than $a$, is numerically more important in these cases; see Eqs.~\eqref{eq:SU6eq}.
Additionally, the Savage-Wise coefficients $c_i$ can be computed by inverting the relations in Eqs.~\eqref{eq:SU3eq}, and the resulting values are presented in the last rows of Table~\ref{tab:abSU3coeff} and in the lower panels of Fig.~\ref{fig:abSU3coeff}.
Due to large uncertainties in the natural case, no conclusions can be made regarding the relative size of the coefficients. In the unnatural case and at the chosen value of the renormalization scale, the $c_5$ coefficient has a larger value than the rest of the coefficients. 
The relative importance of $c_5$ is a remnant of an accidental approximate $SU(16)$ symmetry of $S$-wave two-baryon interactions that is more pronounced in the $SU(3)_f$-symmetric study of Ref.~\cite{Wagman:2017tmp}.
It will be interesting to explore whether the remnant of this symmetry remains visible in studies closer to the physical quark masses.

\begin{table}[t!]
\centering
\caption{Predicted $SU(3)_f$ LECs, $c^{(\text{irrep})}$, as well as the Savage-Wise coefficients, $c_i$, obtained from the Kaplan-Savage $SU(6)$ coefficients $a$ and $b$ using the relations in Eqs.~\eqref{eq:SU6eq} and~\eqref{eq:SU3eq}. Quantities are expressed in units of $[\frac{2\pi}{M_{B}}]$, where $M_B$ is the centroid of the octet-baryon masses.}
\label{tab:abSU3coeff}
\renewcommand{\arraystretch}{1.5}
\begin{tabular}{cccccccc}
\toprule
$\mu$	&	Method	&	$c^{(27)}$	&	$c^{(8_s)}$	&	$c^{(1)}$	&	$c^{(\overline{10})}$	&	$c^{(10)}$	&	$c^{(8_a)}$	\\ \midrule
\multirow{2}{*}{$0$}	&	I	&	$-35_{(-12)}^{(+12)}$	&	$17_{(-73)}^{(+73)}$	&	$-76_{(-73)}^{(+73)}$	&	$-35_{(-12)}^{(+12)}$	&	$7_{(-57)}^{(+57)}$	&	$-24_{(-12)}^{(+12)}$	\\
	&	II	&	$-25_{(-3)}^{(+3)}$	&	$6_{(-17)}^{(+17)}$	&	$-50_{(-17)}^{(+17)}$	&	$-25_{(-3)}^{(+3)}$	&	$0_{(-13)}^{(+14)}$	&	$-19_{(-3)}^{(+3)}$	\\
\multirow{2}{*}{$m_{\pi}$}	&	I	&	$9.5_{(-1.2)}^{(+1.2)}$	&	$18.0_{(-7.5)}^{(+7.9)}$	&	$2.7_{(-7.8)}^{(+7.6)}$	&	$9.5_{(-1.2)}^{(+1.2)}$	&	$16.2_{(-5.9)}^{(+6.2)}$	&	$11.2_{(-1.2)}^{(+1.2)}$	\\
	&	II	&	$10.7_{(-0.7)}^{(+0.6)}$	&	$21.1_{(-4.3)}^{(+4.2)}$	&	$2.4_{(-4.2)}^{(+4.3)}$	&	$10.7_{(-0.7)}^{(+0.6)}$	&	$19.0_{(-3.3)}^{(+3.3)}$	&	$12.8_{(-0.7)}^{(+0.6)}$	\\ \bottomrule\toprule
$\mu$	&	Method	&	$c_1$	&	$c_2$	&	$c_3$	&	$c_4$	&	$c_5$	&	$c_6$	\\ \midrule
\multirow{2}{*}{$0$}	&	I	&	$-18_{(-28)}^{(+28)}$	&	$8_{(-12)}^{(+12)}$	&	$9_{(-13)}^{(+13)}$	&	$-12_{(-19)}^{(+19)}$	&	$1_{(-24)}^{(+24)}$	&	$-8_{(-12)}^{(+12)}$	\\
	&	II	&	$-11_{(-7)}^{(+6)}$	&	$5_{(-3)}^{(+3)}$	&	$5_{(-3)}^{(+3)}$	&	$-7_{(-4)}^{(+4)}$	&	$-2_{(-6)}^{(+6)}$	&	$-5_{(-3)}^{(+3)}$	\\
\multirow{2}{*}{$m_{\pi}$}	&	I	&	$-3.0_{(-3.0)}^{(+2.9)}$	&	$1.3_{(-1.3)}^{(+1.3)}$	&	$1.4_{(-1.4)}^{(+1.4)}$	&	$-2.0_{(-2.0)}^{(+1.9)}$	&	$7.7_{(-2.5)}^{(+2.7)}$	&	$-1.3_{(-1.3)}^{(+1.3)}$	\\
	&	II	&	$-3.6_{(-1.6)}^{(+1.7)}$	&	$1.6_{(-0.7)}^{(+0.7)}$	&	$1.7_{(-0.8)}^{(+0.8)}$	&	$-2.4_{(-1.1)}^{(+1.1)}$	&	$9.0_{(-1.4)}^{(+1.4)}$	&	$-1.6_{(-0.7)}^{(+0.7)}$	\\\bottomrule
\end{tabular}
\end{table}

\begin{table}[tb!]
\centering
\caption{Predicted inverse scattering lengths, $a^{-1}$, for the systems where an ERE fit was not possible, using the $SU(6)$ LECs $a$ and $b$. Quantities are expressed in lattice units.}
\label{tab:predainv}
\renewcommand{\arraystretch}{1.5}
\begin{tabular}{c @{\qquad \quad} c @{\qquad \quad} r @{\qquad \quad} r}
\toprule
$\mu$ & Method & \multicolumn{1}{c @{\qquad \quad}}{$a^{-1}_{\Sigma N \;(\3s1)}$} & \multicolumn{1}{c}{$a^{-1}_{\Xi \Xi \;(\3s1)}$} \\ \midrule
\multirow{2}{*}{$0$} & I & $-0.02_{(-07)}^{(+11)}$ & $-0.02_{(-06)}^{(+10)}$ \\
& II & $0.06_{(-44)}^{(+33)}$ & $0.05_{(-40)}^{(+30)}$ \\
\multirow{2}{*}{$m_{\pi}$} & I & $0.14_{(-07)}^{(+04)}$ & $0.15_{(-06)}^{(+03)}$ \\
& II & $0.16_{(-02)}^{(+02)}$ & $0.17_{(-01)}^{(+02)}$ \\ \bottomrule
\end{tabular}
\end{table}

The values of the $SU(6)$ coefficients $a$ and $b$ allow predictions to be made for the scattering lengths of the systems that could not be constrained in our study by an ERE fit, namely the $\Sigma N \; (\3s1)$ and $\Xi \Xi \; (\3s1)$ channels.
Using the $c^{(\text{irrep})}$ coefficients computed previously, the relations in Eq.~\eqref{eq:scattparam1} can be inverted to obtain $a^{-1}$, assuming that the values of $\bm{c}^{\chi}_i$ are negligible compared to those of $c^{(\text{irrep})}$ (an observation that is only confirmed for given linear combinations of these LECs but is assumed to hold in general given the hints of an approximate $SU(3)_f$ symmetry in this study).
This exercise leads to consistent results for the inverse scattering length for systems for which the ERE allowed a direct extraction of this parameter, while it provides predictions for the channels shown in Table~\ref{tab:predainv}.
For the case of natural interactions, the scattering lengths are not constrained well, although they are consistent within uncertainties with those in the unnatural case, demonstrating the renormalization-scale independence of the scattering length. For the unnatural case, both methods are consistent and give rise to inverse scattering lengths that are positive and larger than those obtained for the rest of the systems studied in this work.
This is in agreement with the parameters found when fitting the results for $k^*\cot\delta$ in these channels beyond the $t$-channel cut, see Table~\ref{tab:scattPar-beyond}.

% Chapter 2 - BB interaction *******************************************

\chapter{Summary and conclusions}\label{chap:summary}

Nuclear and hypernuclear interactions are key inputs into investigations of the properties of matter, and their knowledge continues to be limited in systems with multiple neutrons or when hyperons are present. As can be seen in the summary of the present status of the field, the major limitations come from the lack of experimental data, resulting in contradictory predictions for the scattering parameters obtained with different theoretical models.
In recent years, LQCD has reached the stage where controlled first-principles studies of nuclei are feasible, and may soon constrain nuclear and hypernuclear few-body interactions in nature. The present work demonstrates such a capability in the case of two-baryon interactions, albeit at an unphysically large value of the quark masses corresponding to a pion mass of $\sim 450$ MeV and at only one value of the lattice spacing.
It illustrates how Euclidean two-point correlation functions of systems with the quantum numbers of two baryons computed with LQCD can be used to constrain a wealth of quantities, from scattering phase shifts to low-energy scattering parameters and binding energies, to EFTs of forces, or precisely the LECs describing the interactions of two baryons.
The same type of calculations are foreseen in the near future, with values of the quark masses closer to the physical ones, and their output, in form of both finite-volume energy spectra and constrained EFT interactions, is expected to serve as input of quantum many-body studies of larger isotopes, at both unphysical and physical values of quark masses; see e.g., Refs.~\cite{Barnea:2013uqa,Contessi:2017rww,Bansal:2017pwn} for previous studies in the nuclear sector.
By supplementing the missing experimental input for scattering and spectra of two-baryon systems, such LQCD analyses can constrain phenomenological models and EFTs of hypernuclear forces. 

In~\cref{chap:2}, we have introduced the method of lattice QCD. After a brief description of the fundamental theory in the continuum, QCD, the appropriate changes are applied to allow for the computation of observables using Monte Carlo techniques. These observables are mainly the energies of the system and its interactions with external currents. While in this thesis we have only studied the first type of observables via the construction of two-point correlation functions, a short explanation has been given on how three-point correlation functions behave and some relevant results for nuclear physics have been presented as well. Also, in this chapter, we have shown how the finite-volume energies can be mapped to the scattering phase-shifts using the Lüscher's formalism~\cite{Luscher:1986pf, Luscher:1990ux}, with some detail on the group theory involved when systems are put in a finite volume and the subsequent modification of the angular momentum. Using the same formalism, we have also presented how the ground-state energies obtained with different volumes can be extrapolated to infinite volume, leading to the binding energy of the system.

In~\cref{chap:3}, we have discussed how to extract the energy levels of two-point correlation functions, with special emphasis in systems where noise is significant, as it is the case of few-baryon systems. First, we described a method that has been developed to robustly fit lattice data to a sum of exponentials, where both the fitting range and the number of exponentials were varied to obtain a reliable estimation of the systematic uncertainty~\cite{Beane:2020ycc}. The discussion then focused on how to remove excited-state contamination at early times, with the description of three methods: the generalized pencil-of-functions~\cite{Aubin:2010jc}, the matrix-Prony~\cite{Beane:2009kya}, and the variational~\cite{Michael:1982gb,Luscher:1990ck} method. We have also detailed the different ways in which the errors in the correlation functions can be estimated, using the jackknife and bootstrap resampling methods, but also more robust estimators, like the Hodges–Lehmann estimator.

In~\cref{chap:4}, we have presented the experimental and theoretical status of the baryon-baryon interaction, listing all relevant contributions, and comparing all theoretical predictions of the scattering parameters. We have also discussed and compared the different LQCD methods and results of interest to the present  work, namely the direct method and the potential method. We then moved on to describe the different EFTs that have been constrained with the lattice data. Since our focus is on the low-energy physics regime, we considered pionless EFTs (consisting only of contact interactions) assuming $SU(3)$ flavor~\cite{Savage:1995kv,Petschauer:2013uua} or $SU(6)$ spin-flavor symmetry~\cite{Kaplan:1995yg}. We have complemented this section with group-theoretical arguments to justify the number of operators included at each order. In the last part of this section, we addressed the issue of naturalness of the two-baryon system, as well as its implications in the determination of the LECs of the EFTs from the LQCD results.

The last section of~\cref{chap:4} includes the computation of the lowest-lying spectra of several two-octet baryon systems with strangeness ranging from $0$ to $-4$.
These results have been computed in three different volumes, using a single lattice spacing, and with unphysical values of the light-quark masses~\cite{Illa:2020nsi}. 
Assuming small discretization artifacts given the improved LQCD action that is employed, the results reveal interesting features on the nature of two-baryon forces with larger-than-physical values of the quark masses.
In particular, the determination of scattering parameters of two-baryon systems at low energies has enabled us to constrain the LO and NLO interactions of a pionless EFT, for both the $SU(3)$ flavor-symmetric and the symmetry-breaking interactions. The most precise results are obtained using the binding energy to fix the scattering length, and due to the large uncertainties in the effective ranges, the NLO $SU(3)_f$ symmetric LECs are mostly consistent with zero.  
While the two-baryon channels studied in this work only allowed two sets of leading $SU(3)_f$ symmetry-breaking LECs to be constrained, and those values are seen to be consistent with zero, the present study is the first of this type of analysis to access these interactions, extending the previous EFT matching presented in Ref.~\cite{Wagman:2017tmp} at an $SU(3)$ flavor-symmetric point with $m_\pi = m_K \sim 806$ MeV.
Given the limited knowledge of flavor-symmetry-breaking effects in the two-baryon sector in nature, this demonstrates the potential of LQCD to improve the situation. 
Finally, the observation of an approximate $SU(3)_f$ symmetry in the two-baryon systems included in the present work led to an investigation of the large-$N_c$ predictions of Ref.~\cite{Kaplan:1995yg}, through matching the LQCD results for scattering amplitudes to the EFT.
In particular, the $S$-wave interactions at LO have been found to exhibit an $SU(6)$ spin-flavor symmetry at this pion mass, as also observed in Ref.~\cite{Wagman:2017tmp} at a larger value of the pion mass. Both of the two independent spin-flavor-symmetric interactions at LO have been found to contribute to the amplitude. Nonetheless, the extracted values of the coefficients of the LO $SU(3)$ flavor-symmetric EFT suggest a remnant of an approximate accidental $SU(16)$ symmetry, observed in the $SU(3)$ flavor-symmetric study at $m_\pi \sim 806$~MeV~\cite{Wagman:2017tmp}. It will be interesting to examine these symmetry considerations in the hypernuclear forces at the physical values of the quark masses, particularly given the conjectured connections between the nature of forces in nuclear physics and the quantum entanglement in the underlying systems~\cite{Beane:2018oxh}.
Regarding the binding energy of the systems, we find that most of the channels are bound, except for the spin-triplet $\Sigma N$ and $\Xi \Xi$.
While no attempt is made in the current work to constrain forces within the EFTs at the physical point, a naive extrapolation has been performed using the results of this work and those at $m_\pi \sim 806$ MeV, with simple extrapolation functions, to make predictions of the binding energies of several two-baryon channels.
The obtained results for ground-state energies of two-nucleon systems are found to be compatible with the experimental values. Furthermore, stronger evidence for the existence of bound states in the $\Xi \Xi \; (\1s0)$ and $\Xi N \; (\3s1)$ channels is observed in comparison with other two-baryon systems.
Such predictions are in agreement with current phenomenological models and EFT predictions. However, conclusive results can only be reached by performing LQCD studies of multi-baryon systems at or near the physical values of the quark masses, and upon taking the continuum limit using multiple values of lattice spacing, a program that will be pursued in the upcoming years.

%%%%%%%%%%%%%%%%%%%%%%%%%%%%%%%%%%%%%%%%%%%%%%%%%%%%%
\appendix

% Appendix 2 - *******************************************

\chapter{Group theory tables}\label{appen:GTtables}

This appendix contains the relevant tables for the group theory calculations needed in~\cref{sec:scatteringFV}. These are the character tables for the groups $O^D_h$ (Table~\ref{tab:Oh_char}), $D^D_{4h}$ (Table~\ref{tab:D4_char}), and $C^D_{4v}$ (Table~\ref{tab:C4_char}); the decomposition of the angular momentum from the continuum to finite volume (and vice versa) for the groups $O^D_h$ (Tables~\ref{tab:spin_reduction_o} and~\ref{tab:spin_reduction_o_2}), $D^D_{4h}$ (Tables~\ref{tab:spin_reduction_d} and~\ref{tab:spin_reduction_d_2}), and $C^D_{4v}$ (Tables~\ref{tab:spin_reduction_c} and~\ref{tab:spin_reduction_c_2}); and the basis vectors needed to diagonalize Lüscher's QC into the different irreps for the groups $O^D_h$ (Table~\ref{tab:vectors_o}), $D^D_{4h}$ (Table~\ref{tab:vectors_d}), and $C^D_{4v}$ (Table~\ref{tab:vectors_c}).

\begin{table}[ht]
\centering
\caption{Character table for the double group $O^D_h$.}
\label{tab:Oh_char}
\renewcommand{\arraystretch}{1.2}
\resizebox{\columnwidth}{!}{
\begin{tabular}{c|cccccccccccccccc}
\toprule
\multirow{2}{*}{$O^D_h$} & \multirow{2}{*}{$E$} & \multirow{2}{*}{$\overline{E}$} & \multirow{2}{*}{$8C_3$} & \multirow{2}{*}{$8\overline{C}_3$} & $3C_2$ &  \multirow{2}{*}{$6C_4$} & \multirow{2}{*}{$6\overline{C}_4$} & $6C'_2$ & \multirow{2}{*}{$I$} & \multirow{2}{*}{$\overline{I}$} & \multirow{2}{*}{$8S_6$} & \multirow{2}{*}{$8\overline{S}_6$} & $3\sigma_h$ & \multirow{2}{*}{$6S_4$} & \multirow{2}{*}{$6\overline{S}_4$} &  $6\sigma_d$\\
   & & & & & $3\overline{C}_2$ & & & $6\overline{C}'_2$ & & & & & $3\overline{\sigma}_h$ & & & $6\overline{\sigma}_d$ \\ \midrule
 $A^+_1$  & 1 & 1 & 1 & 1 & 1 & 1 & 1 & 1 & 1 & 1 & 1 & 1 & 1 & 1 & 1 & 1  \\
 $A^+_2$  & 1 & 1 & 1 & 1 & 1 & -1 & -1 & -1 & 1 & 1 & 1 & 1 & 1 & -1 & -1 & -1   \\
 $E^+$  	  & 2 & 2 & -1 & -1 & 2 & 0 & 0 & 0 & 2 & 2 & -1 & -1 & 2 & 0 & 0 & 0  \\
 $T^+_1$  & 3 & 3 & 0 & 0 & -1 & 1 & 1 & -1 & 3 & 3 & 0 & 0 & -1 & 1 & 1 & -1  \\
 $T^+_2$  & 3 & 3 & 0 & 0 & -1 & -1 & -1 & 1 & 3 & 3 & 0 & 0 & -1 & -1 & -1 & 1  \\
  $A^-_1$  & 1 & 1 & 1 & 1 & 1 & 1 & 1 & 1 & -1 & -1 & -1 & -1 & -1 & -1 & -1 & -1  \\
 $A^-_2$  & 1 & 1 & 1 & 1 & 1 & -1 & -1 & -1 & -1 & -1 & -1 & -1 & -1 & 1 & 1 & 1   \\
 $E^-$  	  & 2 & 2 & -1 & -1 & 2 & 0 & 0 & 0 & -2 & -2 & 1 & 1 & -2 & 0 & 0 & 0  \\
 $T^-_1$  & 3 & 3 & 0 & 0 & -1 & 1 & 1 & -1 & -3 & -3 & 0 & 0 & 1 & -1 & -1 & 1  \\
 $T^-_2$  & 3 & 3 & 0 & 0 & -1 & -1 & -1 & 1 & -3 & -3 & 0 & 0 & 1 & 1 & 1 & -1  \\
 $G^+_1$  & 2 & -2 & 1 & -1 & 0 & $\sqrt{2}$ & -$\sqrt{2}$ & 0 & 2 & -2 & 1 & -1 & 0 & $\sqrt{2}$ & -$\sqrt{2}$ & 0  \\
 $G^+_2$  & 2 & -2 & 1 & -1 & 0 & -$\sqrt{2}$ & $\sqrt{2}$ & 0 & 2 & -2 & 1 & -1 & 0 & -$\sqrt{2}$ & $\sqrt{2}$ & 0   \\
 $H^+$  & 4 & -4 & -1 & 1 & 0 & 0 & 0 & 0 & 4 & -4 & -1 & 1 & 0 & 0 & 0 & 0  \\
 $G^-_1$  & 2 & -2 & 1 & -1 & 0 & $\sqrt{2}$ & -$\sqrt{2}$ & 0 & -2 & 2 & -1 & 1 & 0 & -$\sqrt{2}$ & $\sqrt{2}$ & 0  \\
 $G^-_2$  & 2 & -2 & 1 & -1 & 0 & -$\sqrt{2}$ & $\sqrt{2}$ & 0 & -2 & 2 & -1 & 1 & 0 & $\sqrt{2}$ & -$\sqrt{2}$ & 0   \\
 $H^-$  & 4 & -4 & -1 & 1 & 0 & 0 & 0 & 0 & -4 & 4 & 1 & -1 & 0 & 0 & 0 & 0  \\ \midrule          
 $\theta$ & $4\pi$ & $2\pi$ & $2\pi/3$ & $4\pi/3$ & $\pi$ & $\pi/2$ & $3\pi/2$ & $\pi$ & $4\pi$ & $2\pi$ & $2\pi/3$ & $4\pi/3$ & $\pi$ & $\pi/2$ & $3\pi/2$ & $\pi$  \\ \bottomrule
\end{tabular}
}
\end{table}

\begin{table}[ht]
\centering
\caption{Character table for the double group $D^D_{4h}$.}
\label{tab:D4_char}
\renewcommand{\arraystretch}{1.2}
\begin{tabu}{c|cccccccccccccc}
\toprule
\multirow{2}{*}{$D^D_{4h}$} & \multirow{2}{*}{$E$} & \multirow{2}{*}{$\overline{E}$} & \multirow{2}{*}{$2C_4$} & \multirow{2}{*}{$2\overline{C}_4$} & $C_2$ &  $2C'_2$ & $2C''_2$ & \multirow{2}{*}{$I$} & \multirow{2}{*}{$\overline{I}$} & \multirow{2}{*}{$2S_4$} & \multirow{2}{*}{$2\overline{S}_4$} & $\sigma_h$ & $2\sigma_v$ & $2\sigma_d$ \\
   & & & & & $\overline{C}_2$ & $2\overline{C}'_2$ & $2\overline{C}''_2$ & & & & & $\overline{\sigma}_h$ & $2\overline{\sigma}_v$ & $2\overline{\sigma}_d$ \\ \midrule
 $A^+_1$  & 1 & 1 & 1 & 1 & 1 & 1 & 1 & 1 & 1 & 1 & 1 & 1 & 1 & 1  \\
 $A^+_2$  & 1 & 1 & 1 & 1 & 1 & -1 & -1 & 1 & 1 & 1 & 1 & 1 & -1 & -1   \\
 $B^+_1$  & 1 & 1 & -1 & -1 & 1 & 1 & -1 & 1 & 1 & -1 & -1 & 1 & 1 & -1 \\
 $B^+_2$  & 1 & 1 & -1 & -1 & 1 & -1 & 1 & 1 & 1 & -1 & -1 & 1 & -1 & 1  \\
 $E^+$  & 2 & 2 & 0 & 0 & -2 & 0 & 0  & 2 & 2 & 0 & 0 & -2 & 0 & 0    \\
  $A^+_1$  & 1 & 1 & 1 & 1 & 1 & 1 & 1 & -1 & -1 & -1 & -1 & -1 & -1 & -1  \\
 $A^+_2$  & 1 & 1 & 1 & 1 & 1 & -1 & -1 & -1 & -1 & -1 & -1 & -1 & 1 & 1   \\
 $B^+_1$  & 1 & 1 & -1 & -1 & 1 & 1 & -1 & -1 & -1 & 1 & 1 & -1 & -1 & 1 \\
 $B^+_2$  & 1 & 1 & -1 & -1 & 1 & -1 & 1 & -1 & -1 & 1 & 1 & -1 & 1 & -1  \\
 $E^+$  & 2 & 2 & 0 & 0 & -2 & 0 & 0  & -2 & -2 & 0 & 0 & 2 & 0 & 0    \\
 $G^+_1$  & 2 & -2 & $\sqrt{2}$ & $-\sqrt{2}$ & 0 & 0 & 0 & 2 & -2 & $\sqrt{2}$ & $-\sqrt{2}$ & 0 & 0 & 0  \\   
 $G^+_2$  & 2 & -2 & $-\sqrt{2}$ & $\sqrt{2}$ & 0 & 0 & 0 & 2 & -2 & $-\sqrt{2}$ & $\sqrt{2}$ & 0 & 0 & 0   \\   
 $G^-_1$  & 2 & -2 & $\sqrt{2}$ & $-\sqrt{2}$ & 0 & 0 & 0 & -2 & 2 & $-\sqrt{2}$ & $\sqrt{2}$ & 0 & 0 & 0  \\   
 $G^-_2$  & 2 & -2 & $-\sqrt{2}$ & $\sqrt{2}$ & 0 & 0 & 0 & -2 & 2 & $\sqrt{2}$ & $-\sqrt{2}$ & 0 & 0 & 0  \\ \midrule
 $\theta$ & $4\pi$ & $2\pi$ & $\pi/2$ & $3\pi/2$ & $\pi$ & $\pi$ & $\pi$ & $4\pi$ & $2\pi$ & $\pi/2$ & $3\pi/2$ & $\pi$ & $\pi$ & $\pi$   \\ \bottomrule
\end{tabu}
\end{table}

\begin{table}[ht]
\centering
\caption{Character table for the double group $C^D_{4v}$.}
\label{tab:C4_char}
\renewcommand{\arraystretch}{1.2}
\begin{tabu}{c|ccccccc}
\toprule
\multirow{2}{*}{$C^D_{4v}$} & \multirow{2}{*}{$E$} & \multirow{2}{*}{$\overline{E}$} & \multirow{2}{*}{$2C_4$} & \multirow{2}{*}{$2\overline{C}_4$} & $C_2$ & $2\sigma_v$ & $2\sigma_d$ \\
   & & & & & $\overline{C}_2$ & $2\overline{\sigma}_v$ & $2\overline{\sigma}_d$ \\ \midrule
 $A_1$  & 1 & 1 & 1 & 1 & 1 & 1 & 1   \\
 $A_2$  & 1 & 1 & 1 & 1 & 1 & -1 & -1    \\
 $B_1$  & 1 & 1 & -1 & -1 & 1 & 1 & -1  \\
 $B_2$  & 1 & 1 & -1 & -1 & 1 & -1 & 1   \\
 $E$  & 2 & 2 & 0 & 0 & -2 & 0 & 0    \\
 $G_1$  & 2 & -2 & $\sqrt{2}$ & $-\sqrt{2}$ & 0 & 0 & 0  \\   
 $G_2$  & 2 & -2 & $-\sqrt{2}$ & $\sqrt{2}$ & 0 & 0 & 0   \\  \midrule
 $\theta$ & $4\pi$ & $2\pi$ & $\pi/2$ & $3\pi/2$ & $\pi$ & $\pi$ & $\pi$ \\ \bottomrule
\end{tabu}
\end{table}

\begin{table}[ht]
\centering
\caption{Decomposition of angular momentum $J<12$ according to the irreducible representations of the $O^D_h$ group.}
\label{tab:spin_reduction_o}
\renewcommand{\arraystretch}{1.2}
\begin{tabu}{c | l @{\hskip 0.5in}  c | l }
\toprule
$J$ & \multicolumn{1}{c @{\hskip 0.5in}}{$O^D_h$} &$J$ & \multicolumn{1}{c}{$O^D_h$}\\ \cmidrule(r{2.5em}){1-2}\cmidrule(l{-0.5em}){3-4}
 $0$ & $A_1^+$ & $\tfrac{1}{2}$ & $G_1^{\pm}$ \\
 $1$ & $T_1^-$ & $\tfrac{3}{2}$ & $H^{\pm}$ \\
 $2$ & $E^+\oplus T_2^+$ & $\tfrac{5}{2}$ & $G_2^{\pm}\oplus H^{\pm}$ \\
 $3$ & $A_2^-\oplus T^-_1\oplus T^-_2$ & $\tfrac{7}{2}$ & $G_1^{\pm}\oplus G_2^{\pm} \oplus H^{\pm}$ \\ 
 $4$ & $A_1^+\oplus E^+\oplus T^+_1\oplus T^+_2$ & $\tfrac{9}{2}$ & $G_1^{\pm}\oplus 2H^{\pm}$\\
 $5$ & $E^-\oplus 2T^-_1\oplus T^-_2$ & $\tfrac{11}{2}$ & $G_1^{\pm}\oplus G_2^{\pm} \oplus 2H^{\pm}$ \\ 
 $6$ & $A^+_1\oplus A^+_2\oplus E^+\oplus T^+_1\oplus 2 T^+_2$ & $\tfrac{13}{2}$ & $G_1^{\pm}\oplus 2G_2^{\pm} \oplus 2H^{\pm}$ \\
 $7$ & $A^-_2\oplus E^-\oplus  2T^-_1\oplus 2 T^-_2$ & $\tfrac{15}{2}$ & $G_1^{\pm}\oplus G_2^{\pm} \oplus 3H^{\pm}$ \\
 $8$ & $A^+_1\oplus 2E^+\oplus 2T^+_1\oplus 2 T^+_2$ & $\tfrac{17}{2}$ & $2G_1^{\pm}\oplus G_2^{\pm} \oplus 3H^{\pm}$ \\
 $9$ & $A^-_1\oplus A^-_2\oplus E^-\oplus 3 T^-_1\oplus 2 T^-_2$ & $\tfrac{19}{2}$ & $2 G_1^{\pm}\oplus 2G_2^{\pm} \oplus 3 H^{\pm}$ \\
 $10$ & $A^+_1\oplus A^+_2\oplus 2E^+\oplus 2 T^+_1\oplus 3 T^+_2$ &  $\tfrac{21}{2}$ & $G_1^{\pm}\oplus 2G_2^{\pm} \oplus 4 H^{\pm}$ \\
 $11$ & $A^-_2 \oplus 2E^-\oplus 3T^-_1\oplus 3 T^-_2$ & $\tfrac{23}{2}$ & $2G_1^{\pm}\oplus 2G_2^{\pm} \oplus 4H^{\pm}$ \\\bottomrule
\end{tabu}
\end{table}

\begin{table}[ht]
\centering
\caption{Decomposition of angular momentum $J<12$ according to the irreducible representations of the $D^D_{4h}$ group.}
\label{tab:spin_reduction_d}
\renewcommand{\arraystretch}{1.2}
\begin{tabu}{c | l @{\hskip 0.5in}  c | l}
\toprule
$J$ & \multicolumn{1}{c @{\hskip 0.5in}}{$D^D_{4h}$} &$J$ & \multicolumn{1}{c}{$D^D_{4h}$}\\ \cmidrule(r{2.5em}){1-2}\cmidrule(l{-0.5em}){3-4}
 $0$ & $A_1^+$ &  $\tfrac{1}{2}$ & $G_1^{\pm}$ \\
 $1$ & $A^-_2\oplus E^-$ & $\tfrac{3}{2}$ & $G_1^{\pm}\oplus G_2^{\pm}$ \\
 $2$ & $A^+_1\oplus B^+_1\oplus B^+_2\oplus E^+$ & $\tfrac{5}{2}$ & $G_1^{\pm}\oplus 2G_2^{\pm}$ \\
 $3$ & $A^-_2\oplus B^-_1\oplus B^-_2\oplus 2 E^-$ & $\tfrac{7}{2}$ & $2G_1^{\pm}\oplus 2G_2^{\pm}$ \\ 
 $4$ & $2A^+_1\oplus A^+_2\oplus B^+_1\oplus B^+_2\oplus 2E^+$ & $\tfrac{9}{2}$ & $3G_1^{\pm}\oplus 2G_2^{\pm}$ \\
 $5$ & $A^-_1\oplus 2A^-_2\oplus B^-_1\oplus B^-_2\oplus 3E^-$ & $\tfrac{11}{2}$ & $3G_1^{\pm}\oplus 3G_2^{\pm}$ \\ 
 $6$ & $2A^+_1\oplus A^+_2\oplus 2B^+_1\oplus 2B^+_2\oplus 3E^+$ & $\tfrac{13}{2}$ & $3G_1^{\pm}\oplus 4G_2^{\pm}$ \\
 $7$ & $A^-_1\oplus 2A^-_2\oplus 2B^-_1\oplus 2B^-_2\oplus 4E^-$ & $\tfrac{15}{2}$ & $4G_1^{\pm}\oplus 4G_2^{\pm}$ \\
 $8$ & $3A^+_1\oplus 2A^+_2\oplus 2B^+_1\oplus 2B^+_2\oplus 4E^+$ & $\tfrac{17}{2}$ & $5G_1^{\pm}\oplus 4G_2^{\pm}$ \\
 $9$ & $2A^-_1\oplus 3A^-_2\oplus 2B^-_1\oplus 2B^-_2\oplus 5E^-$ & $\tfrac{19}{2}$ & $5G_1^{\pm}\oplus 5G_2^{\pm}$ \\
 $10$ & $3A^+_1\oplus 2A^+_2\oplus 3B^+_1\oplus 3B^+_2\oplus 5E^+$ & $\tfrac{21}{2}$ & $5G_1^{\pm}\oplus 6G_2^{\pm}$ \\
 $11$ & $2A^-_1\oplus 3A^-_2\oplus 3B^-_1\oplus 3B^-_2\oplus 6E^-$ & $\tfrac{23}{2}$ & $6G_1^{\pm}\oplus 6G_2^{\pm}$ \\\bottomrule
\end{tabu}
\end{table}

\begin{table}[ht]
\centering
\caption{Decomposition of angular momentum $J<12$ according to the irreducible representations of the $C^D_{4v}$ group.}
\label{tab:spin_reduction_c}
\renewcommand{\arraystretch}{1.2}
\begin{tabu}{c | l @{\hskip 0.5in} c | l}
\toprule
$J$ & \multicolumn{1}{c @{\hskip 0.5in}}{$C^D_{4v}$} & $J$ & \multicolumn{1}{c}{$C^D_{4v}$}\\ \cmidrule(r{2.5em}){1-2}\cmidrule(l{-0.5em}){3-4}
 $0$ & $A_1$ & $\tfrac{1}{2}$ & $G_1$ \\
 $1$ & $A_1\oplus E$ & $\tfrac{3}{2}$ & $G_1\oplus G_2$ \\
 $2$ & $A_1\oplus B_1\oplus B_2\oplus E$ & $\tfrac{5}{2}$ & $G_1\oplus 2G_2$ \\
 $3$ & $A_1\oplus B_1\oplus B_2\oplus 2E$ & $\tfrac{7}{2}$ & $2G_1\oplus 2G_2$ \\ 
 $4$ & $2A_1\oplus A_2\oplus B_1\oplus B_2\oplus 2E$ & $\tfrac{9}{2}$ & $3G_1\oplus 2G_2$ \\
 $5$ & $2A_1\oplus A_2\oplus B_1\oplus B_2\oplus 3E$ & $\tfrac{11}{2}$ & $3G_1\oplus 3G_2$ \\ 
 $6$ & $2A_1\oplus A_2\oplus 2B_1\oplus 2B_2\oplus 3E$ & $\tfrac{13}{2}$ & $3G_1\oplus 4G_2$ \\
 $7$ & $2A_1\oplus A_2\oplus 2B_1\oplus 2B_2\oplus 4E$ & $\tfrac{15}{2}$ & $4G_1\oplus 4G_2$ \\
 $8$ & $3A_1\oplus 2A_2\oplus 2B_1\oplus 2B_2\oplus 4E$ & $\tfrac{17}{2}$ & $5G_1\oplus 4G_2$ \\
 $9$ & $3A_1\oplus 2A_2\oplus 2B_1\oplus 2B_2\oplus 5E$ & $\tfrac{19}{2}$ & $5G_1\oplus 5G_2$ \\
 $10$ & $3A_1\oplus 2A_2\oplus 3B_1\oplus 3B_2\oplus 5E$ & $\tfrac{21}{2}$ & $5G_1\oplus 6G_2$\\
 $11$ & $3A_1\oplus 2A_2\oplus 3B_1\oplus 3B_2\oplus  6E$ & $\tfrac{23}{2}$ & $6G_1\oplus 6G_2$ \\\bottomrule
\end{tabu}
\end{table}

\begin{table}[ht]
\centering
\caption{Inverse decomposition of angular momentum $J<12$ according to the irreducible representations of the $O^D_h$ group (the number in parentheses indicates the multiplicity of the corresponding $J$ in the associated irreducible representation).}
\label{tab:spin_reduction_o_2}
\renewcommand{\arraystretch}{1.2}
\begin{tabu}{c| l c | l }
\toprule
$O^D_h$ & \multicolumn{1}{c}{$J$} & $O^D_h$ & \multicolumn{1}{c}{$J$} \\ \cmidrule(r){1-2}\cmidrule(l){3-4}
$A^+_1$  & $0,4,6,8,10,\ldots$ & $A^-_1$  & $9,\ldots$  \\
 $A^+_2$  & $6,10,\ldots$ &  $A^-_2$  & $3,7,9,11,\ldots$ \\
 $E^+$  	  & $2,4,6,8(2),10(2),\ldots$ & $E^-$  	  & $5,7,9,11(2),\ldots$ \\
 $T^+_1$  & $4,6,8(2),10(2),\ldots$ & $T^-_1$  & $1,3,5(2),7(2),9(3),11(3),\ldots$ \\
 $T^+_2$  & $2,4,6(2),8(2),10(3),\ldots$ & $T^-_2$  & $3,5,7(2),9(2),11(3),\ldots$ \\\midrule
 $O^D_h$ & \multicolumn{3}{c}{$J$} \\ \midrule
 $G^{\pm}_1$  & \multicolumn{3}{l}{$\tfrac{1}{2},\tfrac{7}{2},\tfrac{9}{2},\tfrac{11}{2},\tfrac{13}{2},\tfrac{15}{2},\tfrac{17}{2}(2),\tfrac{19}{2}(2),\tfrac{21}{2},\tfrac{23}{2}(2),\ldots$} \\ 
 $G^{\pm}_2$  & \multicolumn{3}{l}{$\tfrac{5}{2},\tfrac{7}{2},\tfrac{11}{2},\tfrac{13}{2}(2),\tfrac{15}{2},\tfrac{17}{2},\tfrac{19}{2}(2),\tfrac{21}{2}(2),\tfrac{23}{2}(2),\ldots$} \\
 $H^{\pm}$  & \multicolumn{3}{l}{$\tfrac{3}{2},\tfrac{5}{2},\tfrac{7}{2},\tfrac{9}{2}(2),\tfrac{11}{2}(2),\tfrac{13}{2}(2),\tfrac{15}{2}(3),\tfrac{17}{2}(3),\tfrac{19}{2}(3),\tfrac{21}{2}(4),\tfrac{23}{2}(4),\ldots$} \\\bottomrule
\end{tabu}
\end{table}

\begin{table}[ht]
\centering
\caption{Same as Table~\ref{tab:spin_reduction_o_2} for the $D^D_{4h}$ group.}
\label{tab:spin_reduction_d_2}
\renewcommand{\arraystretch}{1.2}
\begin{tabu}{c| l c | l}
\toprule
$D^D_{4h}$ & \multicolumn{1}{c}{$J$} & $D^D_{4h}$ & \multicolumn{1}{c}{$J$} \\ \cmidrule(r){1-2}\cmidrule(l){3-4}
$A^+_1$ & $0,2,4(2),6(2),8(2),10(3),\ldots$  & $A^-_1$ & $5,7,9(2),11(2),\ldots$ \\
$A^+_2$ & $4,6,8(2),10(2),\ldots$ & $A^-_2$ & $1,3,5(2),7(2),9(3),11(3),\ldots$ \\
$B^+_1$ & $2,4,6(2),8(2),10(3),\ldots$ & $B^-_1$ & $3,5,7(2),9,(2)11(3),\ldots$ \\
$B^+_2$ & $2,4,6(2),8(2),10(3),\ldots$ & $B^-_2$ & $3,5,7(2),9(2),11(3),\ldots$ \\
$E^+$  & $2,4(2),6(3),8(4),10(5),\ldots$ & $E^-$  & $1,3(2),5(3),7(4),9(5),11(6),\ldots$ \\\midrule
$D^D_{4h}$ & \multicolumn{3}{c}{$J$} \\ \midrule
$G^{\pm}_1$  & \multicolumn{3}{l}{$\tfrac{1}{2},\tfrac{3}{2},\tfrac{5}{2},\tfrac{7}{2}(2),\tfrac{9}{2}(3),\tfrac{11}{2}(3),\tfrac{13}{2}(3),\tfrac{15}{2}(4),\tfrac{17}{2}(5),\tfrac{19}{2}(5),\tfrac{21}{2}(5),\tfrac{23}{2}(6),\ldots$} \\
$G^{\pm}_2$  & \multicolumn{3}{l}{$\tfrac{3}{2},\tfrac{5}{2}(2),\tfrac{7}{2}(2),\tfrac{9}{2}(2),\tfrac{11}{2}(3),\tfrac{13}{2}(5),\tfrac{15}{2}(4),\tfrac{17}{2}(4),\tfrac{19}{2}(5),\tfrac{21}{2}(6),\tfrac{23}{2}(6),\ldots$} \\\bottomrule
\end{tabu}
\end{table}

\begin{table}[ht]
\centering
\caption{Same as Table~\ref{tab:spin_reduction_o_2} for the $C^D_{4v}$ group.}
\label{tab:spin_reduction_c_2}
\renewcommand{\arraystretch}{1.2}
\begin{tabu}{c | l}
\toprule
$C^D_{4h}$ & \multicolumn{1}{c}{$J$} \\ \midrule
$A_1$ & $0,1,2,3,4(2),5(2),6(2),7(2),8(3),9(3),10(3),11(3),\ldots$  \\
$A_2$ & $4,5,6,7,8(2),9(2),10(2),11(2),\ldots$ \\
$B_1$ & $2,3,4,5,6(2),7(2),8(2),9(2),10(3),11(3),\ldots$ \\
$B_2$ & $2,3,4,5,6(2),7(2),8(2),9(2),10(3),11(3),\ldots$ \\
$E$  & $1,2,3(2),4(2),5(3),6(3),7(4),8(4),9(5),10(5),11(6),\ldots$ \\
$G_1$  & $\tfrac{1}{2},\tfrac{3}{2},\tfrac{5}{2},\tfrac{7}{2}(2),\tfrac{9}{2}(3),\tfrac{11}{2}(3),\tfrac{13}{2}(3),\tfrac{15}{2}(4),\tfrac{17}{2}(5),\tfrac{19}{2}(5),\tfrac{21}{2}(5),\tfrac{23}{2}(6),\ldots$ \\
$G_2$  & $\tfrac{3}{2},\tfrac{5}{2}(2),\tfrac{7}{2}(2),\tfrac{9}{2}(2),\tfrac{11}{2}(3),\tfrac{13}{2}(5),\tfrac{15}{2}(4),\tfrac{17}{2}(4),\tfrac{19}{2}(5),\tfrac{21}{2}(6),\tfrac{23}{2}(6),\ldots$ \\\bottomrule
\end{tabu}
\end{table}

\begin{table}[ht]
\centering
\caption{Basis vectors up to $J=4$ for the $O_h$ group~\cite{Lee:2017igf}. The index $\nu$ has been omitted from the table since for $J\leq 4$ the multiplicity is always $N(\Gamma,J)=1$.}
\label{tab:vectors_o}
\renewcommand{\arraystretch}{1.2}
\begin{tabu}{c | c | c | c}
\toprule
$\Gamma$ & $J$ & $\sigma$ & Basis in terms of $|J M \rangle$ \\ \midrule
\multirow{2}{*}{$A_1$} & 0 & 1 & $| 0,0 \rangle$\\ \cline{2-4}
 	  & 4 & 1 & $\frac{\sqrt{21}}{6}| 4,0 \rangle+\frac{\sqrt{30}}{12}\left( | 4,4 \rangle + | 4,-4 \rangle \right) $ \\ \hline
$A_2$ & 3 & 1 & $\frac{1}{\sqrt{2}}\left( | 3,2 \rangle - | 3,-2 \rangle \right)$ \\ \hline
\multirow{4}{*}{$E$} & \multirow{2}{*}{2} & 1 & $| 2,0 \rangle$ \\
	  &   & 2 & $\frac{1}{\sqrt{2}}\left( | 2,2 \rangle + | 2,-2 \rangle \right)$ \\\cline{2-4}
	  & \multirow{2}{*}{4} & 1 & $\frac{1}{\sqrt{2}}\left( | 4,2 \rangle + | 4,-2 \rangle \right)$ \\
	  &   & 2 & $\frac{\sqrt{15}}{6}| 4,0 \rangle-\frac{\sqrt{42}}{12}\left( | 4,4 \rangle + | 4,-4 \rangle \right)$ \\ \hline
\multirow{9}{*}{$T_1$} & \multirow{3}{*}{1} & 1 & $| 1,0 \rangle$ \\
	  &   & 2 & $\frac{1}{\sqrt{2}}\left( | 1,1 \rangle - | 1,-1 \rangle \right)$ \\
	  &   & 3 & $\frac{\imag}{\sqrt{2}}\left( | 1,1 \rangle + | 1,-1 \rangle \right)$ \\\cline{2-4}
	  & \multirow{3}{*}{3} & 1 & $| 3,0 \rangle$ \\
	  &   & 2 & $\frac{\sqrt{3}}{4}\left( | 3,1 \rangle - | 3,-1 \rangle \right)-\frac{\sqrt{5}}{4}\left( | 3,3 \rangle - | 3,-3 \rangle \right)$ \\
	  &  & 3 & $-\frac{\imag\sqrt{3}}{4}\left( | 3,1 \rangle + | 3,-1 \rangle \right)-\frac{\imag\sqrt{5}}{4}\left( | 3,3 \rangle + | 3,-3 \rangle \right)$ \\\cline{2-4}
	  & \multirow{3}{*}{4} & 1 & $-\frac{1}{\sqrt{2}}\left( | 4,4 \rangle - | 4,-4 \rangle \right)$ \\
	  &   & 2 & $-\frac{1}{4}\left( | 4,3 \rangle + | 4,-3 \rangle \right)-\frac{\sqrt{7}}{4}\left( | 4,1 \rangle + | 4,-1 \rangle \right)$ \\
	  &   & 3 & $-\frac{\imag}{4}\left( | 4,3 \rangle - | 4,-3 \rangle \right)+\frac{\imag\sqrt{7}}{4}\left( | 4,1 \rangle - | 4,-1 \rangle \right)$ \\ \hline
\multirow{9}{*}{$T_2$} & \multirow{3}{*}{2} & 1 & $-\frac{1}{\sqrt{2}}\left( | 2,1 \rangle + | 2,-1 \rangle \right)$ \\
	  &   & 2 & $-\frac{i}{\sqrt{2}}\left( | 2,1 \rangle - | 2,-1 \rangle \right)$ \\
	  &   & 3 & $\frac{1}{\sqrt{2}}\left( | 2,2 \rangle - | 2,-2 \rangle \right)$ \\\cline{2-4}
	  & \multirow{3}{*}{3} & 1 & $\frac{1}{\sqrt{2}}\left( | 3,2 \rangle + | 3,-2 \rangle \right)$ \\
	  &   & 2 & $\frac{\sqrt{5}}{4}\left( | 3,1 \rangle - | 3,-1 \rangle \right)+\frac{\sqrt{3}}{4}\left( | 3,3 \rangle - | 3,-3 \rangle \right)$ \\
	  &   & 3 & $\frac{\imag\sqrt{5}}{4}\left( | 3,1 \rangle + | 3,-1 \rangle \right)-\frac{\imag\sqrt{3}}{4}\left( | 3,3 \rangle + | 3,-3 \rangle \right)$ \\\cline{2-4}
	  & \multirow{3}{*}{4} & 1 & $-\frac{1}{\sqrt{2}}\left( | 4,2 \rangle - | 4,-2 \rangle \right)$ \\	
	  &   & 2 & $-\frac{1}{4}\left( | 4,1 \rangle + | 4,-1 \rangle \right)+\frac{\sqrt{7}}{4}\left( | 4,3 \rangle + | 4,-3 \rangle \right)$ \\
	  &   & 3 & $-\frac{\imag}{4}\left( | 4,1 \rangle - | 4,-1 \rangle \right)-\frac{\imag\sqrt{7}}{4}\left( | 4,3 \rangle - | 4,-3 \rangle \right)$ \\ \bottomrule
\end{tabu}
\end{table}

\begin{table}[ht]
\centering
\caption{Same as Table~\ref{tab:vectors_o} for the $D_{4h}$ group~\cite{Lee:2017igf}.}
\label{tab:vectors_d}
\renewcommand{\arraystretch}{1.2}
\begin{tabu}{c | c | c | c}
\toprule
$\Gamma$ & $J$ & $\sigma$ & Basis in terms of $|J M \rangle$ \\ \midrule
\multirow{2}{*}{$A_1$} & 0 & 1 & $| 0,0 \rangle$\\\cline{2-4}
	  & 2 & 1 & $| 2,0 \rangle$ \\\hline
$A_2$ & 1 & 1 & $| 1,0 \rangle$ \\\hline
$B_1$ & 2 & 1 & $\frac{1}{\sqrt{2}}\left( | 2,2 \rangle + | 2,-2 \rangle \right)$ \\\hline
$B_2$ & 2 & 1 & $\frac{1}{\sqrt{2}}\left( | 2,2 \rangle - | 2,-2 \rangle \right)$ \\\hline
\multirow{4}{*}{$E$} & \multirow{2}{*}{1} & 1 & $\frac{1}{\sqrt{2}}\left( | 1,1 \rangle - | 1,-1 \rangle \right)$ \\
      &   & 2 & $\frac{1}{\sqrt{2}}\left( | 1,1 \rangle + | 1,-1 \rangle \right)$ \\\cline{2-4}
      & \multirow{2}{*}{2} & 1 & $\frac{1}{\sqrt{2}}\left( | 2,1 \rangle + | 2,-1 \rangle \right) $ \\
	  &   & 2 & $\frac{1}{\sqrt{2}}\left( | 2,1 \rangle - | 2,-1 \rangle \right)$ \\\bottomrule
\end{tabu}
\end{table}

\begin{table}[ht]
\centering
\caption{Same as Table~\ref{tab:vectors_o} for the $C_{4h}$ group~\cite{Gockeler:2012yj}.}
\label{tab:vectors_c}
\renewcommand{\arraystretch}{1.2}
\begin{tabu}{c | c | c | c}
\toprule
$\Gamma$ & $J$ & $\sigma$ & Basis in terms of $|J M \rangle$ \\ \midrule
\multirow{3}{*}{$A_1$} & 0 & 1 & $| 0,0 \rangle$\\\cline{2-4}
	  & 1 & 1 & $| 1,0 \rangle$\\\cline{2-4}
	  & 2 & 1 & $| 2,0 \rangle$ \\ \hline
$B_1$ & 2 & 1 & $\frac{1}{\sqrt{2}}\left( | 2,2 \rangle + | 2,-2 \rangle \right)$\\ \hline
$B_2$ & 2 & 1 & $\frac{1}{\sqrt{2}}\left( -| 2,2 \rangle + | 2,-2 \rangle \right)$\\ \hline
\multirow{4}{*}{$E$} & \multirow{2}{*}{1} & 1 & $-\frac{1+\imag}{2} | 1,1 \rangle + \frac{1-\imag}{2}| 1,-1 \rangle$ \\
	  &   & 2 & $\frac{1}{\sqrt{2}}\left( | 1,1 \rangle - \imag| 1,-1 \rangle \right)$ \\\cline{2-4}
	  & \multirow{2}{*}{2} & 1 & $\frac{1}{\sqrt{2}}\left( -\imag| 2,1 \rangle + | 2,-1 \rangle \right)$ \\
	  &   & 2 & $\frac{1+\imag}{2} | 2,1 \rangle + \frac{1-\imag}{2}| 2,-1 \rangle$ \\\bottomrule
\end{tabu}
\end{table}
 %Group theory tables

% Appendix 1 - *******************************************

\chapter{Evaluation of the \texorpdfstring{$\mathcal{Z}$}{Z}-function}\label{appen:Zfun}

There are several ways to rewrite Eq.~\eqref{eq:zeta_fun} to make it more suitable for numerical evaluation. In his papers~\cite{Luscher:1986pf,Luscher:1990ux}, M.~Lüscher already proposed an integral version of the $\mathcal{Z}$-function (a generalization to boosted systems with unequal masses was presented in Ref.~\cite{Rummukainen:1995vs}). 
In the following we will discuss three versions of the $\mathcal{Z}$-function that can be found in the literature~\cite{Leskovec:2012gb,Yamazaki:2004qb,Beane:2011sc}, and how they relate to each other. We will work the more general case of boosted hadrons with unequal masses (it is easy to take the limit to zero boost and equal masses). We should note though that each of these versions were preceded by works where unboosted systems~\cite{Lellouch:2011qw}, asymmetric lattices~\cite{Feng:2004ua,Detmold:2004qn}, or equal masses~\cite{Feng:2011ah} were considered.
The starting point is Eq.~\eqref{eq:zeta_fun},
\begin{equation}
    \mathcal{Z}^{\bm{d}}_{lm}(s;q^2)=\sum_{\bm{r}\in P_d}\frac{\mathcal{Y}_{lm}(\bm{r})}{\left( |\bm{r}|^2-q^2\right) ^s}\, .
\label{eq:zfun_appen}
\end{equation}
The first step in all the derivations is to rewrite the denominator of Eq.~\eqref{eq:zfun_appen}, $1/\left( |\bm{r}|^2-q^2\right) ^s$, using the integral form of the Gamma function $\Gamma(s)/x^s=\int_0^{\infty}\,dt\, t^{s-1}e^{tx}$,
\begin{equation}
\begin{aligned}
    \mathcal{Z}^{\bm{d}}_{lm}(s;q^2)&=\frac{1}{\Gamma(s)}\sum_{\bm{r}\in P_d} \mathcal{Y}_{lm}(\bm{r})\int_0^{\infty}dt\,t^{s-1}e^{-t( |\bm{r}|^2-q^2)}\\
    &=\frac{1}{\Gamma(s)}\sum_{\bm{r}\in P_d} \mathcal{Y}_{lm}(\bm{r})\left[\int_0^1 dt\,t^{s-1}e^{-t( |\bm{r}|^2-q^2)}+\int_1^{\infty} dt\,t^{s-1}e^{-t( |\bm{r}|^2-q^2)}\right]\, .
\label{eq:two_integrals}
\end{aligned}
\end{equation}
The second integral can be explicitly evaluated for $s=1$,
\begin{equation}
    \frac{1}{\Gamma(s)}\sum_{\bm{r}\in P_d} \mathcal{Y}_{lm}(\bm{r})\int_1^{\infty} dt\,t^{s-1}e^{-t( |\bm{r}|^2-q^2)}\;\overset{s\rightarrow 1}{\longrightarrow}\;\sum_{\bm{r}\in P_d}\mathcal{Y}_{lm}(\bm{r})\frac{e^{-( |\bm{r}|^2-q^2)}}{|\bm{r}|^2-q^2}\, .
\end{equation}
To evaluate the first integral, we use the Poisson summation formula,
\begin{equation}
    \sum_{\bm{n}\in\mathbb{Z}^3}f(\bm{n})=\sum_{\bm{n}\in\mathbb{Z}^3}\int d^3\bm{x} \,f(\bm{x}) e^{\imag 2\pi\bm{n}\cdot\bm{x}}\, .
    \label{eq:Poissformula}
\end{equation}
By changing the sum over $P_d$ by the sum over integers, $\sum_{\bm{r}\in P_d}f(\bm{r})=\sum_{\bm{n}\in\mathbb{Z}^3} f(\bm{r}(\bm{n}))$, with $\bm{r}(\bm{n})=\hat{\gamma}^{-1}(\bm{n}+\alpha\bm{d})$, the first integral takes the form
\begin{equation}
    \frac{1}{\Gamma(s)}\int_0^1 dt\, t^{s-1}e^{tq^2}\sum_{\bm{n}\in\mathbb{Z}^3}\int d^3\bm{x}\; \mathcal{Y}_{lm}(\bm{r}(\bm{x})) e^{-t|\bm{r}(\bm{x})|^2}e^{\imag 2\pi\bm{n}\cdot\bm{x}}\, .
    \label{eq:ap1_trick1}
\end{equation}
Changing the integration variables from $\bm{x}$ to $\bm{r}$, such that $\bm{x}=\hat{\gamma}\bm{r}-\alpha\bm{d}$ and $d^3\bm{x}=\gamma d^3\bm{r}$,
\begin{equation}
\begin{aligned}
    \frac{1}{\Gamma(s)}&\int_0^1 dt\, t^{s-1}e^{tq^2}\sum_{\bm{n}\in\mathbb{Z}^3}\int \gamma d^3\bm{r}\;\mathcal{Y}_{lm}(\bm{r}) e^{-t|\bm{r}|^2}e^{\imag 2\pi\bm{n}\cdot(\hat{\gamma}\bm{r}-\alpha\bm{d})}\\
    &=\frac{\gamma}{\Gamma(s)}\int_0^1 dt\, t^{s-1}e^{tq^2}\sum_{\bm{n}\in\mathbb{Z}^3}e^{-\imag 2\pi\alpha\bm{n}\cdot\bm{d}}\int d^3\bm{r} \,|\bm{r}|^l Y_{lm}(\hat{\bm{r}}) e^{-t|\bm{r}|^2}e^{\imag 2\pi\bm{n}\cdot(\hat{\gamma}\bm{r})}\\
    &=\frac{\gamma}{\Gamma(s)}\int_0^1 dt\, t^{s-1}e^{tq^2}\sum_{\bm{n}\in\mathbb{Z}^3}e^{-\imag 2\pi\alpha\bm{n}\cdot\bm{d}}\int_0^{\infty} |\bm{r}|^2 d|\bm{r}| \,e^{-t|\bm{r}|^2} |\bm{r}|^l \int d\Omega_r\, Y_{lm}(\hat{\bm{r}}) e^{\imag\bm{k}\cdot\bm{r}}\, ,
    \label{eq:ap1_trick2}
\end{aligned}
\end{equation}
with $\bm{k}=2\pi\hat{\gamma}\bm{n}$, and where we have rewritten $\bm{n}\cdot(\hat{\gamma}\bm{r})=(\bm{n}_{\parallel}+\bm{n}_{\perp})\cdot(\gamma \bm{r}_{\parallel}+\bm{r}_{\perp})=\gamma \bm{n}_{\parallel}\cdot \bm{r}_{\parallel}+\bm{n}_{\perp}\cdot\bm{r}_{\perp}=(\hat{\gamma}\bm{n})\cdot \bm{r}$. The angular integral can be performed by expanding $e^{i\bm{k}\cdot\bm{r}}$ in spherical harmonics,
\begin{equation}
\begin{aligned}
    \int d\Omega_r\, Y_{lm}(\hat{\bm{r}}) e^{i\bm{k}\cdot\bm{r}}&=\int d\Omega_r\, Y_{lm}(\hat{\bm{r}}) 4\pi \sum_{l'=0}^{\infty}\sum_{m'=-l'}^{l'} \imag^{l'}j_{l'}(|\bm{k}||\bm{r}|) Y_{l'm'}(\hat{\bm{k}})Y^*_{l'm'}(\hat{\bm{r}})\\
    &=4\pi \sum_{l'=0}^{\infty}\sum_{m'=-l'}^{l'} \imag^{l'}j_{l'}(|\bm{k}||\bm{r}|) Y_{l'm'}(\hat{\bm{k}})\underbrace{\int d\Omega_r\, Y_{lm}(\hat{\bm{r}})Y^*_{l'm'}(\hat{\bm{r}})}_{\delta_{ll'}\delta_{mm'}}\\
    &=4\pi\, \imag^l j_{l} (|\bm{k}||\bm{r}|) Y_{lm}(\hat{\bm{k}})\,  ,
\end{aligned}
\end{equation}
with $j_l(x)$ being the spherical Bessel functions. Now, the integral over $|\bm{r}|$ can be analytically computed, which gives
\begin{equation}
    \int_0^{\infty}|\bm{r}|^2 d|\bm{r}| \, e^{-t|\bm{r}|^2}|\bm{r}|^l j_l(|\bm{k}||\bm{r}|)=\frac{1}{4\pi}\left( \frac{\pi}{t}\right)^{3/2}\left( \frac{|\bm{k}|}{2t}\right)^{l}e^{-|\bm{k}|^2/4t}\, .
\end{equation}
Collecting all the terms, the first integral looks like
\begin{equation}
\begin{aligned}
    \frac{\gamma}{\Gamma(s)}&\int_0^1 dt\, t^{s-1}e^{tq^2}\sum_{\bm{n}\in\mathbb{Z}^3}e^{-\imag 2\pi\alpha\bm{n}\cdot\bm{d}} \left( \frac{\pi}{t} \right)^{3/2} \imag^l \underbrace{ \left( \frac{|\bm{k}|}{2t} \right)^{l}  Y_{lm}(\hat{\bm{k}})}_{\mathcal{Y}_{lm}(\bm{k}/2t)} e^{-|\bm{k}|^2/4t} \\
    &=\frac{\gamma}{\Gamma(s)}\int_0^1 dt\, t^{s-1}e^{tq^2}\sum_{\bm{n}\in\mathbb{Z}^3}e^{-\imag 2\pi\alpha\bm{n}\cdot\bm{d}} \left( \frac{\pi}{t} \right)^{3/2} \imag^l\, \mathcal{Y}_{lm}(\pi \hat{\gamma}\bm{n}/t) e^{-\pi^2|\hat{\gamma}\bm{n}|^2/t}\, .
\end{aligned}
\end{equation}
In the case of $s=1$, the integral over $t$ is finite for all $\bm{n}$ except for $\bm{n}=\bm{0}$. In this case, the divergence occurs only for $l=m=0$, since $\mathcal{Y}_{lm}(\bm{n}=\bm{0})\, \propto\, \delta_{l0}\delta_{m0}$. Then, splitting the sum between $\bm{n}=\bm{0}$ and $\bm{n}\neq \bm{0}$, and looking at the $\bm{n}=\bm{0}$ part,
\begin{equation}
\begin{aligned}
    \frac{\gamma}{\Gamma(s)}&\int_0^1 dt\, t^{s-1}e^{tq^2} \left( \frac{\pi}{t}\right)^{3/2}\frac{1}{\sqrt{4\pi}}\delta_{l0}\delta_{m0}=\frac{\gamma}{\Gamma(s)}\frac{\pi}{2}\int_0^1 dt\, t^{s-5/2}e^{tq^2} \delta_{l0}\delta_{m0}\, .
\end{aligned}
\label{eq:divergent_integral}
\end{equation}
Up to here, all derivations follow these steps.\footnote{In the case of Ref.~\cite{Yamazaki:2004qb}, the steps are slightly different, but the final result is the same.} The differences between the different versions appear in the way this divergent integral is computed:
\begin{itemize}
    \item In Ref.~\cite{Leskovec:2012gb}, by adding and subtracting $t^{s-5/2}$ into the integral,
    \begin{equation}
        \int_0^1 dt\, t^{s-5/2}e^{tq^2} = \int_0^1 dt\, t^{s-5/2}\left(e^{tq^2}-1\right) +\int_0^1 dt\, t^{s-5/2} \, ,
    \end{equation}
    the first integral becomes finite for $s=1$, while the second integral, which comes out finite only for $s > 3/2$, $\int_0^1 dt\, t^{s-5/2} \overset{s>3/2}{=}\frac{1}{s-3/2}$, can be analytically continued to $s = 1$, $\frac{1}{s-3/2}\overset{s\rightarrow 1}{\longrightarrow}-2$.
    \item In Ref.~\cite{Yamazaki:2004qb}, the exponential is replaced by its Taylor expansion,
    \begin{equation}
        \int_0^1 dt\, t^{s-5/2}e^{tq^2} = \int_0^1 dt\, t^{s-5/2} \sum_{\rho=0}^{\infty}\frac{(tq^2)^\rho}{\rho!}=\sum_{\rho=0}^{\infty} \frac{(q^2)^\rho}{\rho!} \int_0^1 dt\, t^{s+\rho-5/2} \, .
    \end{equation}
    Again, the integral is finite only for $s > 3/2$, but it is analytically continued to $s = 1$,
    \begin{equation}
        \sum_{\rho=0}^{\infty} \frac{(q^2)^\rho}{\rho!} \frac{1}{s+\rho-3/2}\;\overset{s\rightarrow 1}{\longrightarrow}\; \sum_{\rho=0}^{\infty}\frac{(q^2)^\rho}{\rho!}\frac{1}{\rho-1/2}\, .
    \end{equation}
    \item In Ref.~\cite{Beane:2011sc}, the divergent integral is integrated by parts, and in the last step it is analytically continued to $s = 1$,
    \begin{equation}
    \begin{aligned}
        \int_0^1 dt\, t^{s-5/2}e^{tq^2} &= \frac{e^{q^2}}{s-3/2}- \int_0^1dt\, q^2e^{tq^2}\frac{t^{s-3/2}}{s-3/2}  \\
        &\overset{s\rightarrow 1}{\longrightarrow} -2e^{q^2}+\int_0^1dt\, e^{tq^2}2t^{-1/2}q^2=-2 e^{q^2}+\int^1_0 dt \frac{e^{tq^2}}{t^{3/2}}2tq^2\, ,
    \end{aligned}    
    \end{equation}
    To recover the result from Ref.~\cite{Detmold:2004qn}, one has to perform another integration by parts,
    \begin{equation}
        -2 e^{q^2}+\int^1_0 dt \,\frac{e^{tq^2}}{t^{3/2}}2tq^2 = 2 e^{q^2}(2q^2-1)-\int^1_0 dt \, \frac{e^{tq^2}}{t^{3/2}} 4t^2q^4 \, .
    \end{equation}
\end{itemize}
Therefore, the final version can be written as
\begin{equation}
\begin{aligned}
    \mathcal{Z}^{\bm{d}}_{lm}(1;q^2)=&\gamma \int_0^1 dt\, e^{tq^2}\sum_{\substack{\bm{n}\in \mathbb{Z}^3\\\bm{n}\neq \bm{0}}} e^{-\imag2\pi\alpha\bm{n}\cdot\bm{d}} \left( \frac{\pi}{t} \right)^{3/2}  \imag^l \, \mathcal{Y}_{lm}(\pi \hat{\gamma}\bm{n}/t) e^{-\pi^2|\hat{\gamma}\bm{n}|^2/t}\\
    &+\sum_{\bm{r}\in P_d} \mathcal{Y}_{lm}(\bm{r})\frac{e^{-( |\bm{r}|^2-q^2)}}{|\bm{r}|^2-q^2}+\gamma\frac{\pi}{2}\begin{Bmatrix}[l]
     \int_0^1 dt\, \frac{1}{t^{3/2}}(e^{tq^2}-1)-2 \\[2ex]
     \sum_{\rho=0}^{\infty}\frac{1}{\rho-1/2}\frac{(q^2)^{\rho}}{\rho!}\\[2ex]
     \int^1_0 dt \frac{e^{tq^2}}{t^{3/2}}2tq^2 -2e^{q^2}
    \end{Bmatrix}\delta_{l0}\delta_{m0}\, ,
\end{aligned}
\end{equation}
where each row inside the curly brackets represents the result of each version. The first and third integral give the same result~\cite{Mathematica},
\begin{equation}
    \int_0^1 dt\, \frac{1}{t^{3/2}}(e^{tq^2}-1)-2 = \int^1_0 dt \frac{e^{tq^2}}{t^{3/2}}2tq^2 -2e^{q^2} = 2e^{q^2}\left[-1+2\sqrt{q^2}D\left(\sqrt{q^2}\right)\right] \, ,
    \label{eq:intDawsonrelation}
\end{equation}
where $D(x)$ is the Dawson integral~\cite{Dawson:1897}, which is defined as $D(x)=e^{-x^2}\int_0^x dy\;e^{y^2}$. For the sum, the result is
\begin{equation}
    \sum_{\rho=0}^{\infty}\frac{1}{\rho-1/2}\frac{(q^2)^{\rho}}{\rho!} = -2\, {}_1F_1\left(-\frac{1}{2};\frac{1}{2};q^2\right)\, ,
    \label{eq:sum1F1relation}
\end{equation}
where ${}_1F_1(a;b;x)$ is the confluent hypergeometric function of the first kind~\cite{abramowitz1964handbook},
\begin{equation}
    {}_1F_1(a;b;x) = \sum_{\rho=0}^{\infty} \frac{(a)_\rho}{(b)_\rho}\frac{x^\rho}{\rho!}\, , \quad \text{with} \;\;(\vartheta)_{\rho} = \Gamma(\vartheta+\rho)/\Gamma(\vartheta)\, .
\end{equation}
In Ref.~\cite{Nijimbere:2017}, the relation between the Dawson integral and ${}_1F_1$ is shown,
\begin{equation}
    D\left(\sqrt{q^2}\right) = \sqrt{q^2} e^{-q^2}{}_1F_1\left(\frac{1}{2};\frac{3}{2};q^2\right)\, .
    \label{eq:Dawson1F1relation}
\end{equation}
The only piece left is to relate ${}_1F_1\left(-\frac{1}{2};\frac{1}{2};q^2\right)$ with ${}_1F_1\left(\frac{1}{2};\frac{3}{2};q^2\right)$, which can be done using the following recurrence relation~\cite{abramowitz1964handbook},
\begin{equation}
    b\, {}_1F_1\left(a;b;x\right)-b\, {}_1F_1\left(a-1;b;x\right)-x\, {}_1F_1\left(a;b+1;x\right)=0\, .
\end{equation}
Setting $a=b=\frac{1}{2}$ and combining Eqs.~\eqref{eq:sum1F1relation} and~\eqref{eq:Dawson1F1relation},
\begin{equation}
\begin{aligned}
    -2\, {}_1F_1\left(-\frac{1}{2};\frac{1}{2};q^2\right) &= -2\, {}_1F_1\left(\frac{1}{2};\frac{1}{2};q^2\right)+4q^2\, {}_1F_1\left(\frac{1}{2};\frac{3}{2};q^2\right)\\
    &=-2 e^{q^2} + 4e^{q^2} \sqrt{q^2} D\left(\sqrt{q^2}\right)\, ,
\end{aligned}
\end{equation}
which is exactly Eq.~\eqref{eq:intDawsonrelation}.

To conclude, just note that an intermediate version between the raw and accelerated form can be found in early 
studies of baryon-baryon interactions~\cite{Beane:2003yx,Beane:2003da}, where the evaluation of the $\mathcal{Z}$-function by brute force takes to the following form,
\begin{equation}
\begin{aligned}
    \mathcal{Z}_{00}(1;q^2)=\lim_{\Lambda\rightarrow \infty} \left(\sum_{\bm{n}=\bm{0}}^{\Lambda} \frac{1}{|\bm{n}|^2-q^2}-4\pi\Lambda\right)\, .
\end{aligned}
\end{equation} %Z-function

% Appendix 3 - *******************************************

\chapter{Summary tables of baryon-baryon channels}\label{appen:BBsummary}

This appendix contains summary tables of the scattering parameters predicted by phenomenological models and EFTs for different two-baryon systems: $\Lambda N$ (Table~\ref{tab:LNscatt}), $\Sigma N$ (Table~\ref{tab:SNscatt}), $\Lambda \Lambda$ or $H$-dibaryon (Table~\ref{tab:LLscatt}), $\Sigma \Sigma$ (Table~\ref{tab:SSscatt}), $\Xi N$ (Table~\ref{tab:XNscatt}), $\Xi \Sigma$ (Table~\ref{tab:XSscatt}), and $\Xi \Xi$ (Table~\ref{tab:XXscatt}). 

This appendix also contains summary tables of the binding energies, as defined in Eq.~\eqref{eq:bind_en}, computed with fully dynamical LQCD calculations: $NN$ (Table~\ref{tab:NNLQCD_summary}), $\Lambda N$ (Table~\ref{tab:LNLQCD_summary}), $\Sigma N$ (Table~\ref{tab:SNLQCD_summary}), $\Lambda \Lambda$ or $H$-dibaryon (Table~\ref{tab:LLLQCD_summary}), $\Sigma \Sigma$ (Table~\ref{tab:SSLQCD_summary}), $\Xi N$ (Table~\ref{tab:XNLQCD_summary}), $\Xi \Sigma$ (Table~\ref{tab:XSLQCD_summary}), and $\Xi \Xi$ (Table~\ref{tab:XXLQCD_summary}). Some results, by virtue of $SU(3)_f$ symmetry, are applicable to different channels belonging to the same irrep (e.g., the CalLat Collaboration is only focused on the $NN$ channels, but since their calculations are performed with $n_f=3$, their results also apply to other channels belonging to the $\mathbf{27}$ irrep). In these tables, there are three additional symbols: $\dag$ indicates that Lüscher's formalism is used to compute the binding energy, $\ddag$ indicates that a continuum extrapolation is performed, and $\S$ that only the potential is shown (no physical quantity, like the energy shift or the phase shift).

The references in the tables are sorted chronologically.

\begin{sidewaystable}
\centering
\renewcommand{\arraystretch}{1.2}
\caption{Scattering parameters for the $\Lambda N$ ($I=1/2$) system}
\label{tab:LNscatt}
\resizebox{\columnwidth}{!}{
\begin{tabular}{lccccc}
\toprule
\multicolumn{2}{l}{Reference}	& $a^{\Lambda N (\1s0)}$ [fm] & $r^{\Lambda N (\1s0)}$ [fm]	&	$a^{\Lambda N (\3s1)}$ [fm] & $r^{\Lambda N (\3s1)}$ [fm] \\\cmidrule(r){1-2}\cmidrule(lr){3-4}\cmidrule(l){5-6}
ND73~\cite{Nagels:1973rq}	&	Nijmegen hard-core model 	&	$-2.46\pm 0.32$	&	$2.03\pm 0.10$	&	$-1.18 \pm 0.06$	&	$2.40\pm 0.09$	\\
ND77~\cite{Nagels:1976xq}	&	Nijmegen hard-core model 	&	$-1.90\pm 0.30$	&	$3.72\pm 0.34$	&	$-1.96 \pm 0.11$	&	$3.24\pm 0.11$	\\
NF78~\cite{Nagels:1978sc}	&	Nijmegen hard-core model 	&	$-2.29$	&	$3.17$	&	$-1.88$	&	$3.36$	\\
Jülich 89~\cite{Holzenkamp:1989tq}	&	Jülich model (models A and B) 	&	$-1.60,  -0.57$	&	$1.33,  7.65$	&	$-1.60, -1.94$	&	$3.15,  2.42$	\\
NSC89~\cite{Maessen:1989sx}	&	Nijmegen soft-core model	&	$-2.78$	&	$2.88$	&	$-1.41$	&	$3.11$	\\
Jülich 94~\cite{Reuber:1993ip}	&	Jülich model (models A, $\tilde{\text{A}}$, B and $\tilde{\text{B}}$) 	&	$-2.04\cdots -0.40$	&	$0.64\cdots 12.28$	&	$-2.12\cdots -1.33$	&	$2.43\cdots 3.91$	\\
Fujiwara \textit{et al.} 96~\cite{Fujiwara:1996qj}	&	$SU_6$ quark model (RGM-H, FSS and RGM-F)	&	$-5.34,-5.39,-2.03$	&	$2.46,2.26,3.05$	&	$-1.04,-1.02,-1.66$	&	$4.92,4.20,3.26$	\\
Ehime 98~\cite{Tamotsu:1998,Tominaga:1998}	&	Ehime model (sets 2, A and B)	&	$-2.76,-2.71,-2.65$	&	$3.19,3.21,3.24$	&	$-2.064,-1.95,-1.80$	&	$3.46,3.56,3.71$\\
NSC97~\cite{Rijken:1998yy, Stoks:1999bz}	& Nijmegen soft-core model (models a$\cdots$f)	&	$-2.51 \cdots -0.71$	&	$3.03 \cdots 5.86$	&	$-2.18 \cdots -1.75$	&	$2.76 \cdots 3.32$	\\
Korpa \textit{et al.} 02~\cite{Korpa:2001au}	& EFT NLO	& $-2.50$	&	$1.61$	& $-1.78$	&	$1.42$	\\
ESC03~\cite{RIJKEN200527}	&	Nijmegen extended-soft-core model	&	$-2.119$	&	$3.177$	&	$-1.824$	&	$2.846$	\\
Jülich 04~\cite{Haidenbauer:2005zh}	&	Jülich model	&	$-2.56$	&	$2.75$	&	$-1.66$	&	$2.93$	\\
ESC04~\cite{Rijken:2006ep}	&	Nijmegen extended-soft-core model (models a$\cdots$d)	&	$-2.073 \cdots -1.941$	&	$2.998 \cdots 3.570$	&	$-1.858 \cdots -1.537$	&	$2.773 \cdots 3.133$	\\
Polinder \textit{et al.} 06~\cite{Polinder:2006zh}	&	$\chi$EFT LO $\Lambda=550\cdots 700$ MeV	&	$-1.91\cdots -1.90$	&	$1.35 \cdots 1.44$	&	$-1.23\cdots -1.22$	&	$2.05 \cdots 2.27$	\\
fss2 07~\cite{Fujiwara:2006yh}	&	$SU_6$ quark model (isospin and particle basis)	&	$-2.59,-2.59$	&	$2.83,2.83$	&	$-1.60,-1.60$	&	$3.01,3.00$	\\
ESC08~\cite{Rijken:2010zzb,Nagels08II:2015}	&	Nijmegen extended-soft-core model (models a'' and c)	&	$-2.70, -2.46$	&	$2.97, 3.14$	&	$-1.65, -1.73$	&	$3.63, 3.55$	\\
Haidenbauer \textit{et al.} 13~\cite{Haidenbauer:2013oca}	& $\chi$EFT NLO $\Lambda=450\cdots 700$ MeV	& $-2.91 \cdots -2.90$	&	$2.56 \cdots 2.86$	& $-1.70 \cdots -1.48$	&	$2.62 \cdots 3.44$	\\
ESC16~\cite{Nagels:2015lfa}	&	Nijmegen extended-soft-core model	&	$-1.88$	&	$3.58$	&	$-1.86$	&	$3.37$	\\
Haidenbauer \textit{et al.} 20~\cite{Haidenbauer:2019boi}	& $\chi$EFT NLO $\Lambda=500\cdots 650$ MeV	& $-2.91 \cdots -2.90$	&	$2.65 \cdots 3.10$	& $-1.52 \cdots -1.40$	&	$2.53 \cdots 2.62$	\\
Ren \textit{et al.} 20~\cite{Ren:2019qow}	&	Relativistic $\chi$EFT LO $\Lambda=20$ GeV	&	$-2.94$	&	$-1.44$	&	$-1.41$	&	$1.61$	\\
Le \textit{et al.} 20~\cite{Le:2019gjp}	&	$\chi$EFT NLO $\Lambda=600$ MeV	&	$-4,-4.5,-5$	&	-	&	$-1.22,-1.15,1.09$	&	-	\\ \midrule
Alexander \textit{et al.} 68~\cite{Alexander:1969cx}	&	ERE approx.\ (CERN, Saclay bubble chamber)	&	$-1.8^{+2.3}_{-4.2}$	&	$2.8^{+14}_{-2.8}$	&	$-1.6^{+1.1}_{-0.8}$	&	$3.3^{+13.7}_{-2.3}$	\\
Sechi-Zorn \textit{et al.} 68~\cite{SechiZorn:1969hk}	&	ERE approx.\ (CERN, Saclay bubble chamber)	&	$-2.0^{+2.0}_{-13.0}$	&	$5.0^{+10.0}_{-5.0}$	&	$-2.2^{+1.6}_{-1.0}$	&	$3.5^{+11.5}_{-1.0}$	\\
Tan 69~\cite{PhysRevLett.23.395}	&	ERE approx.\ (BNL)	&	-	&	-	&	$-2.0\pm 0.5$	&	$3.0\pm 1.0$	\\
HIRES 10~\cite{Budzanowski:2010ib}	&	ERE approx.\ (COSY)	&	$-2.43^{+0.16}_{-0.25}\pm 0.4$	&	$2.21^{+0.16}_{-0.36}\pm >0.4$	&	$-1.56^{+0.19}_{-0.22}\pm 0.4$	&	$3.7^{+0.6}_{-0.6}\pm >0.4$	\\
COSY-TOF 17~\cite{Hauenstein:2016zys}	&	Gasparyan \textit{ et al.} procedure (COSY)	&	-	&	-	&	$-2.55^{+0.72}_{-1.39}\pm 0.6\pm 0.3$	&	-	\\\bottomrule
\end{tabular} }
\end{sidewaystable}

\begin{sidewaystable}
\centering
\renewcommand{\arraystretch}{1.2}
\caption{Scattering parameters for the $\Sigma N$ ($I=3/2$) system.}
\label{tab:SNscatt}
\resizebox{\columnwidth}{!}{
\begin{tabular}{lccccc}
\toprule
\multicolumn{2}{l}{Reference}	& $a^{\Sigma N (\1s0)}$ [fm] & $r^{\Sigma N (\1s0)}$ [fm] & $a^{\Sigma N (\3s1)}$ [fm] & $r^{\Sigma N (\3s1)}$ [fm] \\\cmidrule(r){1-2}\cmidrule(lr){3-4}\cmidrule(l){5-6}
ND73~\cite{Nagels:1973rq}	&	Nijmegen hard-core model 	&	$-2.79\pm 0.45$	&	$3.55\pm 0.32$	&	$0.63$	&	$-0.76$	\\
ND77~\cite{Nagels:1976xq}	&	Nijmegen hard-core model 	&	$-4.61\pm 0.60$	&	$3.69\pm 0.27$	&	$0.32 \pm 0.01$	&	$-6.01\pm 0.12$	\\
NF78~\cite{Nagels:1978sc}	&	Nijmegen hard-core model 	&	$-3.84$	&	$4.03$	&	$0.62$	&	$-1.91$	\\
Jülich 89~\cite{Holzenkamp:1989tq}	&	Jülich model (models A and B) 	&	$-2.28, -1.10$	&	$5.15,  10.11$	&	$-0.78,-0.90$	&	$1.00,  -1.14$	\\
NSC89~\cite{Maessen:1989sx}	&	Nijmegen soft-core model	 &	$-4.71$	&	$3.36$	&	$0.247$	&	$-26.86$	\\
Fujiwara \textit{et al.} 96~\cite{Fujiwara:1996qj}	&	$SU_6$ quark model (RGM-H, FSS and RGM-F w/Coulomb)	&	$-4.21,-2.15,-2.26$	&	$3.28,4.93,2.70$	&	$0.84,0.95,0.79$	&	$0.59,-0.83,-0.66$	\\
Jülich 94~\cite{Reuber:1993ip}	&	Jülich model (models A, $\tilde{\text{A}}$, B and $\tilde{\text{B}}$) 	&	$-2.28\cdots -1.09$	&	$4.96\cdots 10.20$	&	$-0.93\cdots -0.76$	&	$-1.24\cdots 2.50$	\\
NSC97~\cite{Rijken:1998yy, Stoks:1999bz}	& Nijmegen soft-core model (models a$\cdots$f)	&	$-6.16 \cdots -5.89$	&	$3.24 \cdots 3.29$	&	$-0.33 \cdots -0.18$	&	$-40.27 \cdots -11.29$	\\
Korpa \textit{et al.} 02~\cite{Korpa:2001au}	& EFT NLO	& $0.55$	&	$0.36$	& $0.94$	&	$0.35$	\\
ESC03~\cite{RIJKEN200527}	&	Nijmegen extended-soft-core model (w/Coulomb)	&	$-3.18$	&	$3.95$	&	$-3.18$	&	$1.30$	\\
Jülich 04~\cite{Haidenbauer:2005zh}	&	Jülich model	&	$-4.71$	&	$3.31$	&	$0.29$	&	$-11.54$	\\
ESC04~\cite{Rijken:2006ep}	&	Nijmegen extended-soft-core model (models a$\cdots$d w/Coulomb)	&	$-4.09\cdots -3.43$	&	$3.49 \cdots 4.10$	&	$-0.020 \cdots 0.217$	&	$-3356 \cdots -28.94$	\\
Polinder \textit{et al.} 06~\cite{Polinder:2006zh}	&	$\chi$EFT LO $\Lambda=550\cdots 700$ MeV	&	$-2.36\cdots -2.24$	&	$3.53 \cdots 3.74$	&	$0.56\cdots 0.70$	&	$-4.36 \cdots -2.14$	\\
fss2 07~\cite{Fujiwara:2006yh}	&	$SU_6$ quark model (isospin and particle basis)	&	$-2.51,-2.48$	&	$4.92,5.03$	&	$0.729,0.727$	&	$-1.22,-1.31$	\\
ESC08~\cite{Rijken:2010zzb,Nagels08II:2015}	&	Nijmegen extended-soft-core model (models a'' and c w/Coulomb)	&	$-3.85, -3.91$	&	$3.40, 3.41$	&	$0.62, 0.61$	&	$-2.13, -2.35$	\\
Haidenbauer \textit{et al.} 13~\cite{Haidenbauer:2013oca}	& $\chi$EFT NLO $\Lambda=450\cdots 700$ MeV	& $-3.60 \cdots -3.46$	&	$3.45 \cdots 3.59$	& $0.48 \cdots 0.49$	&	$-5.41 \cdots -4.98$	\\
ESC16~\cite{Nagels:2015lfa}	&	Nijmegen extended-soft-core model (w/Coulomb)	&	$-4.30$	&	$3.25$	&	$0.57$	&	$-3.11$	\\
Haidenbauer \textit{et al.} 20~\cite{Haidenbauer:2019boi}	& $\chi$EFT NLO $\Lambda=500\cdots 650$ MeV	& $-3.90 \cdots -3.43$	&	$3.50 \cdots 3.55$	& $0.42 \cdots 0.48$	&	$-6.49 \cdots -5.69$	\\\bottomrule
\end{tabular} }
\end{sidewaystable}

\begin{sidewaystable}
\centering
\renewcommand{\arraystretch}{1.2}
\caption{Scattering parameters for the $\Lambda \Lambda$ ($I=0$) system.}
\label{tab:LLscatt}
\resizebox{\columnwidth}{!}{
\begin{tabular}{lccc}
\toprule
\multicolumn{2}{l}{Reference}	& $a^{\Lambda \Lambda (\1s0)}$ [fm]	& $r^{\Lambda \Lambda (\1s0)}$ [fm]	\\\cmidrule(r){1-2}\cmidrule(l){3-4}
ND~\cite{Morita:2014kza}	&	Nijmegen hard-core model	&	$-10.629\cdots 14.394$	&	$1.300 \cdots 6.863$ \\
NF~\cite{Morita:2014kza}	&	Nijmegen hard-core model	&	$-3.960\cdots 23.956$	&	$0.975\cdots 8.828$ \\
NSC89~\cite{Morita:2014kza}	&	Nijmegen soft-core model	&	$-0.25, -2.10, -1.11$	&	$7.2, 1.9, 3.2$ \\
Ehime 98~\cite{Tominaga:1998}	&	Ehime model (sets 2, A and B)	&	$-3.09,-3.64,-3.40$	&	$2.89,2.73,2.79$	\\
NSC97~\cite{Stoks:1999bz}	& Nijmegen soft-core model (models a$\cdots$f)	&	$-0.53 \cdots -0.27$	&	$7.43 \cdots 15.00$	\\
Filikhin \textit{et al.} 02~\cite{Filikhin:2002wm}	&	Nijmegen models (ND, NSC and ESC00)	&	$-0.31\cdots -10.60$	&	$2.23\cdots 16.6$	\\
Hiyama \textit{et al.} 02~\cite{Hiyama:2002yj}	&	NF+NAGARA	 event	&	$-0.575$	&	$6.45$	\\
ESC03~\cite{RIJKEN200527}	&	Nijmegen extended-soft-core model	&	$-2.94$	&	$2.53$	\\
ESC04~\cite{Rijken:2006ep}	&	Nijmegen extended-soft-core model (models a$\cdots$d)	&	$-1.323\cdots -1.081$	&	$4.401 \cdots 4.482$\\
ESC04~\cite{RijkenIII:2006}	&	Nijmegen extended-soft-core model (models a, d)	&	$-3.804, -1.555$	&	$2.420, 3.617$\\
fss2 07~\cite{Fujiwara:2006yh}	&	$SU_6$ quark model (isospin and particle basis)	&	$-0.821,-0.808$	&	$3.78,3.83$	\\
Polinder \textit{et al.} 07~\cite{Polinder:2007mp}	&	$\chi$EFT LO $\Lambda=550\cdots 700$ MeV	&	$-1.67\cdots -1.52$	&	$0.31 \cdots 0.82$	\\
ESC08~\cite{Rijken:2010zzb, Rijken:2013wxa, Nagels08III:2015}	&	Nijmegen extended-soft-core model (models a'', c1 and c)	&	$-0.88,-0.97,-0.853$	&	$4.34,3.86, 5.126$	\\
Valcarce \textit{et al.} 10~\cite{Valcarce:2010kz}	&	Chiral constituent quark model	&	$-2.54$	&	-	\\
Haidenbauer \textit{et al.} 16~\cite{Haidenbauer:2015zqb}	&	$\chi$EFT NLO $\Lambda=500\cdots 650$ MeV	&	$-0.70\cdots -0.61$	&	$4.56 \cdots 6.95$	\\
Li \textit{et al.} 18~\cite{Li:2018tbt}	&	Relativistic $\chi$EFT LO + HAL QCD potential	&	$-0.60$	&	$3.73$	\\
ESC16~\cite{Nagels:2020oqo}	&	Nijmegen extended-soft-core model	&	$-0.439$	&	$9.533$		\\\midrule
Yoon \textit{et al.} 10~\cite{Yoon:2008zz}	&	Watson’s procedure (KEK-PS E522)	&	$-0.10^{+0.45}_{-2.37}\pm 0.04$	&	$13.90^{>+16.10}_{<-13.90}\pm 9.48$	\\
STAR 15~\cite{Adamczyk:2014vca}	&	Femtoscopy (Au-Au, RHIC) (free $\lambda$)	&	$1.10\pm 0.37^{+0.68}_{-0.08}$	&	$8.52\pm2.56^{+2.09}_{-0.74}$	\\
Morita \textit{et al.} 15~\cite{Morita:2014kza}	&	Femtoscopy (Au-Au, RHIC) (w/o and w/$\Sigma^0$-decay)	&	$-1.25\cdots -0.56, -1.25\cdots 0$	&	$3.5\cdots 7, -\infty\cdots \infty$	\\
Ohnishi \textit{et al.} 16~\cite{Ohnishi:2016elb}	&	Femtoscopy (Au-Au, RHIC) (fixed and free $\lambda$)	&	$-0.79\pm 0.47 [\pm 0.11],  1.10\pm 0.24$	&	$1.76 \pm 11.62 [\pm 0.86], 8.51\pm 2.14$	\\
ALICE 19~\cite{Acharya:2019yvb}	&	Femtoscopy (p-p and p-Pb, LHC) 	&	\multicolumn{2}{c}{see Fig.~\ref{fig:ar_BB_summary} for $1\sigma$ region}   \\\bottomrule
\end{tabular}  }
\end{sidewaystable}

\begin{sidewaystable}
\centering
\renewcommand{\arraystretch}{1.2}
\caption{Scattering parameters for the $\Sigma\Sigma$ ($I=2$) system.}
\label{tab:SSscatt}
\begin{tabular}{lccc}
\toprule
\multicolumn{2}{l}{Reference}	& $a^{\Sigma \Sigma (\1s0)}$ [fm] & $r^{\Sigma \Sigma (\1s0)}$ [fm] \\\cmidrule(r){1-2}\cmidrule(l){3-4}
NSC97~\cite{Stoks:1999bz}	& Nijmegen soft-core model (models a$\cdots$f w/Coulomb)	&	$6.98 \cdots 10.32$	&	$1.46 \cdots 1.60$	\\
fss2 07~\cite{Fujiwara:2006yh}	&	$SU_6$ quark model (isospin and particle basis)	&	$-85.3,-63.7$	&	$2.34,2.37$	\\
Polinder \textit{et al.} 07~\cite{Polinder:2007mp}	&	$\chi$EFT LO $\Lambda=550\cdots 700$ MeV	&	$-9.42\cdots -6.23$	&	$1.88 \cdots 2.17$	\\
ESC08~\cite{Nagels08III:2015}	&	Nijmegen extended-soft-core model (model c)	&	$-0.65$	&	$19.97$\\
Valcarce \textit{et al.} 10~\cite{Valcarce:2010kz}	&	Chiral constituent quark model	&	$0.523$	&	-	\\
Haidenbauer \textit{et al.} 16~\cite{Haidenbauer:2015zqb}	&	$\chi$EFT NLO $\Lambda=500\cdots 650$ MeV	&	$-2.19\cdots -1.82$	&	$5.67 \cdots 6.05$	\\
Li \textit{et al.} 18~\cite{Li:2018tbt}	&	Relativistic $\chi$EFT LO + HAL QCD potential	&	$-0.80$	&	$13.3$	\\
ESC16~\cite{Nagels:2020oqo}	&	Nijmegen extended-soft-core model	&	$0.495$	&	$11.943$	\\\bottomrule
\end{tabular}

\phantom{\resizebox{\columnwidth}{!}{
\begin{tabular}{lccccc}
\toprule
\multicolumn{2}{l}{Reference}	& $a^{\Xi \Xi (\1s0)}$ [fm] & $r^{\Xi \Xi (\1s0)}$ [fm] & $a^{\Xi \Xi (\3s1)}$ [fm] & $r^{\Xi \Xi (\3s1)}$ [fm]	\\
\end{tabular} }}

\caption{Scattering parameters for the $\Xi N$  ($I=0$) system.}
\label{tab:XNscatt}
\begin{tabular}{lccc}
\toprule
\multicolumn{2}{l}{Reference}	& $a^{\Xi N (\3s1)}$ [fm] & $r^{\Xi N (\3s1)}$ [fm] \\\cmidrule(r){1-2}\cmidrule(l){3-4}
Ehime 98~\cite{Tominaga:1998}	&	Ehime model (set B)	&	$-0.352$	&	$17.4$	\\
Ehime 01~\cite{Yamaguchi:2001ip}	&	Ehime model ($g_{\Xi\Xi\sigma}=1.82$ and $17776$)	&	$-0.43,-0.35$	&	$13,17$	\\
ESC04~\cite{RijkenIII:2006}	&	Nijmegen extended-soft-core model (models a and d)	&	$-1.672, 122.5$	&	$2.704, 2.083$	\\
Polinder \textit{et al.} 07~\cite{Polinder:2007mp}	&	$\chi$EFT LO $\Lambda=550\cdots 700$ MeV	&	$-0.34\cdots -0.15$	&	$-8.27 \cdots 16.3$	\\
ESC08~\cite{Rijken:2010zzb, Rijken:2013wxa, Nagels08III:2015}	&	Nijmegen extended-soft-core model (models a'', c1 and c)	&	$6.90,8.86,-5.357$	&	$1.18,1.12, 1.434$	\\
Valcarce \textit{et al.} 10~\cite{Valcarce:2010kz}	&	Chiral constituent quark model	&	$0.28$	&	-	\\
Haidenbauer \textit{et al.} 16~\cite{Haidenbauer:2015zqb}	&	$\chi$EFT NLO $\Lambda=500\cdots 650$ MeV	&	$-0.34\cdots -0.20$	&	$2.93 \cdots 8.36$	\\
Li \textit{et al.} 18~\cite{Li:2018tbt}	&	Relativistic $\chi$EFT LO + HAL QCD potential	&	$-0.14$	&	$-14.0$	\\
Haidenbauer \textit{et al.} 19~\cite{Haidenbauer:2018gvg}	&	$\chi$EFT NLO $\Lambda=500\cdots 650$ MeV	&	$-0.85\cdots -0.33$	&	$-6.86 \cdots 1.42$	\\
ESC16~\cite{Nagels:2020oqo}	&	Nijmegen extended-soft-core model	&	$-0.269$	&	$-10.250$	\\\bottomrule
\end{tabular}
\end{sidewaystable}

\begin{sidewaystable}
\centering
\renewcommand{\arraystretch}{1.2}
\caption{Scattering parameters for the $\Xi \Sigma$ ($I=3/2$) system.}
\label{tab:XSscatt}
\begin{tabular}{lccc}
\toprule
\multicolumn{2}{l}{Reference}	& $a^{\Xi \Sigma (\1s0)}$ [fm] & $r^{\Xi \Sigma (\1s0)}$ [fm] \\\cmidrule(r){1-2}\cmidrule(l){3-4}
NSC97~\cite{Stoks:1999bz}	& Nijmegen soft-core model (models a$\cdots$f)	&	$2.32 \cdots 4.13$	&	$1.17 \cdots 1.46$ \\
fss2 07~\cite{Fujiwara:2006yh}	&	$SU_6$ quark model (isospin and particle basis)	&	$-4.63,-4.70$	&	$2.39,2.37$\\
Haidenbauer \textit{et al.} 10~\cite{Haidenbauer:2009qn}	&	$\chi$EFT LO $\Lambda=550\cdots 700$ MeV	&	$2.74\cdots 4.28$	&	$0.81\cdots 0.96$\\
ESC08~\cite{Rijken:2013wxa}	&	Nijmegen extended-soft-core model (model c1)	&	$-2.80$	&	$2.45$\\\bottomrule
\end{tabular}

\phantom{\resizebox{\columnwidth}{!}{
\begin{tabular}{lccccc}
\toprule
\multicolumn{2}{l}{Reference}	& $a^{\Xi \Xi (\1s0)}$ [fm] & $r^{\Xi \Xi (\1s0)}$ [fm] & $a^{\Xi \Xi (\3s1)}$ [fm] & $r^{\Xi \Xi (\3s1)}$ [fm]	\\\cmidrule(r){1-2}\cmidrule(l){3-4}
NSC97~\cite{Stoks:1999bz}	& Nijmegen soft-core model (models a$\cdots$f)	&	$2.38 \cdots 17.28$	&	$1.29 \cdots 1.85$	&	$0.27 \cdots 0.48$	&	$2.80 \cdots 10.18$	\\
\end{tabular} }}

\caption{Scattering parameters for the $\Xi \Xi$ ($I=1$ for $\1s0$ and $I=0$ for $\3s1$) system.}
\label{tab:XXscatt}
\resizebox{\columnwidth}{!}{
\begin{tabular}{lccccc}
\toprule
\multicolumn{2}{l}{Reference}	& $a^{\Xi \Xi (\1s0)}$ [fm] & $r^{\Xi \Xi (\1s0)}$ [fm] & $a^{\Xi \Xi (\3s1)}$ [fm] & $r^{\Xi \Xi (\3s1)}$ [fm]	\\\cmidrule(r){1-2}\cmidrule(lr){3-4}\cmidrule(l){5-6}
NSC97~\cite{Stoks:1999bz}	& Nijmegen soft-core model (models a$\cdots$f)	&	$2.38 \cdots 17.28$	&	$1.29 \cdots 1.85$	&	$0.27 \cdots 0.48$	&	$2.80 \cdots 10.18$	\\
Ehime 01~\cite{Yamaguchi:2001ip}	&	Ehime model ($g_{\Xi\Xi\sigma}=1.82$ and $1.7776$)	&	$7.41,7.41$	&	$1.73,1.75$	&	$1.83,1.83$	&	$0.66,0.65$		\\
fss2 07~\cite{Fujiwara:2006yh}	&	$SU_6$ quark model (isospin and particle basis)	&	$-1.43,-1.43$	&	$3.20,3.17$	&	$3.20,3.20$	&	$0.218,0.218$\\
Haidenbauer \textit{et al.} 10~\cite{Haidenbauer:2009qn}	&	$\chi$EFT LO $\Lambda=550\cdots 700$ MeV	&	$2.47\cdots 3.92$	&	$0.75\cdots 0.92$	&	$0.52\cdots 0.63$	&	$1.04\cdots 1.11$\\
ESC08~\cite{Rijken:2013wxa}	&	Nijmegen extended-soft-core model (model c1)	&	$-7.25$	&	$2.00$	&	$0.53$	&	$1.63$\\\bottomrule
\end{tabular} }
\end{sidewaystable}

\begin{table}[ht]
\centering
\caption{Summary of the binding energy $B$ for the $NN$ systems obtained from LQCD calculations. An asterisk means that the corresponding reference does not find a bound state.}
\label{tab:NNLQCD_summary}
\renewcommand{\arraystretch}{1.2}
\resizebox{\columnwidth}{!}{
\begin{tabular}{lccccccc}
\toprule
Reference & $n_f$ & $b_s$ [fm] & $m_{\pi}$ [MeV] & $L$ [fm] & $V\rightarrow\infty$ & $B^{NN(\1s0)}$ [MeV] & $B^{NN(\3s1)}$ [MeV] \\\midrule
NPLQCD 06~\cite{Beane:2006mx} & $2+1$ & 0.125 & [593-354] & 2.5 & no & $\Asterisk$ & $\Asterisk$ \\
NPLQCD 10~\cite{Beane:2009py} & $2+1$ & 0.123 & 390 & 2.5 & no & $\Asterisk$ & $\Asterisk$ \\
HAL QCD 10~\cite{Inoue:2010hs} & 3 & 0.121 & [1014-834] & 1.93 & no & $\Asterisk$\rlap{$^\S$} & $\Asterisk$\rlap{$^\S$} \\
NPLQCD 12~\cite{Beane:2011iw} & $2+1$ & 0.123 & 390 & [2.0-4.0] & yes & $7.1(5.2)(7.3)$ & $11(5)(12)$ \\
HAL QCD 12~\cite{Inoue:2011ai} & 3 & 0.121 & [1171-469] & 3.9 & no & $\Asterisk$ & $\Asterisk$ \\
PACS-CS 12~\cite{Yamazaki:2012hi} & $2+1$ & 0.09 & 510 & [2.9-5.8] & yes & $7.4(1.3)(0.6)$ & $11.5(1.1)(0.6)$  \\
NPLQCD 13~\cite{Beane:2012vq} & 3 & 0.145 & 806 & [3.5-7.0] & yes & $15.9(2.7)(2.7)$ & $19.5(3.6)(3.1)$ \\
HAL QCD 13~\cite{Ishii:2013cta} & $2+1$ & 0.091 & [701-411] & 2.9 & no & $\Asterisk$ & $\Asterisk$\rlap{$^\S$} \\
PACS-CS 15~\cite{Yamazaki:2015asa} & $2+1$ & 0.09 & 300 & [4.3-5.8] & yes & $8.5(0.7)_{(-0.5)}^{(+1.6)}$ & $14.5(0.7)_{(-0.8)}^{(+2.4)}$ \\
NPLQCD 15~\cite{Orginos:2015aya} & $2+1$ & 0.117 & 450 & [2.8-5.6] & yes & $12.5_{(-1.9)(-4.5)}^{(+1.7)(+2.5)}$ & $14.4_{(-1.8)(-1.8)}^{(+1.6)(+2.7)}$ \\
CalLat 17~\cite{Berkowitz:2015eaa} & 3 & 0.145 & 806 & [3.5-4.6] & yes & $21.8_{(-5.1)(-2.8)}^{(+3.2)(+0.8)}$ & $30.7_{(-2.5)(-1.6)}^{(+2.4)(+0.5)}$ \\
NPLQCD 17~\cite{Wagman:2017tmp} & 3 & 0.145 & 806 & [3.5-7.0] & yes & $20.6_{(-2.4)(-1.6)}^{(+1.8)(+2.8)}$ & $27.9_{(-2.3)(-1.4)}^{(+3.1)(+2.2)}$ \\
CalLat 21~\cite{Horz:2020zvv} & 3 & 0.086 & 714 & 4.1 & no\rlap{$^\dag$} & $\Asterisk$ & $\Asterisk$ \\
NPLQCD+QCDSF 21~\cite{Beane:2020ycc} & $1+2$ & 0.068 & 449 & [2.2-3.3] & no & $\Asterisk$ & $\Asterisk$ \\
NPLQCD 21~\cite{Illa:2020nsi} & $2+1$ & 0.117 & 450 & [2.8-5.6] & yes & $13.1_{(-3.1)(-0.4)}^{(+2.0)(+2.3)}$ & $12.7_{(-2.4)(-1.7)}^{(+2.4)(+1.5)}$ \\
NPLQCD 21~\cite{Amarasinghe:2021} & $3$ & 0.145 & 806 & 4.6 & no & $5.0(2.4)$ & $2.0(2.5)$ \\
\bottomrule
\end{tabular} }
\end{table}

\begin{table}[ht]
\centering
\caption{Summary of the binding energy $B$ for the $\Lambda N$ systems obtained from LQCD calculations. An asterisk means that the corresponding reference does not find a bound state.}
\label{tab:LNLQCD_summary}
\renewcommand{\arraystretch}{1.2}
\resizebox{\columnwidth}{!}{
\begin{tabular}{lccccccc}
\toprule
Reference & $n_f$ & $b_s$ [fm] & $m_{\pi}$ [MeV] & $L$ [fm] & $V\rightarrow\infty$ & $B^{\Lambda N(\1s0)}$ [MeV] & $B^{\Lambda N(\3s1)}$ [MeV] \\\midrule
NPLQCD 07~\cite{Beane:2006gf} & $2+1$ & 0.125 & 592 & 2.5 & no & $9(8)(20)$ & $13(13)(8)$ \\ 
 & & & 493 & 2.5 & no & $\Asterisk$ & $4(13)(14)$ \\ 
 & & & 354 & 2.5 & no & $\Asterisk$ & $\Asterisk$ \\
PACS-CS 09~\cite{Nemura:2009kc} & $2+1$ & 0.091 & 300 & 2.9 & no & $1.9(1.3)$ & $1.3(1.2)$ \\
NPLQCD 10~\cite{Beane:2009py} & $2+1$ & 0.123 & 390 & 2.5 & no & $\Asterisk$ & $\Asterisk$ \\
 \bottomrule
\end{tabular} }
\end{table}

\begin{table}[ht]
\centering
\caption{Summary of the binding energy $B$ for the $\Sigma N\; (I=3/2)$ systems obtained from LQCD calculations. An asterisk means that the corresponding reference does not find a bound state.}
\label{tab:SNLQCD_summary}
\renewcommand{\arraystretch}{1.2}
\resizebox{\columnwidth}{!}{
\begin{tabular}{lccccccc}
\toprule
Reference & $n_f$ & $b_s$ [fm] & $m_{\pi}$ [MeV] & $L$ [fm] & $V\rightarrow\infty$ & $B^{\Sigma N(\1s0)}$ [MeV] & $B^{\Sigma N(\3s1)}$ [MeV] \\\midrule
NPLQCD 07~\cite{Beane:2006gf} & $2+1$ & 0.125 & 592 & 2.5 & no & $17(11)(27)$ & $\Asterisk$ \\
 & & & 493 & 2.5 & no & $\Asterisk$ & $\Asterisk$ \\
HAL QCD 10~\cite{Inoue:2010hs} & 3 & 0.121 & [1014-834] & 1.93 & no & $\Asterisk$\rlap{$^\S$} & $\Asterisk$\rlap{$^\S$} \\
HAL QCD 12~\cite{Inoue:2011ai} & 3 & 0.121 & [1171-469] & 3.9 & no & $\Asterisk$ & $\Asterisk$\rlap{$^\S$} \\
NPLQCD 13~\cite{Beane:2012vq} & 3 & 0.145 & 806 & [3.4-6.7] & yes & $15.9(2.7)(2.7)$ & $19.5(3.6)(3.1)$ \\
CalLat 17~\cite{Berkowitz:2015eaa} & 3 & 0.145 & 806 & [3.5-4.6] & yes & $21.8_{(-5.1)(-2.8)}^{(+3.2)(+0.8)}$ & $-$ \\
NPLQCD 17~\cite{Wagman:2017tmp} & 3 & 0.145 & 806 & [3.4-6.7] & yes & $20.6_{(-2.4)(-1.6)}^{(+1.8)(+2.8)}$ & $6.7_{(-1.9)(-6.2)}^{(+3.3)(+1.8)}$ \\
CalLat 21~\cite{Horz:2020zvv} & 3 & 0.086 & 714 & 4.1 & no\rlap{$^\dag$} & $\Asterisk$ & $-$ \\
NPLQCD 21~\cite{Illa:2020nsi} & $2+1$ & 0.117 & 450 & [2.8-5.6] & yes & $14.3_{(-3.0)(-2.8)}^{(+3.1)(+0.9)}$ & $\Asterisk$ \\
NPLQCD 21~\cite{Amarasinghe:2021} & $3$ & 0.145 & 806 & 4.6 & no & $5.0(2.4)$ & - \\
 \bottomrule
\end{tabular} }
\end{table}

\begin{table}[ht]
\centering
\caption{Summary of the binding energy $B$ for the $\Lambda \Lambda$ system ($H$-dibaryon) obtained from LQCD calculations. An asterisk means that the corresponding reference does not find a bound state.}
\label{tab:LLLQCD_summary}
\renewcommand{\arraystretch}{1.2}
\begin{tabular}{lccccccc}
\toprule
Reference & $n_f$ & $b_s$ [fm] & $m_{\pi}$ [MeV] & $L$ [fm] & $V\rightarrow\infty$ & $B^{\Lambda \Lambda(\1s0)}$ [MeV] \\\midrule
NPLQCD 10~\cite{Beane:2009py} & $2+1$ & 0.123 & 390 & 2.5 & no & $4.1(1.2)(1.4)$  \\
NPLQCD 11~\cite{Beane:2010hg} & $2+1$ & 0.123 & 390 & [2.0-4.0] & yes & $16.6(2.1)(4.6)$ \\
HAL QCD 11~\cite{Inoue:2010es} & 3 & 0.121 & 1015 & [1.9-3.9] & no & $32.9(4.5)(6.6)$ \\
 &  &  & 837 & [1.9-3.9] & no & $37.4(4.4)(7.3)$ \\
 &  &  & 673 & [1.9-3.9] & no & $35.6(7.4)(4.0)$ \\
NPLQCD 11~\cite{Beane:2011zpa} & $2+1$ & 0.123 & 230 & 4.0 & no & $-0.6(8.9)(10.3)$ \\
NPLQCD 12~\cite{Beane:2011iw} & $2+1$ & 0.123 & 390 & [2.0-4.0] & yes & $13.2(1.8)(4.0)$ \\
HAL QCD 12~\cite{Inoue:2011ai} & 3 & 0.121 & 1171 & 3.9 & no & $49.1(3.4)(5.5)$ \\
 &  &  & 1015 & 3.9 & no & $37.2(3.7)(2.4)$ \\
 &  &  & 837 & 3.9 & no & $37.8(3.1)(4.2)$ \\
 &  &  & 672 & 3.9 & no & $33.6(4.8)(3.5)$ \\
 &  &  & 469 & 3.9 & no & $26.0(4.4)(4.8)$ \\
NPLQCD 13~\cite{Beane:2012vq} & 3 & 0.123 & 806 & 4.0 & yes & $-0.6(8.9)(10.3)$ \\
Mainz 19~\cite{Francis:2018qch} & 2 & 0.066 & 960 & 2.1 & no & $39.0(2.2)$ \\
& & & 960 & 2.1 & no\rlap{$^\dag$} & $19(10)$ \\
& & & 436 & 2.1 & no & $18.8(5.5)$\\
HAL QCD 20~\cite{Sasaki:2019qnh} & $2+1$ & 0.085 & 146 & 8.1 & no & $\Asterisk$ \\
Mainz 21~\cite{Green:2021qol} & 3 & [0.050-0.099]\rlap{$^\ddag$} & 420 & [2.1-3.1] & yes\rlap{$^\dag$} & $3.97(1.16)(0.86)$ \\\bottomrule
\end{tabular}
\end{table}

\begin{table}[ht]
\centering
\caption{Summary of the binding energy $B$ for the $\Sigma \Sigma\; (I=2)$ systems obtained from LQCD calculations. An asterisk means that the corresponding reference does not find a bound state.}
\label{tab:SSLQCD_summary}
\renewcommand{\arraystretch}{1.2}
\begin{tabular}{lccccccc}
\toprule
Reference & $n_f$ & $b_s$ [fm] & $m_{\pi}$ [MeV] & $L$ [fm] & $V\rightarrow\infty$ & $B^{\Sigma \Sigma(\1s0)}$ [MeV] \\\midrule
NPLQCD 10~\cite{Beane:2009py} & $2+1$ & 0.123 & 390 & 2.5 & no & $\Asterisk$ \\
HAL QCD 10~\cite{Inoue:2010hs} & 3 & 0.121 & [1014-834] & 1.93 & no & $\Asterisk$\rlap{$^\S$} \\
NPLQCD 12~\cite{Beane:2011iw} & $2+1$ & 0.123 & 390 & [2.0-4.0] & no & $\Asterisk$ \\
HAL QCD 12~\cite{Inoue:2011ai} & 3 & 0.121 & [1171-469] & 3.9 & no & $\Asterisk$ \\
NPLQCD 13~\cite{Beane:2012vq} & 3 & 0.145 & 806 & [3.4-6.7] & yes & $15.9(2.7)(2.7)$ \\
CalLat 17~\cite{Berkowitz:2015eaa} & 3 & 0.145 & 806 & [3.5-4.6] & yes & $21.8_{(-5.1)(-2.8)}^{(+3.2)(+0.8)}$ \\
NPLQCD 17~\cite{Wagman:2017tmp} & 3 & 0.145 & 806 & [3.4-6.7] & yes & $20.6_{(-2.4)(-1.6)}^{(+1.8)(+2.8)}$ \\
CalLat 21~\cite{Horz:2020zvv} & 3 & 0.086 & 714 & 4.1 & no\rlap{$^\dag$} & $\Asterisk$ \\
NPLQCD 21~\cite{Illa:2020nsi} & $2+1$ & 0.117 & 450 & [2.8-5.6] & yes & $10.2_{(-1.9)(-2.3)}^{(+2.1)(+2.0)}$ \\
NPLQCD 21~\cite{Amarasinghe:2021} & $3$ & 0.145 & 806 & 4.6 & no & $5.0(2.4)$ \\\bottomrule
\end{tabular}
\end{table}

\begin{table}[ht]
\centering
\caption{Summary of the binding energy $B$ for the $\Xi N\; (\3s1,\, I=0)$ systems obtained from LQCD calculations. An asterisk means that the corresponding reference does not find a bound state.}
\label{tab:XNLQCD_summary}
\renewcommand{\arraystretch}{1.2}
\begin{tabular}{lccccccc}
\toprule
Reference & $n_f$ & $b_s$ [fm] & $m_{\pi}$ [MeV] & $L$ [fm] & $V\rightarrow\infty$ & $B^{\Xi N(\3s1)}$ [MeV]\\\midrule
NPLQCD 10~\cite{Beane:2009py} & $2+1$ & 0.123 & 390 & 2.5 & no & $\Asterisk$ \\
HAL QCD 10~\cite{Inoue:2010hs} & 3 & 0.121 & [1014-834] & 1.93 & no & $\Asterisk$\rlap{$^\S$} \\
HAL QCD 12~\cite{Inoue:2011ai} & 3 & 0.121 & [1171-469] & 3.9 & no & $\Asterisk$\rlap{$^\S$} \\
NPLQCD 13~\cite{Beane:2012vq} & 3 & 0.145 & 806 & [3.4-6.7] & yes & $37.7(3.0)(2.7)$ \\
NPLQCD 17~\cite{Wagman:2017tmp} & 3 & 0.145 & 806 & [3.4-6.7] & yes & $40.7_{(-3.2)(-1.4)}^{(+2.1)(+2.4)}$ \\
HAL QCD 20~\cite{Sasaki:2019qnh} & $2+1$ & 0.085 & 146 & 8.1 & no & $\Asterisk$ \\
NPLQCD 21~\cite{Illa:2020nsi} & $2+1$ & 0.117 & 450 & [2.8-5.6] & yes & $25.3_{(-1.5)(-2.2)}^{(+1.5)(+2.2)}$ \\\bottomrule
\end{tabular}
\end{table}

\begin{table}[ht]
\centering
\caption{Summary of the binding energy $B$ for the $\Xi \Sigma\; (\1s0,\, I=3/2)$ system obtained from LQCD calculations. An asterisk means that the corresponding reference does not find a bound state.}
\label{tab:XSLQCD_summary}
\renewcommand{\arraystretch}{1.2}
\begin{tabular}{lccccccc}
\toprule
Reference & $n_f$ & $b_s$ [fm] & $m_{\pi}$ [MeV] & $L$ [fm] & $V\rightarrow\infty$ & $B^{\Xi \Sigma (\1s0)}$ [MeV] \\\midrule
HAL QCD 10~\cite{Inoue:2010hs} & 3 & 0.121 & [1014-834] & 1.93 & no & $\Asterisk$\rlap{$^\S$} \\
HAL QCD 12~\cite{Inoue:2011ai} & 3 & 0.121 & [1171-469] & 3.9 & no & $\Asterisk$ \\
NPLQCD 13~\cite{Beane:2012vq} & 3 & 0.145 & 806 & [3.4-6.7] & yes & $15.9(2.7)(2.7)$ \\
CalLat 17~\cite{Berkowitz:2015eaa} & 3 & 0.145 & 806 & [3.5-4.6] & yes & $21.8_{(-5.1)(-2.8)}^{(+3.2)(+0.8)}$ \\
NPLQCD 17~\cite{Wagman:2017tmp} & 3 & 0.145 & 806 & [3.4-6.7] & yes & $20.6_{(-2.4)(-1.6)}^{(+1.8)(+2.8)}$ \\
CalLat 21~\cite{Horz:2020zvv} & 3 & 0.086 & 714 & 4.1 & no\rlap{$^\dag$} & $\Asterisk$ \\
NPLQCD 21~\cite{Illa:2020nsi} & $2+1$ & 0.117 & 450 & [2.8-5.6] & yes & $12.8_{(-1.6)(-2.2)}^{(+2.1)(+1.6)}$ \\
NPLQCD 21~\cite{Amarasinghe:2021} & $3$ & 0.145 & 806 & 4.6 & no & $5.0(2.4)$ \\\bottomrule
\end{tabular}
\end{table}

\begin{table}[ht]
\centering
\caption{Summary of the binding energy $B$ for the $\Xi \Xi\; (\1s0,\, I=2)$ and $\Xi \Xi\; (\3s1,\, I=0)$ systems obtained from LQCD calculations. An asterisk means that the corresponding reference does not find a bound state.}
\label{tab:XXLQCD_summary}
\renewcommand{\arraystretch}{1.2}
\resizebox{\columnwidth}{!}{
\begin{tabular}{lccccccc}
\toprule
Reference & $n_f$ & $b_s$ [fm] & $m_{\pi}$ [MeV] & $L$ [fm] & $V\rightarrow\infty$ & $B^{\Xi \Xi(\1s0)}$ [MeV] & $B^{\Xi \Xi(\3s1)}$ [MeV] \\\midrule
NPLQCD 10~\cite{Beane:2009py} & $2+1$ & 0.123 & 390 & 2.5 & no & $\Asterisk$ & $-$ \\
HAL QCD 10~\cite{Inoue:2010hs} & 3 & 0.121 & [1014-834] & 1.93 & no & $\Asterisk$\rlap{$^\S$} & $\Asterisk$\rlap{$^\S$} \\
NPLQCD 12~\cite{Beane:2011iw} & $2+1$ & 0.123 & 390 & [2.0-4.0] & yes & $14.0(1.4)(6.7)$ & $-$ \\
HAL QCD 12~\cite{Inoue:2011ai} & 3 & 0.121 & [1171-469] & 3.9 & no & $\Asterisk$ & $\Asterisk$\rlap{$^\S$} \\
NPLQCD 13~\cite{Beane:2012vq} & 3 & 0.145 & 806 & [3.4-6.7] & yes & $15.9(2.7)(2.7)$ & $5.5(3.4)(3.7)$ \\
CalLat 17~\cite{Berkowitz:2015eaa} & 3 & 0.145 & 806 & [3.5-4.6] & yes & $21.8_{(-5.1)(-2.8)}^{(+3.2)(+0.8)}$ & $-$ \\
NPLQCD 17~\cite{Wagman:2017tmp} & 3 & 0.145 & 806 & [3.4-6.7] & yes & $20.6_{(-2.4)(-1.6)}^{(+1.8)(+2.8)}$ & $6.7_{(-1.9)(-6.2)}^{(+3.3)(+1.8)}$ \\
CalLat 21~\cite{Horz:2020zvv} & 3 & 0.086 & 714 & 4.1 & no\rlap{$^\dag$} & $\Asterisk$ & $-$ \\
NPLQCD 21~\cite{Illa:2020nsi} & $2+1$ & 0.117 & 450 & [2.8-5.6] & yes & $14.9_{(-1.0)(-1.8)}^{(+1.5)(+1.4)}$ & $\Asterisk$ \\
NPLQCD 21~\cite{Amarasinghe:2021} & $3$ & 0.145 & 806 & 4.6 & no & $5.0(2.4)$ & - \\
\bottomrule
\end{tabular} }
\end{table} %Summary tables for BB

% Appendix 6 - *******************************************

\chapter{\texorpdfstring{$SU(3)$}{SU(3)} flavor decomposition of two-baryon channels}\label{appen:SU3chann}

In this appendix we tabulate all the octet-octet baryon channels in their corresponding $SU(3)_f$ irreducible representations: $\mathbf{27}$ (Fig~\ref{fig:27irrep_chann}), $\mathbf{8}_s$ (Fig~\ref{fig:8sirrep_chann}), $\mathbf{1}$ (Fig~\ref{fig:1irrep_chann}), $\overline{\mathbf{10}}$ (Fig~\ref{fig:10birrep_chann}), $\mathbf{10}$ (Fig~\ref{fig:10irrep_chann}), and  $\mathbf{8}_a$ (Fig~\ref{fig:8airrep_chann}), obtained by following the procedure explained in~\cref{subsec:SU3EFT}. A similar decomposition can be found in Ref.~\cite{Wagman:2017tmp}, however, we have noticed that in the $\mathbf{8}_a$ irrep (Fig~\ref{fig:8airrep_chann}), the channels ${\color{cc6} \mathbf{3}}-{\color{cc6}\mathbf{5}}$ do not include $\Lambda\Sigma$ contributions (cf.\ Fig. 18 from Ref.~\cite{Wagman:2017tmp}).

For display purposes, the $SU(3)$ weight diagrams are also shown, with the horizontal axis representing the third component of the isospin $I_3$, and the vertical axis representing the hypercharge $Y=B+S$ (with $B$ being baryon number and $S$ strangeness).

The states that are colored {\bf\color{cc0}blue} do not mix with other flavor channels, while mixing happens between states with the same color code and the same isospin, hypercharge, and total angular momentum quantum numbers.

In the literature different combinations can be obtained, and they originate in a different definition of the baryon matrix, Eq.~\eqref{eq:Bmatrix}, where, for example, an additional minus sign in the $\Xi^-$ entry may be present.

\newpage

\begin{figure}
\centering
\begin{tikzpicture}[scale=1.20]
\node (00) at (0,0) {}; 

\foreach \c in {1,...,2}{
\foreach \i in {0,...,5}{
\pgfmathtruncatemacro\j{\c*\i} 
\node[circle,minimum width=4pt,inner sep=0pt] (\c;\j) at (60*\i:\c) {}; } }

\foreach \i in {0,2,...,10}{
\pgfmathtruncatemacro\j{mod(\i+2,12)}
\pgfmathtruncatemacro\k{\i+1}
\node (2\k) at ($(2;\i)!.5!(2;\j)$) {};}

\draw[line width=1.7pt,gray!50] ($(2;4)$)--($(2;2)$);
\draw[line width=1.7pt,gray!50] ($(2;2)$)--($(2;0)$);
\draw[line width=1.7pt,gray!50] ($(2;0)$)--($(2;10)$);
\draw[line width=1.7pt,gray!50] ($(2;10)$)--($(2;8)$);
\draw[line width=1.7pt,gray!50] ($(2;8)$)--($(2;6)$);
\draw[line width=1.7pt,gray!50] ($(2;6)$)--($(2;4)$);

\fill[cc0](2;4) circle (0.1);
\node[text=cc0, font=\bf] at (2;4) [yshift=+2.5ex] {1};
\fill[cc0](23) circle (0.1);
\node[text=cc0, font=\bf] at (23) [yshift=+2.5ex] {2};
\fill[cc0](2;2) circle (0.1);
\node[text=cc0, font=\bf] at (2;2) [yshift=+2.5ex] {3};

\fill[cc0](25) circle (0.1);
\node[text=cc0, font=\bf] at (25) [yshift=+2.5ex] {4};
\fill[cc0](1;2) circle (0.1);
\node[text=cc0, font=\bf] at (1;2) [yshift=+3ex] {5};
\fill[cc0](1;1) circle (0.1);
\node[text=cc0, font=\bf] at (1;1) [yshift=+3ex] {6};
\fill[cc0](21) circle (0.1);
\node[text=cc0, font=\bf] at (21) [yshift=+2.5ex] {7};

\fill[cc0](2;6) circle (0.1);
\node[text=cc0, font=\bf] at (2;6) [xshift=-2.5ex] {8};
\fill[cc0](1;3) circle (0.1);
\node[text=cc0, font=\bf] at (1;3) [yshift=+3ex] {9};
\fill[cc0](00) circle (0.1);
\node[text=cc0, font=\bf] at (00) [yshift=+3.8ex] {10};
\fill[cc0](1;0) circle (0.1);
\node[text=cc0, font=\bf] at (1;0) [yshift=+3ex] {11};
\fill[cc0](2;0) circle (0.1);
\node[text=cc0, font=\bf] at (2;0) [xshift=+2.5ex] {12};

\fill[cc0](27) circle (0.1);
\node[text=cc0, font=\bf] at (27) [yshift=-2.5ex] {13};
\fill[cc0](1;4) circle (0.1);
\node[text=cc0, font=\bf] at (1;4) [yshift=-3ex] {14};
\fill[cc0](1;5) circle (0.1);
\node[text=cc0, font=\bf] at (1;5) [yshift=-3ex] {15};
\fill[cc0](211) circle (0.1);
\node[text=cc0, font=\bf] at (211) [yshift=-2.5ex] {16};

\fill[cc0](2;8) circle (0.1);
\node[text=cc0, font=\bf] at (2;8) [yshift=-2.5ex] {17};
\fill[cc0](29) circle (0.1);
\node[text=cc0, font=\bf] at (29) [yshift=-2.5ex] {18};
\fill[cc0](2;10) circle (0.1);
\node[text=cc0, font=\bf] at (2;10) [yshift=-2.5ex] {19};

\draw[cc1,line width=1.3pt](1;2) circle (0.2);
\node[text=cc1, font=\bf] at (1;2) [xshift=-3.5ex] {20};
\draw[cc1,line width=1.3pt](1;1) circle (0.2);
\node[text=cc1, font=\bf] at (1;1) [xshift=3.5ex] {21};

\draw[cc2,line width=1.3pt](1;3) circle (0.2);
\node[text=cc2, font=\bf] at (1;3) [xshift=-3.5ex] {22};
\draw[cc2,line width=1.3pt](00) circle (0.2);
\node[text=cc2, font=\bf] at (00) [xshift=-4ex] {23};
\draw[cc2,line width=1.3pt](1;0) circle (0.2);
\node[text=cc2, font=\bf] at (1;0) [xshift=3.5ex] {24};

\draw[cc3,line width=1.3pt](1;4) circle (0.2);
\node[text=cc3, font=\bf] at (1;4) [xshift=-3.5ex] {25};
\draw[cc3,line width=1.3pt](1;5) circle (0.2);
\node[text=cc3, font=\bf] at (1;5) [xshift=3.5ex] {26};

\draw[cc4,line width=1.3pt](00) circle (0.3);
\node[text=cc4, font=\bf] at (00) [xshift=4ex] {27};

\draw[decoration={markings,mark=at position 1 with
    {\arrow[scale=2,>=stealth]{>}}},postaction={decorate}, line width=1.2pt] (-3,-2.6)--(3,-2.6) node[right]{$I_3$};
\draw[decoration={markings,mark=at position 1 with
    {\arrow[scale=2,>=stealth]{>}}},postaction={decorate}, line width=1.2pt] (-3,-2.6)--(-3,2.6) node[above]{$Y$};

\foreach \x in {2,...,-2} {%
    \draw ($(\x,-2.6) + (0,-3pt)$) -- ($(\x,-2.6) + (0,3pt)$)
        node [below, yshift=-2ex] {$\x$};
}

\foreach \x in {1.5,0.5} {%
\pgfmathtruncatemacro\k{\x*2}
    \draw ($(\x,-2.6) + (0,-3pt)$) -- ($(\x,-2.6) + (0,3pt)$)
        node [below, yshift=-3ex] {$\frac{\k}{2}$};
}

\foreach \x in {-0.5,-1.5} {%
\pgfmathtruncatemacro\k{\x*-2}
\draw ($(\x,-2.6) + (0,-3pt)$) -- ($(\x,-2.6) + (0,3pt)$) node [below, yshift=-3ex] {$-\frac{\k}{2}$};
}

\foreach \y in {2,...,-2} {%
\pgfmathsetmacro\yy{\y*0.8660254}
\draw ($(-3,\yy) + (-3pt,0)$) -- ($(-3,\yy) + (3pt,0)$) node [left, xshift=-2ex] {$\y$};
}

\node at (2.5,2.2)  {$\mathbf{27}$ irrep};
\end{tikzpicture}\vspace*{-15pt}

\noindent\makebox[\textwidth][c]{ 
\begin{minipage}[t]{0.45\textwidth}
\begin{table}[H]
\begin{tabular}{>{\bfseries\color{cc0}}l c}
1 & $nn$ \\ 
2 & $\sqrt{\frac{1}{2}}(np + pn)$\\
3 & $pp$ 
\end{tabular}
\end{table}
\vspace{-1em}
\begin{table}[H]
\begin{tabular}{>{\bfseries\color{cc0}}lc}
4 & $\Sigma^-n$\\ 
5 & $\sqrt{\frac{2}{3}}\Sigma^0n + \sqrt{\frac{1}{3}}\Sigma^-p$ \\
6 & $-\sqrt{\frac{1}{3}}\Sigma^+n + \sqrt{\frac{2}{3}}\Sigma^0 p$ \\
7 & $\Sigma^+p$
\end{tabular}
\end{table}
\vspace{-1em}
\begin{table}[H]
\begin{tabular}{>{\bfseries\color{cc0}}lc}
8 & $\Sigma^-\Sigma^-$ \\
9 & $\sqrt{\frac{1}{2}}(\Sigma^-\Sigma^0 + \Sigma^0\Sigma^-)$ \\
10 & $\sqrt{\frac{1}{6}}(\Sigma^-\Sigma^+ - 2\Sigma^0\Sigma^0 + \Sigma^+\Sigma^-) $ \\
11 & $\sqrt{\frac{1}{2}}(\Sigma^0\Sigma^++\Sigma^+ \Sigma^0)$ \\
12 & $\Sigma^+ \Sigma^+$
\end{tabular}
\end{table}
\vspace{-1em}
\begin{table}[H]
\begin{tabular}{>{\bfseries\color{cc0}}lc}
13 & $\Xi^-\Sigma^-$ \\
14 & $-\sqrt{\frac{2}{3}}\Sigma^0\Xi^- + \sqrt{\frac{1}{3}}\Sigma^-\Xi^0$ \\ 
15 & $\sqrt{\frac{1}{3}}\Sigma^+\Xi^- + \sqrt{\frac{2}{3}}\Sigma^0\Xi^0$ \\
16 & $\Sigma^+\Xi^0$ 
\end{tabular}
\end{table}
\end{minipage}
\begin{minipage}[t]{0.45\textwidth}
\begin{table}[H]
\begin{tabular}{>{\bfseries\color{cc0}}lc}
17 & $\Xi^-\Xi^-$ \\
18 & $\sqrt{\frac{1}{2}}(\Xi^-\Xi^0 + \Xi^0\Xi^-)$ \\
19 & $\Xi^0\Xi^0$
\end{tabular}
\end{table}
\vspace{-1em}
\begin{table}[H]
\begin{tabular}{>{\bfseries\color{cc1}}lc}
20 &  $\sqrt{\frac{1}{2}}\Lambda n - \sqrt{\frac{1}{6}}\Sigma^0n + \sqrt{\frac{1}{3}}\Sigma^-p$ \\
21 & $\sqrt{\frac{1}{2}}\Lambda p + \sqrt{\frac{1}{3}}\Sigma^+n +\sqrt{\frac{1}{6}}\Sigma^0p$
\end{tabular}
\end{table}
\vspace{-1em}
\begin{table}[H]
\begin{tabular}{>{\bfseries\color{cc2}}lc}
22 & $-\sqrt{\frac{3}{5}}\Lambda\Sigma^- + \sqrt{\frac{2}{5}}\Xi^-n$ \\
23 & $-\sqrt{\frac{3}{5}}\Lambda\Sigma^0 + \sqrt{\frac{1}{5}}(\Xi^-p - \Xi^0n)$ \\
24 & $\sqrt{\frac{3}{5}}\Lambda\Sigma^+ - \sqrt{\frac{2}{5}}\Xi^0p$
\end{tabular}
\end{table}
\vspace{-1em}
\begin{table}[H]
\begin{tabular}{>{\bfseries\color{cc3}}lc}
25 & $\sqrt{\frac{9}{10}}\Lambda\Xi^- + \sqrt{\frac{1}{30}}\Sigma^0\Xi^- + \sqrt{\frac{1}{15}}\Sigma^-\Xi^0$ \\
26 & $\sqrt{\frac{3}{10}}\Lambda\Xi^0 + \sqrt{\frac{1}{15}}\Sigma^+\Xi^- - \sqrt{\frac{1}{30}}\Sigma^0\Xi^0$
\end{tabular}
\end{table}
\vspace{-1em}
\begin{table}[H]
\begin{tabular}{>{\bfseries\color{cc4}}lc}
27 & $-\sqrt{\frac{1}{120}}(\Sigma^+\Sigma^- + \Sigma^0 \Sigma^0 + \Sigma^- \Sigma^+)$\\
& $+ \sqrt{\frac{3}{20}}(\Xi^0n + \Xi^-p) - \sqrt{\frac{27}{40}}\Lambda\Lambda$
\end{tabular}
\end{table}
\end{minipage}}
\caption{Weight diagram and two-baryon channels for the $\mathbf{27}$ irrep.}
\label{fig:27irrep_chann}
\end{figure}

\begin{figure}
\centering
\begin{tikzpicture}[scale=1.20]
\node (00) at (0,0) {}; 

\foreach \c in {1,...,1}{
\foreach \i in {0,...,5}{
\pgfmathtruncatemacro\j{\c*\i} 
\node[circle,minimum width=4pt,inner sep=0pt] (\c;\j) at (60*\i:\c) {}; } }

% \foreach \i in {0,2,...,10}{
% \pgfmathtruncatemacro\j{mod(\i+2,12)}
% \pgfmathtruncatemacro\k{\i+1}
% \node (2\k) at ($(2;\i)!.5!(2;\j)$) {};}

\draw[line width=1.7pt,gray!50] ($(1;2)$)--($(1;1)$);
\draw[line width=1.7pt,gray!50] ($(1;1)$)--($(1;0)$);
\draw[line width=1.7pt,gray!50] ($(1;0)$)--($(1;5)$);
\draw[line width=1.7pt,gray!50] ($(1;5)$)--($(1;4)$);
\draw[line width=1.7pt,gray!50] ($(1;4)$)--($(1;3)$);
\draw[line width=1.7pt,gray!50] ($(1;3)$)--($(1;2)$);

\fill[cc1](1;2) circle (0.1);
\node[text=cc1, font=\bf] at (1;2) [yshift=2.5ex] {1};
\fill[cc1](1;1) circle (0.1);
\node[text=cc1, font=\bf] at (1;1) [yshift=2.5ex] {2};

\fill[cc2](1;3) circle (0.1);
\node[text=cc2, font=\bf] at (1;3) [xshift=-2.5ex] {3};
\fill[cc2](00) circle (0.1);
\node[text=cc2, font=\bf] at (00) [xshift=-3ex] {4};
\fill[cc2](1;0) circle (0.1);
\node[text=cc2, font=\bf] at (1;0) [xshift=2.5ex] {5};

\fill[cc3](1;4) circle (0.1);
\node[text=cc3, font=\bf] at (1;4) [yshift=-2.5ex] {6};
\fill[cc3](1;5) circle (0.1);
\node[text=cc3, font=\bf] at (1;5) [yshift=-2.5ex] {7};

\draw[cc4,line width=1.3pt](00) circle (0.2);
\node[text=cc4, font=\bf] at (00) [xshift=3ex] {8};

\draw[decoration={markings,mark=at position 1 with
    {\arrow[scale=2,>=stealth]{>}}},postaction={decorate}, line width=1.2pt] (-2,-1.7)--(2,-1.7) node[right]{$I_3$};
\draw[decoration={markings,mark=at position 1 with
    {\arrow[scale=2,>=stealth]{>}}},postaction={decorate}, line width=1.2pt] (-2,-1.7)--(-2,1.7) node[above]{$Y$};

\foreach \x in {1,...,-1} {%
    \draw ($(\x,-1.7) + (0,-3pt)$) -- ($(\x,-1.7) + (0,3pt)$)
        node [below, yshift=-2ex] {$\x$};
}

\foreach \x in {0.5} {%
\pgfmathtruncatemacro\k{\x*2}
    \draw ($(\x,-1.7) + (0,-3pt)$) -- ($(\x,-1.7) + (0,3pt)$)
        node [below, yshift=-3ex] {$\frac{\k}{2}$};
}

\foreach \x in {-0.5} {%
\pgfmathtruncatemacro\k{\x*-2}
\draw ($(\x,-1.7) + (0,-3pt)$) -- ($(\x,-1.7) + (0,3pt)$) node [below, yshift=-3ex] {$-\frac{\k}{2}$};
}

\foreach \y in {1,...,-1} {%
\pgfmathsetmacro\yy{\y*0.8660254}
\draw ($(-2,\yy) + (-3pt,0)$) -- ($(-2,\yy) + (3pt,0)$) node [left, xshift=-2ex] {$\y$};
}

\node at (1.8,1.3) {$\mathbf{8}_s$ irrep};
\end{tikzpicture} \vspace*{-15pt}

\noindent\makebox[\textwidth][c]{
\begin{minipage}[t]{0.45\textwidth}
\begin{table}[H]
\begin{tabular}{>{\bfseries\color{cc1}}lc}
1 & $-\sqrt{\frac{9}{10}}\Lambda n - \sqrt{\frac{1}{30}}\Sigma^0 n + \sqrt{\frac{2}{30}}\Sigma^- p$ \\
2 & $\sqrt{\frac{9}{10}}\Lambda p + \sqrt{\frac{2}{30}}\Sigma^+ n + \sqrt{\frac{1}{30}}\Sigma^0 p$ \\
\end{tabular}
\end{table}
\vspace{-1em}
\begin{table}[H]
\begin{tabular}{>{\bfseries\color{cc2}}lc}
3 & $\sqrt{\frac{2}{5}}\Lambda\Sigma^- + \sqrt{\frac{3}{5}}\Xi^-n$ \\
4 & $\sqrt{\frac{2}{5}}\Lambda\Sigma^0 + \sqrt{\frac{3}{5}}(\Xi^-p - \Xi^0n)$ \\
5 & $\sqrt{\frac{2}{5}}\Lambda\Sigma^+ + \sqrt{\frac{3}{5}}\Xi^0p$ \\
\end{tabular}
\end{table}
\end{minipage}
\begin{minipage}[t]{0.5\textwidth}
\begin{table}[H]
\begin{tabular}{>{\bfseries\color{cc3}}lc}
6 & $-\sqrt{\frac{1}{10}}\Lambda\Xi^- + \sqrt{\frac{3}{10}}\Sigma^0\Xi^- + \sqrt{\frac{6}{10}}\Sigma^-\Xi^0$ \\
7 & $\sqrt{\frac{1}{10}}\Lambda\Xi^0 - \sqrt{\frac{6}{10}}\Sigma^+\Xi^- + \sqrt{\frac{3}{10}}\Sigma^0\Xi^0$ \\
\end{tabular}
\end{table}
\vspace{-1em}
\begin{table}[H]
\begin{tabular}{>{\bfseries\color{cc4}}lc}
8 & $-\sqrt{\frac{1}{5}}(\Sigma^+\Sigma^- + \Sigma^0 \Sigma^0 + \Sigma^- \Sigma^+)$\\ 
& $+ \sqrt{\frac{1}{10}}(\Xi^0n + \Xi^-p) - \sqrt{\frac{1}{5}}\Lambda\Lambda$ \\
\end{tabular}
\end{table}
\end{minipage}}
\caption{Weight diagram and two-baryon channels for the $\mathbf{8}_s$ irrep.}
\label{fig:8sirrep_chann}
\end{figure}
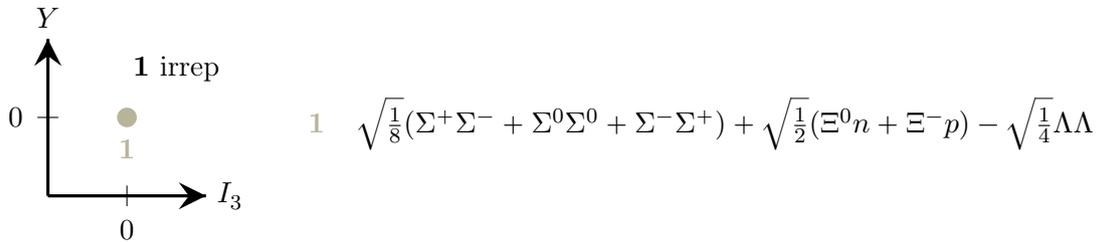

\begin{figure}
\noindent\makebox[\textwidth][c]{
\begin{minipage}[c]{0.25\textwidth}
\begin{figure}[H]
\centering
\begin{tikzpicture}[scale=1.30,baseline={([yshift=0ex]current bounding box.center)}]
\node (00) at (0,0) {}; 

\fill[cc4](00) circle (0.1);
\node[text=cc4, font=\bf] at (00) [yshift=-2.5ex] {1};

\draw[decoration={markings,mark=at position 1 with
    {\arrow[scale=2,>=stealth]{>}}},postaction={decorate}, line width=1.2pt] (-0.8,-0.8)--(0.8,-0.8) node[right]{$I_3$};
\draw[decoration={markings,mark=at position 1 with
    {\arrow[scale=2,>=stealth]{>}}},postaction={decorate}, line width=1.2pt] (-0.8,-0.8)--(-0.8,0.8) node[above]{$Y$};

\draw ($(0,-0.8) + (0,-3pt)$) -- ($(0,-0.8) + (0,3pt)$) node [below, yshift=-2ex] {$0$};

\draw ($(-0.8,0) + (-3pt,0)$) -- ($(-0.8,0) + (3pt,0)$) node [left, xshift=-2ex] {$0$};

\node at (0.5,0.5) {$\mathbf{1}$ irrep};
\end{tikzpicture} 
\end{figure}
\end{minipage}
\begin{minipage}{0.75\textwidth}
\begin{table}[H]
\begin{tabular}{>{\bfseries\color{cc4}}lc}
1 & $\sqrt{\frac{1}{8}}(\Sigma^+\Sigma^- + \Sigma^0 \Sigma^0 + \Sigma^- \Sigma^+) + \sqrt{\frac{1}{2}}(\Xi^0n + \Xi^-p) - \sqrt{\frac{1}{4}}\Lambda\Lambda$ \\
\end{tabular}
\end{table}
\end{minipage}}
\caption{Weight diagram and two-baryon channels for the $\mathbf{1}$ irrep.}
\label{fig:1irrep_chann}
\end{figure}

\begin{figure}
\centering
\begin{tikzpicture}[scale=1.20]
\node (00) at (0,0) {}; 

\foreach \c in {1,...,2}{
\foreach \i in {0,...,5}{
\pgfmathtruncatemacro\j{\c*\i} 
\node[circle,minimum width=4pt,inner sep=0pt] (\c;\j) at (60*\i:\c) {}; } }

\foreach \i in {0,2,...,10}{
\pgfmathtruncatemacro\j{mod(\i+2,12)}
\pgfmathtruncatemacro\k{\i+1}
\node (2\k) at ($(2;\i)!.5!(2;\j)$) {};}

\draw[line width=1.7pt,gray!50] ($(2;6)!.5!(2;8)$)--($(2;0)!.5!(2;10)$);
\draw[line width=1.7pt,gray!50] ($(2;6)!.5!(2;8)$)--($(2;4)!.5!(2;2)$);
\draw[line width=1.7pt,gray!50] ($(2;4)!.5!(2;2)$)--($(2;0)!.5!(2;10)$);

\fill[cc0](23) circle (0.1);
\node[text=cc0, font=\bf] at (23) [yshift=+2.5ex] {1};

\fill[cc5](1;2) circle (0.1);
\node[text=cc5, font=\bf] at (1;2) [xshift=-2.5ex] {2};
\fill[cc5](1;1) circle (0.1);
\node[text=cc5, font=\bf] at (1;1) [xshift=+2.5ex] {3};

\fill[cc6](1;3) circle (0.1);
\node[text=cc6, font=\bf] at (1;3) [xshift=-2.5ex] {4};
\fill[cc6](00) circle (0.1);
\node[text=cc6, font=\bf] at (00) [yshift=+2.5ex] {5};
\fill[cc6](1;0) circle (0.1);
\node[text=cc6, font=\bf] at (1;0) [xshift=+2.5ex] {6};

\fill[cc0](27) circle (0.1);
\node[text=cc0, font=\bf] at (27) [yshift=-2.5ex] {7};
\fill[cc0](1;4) circle (0.1);
\node[text=cc0, font=\bf] at (1;4) [yshift=-2.5ex] {8};
\fill[cc0](1;5) circle (0.1);
\node[text=cc0, font=\bf] at (1;5) [yshift=-2.5ex] {9};
\fill[cc0](211) circle (0.1);
\node[text=cc0, font=\bf] at (211) [yshift=-2.5ex] {10};

\draw[decoration={markings,mark=at position 1 with
    {\arrow[scale=2,>=stealth]{>}}},postaction={decorate}, line width=1.2pt] (-2.5,-1.8)--(2.5,-1.8) node[right]{$I_3$};
\draw[decoration={markings,mark=at position 1 with
    {\arrow[scale=2,>=stealth]{>}}},postaction={decorate}, line width=1.2pt] (-2.5,-1.8)--(-2.5,2.6) node[above]{$Y$};

\foreach \x in {1,...,-1} {%
    \draw ($(\x,-1.8) + (0,-3pt)$) -- ($(\x,-1.8) + (0,3pt)$)
        node [below, yshift=-2ex] {$\x$};
}

\foreach \x in {1.5,0.5} {%
\pgfmathtruncatemacro\k{\x*2}
    \draw ($(\x,-1.8) + (0,-3pt)$) -- ($(\x,-1.8) + (0,3pt)$)
        node [below, yshift=-3ex] {$\frac{\k}{2}$};
}

\foreach \x in {-0.5,-1.5} {%
\pgfmathtruncatemacro\k{\x*-2}
\draw ($(\x,-1.8) + (0,-3pt)$) -- ($(\x,-1.8) + (0,3pt)$) node [below, yshift=-3ex] {$-\frac{\k}{2}$};
}

\foreach \y in {2,...,-1} {%
\pgfmathsetmacro\yy{\y*0.8660254}
\draw ($(-2.5,\yy) + (-3pt,0)$) -- ($(-2.5,\yy) + (3pt,0)$) node [left, xshift=-2ex] {$\y$};
}

\node at (2.5,2.2) {$\overline{\mathbf{10}}$ irrep};
\end{tikzpicture} \vspace*{-15pt}

\noindent\makebox[\textwidth][c]{
\begin{minipage}[t]{0.35\textwidth}
\begin{table}[H]
\begin{tabular}{>{\bfseries\color{cc0}}lc}
1 & $\sqrt{\frac{1}{2}}(np-pn)$
\end{tabular}
\end{table}
\vspace{-1em}
\begin{table}[H]
\begin{tabular}{>{\bfseries\color{cc0}}lc}
7 & $\Sigma^-\Xi^-$ \\
8 & $-\sqrt{\frac{2}{3}}\Sigma^0\Xi^- + \sqrt{\frac{1}{3}}\Sigma^-\Xi^0$ \\
9 & $\sqrt{\frac{1}{3}}\Sigma^+\Xi^- + \sqrt{\frac{2}{3}}\Sigma^0\Xi^0$ \\
10 & $\Sigma^+\Xi^0$
\end{tabular}
\end{table}
\end{minipage}
\begin{minipage}[t]{0.65\textwidth}
\begin{table}[H]
\begin{tabular}{>{\bfseries\color{cc5}}lc}
2 & $-\sqrt{\frac{1}{6}}\Sigma^0n + \sqrt{\frac{1}{3}}\Sigma^-p - \sqrt{\frac{1}{2}}\Lambda n$ \\
3 & $\sqrt{\frac{1}{3}}\Sigma^+n - \sqrt{\frac{1}{6}}\Sigma^0p + \sqrt{\frac{1}{2}}\Lambda p$
\end{tabular}
\end{table}
\vspace{-1em}
\begin{table}[H]
\begin{tabular}{>{\bfseries\color{cc6}}lc}
4 & $\sqrt{\frac{1}{12}}(\Sigma^-\Sigma^0 - \Sigma^0\Sigma^-) + \sqrt{\frac{1}{3}}\Xi^-n - \sqrt{\frac{1}{2}}\Lambda\Sigma^-$ \\
5 & $-\sqrt{\frac{1}{12}}(\Sigma^-\Sigma^+ - \Sigma^+\Sigma^-) + \sqrt{\frac{1}{6}}(\Xi^-p - \Xi^0n) - \sqrt{\frac{1}{2}}\Lambda\Sigma^0$ \\
6 & $-\sqrt{\frac{1}{12}}(\Sigma^0\Sigma^+ - \Sigma^+\Sigma^0) - \sqrt{\frac{1}{3}}\Xi^0p + \sqrt{\frac{1}{2}}\Lambda\Sigma^+$
\end{tabular}
\end{table}
\end{minipage}}
\caption{Weight diagram and two-baryon channels for the $\overline{\mathbf{10}}$ irrep.}
\label{fig:10birrep_chann}
\end{figure}
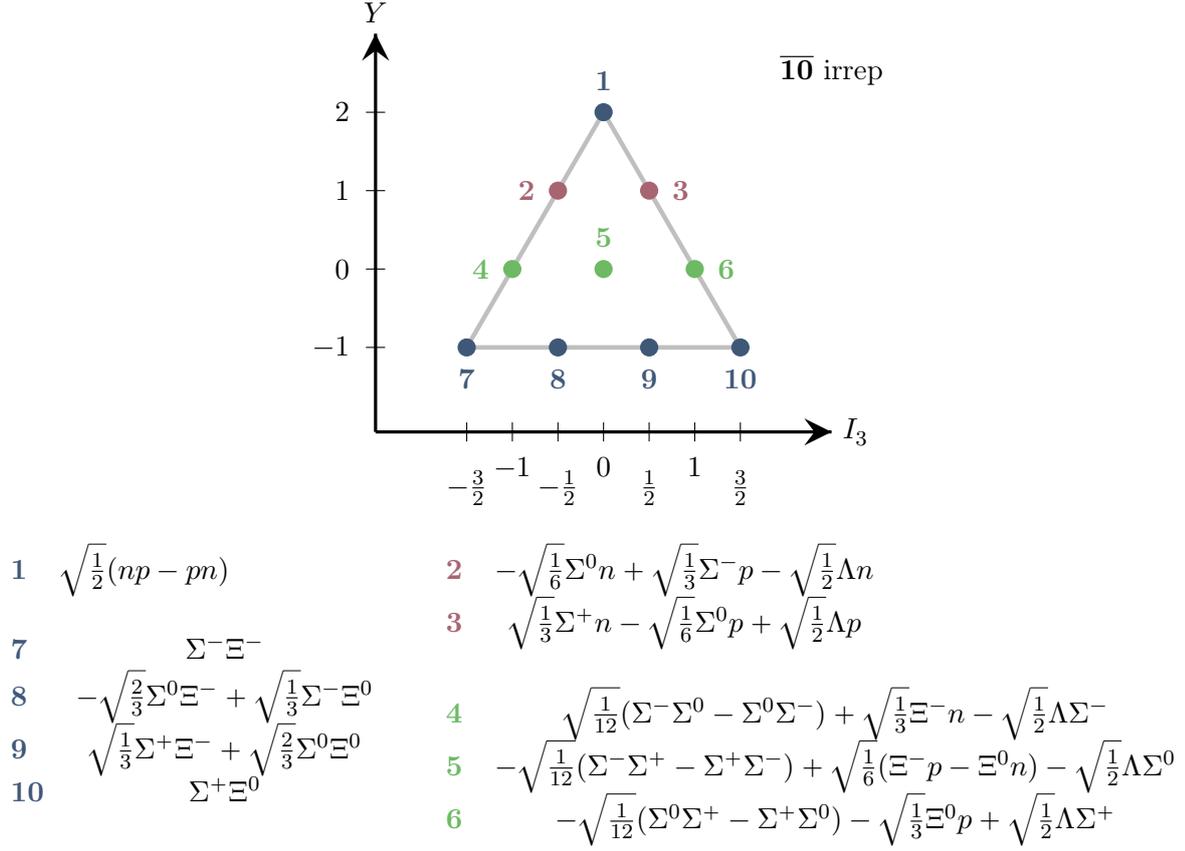

\begin{figure}
\centering
\begin{tikzpicture}[scale=1.20]
\node (00) at (0,0) {}; 

\foreach \c in {1,...,2}{
\foreach \i in {0,...,5}{
\pgfmathtruncatemacro\j{\c*\i} 
\node[circle,minimum width=4pt,inner sep=0pt] (\c;\j) at (60*\i:\c) {}; } }

\foreach \i in {0,2,...,10}{
\pgfmathtruncatemacro\j{mod(\i+2,12)}
\pgfmathtruncatemacro\k{\i+1}
\node (2\k) at ($(2;\i)!.5!(2;\j)$) {};}

\draw[line width=1.7pt,gray!50] ($(2;6)!.5!(2;4)$)--($(2;0)!.5!(2;2)$);
\draw[line width=1.7pt,gray!50] ($(2;6)!.5!(2;4)$)--($(2;8)!.5!(2;10)$);
\draw[line width=1.7pt,gray!50] ($(2;8)!.5!(2;10)$)--($(2;0)!.5!(2;2)$);

\fill[cc0](25) circle (0.1);
\node[text=cc0, font=\bf] at (25) [yshift=+2.5ex] {1};
\fill[cc0](1;2) circle (0.1);
\node[text=cc0, font=\bf] at (1;2) [yshift=+2.5ex] {2};
\fill[cc0](1;1) circle (0.1);
\node[text=cc0, font=\bf] at (1;1) [yshift=+2.5ex] {3};
\fill[cc0](21) circle (0.1);
\node[text=cc0, font=\bf] at (21) [yshift=+2.5ex] {4};

\fill[cc6](1;3) circle (0.1);
\node[text=cc6, font=\bf] at (1;3) [xshift=-2.5ex] {5};
\fill[cc6](00) circle (0.1);
\node[text=cc6, font=\bf] at (00) [yshift=+2.5ex] {6};
\fill[cc6](1;0) circle (0.1);
\node[text=cc6, font=\bf] at (1;0) [xshift=+2.5ex] {7};

\fill[cc7](1;4) circle (0.1);
\node[text=cc7, font=\bf] at (1;4) [xshift=-2.5ex] {8};
\fill[cc7](1;5) circle (0.1);
\node[text=cc7, font=\bf] at (1;5) [xshift=+2.5ex] {9};

\fill[cc0](29) circle (0.1);
\node[text=cc0, font=\bf] at (29) [yshift=-2.5ex] {10};

\draw[decoration={markings,mark=at position 1 with
    {\arrow[scale=2,>=stealth]{>}}},postaction={decorate}, line width=1.2pt] (-2.5,-2.6)--(2.5,-2.6) node[right]{$I_3$};
\draw[decoration={markings,mark=at position 1 with
    {\arrow[scale=2,>=stealth]{>}}},postaction={decorate}, line width=1.2pt] (-2.5,-2.6)--(-2.5,1.8) node[above]{$Y$};

\foreach \x in {1,...,-1} {%
    \draw ($(\x,-2.6) + (0,-3pt)$) -- ($(\x,-2.6) + (0,3pt)$)
        node [below, yshift=-2ex] {$\x$};
}

\foreach \x in {1.5,0.5} {%
\pgfmathtruncatemacro\k{\x*2}
    \draw ($(\x,-2.6) + (0,-3pt)$) -- ($(\x,-2.6) + (0,3pt)$)
        node [below, yshift=-3ex] {$\frac{\k}{2}$};
}

\foreach \x in {-0.5,-1.5} {%
\pgfmathtruncatemacro\k{\x*-2}
\draw ($(\x,-2.6) + (0,-3pt)$) -- ($(\x,-2.6) + (0,3pt)$) node [below, yshift=-3ex] {$-\frac{\k}{2}$};
}

\foreach \y in {1,...,-2} {%
\pgfmathsetmacro\yy{\y*0.8660254}
\draw ($(-2.5,\yy) + (-3pt,0)$) -- ($(-2.5,\yy) + (3pt,0)$) node [left, xshift=-2ex] {$\y$};
}

\node at (2.5,1.5) {$\mathbf{10}$ irrep};
\end{tikzpicture} \vspace*{-15pt}

\noindent\makebox[\textwidth][c]{
\begin{minipage}[t]{0.35\textwidth}
\begin{table}[H]
\begin{tabular}{>{\bfseries\color{cc0}}lc}
1 & $\Sigma^-n$ \\
2 & $\sqrt{\frac{2}{3}}\Sigma^0n + \sqrt{\frac{1}{3}}\Sigma^-p$ \\
3 & $-\sqrt{\frac{1}{3}}\Sigma^+n + \sqrt{\frac{2}{3}}\Sigma^0p$ \\
4 & $\Sigma^+p$
\end{tabular}
\end{table}
\vspace{-1em}
\begin{table}[H]
\begin{tabular}{>{\bfseries\color{cc0}}lc}
10 & $\sqrt{\frac{1}{2}}(\Xi^0\Xi^- - \Xi^-\Xi^0)$
\end{tabular}
\end{table}
\end{minipage}
\begin{minipage}[t]{0.65\textwidth}
\begin{table}[H]
\begin{tabular}{>{\bfseries\color{cc6}}lc}
5 & $\sqrt{\frac{1}{12}}(\Sigma^-\Sigma^0 - \Sigma^0\Sigma^-) + \sqrt{\frac{1}{3}}\Xi^-n + \sqrt{\frac{1}{2}}\Lambda\Sigma^-$ \\
6 & $-\sqrt{\frac{1}{12}}(\Sigma^-\Sigma^+ - \Sigma^+\Sigma^-) + \sqrt{\frac{1}{6}}(\Xi^-p - \Xi^0n) +\sqrt{ \frac{1}{2}}\Lambda\Sigma^0$ \\
7 & $\sqrt{\frac{1}{12}}(\Sigma^0\Sigma^+ - \Sigma^+\Sigma^0) + \sqrt{\frac{1}{3}}\Xi^0p + \sqrt{\frac{1}{2}}\Lambda\Sigma^+$
\end{tabular}
\end{table}
\vspace{-1em}
\begin{table}[H]
\begin{tabular}{>{\bfseries\color{cc7}}lc}
8 & $\sqrt{\frac{1}{6}}\Sigma^0\Xi^- + \sqrt{\frac{1}{3}}\Sigma^-\Xi^0 - \sqrt{\frac{1}{2}}\Lambda\Xi^-$ \\
9 & $-\sqrt{\frac{1}{3}}\Sigma^+\Xi^- + \sqrt{\frac{1}{6}}\Sigma^0\Xi^0 + \sqrt{\frac{1}{2}}\Lambda\Xi^0$
\end{tabular}
\end{table}
\end{minipage}}
\caption{Weight diagram and two-baryon channels for the $\mathbf{10}$ irrep.}
\label{fig:10irrep_chann}
\end{figure}

\begin{figure}
\centering
\begin{tikzpicture}[scale=1.20]
\node (00) at (0,0) {}; 

\foreach \c in {1,...,1}{
\foreach \i in {0,...,5}{
\pgfmathtruncatemacro\j{\c*\i} 
\node[circle,minimum width=4pt,inner sep=0pt] (\c;\j) at (60*\i:\c) {}; } }

\draw[line width=1.7pt,gray!50] ($(1;2)$)--($(1;1)$);
\draw[line width=1.7pt,gray!50] ($(1;1)$)--($(1;0)$);
\draw[line width=1.7pt,gray!50] ($(1;0)$)--($(1;5)$);
\draw[line width=1.7pt,gray!50] ($(1;5)$)--($(1;4)$);
\draw[line width=1.7pt,gray!50] ($(1;4)$)--($(1;3)$);
\draw[line width=1.7pt,gray!50] ($(1;3)$)--($(1;2)$);

\fill[cc5](1;2) circle (0.1);
\node[text=cc5, font=\bf] at (1;2) [yshift=2.5ex] {1};
\fill[cc5](1;1) circle (0.1);
\node[text=cc5, font=\bf] at (1;1) [yshift=2.5ex] {2};

\fill[cc6](1;3) circle (0.1);
\node[text=cc6, font=\bf] at (1;3) [xshift=-2.5ex] {3};
\fill[cc6](00) circle (0.1);
\node[text=cc6, font=\bf] at (00) [xshift=-3ex] {4};
\fill[cc6](1;0) circle (0.1);
\node[text=cc6, font=\bf] at (1;0) [xshift=2.5ex] {5};

\fill[cc7](1;4) circle (0.1);
\node[text=cc7, font=\bf] at (1;4) [yshift=-2.5ex] {6};
\fill[cc7](1;5) circle (0.1);
\node[text=cc7, font=\bf] at (1;5) [yshift=-2.5ex] {7};

\draw[cc0,line width=1.3pt](00) circle (0.2);
\node[text=cc0, font=\bf] at (00) [xshift=3ex] {8};

\draw[decoration={markings,mark=at position 1 with
    {\arrow[scale=2,>=stealth]{>}}},postaction={decorate}, line width=1.2pt] (-2,-1.7)--(2,-1.7) node[right]{$I_3$};
\draw[decoration={markings,mark=at position 1 with
    {\arrow[scale=2,>=stealth]{>}}},postaction={decorate}, line width=1.2pt] (-2,-1.7)--(-2,1.7) node[above]{$Y$};

\foreach \x in {1,...,-1} {%
    \draw ($(\x,-1.7) + (0,-3pt)$) -- ($(\x,-1.7) + (0,3pt)$)
        node [below, yshift=-2ex] {$\x$};
}

\foreach \x in {0.5} {%
\pgfmathtruncatemacro\k{\x*2}
    \draw ($(\x,-1.7) + (0,-3pt)$) -- ($(\x,-1.7) + (0,3pt)$)
        node [below, yshift=-3ex] {$\frac{\k}{2}$};
}

\foreach \x in {-0.5} {%
\pgfmathtruncatemacro\k{\x*-2}
\draw ($(\x,-1.7) + (0,-3pt)$) -- ($(\x,-1.7) + (0,3pt)$) node [below, yshift=-3ex] {$-\frac{\k}{2}$};
}

\foreach \y in {1,...,-1} {%
\pgfmathsetmacro\yy{\y*0.8660254}
\draw ($(-2,\yy) + (-3pt,0)$) -- ($(-2,\yy) + (3pt,0)$) node [left, xshift=-2ex] {$\y$};
}

\node at (1.8,1.3) {$\mathbf{8}_a$ irrep};
\end{tikzpicture} \vspace*{-15pt}

\noindent\makebox[\textwidth][c]{
\begin{minipage}[t]{0.5\textwidth}
\begin{table}[H]
\begin{tabular}{>{\bfseries\color{cc5}}lc}
1 & $-\sqrt{\frac{1}{6}}\Sigma^0n + \sqrt{\frac{1}{3}}\Sigma^-p + \sqrt{\frac{1}{2}}\Lambda n$ \\
2 & $\sqrt{\frac{1}{3}}\Sigma^+n + \sqrt{\frac{1}{6}}\Sigma^0p + \sqrt{\frac{1}{2}}\Lambda p$
\end{tabular}
\end{table}
\vspace{-1em}
\begin{table}[H]
\begin{tabular}{>{\bfseries\color{cc6}}lc}
3 & $-\sqrt{\frac{1}{3}}(\Sigma^-\Sigma^0 - \Sigma^0\Sigma^-) + \sqrt{\frac{1}{3}}\Xi^-n$ \\
4 & $\sqrt{\frac{1}{3}}(\Sigma^-\Sigma^+ - \Sigma^+\Sigma^-) + \sqrt{\frac{1}{6}}(\Xi^-p-\Xi^0n)$ \\
5 & $-\sqrt{\frac{1}{3}}(\Sigma^0\Sigma^+ - \Sigma^+\Sigma^0) + \sqrt{\frac{1}{3}}\Xi^0p$
\end{tabular}
\end{table}
\end{minipage}
\begin{minipage}[t]{0.45\textwidth}
\begin{table}[H]
\begin{tabular}{>{\bfseries\color{cc7}}lc}
6 & $\sqrt{\frac{1}{6}}\Sigma^0\Xi^- + \sqrt{\frac{1}{3}}\Sigma^-\Xi^0 + \sqrt{\frac{1}{2}}\Lambda\Xi^-$ \\
7 & $\sqrt{\frac{1}{3}}\Sigma^+\Xi^- - \sqrt{\frac{1}{6}}\Sigma^0\Xi^0 + \sqrt{\frac{1}{2}}\Lambda\Xi^0$
\end{tabular}
\end{table}
\vspace{-1em}
\begin{table}[H]
\begin{tabular}{>{\bfseries\color{cc0}}lc}
8 & $\sqrt{\frac{1}{2}}(\Xi^0n + \Xi^-p)$
\end{tabular}
\end{table}
\end{minipage}}
\caption{Weight diagram and two-baryon channels for the $\mathbf{8}_a$ irrep.}
\label{fig:8airrep_chann}
\end{figure}
 %SU(3)f decomposition

% Appendix 8 - *******************************************

\chapter{Partial-wave analysis of the NLO terms in the EFT}\label{appen:NLOEFT}

In this appendix we classify all the 19 $SU(3)_f$-symmetric terms appearing at NLO in the Lagrangian of Ref.~\cite{Petschauer:2013uua}. Instead of derivatives, we will work with momenta, with $\bm{p}$ and $\bm{p}'$ being the momentum of the initial and final baryon in the c.m.\ frame, respectively. Keeping only the large components of the spinor, we get several combinations of the momenta and Pauli matrices: $\bm{p}^2$, $\bm{p}'^2$, $\bm{p}\cdot\bm{p}'$, $(\bm{p}\cdot\sigma)(\bm{p}'\cdot\sigma')$, $(\bm{p}\cdot\sigma)(\bm{p}\cdot\sigma')$, and  $(\bm{p}\times\bm{p}')\cdot\sigma$, where the prime in the $\sigma$ is to indicate that it acts on a different baryon field. Following Ref.~\cite{Varshalovich:1988}, in order to disentangle and identify all the partial wave contributions in each operator, we expand it in spherical harmonics,
\begin{equation}
    O(\bm{p},\bm{p}')=\sum_{L M_L L' M'_L}O_{L M_L; L' M'_L}(|\bm{p}|,|\bm{p}'|)Y_{L M_L}(\hat{\bm{p}}) Y_{L' M'_L}(\hat{\bm{p}}')\, ,
\end{equation}
where the coefficients $O_{L M_L; L' M'_L}$ are given by
\begin{equation}
    O_{L M_L; L' M'_L}(|\bm{p}|,|\bm{p}'|)=\int d \hat{\bm{p}}\, d \hat{\bm{p}}'\, Y^*_{L M_L}(\hat{\bm{p}}) Y^*_{L' M'_L}(\hat{\bm{p}}')O(\bm{p},\bm{p}')\, .
\end{equation}
We have labeled $L$ (and $M_L$) the angular momentum of the initial baryons, and $L'$ (and $M'_L$) the angular momentum of the final baryons.
Since we just want to see which $L$ (and $L'$) has non-zero contribution to the scattering, we do not have to find the exact form of $O_{L M_L; L' M'_L}$, we only need to look for the non-zero coefficients. In addition, to expand these terms, it is useful to use the spherical basis,
\begin{equation}
    \begin{Bmatrix}
    A_{+1}=-\frac{1}{\sqrt{2}}(A_x+\imag A_y) \\
    A_{0}=A_z \\
    A_{-1}=\frac{1}{\sqrt{2}}(A_x-\imag A_y) \\
    \end{Bmatrix} \quad \Rightarrow \quad A_{m}=\sqrt{\frac{4\pi}{3}}|A|Y_{1m}(\hat{A})\, ,\quad \text{with }\;  m\in\{\pm 1,0\}\, .
\end{equation}
The relation between the upper and lower indices components is $A_{m}=(-1)^{m}A^{-m}$, and the product of two vectors is $A\cdot B=\sum_{m}A_{m}B^{m}=\sum_{m}(-1)^{m}A_{m}B_{-m}$ for the scalar product and $(A \times B)_{m}=-\imag\sqrt{2}\sum_{m_1m_2}C^{1 m}_{1 m_1;1 m_2}A_{m_1}B^{m_2}$ for the vector product (where $C^{JM}_{j_1m_1;j_2 m_2}$ is the corresponding Clebsch-Gordan coefficient). Since here we are dealing with vectors ($\bm{p}$ and $\sigma$), these are rank 1 tensors, so we can couple them, $A_{m_1}B_{m_2}=\sum_{J M}C^{J M}_{1 m_1;1 m_2}[A\otimes B]_{J M}$, with $0\leq J \leq 2$ (the same for the inverse relation, $[A\otimes B]_{J M}=\sum_{m_1 m_2}C^{J M}_{1m_1;1 m_2}A_{m_1}B_{m_2}$, with $[A\otimes B]_{00}=-\frac{1}{\sqrt{3}}A\cdot B$ or $[A\otimes B]_{1 M}=\frac{\imag}{\sqrt{2}}(A\times B)_{M}$, for example). Let us look at each term:
\begin{enumerate}[label=(\roman*)]
    \item For the term $\bm{p}^2$,
    \begin{equation}
        \displayindent0pt
        \displaywidth\textwidth
        \begin{aligned}
        &\int d \hat{\bm{p}}\, d \hat{\bm{p}}' \,Y^*_{L M_L}(\hat{\bm{p}}) Y^*_{L' M'_L}(\hat{\bm{p}}') \bm{p}^2= \int d \hat{\bm{p}}\, d \hat{\bm{p}}' \, Y^*_{L M_L}(\hat{\bm{p}}) Y^*_{L' M'_L}(\hat{\bm{p}}') \bm{p}^2 \sqrt{4\pi}Y_{00}(\hat{\bm{p}})\sqrt{4\pi}Y_{00}(\hat{\bm{p}}')\\
        &=4\pi \bm{p}^2 \int d \hat{\bm{p}}\, Y^*_{L M_L}(\hat{\bm{p}}) Y_{00}(\hat{\bm{p}})\int d \hat{\bm{p}}'\,  Y^*_{L' M'_L}(\hat{\bm{p}}') Y_{00}(\hat{\bm{p}}')\; \propto\; \bm{p}^2 \delta_{L0}\delta_{M_L 0} \delta_{L' 0}\delta_{M'_L 0}\, .
        \end{aligned}
    \end{equation}
    Therefore, terms which contain only $\bm{p}^2$ or $\bm{p}'^2$ will contribute exclusively to the $S$ wave.
    \item For the term $\bm{p}\cdot\bm{p}'$,
    \begin{equation}
        \displayindent0pt
        \displaywidth\textwidth
        \begin{aligned}
        &\int d \hat{\bm{p}}\, d \hat{\bm{p}}'\, Y^*_{L M_L}(\hat{\bm{p}}) Y^*_{L' M'_L}(\hat{\bm{p}}') (\bm{p}\cdot\bm{p}')=\sum_{m}(-1)^{m} \int d \hat{\bm{p}}\, Y^*_{L M_L}(\hat{\bm{p}}) \bm{p}_{m}\int d \hat{\bm{p}}'\,  Y^*_{L' M'_L}(\hat{\bm{p}}') \bm{p}'_{-m}\\
        &=\frac{4\pi}{3} |\bm{p}||\bm{p}'|\sum_{m}  (-1)^{m}\int d \hat{\bm{p}}\, Y^*_{L M_L}(\hat{\bm{p}}) Y_{1 m}(\hat{\bm{p}})\int d \hat{\bm{p}}'\, Y^*_{L' M'_L}(\hat{\bm{p}}') Y_{1 -m}(\hat{\bm{p}}')\; \propto\; |\bm{p}||\bm{p}'| \delta_{L1}\delta_{L'1}\, .
        \end{aligned}
    \label{eq:ppPwave}
    \end{equation}
    This term contributes to the $P$ wave.
    \item For the term $(\bm{p}\cdot\sigma)(\bm{p}'\cdot\sigma')$,
    \begin{equation}
        \displayindent0pt
        \displaywidth\textwidth
        \begin{aligned}
        (\bm{p}\cdot\sigma)(\bm{p}'\cdot\sigma') &= \sum_{m_1,m_2} (-1)^{m_1+m_2} \bm{p}_{m_1} \sigma_{-m_1} \bm{p}'_{m_2} \sigma'_{-m_2} \\
        &=\sum_{m_1,m_2}\sum_{J,M}(-1)^{m_1+m_2}C^{J M}_{1m_1;1 m_2 }\sigma_{-m_1}\sigma'_{-m_2}[\bm{p}\otimes \bm{p}']_{J M}\\
        &=\sum_{m_1,m_2}\sum_{J,M}(-1)^{m_1+m_2}(-1)^{1+1-J}C^{J -M}_{1-m_1;1 -m_2 }\sigma_{-m_1}\sigma'_{-m_2}[\bm{p}\otimes \bm{p}']_{J M}\\
        &=\sum_{m_1,m_2}\sum_{J,M}(-1)^{M-J}C^{J -M}_{1-m_1;1 -m_2 }\sigma_{-m_1}\sigma'_{-m_2}[\bm{p}\otimes \bm{p}']_{J M}\\
        &=\sum_{J M}(-1)^{M-J}[\sigma\otimes \sigma']_{J -M}[\bm{p}\otimes \bm{p}']_{J M}\, .
        \end{aligned}
    \end{equation}
    The first contribution will come from $J=0$ (and $M=0$),
    \begin{equation}
        \displayindent0pt
        \displaywidth\textwidth
        (\bm{p}\cdot\sigma)(\bm{p}'\cdot\sigma') \xrightarrow{J=0} [\sigma\otimes \sigma']_{00} [\bm{p}\otimes \bm{p}']_{00} = \frac{1}{3}(\sigma \cdot \sigma') (\bm{p}\cdot \bm{p}')\, ,
    \end{equation}
    which contributes to the $P$ wave, as seen in Eq.~\eqref{eq:ppPwave}. Next, the $J=1$ term,
    \begin{equation}
        \displayindent0pt
        \displaywidth\textwidth
        \begin{aligned}
        (\bm{p}\cdot\sigma)(\bm{p}'\cdot\sigma')& \xrightarrow{J=1}  \sum_M (-1)^{M-1}[\sigma\otimes \sigma']_{1-M}[\bm{p}\otimes \bm{p}']_{1M} \\
        &= \sum_M(-1)^{M-1} \sum_{m_1,m_2} C^{1-M}_{1m_1;1m_2} \sigma_{m_1} \sigma'_{m_2} \sum_{m_3,m_4}C^{1M}_{1m_3;1m_4} \bm{p}_{m_3} \bm{p}'_{m_4}\\
        &= \sum_{\substack{m_1,m_2\\m_3,m_4}} \left[\sum_M (-1)^{M-1} C^{1-M}_{1m_1;1m_2}C^{1M}_{1m_3;1m_4} \right] \sigma_{m_1} \sigma'_{m_2}\bm{p}_{m_3} \bm{p}'_{m_4} \\
        &= \sum_{\substack{m_1,m_2\\m_3,m_4}} \left[\sum_M (-1)^M C^{1M}_{1-m_1;1-m_2}C^{1M}_{1m_3;1m_4} \right] \sigma_{m_1} \sigma'_{m_2}\bm{p}_{m_3} \bm{p}'_{m_4}\\
        &=|\bm{p}||\bm{p}'|\frac{4\pi}{3}\sum_{\substack{m_1,m_2\\m_3,m_4}} \left[\sum_M (-1)^M C^{1M}_{1-m_1;1-m_2}C^{1M}_{1m_3;1m_4} \right] \sigma_{m_1} \sigma'_{m_2}Y_{1m_3}(\hat{\bm{p}}) Y_{1m_4}(\hat{\bm{p}}')\, .
        \end{aligned}
    \end{equation}
    This also contributes to the $P$ wave. Finally, the $J=2$ term,
    \begin{equation}
        \displayindent0pt
        \displaywidth\textwidth
        \begin{aligned}
        (\bm{p}\cdot\sigma)(\bm{p}'\cdot\sigma')& \xrightarrow{J=2}  \sum_M (-1)^{M-2} [\sigma\otimes \sigma']_{2-M} [\bm{p}\otimes \bm{p}']_{2M} \\
        &= \sum_M(-1)^{M-2} \sum_{m_1,m_2} C^{2-M}_{1m_1;1m_2} \sigma_{m_1} \sigma'_{m_2} \sum_{m_3,m_4}C^{2M}_{1m_3;1m_4} \bm{p}_{m_3} \bm{p}'_{m_4}\\
        &=\sum_{\substack{m_1,m_2\\m_3,m_4}} \left[\sum_M (-1)^{M-2} C^{2-M}_{1m_1;1m_2}C^{2M}_{1m_3;1m_4} \right] \sigma_{m_1} \sigma'_{m_2}\bm{p}_{m_3} \bm{p}'_{m_4} \\
        &= \sum_{\substack{m_1,m_2\\m_3,m_4}} \left[\sum_M (-1)^M C^{1M}_{1-m_1;1-m_2}C^{1M}_{1m_3;1m_4} \right] \sigma_{m_1} \sigma'_{m_2}\bm{p}_{m_3} \bm{p}'_{m_4}\\
        &=|\bm{p}||\bm{p}'|\frac{4\pi}{3}\sum_{\substack{m_1,m_2\\m_3,m_4}} \left[ \sum_M (-1)^M C^{2M}_{1-m_1;1-m_2}C^{2M}_{1m_3;1m_4} \right] \sigma_{m_1} \sigma'_{m_2}Y_{1m_3}(\hat{\bm{p}}) Y_{1m_4}(\hat{\bm{p}}')\, ,
        \end{aligned}
    \end{equation}
    which also contributes to the $P$ wave.
    \item For the term $(\bm{p}\cdot\sigma)(\bm{p}\cdot\sigma')$, which is similar to the previous one except for the exchange $\bm{p}'\rightarrow \bm{p}$,
    \begin{equation}
        \displayindent0pt
        \displaywidth\textwidth
        (\bm{p}\cdot\sigma)(\bm{p}\cdot\sigma') = \sum_{J,M}(-1)^{M-J} [\sigma\otimes \sigma']_{J -M} [\bm{p}\otimes \bm{p}]_{J M}\, .
    \end{equation}
    The first contribution will come from $J=0$ (and $M=0$),
    \begin{equation}
        \displayindent0pt
        \displaywidth\textwidth
        (\bm{p}\cdot\sigma)(\bm{p}\cdot\sigma') \xrightarrow{J=0} [\sigma\otimes \sigma']_{00}[\bm{p}\otimes \bm{p}]_{00}=\frac{1}{3}(\sigma\cdot\sigma')\bm{p}^2\, .
    \end{equation}
    This term contributes to the $S$ wave.
    Next, the $J=1$ term vanishes, since $\bm{p}\times\bm{p}=0$, and the $J=2$ term,
    \begin{equation}
        \displayindent0pt
        \displaywidth\textwidth
        (\bm{p}\cdot\sigma)(\bm{p}\cdot\sigma') \xrightarrow{J=2} \bm{p}^2 \frac{4\pi}{3} \sum_M (-1)^{M-2} [\sigma\otimes \sigma']_{2-M} [Y_{1}(\hat{\bm{p}})\otimes Y_1(\hat{\bm{p}})]_{2M}\, .
    \end{equation}
    The last object can be written as
    \begin{equation}
        \displayindent0pt
        \displaywidth\textwidth
        [Y_{l_1}(\hat{\bm{p}})\otimes Y_{l_2}(\hat{\bm{p}})]_{LM}=\sqrt{\frac{(2l_1+1)(2l_2+1)}{4\pi(2L+1)}}C^{L0}_{l_10;l_20}Y_{LM}(\hat{\bm{p}})\, .
    \end{equation}
    Then,
    \begin{equation}
        \displayindent0pt
        \displaywidth\textwidth
        \begin{aligned}
        &\bm{p}^2 \frac{4\pi}{3}\sum_M (-1)^{M-2}[\sigma\otimes \sigma']_{2-M}[Y_{1}(\hat{\bm{p}})\otimes Y_1(\hat{\bm{p}})]_{2M} = \bm{p}^2 \frac{4\pi}{3}\sqrt{\frac{3}{10\pi}} \sum_M (-1)^{M}[\sigma\otimes \sigma']_{2-M}Y_{2M}(\hat{\bm{p}})\\
        &=\bm{p}^2 \sqrt{\frac{8\pi }{15}} \sum_M (-1)^{M-2}[\sigma\otimes \sigma']_{2-M}Y_{2M}(\hat{\bm{p}})=\bm{p}^2 \sqrt{\frac{8\pi }{15}} \sum_M\sum_{m_1,m_2} (-1)^{M} C^{2-M}_{1m_1;1m_2} \sigma_{m_1} \sigma'_{m_2} Y_{2M}(\hat{\bm{p}})\\
        &= \bm{p}^2 \sqrt{\frac{8\pi }{15}} \sum_{m_1,m_2} (-1)^{m_1+m_2} C^{2\, m_1+m_2}_{1m_1;1m_2} \sigma_{m_1} \sigma'_{m_2} Y_{2\, -m_1-m_2}(\hat{\bm{p}}) \\
        &=\bm{p}^2 \sqrt{\frac{8\pi }{15}} \sum_{m_1,m_2} C^{2\, m_1+m_2}_{1m_1;1m_2} \sigma_{m_1} \sigma'_{m_2} Y^*_{2\, m_1+m_2}(\hat{\bm{p}})\, .
        \end{aligned}
    \end{equation}
    This term is responsible for the $\3s1 \rightarrow {}^3 \hskip -0.025in D_1$ transition.
    \item For the term $(\bm{p}\times\bm{p}')\cdot\sigma$,
    \begin{equation}
        \displayindent0pt
        \displaywidth\textwidth
        \begin{aligned}
        (\bm{p}\times\bm{p}')\cdot\sigma &= -\imag\sqrt{2}\sum_{m_1,m_2,m_3} (-1)^{m_3}C^{1\, m_3}_{1m_1;1m_2}\bm{p}_{m_1}\bm{p}'_{m_2}\sigma_{-m_3}\\
        &=-\imag\sqrt{2}|\bm{p}||\bm{p}'|\frac{4\pi}{3}\sum_{m_1,m_2,m_3} (-1)^{m_3} C^{1\, m_3}_{1m_1;1m_2}Y_{1m_1}(\hat{\bm{p}})Y_{1m_2}(\hat{\bm{p}}')\sigma_{-m_3}\, ,
        \end{aligned}
    \end{equation}
    which is responsible for the $^1 \hskip -0.025in P_1 \rightarrow {}^3 \hskip -0.025in P_1$ transition.
\end{enumerate}
After analyzing all the terms, the ones that contribute to the $S$ wave contain $\bm{p}^2$ and $\bm{p}'^2$, so the Lagrangian will only contain second derivatives, as written in Eq.~\eqref{eq:LagNLO1}. %Partial-wave NLO

% Appendix 7 - *******************************************

\chapter{On leading flavor-symmetry breaking coefficients in the EFT}\label{appen:LECsEFT}

Table 10 of Ref.~\cite{Petschauer:2013uua} lists the $SU(3)_f$ symmetry-breaking LECs $c^i_{\chi}$ for all of the two-(octet) baryon channels. These coefficients are a combination of different terms in the Lagrangian shown in Table 9 of the same reference (terms 29-40). The relations between the $c^i_{\chi}$ from Ref.~\cite{Petschauer:2013uua} and the ones in Eq.~\eqref{eq:LagNLO2} introduced in the present work are presented in Table~\ref{tab:coeff_petshauervsme}. Instead of the $(^{2s+1} L_J,I)$ notation, the channels are labeled as $(^{2I+1}_{2s+1})$ for brevity, as $L=0$ in all cases.

In Table~\ref{tab:full_eft_coeff}, a list of the two-baryon channels that are required to obtain independently all the LECs of this work is provided. There are 6 LO and 12 NLO symmetry-breaking coefficients that are referred to as momentum independent in this paper, as well as 6 NLO momentum-dependent coefficients, making a total of 24 parameters that would need to be constrained in a potentially more exhaustive future study. For the momentum-independent coefficients, the choice of the systems is not unique, as there are 37 different channels that can be used to constrain only 18 parameters (assuming $SU(2)_f$ symmetry and no electromagnetic interaction). For the momentum-dependent coefficients, no extra channels are needed besides those used for the momentum-independent ones. For simplicity, only channels that do not change the baryon content are used (e.g., $\Sigma N \rightarrow \Sigma N$, denoted as $\Sigma N$ in short).

\begin{table}[h!]
\caption{Comparison between the symmetry-breaking LECs of this work and those in Ref.~\cite{Petschauer:2013uua} for the two-baryon channels for which only one $c^i_{\chi}$ appears in that reference.}
\label{tab:coeff_petshauervsme}
\centering
\renewcommand{\arraystretch}{1.3}
\resizebox{\columnwidth}{!}{
\begin{tabular}{ccc}
\toprule
Channel $(^{2I+1}_{2s+1})$	&	Ref.~\cite{Petschauer:2013uua} & Coefficients in Eq.~\eqref{eq:LagNLO2}	\\ \midrule
$NN\rightarrow NN \; (^3_1)$ & $c^1_{\chi}/2$ & $4(c^{\chi}_3 - c^{\chi}_4)$ \\
$\Lambda N\rightarrow \Lambda N \; (^2_1)$	&	$c^2_{\chi}$	&	$\frac{1}{3}(4c^{\chi}_1-4c^{\chi}_2+9c^{\chi}_3-9c^{\chi}_4-4c^{\chi}_5+4c^{\chi}_6-c^{\chi}_9+c^{\chi}_{10}+4c^{\chi}_{11}-4c^{\chi}_{12})$ 	\\
$\Lambda N\rightarrow \Sigma N	 \; (^2_1)$	&	$-c^3_{\chi}$	&	$c^{\chi}_3-c^{\chi}_4+2c^{\chi}_5-2c^{\chi}_6+c^{\chi}_9-c^{\chi}_{10}$	\\
$\Sigma N\rightarrow \Sigma N \; (^2_1)$	&	$c^4_{\chi}$	&	$-c^{\chi}_3+c^{\chi}_4-3c^{\chi}_9+3c^{\chi}_{10}$	\\
$\Lambda \Lambda\rightarrow \Lambda \Lambda \; (^1_1)$	&	$\frac{c^5_{\chi}}{2}$	& $\frac{8}{9}(2c^{\chi}_1-2c^{\chi}_2+2c^{\chi}_3-2c^{\chi}_4-4c^{\chi}_5+4c^{\chi}_6-2c^{\chi}_7+2c^{\chi}_8-2c^{\chi}_9+2c^{\chi}_{10}+3c^{\chi}_{11}-3c^{\chi}_{12})$	\\
$\Xi N\rightarrow \Xi N \; (^3_1)$	&	$c^6_{\chi}$	&	$2(-2c^{\chi}_5+2c^{\chi}_6+c^{\chi}_{11}-c^{\chi}_{12})$	\\
$NN\rightarrow NN \; (^1_3)$	&	$c^7_{\chi}/2$	&	$4(c^{\chi}_3+ c^{\chi}_4)$	\\
$\Lambda N\rightarrow \Lambda N \; (^2_3)$	&	$c^8_{\chi}$	&	$\frac{1}{3}(4c^{\chi}_1+4c^{\chi}_2+7c^{\chi}_3+7c^{\chi}_4+12c^{\chi}_5+12c^{\chi}_6+9c^{\chi}_9+9c^{\chi}_{10}+4c^{\chi}_{11}+4c^{\chi}_{12})$	\\
$\Lambda N\rightarrow \Sigma N \; (^2_3)$	&	$-c^9_{\chi}$	&	$-c^{\chi}_3-c^{\chi}_4+2c^{\chi}_5+2c^{\chi}_6+3c^{\chi}_9+3c^{\chi}_{10}$ \\
$\Sigma N\rightarrow \Sigma N \; (^2_3)$	&	$c^{10}_{\chi}$	&	$c^{\chi}_3+c^{\chi}_4+3c^{\chi}_9+3c^{\chi}_{10}$ 	\\ 
$\Xi N\rightarrow \Xi N \; (^1_3)$	&	$c^{11}_{\chi}$	&	 $2(2c^{\chi}_5+2c^{\chi}_6+2c^{\chi}_7+2c^{\chi}_8+2c^{\chi}_9+2c^{\chi}_{10}+c^{\chi}_{11}+c^{\chi}_{12})$	\\ 
$\Xi N\rightarrow \Xi N \; (^3_3)$	&	$c^{12}_{\chi}$	&	$2(2c^{\chi}_5+2c^{\chi}_6+c^{\chi}_{11}+c^{\chi}_{12})$ \\\bottomrule
\end{tabular}}
\end{table}
\begin{table}[!htbp]
\renewcommand{\arraystretch}{1.6}
\caption{Combinations of two-baryon channels necessary to constrain independently all of the LO+NLO EFT LECs introduced in~\cref{subsec:SU3EFT}.}
\label{tab:full_eft_coeff}
\resizebox{\columnwidth}{!}{
\begin{tabular}{c >{\footnotesize}c c >{\footnotesize}c}
\toprule
Coefficient & \multicolumn{3}{c}{Channels $(^{2I+1}_{2s+1})$}\\
\midrule
$c^{(27)}$ & \multicolumn{3}{ >{\footnotesize}c}{$2 \Xi \Sigma (^4_1)- \Xi\Xi (^3_1)$} \\
$c^{(8_s)}$ & \multicolumn{3}{ >{\footnotesize}c}{$\frac{15}{4} \Lambda \Lambda (^1_1)+\frac{35}{36} \Sigma\Sigma (^1_1)-5\Xi\Lambda (^2_1)-\frac{5}{3}\Xi N (^1_1)+\frac{5}{9}\Xi \Sigma (^2_1)+\frac{1}{3}\Xi \Sigma (^4_1) +\frac{37}{18}\Xi\Xi (^3_1)$} \\
$c^{(1)}$ & \multicolumn{3}{ >{\footnotesize}c}{$-6 \Lambda \Lambda (^1_1)+\frac{10}{9} \Sigma\Sigma (^1_1)+8\Xi\Lambda (^2_1) +\frac{8}{3}\Xi N (^1_1)-\frac{8}{9}\Xi \Sigma (^2_1)-\frac{2}{3}\Xi \Sigma (^4_1) -\frac{29}{9}\Xi\Xi (^3_1)$} \\
$c^{(\overline{10})}$ & \multicolumn{3}{ >{\footnotesize}c}{$\frac{1}{3}NN(^1_3)+\frac{8}{9}\Sigma N (^2_3)+\frac{2}{9}\Sigma N (^4_3)+\frac{4}{3}\Xi\Lambda (^2_3)-\frac{2}{3}\Xi N (^1_3)-\frac{2}{3}\Xi N (^3_3)-\frac{4}{9}\Xi\Sigma (^2_3)+\frac{4}{9}\Xi\Sigma (^4_3)-\frac{4}{9}\Xi\Xi (^1_3)$}\\
$c^{(10)}$ & \multicolumn{3}{ >{\footnotesize}c}{$\frac{1}{3}NN(^1_3)-\frac{4}{9}\Sigma N (^2_3)+\frac{8}{9}\Sigma N (^4_3)-\frac{2}{3}\Xi\Lambda (^2_3)+\frac{1}{3}\Xi N (^1_3)+\frac{1}{3}\Xi N (^3_3)+\frac{2}{9}\Xi\Sigma (^2_3)-\frac{2}{9}\Xi\Sigma (^4_3)+\frac{2}{9}\Xi\Xi (^1_3)$} \\
$c^{(8_a)}$ & \multicolumn{3}{ >{\footnotesize}c}{$-\frac{11}{12}NN(^1_3)+\frac{17}{9}\Sigma N (^2_3)-\frac{7}{9}\Sigma N (^4_3)-\frac{1}{6}\Xi\Lambda (^2_3)-\frac{17}{12}\Xi N (^1_3)+\frac{19}{12}\Xi N (^3_3)+\frac{37}{18}\Xi\Sigma (^2_3)-\frac{5}{9}\Xi\Sigma (^4_3)-\frac{25}{36}\Xi\Xi (^1_3)$} \\[1.5ex]
$c^{\chi}_1$ & \multicolumn{3}{ >{\footnotesize}c}{$\begin{aligned}
&\tfrac{9}{16}\Lambda\Lambda (^1_1)-\tfrac{1}{48}NN(^1_3) -\tfrac{5}{36}\Sigma N (^2_3)-\tfrac{1}{2}\Sigma N (^4_1)+\tfrac{1}{36}\Sigma N (^4_3)+\tfrac{7}{48}\Sigma\Sigma (^1_1)-\tfrac{5}{24}\Xi \Lambda (^2_3)-\tfrac{1}{4}\Xi N (^1_1)\\
&+\tfrac{5}{48}\Xi N (^1_3)-\tfrac{1}{4}\Xi N (^3_1)+\tfrac{5}{48}\Xi N (^3_3)+\tfrac{5}{72}\Xi\Sigma (^2_3)+\tfrac{5}{6}\Xi \Sigma (^4_1)+\tfrac{1}{18}\Xi \Sigma (^4_3)+\tfrac{1}{144}\Xi \Xi (^1_3)-\tfrac{13}{24}\Xi \Xi (^3_1)
\end{aligned}$} \\[3.5ex]
$c^{\chi}_2$ & \multicolumn{3}{ >{\footnotesize}c}{$\begin{aligned}
&-\tfrac{9}{16}\Lambda\Lambda (^1_1)-\tfrac{1}{48}NN(^1_3) -\tfrac{5}{36}\Sigma N (^2_3)+\tfrac{1}{2}\Sigma N (^4_1)+\tfrac{1}{36}\Sigma N (^4_3)-\tfrac{7}{48}\Sigma\Sigma (^1_1)-\tfrac{5}{24}\Xi \Lambda (^2_3)+\tfrac{1}{4}\Xi N (^1_1)\\
&+\tfrac{5}{48}\Xi N (^1_3)+\tfrac{1}{4}\Xi N (^3_1)+\tfrac{5}{48}\Xi N (^3_3)+\tfrac{5}{72}\Xi\Sigma (^2_3)-\tfrac{5}{6}\Xi \Sigma (^4_1)+\tfrac{1}{18}\Xi \Sigma (^4_3)+\tfrac{1}{144}\Xi \Xi (^1_3)+\tfrac{13}{24}\Xi \Xi (^3_1)
\end{aligned}$} \\[3.5ex]
$c^{\chi}_3$ & \multicolumn{3}{ >{\footnotesize}c}{$\begin{aligned}
&\tfrac{1}{12}NN(^1_3)-\tfrac{1}{9}\Sigma N (^2_3)+\tfrac{1}{4}\Sigma N (^4_1)-\tfrac{1}{36}\Sigma N (^4_3)-\tfrac{1}{6}\Xi\Lambda (^2_3)+\tfrac{1}{12}\Xi N (^1_3)\\
&+\tfrac{1}{12}\Xi N (^3_3)+\tfrac{1}{18}\Xi\Sigma (^2_3)-\tfrac{1}{2}\Xi\Sigma (^4_1)-\tfrac{1}{18}\Xi\Sigma (^4_3)+\tfrac{1}{18}\Xi\Xi (^1_3)+\tfrac{1}{4}\Xi\Xi (^3_1)
\end{aligned}$} \\[3.5ex]
$c^{\chi}_4$ & \multicolumn{3}{ >{\footnotesize}c}{$\begin{aligned}
&\tfrac{1}{12}NN(^1_3)-\tfrac{1}{9}\Sigma N (^2_3)-\tfrac{1}{4}\Sigma N (^4_1)-\tfrac{1}{36}\Sigma N (^4_3)-\tfrac{1}{6}\Xi\Lambda (^2_3)+\tfrac{1}{12}\Xi N (^1_3)\\
&+\tfrac{1}{12}\Xi N (^3_3)+\tfrac{1}{18}\Xi\Sigma (^2_3)+\tfrac{1}{2}\Xi\Sigma (^4_1)-\tfrac{1}{18}\Xi\Sigma (^4_3)+\tfrac{1}{18}\Xi\Xi (^1_3)-\tfrac{1}{4}\Xi\Xi (^3_1)
\end{aligned}$} \\[3.5ex]
$c^{\chi}_5$ & \multicolumn{3}{ >{\footnotesize}c}{$\begin{aligned}
&\tfrac{1}{24}NN(^1_3)-\tfrac{1}{18}\Sigma N (^2_3)+\tfrac{1}{4}\Sigma N (^4_1)+\tfrac{1}{36}\Sigma N (^4_3)-\tfrac{3}{8}\Xi\Lambda (^2_1)+\tfrac{1}{24}\Xi\Lambda (^2_3)+\tfrac{1}{24}\Xi N (^1_3)\\
&+\tfrac{1}{24}\Xi N (^3_3)+\tfrac{1}{24}\Xi\Sigma (^2_1)-\tfrac{7}{72}\Xi\Sigma (^2_3)-\tfrac{5}{12}\Xi\Sigma (^4_1)-\tfrac{1}{36}\Xi\Sigma (^4_3)-\tfrac{1}{72}\Xi\Xi (^1_3)+\tfrac{1}{2}\Xi\Xi (^3_1)
\end{aligned}$} \\[3.5ex]
$c^{\chi}_6$ & \multicolumn{3}{ >{\footnotesize}c}{$\begin{aligned}
&\tfrac{1}{24}NN(^1_3)-\tfrac{1}{18}\Sigma N (^2_3)-\tfrac{1}{4}\Sigma N (^4_1)+\tfrac{1}{36}\Sigma N (^4_3)+\tfrac{3}{8}\Xi\Lambda (^2_1)+\tfrac{1}{24}\Xi\Lambda (^2_3)+\tfrac{1}{24}\Xi N (^1_3)\\
&+\tfrac{1}{24}\Xi N (^3_3)-\tfrac{1}{24}\Xi\Sigma (^2_1)-\tfrac{7}{72}\Xi\Sigma (^2_3)+\tfrac{5}{12}\Xi\Sigma (^4_1)-\tfrac{1}{36}\Xi\Sigma (^4_3)-\tfrac{1}{72}\Xi\Xi (^1_3)-\tfrac{1}{2}\Xi\Xi (^3_1)
\end{aligned}$} \\[3.5ex]
$c^{\chi}_7$ & \multicolumn{3}{ >{\footnotesize}c}{$\begin{aligned}
\tfrac{1}{12}&NN(^1_3)-\tfrac{1}{9}\Sigma N (^2_3)+\tfrac{1}{2}\Sigma N (^4_1)+\tfrac{1}{18}\Sigma N (^4_3)-\tfrac{3}{4}\Xi\Lambda (^2_1)+\tfrac{1}{12}\Xi\Lambda (^2_3)+\tfrac{1}{12}\Xi N (^1_3)+\tfrac{1}{4}\Xi N (^3_1)\\
&-\tfrac{1}{6}\Xi N (^3_3)-\tfrac{1}{12}\Xi\Sigma (^2_1)-\tfrac{1}{36}\Xi\Sigma (^2_3)-\tfrac{11}{12}\Xi\Sigma (^4_1)+\tfrac{1}{36}\Xi\Sigma (^4_3)-\tfrac{1}{36}\Xi\Xi (^1_3)+\Xi\Xi (^3_1)
\end{aligned}$} \\[3.5ex]
$c^{\chi}_8$ & \multicolumn{3}{ >{\footnotesize}c}{$\begin{aligned}
\tfrac{1}{12}&NN(^1_3)-\tfrac{1}{9}\Sigma N (^2_3)-\tfrac{1}{2}\Sigma N (^4_1)+\tfrac{1}{18}\Sigma N (^4_3)+\tfrac{3}{4}\Xi\Lambda (^2_1)+\tfrac{1}{12}\Xi\Lambda (^2_3)+\tfrac{1}{12}\Xi N (^1_3)-\tfrac{1}{4}\Xi N (^3_1)\\
&-\tfrac{1}{6}\Xi N (^3_3)+\tfrac{1}{12}\Xi\Sigma (^2_1)-\tfrac{1}{36}\Xi\Sigma (^2_3)+\tfrac{11}{12}\Xi\Sigma (^4_1)+\tfrac{1}{36}\Xi\Sigma (^4_3)-\tfrac{1}{36}\Xi\Xi (^1_3)-\Xi\Xi (^3_1)
\end{aligned}$} \\[3.5ex]
$c^{\chi}_9$ & \multicolumn{3}{ >{\footnotesize}c}{$ \begin{aligned}
-\tfrac{9}{16}&\Lambda\Lambda (^1_1)+\tfrac{1}{48}NN(^1_3)-\tfrac{1}{36}\Sigma N (^2_3)-\tfrac{1}{2}\Sigma N (^4_1)+\tfrac{1}{18}\Sigma N (^4_3)+\tfrac{1}{48}\Sigma\Sigma (^1_1)+\tfrac{3}{2}\Xi \Lambda (^2_1)-\tfrac{1}{24}\Xi \Lambda (^2_3)+\tfrac{1}{8}\Xi N (^1_1)\\
&+\tfrac{7}{48}\Xi N (^1_3)-\tfrac{1}{8}\Xi N (^3_1)-\tfrac{5}{48}\Xi N (^3_3)-\tfrac{11}{72}\Xi\Sigma (^2_3)+\tfrac{5}{6}\Xi \Sigma (^4_1)+\tfrac{1}{36}\Xi \Sigma (^4_3)+\tfrac{11}{144}\Xi \Xi (^1_3)-\tfrac{31}{24}\Xi \Xi (^3_1)
\end{aligned}$} \\[3.5ex]
$c^{\chi}_{10}$ & \multicolumn{3}{ >{\footnotesize}c}{$\begin{aligned}
\tfrac{9}{16}&\Lambda\Lambda (^1_1)+\tfrac{1}{48}NN(^1_3)-\tfrac{1}{36}\Sigma N (^2_3)+\tfrac{1}{2}\Sigma N (^4_1)+\tfrac{1}{18}\Sigma N (^4_3)-\tfrac{1}{48}\Sigma\Sigma (^1_1)-\tfrac{3}{2}\Xi \Lambda (^2_1)-\tfrac{1}{24}\Xi \Lambda (^2_3)-\tfrac{1}{8}\Xi N (^1_1)\\
&+\tfrac{7}{48}\Xi N (^1_3)+\tfrac{1}{8}\Xi N (^3_1)-\tfrac{5}{48}\Xi N (^3_3)-\tfrac{11}{72}\Xi\Sigma (^2_3)-\tfrac{5}{6}\Xi \Sigma (^4_1)+\tfrac{1}{36}\Xi \Sigma (^4_3)+\tfrac{11}{144}\Xi \Xi (^1_3)+\tfrac{31}{24}\Xi \Xi (^3_1)
\end{aligned}$} \\[3.5ex]
$c^{\chi}_{11}$ & \multicolumn{3}{ >{\footnotesize}c}{$\begin{aligned}
&-\tfrac{9}{16}\Lambda\Lambda (^1_1)-\tfrac{1}{16}NN(^1_3)-\tfrac{1}{12}\Sigma N (^2_3)+\tfrac{1}{2}\Sigma N (^4_1)-\tfrac{1}{12}\Sigma N (^4_3)-\tfrac{7}{48}\Sigma\Sigma (^1_1)-\tfrac{1}{8}\Xi \Lambda (^2_3)+\tfrac{1}{4}\Xi N (^1_1)\\
&+\tfrac{1}{16}\Xi N (^1_3)+\tfrac{1}{4}\Xi N (^3_1)+\tfrac{1}{16}\Xi N (^3_3)+\tfrac{1}{24}\Xi\Sigma (^2_3)-\tfrac{13}{12}\Xi \Sigma (^4_1)+\tfrac{1}{12}\Xi \Sigma (^4_3)+\tfrac{5}{48}\Xi \Xi (^1_3)+\tfrac{19}{24}\Xi \Xi (^3_1)
\end{aligned}$} \\[3.5ex]
$c^{\chi}_{12}$ & \multicolumn{3}{ >{\footnotesize}c}{$\begin{aligned}
&\tfrac{9}{16}\Lambda\Lambda (^1_1)-\tfrac{1}{16}NN(^1_3)-\tfrac{1}{12}\Sigma N (^2_3)-\tfrac{1}{2}\Sigma N (^4_1)-\tfrac{1}{12}\Sigma N (^4_3)+\tfrac{7}{48}\Sigma\Sigma (^1_1)-\tfrac{1}{8}\Xi \Lambda (^2_3)-\tfrac{1}{4}\Xi N (^1_1)\\
&+\tfrac{1}{16}\Xi N (^1_3)-\tfrac{1}{4}\Xi N (^3_1)+\tfrac{1}{16}\Xi N (^3_3)+\tfrac{1}{24}\Xi\Sigma (^2_3)+\tfrac{13}{12}\Xi \Sigma (^4_1)+\tfrac{1}{12}\Xi \Sigma (^4_3)+\tfrac{5}{48}\Xi \Xi (^1_3)-\tfrac{19}{24}\Xi \Xi (^3_1)
\end{aligned}$} \\[2.5ex]
\midrule
$\tilde{c}^{(27)}$ & $\phantom{\Lambda\Lambda\Lambda\Lambda (^1_1)} \Xi\Xi (^3_1) \phantom{\Lambda\Lambda\Lambda\Lambda (^1_1)}$ & $\tilde{c}^{(\overline{10})}$ & $NN(^1_3)$\\
$\tilde{c}^{(8_s)}$ & $\phantom{\Lambda\Lambda\Lambda\Lambda (^1_1)} \tfrac{1}{3}\Lambda\Lambda (^1_1)+2\Sigma\Sigma (^1_1)-\tfrac{5}{3}\Xi N (^1_1) \phantom{\Lambda\Lambda\Lambda\Lambda (^1_1)}$ & $\tilde{c}^{(10)}$ & $\Xi \Xi (^1_3)$\\
$\tilde{c}^{(1)}$ & $\phantom{\Lambda\Lambda\Lambda\Lambda (^1_1)} -\tfrac{7}{6}\Lambda\Lambda (^1_1)-\tfrac{1}{2}\Sigma\Sigma (^1_1)+\tfrac{8}{3}\Xi N (^1_1) \phantom{\Lambda\Lambda\Lambda\Lambda (^1_1)}$ & $\tilde{c}^{(8_a)}$ & $\Xi N (^1_3)$ \\ \bottomrule
\end{tabular}}
\end{table}
 %SU(3)f breaking

% Appendix 4 - *******************************************

\chapter{Comparison with previous results and low-energy theorems}\label{appen:vs2015}

A subset of the correlation functions used in this work has already been analyzed in Ref.~\cite{Orginos:2015aya}, where the $NN\; (\1s0)$ and $NN\; (\3s1)$ channels were studied. In this appendix, we present the outcome of a careful comparison of the results obtained using both analyses, along with a comparison of the updated scattering parameters obtained in this work and those obtained from low-energy theorems in Ref.~\cite{Baru:2016evv}.

%%%%%%%%%%%%%%%%%%%%%%%%%%%%%%%%%%%%%%%%%%%
\subsubsection{Differences in the fitting strategy}

\begin{figure}[b!]
\centering
\includegraphics[width=0.8\textwidth]{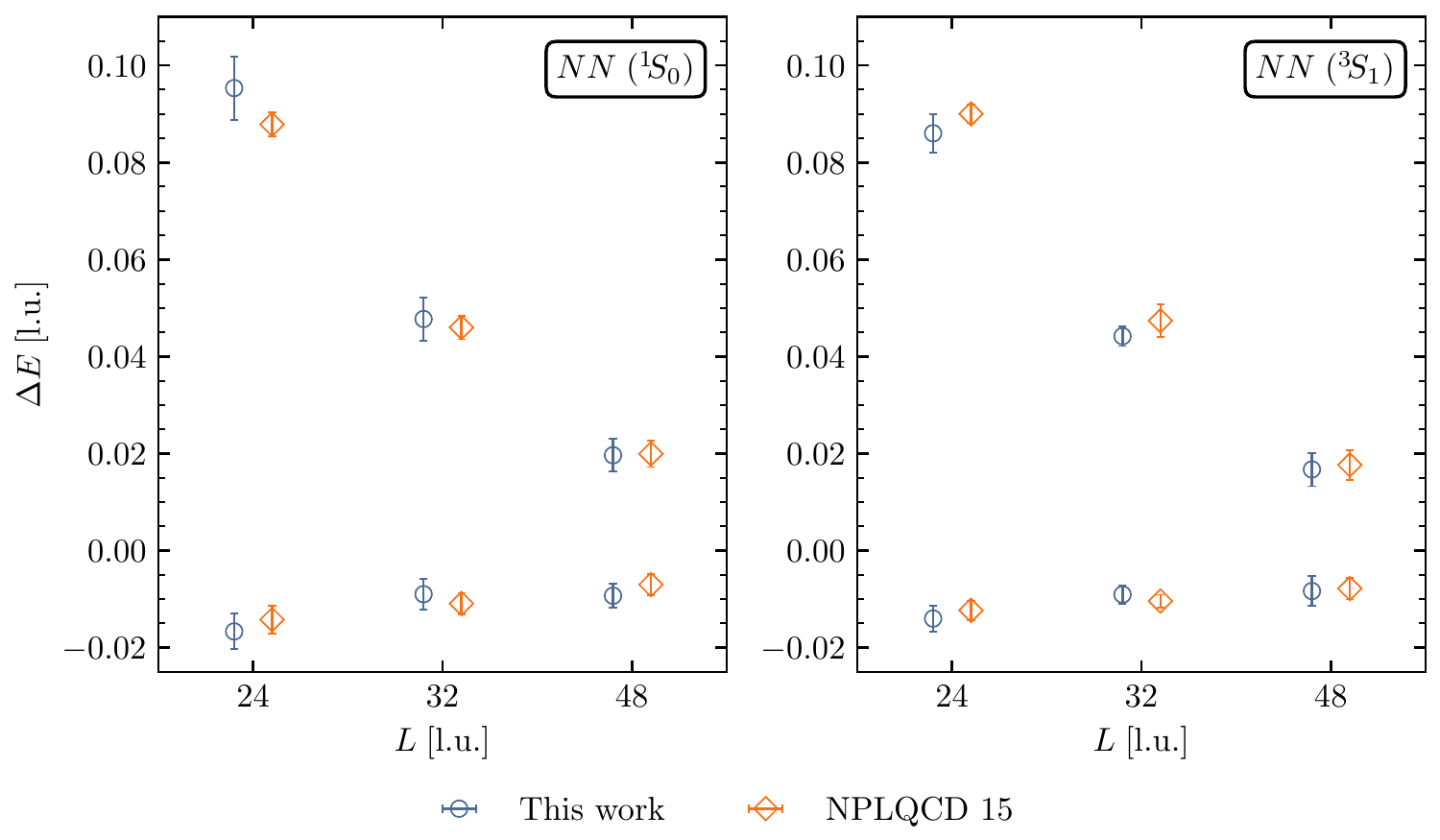}
\caption{ Comparison of the ground-state and first excited-state energies obtained in this work (blue circles) and from Ref.~\cite{Orginos:2015aya} (orange diamonds), labeled as NPLQCD 15. The figure shows results with statistical and systematic uncertainties combined in quadrature. Quantities are expressed in lattice units.}
\label{fig:20152020_energiescompar}
\end{figure}

The ground-state and first excited-state energies obtained in this work and those from Ref.~\cite{Orginos:2015aya} are shown in Fig.~\ref{fig:20152020_energiescompar}.
While all numbers are in agreement within uncertainties, it is clear that, in general, the analysis performed in Ref.~\cite{Orginos:2015aya} led to smaller uncertainties (one exception is the $NN\; (\3s1)$ first excited state with $L=32$).
That analysis consisted of the following: i) taking linear combinations of the SP and SS correlation functions (except for the $L=48$ ensemble, where only SP correlation functions were computed); ii) the use of the Hodges-Lehmann (HL) robust estimator under bootstrap resampling to estimate the ensemble-averaged correlation functions; and iii) fitting constants to the effective (mass) energy functions built from the combinations mentioned above.
In the present analysis, multi-exponential fits are performed to both SP and SS correlation functions in a correlated way (when available), using the mean under bootstrap resampling. 

\begin{figure}[t!]
\includegraphics[width=\textwidth]{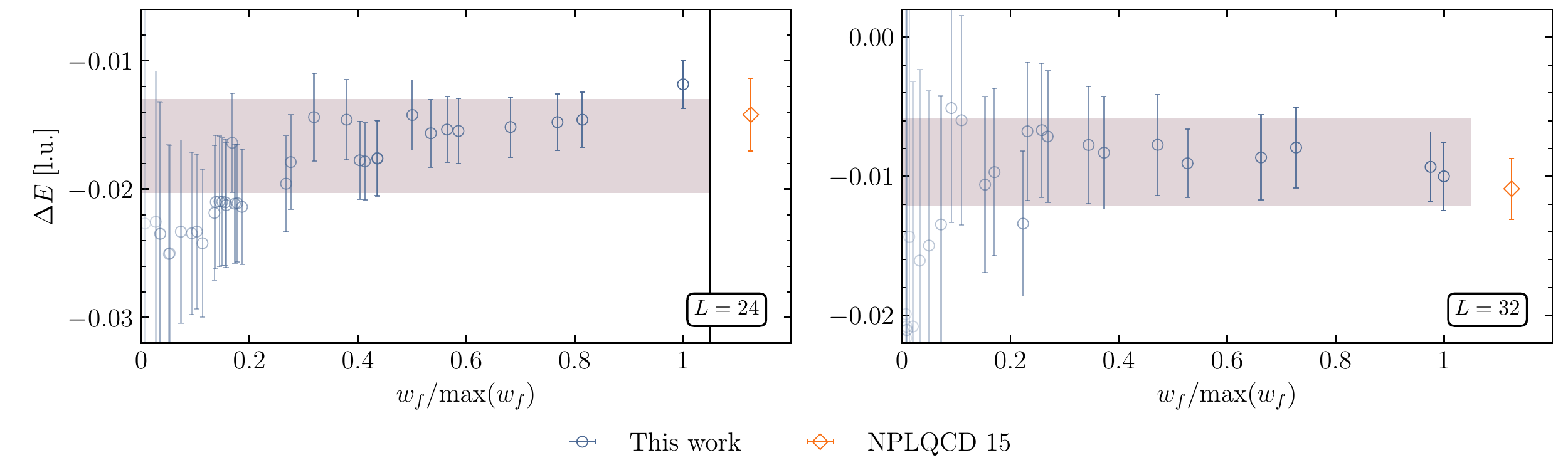}
\caption{Ground-state energies for the $NN\; (\1s0)$ system computed on ensembles with $L=24$ (left) and $L=32$ (right), sorted by their weight. The weight of each individual fit is indicated by the level of transparency of each point (darker points have larger weight). The band shows the final result, obtained by combining the individual points with the corresponding weight according to Eq.~\eqref{eq:weights_fitting}, with statistical and systematic uncertainties combined in quadrature. To facilitate the comparison, the orange point in the right panel of each figure shows the result of Ref.~\cite{Orginos:2015aya}, labeled as NPLQCD 15.}
\label{fig:nn1s0_example}
\end{figure}

Taking a closer look at how the statistical and systematic uncertainties are computed, it is worth examining the individual fits from all accepted time windows.
These are shown in Fig.~\ref{fig:nn1s0_example} for the $NN\; (\1s0)$ $L=24$ and $L=32$ ground states, sorted by their weight, $w_f$, as defined in Eq.~\eqref{eq:weights}. As can be seen, there are cases for which the size of the uncertainty is similar to or smaller than that presented in Ref.~\cite{Orginos:2015aya}.
However, the final combined uncertainty, represented by the band in Fig.~\ref{fig:nn1s0_example}, is larger. This can be understood as using a more conservative procedure for quantifying the systematic uncertainty, as well as a more thorough one: not only are variations of the fitting range considered, but also variations in the fitting form, including forms with multiple exponentials, see~\cref{subsec:2ptfitting}.

\begin{figure}[t!]
\centering
\renewcommand{\arraystretch}{1.5}
\begin{tabular}{rrccccl}
& & \multicolumn{2}{c}{One input (SS)} & \multicolumn{2}{c}{Multiple inputs (SS and SP)} & \\
& & Effective energy & Corr. & Effective energy & Corr. & \multirow{3}{*}[-19ex]{\includegraphics[width=0.45cm]{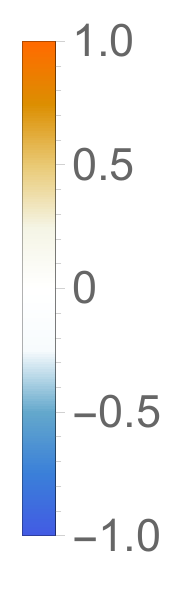}}\\
\multirow{2}{*}{\rotatebox{90}{$L=24$}} & \rotatebox{90}{Mean} & \includegraphics[width=3cm]{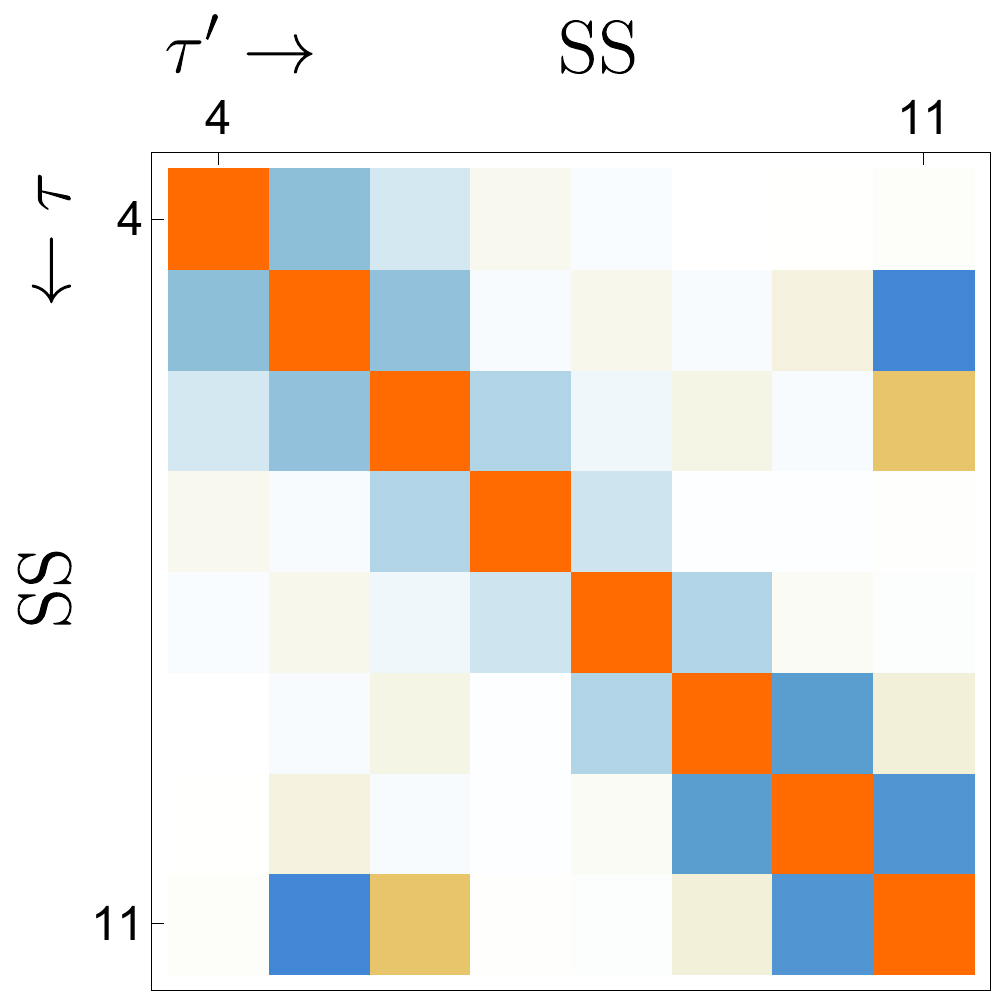} & \includegraphics[width=3cm]{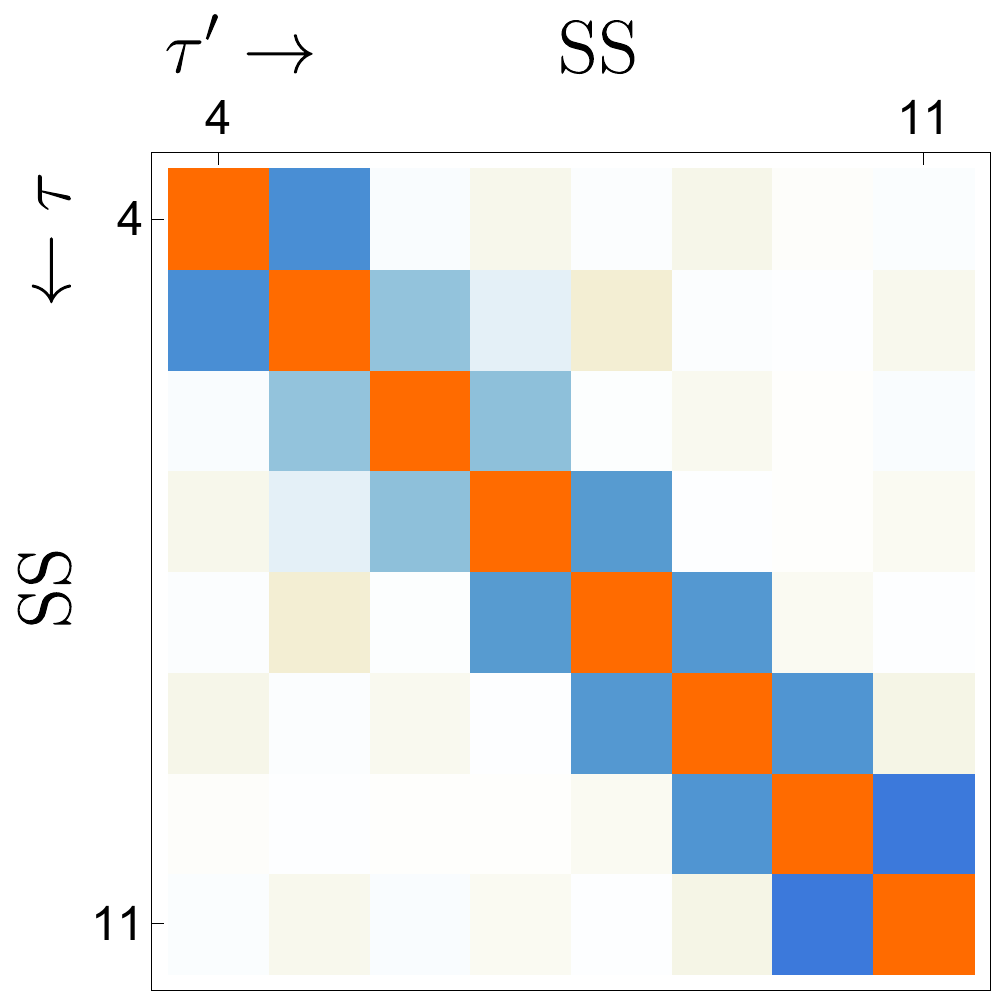} & \includegraphics[width=3cm]{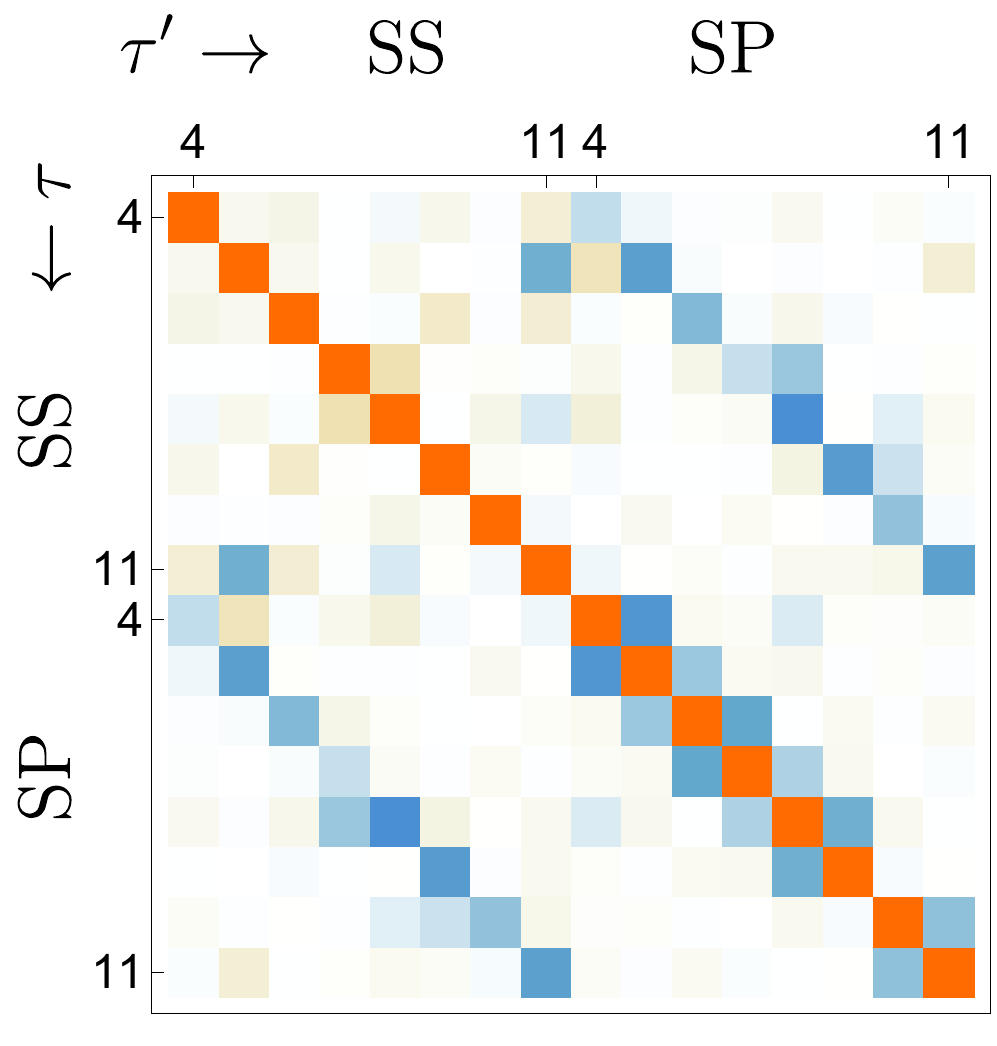} & \includegraphics[width=3cm]{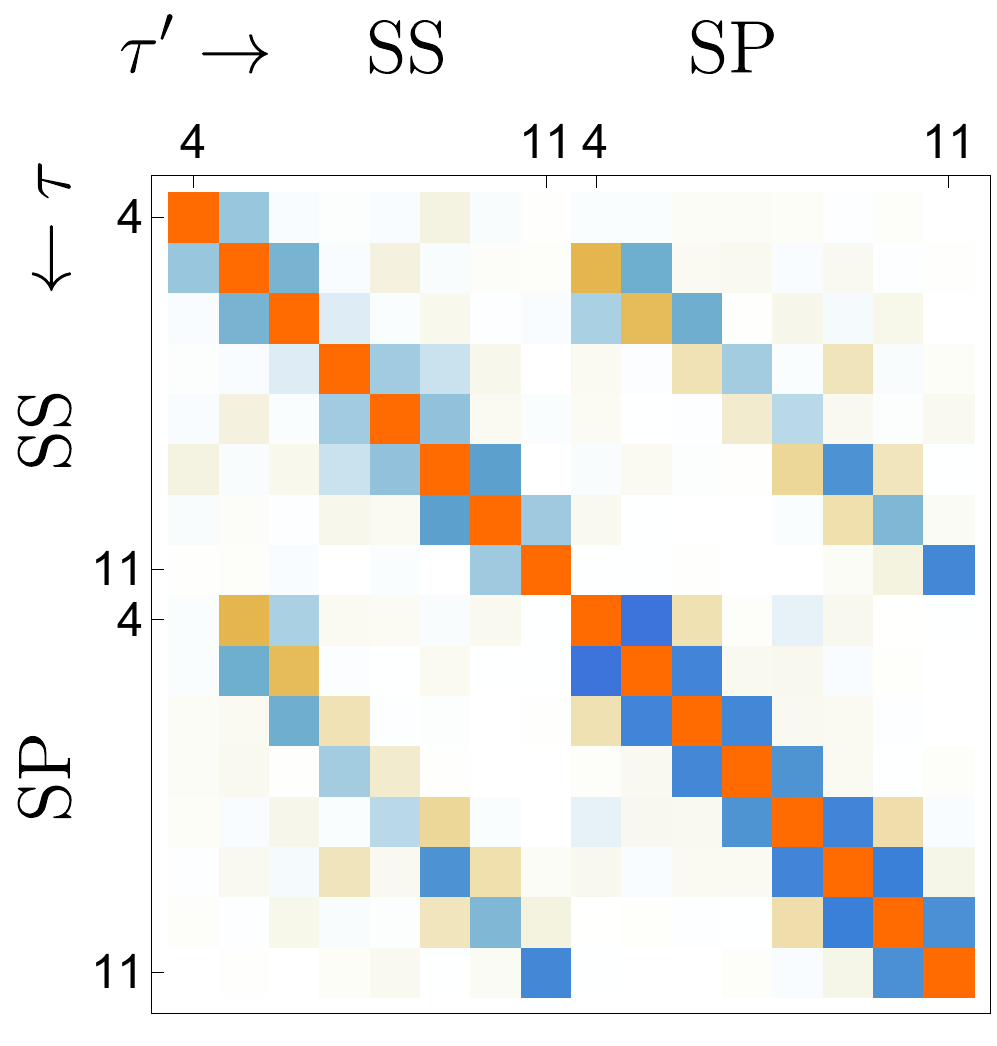} & \\ 
& \rotatebox{90}{HL estimator} & \includegraphics[width=3cm]{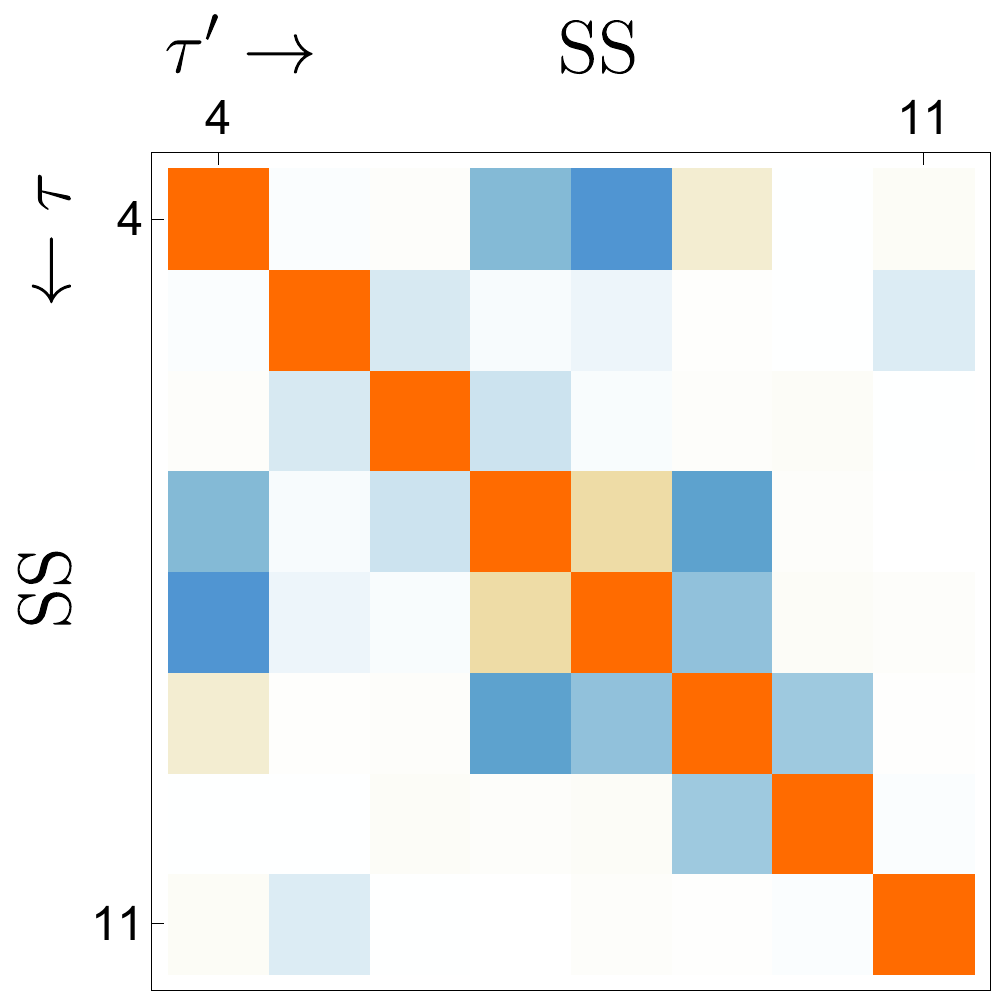} & \includegraphics[width=3cm]{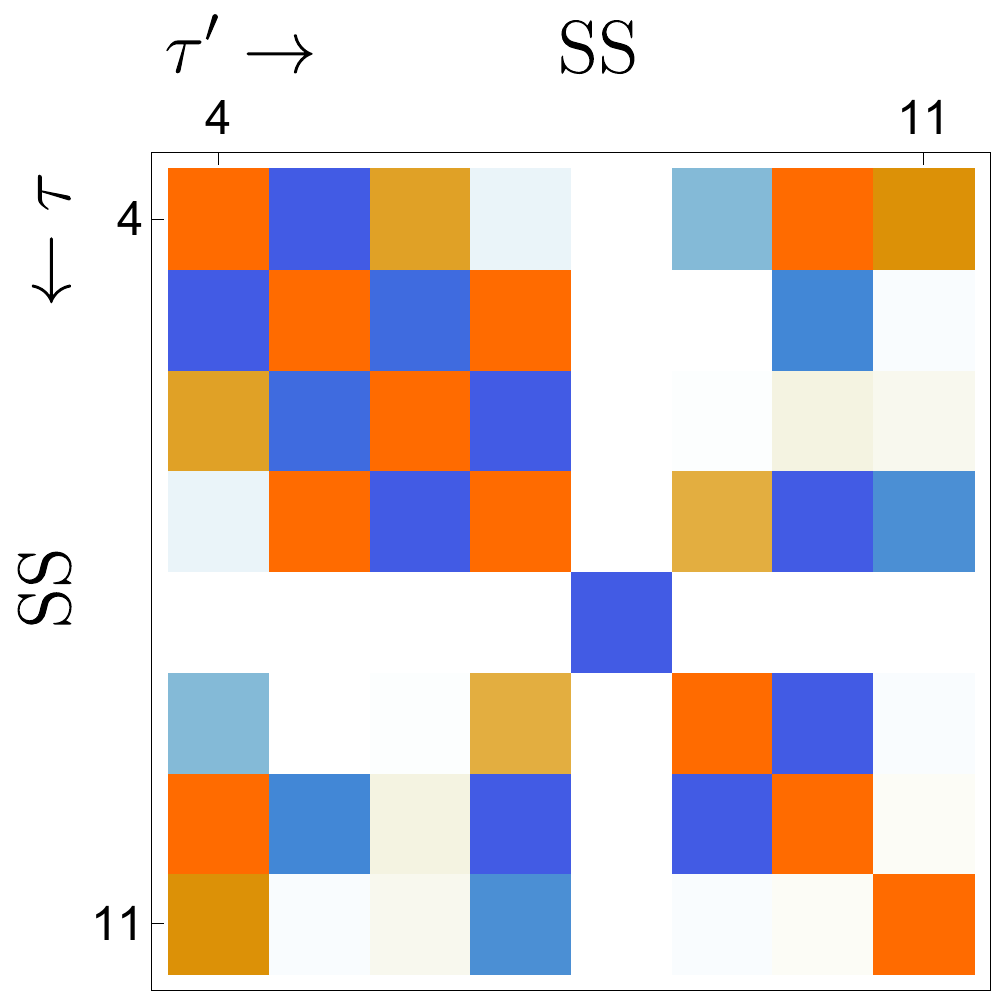} & \includegraphics[width=3cm]{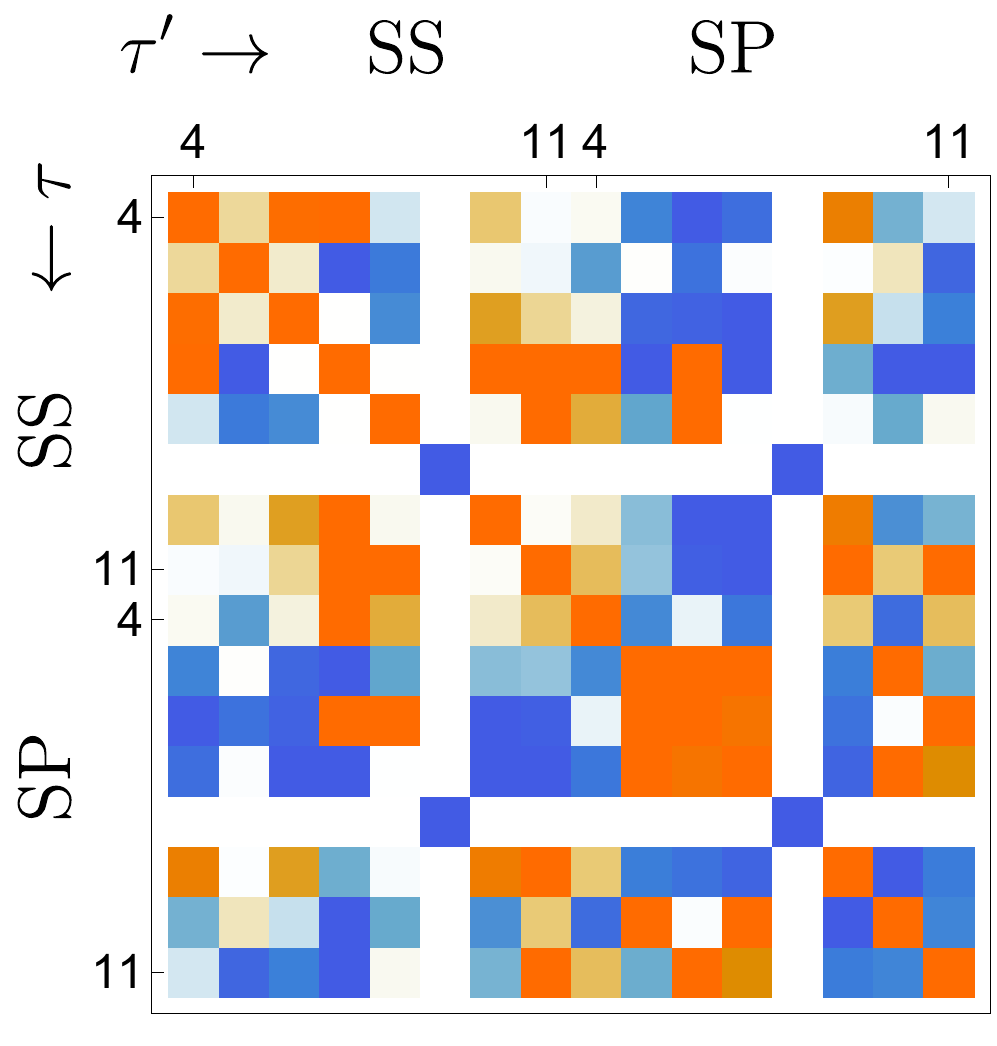} & \includegraphics[width=3cm]{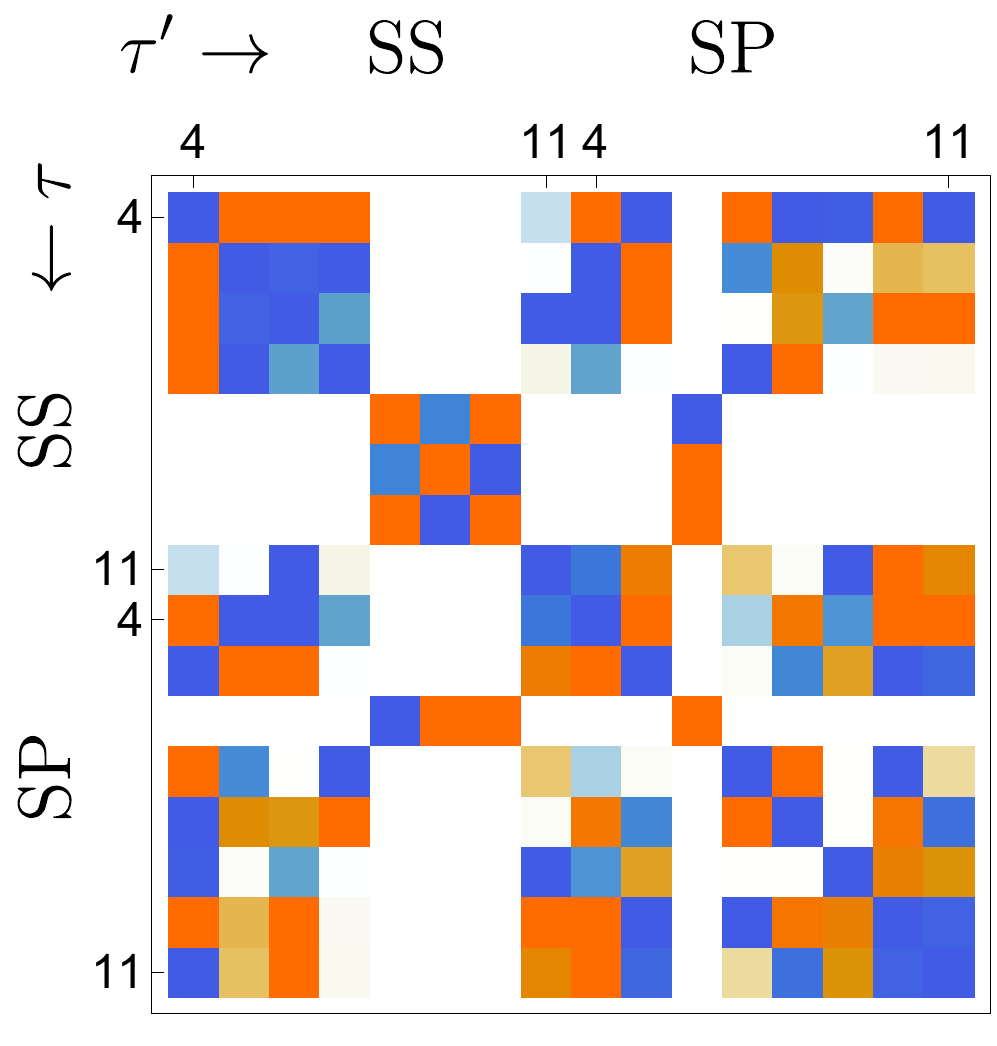} & \\
\multirow{2}{*}{\rotatebox{90}{$L=32$}} & \rotatebox{90}{Mean} & \includegraphics[width=3cm]{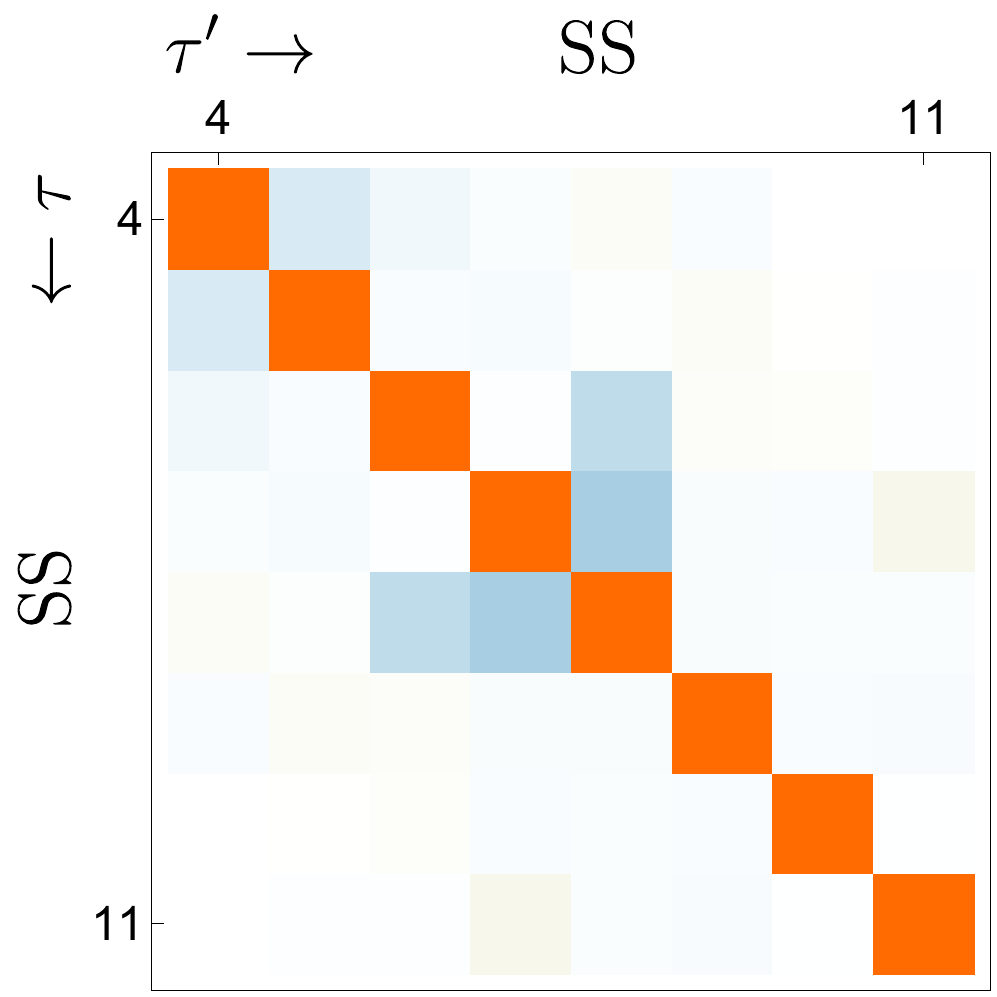} & \includegraphics[width=3cm]{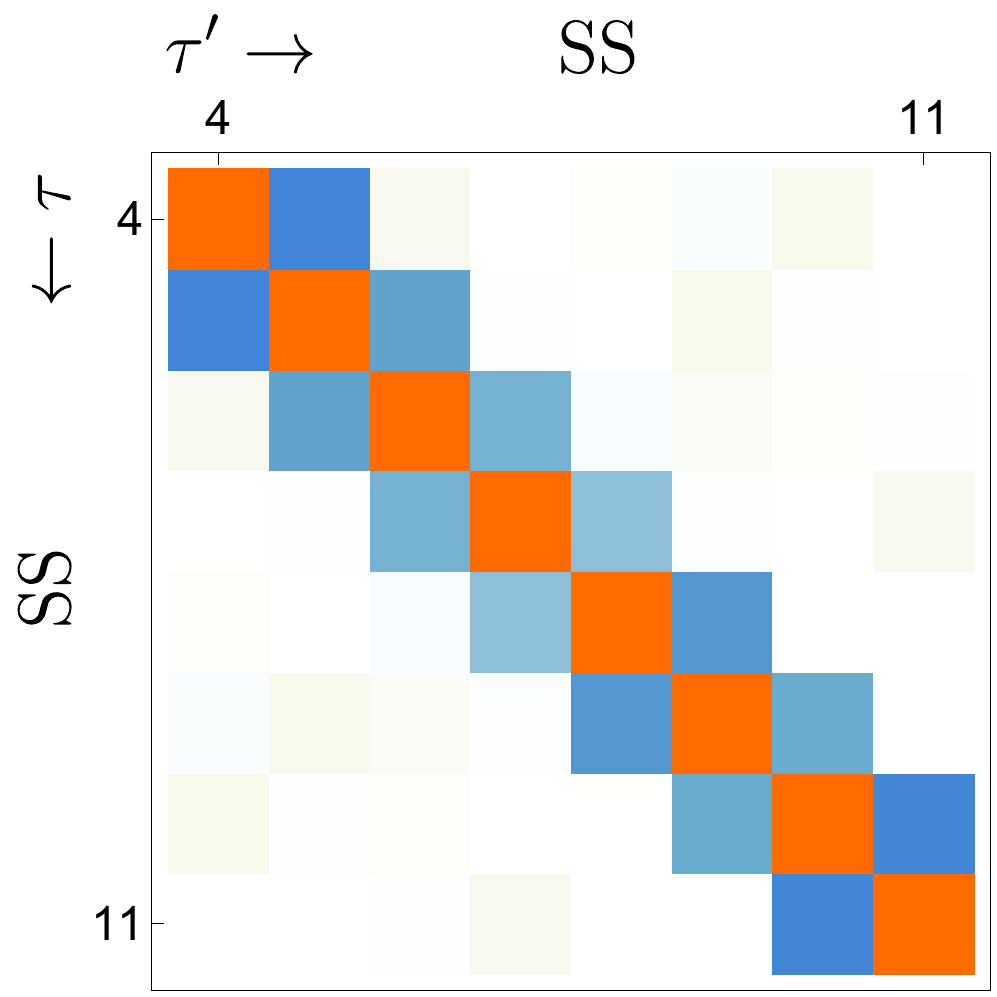} & \includegraphics[width=3cm]{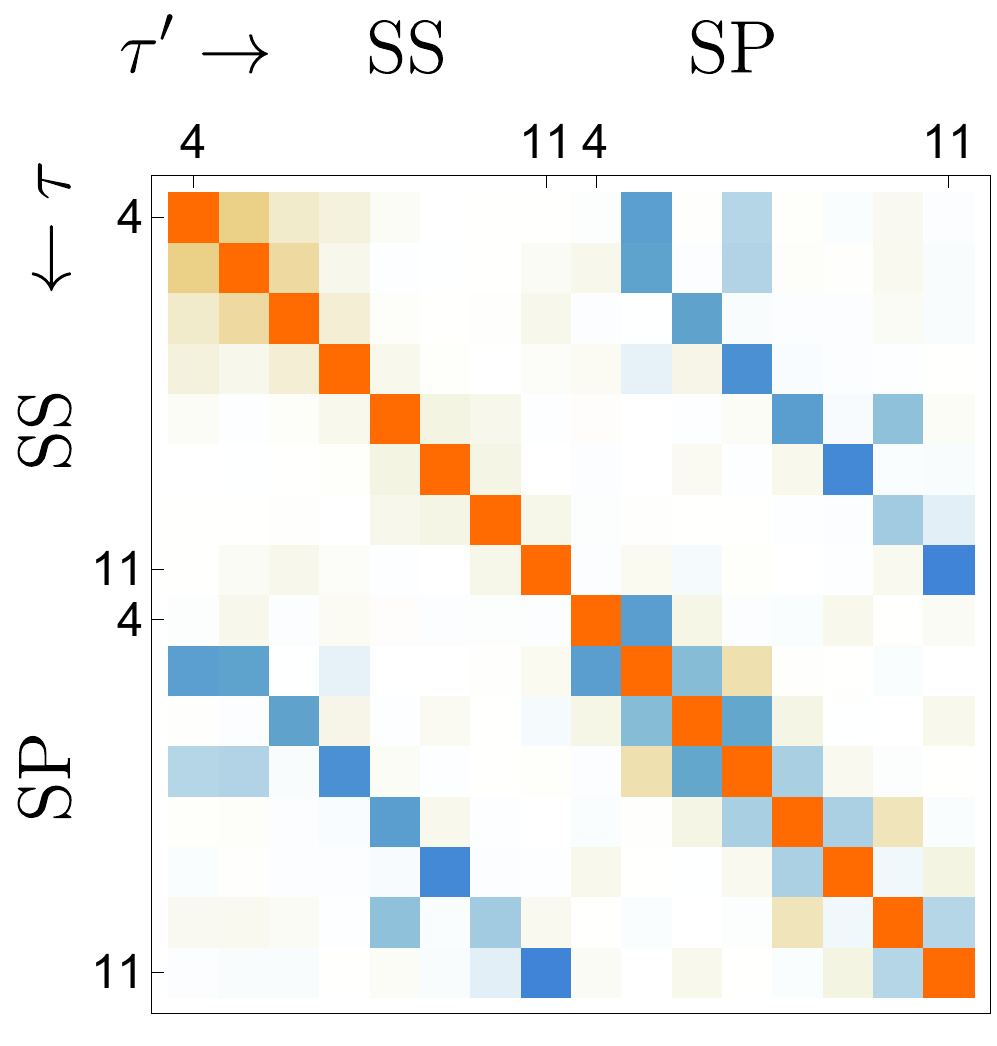} & \includegraphics[width=3cm]{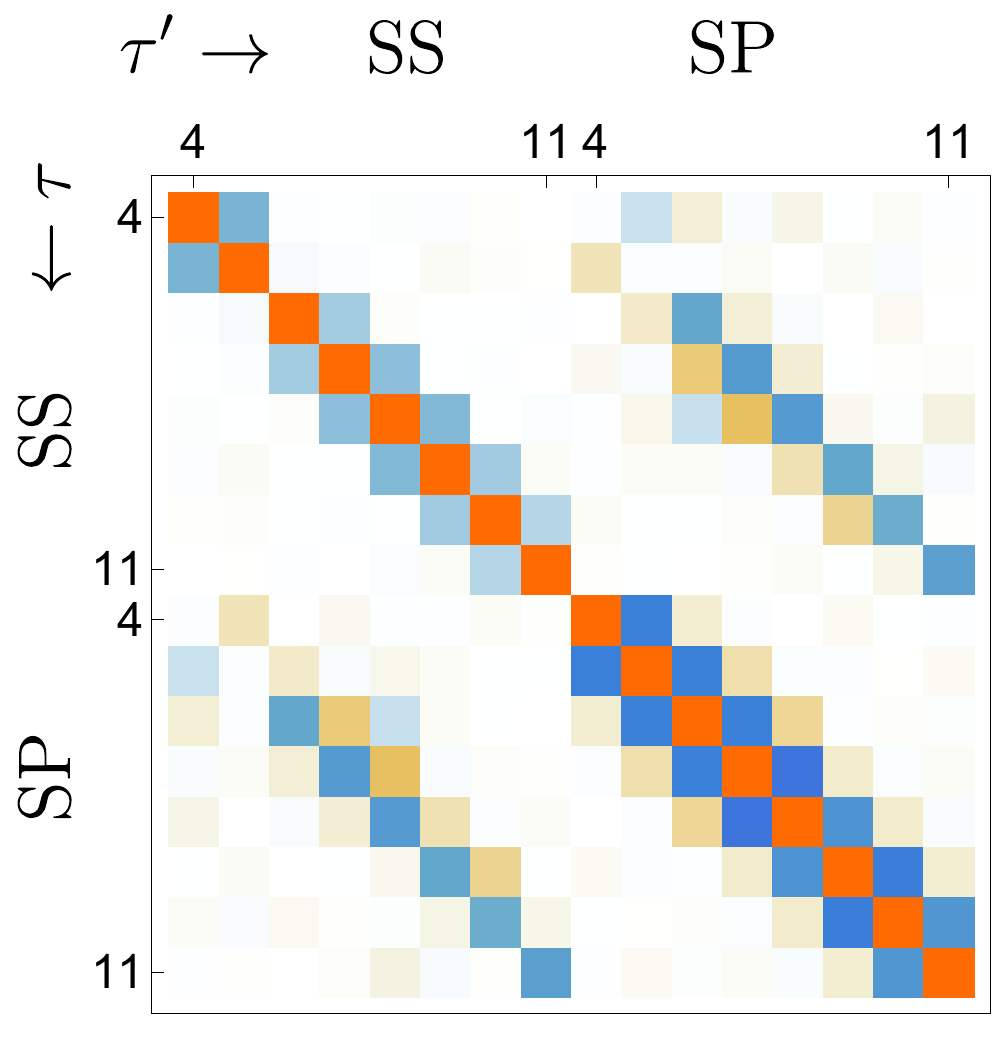} & \multirow{2}{*}[0ex]{\includegraphics[width=0.45cm]{figures/appendix4/legend.pdf}}\\ 
& \rotatebox{90}{HL estimator} & \includegraphics[width=3cm]{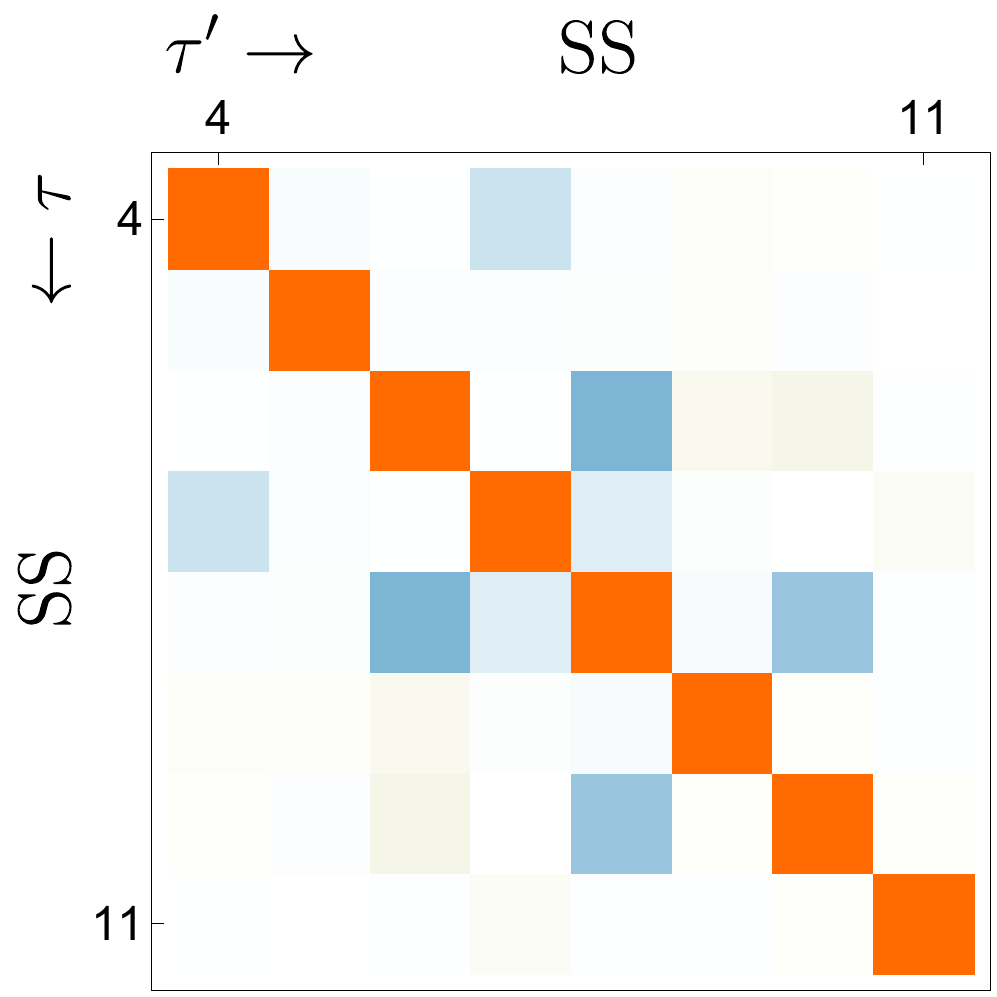} & \includegraphics[width=3cm]{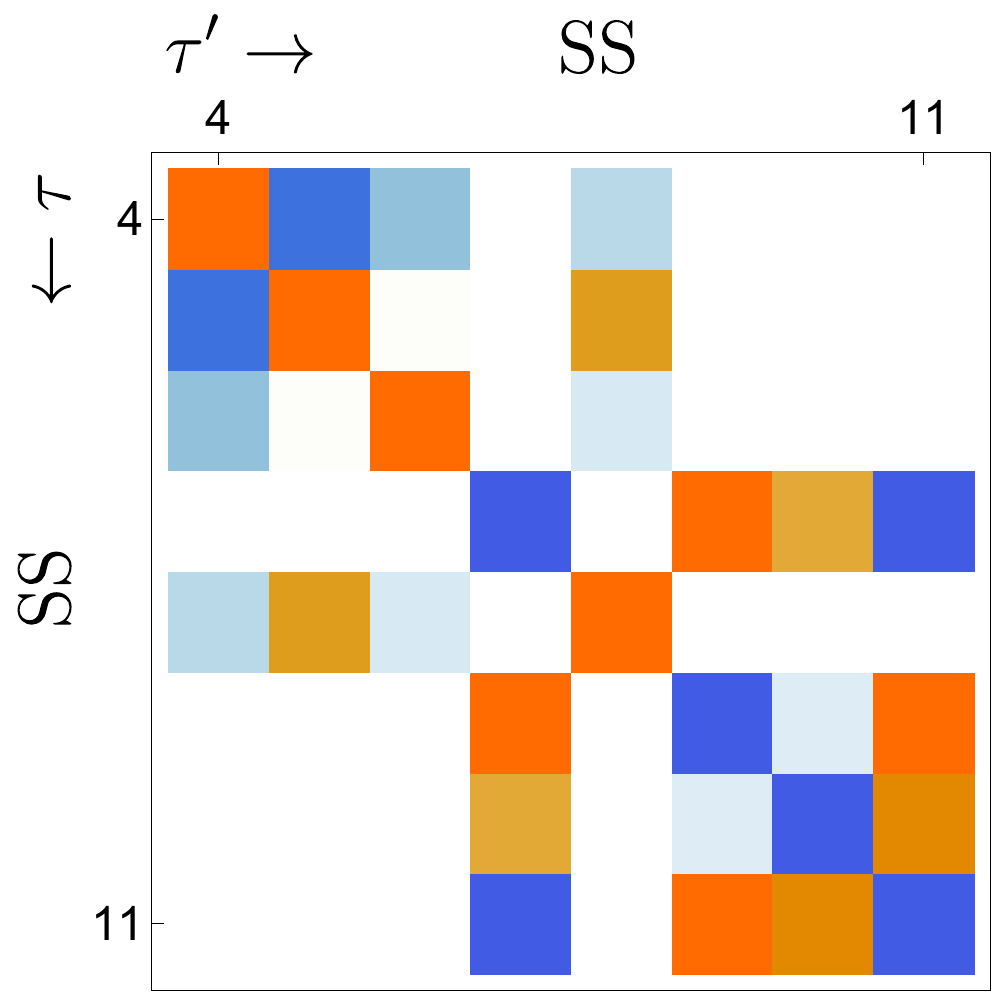} & \includegraphics[width=3cm]{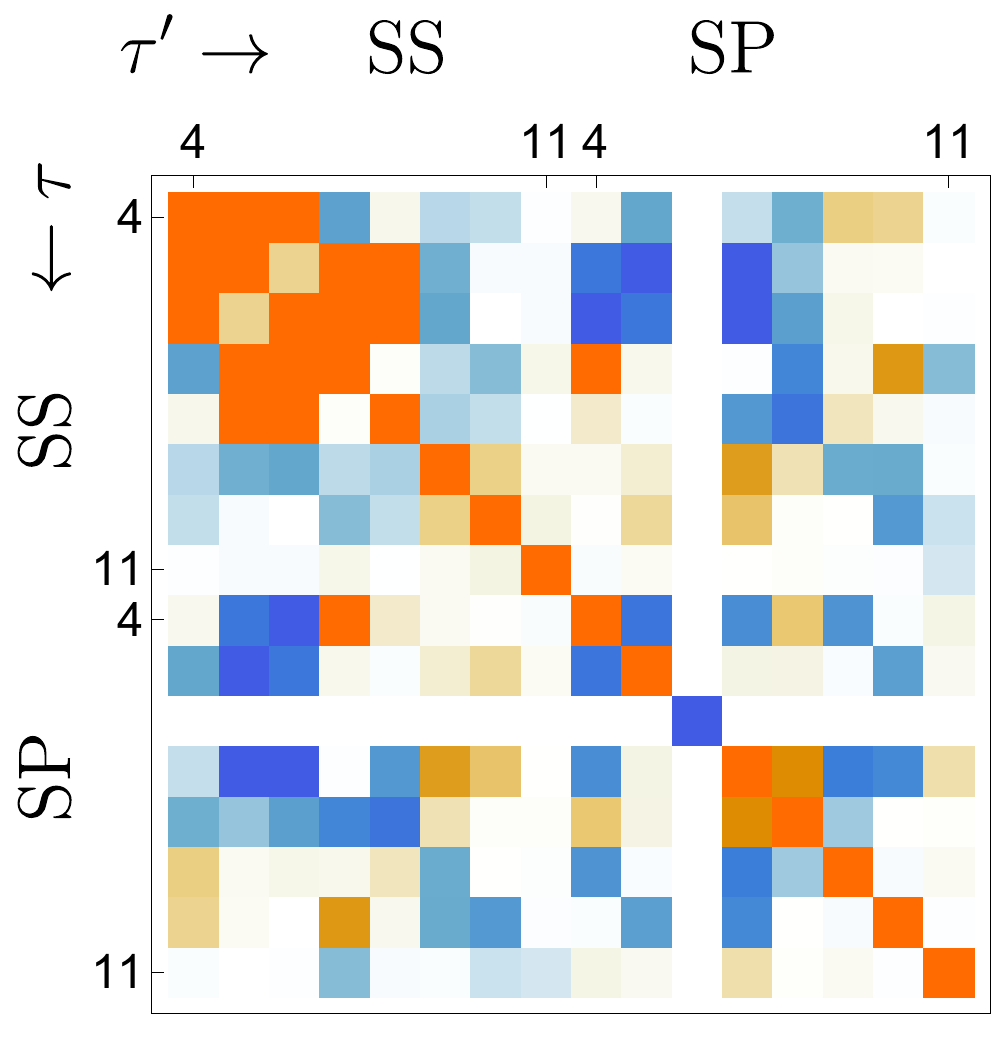} & \includegraphics[width=3cm]{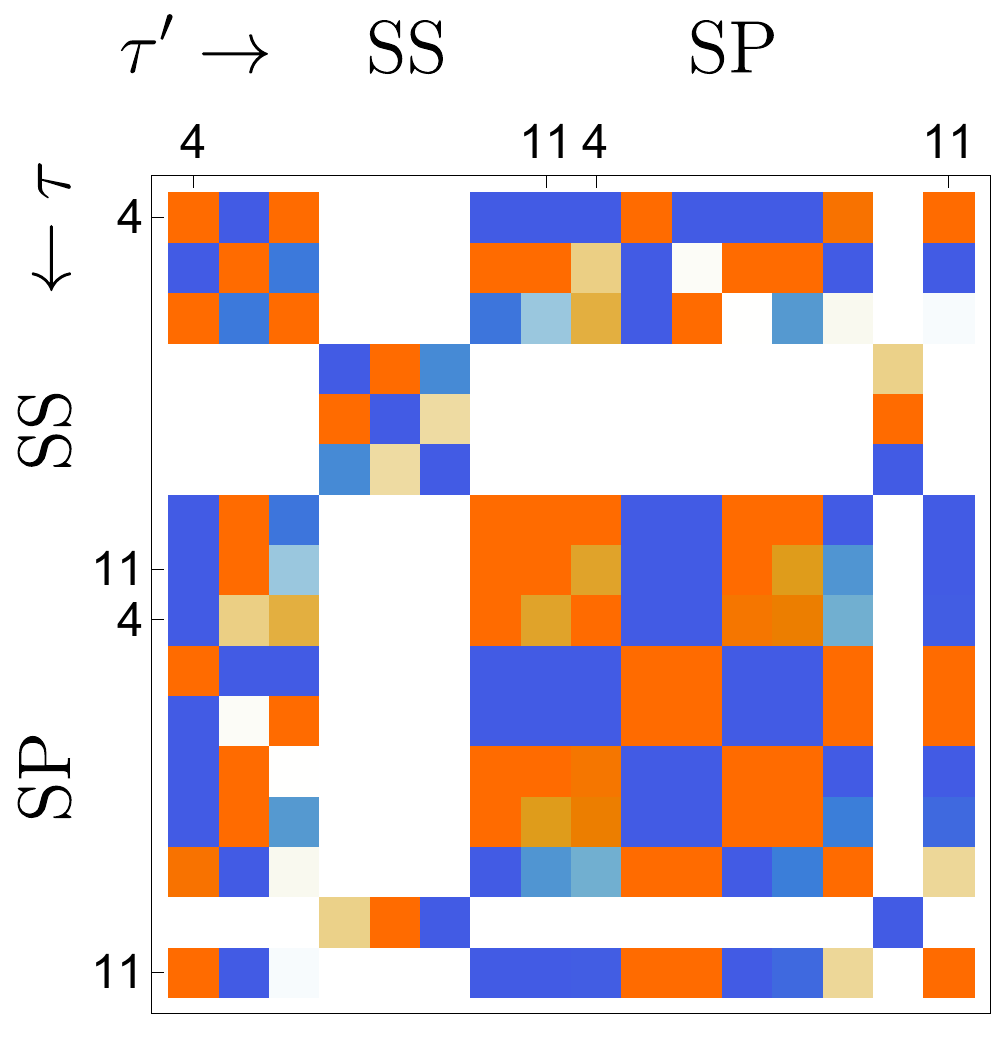} &
\end{tabular}
\caption{Normalized inverse covariance matrices computed for the $NN\; (\1s0)$ ground state with $L=24$ (top) and $L=32$ (bottom) for $\tau\in [4,11]$ l.u.\ using the mean and HL estimators applied to the effective energy function and correlation function. }
\label{fig:covmatrices}
\end{figure}

Next, the implications of using the HL estimator (instead of the mean) on the individual SP and SS correlation functions are analyzed. When correlations are fully taken into account, the covariance matrix associated with the HL estimator is computed with the median absolute deviation (MAD), as explained in~\cref{subsec:HL}. However, in some cases the resulting covariance matrix is found not to be positive semi-definite, and it only becomes well behaved when a single type of correlation function is used (or a linear combination of several) in the form of an effective (mass) energy function.
To illustrate this, Fig.~\ref{fig:covmatrices} shows the normalized inverse covariance matrix, $\mathcal{C}^{-1}(\tau,\tau')/\sqrt{\mathcal{C}^{-1}(\tau,\tau)\mathcal{C}^{-1}(\tau',\tau')}$, for the $NN\; (\1s0)$ ground state with $L=24$ and $L=32$ for all possible choices, i.e., HL estimator versus mean and correlation function versus the effective energy function.

\begin{figure}[t!]
\includegraphics[width=\textwidth]{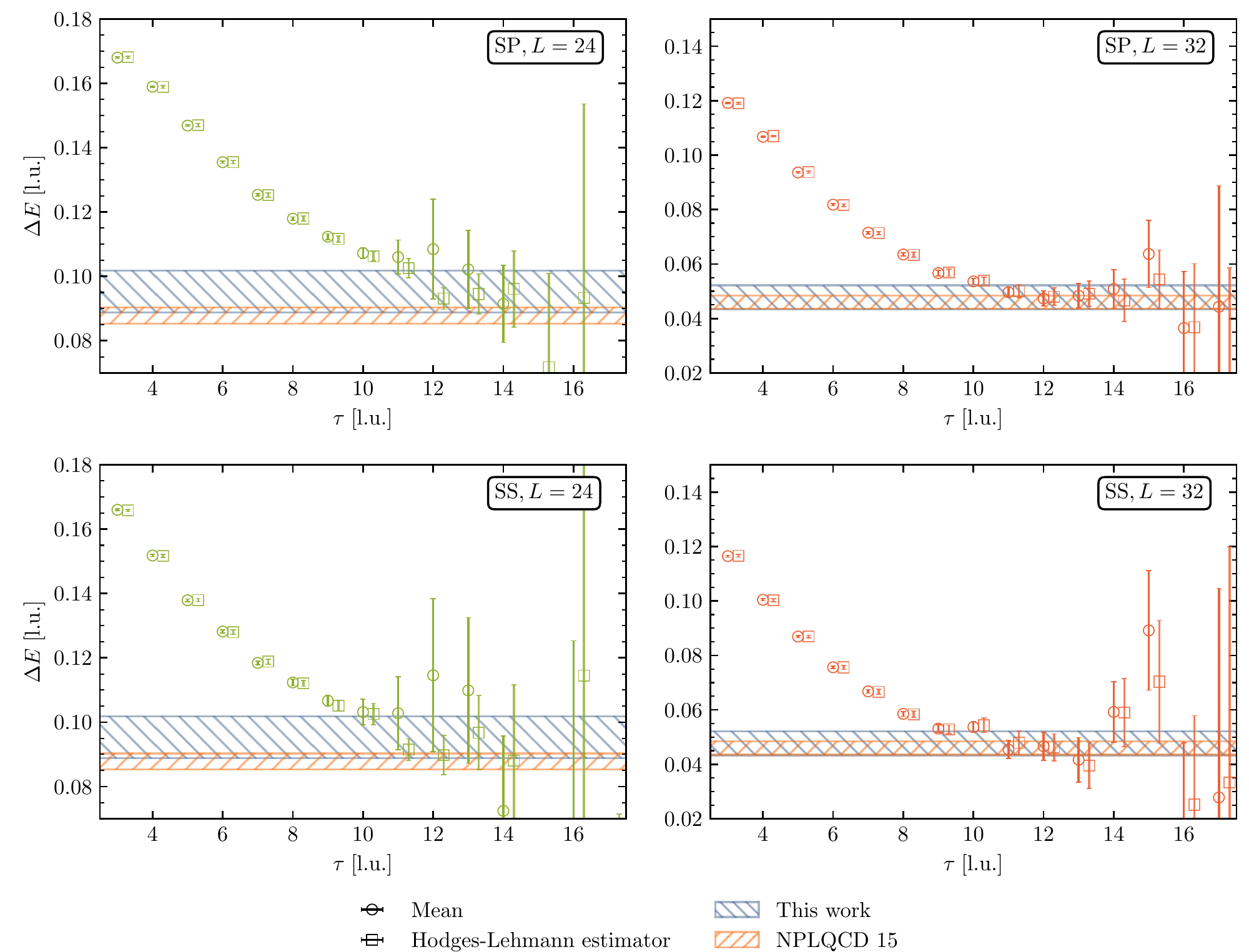}
\caption{Comparison of the effective energy-shift plots of the SP and SS correlation functions for the $NN\; (\1s0)$ $L=24$ (left panel) and $L=32$ (right panel) first excited states computed using the mean (dark green/red circles) and the HL estimator (light green/red squares, shifted horizontally for clarity). The bands show the results of this work and of Ref.~\cite{Orginos:2015aya}, labeled as NPLQCD 15.}
\label{fig:nn1s0EMP_24vs32}
\end{figure}

Therefore, in order to incorporate the HL estimator into the fitting strategy used here, only the fully uncorrelated covariance matrix can be used, and this leads to results which are compatible with the ones presented here using the mean.
In Fig.~\ref{fig:nn1s0EMP_24vs32}, the effective energy functions computed with the mean and HL are compared for the $NN\; (\1s0)$ first excited states, showing agreement within uncertainties.

\begin{figure}[t!]
\includegraphics[width=\textwidth]{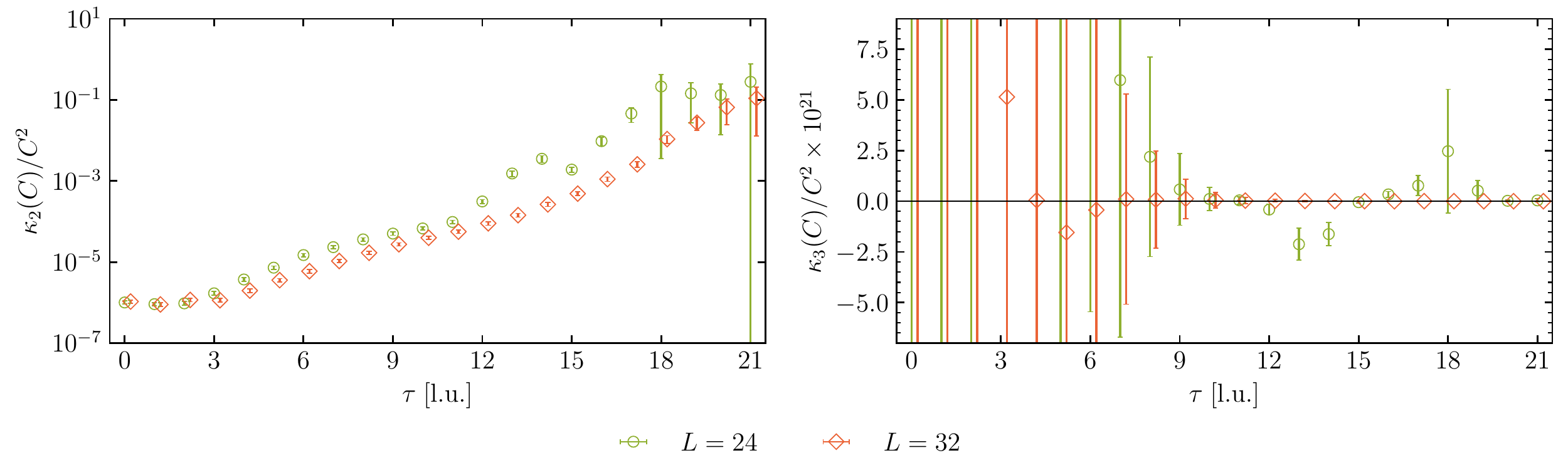}
\caption{The bootstrap estimates of the variance $\kappa_2[C(\tau)]/C^2(\tau)$ and skewness $\kappa_3[C(\tau)]/C^2(\tau)$ for the SS correlation functions corresponding to the $NN\; (\1s0)$ first excited state with $L=24$ (green circles) and $L=32$ (red diamonds). The $L=32$ points have been shifted slightly along the $\tau$ axis for clarity.}
\label{fig:var_skew}
\end{figure}

To understand the ill-behaved behavior of some of the HL correlation functions, it is important to recall that baryonic correlation functions exhibit distributions that are largely non-Gaussian with heavy tails, and the mean becomes Gaussian only in the limit of large statistics.
However, at late times, the signal-to-noise degradation worsens, and outliers occur more frequently in the distribution. For the $L=32$ and $L=48$ cases, the point at which the HL estimator gives different results compared with the usual estimator (mean and standard deviation), which would indicate a deviation from Gaussian behavior, occurs at a much later time compared with the maximum time included in the fits using the automated fitter of this work.
For the $L=24$ case, the data are more noisy than on the other two ensembles, showing non-Gaussianity at earlier times. To illustrate the different behavior between the $L=24$ and $L=32$ ensembles, the second and third cumulants of $C(\tau)$, defined as
\begin{equation}
    \kappa_n[C(\tau)]=\left\langle C(\tau)^n \right\rangle -\sum_{m=1}^{n-1} \binom{n-1}{m-1} \kappa_m[C(\tau)] \left\langle C(\tau)^{n-m} \right\rangle\, ,
\end{equation}
with $n\in\{2,3\}$, respectively, are shown in Fig.~\ref{fig:var_skew} for the two ensembles in the case of the $NN\; (\1s0)$ first excited state.
Looking at the second cumulant (variance), $\kappa_2$, it is clear that $L=24$ is more noisy than $L=32$, and looking at the third cumulant (skewness), $\kappa_3$, it is clear that $L=24$ deviates from zero, an indication of the non-Gaussian behavior. The use of robust estimators is, therefore, questionable in this case. This is the main reason for abandoning the use of the HL estimator in the analysis of correlation functions in the present study.

\subsubsection{Differences in the scattering parameters}

\begin{figure}[b!]
\includegraphics[width=\textwidth]{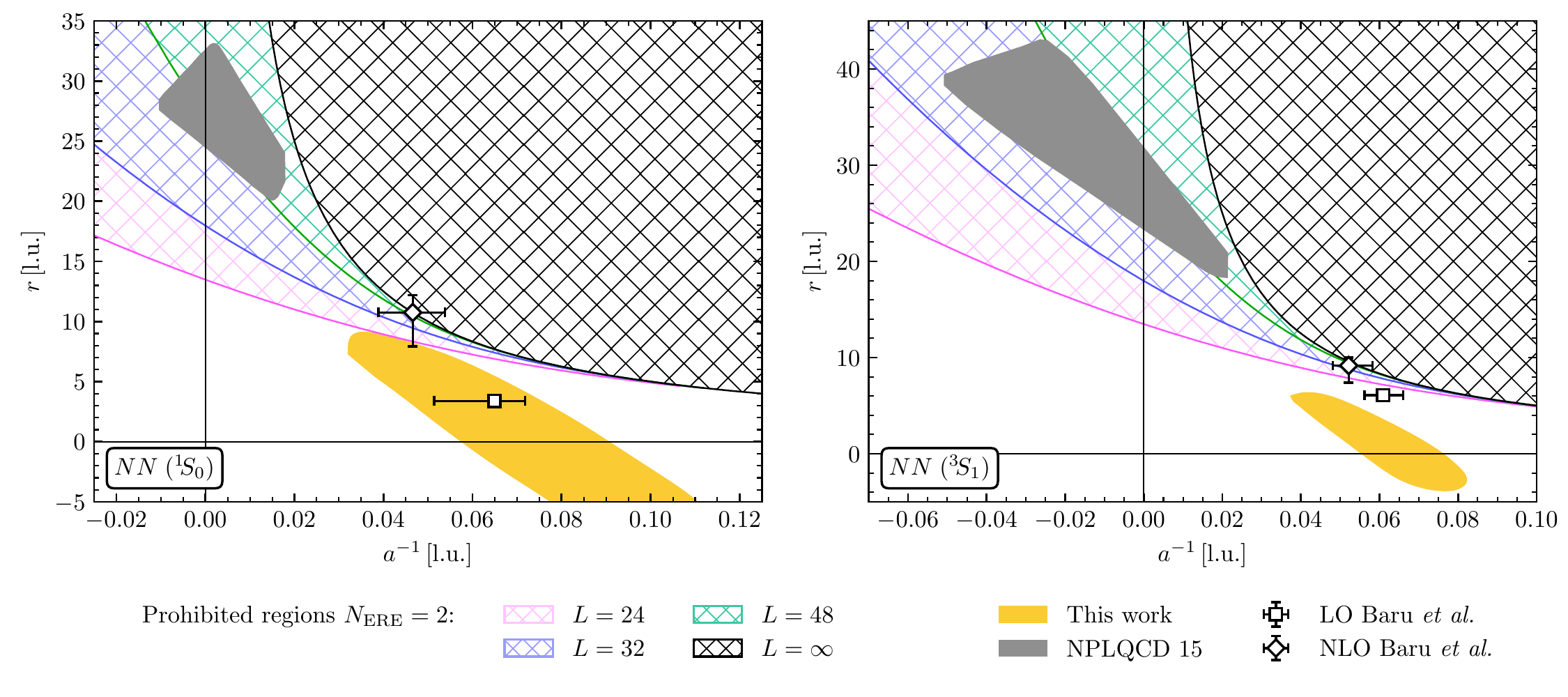}
\caption{Comparison of the 68\% confidence region of the scattering parameters obtained in this work (yellow area), from Ref.~\cite{Orginos:2015aya} (gray area, labeled as NPLQCD 15), and predictions of low-energy theorems from Ref.~\cite{Baru:2016evv} (LO and NLO results). The regions include both statistical and systematic uncertainties combined in quadrature. The prohibited regions where the two-parameter ERE does not cross the $\mathcal{Z}$-functions at given volumes or in the infinite-volume limit are denoted as hashed areas. Quantities are expressed in lattice units.}
\label{fig:20152020Baru_EREcompar}
\end{figure}

The 68\% confidence region of the scattering parameters from a two-parameter ERE extracted in this work and in Ref.~\cite{Orginos:2015aya} are shown in Fig.~\ref{fig:20152020Baru_EREcompar}.
It can be seen that the values of the parameters obtained in the two analyses do not fully agree at the $1\sigma$ level, although the uncertainties are rather large.

\begin{figure}[t!]
\centering
\includegraphics[width=0.94\textwidth]{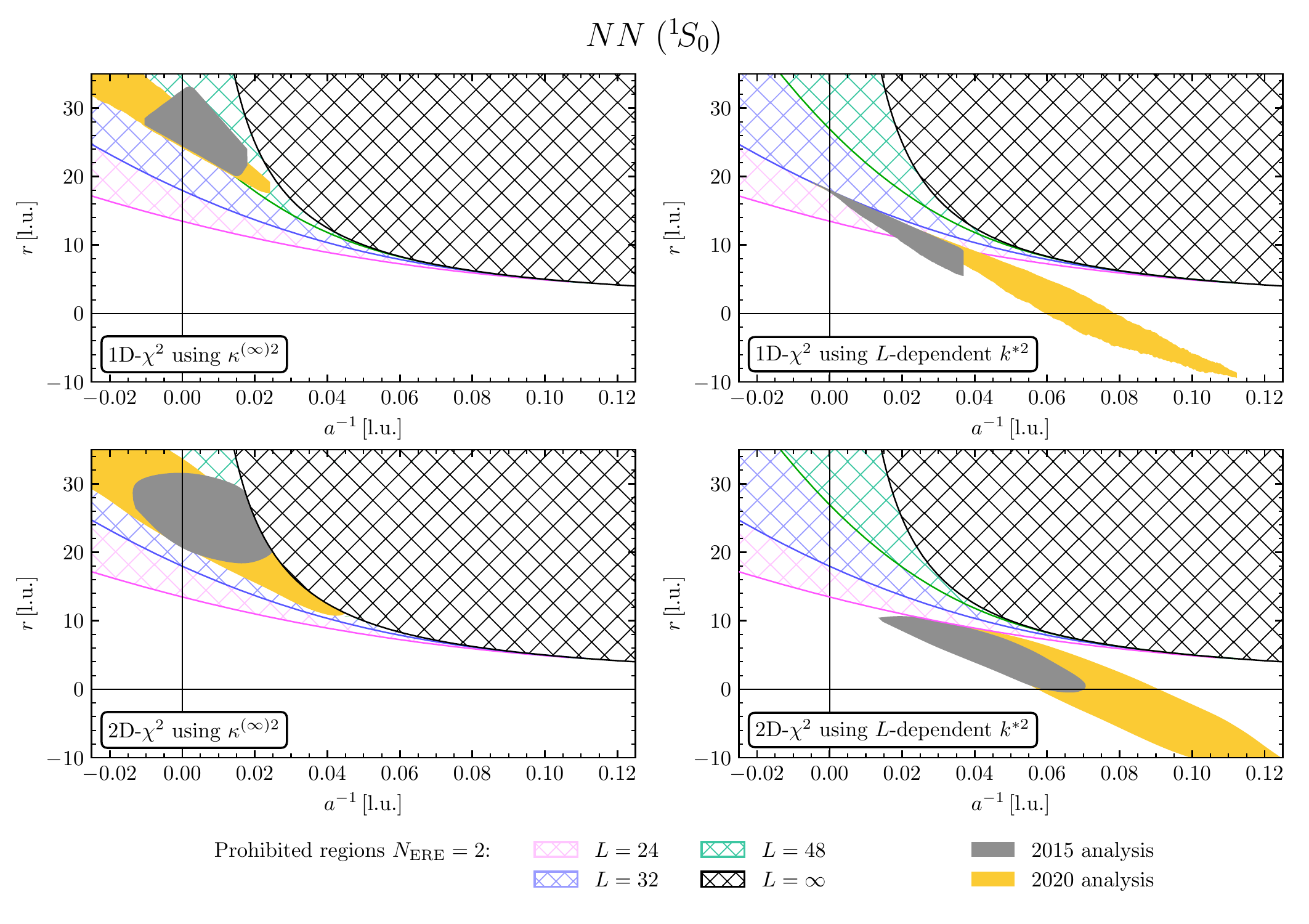}
\includegraphics[width=0.94\textwidth]{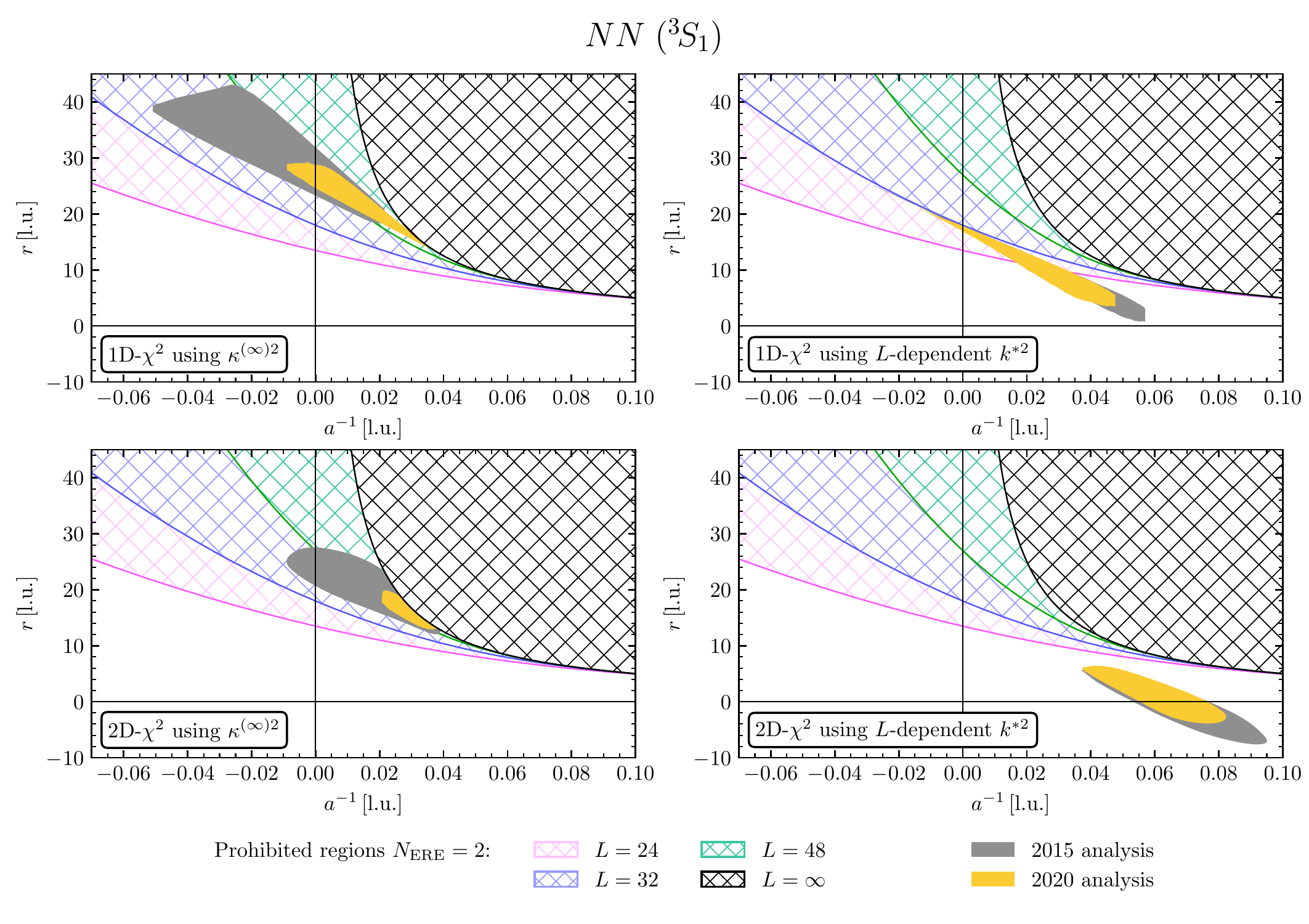}
\caption{Comparison of the 68\% confidence region of the scattering parameters obtained using the energy levels extracted in this work (yellow area) and from Ref.~\cite{Orginos:2015aya} (gray area) with four different analyses. The regions include both statistical and systematic uncertainties combined in quadrature. The prohibited regions where the two-parameter ERE does not cross the $\mathcal{Z}$-function are the crossed areas. Quantities are expressed in lattice units.}
\label{fig:20152020_EREcompar_tests}
\end{figure}

There are two significant differences between the two analyses: i) the use of the new definition for the $\chi^2$ function (2D-$\chi^2$) in the present work, as opposed to the usual $\chi^2$ function (1D-$\chi^2$) used in Ref.~\cite{Orginos:2015aya}, and ii) the use of the $L$-dependent ground-state $k^{*2}$ values in the fits to ERE in the present work, instead of using only the infinite-volume extrapolated value, $\kappa^{(\infty)}$, used in Ref.~\cite{Orginos:2015aya}. To see the effects of each, a comprehensive analysis has been performed, the results of which are shown in Fig.~\ref{fig:20152020_EREcompar_tests}. Here, four different possibilities, corresponding to the types of the $\chi^2$ function (1D or 2D) and the use of ground-state $k^{*2}$ data ($L$-dependent or extrapolated), are tested using the lowest-lying spectra obtained in Ref.~\cite{Orginos:2015aya} and those in the present work.

From these tests, several interesting features are observed. First, the use of the 1D-$\chi^2$, either with the $L$-dependent $k^{*2}$ or the extrapolated one, is insensitive to the conditions imposed by Lüscher's QC, and as a result, the confidence regions of the scattering parameters could lie on top of the prohibited regions. This is because the distance minimized in the 1D-$\chi^2$ is the vertical one, and not the one along the $\mathcal{Z}$-function, so the ERE is not forced to cross it. Second, when the 2D-$\chi^2$ is used with the extrapolated $k^{*2}$ value, $\kappa^{(\infty)2}$, the only region that is avoided is the one corresponding to $L=\infty$ in the figures, which is expected: with the value of the pole position given by Eq.~\eqref{eq:kinf3}, the function $k^*\cot\delta|_{k^* = i \kappa^{(\infty)}}$ equals $-\sqrt{-k^{*2}}$ and the ERE crosses the $-\sqrt{-k^{*2}}$ function, imposing the $r/a<1/2$ constraint on the scattering parameters. Third, it is reassuring that the regions obtained using the two different energy inputs, from this work or from Ref.~\cite{Orginos:2015aya}, are always overlapping.

Perhaps the most significant observation is that the choice of including the points in the negative $k^{*2}$ region in the fit, i.e., the infinite-volume extrapolated value of the momenta versus the $L$-dependent values, has far more impact on the differences observed than which $\chi^2$ function is used. What the new $\chi^2$ function does is to move the scattering parameters to the allowed region by the $\mathcal{Z}$-functions. Furthermore, with the new fitting methodology, several questions raised about the validity of the ERE fits are addressed, as was presented in~\cref{subsec:validityLQCD}. An important one is that the updated results of this work recover the position of the bound state pole obtained via the infinite-volume extrapolation of the energies, and do not yield a second pole near threshold, which would be incompatible with the use of the ERE. As a final remark, it should be noted that the data fitted to extract these parameters are highly non-Gaussian, as can be seen from the correlation between $k^{*2}$ and $k\cot\delta$ in Fig.~\ref{fig:kcotdelta}, and exhibit large uncertainties. This can be compared with the results of Refs.~\cite{Beane:2013br,Wagman:2017tmp} at $m_{\pi}\sim 806$ MeV, where more finite-volume energy eigenvalues, with better precision, could be used in the ERE fitting. As a result, it has been verified that either the $L$-dependent or the infinite-volume extrapolated value of $k^{*2}$ in the ERE fitting gives compatible scattering parameters.

In Ref.~\cite{Baru:2016evv}, low-energy theorems~\cite{Baru:2015ira} were used to compute the scattering parameters from the binding energies of the $NN$ systems obtained in Ref.~\cite{Orginos:2015aya}, and it was pointed out that there were some tensions with the scattering parameters obtained from the LQCD data using Lüscher's method, i.e., those reported in Ref.~\cite{Orginos:2015aya}. Since the binding energies obtained in this work are in full agreement with those obtained in Ref.~\cite{Orginos:2015aya}, the results obtained in Ref.~\cite{Baru:2016evv} can be compared with the updated scattering parameters of this work. As is depicted in Fig.~\ref{fig:20152020Baru_EREcompar}, the tension has reduced considerably. For the two-parameter ERE results, the scattering length is now completely consistent with the low-energy theorem predictions, at both LO and NLO. For the effective range, since the NLO predictions of the low-energy theorems enter the prohibited region for the two-parameter ERE, the comparison may only be made with the LO results. As is seen, for both the $\1s0$ and $\3s1$ channels, the effective ranges are also in agreement (with the $\1s0$ state having a better overlap). %Comparison old results

% Appendix 5 - *******************************************

\chapter{Supplementary figures and tables of Section \texorpdfstring{\ref{sec:450results}}{4.3}}\label{appen:figtab}

\begin{figure}[hbt!]
\includegraphics[width=\textwidth]{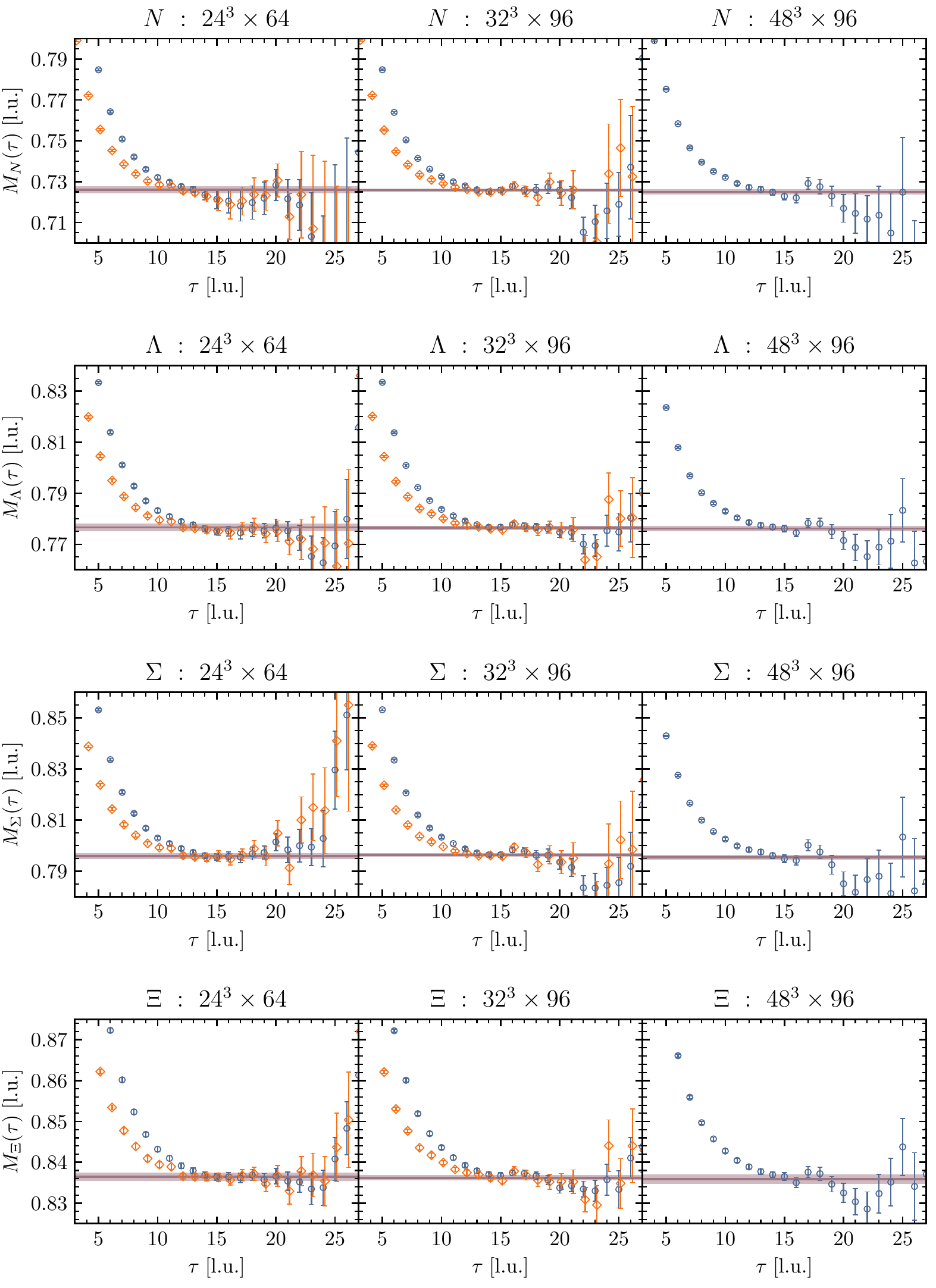}
\caption{Single-baryon EMPs for the SP (blue squares) and SS (orange diamonds) source-sink combinations. The SS points have been slightly shifted along the horizontal axis for clarity.}
\label{fig:B1_EMP}
\end{figure}

\begin{figure}[hbt!]
\includegraphics[width=0.95\textwidth]{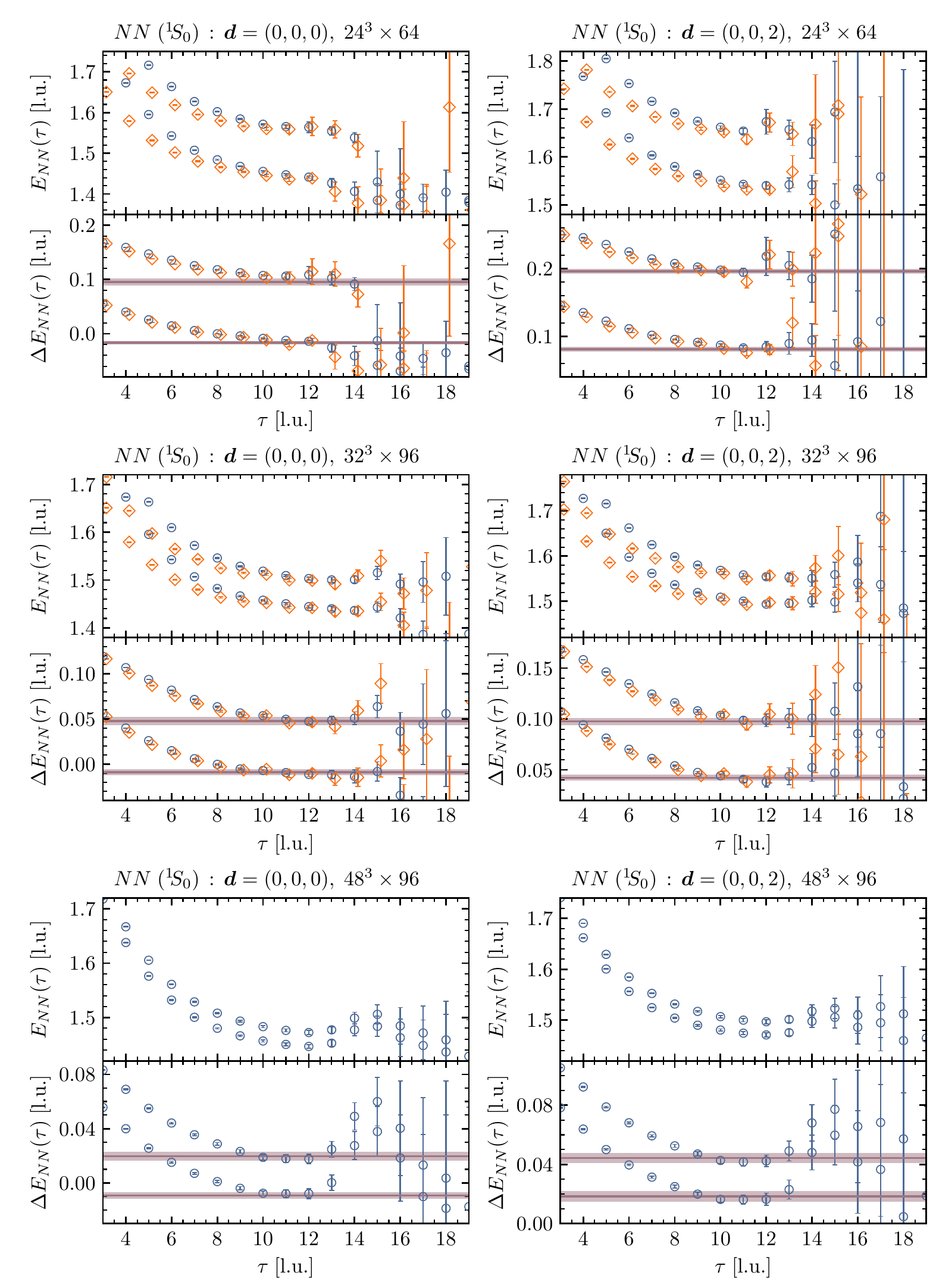}
\caption{The effective energy plots (upper panel of each segment) and the effective energy-shift plots (lower panel of each segment) for the $NN\; (\1s0)$ system at rest (left panels) and with boost $\bm{d}=(0,0,2)$ (right panels) for the SP (blue circles) and SS (orange diamonds) source-sink combinations.}
\label{fig:NN1s0_EMP}
\end{figure}

\begin{figure}[hbt!]
\includegraphics[width=0.95\textwidth]{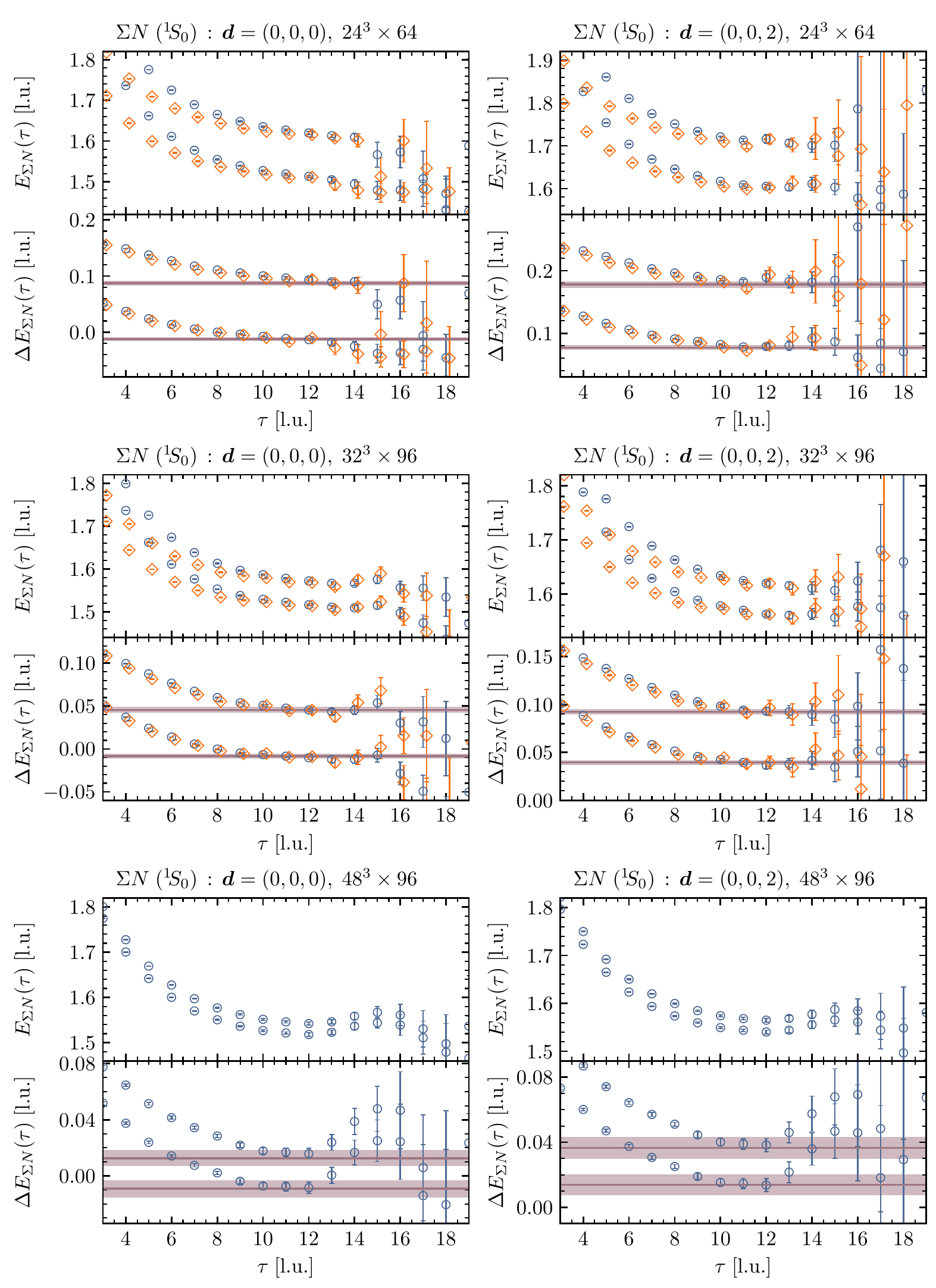}
\caption{The effective energy plots (upper panel of each segment) and the effective energy-shift plots (lower panel of each segment) for the $\Sigma N(^1S_0)$ system at rest (left panels) and with boost $\bm{d}=(0,0,2)$ (right panels) for the SP (blue circles) and SS (orange diamonds) source-sink combinations.}
\label{fig:SN1s0_EMP}
\end{figure}

\begin{figure}[hbt!]
\includegraphics[width=0.95\textwidth]{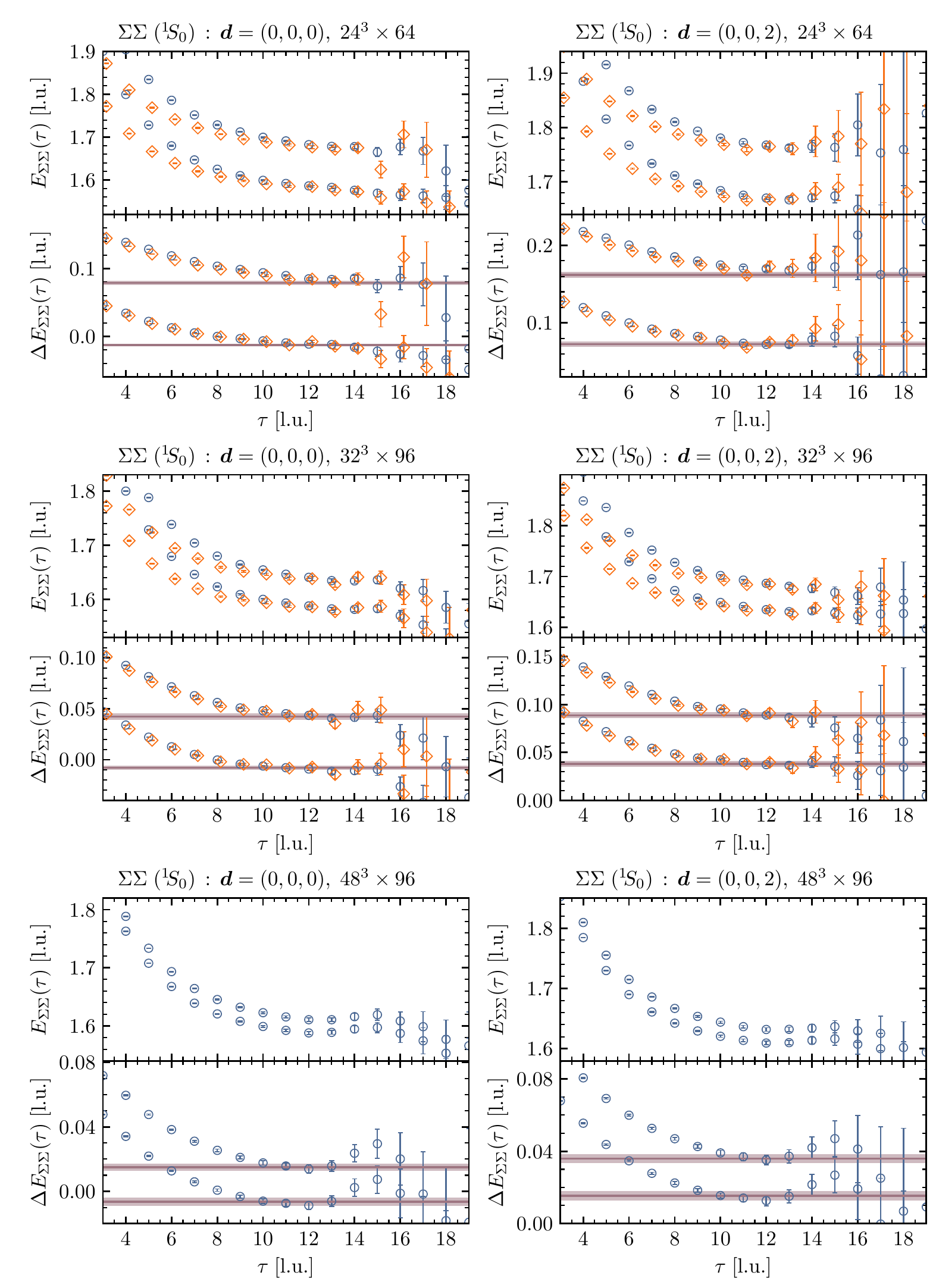}
\caption{The effective energy plots (upper panel of each segment) and the effective energy-shift plots (lower panel of each segment) for the $\Sigma \Sigma (^1S_0)$ system at rest (left panels) and with boost $\bm{d}=(0,0,2)$ (right panels) for the SP (blue circles) and SS (orange diamonds) source-sink combinations.}
\label{fig:SS1s0_EMP}
\end{figure}

\begin{figure}[hbt!]
\includegraphics[width=0.95\textwidth]{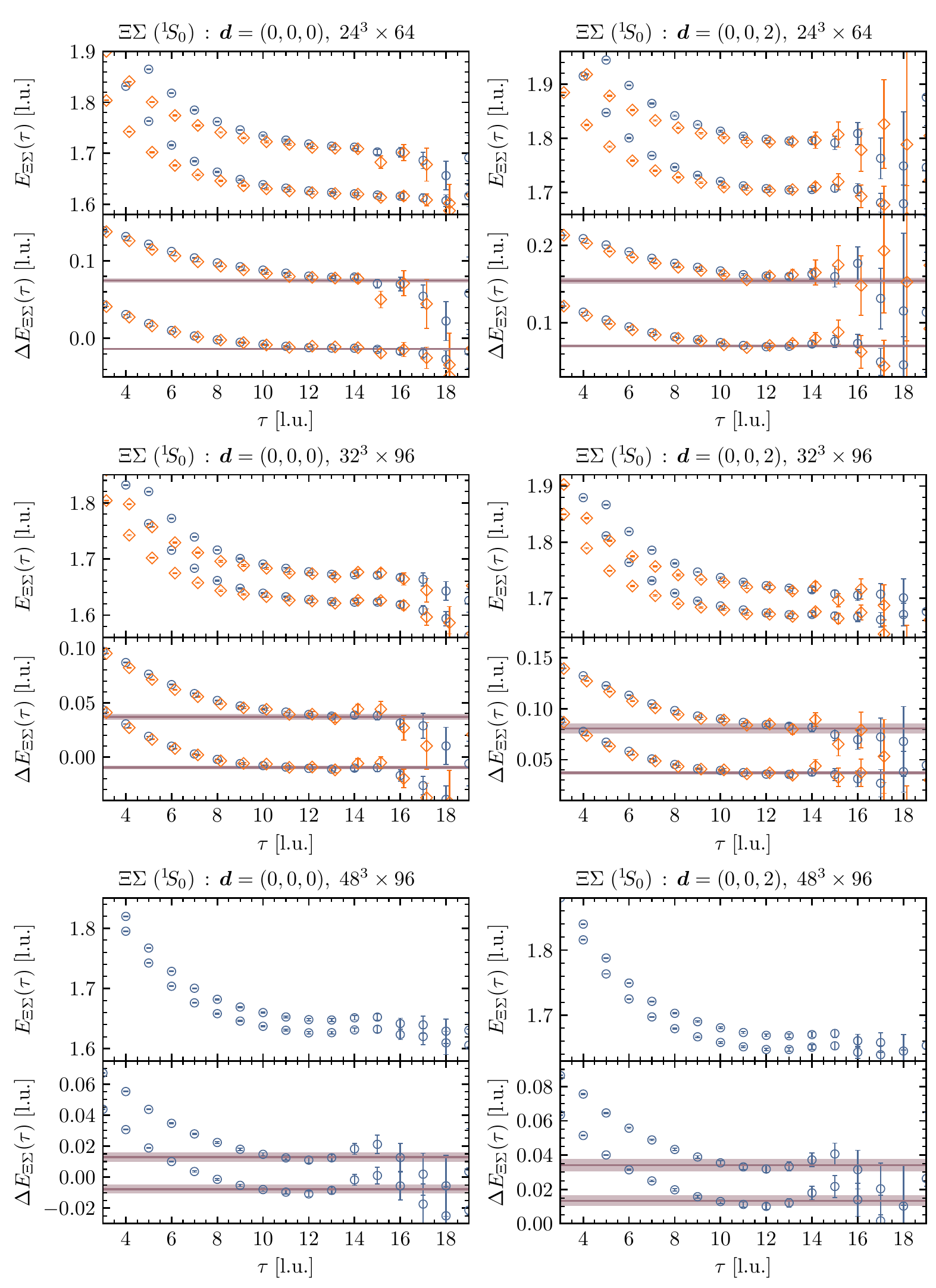}
\caption{The effective energy plots (upper panel of each segment) and the effective energy-shift plots (lower panel of each segment) for the $\Xi \Sigma (^1S_0)$ system at rest (left panels) and with boost $\bm{d}=(0,0,2)$ (right panels) for the SP (blue circles) and SS (orange diamonds) source-sink combinations.}
\label{fig:XS1s0_EMP}
\end{figure}

\begin{figure}[hbt]
\includegraphics[width=0.95\textwidth]{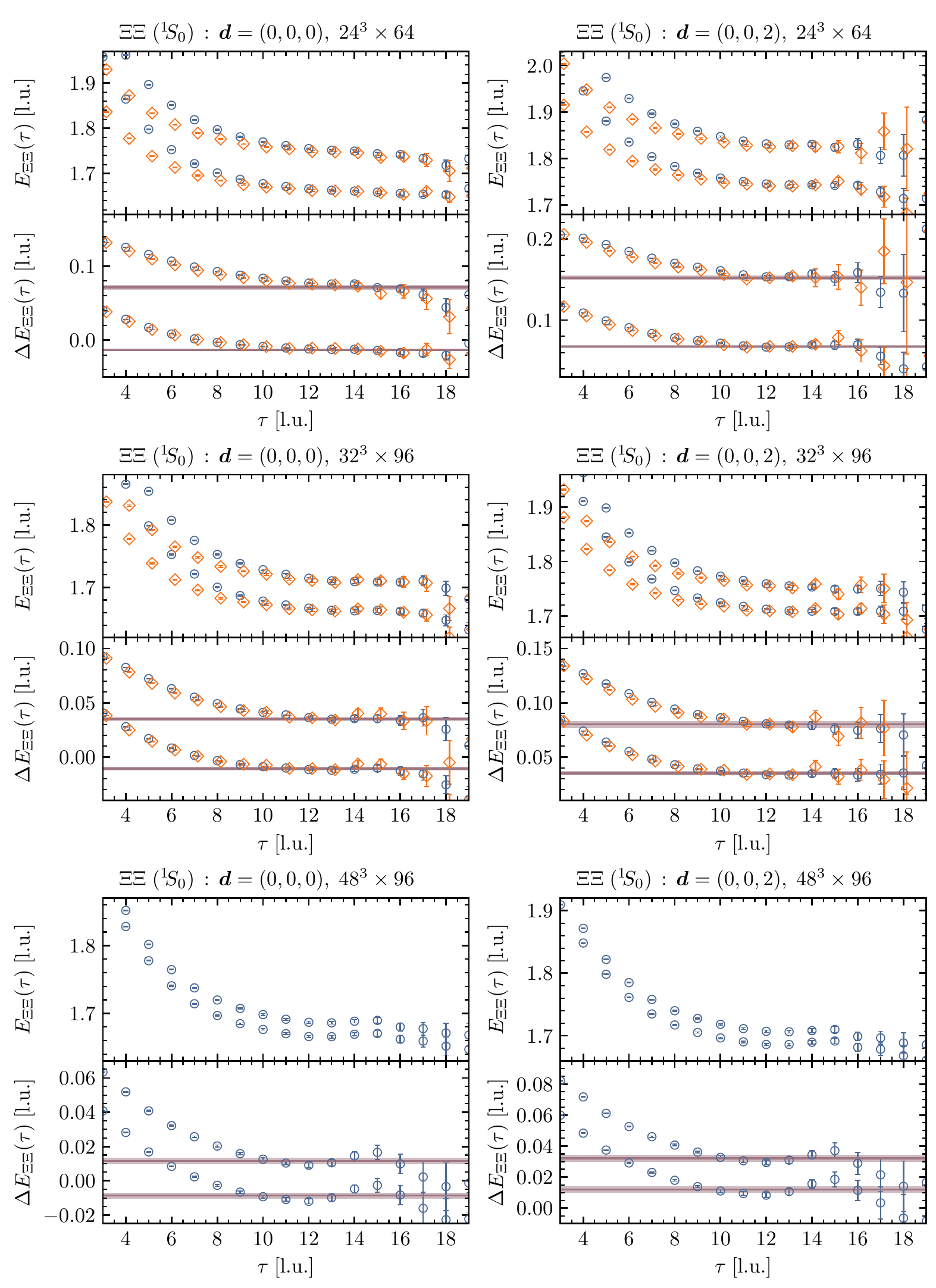}
\caption{The effective energy plots (upper panel of each segment) and the effective energy-shift plots (lower panel of each segment) for the $\Xi \Xi (^1S_0)$ system at rest (left panels) and with boost $\bm{d}=(0,0,2)$ (right panels) for the SP (blue circles) and SS (orange diamonds) source-sink combinations.}
\label{fig:XX1s0_EMP}
\end{figure}

\begin{figure}[hbt!]
\includegraphics[width=0.95\textwidth]{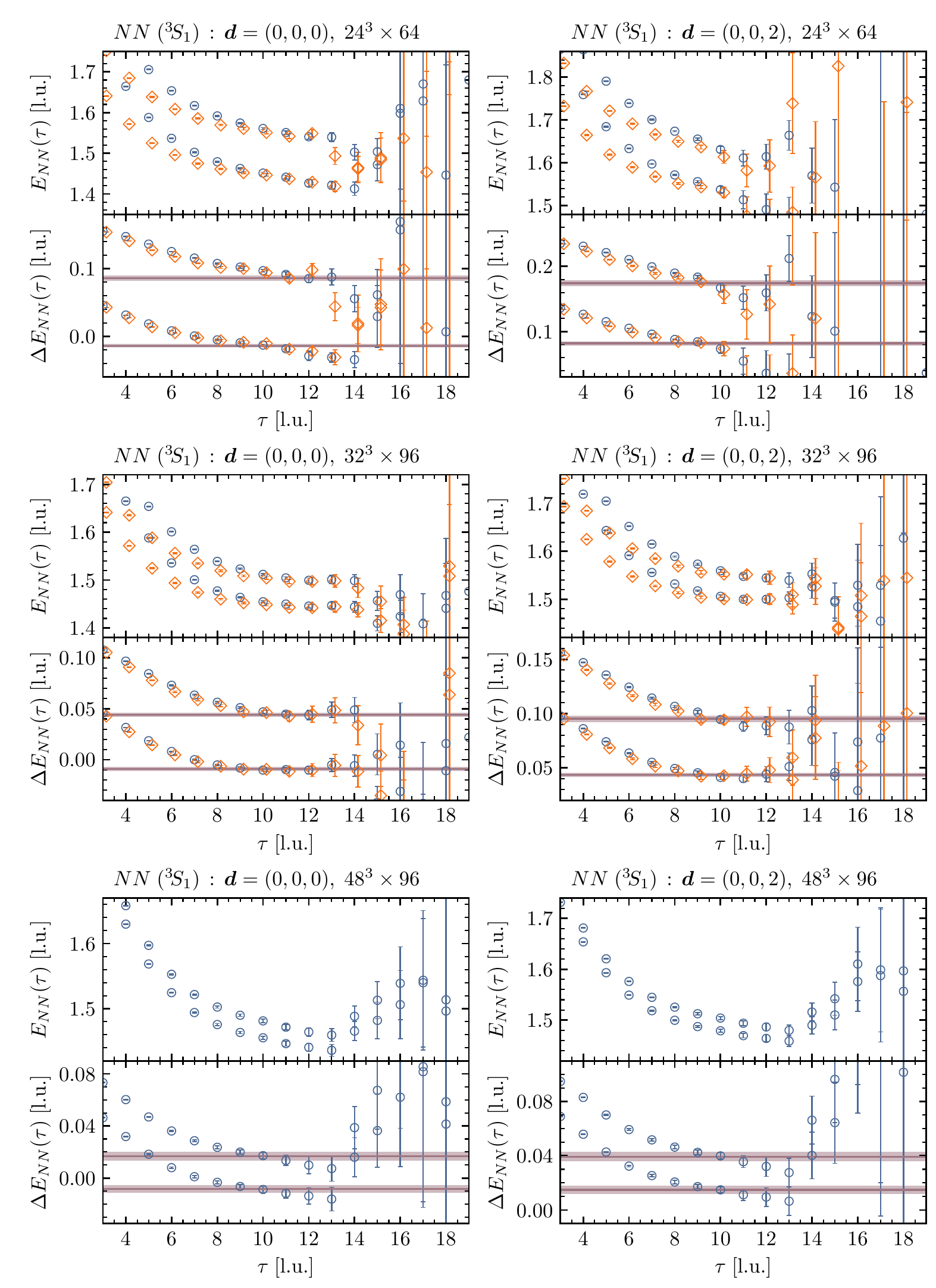}
\caption{The effective energy plots (upper panel of each segment) and the effective energy-shift plots (lower panel of each segment) for the $NN(^3S_1)$ system at rest (left panels) and with boost $\bm{d}=(0,0,2)$ (right panels) for the SP (blue circles) and SS (orange diamonds) source-sink combinations.}
\label{fig:NN3s1_EMP}
\end{figure}

\begin{figure}[hbt!]
\includegraphics[width=0.95\textwidth]{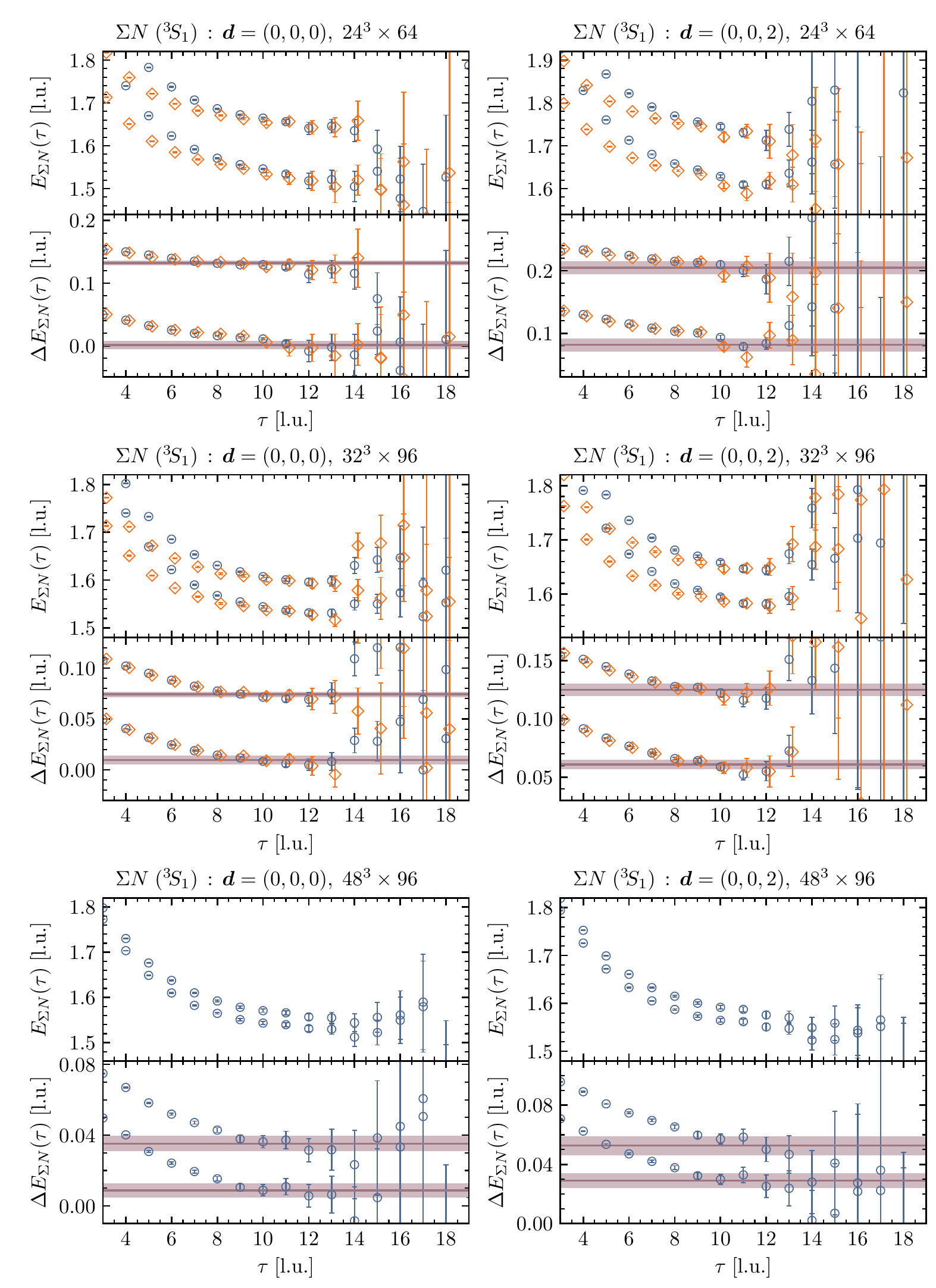}
\caption{The effective energy plots (upper panel of each segment) and the effective energy-shift plots (lower panel of each segment) for the $\Sigma N(^3S_1)$ system at rest (left panels) and with boost $\bm{d}=(0,0,2)$ (right panels) for the SP (blue circles) and SS (orange diamonds) source-sink combinations.}
\label{fig:SN3s1_EMP}
\end{figure}

\begin{figure}[hbt!]
\includegraphics[width=0.95\textwidth]{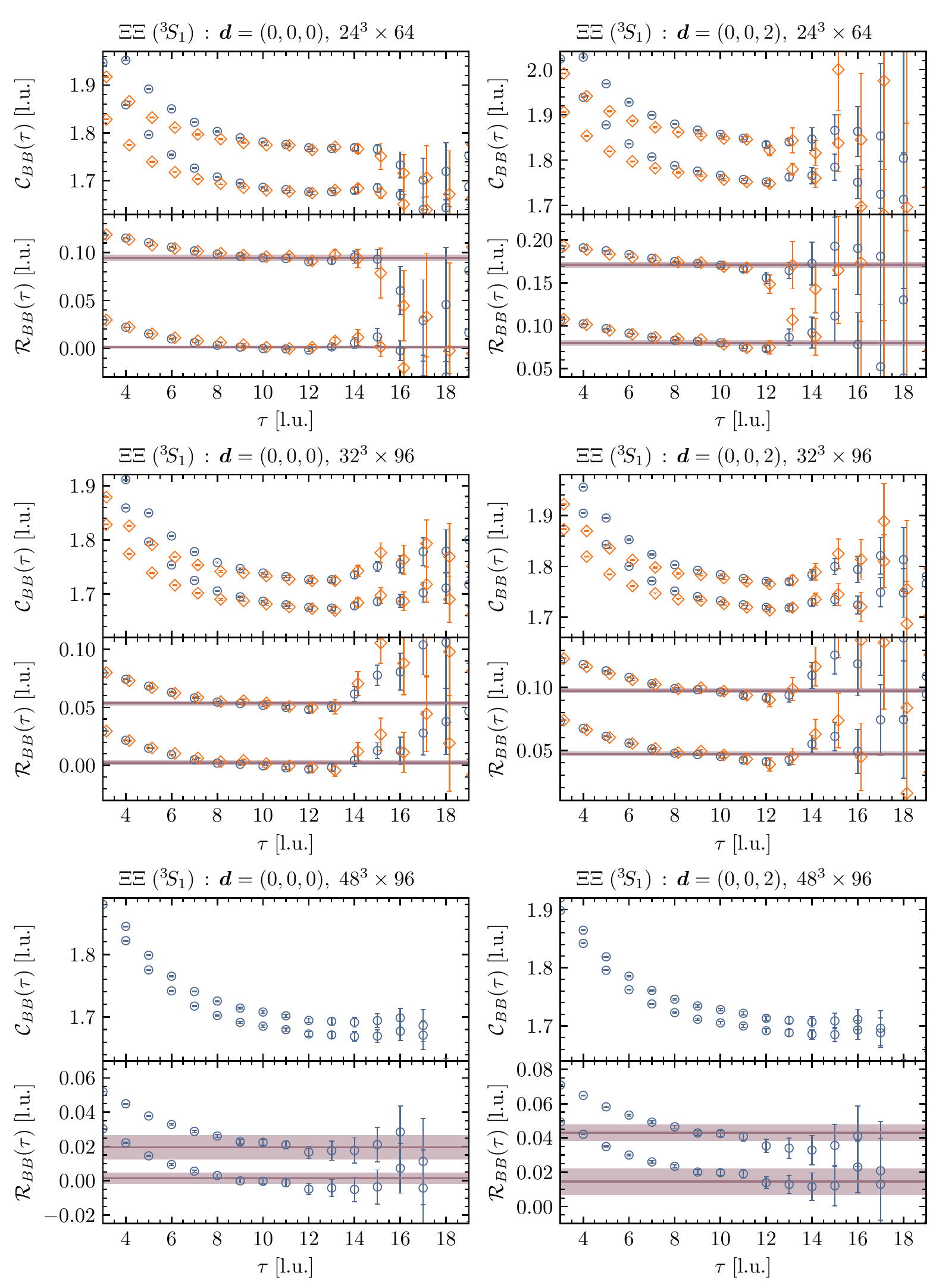}
\caption{The effective energy plots (upper panel of each segment) and the effective energy-shift plots (lower panel of each segment) for the $\Xi \Xi (^3S_1)$ system at rest (left panels) and with boost $\bm{d}=(0,0,2)$ (right panels) for the SP (blue circles) and SS (orange diamonds) source-sink combinations.}
\label{fig:XX3s1_EMP}
\end{figure}

\begin{figure}[hbt!]
\includegraphics[width=0.95\textwidth]{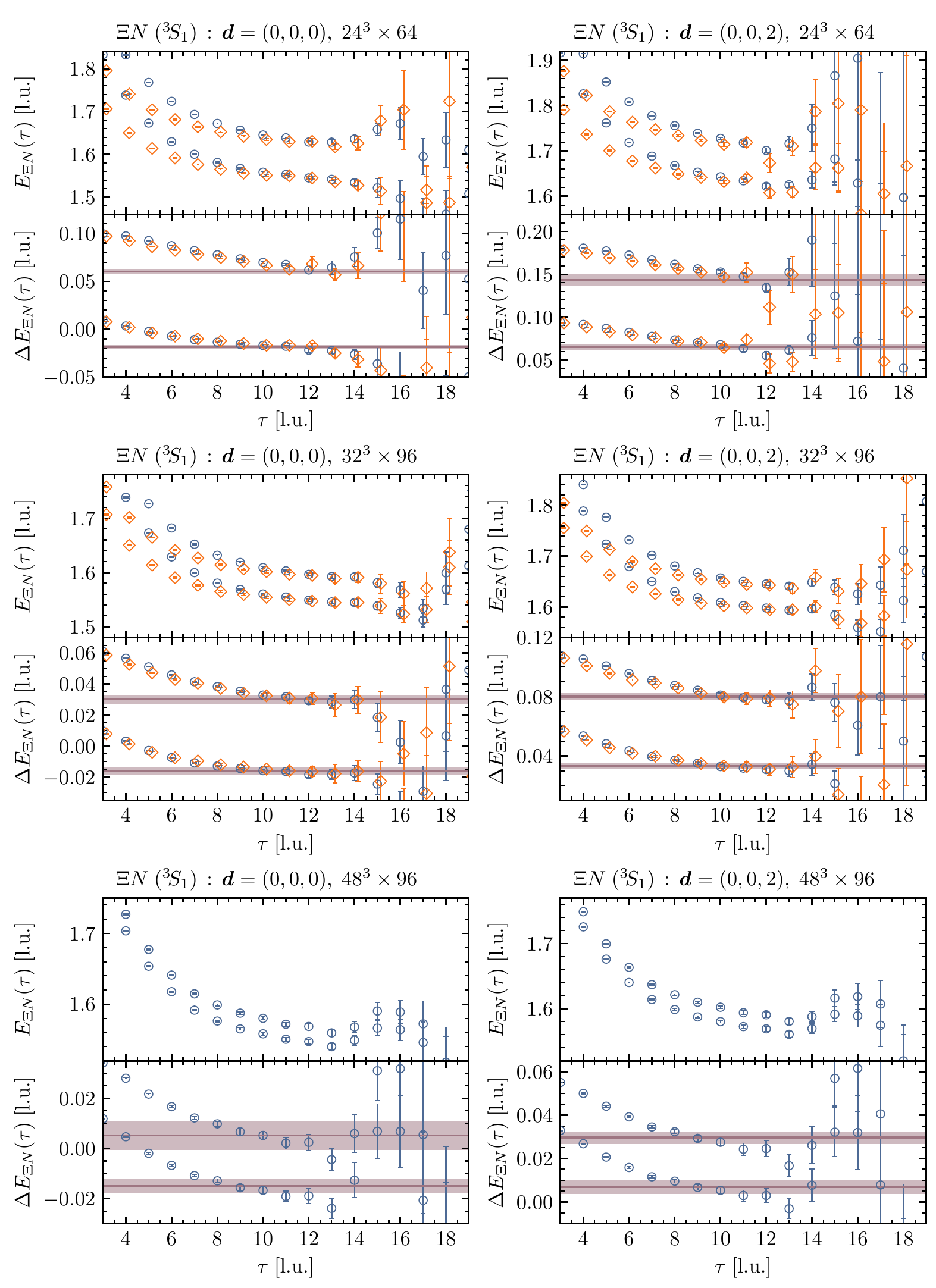}
\caption{The effective energy plots (upper panel of each segment) and the effective energy-shift plots (lower panel of each segment) for the $\Xi N(^3S_1)$ system at rest (left panels) and with boost $\bm{d}=(0,0,2)$ (right panels) for the SP (blue circles) and SS (orange diamonds) source-sink combinations.}
\label{fig:XN3s1_EMP}
\end{figure}

This appendix contains all the figures omitted from the main body of the paper for ease of presentation. These include the effective mass plots of the single baryons in Fig.~\ref{fig:B1_EMP}, and the effective energy and effective energy-shift plots for the two-baryon systems $NN\;(\1s0)$ (Fig.~\ref{fig:NN1s0_EMP}), $\Sigma N\;(\1s0)$ (Fig.~\ref{fig:SN1s0_EMP}), $\Sigma\Sigma\;(\1s0)$ (Fig.~\ref{fig:SS1s0_EMP}), $\Xi \Sigma\;(\1s0)$ (Fig.~\ref{fig:XS1s0_EMP}), $\Xi \Xi\;(\1s0)$ (Fig.~\ref{fig:XX1s0_EMP}), $NN\;(\3s1)$ (Fig.~\ref{fig:NN3s1_EMP}), $\Sigma N\;(\3s1)$ (Fig.~\ref{fig:SN3s1_EMP}), $\Xi \Xi\;(\3s1)$ (Fig.~\ref{fig:XX3s1_EMP}), and $\Xi N\;(\3s1)$ (Fig.~\ref{fig:XN3s1_EMP}). In Fig.~\ref{fig:B1_EMP}, the thin horizontal line and the horizontal band surrounding it represent, respectively, the central value of the baryon mass at each volume and the associated statistical and systematic uncertainties combined in quadrature, obtained with the fitting procedure described in~\cref{subsec:2ptfitting}. Similarly, in Figs.~\ref{fig:NN1s0_EMP}-\ref{fig:XN3s1_EMP} the line and the band represent, respectively, the central value of the two-baryon energy shifts compared to non-interacting baryons at rest (bottom panels) for each volume, and the associated statistical and systematic uncertainties combined in quadrature.

The appendix also contains the numerical results that were omitted from the main body. These include the energy shifts $\Delta E$ of the two-baryon systems, the c.m.\ momenta $k^{*2}$, and the value of $k^*\cot\delta$ for all the systems: $NN\;(\1s0)$ (Table~\ref{tab:eshift_ini}), $\Sigma N\;(\1s0)$ (Table~\ref{tab:eshift_SN1s0}), $\Sigma\Sigma\;(\1s0)$ (Table~\ref{tab:eshift_SS1s0}), $\Xi \Sigma\;(\1s0)$ (Table~\ref{tab:eshift_XS1s0}), $\Xi \Xi\;(\1s0)$ (Table~\ref{tab:eshift_XX1s0}), $NN\;(\3s1)$ (Table~\ref{tab:eshift_NN3s1}), $\Sigma N\;(\3s1)$ (Table~\ref{tab:eshift_SN3s1}), $\Xi \Xi\;(\3s1)$ (Table~\ref{tab:eshift_XX3s1}), and $\Xi N\;(\3s1)$ (Table~\ref{tab:eshift_fin}). In these tables, the values in the first and second parentheses correspond to statistical and systematic uncertainties, respectively, while those in the upper and lower parentheses are, respectively, the right and left uncertainties when the error bars are asymmetric, as is generally the case for the $k^*\cot\delta$ values. When there is a dash sign in the tables, it indicates that the quantity $k^*\cot\delta$ diverges due to the singularities in the $\mathcal{Z}^{\bm{d}}_{00}$ function.

All quantities in the plots and tables are expressed in lattice units.

\begin{table}[hbt!]
\centering
\caption{The values of the energy shift $\Delta E$, the c.m.\ momentum $k^{*2}$, and $k^*\cot\delta$ for the $NN \; (\1s0)$ channel.}
\label{tab:eshift_ini}
\renewcommand{\arraystretch}{1.5}
\begin{tabular}{cccrrr}
\toprule
Ensemble & Boost vector & State & \multicolumn{1}{c}{$\Delta E$ [l.u.]} & \multicolumn{1}{c}{$k^{*2}$ [l.u.]} & \multicolumn{1}{c}{$k^*\cot\delta$ [l.u.]} \\
\midrule
\multirow{4}{*}{$24^3\times 64$}	&	\multirow{2}{*}{$(0,0,0)$}	&	$n=1$	&	$-0.0166(19)(31)$	&	$-0.0120(13)(22)$	&	$-0.078_{(-11)(-17)}^{(+13)(+25)}$	\\
	&		&	$n=2$	&	$ 0.0953(23)(61)$	&	$ 0.0715(17)(47)$	&	\multicolumn{1}{c}{-}	\\
	&	\multirow{2}{*}{$(0,0,2)$}	&	$n=1$	&	$ 0.0812(16)(28)$	&	$-0.0079(12)(21)$	&	$-0.033_{(-18)(-27)}^{(+23)(+51)}$	\\
	&		&	$n=2$	&	$ 0.1960(16)(35)$	&	$ 0.0833(13)(29)$	&	$-0.233_{(-34)(-93)}^{(+32)(+65)}$	\\\midrule
\multirow{4}{*}{$32^3\times 96$}	&	\multirow{2}{*}{$(0,0,0)$}	&	$n=1$	&	$-0.0090(25)(20)$	&	$-0.0065(18)(14)$	&	$-0.056_{(-19)(-15)}^{(+29)(+27)}$	\\
	&		&	$n=2$	&	$ 0.0477(37)(24)$	&	$ 0.0352(28)(17)$	&	$0.7_{(-0.3)(-0.1)}^{(+3.4)(+32.0)}$	\\
	&	\multirow{2}{*}{$(0,0,2)$}	&	$n=1$	&	$ 0.0422(20)(21)$	&	$-0.0075(15)(16)$	&	$-0.068_{(-13)(-13)}^{(+18)(+21)}$	\\
	&		&	$n=2$	&	$ 0.0976(22)(27)$	&	$ 0.0347(18)(21)$	&	\multicolumn{1}{c}{-}	\\\midrule
\multirow{4}{*}{$48^3\times 96$}	&	\multirow{2}{*}{$(0,0,0)$}	&	$n=1$	&	$-0.0093(22)(11)$	&	$-0.0067(16)(08)$	&	$-0.079_{(-10)(-05)}^{(+13)(+06)}$	\\
	&		&	$n=2$	&	$ 0.0197(25)(23)$	&	$ 0.0143(18)(16)$	&	$0.2_{(-0.1)(-0.1)}^{(+0.5)(+1.8)}$	\\
	&	\multirow{2}{*}{$(0,0,2)$}	&	$n=1$	&	$ 0.0183(25)(26)$	&	$-0.0038(18)(20)$	&	$-0.051_{(-19)(-17)}^{(+38)(+80)}$	\\
	&		&	$n=2$	&	$ 0.0444(25)(24)$	&	$ 0.0156(19)(19)$	&	\multicolumn{1}{c}{-}	\\\bottomrule
\end{tabular}
\end{table}

\begin{table}[hbt!]
\centering
\caption{The values of the energy shift $\Delta E$, the c.m.\ momentum $k^{*2}$, and $k^*\cot\delta$ for the $\Sigma N \; (\1s0)$ channel.}
\label{tab:eshift_SN1s0}
\renewcommand{\arraystretch}{1.5}
\begin{tabular}{cccrrr}
\toprule
Ensemble & Boost vector & State & \multicolumn{1}{c}{$\Delta E$ [l.u.]} & \multicolumn{1}{c}{$k^{*2}$ [l.u.]} & \multicolumn{1}{c}{$k^*\cot\delta$ [l.u.]} \\
\midrule
\multirow{4}{*}{$24^3\times 64$}	&	\multirow{2}{*}{$(0,0,0)$}	&	$n=1$	&	$-0.0122(13)(26)$	&	$-0.0093(10)(20)$	&	$-0.048_{(-12)(-22)}^{(+14)(+35)}$	\\
	&		&	$n=2$	&	$ 0.0873(19)(32)$	&	$ 0.0682(15)(26)$	&	\multicolumn{1}{c}{-}	\\
	&	\multirow{2}{*}{$(0,0,2)$}	&	$n=1$	&	$ 0.0771(17)(35)$	&	$-0.0083(14)(27)$	&	$-0.040_{(-18)(-31)}^{(+24)(+60)}$	\\
	&		&	$n=2$	&	$ 0.1780(18)(48)$	&	$ 0.0747(16)(40)$	&	$-0.7_{(-0.2)(-1.7)}^{(+0.2)(+0.3)}$	\\ \midrule
\multirow{4}{*}{$32^3\times 96$}	&	\multirow{2}{*}{$(0,0,0)$}	&	$n=1$	&	$-0.0082(21)(16)$	&	$-0.0063(16)(12)$	&	$-0.052_{(-18)(-13)}^{(+26)(+25)}$	\\
	&		&	$n=2$	&	$ 0.0456(31)(16)$	&	$ 0.0351(24)(13)$	&	$0.6_{(-0.3)(-0.1)}^{(+1.6)(+2.2)}$	\\
	&	\multirow{2}{*}{$(0,0,2)$}	&	$n=1$	&	$ 0.0396(18)(17)$	&	$-0.0080(14)(13)$	&	$-0.073_{(-12)(-10)}^{(+14)(+16)}$	\\
	&		&	$n=2$	&	$ 0.0924(18)(23)$	&	$ 0.0339(15)(19)$	&	\multicolumn{1}{c}{-}	\\ \midrule
\multirow{4}{*}{$48^3\times 96$}	&	\multirow{2}{*}{$(0,0,0)$}	&	$n=1$	&	$-0.0092(48)(38)$	&	$-0.0070(36)(29)$	&	$-0.081_{(-21)(-16)}^{(+36)(+43)}$	\\
	&		&	$n=2$	&	$ 0.0126(48)(30)$	&	$ 0.0096(36)(23)$	&	$0.03_{(-06)(-03)}^{(+11)(+11)}$	\\
	&	\multirow{2}{*}{$(0,0,2)$}	&	$n=1$	&	$ 0.0139(50)(40)$	&	$-0.0066(38)(30)$	&	$-0.078_{(-23)(-18)}^{(+42)(+66)}$	\\
	&		&	$n=2$	&	$ 0.0366(51)(44)$	&	$ 0.0110(39)(34)$	&	\multicolumn{1}{c}{-}	\\\bottomrule
\end{tabular}
\end{table}

\begin{table}[hbt!]
\centering
\caption{The values of the energy shift $\Delta E$, the c.m.\ momentum $k^{*2}$, and $k^*\cot\delta$ for the $\Sigma \Sigma \; (\1s0)$ channel.}
\label{tab:eshift_SS1s0}
\renewcommand{\arraystretch}{1.5}
\begin{tabular}{cccrrr}
\toprule
Ensemble & Boost vector & State & \multicolumn{1}{c}{$\Delta E$ [l.u.]} & \multicolumn{1}{c}{$k^{*2}$ [l.u.]} & \multicolumn{1}{c}{$k^*\cot\delta$ [l.u.]} \\
\midrule
\multirow{4}{*}{$24^3\times 64$}	&	\multirow{2}{*}{$(0,0,0)$}	&	$n=1$	&	$-0.0126(13)(15)$	&	$-0.0100(10)(12)$	&	$-0.057_{(-11)(-12)}^{(+13)(+17)}$	\\
	&		&	$n=2$	&	$ 0.0788(21)(26)$	&	$ 0.0643(18)(21)$	&	$1.3_{(-0.4)(-0.3)}^{(+0.9)(+2.1)}$	\\
	&	\multirow{2}{*}{$(0,0,2)$}	&	$n=1$	&	$ 0.0725(14)(29)$	&	$-0.0095(11)(24)$	&	$-0.054_{(-13)(-25)}^{(+15)(+39)}$	\\
	&		&	$n=2$	&	$ 0.1618(34)(17)$	&	$ 0.0668(30)(15)$	&	\multicolumn{1}{c}{-}	\\ \midrule
\multirow{4}{*}{$32^3\times 96$}	&	\multirow{2}{*}{$(0,0,0)$}	&	$n=1$	&	$-0.0080(12)(14)$	&	$-0.0063(10)(12)$	&	$-0.053_{(-12)(-13)}^{(+14)(+20)}$	\\
	&		&	$n=2$	&	$ 0.0424(19)(25)$	&	$ 0.0342(16)(20)$	&	$0.50_{(-15)(-15)}^{(+31)(+71)}$	\\
	&	\multirow{2}{*}{$(0,0,2)$}	&	$n=1$	&	$ 0.0381(19)(23)$	&	$-0.0079(15)(19)$	&	$-0.071_{(-13)(-14)}^{(+17)(+25)}$	\\
	&		&	$n=2$	&	$ 0.0889(22)(22)$	&	$ 0.0341(18)(19)$	&	\multicolumn{1}{c}{-}	\\ \midrule
\multirow{4}{*}{$48^3\times 96$}	&	\multirow{2}{*}{$(0,0,0)$}	&	$n=1$	&	$-0.0065(19)(17)$	&	$-0.0051(15)(14)$	&	$-0.066_{(-12)(-10)}^{(+17)(+18)}$	\\
	&		&	$n=2$	&	$ 0.0150(19)(11)$	&	$ 0.0119(16)(09)$	&	$0.083_{(-42)(-23)}^{(+66)(+47)}$	\\
	&	\multirow{2}{*}{$(0,0,2)$}	&	$n=1$	&	$ 0.0154(19)(20)$	&	$-0.0049(16)(16)$	&	$-0.063_{(-13)(-13)}^{(+19)(+24)}$	\\
	&		&	$n=2$	&	$ 0.0359(19)(16)$	&	$ 0.0117(15)(14)$	&	$0.077_{(-39)(-31)}^{(+64)(+72)}$	\\\bottomrule
\end{tabular}
\end{table}

\begin{table}[hbt!]
\centering
\caption{The values of the energy shift $\Delta E$, the c.m.\ momentum $k^{*2}$, and $k^*\cot\delta$ for the $\Xi \Sigma \; (\1s0)$ channel.}
\label{tab:eshift_XS1s0}
\renewcommand{\arraystretch}{1.5}
\begin{tabular}{cccrrr}
\toprule
Ensemble & Boost vector & State & \multicolumn{1}{c}{$\Delta E$ [l.u.]} & \multicolumn{1}{c}{$k^{*2}$ [l.u.]} & \multicolumn{1}{c}{$k^*\cot\delta$ [l.u.]} \\
\midrule
\multirow{4}{*}{$24^3\times 64$}	&	\multirow{2}{*}{$(0,0,0)$}	&	$n=1$	&	$-0.0137(10)(09)$	&	$-0.0112(08)(08)$	&	$-0.070_{(-08)(-07)}^{(+09)(+08)}$	\\
	&		&	$n=2$	&	$ 0.0745(17)(22)$	&	$ 0.0621(14)(19)$	&	$0.81_{(-16)(-18)}^{(+25)(+44)}$	\\
	&	\multirow{2}{*}{$(0,0,2)$}	&	$n=1$	&	$ 0.0701(12)(17)$	&	$-0.0101(10)(15)$	&	$-0.062_{(-11)(-15)}^{(+13)(+19)}$	\\
	&		&	$n=2$	&	$ 0.1541(29)(21)$	&	$ 0.0631(25)(19)$	&	\multicolumn{1}{c}{-}	\\ \midrule
\multirow{4}{*}{$32^3\times 96$}	&	\multirow{2}{*}{$(0,0,0)$}	&	$n=1$	&	$-0.0096(12)(13)$	&	$-0.0078(10)(10)$	&	$-0.070_{(-09)(-09)}^{(+10)(+12)}$	\\
	&		&	$n=2$	&	$ 0.0370(20)(18)$	&	$ 0.0306(17)(15)$	&	$0.233_{(-60)(-46)}^{(+83)(+98)}$	\\
	&	\multirow{2}{*}{$(0,0,2)$}	&	$n=1$	&	$ 0.0371(10)(14)$	&	$-0.0079(09)(11)$	&	$-0.071_{(-08)(-09)}^{(+09)(+13)}$	\\
	&		&	$n=2$	&	$ 0.0806(34)(34)$	&	$ 0.0288(29)(29)$	&	$0.19_{(-08)(-06)}^{(+16)(+25)}$	\\ \midrule
\multirow{4}{*}{$48^3\times 96$}	&	\multirow{2}{*}{$(0,0,0)$}	&	$n=1$	&	$-0.0078(26)(12)$	&	$-0.0063(22)(09)$	&	$-0.075_{(-14)(-05)}^{(+20)(+08)}$	\\
	&		&	$n=2$	&	$ 0.0129(26)(13)$	&	$ 0.0106(22)(11)$	&	$0.045_{(-41)(-18)}^{(+65)(+41)}$	\\
	&	\multirow{2}{*}{$(0,0,2)$}	&	$n=1$	&	$ 0.0133(28)(14)$	&	$-0.0063(23)(10)$	&	$-0.075_{(-15)(-07)}^{(+22)(+11)}$	\\
	&		&	$n=2$	&	$ 0.0340(32)(18)$	&	$ 0.0109(26)(14)$	&	$0.06_{(-05)(-03)}^{(+11)(+08)}$	\\\bottomrule
\end{tabular}
\end{table}

\begin{table}[hbt!]
\centering
\caption{The values of the energy shift $\Delta E$, the c.m.\ momentum $k^{*2}$, and $k^*\cot\delta$ for the $\Xi \Xi \; (\1s0)$ channel.}
\label{tab:eshift_XX1s0}
\renewcommand{\arraystretch}{1.5}
\begin{tabular}{cccrrr}
\toprule
Ensemble & Boost vector & State & \multicolumn{1}{c}{$\Delta E$ [l.u.]} & \multicolumn{1}{c}{$k^{*2}$ [l.u.]} & \multicolumn{1}{c}{$k^*\cot\delta$ [l.u.]} \\
\midrule
\multirow{4}{*}{$24^3\times 64$}	&	\multirow{2}{*}{$(0,0,0)$}	&	$n=1$	&	$-0.0134(09)(15)$	&	$-0.0112(07)(12)$	&	$-0.070_{(-07)(-11)}^{(+08)(+14)}$	\\
	&		&	$n=2$	&	$ 0.0712(14)(30)$	&	$ 0.0609(12)(26)$	&	$0.66_{(-10)(-17)}^{(+13)(+43)}$	\\
	&	\multirow{2}{*}{$(0,0,2)$}	&	$n=1$	&	$ 0.0675(10)(13)$	&	$-0.0110(09)(11)$	&	$-0.070_{(-08)(-10)}^{(+09)(+13)}$	\\
	&		&	$n=2$	&	$ 0.1519(14)(27)$	&	$ 0.0643(12)(25)$	&	\multicolumn{1}{c}{-}	\\ \midrule
\multirow{4}{*}{$32^3\times 96$}	&	\multirow{2}{*}{$(0,0,0)$}	&	$n=1$	&	$-0.0109(11)(14)$	&	$-0.0090(09)(12)$	&	$-0.081_{(-07)(-09)}^{(+08)(+11)}$	\\
	&		&	$n=2$	&	$ 0.0349(12)(16)$	&	$ 0.0295(10)(14)$	&	$0.195_{(-31)(-38)}^{(+38)(+58)}$	\\
	&	\multirow{2}{*}{$(0,0,2)$}	&	$n=1$	&	$ 0.0349(10)(17)$	&	$-0.0091(08)(14)$	&	$-0.082_{(-06)(-10)}^{(+07)(+13)}$	\\
	&		&	$n=2$	&	$ 0.0800(11)(30)$	&	$ 0.0299(10)(26)$	&	$0.23_{(-04)(-08)}^{(+04)(+17)}$	\\ \midrule
\multirow{4}{*}{$48^3\times 96$}	&	\multirow{2}{*}{$(0,0,0)$}	&	$n=1$	&	$-0.0087(12)(13)$	&	$-0.0072(10)(11)$	&	$-0.082_{(-06)(-07)}^{(+07)(+08)}$	\\
	&		&	$n=2$	&	$ 0.0115(13)(14)$	&	$ 0.0096(11)(12)$	&	$0.026_{(-19)(-19)}^{(+22)(+27)}$	\\
	&	\multirow{2}{*}{$(0,0,2)$}	&	$n=1$	&	$ 0.0120(13)(16)$	&	$-0.0071(11)(13)$	&	$-0.081_{(-07)(-08)}^{(+08)(+10)}$	\\
	&		&	$n=2$	&	$ 0.0321(13)(17)$	&	$ 0.0099(11)(15)$	&	$0.033_{(-20)(-26)}^{(+27)(+37)}$	\\\bottomrule
\end{tabular}
\end{table}

\begin{table}[hbt!]
\centering
\caption{The values of the energy shift $\Delta E$, the c.m.\ momentum $k^{*2}$, and $k^*\cot\delta$ for the $NN \; (\3s1)$ channel.}
\label{tab:eshift_NN3s1}
\renewcommand{\arraystretch}{1.5}
\begin{tabular}{cccrrr}
\toprule
Ensemble & Boost vector & State & \multicolumn{1}{c}{$\Delta E$ [l.u.]} & \multicolumn{1}{c}{$k^{*2}$ [l.u.]} & \multicolumn{1}{c}{$k^*\cot\delta$ [l.u.]} \\
\midrule
\multirow{4}{*}{$24^3\times 64$}	&	\multirow{2}{*}{$(0,0,0)$}	&	$n=1$	&	$-0.0140(18)(19)$	&	$-0.0101(13)(14)$	&	$-0.058_{(-14)(-13)}^{(+17)(+20)}$	\\
	&		&	$n=2$	&	$ 0.0860(30)(26)$	&	$ 0.0643(22)(20)$	&	$1.3_{(-0.5)(-0.3)}^{(+1.4)(+2.8)}$	\\
	&	\multirow{2}{*}{$(0,0,2)$}	&	$n=1$	&	$ 0.0819(18)(27)$	&	$-0.0074(14)(20)$	&	$-0.023_{(-22)(-29)}^{(+31)(+55)}$	\\
	&		&	$n=2$	&	$ 0.1744(25)(36)$	&	$ 0.0658(20)(29)$	&	\multicolumn{1}{c}{-}	\\ \midrule
\multirow{4}{*}{$32^3\times 96$}	&	\multirow{2}{*}{$(0,0,0)$}	&	$n=1$	&	$-0.0090(15)(12)$	&	$-0.0065(11)(08)$	&	$-0.056_{(-12)(-09)}^{(+15)(+13)}$	\\
	&		&	$n=2$	&	$ 0.0442(16)(12)$	&	$ 0.0326(12)(10)$	&	$0.340_{(-70)(-45)}^{(+97)(+98)}$	\\
	&	\multirow{2}{*}{$(0,0,2)$}	&	$n=1$	&	$ 0.0434(18)(12)$	&	$-0.0066(13)(08)$	&	$-0.057_{(-14)(-08)}^{(+20)(+14)}$	\\
	&		&	$n=2$	&	$ 0.0952(20)(23)$	&	$ 0.0328(15)(18)$	&	$0.44_{(-13)(-12)}^{(+30)(+75)}$	\\ \midrule
\multirow{4}{*}{$48^3\times 96$}	&	\multirow{2}{*}{$(0,0,0)$}	&	$n=1$	&	$-0.0083(27)(15)$	&	$-0.0060(19)(11)$	&	$-0.073_{(-13)(-07)}^{(+18)(+12)}$	\\
	&		&	$n=2$	&	$ 0.0167(30)(17)$	&	$ 0.0122(22)(12)$	&	$0.09_{(-06)(-03)}^{(+13)(+11)}$	\\
	&	\multirow{2}{*}{$(0,0,2)$}	&	$n=1$	&	$ 0.0149(29)(15)$	&	$-0.0063(21)(11)$	&	$-0.076_{(-14)(-07)}^{(+19)(+12)}$	\\
	&		&	$n=2$	&	$ 0.0393(31)(16)$	&	$ 0.0117(23)(12)$	&	$0.08_{(-06)(-02)}^{(+12)(+09)}$	\\\bottomrule
\end{tabular}
\end{table}

\begin{table}[hbt!]
\centering
\caption{The values of the energy shift $\Delta E$, the c.m.\ momentum $k^{*2}$, and $k^*\cot\delta$ for the $\Sigma N \; (\3s1)$ channel.}
\label{tab:eshift_SN3s1}
\renewcommand{\arraystretch}{1.5}
\begin{tabular}{cccrrr}
\toprule
Ensemble & Boost vector & State & \multicolumn{1}{c}{$\Delta E$ [l.u.]} & \multicolumn{1}{c}{$k^{*2}$ [l.u.]} & \multicolumn{1}{c}{$k^*\cot\delta$ [l.u.]} \\
\midrule
\multirow{4}{*}{$24^3\times 64$}	&	\multirow{2}{*}{$(0,0,0)$}	&	$n=1$	&	$ 0.0012(57)(41)$	&	$ 0.0009(44)(31)$	&	\multicolumn{1}{c}{-}	\\
	&		&	$n=2$	&	$ 0.1325(28)(27)$	&	$ 0.1050(24)(22)$	&	$0.102_{(-40)(-36)}^{(+42)(+40)}$	\\
	&	\multirow{2}{*}{$(0,0,2)$}	&	$n=1$	&	$ 0.0816(57)(88)$	&	$-0.0048(46)(69)$	&	\multicolumn{1}{c}{-}	\\
	&		&	$n=2$	&	$ 0.2047(84)(60)$	&	$ 0.0975(71)(50)$	&	$0.03_{(-12)(-10)}^{(+13)(+09)}$	\\ \midrule
\multirow{4}{*}{$32^3\times 96$}	&	\multirow{2}{*}{$(0,0,0)$}	&	$n=1$	&	$ 0.0096(32)(33)$	&	$ 0.0073(24)(25)$	&	$-0.107_{(-37)(-53)}^{(+27)(+26)}$	\\
	&		&	$n=2$	&	$ 0.0742(22)(13)$	&	$ 0.0578(18)(10)$	&	$0.047_{(-38)(-18)}^{(+40)(+24)}$	\\
	&	\multirow{2}{*}{$(0,0,2)$}	&	$n=1$	&	$ 0.0609(38)(16)$	&	$ 0.0088(30)(11)$	&	$-0.087_{(-40)(-17)}^{(+29)(+11)}$	\\
	&		&	$n=2$	&	$ 0.1252(46)(25)$	&	$ 0.0604(38)(19)$	&	$0.15_{(-10)(-05)}^{(+14)(+08)}$	\\ \midrule
\multirow{4}{*}{$48^3\times 96$}	&	\multirow{2}{*}{$(0,0,0)$}	&	$n=1$	&	$ 0.0087(32)(25)$	&	$ 0.0066(24)(18)$	&	$-0.021_{(-33)(-28)}^{(+37)(+31)}$	\\
	&		&	$n=2$	&	$ 0.0352(34)(27)$	&	$ 0.0270(26)(19)$	&	$0.08_{(-09)(-06)}^{(+14)(+13)}$	\\
	&	\multirow{2}{*}{$(0,0,2)$}	&	$n=1$	&	$ 0.0291(35)(37)$	&	$ 0.0051(27)(29)$	&	$-0.041_{(-45)(-91)}^{(+39)(+44)}$	\\
	&		&	$n=2$	&	$ 0.0527(38)(52)$	&	$ 0.0235(30)(41)$	&	$-0.03_{(-14)(-65)}^{(+11)(+18)}$	\\\bottomrule
\end{tabular}
\end{table}

\begin{table}[hbt!]
\centering
\caption{The values of the energy shift $\Delta E$, the c.m.\ momentum $k^{*2}$, and $k^*\cot\delta$ for the $\Xi \Xi \; (\3s1)$ channel.}
\label{tab:eshift_XX3s1}
\renewcommand{\arraystretch}{1.5}
\label{tab:eshift_xx3s1}
\begin{tabular}{cccrrr}
\toprule
Ensemble & Boost vector & State & \multicolumn{1}{c}{$\Delta E$ [l.u.]} & \multicolumn{1}{c}{$k^{*2}$ [l.u.]} & \multicolumn{1}{c}{$k^*\cot\delta$ [l.u.]} \\
\midrule
\multirow{4}{*}{$24^3\times 64$}	&	\multirow{2}{*}{$(0,0,0)$}	&	$n=1$	&	$ 0.0012(11)(09)$	&	$ 0.0010(10)(08)$	&	\multicolumn{1}{c}{-}	\\
	&		&	$n=2$	&	$ 0.0944(28)(16)$	&	$ 0.0812(25)(14)$	&	$-0.38_{(-12)(-08)}^{(+09)(+05)}$	\\
	&	\multirow{2}{*}{$(0,0,2)$}	&	$n=1$	&	$ 0.0797(13)(28)$	&	$-0.0003(12)(24)$	&	\multicolumn{1}{c}{-}	\\
	&		&	$n=2$	&	$ 0.1712(19)(29)$	&	$ 0.0820(18)(26)$	&	$-0.29_{(-06)(-10)}^{(+05)(+07)}$	\\ \midrule
\multirow{4}{*}{$32^3\times 96$}	&	\multirow{2}{*}{$(0,0,0)$}	&	$n=1$	&	$ 0.0024(15)(14)$	&	$ 0.0020(12)(11)$	&	$-0.27_{(-30)(-95)}^{(+08)(+06)}$	\\
	&		&	$n=2$	&	$ 0.0535(14)(13)$	&	$ 0.0455(12)(12)$	&	$-0.297_{(-77)(-89)}^{(+57)(+53)}$	\\
	&	\multirow{2}{*}{$(0,0,2)$}	&	$n=1$	&	$ 0.0471(13)(15)$	&	$ 0.0014(11)(13)$	&	\multicolumn{1}{c}{-}	\\
	&		&	$n=2$	&	$ 0.0976(14)(13)$	&	$ 0.0454(12)(11)$	&	$-0.262_{(-73)(-72)}^{(+52)(+44)}$	\\ \midrule
\multirow{4}{*}{$48^3\times 96$}	&	\multirow{2}{*}{$(0,0,0)$}	&	$n=1$	&	$ 0.0014(17)(29)$	&	$ 0.0012(14)(23)$	&	\multicolumn{1}{c}{-}	\\
	&		&	$n=2$	&	$ 0.0195(67)(25)$	&	$ 0.0164(56)(21)$	&	\multicolumn{1}{c}{-}	\\
	&	\multirow{2}{*}{$(0,0,2)$}	&	$n=1$	&	$ 0.0145(78)(06)$	&	$-0.0049(66)(13)$	&	\multicolumn{1}{c}{-}	\\
	&		&	$n=2$	&	$ 0.0430(19)(45)$	&	$ 0.0192(16)(39)$	&	\multicolumn{1}{c}{-}	\\\bottomrule
\end{tabular}
\end{table}

\begin{table}[hbt!]
\centering
\caption{The values of the energy shift $\Delta E$, the c.m.\ momentum $k^{*2}$, and $k^*\cot\delta$ for the $\Xi N \; (\3s1)$ channel.}
\renewcommand{\arraystretch}{1.5}
\label{tab:eshift_fin}
\begin{tabular}{cccrrr}
\toprule
Ensemble & Boost vector & State & \multicolumn{1}{c}{$\Delta E$ [l.u.]} & \multicolumn{1}{c}{$k^{*2}$ [l.u.]} & \multicolumn{1}{c}{$k^*\cot\delta$ [l.u.]} \\
\midrule
\multirow{4}{*}{$24^3\times 64$}	&	\multirow{2}{*}{$(0,0,0)$}	&	$n=1$	&	$-0.0186(16)(14)$	&	$-0.0144(12)(11)$	&	$-0.097_{(-08)(-07)}^{(+09)(+09)}$	\\
	&		&	$n=2$	&	$ 0.0602(21)(18)$	&	$ 0.0478(17)(15)$	&	$0.165_{(-26)(-21)}^{(+29)(+27)}$	\\
	&	\multirow{2}{*}{$(0,0,2)$}	&	$n=1$	&	$ 0.0651(19)(35)$	&	$-0.0166(15)(28)$	&	$-0.113_{(-08)(-14)}^{(+09)(+19)}$	\\
	&		&	$n=2$	&	$ 0.1436(23)(60)$	&	$ 0.0486(19)(51)$	&	\multicolumn{1}{c}{-}	\\ \midrule
\multirow{4}{*}{$32^3\times 96$}	&	\multirow{2}{*}{$(0,0,0)$}	&	$n=1$	&	$-0.0161(13)(23)$	&	$-0.0124(10)(17)$	&	$-0.104_{(-06)(-09)}^{(+06)(+11)}$	\\
	&		&	$n=2$	&	$ 0.0302(14)(25)$	&	$ 0.0237(11)(20)$	&	$0.067_{(-16)(-27)}^{(+17)(+35)}$	\\
	&	\multirow{2}{*}{$(0,0,2)$}	&	$n=1$	&	$ 0.0331(13)(16)$	&	$-0.0124(10)(13)$	&	$-0.104_{(-06)(-07)}^{(+06)(+08)}$	\\
	&		&	$n=2$	&	$ 0.0801(14)(17)$	&	$ 0.0255(11)(14)$	&	$0.122_{(-26)(-29)}^{(+30)(+45)}$	\\ \midrule
\multirow{4}{*}{$48^3\times 96$}	&	\multirow{2}{*}{$(0,0,0)$}	&	$n=1$	&	$-0.0151(15)(25)$	&	$-0.0117(11)(19)$	&	$-0.107_{(-05)(-09)}^{(+06)(+10)}$	\\
	&		&	$n=2$	&	$ 0.0053(49)(31)$	&	$ 0.0041(38)(24)$	&	\multicolumn{1}{c}{-}	\\
	&	\multirow{2}{*}{$(0,0,2)$}	&	$n=1$	&	$ 0.0068(16)(27)$	&	$-0.0118(13)(21)$	&	$-0.108_{(-06)(-09)}^{(+06)(+11)}$	\\
	&		&	$n=2$	&	$ 0.0296(18)(22)$	&	$ 0.0062(14)(17)$	&	$-0.026_{(-20)(-25)}^{(+21)(+27)}$	\\\bottomrule
\end{tabular}
\vspace{12.5cm}
\end{table} %Tables and Figs Sec. 4.3

%%%%%%%%%%%%%%%%%%%%%%%%%%%%%%%%%%%%%%%%%%%%%%%%%%%%
\backmatter

\bibliography{Bibliography}

\end{document}